\documentclass[english,floatfix,twoside]{toraphd}
\usepackage[dvips]{graphicx}
\usepackage{eucal,calrsfs,bbm,sans,amsmath,amssymb,amsbsy,subfigure,stmaryrd,pifont}
\usepackage{hyperref}
\usepackage{color}

\usepackage[latin1]{inputenc}
\usepackage[T1]{fontenc}

 \newtheorem{prop}{Proposition}
 \newtheorem{defn}{Definition}

\newcommand{\arcsinh}{{\rm arcsinh}}

 \newcommand{\ham}{\op{\CMcal{H}}}
 \newcommand{\comm}[2]{\left[#1,#2\right]}
\newcommand{\adj}[1]{{#1}^{\dag}}
 \newcommand{\norm}[1]{\left\Vert#1\right\Vert}
\newcommand{\qph}{e--print quant-ph/}
 \newcommand{\abs}[1]{\left\vert#1\right\vert}
\newcommand{\ro}{\varrho}
\newcommand{\R}{\mathbbm{R}}

\newcommand{\rr}{\mathbbm{R}}

\newcommand{\id}{\mathbbm{1}}
\newcommand{\Id}[1]{\mathbbm{1}_{#1}}
\newcommand{\op}[1]{\hat{#1}}
 \newcommand{\dd}{\mathcal{D}}
 \newcommand{\set}[1]{\left\{#1\right\}}
\newcommand{\n}[1]{\;\!\diagup \!\!\!\!\!#1}
 \newcommand{\conc}{\CMcal{C}}

\newcommand{\diag}{\,{\rm diag}\,}

\newcommand{\sy}[1]{Sp_{(#1,\R)}}

\newcommand{\tr}{{\rm Tr}\,}
\newcommand{\Tr}[1]{{\rm Tr}\,\left[#1\right]}
\newcommand{\ptr}[2]{{\rm Tr}_{#1}\, #2}
\renewcommand{\det}{{\rm Det}\,}
\newcommand{\gr}[1]{\boldsymbol{#1}}
\newcommand{\be}{\begin{equation}}
\newcommand{\ee}{\end{equation}}
\newcommand{\bea}{\begin{eqnarray}}
\newcommand{\eea}{\end{eqnarray}}
\newcommand{\ket}[1]{|#1\rangle}
\newcommand{\bra}[1]{\langle#1|}

\newcommand{\ketbra}[2]{\vert #1 \rangle \! \langle #2 \vert}
 \newcommand{\hh}{\mathcal{H}}
  \newcommand{\W}{\op{\CMcal{W}}}
 \newcommand{\s}{\CMcal{S}}
\newcommand{\avr}[1]{\langle#1\rangle}

\newcommand{\N}{{\CMcal N}}
\newcommand{\D}{\Delta}
\newcommand{\sig}{\gr{\sigma}}
\newcommand{\gam}{\gr{\gamma}}
\newcommand{\eps}{\gr{\varepsilon}}
\newcommand{\bet}{\gr{\beta}}
\newcommand{\alp}{\gr{\alpha}}
\renewcommand{\abs}[1]{\left\vert#1\right\vert}
\newcommand{\eq}[1]{Eq.~{\rm(\ref{#1})}}
\newcommand{\ineq}[1]{Ineq.~{\rm(\ref{#1})}}
\newcommand{\rref}[1]{{\rm\ref{#1}}}
\newcommand{\pref}[1]{{\rm(\ref{#1})}}
\newcommand{\eg}{\emph{e.g.}~}
\newcommand{\ie}{\emph{i.e.}~}

\newcommand{\ps}{\ket{\psi}}
\newcommand{\ph}{\varphi}
\newcommand{\T}{^{\sf T}}
\newcommand{\PT}[1]{^{{\sf T\!}_#1}}

\begin{document}
\pagestyle{empty} \frontmatter

\title{{\sc{\textbf{\sf \textbf{
%\HUGE S{\Huge TRUCTURE AND} \\ \vspace*{-2.1cm}
\HUGE E{\Huge NTANGLEMENT OF}\\ \vspace*{.5cm} G{\Huge AUSSIAN}
S{\Huge TATES}}}}}}
\date{\normalsize{\vspace*{0.9cm}\sf V Ciclo Nuova Serie (2003--2006) \\
\vspace*{0.1cm} Coordinator: {\sc \large Prof. G. Vilasi}}}
\author{\vspace*{0.3cm}\Huge Gerardo Adesso\vspace*{0.8cm}}

\department{Dipartimento di Fisica ``E. R. Caianiello'' \\
\vspace*{0.1cm}{\large Facolt\`a di Scienze Matematiche Fisiche e
Naturali} }

\director{Dr. Fabrizio Illuminati}

\maketitle

\cleardoublepage

\begin{flushright}
\includegraphics[width=4cm]{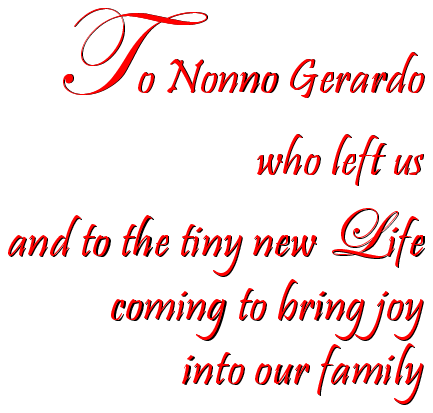}
\end{flushright}

\vspace*{3cm}

\vfil \vfil \vfil
\begin{flushleft} {\vspace*{1cm}
\hspace*{0.1cm}\includegraphics[width=6cm]{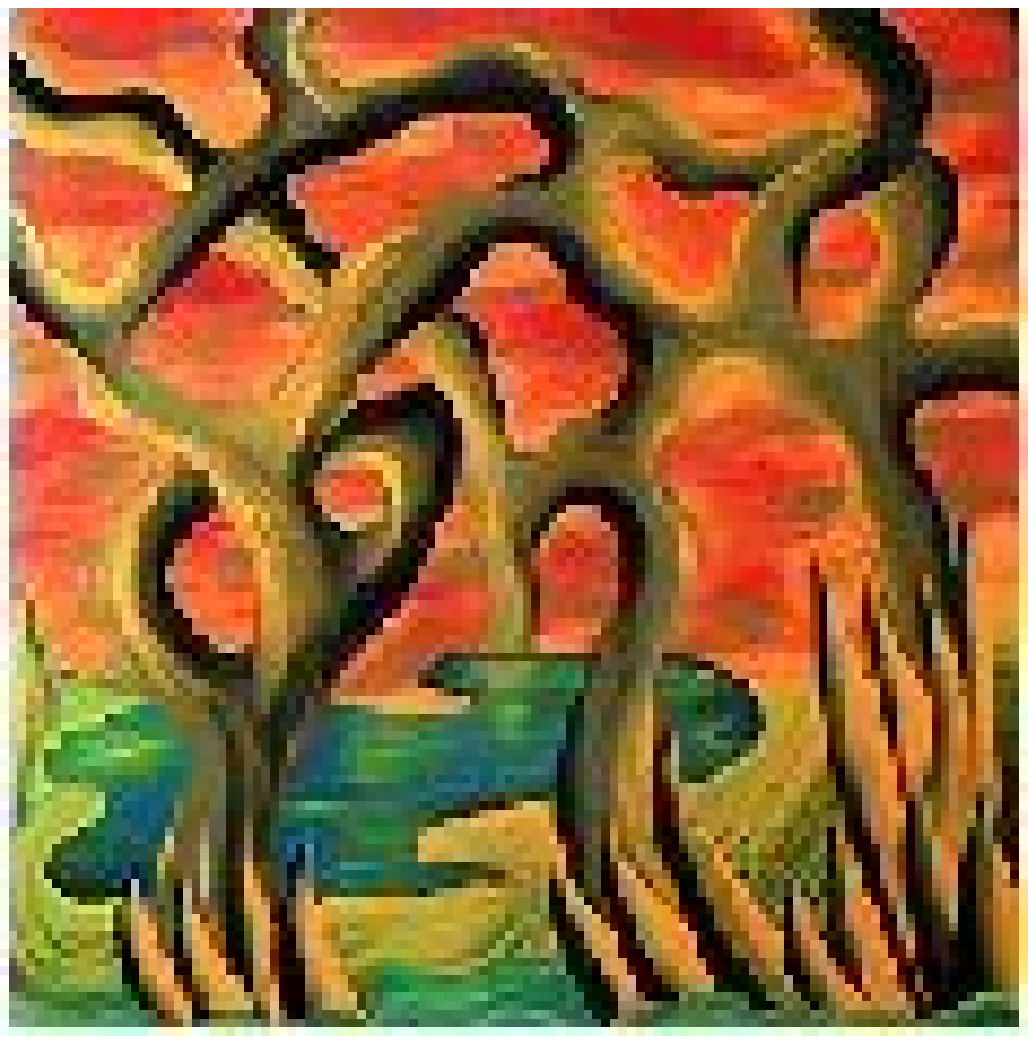} \\
\vspace*{0.6cm} {\rm \normalsize  {\em Life's Entanglement.}  CR
Studio Inc., 2002.
\\ \vspace{0.05cm}
\hspace*{0.2cm}\texttt{\footnotesize
http://www.crstudioinc.com/paint/54.html}}} \label{dedica}
\end{flushleft}
%\vfil

%\begin{figure}[h!]
%\includegraphics[width=5.349in]{entangledmosaic.eps}%
%\caption{\label{dummy2} eccolo.}
%\\

%asdg
%\end{figure}

\cleardoublepage

\chapter*{Foreword} %Acknowledgments

{\footnotesize{\it Three years have passed quickly but at the same
time filled with entertainment, work and new experiences which have
left an indelible footprint on my formation. I have shared this
adventure with many people, and it is a pity I cannot list them all
here for brevity reasons.

Many thanks with sincere appreciation and esteem to my advisor Dr.
Fabrizio Illuminati for all the simultaneously funny and serious
time, for having taught me how to make top-quality research and how
to write papers in a unique style. We had many animated discussions,
and he was always prompt to give me wise advices without ever
imposing his point of view, or preventing me to take my own
decisions. Thanks also for his breathtaking attributes when
referring to me in letters of support. I am grateful to Prof. Silvio
De Siena, who leads with Fabrizio the Quantum Theory and Quantum
Technologies Group in Salerno: his patience, experience and
friendliness have been useful in several critical moments. It is
impossible not to smile while thinking of, and thanking my friend
and colleague Dr. Alessio Serafini (now at Imperial College,
London), arguably the only person with whom I was able to establish
a `proper' collaboration (and who managed to have me doing
calculations with pencil and paper): he is to be blamed for
introducing me to continuous variable systems and Gaussian states. I
hope to keep  sharing ideas with him, even though the shiny
``basset-crew'' trio (me, him and Fabrizio) is now spatially
separated.

During my PhD I had the luck of spending one year in the Centre for
Quantum Computation (CQC) of the University of Cambridge. I cannot
recall how many interesting discussions (mostly at lunch-time) arose
there, and an estimate of how much the Cambridge experience was good
for myself (on both professional and personal grounds) would surely
result in a lower bound. Hence I express all my gratitude to my host
Dr. Marie Ericsson for the joy of making physics together, for her
enthusiastic attitude towards work, and  for the friendship born
during my year at CQC. Accordingly, I thank Prof. Artur Ekert,
director of the CQC, who accepted me as a full member, and whose
(quantum) capacity of efficiently transmitting advices in short-time
interactions is probably close to infinity.

I acknowledge collaborating with several researchers during these
years. First, thanks to the experimental group of Prof. Claude Fabre
(Lab. Kastler-Brossel, Paris), and specifically to Dr. Julien Laurat
(now at Caltech), Dr. Thomas Coudreau, Ga{\"e}lle Keller and Jos\'e
Augusto Oliveira-Huguenin, for the fruitful exchanges and for
letting me see what a beam-splitter is in reality. Thanks also to
Tohya Hiroshima for welcoming our contribution to the proof of the
monogamy of entanglement in all Gaussian states, and to Ivette
Fuentes-Schuller for the relativistic insights coming from our joint
project on continuous variable entanglement in non-inertial
reference frames. I am furthermore deeply grateful to a series of
colleagues, not enumerated here, whom I met in various venues and
with whom I discussed for brief or long time, learning new things
and understanding old ones.

Thanks to Alfonso and to all my friends, colleagues and officemates
both in Salerno and in Cambridge: time without them would have been
duller and less productive. Thanks to my parents, my sister and all
close members of my family, for their neverending encouragement and
for the trust they placed in me. Special thanks to
{\texttt{www.enigmi.net}} for keeping my brain active during this
last year.

My unlimited thanks and love to Samanta, whom I had the pleasure and
honor to marry in the middle of our PhD. Thanks for all her support,
hints, criticisms, proofreading, thanks for reminding me of the
sentence quoted in the caption of Fig.~\ref{figbasset}, thanks for
all her delicious dinners, her sweetness, her patience, her kisses,
her love, for all her being herself. Thanks for making a better man
out of me... and for the miracle of life which blessed her and me.

\vspace*{0.3cm}

\begin{flushright}

Gerardo

\end{flushright}

} }

\begin{abstract}
\small{This Dissertation collects results of my own work on the
interpretation, characterization, quantification and application of
bipartite and multipartite entanglement in Gaussian states of
continuous variable (CV)  systems.\footnote{\scriptsize{Most of the
research achievements presented here are published in (or  under
consideration for) scientific papers, as listed on page {\bf
\pageref{listofpubs}}. My publications will be quoted as {\rm [GAx]}
throughout the Dissertation. Some results, excluding the most recent
advances, are also summarized in a book chapter \cite{adebook}. The
structural and theoretical parts of this Dissertation are the basis
for a review article \cite{benarev}.}}

In the context of investigating connections between bipartite
entanglement and global and local degrees of information
\cite{AdessoPRA,polacchi}, we show how entanglement of two-mode
Gaussian states can be accurately quantified in terms of the global
and local amounts of mixedness \cite{prl,extremal}, and efficiently
estimated experimentally by direct measurements of the associated
purities \cite{prl,francamentemeneinfischio}. More generally, we
discuss different measures of bipartite entanglement and show their
inequivalence in ordering two-mode Gaussian states \cite{ordering}.
For multimode Gaussian states endowed with local symmetry with
respect to a given bipartition, we show how the multimode block
entanglement can be completely and reversibly localized onto a
single pair of modes by local, unitary operations
\cite{adescaling,unitarily}.

We then analyze the distribution of entanglement among multiple
parties  in multimode Gaussian states \cite{pisa}, introducing a new
entanglement monotone, the `contangle', adapted to a CV scenario
\cite{contangle}. We prove that, in all Gaussian states of an
arbitrary number of modes, entanglement distributes (as already
observed for qubit systems) according to a monogamy law
\cite{contangle,hiroshima}. Focusing on three-mode Gaussian states,
we study their genuine tripartite entanglement by means of the
residual contangle \cite{contangle}, we discuss their usefulness for
quantum communication implementations \cite{3mj}, and we investigate
in detail their distributed entanglement structure \cite{3mpra},
evidencing how, under a strong symmetry, an arbitrary tripartite
entanglement coexists with a limited, nonzero bipartite
entanglement: a feature named `promiscuous' entanglement sharing
\cite{contangle}. We then unfold how in four-mode Gaussian states
with more relaxed symmetry constraints, entanglement can be
infinitely promiscuous (at variance with the corresponding states of
qubits), with a coexistence of an unlimited four-partite
entanglement and an unlimited residual bipartite entanglement in two
pairs of modes \cite{unlim}.
%Such a
%result may lead to novel developments in quantum communication and
%computation with CV systems.

We moreover study entanglement distribution in harmonic lattices
with an underlying `valence bond' structure \cite{gvbs,minsk}, and,
in the general case of pure {\em N}-mode Gaussian states, we provide
standard forms under local operations \cite{sformato}, which yield
an efficient characterization of generic entanglement, together with
an optimal scheme to engineer such states in the lab with minimal
resources \cite{generic}. Operationally, multipartite entanglement
in symmetric {\em N}-mode Gaussian resources is qualitatively and
quantitatively proven to be equivalent to the success of multiparty
CV quantum teleportation networks \cite{telepoppate}. We conclude
with an application of our machinery to a relativistic setting:
namely, we study Gaussian entanglement sharing between modes of a
free scalar field from the perspective of observers in relative
acceleration, interpreting the entanglement loss due to the Unruh
effect in the light of a redistribution of entanglement between
accessible and unaccessible causally disconnected modes
\cite{ivette}. Such studies are of relevance in the context of the
information loss paradox in black holes \cite{myletter}.

}

\end{abstract} \pagestyle{headings}

 \tableofcontents

%\listoffigures

%\def\bibname{List of Publications}
%\begin{thebibliography}{99}
%\bibitem{pero}
%G. Adess
%\end{thebibliography}
%\def\bibname{Bibliography}
%\bibliography[List of Publications]{torabib}[List of Publications]

\mainmatter
\part{Preliminaries}
{\vspace*{1cm}
\includegraphics[width=9cm]{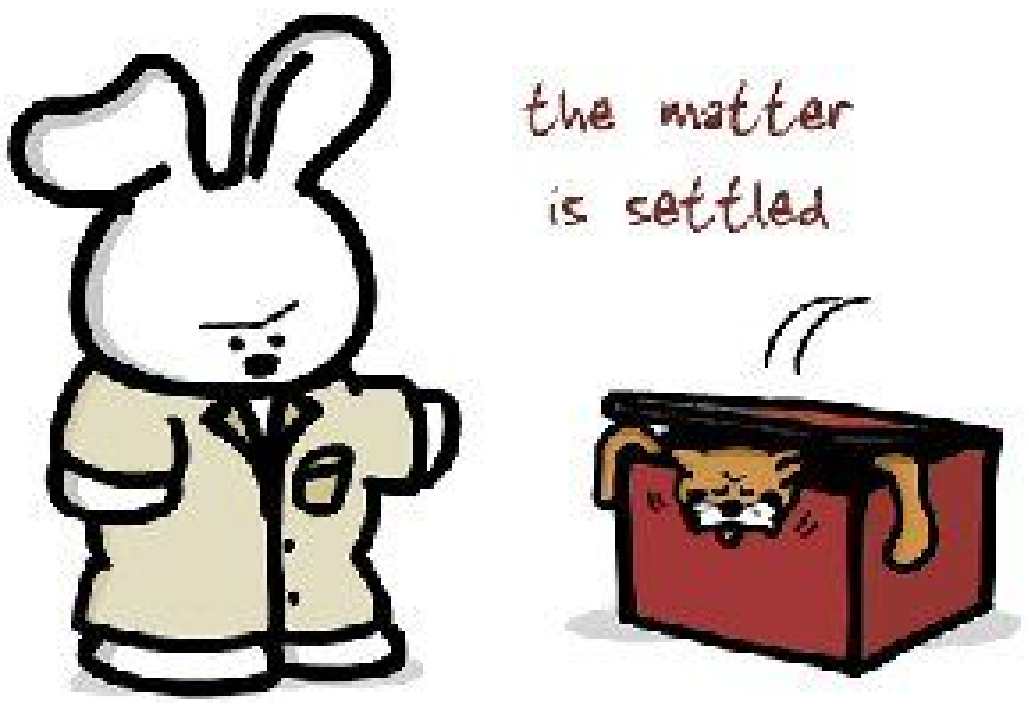} \\
\vspace*{0.6cm} {\rm \normalsize  {\em Brian's guide to quantum
mechanics.} John Walker, 2006.
\\ \vspace*{-0.4cm} \texttt{\footnotesize http://www.briansguide.net/index.cgi?ID=1140392885-C-CSI}}}
\label{PartPrelim}

\chapter*{Introduction} \label{ChapIntro}

{\sf

About eighty years after their inception, quantum mechanics and
quantum theory are still an endless source of new and precious
knowledge on the physical world and at the same time keep evolving
in their mathematical structures, conceptual foundations, and
intellectual and cultural implications. This is one of the reasons
why quantum physics is still so specially fascinating to all those
that approach it for the first time and never ceases to be so for
those that are professionally involved with it. In particular, since
the early nineties of the last century and in the last ten-fifteen
years, a quiet revolution has taken place in the quantum arena. This
revolution has progressively indicated and clarified that aspects
once thought to be problematic, such as quantum non-separability and
``spooky'' actions at a distance, are actually not only the origin
of paradoxes but rather some of the key ingredients that are
allowing a deeper understanding of quantum mechanics, its
applications to new and exciting fields of research (such as quantum
information and quantum computation), and tremendous progress in the
development of its mathematical and conceptual foundations. Among
the key elements of the current re-foundation of quantum theory,
{\em entanglement} certainly plays a very important role, also
because it is a concept that can be mathematically qualified and
quantified in a way that allows it to provide new and general
characterizations of quantum properties, operations, and states.

The existence of entangled states, stemming directly from the
superposition principle, can be regarded as a founding feature, or
better ``the characteristic trait'' (according to Schr\"odinger) of
quantum mechanics itself.
 Entanglement arises when the state
of two or more subsystems of a compound quantum system cannot be
factorized into pure local states of the subsystems. The subsystems
thus share quantum correlations which can be stronger than any
classical correlation.  Quantum information science was born upon
the key observation that the exploitation of such nonclassical
correlations enables encoding, processing and distribution of
information in ways impossible, or very inefficient, with classical
means. Hence the possibility of implementing entangled resources
resulted in futuristic proposals (quantum teleportation, quantum
cryptography, quantum computation, ...) which are now made, to a
certain extent, into reality. On a broader perspective, it is now
recognized that entanglement plays a fundamental role in the physics
of many-body systems, in particular in critical phenomena like
quantum phase transitions, and in the description of the
interactions between complex systems at the quantum scale.

Despite its prominent role in the physics of microscopic but also
macroscopic systems, it still stands as an open issue to achieve a
conclusive characterization and quantification of bipartite
entanglement for mixed states, and especially to provide a
definition and interpretation of multipartite entanglement both for
pure states and in the presence of mixedness. While important
insights have been gained on these issues in the context of qubit
systems (two-level quantum systems traditionally employed as the
main logical units for quantum computing and quantum information in
general), a less satisfactory understanding has been achieved until
recent times on higher-dimensional systems, as the structure of
entangled states in Hilbert spaces of high dimensionality scales
exhibiting a formidable degree of complexity.

However, and quite remarkably, in systems endowed with
infinite-dimensional Hilbert spaces (where entanglement can arise
between degrees of freedom with continuous spectra), recent advances
have been recorded for what concerns the understanding and the
quantification of the entanglement properties of a restricted class
of states, the so-called {\em Gaussian states}.  Gaussian states
distinctively stand out of the infinite variety of continuous
variable systems, because on one hand they allow a clean framework
for the analytical study of the structure of non-local correlations,
and on the other hand they are of great practical relevance in
applications to quantum optics and quantum information. Two-mode and
multimode coherent and squeezed Gaussian states are indeed key
resources, producible and manipulatable in the lab with a high
degree of control, for a plethora of two-party and multi-party
quantum communication protocols, ranging from deterministic
teleportation and secure key distribution, to quantum data storage
and cluster computation.

\medskip

This PhD Dissertation collects my personal contributions to the
understanding, qualification, quantification, structure, production,
operational interpretation, and applications of entanglement in
Gaussian states of continuous variable systems. Let us briefly
mention some of the most important results, the majority of which
have appeared in
Refs.~{\rm[\href{#cite.prl}{GA2}---\href{#cite.ivette}{GA20}]}.

In the first place we enriched the well-established theory of
bipartite entanglement in two-mode Gaussian states, providing new
physically insightful connections between the entanglement and the
degrees of information associated with the global system and its
subsystems. We thus showed that the negativity (an entanglement
monotone) can be accurately qualified and quantitatively estimated
in those states by direct purity measurements. We also proved that
different entanglement quantifiers (negativities and Gaussian
entanglement measures) induce inequivalent orderings on the set of
entangled, nonsymmetric two-mode Gaussian states. We then extended
our scope to investigate multimode, bipartite entanglement in
$N$-mode  Gaussian states endowed with some symmetry constraints,
and its scaling with the number of the modes; this study was enabled
by our central observation that entanglement in such states is
unitarily localizable to an effective two-mode entanglement. We
could thus extend the validity of the necessary and sufficient
positive-partial-transposition condition for separability to
bisymmetric Gaussian states of an arbitrary number of modes, and
exactly quantify the block entanglement between different mode
partitions, revealing signatures of a genuine multipartite
entanglement arising among all modes. Under these premises, we
developed {\em ex novo} a theory of multipartite entanglement for
Gaussian states, based on the crucial fact that entanglement cannot
be freely shared and its distribution is constrained to a monogamy
inequality, which we proved to hold for all (pure and mixed)
$N$-mode Gaussian states distributed among $N$ parties. To this aim,
we introduced new continuous variable entanglement monotones, namely
(Gaussian) `contangle' and Gaussian tangle, for the quantification
of entanglement sharing in Gaussian states. Implications of this
analysis include, in particular, the definition of the `residual
contangle' as the first {\em bona fide} measure of genuine
multipartite (specifically, tripartite) entanglement in a continuous
variable setting, a complete quantitative analysis of multipartite
entanglement in the paradigmatic instance of three-mode Gaussian
states, the discovery of the promiscuous nature of entanglement
sharing in symmetric `GHZ/$W$' Gaussian states, and the
demonstration of the possible coexistence of unlimited bipartite and
multipartite entanglement in states of at least four modes.

Our investigation was not confined to theoretical and structural
aspects of entanglement only. Along parallel lines, we got
interested on one hand in how to produce  bipartite and/or
multipartite entanglement in the lab with efficient means, and on
the other hand in how to optimally employ such entanglement for
practical applications, endowing the entanglement itself with an
operational interpretation. In the case of two-mode Gaussian states,
we joined an experiment concerning production, characterization and
manipulation of entanglement in the context of quantum optics. In
the three- and four-mode instances, we proposed several schemes to
efficiently engineer family of Gaussian states with relevant
entanglement properties. In general, we devised an optimal scheme to
produce generic pure $N$-mode Gaussian states in a standard form not
encoding direct correlations between position and momentum operators
(and so encompassing all the instances of multimode Gaussian states
commonly employed in practical implementations); such an analysis
allows to interpret entanglement in this subclass of Gaussian states
entirely in terms of the two-point correlations between any pair of
modes. In this respect,  one theoretical result of direct interest
for the characterization of entanglement in Gaussian states, is the
qualitative and quantitative equivalence we established between the
presence of bipartite (multipartite) entanglement in two-mode
($N$-mode) fully symmetric Gaussian states shared as resources for a
two-party teleportation experiment ($N$-party teleportation
network), and the maximal fidelity of the protocol, optimized over
local single-mode unitary operations performed on the shared
resource. In the special case of three-mode, pure GHZ/$W$ states,
this optimal fidelity is a monotonically increasing function of the
residual contangle (which quantifies genuine tripartite
entanglement), providing the latter with a strong operational
significance. Based on this equivalence, we presented a proposal to
experimentally verify the promiscuous sharing structure of
tripartite Gaussian entanglement in such states in terms of the
success of two-party and three-party teleportation experiments.
Telecloning with three-mode Gaussian resources was also thoroughly
investigated.

We finally considered two applications of the Gaussian machinery to
the companion areas of condensed matter/statistical mechanics, and
relativity theory. Concerning the former, we studied entanglement
distribution in ground states of translationally invariant many-body
harmonic lattice systems endowed with a Gaussian `valence bond'
structure. We characterized the range of correlations in such
harmonic models, connecting it to the degree of entanglement in a
smaller Gaussian structure, named `building block', which enters in
the valence bond construction. We also discussed the experimental
production of Gaussian valence bond states of an arbitrary number of
modes, and their usefulness for multiparty telecloning of coherent
states. On the other hand, in a relativistic setting we studied the
distribution of entanglement between modes of a free scalar field
from the perspective of observers in relative acceleration. The
degradation of entanglement due to the Unruh effect was analytically
characterized for two parties sharing a two-mode Gaussian state in
an inertial frame, in the cases of either one or both observers
undergoing uniform acceleration.  Within the monogamy framework, we
precisely explained the loss of entanglement as a redistribution of
the inertial entanglement into multipartite quantum correlations
among accessible and unaccessible modes from a non-inertial
perspective.

\medskip

This Dissertation is organized as follows.

\smallskip

Part \ref{PartPrelim} is devoted to introducing the main ingredients
of our analysis, entanglement on one side, and Gaussian states on
the other. In particular, Chapter \ref{ChapEnt} contains the basics
of entanglement theory: how to quantify quantum information, the
separability problem, different entanglement measures, and a
discussion on entanglement sharing. In Chapter \ref{ChapGauss} we
give a self-contained introduction to phase-space and symplectic
methods in the study of Gaussian states of infinite-dimensional
bosonic systems, we discuss the covariance matrix formalism, and we
provide a classification of pure and mixed Gaussian states according
to the various standard forms that the associated covariance
matrices can take.

We collect in Part \ref{PartBip} all results concerning bipartite
entanglement of Gaussian states with two or more modes. In Chapter
\ref{ChapEntGauss} we illustrate the machinery of bipartite
entanglement qualification and quantification in Gaussian states.
The massive Chapter \ref{Chap2M} contains our specific results on
two-mode Gaussian states, including the existence of extremally
(minimally and maximally) entangled states
 at given degrees of mixedness, and the
different orderings induced on entangled states by different
measures of entanglement. In Chapter \ref{ChapUniloc}  we describe
the unitary (and therefore reversible) localization of bipartite
multimode entanglement to a  bipartite two-mode entanglement in
fully symmetric and bisymmetric multimode Gaussian states, and its
scaling with the number of modes.

Multipartite entanglement of Gaussian states is the topic of Part
\ref{PartMulti}. In Chapter \ref{ChapMonoGauss} we present our
crucial advances in the understanding of entanglement sharing in
multimode Gaussian states, including the proof of the monogamy
inequality on distributed entanglement for all Gaussian states.
Multipartite entanglement of three-mode Gaussian states is analyzed
in Chapter \ref{Chap3M} by discussing the structural properties of
such states, and the main consequences of the monogamy inequality,
such as the quantification of genuine tripartite entanglement, and
the promiscuous nature of entanglement sharing in Gaussian states
with symmetry properties. Chapter \ref{ChapUnlim} deals with the
remarkable property of multipartite entanglement in Gaussian states
(as opposed to low-dimensional systems), to coexist to an arbitrary
extent with bipartite entanglement, in simple families of states of
at least four modes, within the holding of the monogamy inequality.

In Part \ref{PartEngi} we show how to engineer multimode Gaussian
resources with optical means. Chapter \ref{Chap2MExp} contains
schemes for the production of extremally entangled two-mode states,
as well as experimental results on the production, characterization
and manipulation of two-mode entanglement with a novel optical
setup. In Chapter \ref{Chap3M4Mengi} we provide a systematic
investigation on the preparation of several families of three- and
four-mode Gaussian states with peculiar entanglement properties,
providing efficient schemes. Chapter \ref{ChapGeneric} deals instead
with the general instance of pure $N$-mode Gaussian states, in which
case for the relevant family of `block-diagonal' states an optimal
state engineering recipe is proposed, which enables to connect
generic entanglement to operationally meaningful resources.

The companion Part \ref{Part4} looks at entanglement from a
practical perspective for quantum information and communication
implementations, and other less conventional applications. Bipartite
and multipartite teleportation-based communication (including
telecloning) with Gaussian states is studied in Chapter
\ref{ChapCommun}, where the equivalence between optimal
teleportation fidelity and shared entanglement is established,
together with an experimentally testable connection between
teleportation efficiency, multipartite entanglement, and promiscuous
sharing structure. Entanglement distribution and the investigation
of the correlation range in many-body harmonic rings with a Gaussian
valence bond structure is addressed in Chapter \ref{ChapGVBS}. The
degradation of Gaussian entanglement as detected by accelerated
observers is instead investigated in Chapter \ref{ChapIvette}, and
interpreted in terms of an entanglement re-distribution in
multipartite form among accessible and unaccessible modes from a
non-inertial perspective.

Part \ref{PartConcl} concludes this Dissertation with a summary on
the various applications of Gaussian entanglement not covered by our
personal research, a brief discussion about recent advances in the
qualification and quantification of entanglement in non-Gaussian
states --- a field of investigation that is to a large extent yet to
be fully explored ---  and an overview on open problems and current
research directions. Appendix \ref{ChapAppendixSF} contains some
tools of symplectic analysis necessary for the structural
characterization of the covariance matrix of pure Gaussian states.

}

\chapter{Characterizing entanglement} \label{ChapEnt}

%\begin{figure}[t!]
%\includegraphics[width=4cm]{dummy.eps}%
%\caption{\label{dummy2} eccolo.}
%\end{figure}

{\sf According to Erwin Schr\"odinger, quantum entanglement is not
\emph{``one but rather the characteristic trait of quantum
mechanics, the one that enforces its entire departure from classical
lines of thought''} \cite{schr}. Entanglement has been widely
recognized as a fundamental aspect of quantum theory, stemming
directly from the superposition principle and quantum
non-factorizability. Remarkably, it is now also acknowledged as a
fundamental physical resource, much on the same status as energy and
entropy, and as a key factor in the realization of information
processes otherwise impossible to implement on classical systems.
Thus the degree of entanglement and information are the crucial
features of a quantum state from the point of view of Quantum
Information Theory \cite{chuaniels,heiss}. Indeed, the search for
proper mathematical frameworks to quantify such features in general
(mixed) quantum states cannot be yet considered accomplished. In
view of such considerations, it is clear that the full understanding
of the relationships between the quantum correlations contained in a
multipartite state and the global and local ({\em i.e.~}referring to
the reduced states of the subsystems) degrees of information of the
state, is of critical importance. In particular, it would represent
a relevant step towards the clarification of the nature of quantum
correlations and, possibly, of the distinction between quantum and
classical correlations in mixed quantum states
\cite{henvedral01,audenplenio,berry}.

We open this Chapter with a discussion about the interpretation and
measures of information in quantum systems. We then move to a
detailed discussion about quantum entanglement, its definition,
qualification and quantification in the bipartite and, to some
extent, in the multipartite setting. Special attention will be
devoted to the fundamental property of entanglement to be
distributed in a so-called ``monogamous'' way, and the implications
of such feature for the characterization of many-body entanglement
sharing.

\section{Information contained in a quantum state}\label{ParInfo}

Both discrete-variable and continuous-variable systems, endowed
respectively with a finite-dimensional and a infinite-dimensional
Hilbert space, can be efficiently employed in quantum information
theory for the encoding, manipulation and transmission of
information. In this context, it is natural to question how much
information a quantum state contains.

Suppose we have prepared a physical system in a certain state and we
would like to test the system somehow, for instance with a
measurement. Before performing the experimental verification, we can
only predict the probabilities  $p_1,\ldots,p_N$ associated to the
$N$ possible outcomes. After the measurement, one of these outcomes
will have occurred and we will possess a complete information
(certainty) about the state of our system.

The degree of {\em information} contained in a state corresponds to
how much certainty we possess {\em a priori} on predicting the
outcome of any test performed on the state \cite{BPeres}.

\subsection{Purity and linear entropy}

 The quantification of information will in general depend not only
 on the state preparation procedure, but also on the choice of the
measurement with its associated probabilities $\{p_k\}$. If for any
test one has a complete ignorance (uncertainty), \ie for a system in
a $N$-dimensional Hilbert space one finds  $p_k=1/N\,\,\forall\,k$,
then the state is {\em maximally mixed}, in other words prepared in
a totally random mixture, with density matrix proportional to the
identity, $\ro_m={\Id N}/{N}$. For instance, a photon emitted by a
thermal source is called `unpolarized', reflecting the fact that,
with respect to any unbiased polarization measurement, the two
outcomes (horizontal/vertical) have the same probability. The
opposite case is represented by {\em pure quantum states}, whose
density matrix is a projector $\ro_p=\ketbra{\psi}{\psi}$, such that
$\ro_p^2 = \ro_p\,$. A pure state of a quantum system contains the
maximum information one has at disposal on the preparation of the
system. All the intermediate instances correspond to a partial
information encoded in the state of the system under consideration.

A hint on how to quantify this information comes from the general
properties of a quantum density operator. We recall that
\[
\tr{\ro^2}\quad \left\{
\begin{array}{lll}
=1 &\Leftrightarrow & \ro\ {\rm pure\ state}\,; \\
<1 &\Leftrightarrow & \ro\ {\rm mixed\ state}\,.
\end{array}
\right.
\]
It is thus natural to address the trace of  $\ro^2$ as {\em purity}
$\mu$ of a state $\ro$,
\begin{equation}\label{QM:purity}
\mu(\ro) = \tr{\ro^2}\,.
\end{equation}
The purity is a measure of information. For states of a Hilbert
space $\hh$ with $\dim\hh=N$, the purity varies in the range
$$
\frac1N \le \mu \le 1\,,
$$
reaching its minimum on the totally random mixture, and equating
unity of course on pure states. In the limit of continuous variable
systems $(N \rightarrow \infty)$, the minimum purity tends
asymptotically to zero.

Accordingly, the ``impurity'' or degree of {\em mixedness} of a
quantum state $\ro$, which characterizes our ignorance before
performing any quantum test on $\ro$, can be quantified via the
functional
\begin{equation}\label{QM:SL}
S_L(\ro) = \frac{N}{N-1} \left( 1- \mu \right) = \frac{N}{N-1}
\left( 1- \tr{\ro^2} \right)\,.
\end{equation}
The quantity $S_L$ (ranging between $0$ and $1$) defined by
\eq{QM:SL} is known as {\em linear entropy} and it is a very useful
measure of mixedness in quantum information theory due to its direct
connection with the purity and the effective simplicity in its
computation.  Actually, the name `linear entropy' follows from the
observation that $S_L$ can be interpreted as a first-order
approximation of the canonical measure of lack-of-information in
quantum theory, that is {\em Von Neumann entropy}. In practice, the
two quantities are not exactly equivalent, and differences between
the two will be singled out in the context of characterizing
entanglement, as we will see in the next Part.

\subsection{Shannon--Von Neumann entropy}

Let us go back to our physical system and to our ensemble of {\em a
priori} known probabilities  $p_1,\ldots,p_N$, associated to the
possible outcomes of a particular measurement we are going to
perform on the system. If we imagine to repeat the measurement on
$n$ copies of the system, all prepared in the same state, with $n$
arbitrarily large, we can expect the outcome ``system in state $j$''
be obtained  $\sim n_j = n\,p_j$ times. Based on our knowledge on
the preparation of the system, we are in the position to predict the
statistical frequencies corresponding to the different outcomes, but
not the order in which the single outcomes will be obtained.
Assuming that, on $n$ measurement runs, outcome $1$ is obtained
$n_1$ times, outcome $2$ $n_2$ times, and so on, the total number of
permutations of the $n$ outcomes is given by $\left({n!}/{\prod_k
n_k!}\right)\,. $ For $n \rightarrow \infty$, also the individual
frequencies will diverge,  $n_j = n\, p_j \rightarrow \infty$, so
that by using Stirling's formula one finds
 $$
\log \frac{n!}{\prod_k n_k!} \simeq n \log n - n - \sum_k (n_k \log
n_k - n_k) = -n \sum_k p_k \log p_k\,.
$$
The expression
\begin{equation}\label{QM:entropy}
S = - \sum_{k=1}^N p_k \log p_k
\end{equation}
is named  \emph{entropy} associated to the probability distribution
$\{p_1,\ldots,p_N\}$: it is a measure of our ignorance prior to the
measurement.

The notion of entropy, originating from thermodynamics, has been
reconsidered in the context of classical information theory by
Shannon  \cite{Shannon48}.  In quantum information theory the
probabilities  $\{p_k\}$  of \eq{QM:entropy} are simply the
eigenvalues of the density matrix $\ro$, and Shannon entropy is
substituted by \emph{Von Neumann entropy} \cite{BVonNeumann}
\begin{equation}\label{QM:SV}
S_V = - \Tr{\ro\, \log \ro} = - \sum_k p_k \log p_k\,.
\end{equation}

Purity $\mu$, linear entropy $S_L$ and Von Neumann entropy $S_V$ of
a quantum state $\ro$ are all invariant quantities under unitary
transformations, as they depend only on the eigenvalues of  $\ro$.
Moreover, Von Neumann entropy  $S_V (\ro)$ satisfies a series of
important mathematical properties, each reflecting a well-defined
physical requirement \cite{Wehrl78}. Some of them are listed as
follows.

\begin{itemize}
    \item \emph{Concavity.}
    \begin{equation}\label{QM:SVconcav}
    S_V(\lambda_1 \ro_1 + \ldots + \lambda_n \ro_n) \ge \lambda_1
    S_V(\ro_1) + \ldots + \lambda_n S_V(\ro_n)\,,
    \end{equation}
    with $\lambda_i \ge 0,\,\sum_i \lambda_i = 1$.
    \eq{QM:SVconcav} means that Von Neumann entropy increases by mixing states, \ie is greater
    if we are more ignorant about the preparation of the system.
    This property follows from the concavity of the log function.

    \item \emph{Subadditivity.} Consider a bipartite system $\s$
    (described by the Hilbert space  $\hh = \hh_1 \otimes
    \hh_2$) in the state  $\ro$. Then
    \begin{equation}\label{QM:SVsubadd}
    S_V(\ro) \le S_V(\ro_1) + S_V(\ro_2)\,,
    \end{equation}
    where $\ro_{1,2}$ are the reduced density matrices $\ro_{1,2} = \ptr{2,1}{\ro}$
    associated to subsystems  $\s_{1,2}$.
    For states of the form $\ro^{\otimes} = \ro_1 \otimes
    \ro_2$, \eq{QM:SVsubadd} is saturated, yielding that  Von Neumann entropy is
    \emph{additive} on tensor product states:
    \begin{equation}\label{QM:SVaddprod}
    S_V(\ro_1 \otimes \ro_2) = S_V(\ro_1) + S_V(\ro_2)\,.
    \end{equation}
    The purity, \eq{QM:purity}, is instead \emph{multiplicative} on
    product states, as the trace of a product equates the product of
    the traces:
    \begin{equation}\label{QM:mumultprod}
    \mu(\ro_1 \otimes \ro_2) = \mu(\ro_1) \cdot \mu(\ro_2)\,.
    \end{equation}
    %    \item \emph{Subadditività forte.}
    \item \emph{Araki--Lieb inequality \cite{ArakiLieb70}.}
    In a bipartite system,
    \begin{equation}\label{QM:SVarale}
    S_V(\ro) \ge \abs{S_V(\ro_1) - S_V(\ro_2)}\,.
    \end{equation}
    Properties \pref{QM:SVsubadd} and \pref{QM:SVarale} are typically grouped
    in the so-called  \emph{triangle inequality}
    \begin{equation}\label{QM:SVtriangle}
    \abs{S_V(\ro_1) - S_V(\ro_2)} \le S_V(\ro) \le S_V(\ro_1) + S_V(\ro_2)\,.
    \end{equation}
\end{itemize}

    It is interesting to remark that \ineq{QM:SVsubadd} is in sharp
    contrast with the analogous property of classical Shannon
    entropy, \begin{equation}\label{QM:Shanarale}
    S(X,Y) \ge S(X),\,S(Y)\,.
    \end{equation}
    Shannon entropy of a joint  probability distribution is always greater than
    the Shannon entropy of each marginal probability distribution,
    meaning that there is more information in a global classical
    system than in any of its parts. On the other hand, consider a
    bipartite quantum system in a pure state
     $\ro=\ketbra{\psi}{\psi}$ . We have then for  Von Neumann
     entropies: $S_V(\ro)=0$, while
 $S_V(\ro_1) \overset{_{\sf \tiny (\ref{QM:SVarale})}}{=} S_V(\ro_2) \ge 0$. The global
 state  $\ro$ has been prepared in a well defined way, but if we measure
 local observables on the subsystems, the measurement outcomes are
 unavoidably random and to some extent unpredictable. We cannot
 reconstruct
 the whole information about how the global system was
 prepared in the state $\ro$ (apart from the trivial instance of $\ro$ being a product state $\ro=\ro_1 \otimes \ro_2$),
 by only looking separately at the two
 subsystems. Information is rather encoded in non-local and
 non-factorizable quantum correlations --- {\em entanglement} --- between the two
 subsystems. The comparison between  the relations \eq{QM:SVsubadd} and \eq{QM:Shanarale} clearly
 evidences the difference between classical and quantum information.

\subsection{Generalized entropies}

In general, the degree of mixedness of a quantum state $\varrho$ can
be characterized completely by the knowledge of all the associated
Schatten $p$--norms \cite{bathia} \be \label{schatten} \|\varrho\|_p
= (\,{\rm Tr}\,|\varrho|^p)^{\frac1p} =(\,{\rm
Tr}\,\varrho^p)^{\frac1p}\, , \quad \,{\rm with} \: p \ge 1. \ee In
particular, the case $p=2$ is directly related to the purity $\mu$,
\eq{QM:purity}, as it is essentially equivalent (up to
normalization) to the linear entropy \eq{QM:SL}. The $p$-norms are
multiplicative on tensor product states and thus determine a family
of non-extensive ``generalized entropies'' $S_{p}$
\cite{bastiaans,tsallis}, defined as \be S_{p} = \frac{1-\,{\rm
Tr}\,\varrho^p}{p-1} \; , \quad p > 1. \label{pgen} \ee These
quantities have been introduced independently  Bastiaans in the
context of quantum optics \cite{bastiaans}, and by  Tsallis in the
context of statistical mechanics \cite{tsallis}. In the quantum
arena, they can be interpreted both as quantifiers of the degree of
mixedness of a state $\varrho$ by the amount of information it
lacks, and as measures of the overall degree of coherence of the
state.

The generalized entropies $S_p$'s range from $0$ for pure states to
$1/(p-1)$ for completely mixed states with fully degenerate
eigenspectra.  We also mention that, in the asymptotic limit of
arbitrary large $p$, the function $\tr{\varrho^p}$ becomes a
function only of the largest eigenvalue of $\varrho$: more and more
information about the state is discarded in such an estimate for the
degree of purity; considering for any non-pure state $S_p$ in the
limit $p\rightarrow \infty$, yields a trivial constant null
function, with no information at all about the state under exam. We
also note that, for any given quantum state, $S_{p}$ is a
monotonically decreasing function of $p$.

Finally, another important class of entropic measures includes the
R\'{e}nyi entropies \cite{renyi} \be S_{p}^{R} = \frac{\log \, {\rm
Tr} \, \varrho^{p}}{1-p} \; , \quad p > 1. \ee It can be  shown that
\cite{SeralePHD} \be
\lim_{p\rightarrow1+}S_{p}=\lim_{p\rightarrow1+}S_{p}^{R}= -\,{\rm
Tr}\,(\varrho\log\varrho) \equiv S_{V} \, , \label{genvneu} \ee so
that also the Von Neumann entropy, \eq{QM:SV}, can be defined in
terms of $p$-norms and within the framework of generalized
entropies.

\subsection{Mutual information}

The subadditivity property \pref{QM:SVsubadd} of Von Neumann entropy
is at the heart of the measure typically employed in quantum
information theory to quantify total --- classical and quantum ---
correlations in a quantum state, namely the {\em mutual information}
\cite{berry}
\begin{equation}\label{QM:I}
I(\ro) = S_V(\ro_1) + S_V(\ro_2) - S_V(\ro)\,,
\end{equation}
where $\ro$ is the state of the global system and $\ro_{1,2}$
correspond to the reduced density matrices. Mutual information
quantifies the information we obtain on $\ro$ by looking at the
system in its entirety, minus the information we can extract from
the separate observation of the subsystems. It can in fact be
written as relative entropy between $\ro$ and the corresponding
product state $\ro^\otimes = \ro_1 \otimes \ro_2$,
\begin{equation}\label{QM:IRE}
I(\ro) = S_R(\ro \Vert \ro^{\otimes})\,,
\end{equation}
where the \emph{relative entropy}, a distance-like measure between
two quantum states in terms of information, is defined as
\cite{VedralRMP}
\begin{equation}\label{QM:RE}
S_R(\ro \Vert {\sigma}) = -S_V(\ro) - \Tr{\ro \, \log {\sigma}} =
\Tr{\ro\,(\log\ro - \log {\sigma})}\,.
\end{equation}

If $\varrho$ is a pure quantum state [$S_V(\varrho)=0$], the Von
Neumann entropy of its reduced states
$S_V(\varrho_1)=S_V(\varrho_2)$ quantifies the entanglement between
the two parties, as we will soon show. Being
$I(\varrho)=2S_V(\varrho_{1})=2S_V(\varrho_{2})$ in this case, one
says that the pure state also contains some classical correlations,
equal in content to the quantum part,
$S_V(\varrho_1)=S_V(\varrho_2)$.

In mixed states a more complex scenario emerges. The mere
distinction between classical correlations, \ie producible by means
of local operations and classical communication (LOCC) only, and
entanglement, due to a purely quantum interaction between
subsystems, is a highly nontrivial, and not generally accomplished
yet, task \cite{henvedral01,berry}.

We are now going to summarize the most relevant results to date
concerning the qualitative and quantitative characterization of
entanglement.

\section{Entanglement and non-locality}

%\section{Il problema dell'entanglement}
 From a phenomenological point of view, the phenomenon of
entanglement is fairly simple. When two physical systems come to an
interaction, some correlation of a quantum nature is generated
between the two of them, which persists even when the interaction is
switched off and the two systems are spatially
separated\footnote{\sf Entanglement can be also created without
direct interaction between the subsystems, via the so-called
entanglement swapping \cite{Telep}.}. If we measure a local
observable on the first system, its state collapses of course in an
eigenstate of that observable. Surprisingly, also the state of the
second system, wherever it is (in the ideal case of zero
environmental decoherence), is modified instantly. Responsible for
this ``spooky action at a distance'' \cite{EPR35} is the
non-classical and non-local quantum correlation known as {\em
entanglement}.

Suppose we have a bipartite or multipartite quantum state: well, the
answer to an apparently innocent question like
\begin{quote}
\centering {\it Does this state contain quantum correlations?}
\end{quote}
is extremely hard to be achieved
\cite{heiss,audenplenio,PlenioVirmani}. The first step concerns a
basic understanding of what such a question really means.

One may argue that a system contains quantum correlations if the
observables associated to the different subsystems are correlated,
and their correlations cannot be reproduced with purely classical
means. This implies that some form of inseparability or
non-factorizability is necessary to properly take into account those
correlations. For what concerns globally {\em pure states} of the
composite quantum system, it is relatively easy to check if the
correlations are of genuine quantum nature. In particular, it is
enough to check if a Bell-CHSH inequality \cite{Bell64,CHSH69} is
violated \cite{Gisin91}, to conclude that a pure quantum state is
entangled. There are in fact many different criteria to characterize
entanglement, but all of them are practically based on equivalent
forms of non-locality in pure quantum states.

These equivalences fade when we deal with {\em mixed states}. At
variance with a pure state, a mixture can be prepared in (generally
infinitely) many different ways. Not being able to reconstruct the
original preparation of the state, one cannot extract all the
information it contains. Accordingly, there is not a completely
general and practical criterion to decide whether correlations in a
mixed quantum state are of classical or quantum nature. Moreover,
different manifestations of quantum inseparability are in general
{\em not} equivalent. For instance, one pays more (in units of Bell
singlets) to create an entangled mixed state $\ro$ --- entanglement
cost \cite{Bennett96pra} --- than what one can get back from
reconverting $\ro$ into a product of singlets  --- distillable
entanglement \cite{Bennett96pra} --- via LOCC \cite{Synak}. Another
example is provided by Werner in a seminal work \cite{Werner89},
where he introduced a parametric family of mixed states (known as
Werner states) which, in some range of the parameters, are entangled
(inseparable) without violating any Bell inequality on local
realism, and thus admitting a description in terms of local hidden
variables. It is indeed an open question in quantum information
theory to prove whether any entangled state violates some Bell-type
inequality \cite{QIProb,Terhal00}.

In fact, entanglement and non-locality are different resources
\cite{GisinNJP}. This can be understood within the general framework
of no-signalling theories which exhibit  even more non-local
features than quantum mechanics. Let us briefly recall what is
intended by non-locality according to Bell \cite{BBell}: there
exists in Nature a channel that allows one to distribute
correlations between distant observers, such that  the correlations
are not already established at the source, and  the correlated
random variables can be created in a configuration of space-like
separation, \ie no normal signal (assuming no superluminal
transmission) can be the cause of the correlations
\cite{EPR35,Bell64}. A convenient description of the intriguing
phenomenon of non-locality is already known: quantum mechanics
describes the channel as a pair of entangled particles. But such
interpretation is not the only one. In recent years, there has been
a growing interest in providing other descriptions of this channel,
mainly assuming a form of communication \cite{tapp}, or the usage of
an hypothetical ``non-local machine'' \cite{prbox} able to violate
the CHSH inequality \cite{CHSH69} up to its algebraic value of $4$
(while the local realism threshold is $2$ and the maximal violation
admitted by quantum mechanics is $2\sqrt2$, the Cirel'son bound
\cite{cirel}). Usually, the motivation for looking into these
descriptions does not come from a rejection of quantum mechanics and
the desire to replace it with something else; rather the opposite:
the goal is to quantify how powerful quantum mechanics is by
comparing its achievements to those of other resources. The
interested reader may have a further look at
Ref.~\cite{Gisinonlocal}.

\section{Theory of bipartite entanglement}

\subsection{Pure states: qualification and
quantification}\label{SecPure}

\begin{defn} \label{defEntPure}
A pure quantum state $\ps \in \hh = \hh_1 \otimes \hh_2$ is {\em
separable} if it can be written as a product state, \ie if there
exist  $\ket{\ph}_1 \in \hh_1$ and $\ket{\chi}_2 \in \hh_2$ such
that
\begin{equation}\label{E:ketprod}
\ps = \ket{\ph}_1 \otimes \ket{\chi}_2 \equiv \ket{\ph,\chi}\,.
\end{equation}
Otherwise, $\ps$ is an {\em entangled} state.
\end{defn}

Qualifying entanglement means having an operational criterion which
would allow us to answer our original question, namely if a given
state is entangled or not.

To this aim, it is useful to write a pure quantum state in its
unique Schmidt decomposition \cite{BPreskill},
\begin{equation}\label{E:schmidt}
\ps = \sum_{k=1}^d \lambda_k \ket{u_k,v_k}\,,
\end{equation}
\begin{eqnarray}
&{\rm where}\quad  & d = \min\{d_1,d_2\}\,, \label{E:schmidtnum}\\
& &\lambda_k \ge 0\,,\quad \sum_{k=1}^d \lambda_k^2 = 1\,.
\label{E:schmidtsum}
\end{eqnarray}
The number $d$ of non-zero terms in the expansion \pref{E:schmidt}
is known as Schmidt number, the positive numbers $\{\lambda_k\}$ are
the Schmidt coefficients, and the local bases $\{\ket{u_k}\} \in
\hh_1$ and $\{\ket{v_k}\} \in \hh_2$ are the Schmidt bases.

From the Schmidt decomposition it follows that the reduced density
matrices of $\ps$,
\begin{eqnarray}
  \ro_1 &=& \sum_{k=1}^d \lambda_k^2 \ketbra{u_k}{u_k}\,, \nonumber\\
  \ro_2 &=& \sum_{k=1}^d \lambda_k^2 \ketbra{v_k}{v_k}\,. \label{E:redsch}
\end{eqnarray}
have the same nonzero eigenvalues (equal to the squared Schmidt
coefficients) and their total number is the Schmidt number $d$. One
can then observe that product states $\ps = \ket{\ph,\chi}$ are
automatically written in Schmidt form with $d=1$, \ie the reduced
density matrices correspond to pure states  ($\ro_1 =
\ketbra{\ph}{\ph}\,,\,\ro_2 = \ketbra{\chi}{\chi})$. On the other
hand, if a state admits a Schmidt decomposition with only one
coefficient, then it is necessarily a product state. We can then
formulate an entanglement criterion for pure quantum states. Namely,
a state
 $\ps$ of a bipartite system is entangled if and only if the reduced
 density matrices describe mixed states,
\begin{equation}\label{E:entpure}
\ps\,\,\,entangled \quad\Leftrightarrow\quad d>1\,,
\end{equation}
with $d$ defined by \eq{E:schmidtnum}.

We thus retrieve that bipartite entanglement of pure quantum states
is qualitatively equivalent to the presence of local mixedness, as
intuitively expected. This connection is in fact also quantitative.
The {\em entropy of entanglement} $E_V(\ps)$ of a pure bipartite
state $\ps$ is defined as the Von Neumann entropy, \eq{QM:SV}, of
its reduced density matrices \cite{Bennett96},
\begin{equation}\label{E:E}
E_V(\ps) = S_V(\ro_1) = S_V(\ro_2) = -\sum_{k=1}^d \lambda_k^2 \,
\log \lambda_k^2\,.
\end{equation}
The entropy of entanglement is {\em the} canonical measure of
bipartite entanglement in pure states. It depends only on the
Schmidt coefficients   $\lambda_k$, not on the corresponding bases;
as a consequence it is invariant under local unitary operations
\begin{equation}\label{E:Elocalinv}
E_V\Big((\op{U}_1 \otimes \op{U}_2) \ps\Big) = E_V\Big(\ps\Big).
\end{equation}
It can be shown \cite{Popescu97} that $E_V(\ps)$ cannot increase
under LOCC performed on the state $\ps$: this is a fundamental
physical requirement as it reflects the fact that entanglement
cannot be created via LOCC only \cite{VedralPRK,Vidal00}. It can be
formalized as follows. Let us suppose, starting with a state $\ps$
of the global system $\s$, to perform local measurements on  $\s_1$
and $\s_2$, and to obtain, after the measurement, the state
$\ket{\ph_1}$ with probability $p_1$, the state $\ket{\ph_2}$ with
probability $p_2$, and so on. Then
\begin{equation}\label{E:EnoLO}
E_V(\ps) \ge \sum_k p_k E_V(\ket{\ph_k})\,.
\end{equation}
Note that entanglement cannot increase \emph{on average}, \ie
nothing prevents, for a given $k$, that $E_V(\ket{\ph_k})
> E_V(\ps)$. On this the concept of {\em entanglement distillation} is based
\cite{Bennett96prl,Bennett96pra,Gisin96}: with a probability $p_k$,
it is possible to increase entanglement via LOCC manipulations.

\subsection{Mixed states: entanglement vs separability}

A mixed state $\ro$ can be decomposed as a convex combination of
pure states,
\begin{equation}\label{E:rodec}
\ro = \sum_k p_k \ketbra{\psi_k}{\psi_k}\,.
\end{equation}
\eq{E:rodec} tells us how to create the state described by the
density matrix $\ro$: we have to prepare the state  $\ket{\psi_1}$
with probability  $p_1$, the state $\ket{\psi_2}$ with probability
$p_2$, etc. For instance, we could collect $N$ copies ($N \gg 1$) of
the system, prepare $n_k \simeq N p_k$ of them in the state
$\ket{\psi_k}$, and pick a random system.

The problem is that the decomposition of  \eq{E:rodec} is not
unique: unless $\ro$ is already a pure state, there exist infinitely
many decompositions of a generic $\ro$ in ensembles of pure states,
meaning that the mixed state can be prepared in infinitely many
different ways. One can expect that this has some consequence on the
entanglement. Let us suppose we have a bipartite system and we
detect, by local measurements, the presence of correlations between
the two subsystems. Given the ambiguity on the state preparation, we
cannot know {\em a priori} if those correlations arose from a
quantum interaction between the subsystems (meaning entanglement) or
were induced by means of LOCC (meaning classical correlations). It
is thus clear that a mixed state can be defined separable
(classically correlated) if there exist at least one way of
engineering it by LOCC; on the other hand it is entangled (quantumly
correlated) if, among the infinite possible preparation procedures,
there is no one which relies on LOCC only \cite{Werner89}.

\begin{defn} \label{defEntMix}
A mixed quantum state $\ro$ of a bipartite system, described by the
Hilbert space  $\hh = \hh_1 \otimes \hh_2$, is {\em separable} if
and only if there exist coefficients $\{p_k \,\vert\, p_k \ge
0,\,\sum_k p_k=1\}$, and states $\{{\sigma}_k\} \in \hh_1$ and
$\{{\tau}_k\} \in \hh_2$, such that
\begin{equation}\label{E:rosep}
\ro = \sum_k p_k \, \left( {\sigma}_k \otimes {\tau}_k \right)\,.
\end{equation}
Otherwise, $\ro$ is an {\em entangled} state.
\end{defn}

For pure states, the expansion  \eq{E:rosep} has a single term and
we recover {\rm Def.~\ref{defEntPure}}, \ie the only separable pure
states are product states. For mixed states, not only product states
(containing zero correlations of any form) but in general any convex
combination of product states is separable. This is obvious as the
state of \eq{E:rosep} contains only classical correlations, since it
can be prepared by means of LOCC.

However, {\rm Def.~\ref{defEntMix}} is in all respects impractical.
Deciding separability according to the above definition would imply
checking all the infinitely many decomposition of a state $\ro$ and
looking for at least one of the form  \eq{E:rosep}, to conclude that
the state is not entangled. This is clearly impossible. For this
reason, several {\em operational} criteria have been developed in
order to detect entanglement in mixed quantum states
\cite{CiracPrimer,BrussJMP,BInfo}. Two of them are discussed in the
following.

\subsubsection{Positive Partial Transposition criterion}
\label{SecPPT}

One of the most powerful results to date in the context of
separability criteria is the Peres--Horodecki condition
\cite{Peres96,Horodecki96}. It is based on the operation of {\em
partial transposition} of the density matrix of a bipartite system,
obtained by performing transposition  with respect to the degrees of
freedom of one subsystem only. Peres criterion states that, if a
state $\ro_s$ is separable, then its partial transpose $\ro_s\PT1$
(with respect \eg to subsystem $\s_1$) is a valid density matrix, in
particular positive semidefinite, $\ro_s\PT1 \ge 0.$ The same holds
naturally for $\ro_s\PT2$. Positivity of the partial transpose (PPT)
is therefore a necessary condition for separability \cite{Peres96}.
The converse (\ie $\ro\PT1\ge0 \Rightarrow \ro$ separable) is in
general false. The Horodecki's have proven that it is indeed true
for low-dimensional systems, specifically bipartite systems of
dimensionality $2 \times 2$ and  $2 \times 3$, in which case PPT is
equivalent to separability \cite{Horodecki96}. For higher
dimensional systems, PPT entangled states (with  $\ro\PT1 \ge 0$)
have been shown to exist \cite{Horodecki97}. These states are known
as \emph{bound entangled} \cite{Horodecki98} as their entanglement
cannot be distilled to obtain maximally entangled states. The
existence of bound entangled (undistillable) states with negative
partial transposition is conjectured as well
\cite{Dur00,DiVincenzo00}, yet a fully rigorous analytical proof of
this fact is still lacking \cite{QIProb}.

Recently, PPT criterion has been revisited in the continuous
variable scenario by Simon  \cite{Simon00}, who showed how the
transposition operation acquires in infinite-dimensional Hilbert
spaces an elegant geometric interpretation in terms of time
inversion (mirror reflection of the momentum operator). It follows
that the PPT criterion is again necessary and sufficient for
separability in all $(1+N)$-mode Gaussian states
 of continuous variable systems\footnote{\sf Gaussian states of a
$N$-mode continuous variable system are by definition states whose
characteristic function and quasi-probability distributions are
Gaussian on a $2N$-dimensional real phase space. See
Chapter~\ref{ChapGauss} for a rigorous definition.} with respect to
$1 \times N$ bipartitions \cite{Simon00,Duan00,werewolf}. We have
extended its validity to ``bisymmetric'' $M \times N$ Gaussian
states \cite{unitarily}, \ie invariant under local mode permutations
in the $M$-mode and in the $N$-mode partitions, as detailed in
Sec.~\ref{SecPPTG}.

\subsubsection{Entanglement witnesses} \label{SecWitn}

A state  $\ro$ is entangled if and only if there exists a Hermitian
operator $\W$ such that $\Tr{\W\,\ro}<0$ and $\Tr{\W\,{\sigma}}\ge0$
for any state ${\sigma} \in \dd$, where $\dd \subset \hh$ is the
convex and compact subset of separable states
\cite{Horodecki96,Terhal00}.
%\end{thm}
The operator  $\W$ is the {\em witness} responsible for detecting
entanglement in the state  $\ro$. According to the Hahn-Banach
theorem, given a convex and compact set $\dd$ and given $\ro \not\in
\dd$, there exists an hyperplane which separates $\ro$ from $\dd$.
Optimal entanglement witnesses induce an hyperplane which is tangent
to the set $\dd$ \cite{Lewenstein00}. A sharper detection of
separability can be achieved by means of nonlinear entanglement
witnesses, curved towards the set $\dd$ of separable states
\cite{NorbertoGuhne,illuso}.

Entanglement witnesses are quite powerful tools to distinguish
entangled from separable states, especially in practical contexts.
With some preliminary knowledge about the form of the states one is
willing to engineer or implement in a quantum information
processing, one can systematically find entanglement witnesses in
terms of experimentally accessible observables, to have a direct
tool to test the presence of quantum correlations, as demonstrated
in the lab \cite{Barbieri03,Bourennane03,blattguhne,panguhne}.

\subsection{Mixed states: quantifying
entanglement}\label{SecQuantEnt}

The issue of quantifying bipartite entanglement cannot be considered
accomplished yet. We are assisting to a proliferation of
entanglement measures, each motivated by a special context in which
quantum correlations play a central role, and each accounting for a
different, sometimes inequivalent quantification and ordering of
entangled states. Detailed treatments of the topic can be found \eg
in Refs.~\cite{BrussJMP,QIC01,PlenioVirmani,ChristandlPHD}. In
general, some physical requirements any good entanglement measure
$E$ should satisfy are the following.

\subsubsection{Properties of entanglement monotones}

\begin{itemize}
    \item \emph{Nullification.\quad} $E(\ro)\ge 0$. If $\ro$ is separable,
    then
    $E(\ro)=0$.
    \item \emph{Normalization.\quad} For a maximally entangled
    state  in $d \times d$ dimension, $\ket{\Phi} =
\left(\sum_{i=0}^{d-1}\ket{i,i}\right)/\sqrt{d}$, it should be
\begin{equation}\label{E:Emixmax}
    E(\ketbra{\Phi}{\Phi}) = \log d\,.
    \end{equation}
    \item \emph{Local invariance.\quad} The measure $E(\ro)$ should be invariant under
    local unitary transformations,
\begin{equation}\label{E:Emixlocalinv}
    E\left((\op{U}_1 \otimes \op{U}_2)\, \ro\, (\adj{\op{U}}_1 \otimes \adj{\op{U}}_2)\right) = E(\ro)\,.
    \end{equation}
    \item \emph{LOCC monotonicity.\quad}  The measure $E(\ro)$ should not increase
    on average upon application of LOCC transformations,
\begin{equation}\label{E:EmixnoincrLOCC}
    E(\op{O}_{_{\rm LOCC}}(\ro)) \le E(\ro)\,,
    \end{equation}
    \item \emph{Continuity.\quad} The entanglement difference
    between two density matrices infinitely close in trace norm
    should tend to zero,
    \begin{equation}\label{E:Emixcont}
    \norm{\ro - {\sigma}} \rightarrow 0 \quad \Rightarrow \quad
    E(\ro) - E({\sigma}) \rightarrow 0\,.
\end{equation}
\end{itemize}
Additional requirements (not strictly needed, and actually not
satisfied even by some `good' entanglement measures), include: {\em
additivity} on tensor product states, $E(\ro^{\otimes N}) = N\,
E(\ro)$; {\em convexity},      $E(\lambda\, \ro + (1-\lambda)\,
{\sigma}) \le
    \lambda\,E(\ro) + (1-\lambda)\, E({\sigma})$ with $0\le\lambda\le1$; reduction to the entropy of entanglement \eq{E:E}
    on pure states. The latter constraint is clearly not necessary, as it is enough
    for a quantifier $E'(\ro)$ to be a strictly monotonic and convex function of another measure
     $E''(\ro)$ which satisfies the above listed properties, in order for $E'(\ro)$ to be regarded
     as a good entanglement measure.
     Another interesting property a good entanglement measure should
     satisfy, which becomes crucial in the multipartite setting, is
     {\em monogamy} in the sense of Coffman-Kundu-Wootters
     \cite{CKW}. In a tripartite state $\ro_{ABC}$,
\begin{equation}\label{E:monogamy}
E(\ro_{A|(BC)}) \ge E(\ro_{A|B}) + E(\ro_{A|C})\,.
\end{equation}
A detailed discussion about entanglement sharing and monogamy
constraints \cite{pisa} will be provided in Sec.~\ref{SecPisa}, as
it embodies the central idea behind the results of Part
\ref{PartMulti} of this Dissertation.

\subsubsection{Entanglement measures} \label{SecEntMeas}

We will now recall the definition of some `popular' entanglement
measures, which have special relevance for our results obtained in
the continuous variable scenario. The author is referred to Refs.
\cite{PlenioVirmani,ChristandlPHD} for better and more comprehensive
reviews.

\vspace*{0.3cm}

\begin{itemize}
  \item
{\rm Entanglement of formation.}--- The {\em entanglement of
formation} $E_F(\ro)$ \cite{Bennett96pra} is the convex-roof
extension \cite{OsborneCROOF} of the entropy of entanglement
\eq{E:E}, \ie the weighted average of the pure-state entanglement,
\begin{equation}\label{E:EF}
E_F(\ro) = \min_{\{p_k,\,\ket{\psi_k}\}} \sum_k
p_k\,E_V(\ket{\psi_k})\,,
\end{equation}
minimized over all decompositions of the mixed state $\ro=\sum_k p_k
\ketbra{\psi_k}{\psi_k}$. An explicit solution of such nontrivial
optimization problem is available for two qubits \cite{Wootters98},
for highly symmetric states like Werner states and isotropic states
in arbitrary dimension \cite{Terhal01,VollWerner01}, and for
symmetric two-mode Gaussian states \cite{giedke03}.  The additivity
of the entanglement of formation is currently an open problem
\cite{QIProb}.

\item {\rm Entanglement cost.}---
The {\em entanglement cost} $E_C(\ro)$ \cite{Bennett96pra}
quantifies how much Bell pairs
$\ket{\Phi}=(\ket{00}+\ket{11})/\sqrt2$ one has to spend to create
the entangled state   $\ro$ by means of LOCC. It is defined as the
asymptotic ratio between the minimum number $M$ of used Bell pairs,
and the number $N$ of output copies of $\ro$,
\begin{equation}\label{E:EC}
E_C(\ro) = \min_{_{\{{\rm LOCC}\}}} \lim_{N \rightarrow \infty}
\frac{M^{\rm\, in}}{N^{\rm\, out}}\,.
\end{equation}
The entanglement cost, a difficult quantity to be computed in
general  \cite{Dur02}, is equal to the asymptotic regularization of
the entanglement of formation \cite{Hayden01}, $E_C(\ro) = \lim_{N
\rightarrow \infty} [E_F(\ro^{\otimes N})/{N}]$, and would coincide
with  $E_F(\ro)$ if the additivity of the latter was proven.

\item{\rm Distillable entanglement.}--- The converse of the entanglement cost
is the {\em distillable entanglement} \cite{Bennett96pra}, which is
defined as the asymptotic fraction $M/N$ of Bell pairs which can be
extracted from $N$ copies of the state $\ro$ by using the optimal
LOCC distillation protocol,
\begin{equation}\label{E:ED}
E_D(\ro) = \max_{_{\{{\rm LOCC}\}}} \lim_{N \rightarrow \infty}
\frac{M^{\rm\, out}}{N^{\rm\, in}}\,.
\end{equation}
The distillable entanglement vanishes for bound entangled states.
The quantity $E_C(\ro)-E_D(\ro)$ can be regarded as the
undistillable entanglement.  It is strictly nonzero for all
entangled mixed states \cite{Synak}, meaning that LOCC manipulation
of quantum states is asymptotically irreversible apart from the case
of pure states (one loses entanglement units in the currency
exchange!).

\item{\rm Relative entropy of entanglement.}--- An intuitive way to measure
entanglement is to consider the minimum ``distance'' between the
state $\ro$ and the convex set  $\dd \subset \hh$ of separable
states. In particular, the {\em relative entropy of entanglement}
 $E_R(\ro)$ \cite{Vedral97} is the entropic distance [\ie the quantum relative entropy
 \eq{QM:RE}] between  $\ro$ and the closest separable state
 ${\sigma}^\star$,
\begin{equation}\label{E:ER}
E_R(\ro) = \min_{{\sigma}^\star \in \dd} \Tr{\ro\, (\log\ro -
\log{\sigma}^\star)}\,.
\end{equation}
Note that the closest separable state ${\sigma}^\star$ is typically
not the corresponding product state $\ro^{\otimes} =\ro_1 \otimes
\ro_2$; the relative entropy between $\ro$ and $\ro^{\otimes}$ is
indeed the mutual information \eq{QM:IRE}, a measure of total
correlations which therefore overestimates entanglement (being
nonzero also on separable, classically correlated states).
\end{itemize}

\smallskip

Let us recall that the all the above mentioned entanglement measures
coincide on pure states, $E_F(\ps)=E_C(\ps)=E_R(\ps)=E_D(\ps) \equiv
E_V(\ps)$, while for generic mixed states the following chain of
analytic inequalities holds \cite{Horodecki00,Donald02,QIC01},
\begin{equation}\label{E:EFCRD}
E_F(\ro) \ge E_C(\ro) \ge E_R(\ro) \ge E_D(\ro)\,.
\end{equation}

\smallskip

\begin{itemize}
\item{\rm Negativities.}--- An important class of entanglement measures is constituted by the
negativities,  which quantify the violation of the PPT criterion for
separability (see Sec.~\ref{SecPPT}), \ie how much the partial
transposition of $\ro$ fails to be positive. The {\em negativity}
$\N(\ro)$ \cite{Zyczkowski98,EisertPHD} is defined as
\begin{equation}\label{E:N}
\N(\ro) = \frac{\norm{\ro\PT{i}}_1-1}{2}\,,
\end{equation}
where
\begin{equation}\label{E:tracenorm}
\|\op{O}\|_1 = \tr{\sqrt{\adj{\op{O}}\op{O}}}
\end{equation}
is the trace norm of the operator $\op O$. The negativity is a
computable measure of entanglement, being
\begin{equation}\label{E:Nsum}
\N(\ro)=\max\set{0,-{\sum_k \lambda_k^-}}\,,
\end{equation}
where the $\set{\lambda_k^-}$'s are the negative eigenvalues of the
partial transpose.

In continuous variable systems, the negativity is still a proper
entanglement measure \cite{VidalWerner02}, even though a related
measure is more often used, the  \emph{logarithmic negativity}
$E_\N(\ro)$ \cite{VidalWerner02, EisertPHD, Plenio05},
\begin{equation}\label{E:EN}
E_{\N}(\ro)= \log\|\ro\PT{i}\|_{1} = \log \left[1+2\N(\ro)\right]\,.
\end{equation}
The logarithmic negativity is additive and, despite being
non-convex, is a full entanglement monotone under LOCC
\cite{Plenio05}; it is an upper bound for the distillable
entanglement \cite{EisertPHD}, $E_\N(\ro) \ge E_D(\ro)$, and it is
the exact entanglement cost under operations preserving the
positivity of the partial transpose \cite{Eisert03}. The logarithmic
negativity will be our measure of choice for the quantification of
bipartite entanglement of Gaussian states (see Part~\ref{PartBip})
and on it we will base the definition of a new entanglement measure
for continuous variable systems, the {\em contangle}
\cite{contangle}, which will be exploited in the analysis of
distributed multipartite entanglement of Gaussian states (see
Part~\ref{PartMulti}).

\item{\rm Squashed entanglement.}---
Another interesting entanglement measure is the {\em squashed
entanglement} $E_{sq}(\ro)$ \cite{Squashed} which is defined as
\begin{equation}\label{E:squash}
    E_{sq}(\ro_{AB}) = \inf_{E} \left[ \frac12 I(\ro_{ABE})
    \right]\ : \ {{\rm tr}_E\{\ro_{ABE}\}=\ro_{AB}}\,,
    \end{equation}
where $I(\ro_{ABE})
=S(\ro_{AE})+S(\ro_{BE})-S(\ro_{ABE})-S(\ro_{E})$ is the quantum
conditional mutual information, which is often also denoted as
$I(A;B|E)$. The motivation behind $E_{sq}$ comes from related
quantities in classical cryptography that determine correlations
between two communicating parties and an eavesdropper. The squashed
entanglement is a convex entanglement monotone that is a lower bound
to $E_F(\ro)$ and an upper bound to $E_D(\ro)$, and is hence
automatically equal to $E_V(\ro)$ on pure states. It is also
additive on tensor products, and continuous \cite{AlickiF03}. Cherry
on the cake, it is also monogamous \ie it satisfies \eq{E:monogamy}
for arbitrary systems \cite{KoashiWinter}. The severe drawback which
affects this otherwise ideal measure of entanglement is its
computability: in principle the minimization in \eq{E:squash}  must
be carried out over {\em all} possible extensions, including
infinite-dimensional ones, which is highly nontrivial. Maybe the
task can be simplified, in the case of Gaussian states, by
restricting to Gaussian extensions. This is currently under our
investigation, but no significant progress has been achieved yet.
\end{itemize}

%\vspace*{0.3cm}

\subsubsection{Entanglement-induced ordering of states}
\label{SecOrderDiscuss}

Let us remark that having so many entanglement measures (of which
only a small portion has been recalled here) means in particular
that different {\em orderings} are induced on the set of entangled
states. One can show that any two LOCC-monotone entanglement
measures can only impose the same ordering on the set of entangled
states, if they are actually exactly the same measure
\cite{VirmaniPlenioTheorem}. Therefore there exist in general pairs
of states $\ro_A$ and $\ro_B$ such that $E'(\ro_A)
> E'(\ro_B)$ and $E''(\ro_A) < E''(\ro_B)$, according to two different entanglement
monotones  $E'(\ro)$ and $E''(\ro)$ (see Sec.~\ref{secorder} for an
explicit analysis in the case of two-mode Gaussian states
\cite{ordering}).
 Given the wide range of tasks
that exploit entanglement \cite{chuaniels}, one might understand
that the motivations behind the definitions of entanglement as `that
property which is exploited in such protocols' are manifold. This
means that situations will almost certainly arise where a state
$\ro_A$ is better than another state $\ro_B$ for achieving one task,
but for achieving a different task $\ro_B$ is better than $\ro_A$.
Consequently, the fact that using a task-based approach to
quantifying entanglement will certainly not lead to a single unified
perspective, is somehow expected.

In this respect, it is important to know that (in finite-dimensional
Hilbert spaces) {\em all bipartite entangled states are useful for
quantum information processing} \cite{Masanes05}. For a long time
the quantum information community has used a `negative'
characterization of the term entanglement, essentially defining
entangled states as those that cannot be created by LOCC alone
\cite{PlenioVirmani}. However, remarkably, it has been recently
shown  that for any non-separable state $\ro$ according to {\rm
Def.~\ref{defEntMix}}, one can find another state $\sigma$ whose
teleportation fidelity may be enhanced if $\ro$ is also
present\footnote{\sf We have independently achieved a somehow
similar operational interpretation for (generally multipartite)
continuous-variable entanglement of symmetric Gaussian states in
terms of optimal teleportation fidelity \cite{telepoppate}, as will
be discussed in Sec.~\ref{SecTelepoppy}.}
\cite{Masanes05,Masanes05a,Brandao05}. This is interesting as it
allows us to positively characterize non-separable states as those
possessing a useful resource that is not present in separable
states. The synonymous use of the terms {\em non-separable} and {\em
entangled} is hence justified.

\section{Multipartite entanglement sharing and monogamy constraints}
\label{SecPisa}

It is a central trait of quantum information theory that there exist
limitations to the free sharing of quantum correlations among
multiple parties. Such {\em monogamy constraints} have been
introduced in a landmark paper by Coffman, Kundu and Wootters, who
derived a quantitative inequality expressing a trade-off between the
couplewise and the genuine tripartite entanglement for states of
three qubits \cite{CKW}.  Since then, a lot of efforts have been
devoted to the investigation of distributed entanglement in
multipartite quantum systems. In this Section, based on
Ref.~\cite{pisa}, we report in a unifying framework a bird's eye
view of the most relevant results that have been established so far
on entanglement sharing. We will take off from the domain of $N$
qubits, graze qudits ({\ie}$d$-dimensional quantum systems), and
drop the premises for the fully continuous-variable analysis of
entanglement sharing in Gaussian states which will presented in
Part~\ref{PartMulti}.

\subsection{Coffman-Kundu-Wootters inequality}

The simplest conceivable quantum system in which multipartite
entanglement can arise is a system of three two-level particles
(qubits). Let two of these qubits, say A and B, be in a maximally
entangled state (a Bell state). Then {\em no} entanglement is
possible between each of them and the third qubit C. In fact
entanglement between C and A (or B) would imply A and B being in a
mixed state, which is impossible because they are sharing a pure
Bell state. This simple observation embodies, in its sharpest
version, the {\em monogamy} of quantum entanglement \cite{monogamy},
as opposed to classical correlations which can be freely shared.

We find it instructive to look at this feature as a simple
consequence of the no-cloning theorem \cite{nocloning1,nocloning2}.
In fact, maximal couplewise entanglement in both bipartitions AB and
AC of a three-particle ABC system, would enable perfect $1
\rightarrow 2$ telecloning \cite{telequb} of an unknown input state,
which is impossible due to the linearity of quantum mechanics. The
monogamy constraints thus emerge as fundamental properties enjoyed
by quantum systems involving more than two parties, and play a
crucial role {\em e.g.}~in the security of quantum key distribution
schemes based on entanglement \cite{arturo}, limiting the
possibilities of the malicious eavesdropper. Just like in the
context of cloning, where research is devoted to the problem of
creating the best possible {\em approximate} copies of a quantum
state, one can address the question of entanglement sharing in a
weaker form. If the two qubits A and B are still entangled but not
in a Bell state, one can then ask how much entanglement each of them
is allowed to share with qubit C, and what is the maximum genuine
tripartite entanglement that they may share all together. The answer
is beautifully encoded in the Coffman-Kundu-Wootters (CKW)
inequality \cite{CKW}
\begin{equation}\label{ckw3}
E^{A|(BC)} \, \ge \, E^{A|B \n C} + E^{A|C \n B}\,,
\end{equation}
where $E^{A|(BC)}$ denotes the entanglement between qubit A and
subsystem (BC), globally in a state $\varrho$, while $E^{A|B \n C}$
denotes entanglement between A and B in the reduced state obtained
tracing out qubit C (and similarly for $E^{A|C \n B}$ exchanging the
roles of B and C). \ineq{ckw3} states that the bipartite
entanglement between one single qubit, say A, and all the others, is
greater than the sum of all the possible couplewise entanglements
between A and each other qubit.

\subsection{Which entanglement is shared?}

While originally derived for system of three qubits, it is natural,
due to the above considerations, to assume that \ineq{ckw3} be a
general feature of any three-party quantum system in arbitrary (even
infinite) dimensions. However, before proceeding, the careful reader
should raise an important question, namely {\em how} are we
measuring the bipartite entanglement in the different bipartitions,
and what the symbol $E$ stands for in \ineq{ckw3}.

Even if the system of three qubits is globally in a pure state, its
reductions will be obviously mixed. As seen in
Sec.~\ref{SecQuantEnt}, there is a piebald scenario of several,
inequivalent measures of entanglement for mixed states, and each of
them must be chosen, depending on the problem one needs to address,
and/or on the desired use of the entangled resources. This picture
is consistent, provided that each needed measure is selected out of
the cauldron of {\em bona fide} entanglement measures, at least
positive on inseparable states and monotone under LOCC. Here we are
addressing the problem of entanglement sharing: one should not be so
surprised to discover that not all entanglement measures satisfy
\ineq{ckw3}. In particular, the entanglement of formation,
\eq{E:EF}, fails to fulfill the task, and this fact led CKW to
define, for qubit systems, a new measure of bipartite entanglement
consistent with the quantitative monogamy constraint expressed by
\ineq{ckw3}.

\subsubsection{Entanglement of two qubits} \label{SecEnt2Q}

 For arbitrary states of two qubits, the entanglement of
formation, \eq{E:EF}, has been explicitly computed by Wootters
\cite{Wootters98}, and reads
\begin{equation}
  E_F(\varrho) = {\CMcal F}[\conc(\varrho)]\,, \label{eqwootters}
  \end{equation}
  with ${\CMcal F}(x) = H [(1 + \sqrt{1-x^2})/2]$ and $H(x) = - x \log_2 x - (1-x) \log_2 (1-x)$.
 The quantity $\conc(\varrho)$ is called the
\emph{concurrence} \cite{Wootters97} of the state $\varrho$ and is
defined as \be \label{concurrence}\conc(\varrho) =
\max\{0,\sqrt{\lambda_1}-\sqrt{\lambda_2}-\sqrt{\lambda_3}-\sqrt{\lambda_4}\}\,
,\ee where the $\{\lambda_i\}$'s are the eigenvalues of the matrix
$\varrho (\sigma_y \otimes \sigma_y) \varrho^{\ast} (\sigma_y
\otimes \sigma_y)$ in decreasing order, $\sigma_y$ is the Pauli spin
matrix and the star denotes complex conjugation in the computational
basis $\{\ket{ij}=\ket{i}\otimes\ket{j},\;i,j=0,1\}$. Because
${\CMcal F}(x)$ is a monotonic convex function of $x \in [0,\,1]$,
the concurrence $\conc(\varrho)$ and its square, the \emph{tangle}
\cite{CKW} \be\label{tangle}\tau(\varrho)=\conc^2(\varrho)\,,\ee are
proper entanglement monotones as well. On pure states, they are
monotonically increasing functions of the entropy of entanglement,
\eq{E:E}.

The concurrence coincides (for pure qubit states) with another
entanglement monotone, the {\em negativity} \cite{Zyczkowski98},
defined in \eq{E:N}, which properly quantifies entanglement of two
qubits as PPT criterion \cite{Peres96,Horodecki96} is necessary and
sufficient for separability (see Sec.~\ref{SecPPT}). On the other
hand, the tangle is equal (for pure states $\ket\psi$) to the linear
entropy of entanglement $E_L$, defined as the linear entropy $S_L
(\varrho_{\rm A}) = 1-{\rm Tr}_{\rm A}\varrho_{\rm A}^2$,
\eq{QM:SL}, of the reduced state $\varrho_{\rm A} = {\rm Tr}_{\rm B}
\ketbra{\psi}{\psi}$ of one party.

\subsection{Residual tripartite entanglement} \label{sec3tangle} After this survey, we can now
recall the crucial result that, for three qubits, the desired
measure $E$ such that the CKW inequality \pref{ckw3} is satisfied is
exactly the tangle \cite{CKW} $\tau$, \eq{tangle}. The general
definition of the tangle, needed {\eg}to compute the leftmost term
in \ineq{ckw3} for mixed states, involves a convex roof analogous to
that defined in \eq{E:E}, namely
\begin{equation}\label{mixtangle}
\tau(\varrho) = \min_{\{p_i,\psi_i\}} \sum_i p_i\,\tau
(\ketbra{\psi_i}{\psi_i})\,.
\end{equation}
%For a pure state of $N$ qubits, in particular, the tangle between
%qubit A (described by a reduced density matrix $\varrho_{\rm A}$),
%which enters in the decomposition \eq{mixtangle}, is just equal to
%$4\ \det \varrho_{\rm A}$.
With this general definition, which implies that the tangle is a
convex measure on the set of density matrices, it was sufficient for
CKW to prove \ineq{ckw3} only for pure states of three qubits, to
have it satisfied for free by mixed states as well \cite{CKW}.

Once one has established a monogamy inequality like \ineq{ckw3}, the
following natural step is to study the difference between the
leftmost quantity and the rightmost one, and to interpret this
difference as the {\em residual entanglement}, not stored in
couplewise correlations, that hence quantifies the genuine
tripartite entanglement shared by the three qubits. The emerging
measure
\begin{equation}\label{tau3}
\tau_3^{A|B|C} = \tau^{A|(BC)} - \tau^{A|B \n C} - \tau^{A|C \n
B}\,,
\end{equation}
known as the {\em three-way tangle} \cite{CKW}, has indeed some nice
features. For pure states, it is invariant under permutations of any
two qubits, and more remarkably it has been proven to be a
tripartite entanglement monotone under LOCC \cite{wstates}. However,
no operational interpretation for the three-tangle, possibly
relating it to the optimal distillation rate of some canonical
`multiparty singlet', is currently known. The reason lies probably
in the fact that the notion of a well-defined maximally entangled
state becomes fuzzier when one moves to the multipartite setting. In
this context, it has been shown that there exist two classes of
three-party fully inseparable pure states of three qubits,
inequivalent under stochastic LOCC operations, namely the
Greenberger-Horne-Zeilinger (GHZ) state \cite{GHZ}
\begin{equation}\label{ghzstates}
\ket{\psi_{\rm GHZ}} = \frac1{\sqrt{2}} \left(\ket{000} +
\ket{111}\right)\,,
\end{equation}
 and the $W$ state \cite{wstates}
\begin{equation}\label{wstates}
\ket{\psi_W}=\frac1{\sqrt{3}} \left(\ket{001} + \ket{010} +
\ket{100}\right)\,.
\end{equation}
From the point of view of entanglement, the big difference between
them is that the GHZ state has maximum residual three-party tangle
[$\tau_3(\psi_{\rm GHZ}) = 1$] with zero couplewise quantum
correlations in any two-qubit reduction, while the $W$ state
contains maximum two-party entanglement between any couple of qubits
in the reduced states and it consequently saturates \ineq{ckw3}
[$\tau_3(\psi_W) = 0$]. The full inseparability of the $W$ state can
be however detected by the `Schmidt measure' \cite{schmeasure}.

\subsection{Monogamy inequality for {\em N} parties}

So far we have recalled the known results on the problem of
entanglement sharing in finite-dimensional systems of three parties,
leading to the definition of the residual tangle as a proper measure
of genuine tripartite entanglement for three qubits. However, if the
monogamy of entanglement is really a universal property of quantum
systems, one should aim at finding more general results.

There are two axes along which one can move, pictorially, in this
respect. One direction concerns the investigation on distributed
entanglement in systems of more than three parties, starting with
the simplest case of $N \ge 4$ qubits (thus moving along the
horizontal axis of increasing number of parties). On the other hand,
one should analyze the sharing structure of multipartite
entanglement in higher dimensional systems, like qudits, moving, in
the end, towards continuous variable  systems (thus going along the
vertical axis of increasing Hilbert space dimensions). The final
goal would be to cover the entire square spanned by these two axes,
in order to establish a really complete theory of entanglement
sharing.

Let us start moving to the right. It is quite natural to expect
that, in a $N$-party system, the entanglement between qubit $p_i$
and the rest should be greater than the total two-party entanglement
between qubit $p_i$ and each of the other $N-1$ qubits. So the
generalized version of \ineq{ckw3} reads
\begin{equation}\label{ckwn}
E^{p_i|{\CMcal P}_i} \, \ge \, \sum_{j \ne i} E^{p_i|p_j}\,,
\end{equation}
with ${\CMcal P}_i = (p_1,\ldots,p_{i-1},p_{i+1},\ldots,p_N)$.
Proving \ineq{ckwn} for {\em any} quantum system in arbitrary
dimension, would definitely fill the square; it appears though as a
formidable task to be achieved for a computable entanglement
measure. It is known in fact that squashed entanglement
\cite{Squashed}, \eq{E:squash}, is monogamous for arbitrary
partitions of arbitrary-dimensional systems \cite{KoashiWinter}, yet
its impracticality renders this result of limited relevance.
However, partial encouraging results have been recently obtained
which directly generalize the pioneering work of CKW.

Osborne and Verstraete have shown that the generalized monogamy
inequality \pref{ckwn} holds true for any (pure or mixed) state of a
system of $N$ qubits \cite{osborne}, proving a longstanding
conjecture due to CKW themselves \cite{CKW}. Again, the entanglement
has to be measured by the tangle $\tau$. This is an important
result; nevertheless, one must admit that, if more than three
parties are concerned, it is not so obvious why all the bipartite
entanglements should be decomposed only with respect to a single
elementary subsystem. One has in fact an exponentially increasing
number of ways to arrange blocks of subsystems and to construct
multiple splittings of the whole set of parties, across which the
bipartite (or, even more intriguingly, the multipartite)
entanglements can be compared. This may be viewed as a third,
multifolded axis in our `geometrical' description of the possible
generalizations of \ineq{ckw3}. Leaving aside in the present context
this intricate plethora of additional situations, we stick to the
monogamy constraint of \ineq{ckwn}, obtained decomposing the
bipartite entanglements with respect to a single particle, while
keeping in mind that for more than three particles the residual
entanglement emerging from \ineq{ckwn} is not necessarily {\em the}
measure of multipartite entanglement. Rather, it properly quantifies
the  entanglement not stored in couplewise correlations, and thus
finds interesting applications for instance in the study of quantum
phase transitions and criticality in spin systems
\cite{quphosborne,FazioNAT,verrucchi}.

\subsection{Entanglement sharing among qudits}

The first problem one is faced with when trying to investigate the
sharing of quantum correlations in higher dimensional systems is to
find the correct measure for the quantification of bipartite
entanglement. Several approaches to generalize Wootters' concurrence
and/or tangle have been developed \cite{concext,mintert}. In the
present context, Yu and Song \cite{cinesi} have recently established
a CKW-like monogamy inequality for an arbitrary number of qudits,
employing an entanglement quantifier which is a lower bound to the
tangle for any finite $d$. They define the tangle for mixed states
as  the convex-roof extension \eq{mixtangle} of the linear entropy
of entanglement $E_L$ for pure states. Moreover, the authors claim
that the corresponding residual tangle is a proper measure of
multipartite entanglement. Let us remark however that, at the
present stage in the theory of entanglement sharing, trying to make
sense of a heavy mathematical framework (within which, moreover, a
proof of monotonicity of the $N$-way tangle under LOCC has not been
established yet for $N >3$, not even for qubits) with little, if
any, physical insight, is likely not worth trying. Probably the CKW
inequality is interesting not because of the multipartite measure it
implies, but because it embodies a quantifiable trade-off between
the distribution of bipartite entanglement.

In this respect, it seems relevant to address the following
question, raised by Dennison and Wootters \cite{qudits}. One is
interested in computing the maximum possible bipartite entanglement
between {\em any} couple of parties, in a system of three or more
qudits, and in comparing it with the entanglement capacity $\log_2
d$ of the system. Their ratio $\varepsilon$ would provide an
immediate quantitative bound on the shareable entanglement, stored
in couplewise correlations. Results obtained for $d=2$, $3$ and $7$
(using the entanglement of formation) suggest for three qudits a
general trend of increasing $\varepsilon$ with increasing $d$
\cite{qudits}. While this is only a preliminary analysis, it raises
intriguing questions, pushing the interest in entanglement sharing
towards infinite-dimensional systems. In fact, if $\varepsilon$
saturated to $1$ for $d \rightarrow \infty$, this would entail the
really counterintuitive result that entanglement could be freely
shared in this limit! We notice that, being the entanglement
capacity infinite for $d \rightarrow \infty$, $\varepsilon$ vanishes
if the maximum  couplewise entanglement is not infinite. And this is
the case, because again an infinite shared entanglement between two
two-party reductions would allow perfect $1 \rightarrow 2$
telecloning \cite{telecloning} exploiting Einstein-Podolsky-Rosen
(EPR) \cite{EPR35} correlations, but this is forbidden by quantum
mechanics.

Nevertheless, the study of entanglement sharing in continuous
variable systems yields surprising consequences, as we will show in
Part \ref{PartMulti} of this Dissertation. We will indeed define
proper infinite-dimensional analogues of the tangle
\cite{contangle,hiroshima}, and establish the general monogamy
inequality \pref{ckwn} on entanglement sharing for all $N$-mode
Gaussian states distributed among $N$ parties \cite{hiroshima}. An
original, possibly ``promiscuous'' structure of entanglement sharing
in Gaussian states with some symmetry constraints will be also
elucidated \cite{contangle,3mpra,3mj,unlim}.

}

\chapter{Gaussian states: structural properties} \label{ChapGauss}
{\sf

In this Chapter we will recall the main definitions and set up our
notation for the mathematical treatment of Gaussian states of
continuous variable systems. Some of our results concerning the
evaluation of entropic measures for Gaussian states \cite{extremal}
and the reduction of Gaussian covariance matrices into standard
forms under local operations \cite{sformato} will be included here
as well.

\section{Introduction to continuous variable
systems}\label{secIntroCV}
 A continuous variable (CV) system \cite{brareview,eisplenio,COVAQIAL} of $N$ canonical bosonic modes
is described by a Hilbert space
\be\label{hilfok}\hh=\bigotimes_{k=1}^{N} \hh_{k}\ee resulting from
the tensor product structure of infinite-dimensional Fock spaces
$\hh_{k}$'s. One can think for instance to the quantized
electromagnetic field, whose Hamiltonian describes a system of $N$
harmonic oscillators, the {\em modes} of the field,
\begin{equation}\label{CV:Ham}
\ham = \sum_{k=1}^N \hbar \omega_k \left(\adj{\op{a}}_k\op{a}_k +
\frac12\right)\,.
\end{equation}
Here $\op{a}_k$ and $\adj{\op{a}}_k$ are the  annihilation and
creation operators of a photon in mode $k$ (with frequency
$\omega_k$), which satisfy the bosonic commutation relation
\begin{equation}\label{CV:comm}
\comm{\op{a}_k}{\adj{\op{a}}_{k'}}=\delta_{kk'}\,,\quad
\comm{\op{a}_k^}{\op{a}_{k'}}=\comm{\adj{\op{a}}_k}{\adj{\op{a}}_{k'}}=0\,.
\end{equation}
From now on we will assume for convenience natural units with
$\hbar=2$. The corresponding quadrature phase operators (position
and momentum)  for each mode are defined as
\begin{eqnarray}
  \hat q_{k} &=& (\op a_{k}+\op a^{\dag}_{k})\,, \label{CV:q}\\
  \hat p_{k} &=& (\op a_{k}-\op a^{\dag}_{k})/i \label{CV:p}
\end{eqnarray}
We can group together the canonical operators in the vector
\be\label{CV:R}
\hat{R}=(\hat{q}_1,\hat{p}_1,\ldots,\hat{q}_N,\hat{p}_N)\T\,,\ee
which enables us to write in a compact form the  bosonic commutation
relations between the quadrature phase operators, \be
[\hat{R}_k,\hat{R}_l]=2 i\Omega_{kl} \; ,\label{ccr}\ee where
$\Omega$ is the symplectic form \be
\Omega=\bigoplus_{k=1}^{N}\omega\, , \quad \omega=
\left(\begin{array}{cc}
0&1\\
-1&0
\end{array}\right)\, . \label{symform}
\ee

The space $\hh_k$ is spanned by the Fock basis $\{\ket{n}_k\}$ of
eigenstates of the number operator $\hat{n}_k = \hat a_k^{\dag}\hat
a_k$, representing the Hamiltonian of the non-interacting mode via
\eq{CV:Ham}. The Hamiltonian of any mode is bounded from below, thus
ensuring the stability of the system, so that for any mode a vacuum
state $\ket{0}_k\in \hh_k$ exists, for which $\hat a_k\ket{0}_k=0$.
The vacuum state of the global Hilbert space will be denoted by
$\ket{0}=\bigotimes_k \ket{0}_k$. In the single-mode Hilbert space
${\mathcal H}_k$, the eigenstates of $\hat a_k$ constitute the
important set of {\em coherent states} \cite{BWallsMilburn}, which
is overcomplete in $\hh_k$. Coherent states result from applying the
single-mode Weyl displacement operator $\hat D_k$ to the vacuum
$\ket{0}_k$, $\ket{\alpha}_k= \hat D_k(\alpha)\ket{0}_k$, where
\be\label{CV:Weyl}\hat D_k(\alpha)=\,{\rm e}^{\alpha \hat
a_k^{\dag}-\alpha^{*}\hat a_k}\,,\ee and the coherent amplitude
$\alpha\in{\mathbbm C}$ satisfies $\hat
a_k\ket{\alpha}_k=\alpha\ket{\alpha}_k$. In terms of the Fock basis
of mode $k$ a coherent state reads
\be\label{CV:coh}\ket{\alpha}_k=\,{\rm e}^{-\frac12|\alpha|^2}
\sum_{n=1}^{\infty}\frac{\alpha^{n}}{\sqrt{n!}}\ket{n}_k\,.\ee
Tensor products of coherent states of different modes are obtained
by applying the $N$-mode Weyl operators $\hat D_{\xi}$ to the global
vacuum $\ket{0}$. For future convenience, we define the operators
$\op D_{\xi}$ in terms of the canonical operators $\hat{R}$, \be
\hat D_{\xi} = \,{\rm e}^{i\hat{R}^{\sf T} \Omega \xi}\, , \quad
{\rm with} \quad\xi\in {\mathbbm R}^{2N} \; . \ee One  has then
$\ket{\xi}=\hat D_{\xi}\ket{0}$, which entails $\hat
a_k\ket{\xi}=(\xi_k+i\xi_{k+1})\ket{\xi}$.

\subsection{Quantum phase-space picture}

The states of a CV system are the set of positive trace-class
operators $\{\varrho\}$ on the Hilbert space $\hh$, \eq{hilfok}.
However, the complete description of any quantum state $\varrho$ of
such an infinite-dimensional system can be provided by one of its
$s$-ordered {\em characteristic functions} \cite{barnett} \be \chi_s
(\xi) = \,{\rm Tr}\,[\varrho \hat D_{\xi}] \,{\rm e}^{s\|\xi\|^2/2}
\; , \label{cfs} \ee with $\xi\in\R^{2N}$, $\|\cdot\|$ standing for
the Euclidean norm of $\R^{2N}$. The vector $\xi$ belongs to the
real $2N$-dimensional space $\Gamma=({\mathbbm R}^{2N},\Omega)$,
which is called {\em phase space}, in analogy with classical
Hamiltonian dynamics. One can see from the definition of the
characteristic functions that in the phase space picture, the tensor
product structure is replaced by a direct sum structure, so that the
$N$-mode phase space is $\Gamma = \bigoplus_k \Gamma_k$, where
$\Gamma_k=({\mathbbm R}^{2},\omega)$ is the local phase space
associated with mode $k$.

The family of characteristic functions is in turn related, via
complex Fourier transform, to the {\em quasi-probability
distributions} $W_s$, which constitute another set of complete
descriptions of the quantum states \be
W_s(\xi)=\frac{1}{\pi^2}\int_{{\mathbbm R}^{2N}}\kappa
\chi_s(\kappa) \,{\rm e}^{i\kappa^{\sf T} \Omega \xi}\,{\rm d}^{2N}
\, . \label{qps} \ee As well known, there exist states for which the
function $W_s$ is not a regular probability distribution for any
$s$, because it can in general be singular or assume negative
values. Note that the value $s=-1$ corresponds to the Husimi
`Q-function' \cite{Husimi40}
$W_{-1}(\xi)=\bra{\xi}\varrho\ket{\xi}/\pi$ and thus always yields a
regular probability distribution. The case $s=0$ corresponds to the
so-called `Wigner function' \cite{Wigner32}, which will be denoted
simply by $W$. Likewise, for the sake of simplicity, $\chi$ will
stand for the symmetrically ordered characteristic function
$\chi_0$. Finally, the most singular case $s=1$ brings to the
celebrated `P-representation', which was introduced, independently,
by Glauber \cite{Glauber63} and Sudarshan \cite{Sudarshan63}.

The quasiprobability distributions of integer order $W_{-1}$, $W_0$
and $W_1$ are deeply related to, respectively, the antinormally
ordered, symmetrically ordered and normally ordered expressions of
operators. More precisely, if the operator $\hat{O}$ can be
expressed as $\hat{O}= f(\hat a_k,\hat a^{\dag}_k)$ for
$k=1,\ldots,N$, where $f$ is a, say, symmetrically ordered function
of the field operators, then one has \cite{cahillglau1,cahillglau2}
\[
\,{\rm Tr}[\varrho \hat{O}] =
\int_{\R^{2N}}W_0(\kappa)\bar{f}(\kappa) \,{\rm d}^{2N}\kappa \, ,
\]
where
$\bar{f}(\kappa)=f(\kappa_k+i\kappa_{k+1},\kappa_k-i\kappa_{k+1})$
and $f$ takes the same form as the operatorial function previously
introduced. The same relationship holds between $W_{-1}$ and the
antinormally ordered expressions of the operators, and between
$W_{1}$ and the normally ordered ones. We also recall that the
normally ordered function of a given operator is provided by its
Wigner representation. This entails the following equalities for the
trace \be 1 = \tr\varrho = \int_{\R^{2N}} W(\kappa) \,{\rm
d}^{2N}\kappa = \chi(0) \,\label{wigtr} , \ee and for the purity
 \be \mu = \tr\varrho^2 = \int_{\R^{2N}}
W^2(\kappa) \,{\rm d}^{2N}\kappa = \int_{\R^{2N}} |\chi(\xi)|^2
\,{\rm d}^{2N}\xi    \,\label{wigpr} , \ee of a state $\varrho$,
which will come handy in the following. The various Appendixes of
Ref.~\cite{barnett} contain other practical relations between the
relevant properties of a density matrix and the corresponding
phase-space description.

The (symmetric) Wigner function can be written as follows in terms
of the (non-normalized) eigenvectors $\ket{x}$ of the quadrature
operators $\{\hat q_j\}$ (for which $\hat q_j \ket{x}=q_j \ket{x}$,
$x\in\rr^{N}$, for $j=1,\ldots,N$) \cite{Simon00} \be W(x,p) =
\frac{1}{\pi^{N}} \int_{\rr^{N}} \bra{x-x'}\varrho\ket{x+x'}\,{\rm
e}^{ix'\cdot p} \,{\rm d}^{N}x' \: , \quad x,p\in\rr^{N}\,.
\label{wig} \ee From an operational point of view, the Wigner
function admits a clear interpretation in terms of homodyne
measurements \cite{sculzub}: the marginal integral of the Wigner
function over the variables $p_1,\ldots,p_N,x_1,\ldots,x_{N-1}$,
$$\int_{\rr^{2N-1}}W(x,p)\,{\rm d}^{N}p\,{\rm d}\,x_1\ldots{\rm d}\,x_{N-1}\,,$$
gives the probability of the results of homodyne detections on the
remaining quadrature $x_{N}$ \cite{francamentemeneinfischio} (for
more details see Sec.~\ref{secFrancesi}).

Table \ref{CVtab} is a useful scheme to summarize the properties of
quantum phase spaces. It will be completed in the next Section,
where powerful tools special to Gaussian states and Gaussian
operations will be reviewed.

\begin{table}[t!]   {\rm
\begin{tabular}{c||c|c}
%  \hline
   & Hilbert space $\hh$ & Phase space $\Gamma$ \\
  \hline \hline
   dimension & $\infty$ & $2N$ \\ \hline
   structure & $\bigotimes$ & $\bigoplus$ \\ \hline
   description & $\ro$ & $\chi_s,\,W_s$ \\ \hline
 \end{tabular}}
 \caption{Schematic comparison between Hilbert-space  and  phase-space \newline  pictures
 for $N$-mode continuous variable systems.}
\label{CVtab}\end{table}

\section{Mathematical description of Gaussian states}

The set of {\em Gaussian states} is, by definition, the set of
states with Gaussian characteristic functions and quasi-probability
distributions on the multimode quantum phase space.  Such states are
at the heart of information processing in CV systems
\cite{adebook,benarev,brareview} and are the main subject of this
Dissertation.

\subsection{Covariance matrix formalism}

From the definition it follows that  a Gaussian state $\varrho$ is
completely characterized by the first and second statistical moments
of the quadrature field operators, which will be denoted,
respectively, by the vector of first moments $\bar R =
\left(\langle\hat R_{1} \rangle,\langle\hat
R_{1}\rangle,\ldots,\langle\hat R_{N}\rangle, \langle\hat
R_{n}\rangle\right)$ and by the covariance matrix (CM) $\sig$ of
elements
\begin{equation}
\sig_{ij} = \frac{1}{2}\langle \hat{R}_i \hat{R}_j + \hat{R}_j
\hat{R}_i \rangle - \langle \hat{R}_i \rangle \langle \hat{R}_j
\rangle  \label{covariance}\,.
\end{equation}

First moments can be arbitrarily adjusted by local unitary
operations, namely displacements in phase space, \ie applications of
the single-mode Weyl operator \eq{CV:Weyl} to locally re-center the
reduced Gaussian corresponding to each single mode\footnote{\sf
Recall that the reduced state obtained from a Gaussian state by
partial tracing over a subset of modes is still Gaussian.}. Such
operations leave any informationally relevant property, such as
entropy and entanglement, invariant: therefore, first moments are
unimportant to the whole scope of our analysis and from now on
(unless explicitly stated) we will set them to $0$ without any loss
of generality.

With this position, the Wigner function of a Gaussian state can be
written as follows in terms of phase-space quadrature variables
\begin{equation}
W(R)=\frac{\,{\rm
e}^{-\frac{1}{2}R\boldsymbol{\sigma}^{-1}R\T}}{\pi\sqrt{{\rm
Det}\,\boldsymbol{\sigma}}}{\:,}\label{wigner}
\end{equation}
where $R$ stands for the real phase-space vector
$(q_{1},p_{1},\ldots,q_{N},p_{N})\in\Gamma$. Despite the infinite
dimension of the Hilbert space in which it lives, a {\em complete}
description of an arbitrary Gaussian state (up to local unitary
operations) is therefore encoded in the $2N \times 2N$ CM $\sig$,
which in the following will be assumed indifferently to denote the
matrix of second moments of a Gaussian state, or the state itself.
In the formalism of statistical mechanics, the CM elements are the
two-point truncated correlation functions between the $2N$ canonical
continuous variables. We notice also that the entries of the CM can
be expressed as energies by multiplying them by the quantity $\hbar
\omega_k$, where $\omega_k$ is the frequency of each mode $k$, in
such a way that $\tr{\sig}$ is related to the mean energy of the
state, \ie the average of the non-interacting Hamiltonian
\eq{CV:Ham}. This mean energy is generally unbounded in CV systems.

As the real $\sig$ contains the complete locally-invariant
information on a Gaussian state, we can expect some constraints to
exist to be obeyed by any {\em bona fide} CM, reflecting in
particular the requirements of positive-semidefiniteness of the
associated density matrix $\varrho$. Indeed, such condition together
with the canonical commutation relations
 imply
\begin{equation}
\sig+ i\Omega\ge 0 \; , \label{bonfide}
\end{equation}
\ineq{bonfide} is the only necessary and sufficient constraint the
matrix $\sig$ has to fulfill to be the CM corresponding to a
physical Gaussian state \cite{simon87,simon94}. More in general, the
previous condition is necessary for the CM of {\em any}, generally
non-Gaussian, CV state (characterized in principle by the moments of
any order). We note that such a constraint implies $\sig\ge0$.
\ineq{bonfide} is the expression of the uncertainty principle on the
canonical operators in its strong, Robertson--Schr\"odinger form
\cite{robertson30,schrodinger30,serafozziprl}.

For future convenience, let us define and write down the CM
$\sig_{1\ldots N}$ of an $N$-mode Gaussian state in terms of two by
two submatrices as \be \sig_{1\ldots N} = \left(\begin{array}{cccc}
\sig_{1} & \eps_{1,2}\; & \cdots & \eps_{1,N} \\
&&&\\
\eps_{1,2}^{\sf T}\; & \ddots & \ddots & \vdots \\
&&&\\
\vdots & \ddots & \ddots & \eps_{N-1,N} \\
&&&\\
\eps_{1,N}^{\sf T}& \cdots & \eps_{N-1,N}^{\sf T} & \sig_{N} \\
\end{array}\right) \; . \label{CM}
\ee Each diagonal block $\sig_k$ is respectively the local CM
corresponding to the reduced state of mode $k$, for all
$k=1,\ldots,N$. On the other hand, the off-diagonal matrices
$\eps_{i,j}$ encode the intermodal correlations (quantum and
classical) between subsystems $i$ and $j$. The matrices $\eps_{i,j}$
all vanish for a product state.

In this preliminary overview, let us just mention an important
instance of two-mode Gaussian state, the {\em two-mode squeezed
state} $\ket{\psi^{sq}}_{i,j}=\hat U_{i,j}(r)
\left(\ket{0}_i\!\otimes\ket{0}_j\right)$ with squeezing factor  $r
\in \R$, where  the (phase-free) two-mode squeezing operator is
given by
\begin{equation}\label{tmsU}
\hat U_{i,j}(r) = \exp \left[-\frac{r}{2} (\hat {a}_i^\dag \hat
{a}_j^\dag -\hat {a}_i \hat {a}_j ) \right]\,,
\end{equation}
In the limit of infinite squeezing ($r\to \infty )$, the state
approaches the ideal Einstein-Podolsky-Rosen (EPR) state
\cite{EPR35}, simultaneous eigenstate of total momentum and relative
position of the two subsystems, which thus share {\em infinite
entanglement}. The EPR state is unnormalizable and unphysical:
two-mode squeezed states, being arbitrarily good approximations of
it with increasing squeezing, are therefore key resources for
practical implementations of CV quantum information protocols
\cite{brareview} and play a central role in the subsequent study of
the entanglement properties of general Gaussian states. A two-mode
squeezed state with squeezing degree $r$ (also known in optics as
{\em twin-beam} state \cite{BWallsMilburn}) will be described by a
CM
\begin{equation}\label{tms}
\sig^{sq}_{i,j}(r)= \left(\begin{array}{cccc}
\cosh(2r)&0&\sinh(2r)&0\\
0&\cosh(2r)&0&-\sinh(2r)\\
\sinh(2r)&0&\cosh(2r)&0\\
0&-\sinh(2r)&0&\cosh(2r)
\end{array}\right)\! .
\end{equation}

The CM of $N$-mode coherent states (including the vacuum) is instead
the $2N \times 2N$ identity matrix.

\subsection{Symplectic operations} \label{SecSympl}

A major role in the theoretical and experimental manipulation of
Gaussian states is played by unitary operations which preserve the
Gaussian character of the states on which they act. Such operations
are all those generated by Hamiltonian terms at most quadratic in
the field operators.  As a consequence of the Stone-Von Neumann
theorem, the so-called {\em metaplectic} representation
\cite{simon94} entails that any such unitary operation at the
Hilbert space level corresponds, in phase space, to a symplectic
transformation, {\ie}to a linear transformation $S$ which preserves
the symplectic form $\Omega$, so that \be\label{symplectic}S\T\Omega
S = \Omega\,.\ee Symplectic transformations on a $2N$-dimensional
phase space form the (real) symplectic group $\sy{2N}$
\cite{pramana}. Such transformations act linearly on first moments
and by congruence on CMs, $\sig\mapsto S \sig S\T$. \eq{symplectic}
implies $\det{S}=1$, $\forall\,S\in\sy{2N}$. Ideal beam-splitters,
phase shifters and squeezers are all described by some kind of
symplectic transformation (see \eg \cite{francamentemeneinfischio}).
For instance, the two-mode squeezing operator \eq{tmsU} corresponds
to the symplectic transformation
\begin{equation}\label{tmsS}
S_{i,j}(r)=\left(\begin{array}{cccc}
\cosh r&0&\sinh r&0\\
0&\cosh r&0&-\sinh r\\
\sinh r&0&\cosh r&0\\
0&-\sinh r&0&\cosh r
\end{array}\right)\, ,
\end{equation}
where the matrix is understood to act on the couple of modes $i$ and
$j$. In this way, the two-mode squeezed state, \eq{tms}, can be
obtained as $\sig^{sq}_{i,j}(r) = S_{i,j}(r) \id S_{i,j}\T(r)$
exploiting the fact that the CM of the two-mode vacuum state is the
$4 \times 4$ identity matrix.

Another common symplectic operation is the ideal (phase-free) {\em
beam-splitter}, whose action $\hat{B}_{i,j}$ on a pair of modes $i$
and $j$ is defined as
\begin{equation}\label{bsplit}
\hat{B}_{i,j}(\theta):\left\{
\begin{array}{l}
\hat a_i \rightarrow \hat a_i \cos\theta + \hat a_j\sin\theta \\
\hat a_j \rightarrow \hat a_i \sin\theta - \hat a_j\cos\theta \\
\end{array} \right.\,,
\end{equation}
with $\hat a_l$ being the annihilation operator of mode $k$. A
beam-splitter with transmittivity $\tau$ corresponds to a rotation
of $\theta = \arccos\sqrt{\tau}$ in phase space ($\theta=\pi/4$
amounts to a balanced 50:50 beam-splitter, $\tau=1/2$), described by
a symplectic transformation
\begin{equation}\label{bbs}
B_{i,j}(\tau)=\left(
\begin{array}{cccc}
 \sqrt{\tau } & 0 & \sqrt{1 - \tau } & 0 \\
 0 & \sqrt{\tau } & 0 & \sqrt{1 - \tau } \\
 \sqrt{1 - \tau } & 0 & - \sqrt{\tau } & 0 \\
 0 & \sqrt{1 - \tau } & 0 & - \sqrt{\tau }
\end{array}
\right)\,.
\end{equation}

Single-mode symplectic operations are easily retrieved as well,
being just combinations of planar (orthogonal) rotations and of
single-mode squeezings of the form \be\label{sqz} S_j(r)
=\diag(\,{\rm e}^{r},\,{\rm e}^{-r})\,,\ee acting on mode $j$, for
$r>0$. In this respect, let us mention that the two-mode squeezed
state \eq{tms} can also be obtained indirectly, by individually
squeezing two single modes $i$ and $j$ in orthogonal quadratures,
and by letting them interfere at a 50:50 beam-splitter. The total
transformation realizes what we can call a ``twin-beam box'',
\begin{equation}\label{twin}
T_{i,j}(r) =B_{i,j}(1/2) \cdot (S_{i}(r) \oplus S_j(-r))\,,
\end{equation}
which, if applied to two uncorrelated vacuum modes $i$ and $j$
(whose initial CM is the identity matrix), results in the production
of a pure two-mode squeezed Gaussian state with CM exactly equal to
$T_{i,j}(r) T_{i,j}\T(r) \equiv \sig^{sq}_{i,j}(r)$ from \eq{tms}.

In general, symplectic transformations in phase space are generated
by exponentiation of matrices written as $J\Omega$, where $J$ is
antisymmetric \cite{pramana}. Such generators can be symmetric or
antisymmetric. The operations $B_{ij}(\tau)$, \eq{bbs}, generated by
antisymmetric operators are orthogonal and, acting by congruence on
the CM $\sig$, preserve the value of $\tr{\sig}$. Since $\tr{\sig}$
gives the contribution of the second moments to the average of the
Hamiltonian $\bigoplus_k \hat a_k^{\dag}\hat a_k$, these
transformations are said to be {\em passive} (they belong to the
compact subgroup of $\sy{2N}$). Instead, operations $S_{i,j}(r)$,
\eq{tmsS}, generated by symmetric operators, are not orthogonal and
do not preserve $\tr{\sig}$ (they belong to the non-compact subgroup
of $\sy{2N}$). This mathematical difference between squeezers and
phase-space rotations accounts, in a quite elegant way, for the
difference between {\em active} (energy consuming) and passive
(energy preserving) optical transformations \cite{passive}.

Let us remark that {\em local} symplectic operations belong to the
group ${\sy{2}}^{\oplus N}$. They correspond, on the Hilbert space
level, to tensor products of unitary transformations, each acting on
the space of a single mode. It is useful to notice that the
determinants of each $2\times2$ submatrix of a $N$-mode CM, \eq{CM},
are {\em all} invariant under local symplectic operations $S \in
{\sy{2}}^{\oplus N}$.\footnote{\sf The invariance of the
off-diagonal terms $\det\eps_{i,j}$ follows from Binet's formula for
the determinant of a matrix \cite{bathia}, plus the fact that any
symplectic transformation $S$ has $\det S = 1$.} This mathematical
property reflects the physical requirement that both marginal
informational properties, and correlations between the various
individual subsystems, cannot be altered by local operations only.

\subsubsection{Symplectic eigenvalues and
invariants}\label{SecWilly} A crucial symplectic transformation is
the one realizing the decomposition of a Gaussian state in normal
modes. Through this decomposition, thanks to Williamson theorem
\cite{williamson36}, the CM of a $N$-mode Gaussian state can always
be written in the so-called Williamson normal, or diagonal form
\begin{equation}
\sig=S\T \gr\nu S \; , \label{willia}
\end{equation}
where $S\in Sp_{(2N,\mathbb{R})}$ and $\gr\nu$ is the CM
\begin{equation}
\gr{\nu}=\bigoplus_{k=1}^{N}\left(\begin{array}{cc}
\nu_k&0\\
0&\nu_k
\end{array}\right) \, , \label{therma}
\end{equation}
corresponding to a tensor product state with a diagonal density
matrix $\varrho^{_\otimes}$ given by \be
\varrho^{_\otimes}=\bigotimes_{k}
\frac{2}{\nu_{k}+1}\sum_{n=0}^{\infty}\left(
\frac{\nu_{k}-1}{\nu_{k}+1}\right)\ket{n}_{k}{}_{k}\bra{n}\; ,
\label{thermas} \ee where $\ket{n}_k$ denotes the number state of
order $n$ in the Fock space $\hh_{k}$. In the Williamson form, each
mode with frequency $\omega_k$ is a Gaussian state in  thermal
equilibrium at a temperature $T_k$, characterized by an average
number of thermal photons $\bar n_k$ which obeys  Bose-Einstein
statistics,
\begin{equation}\label{temperature}
\bar n_k = \frac{\nu_k-1}{2} =
\frac{1}{\exp\left(\frac{\hbar\omega_k}{k_B T_k}\right)-1}\,.
\end{equation}

The $N$ quantities $\nu_{k}$'s form the {\em symplectic spectrum} of
the CM $\sig$, and are invariant under the action of global
symplectic transformations on the matrix $\sig$. The symplectic
eigenvalues can be computed as the orthogonal eigenvalues of the
matrix $|i\Omega\sig|$ \cite{SeralePHD} and are thus determined by
$N$ invariants of the characteristic polynomial of such a matrix
\cite{serafozziprl}. One global symplectic invariant is simply the
determinant of the CM (whose invariance is a consequence of the fact
that ${\rm Det}\,S=1$ $\forall S\in \sy{2N}$), which once computed
in the Williamson diagonal form reads
\begin{equation}\label{detsigma}
\det\sig = \prod_{k=1}^N \nu_k^2\,.
\end{equation}
Another important invariant under global symplectic operations is
the so-called {\em seralian} $\Delta$ \cite{polacchi}, defined as
the sum of the determinants of all $2 \times 2$ submatrices of a CM
$\sig$, \eq{CM}, which can be readily computed in terms of its
symplectic eigenvalues as
\begin{equation}\label{seralian}
\Delta(\sig) = \sum_{k=1}^N \nu_k^2\,.
\end{equation}
The invariance of $\Delta_{\sig}$ in the multimode case
\cite{serafozziprl} follows from its invariance in the case of
two-mode states, proved in Ref.~\cite{SymplecticInvariants}, and
from the fact that any symplectic transformation can be decomposed
as the product of two-mode transformations \cite{agarwal94}.

\subsubsection{Symplectic representation of the uncertainty
principle}\label{SecSympHeis}

The symplectic eigenvalues $\nu_{k}$ encode essential information on
the Gaussian state $\gr{\sigma}$ and provide powerful, simple ways
to express its fundamental properties \cite{SymplecticInvariants}.
For instance, let us consider the uncertainty relation
\pref{bonfide}. Since the inverse of a symplectic operation  is
itself symplectic, one has from \eq{symplectic}, ${S^{-1}}\T
{\Omega}S^{-1}= {\Omega}$, so that \ineq{bonfide} is {\em
equivalent} to $\gr{\nu}+i{\Omega}\ge 0$. In terms of the symplectic
eigenvalues $\nu_{k}$ the uncertainty relation then simply reads \be
{\nu}_{k}\ge1 \; . \label{sympheis} \ee Inequality \pref{sympheis}
is completely equivalent to the uncertainty relation \pref{bonfide}
provided that the CM $\sig$ satisfies $\sig\ge 0$.

We can, without loss of generality, rearrange the modes of a
$N$-mode state such that the corresponding symplectic eigenvalues
are sorted in ascending order
$$ \nu_- \equiv \nu_1 \le \nu_2 \le \ldots \le \nu_{N-1} \le \nu_N \equiv \nu_+\,.$$ With this
notation, the uncertainty relation reduces to $\nu_- \ge 1$. We
remark that the full saturation of the uncertainty principle can
only be achieved by {\em pure} $N$-mode Gaussian states, for which
$$\nu_i=1\,\,\forall i=1,\ldots, N\,,$$ meaning that the Williamson
normal form of any pure Gaussian state is the vacuum $\ket0$ of the
$N$-mode Hilbert space $\hh$. Instead, mixed states such that
$\nu_{i\le k}=1$ and $\nu_{i>k}>1$, with $1\le k\le N$, only
partially saturate the uncertainty principle, with partial
saturation becoming weaker with decreasing $k$. Such states are
minimum-uncertainty mixed Gaussian states in the sense that the
phase quadrature operators of the first $k$ modes satisfy the
Robertson-Schr\"odinger minimum uncertainty, while for the remaining
$N-k$ modes the state indeed contains some additional thermal
correlations which are responsible for the global mixedness of the
state.

 We can  define in all generality the {\em symplectic rank}
$\aleph$ of a CM $\sig$ as the number of its symplectic eigenvalues
different from $1$, corresponding to the number of non-vacua normal
modes \cite{generic}. A Gaussian state is pure if and only if
$\aleph=0$, while for mixed $N$-mode states one has $1\le \aleph \le
N$ according to their degree of partial minimum-uncertainty
saturation. This is in analogy with the standard rank of
finite-dimensional (density) matrices, defined as the number of
non-zero eigenvalues; in that case, only pure states
$\varrho=\ketbra{\psi}{\psi}$ have rank $1$, and mixed states have
in general higher rank. As we will now show, {\em all} the
informational properties of Gaussian states can be recast in terms
of their symplectic spectra.

A mnemonic summary of the main structural features of Gaussian
states in the phase-space/CM description (including the definition
of purity given in the next subsection) is provided by Table
\ref{CVtabG}.

\begin{table}[t!]   {\rm
\begin{tabular}{c||c|c}
%  \hline
   & Hilbert space $\hh$ & Phase space $\Gamma$ \\
  \hline \hline
  dimension & $\infty$ & $2N$ \\ \hline
  structure & $\bigotimes$ & $\bigoplus$ \\ \hline
  description & $\ro$ & $\sig$ \\ \hline
  {\em bona fide} & $\ro \ge 0$ & $\sig + i \Omega \ge 0$ \\ \hline
  operations & $\underset{\ro \mapsto U \ro \adj{U}}{\overset{}{U}: \adj U U = \id}$ &
$\underset{\sig \mapsto S \sig S\T}{\overset{}{S}: S\T \Omega S =
\Omega}$
  \\ \hline
  $\overset{^{}}{{\sf spectra}}$ &
  $\underset{0 \le \lambda_k \le 1}{U \ro \adj{U} = {\rm diag}\{\lambda_k\}}$ &
  $\underset{1 \le \nu_k < \infty}{S \sig S\T = {\rm diag}\{\nu_k\}}$
  \\ \hline
  pure states & $\lambda_i = 1,\,\lambda_{j \neq i}=0$ & $\nu_j = 1,\, \forall j=1\ldots N$
  \\ \hline
  purity & $\overset{}{\tr{\ro^2} = \sum_k \lambda_k^2}$ &
  $\overset{}{1/\sqrt{\det{\sig}}=\prod_k \nu_k^{-1}}$ \\ \hline
 \end{tabular}}
 \caption{Schematic comparison between Hilbert-space  and  phase-space  pictures
 for $N$-mode Gaussian states. The first two rows are taken from Table \ref{CVtab} and apply to
 general states of CV systems. The following rows are special to Gaussian states, relying on
 the covariance matrix description and the properties of the symplectic group.}
\label{CVtabG}\end{table}

\section{Degree of information encoded in a Gaussian state}
\label{SecEntroG}

Several entropic measures able to quantify the degree of information
(or lack thereof) encoded in a quantum states and, equivalently, its
degree of purity (or lack thereof) have been introduced in
Sec.~\ref{ParInfo}. Here, based on Ref.~\cite{extremal}, we
illustrate their evaluation for arbitrary Gaussian states.

\subsection{Purity and generalized entropies}
The generalized purities ${\rm Tr}\,\varrho^p$ defined by
\eq{schatten} are invariant under global unitary operations.
Therefore, for any $N$-mode Gaussian state they are only functions
of the symplectic eigenvalues $\nu_{k}$ of $\gr{\sigma}$. In fact, a
symplectic transformation acting on $\gr{\sigma}$ is embodied by a
unitary (trace preserving) operator acting on $\varrho$, so that
${\rm Tr}\,\varrho^{p}$ can be easily computed on the Williamson
diagonal state $\gr\nu$ of \eq{therma}. One obtains \cite{extremal}
\be {\rm Tr}\,\varrho^{p}=\prod_{i=1}^{N}g_{p}(\nu_i)\; ,
\label{pgau} \ee where
\[
g_{p}(x)=\frac{2^p}{(x+1)^p-(x-1)^p} \, .
\]
A first remarkable consequence of \eq{pgau} is that \be
\mu(\varrho)=\frac{1}{\prod_i \nu_{i}}=\frac{1}{\sqrt{{\rm
Det}\,\gr{\sigma}}} \, . \label{purgau} \ee Regardless of the number
of modes, the purity of a Gaussian state is {\em fully determined}
by the global symplectic invariant ${\rm Det}\,\gr{\sigma}$ alone,
\eq{detsigma}. We recall that the purity is related to the linear
entropy $S_L$ via \eq{QM:SL}, which in CV systems simply becomes
$S_L = 1-\mu$. A second consequence of \eq{pgau} is that, together
with Eqs.~\pref{pgen} and \pref{genvneu}, it allows for the
computation of the Von Neumann entropy $S_{V}$, \eq{QM:SV}, of a
Gaussian state $\varrho$, yielding \cite{SymplecticInvariants} \be
S_{V}(\varrho)=\sum_{i=1}^{N}f(\nu_{i}) \; , \label{vneugau} \ee
where \be\label{entfunc} f(x) \equiv
\frac{x+1}{2}\log\left(\frac{x+1}{2}\right)-
\frac{x-1}{2}\log\left(\frac{x-1}{2}\right) \, . \ee Such an
expression for the Von Neumann entropy of a Gaussian state was first
explicitly given in Ref.~\cite{holevo99}. Notice that $S_V$ diverges
on infinitely mixed CV states, while $S_L$ is normalized to $1$. Let
us remark that, clearly, the symplectic spectrum of single-mode
Gaussian states, which consists of only one eigenvalue $\nu_1$, is
fully determined by the invariant ${\rm
Det}\,\gr{\sigma}=\nu_{1}^2$. Therefore, all the entropies $S_{p}$'s
(and $S_{V}$ as well) are just increasing functions of ${\rm
Det}\,\gr{\sigma}$ (\emph{i.e.}~of $S_L$) and induce \emph{the same}
hierarchy of mixedness on the set of one-mode Gaussian states. This
is no longer true for multi-mode states, even for the relevant,
simple instance of two-mode states \cite{extremal}, as we will show
in the following.

Accordingly, for an arbitrary Gaussian state the {\em mutual
information}, \eq{QM:I}, quantifying the total (classical and
quantum) correlations between two subsystems \cite{berry}, can be
computed as well. Namely, for a bipartite Gaussian state with global
CM $\sig_{A|B}$, the mutual information yields
\cite{holevo99,SymplecticInvariants}
\begin{equation}\label{migau}
I(\sig_{A|B}) = S_V(\sig_A) + S_V(\sig_B) - S_V(\sig_{A|B})\,,
\end{equation}
where each Von Neumann entropy can be evaluated from the respective
symplectic spectrum using \eq{vneugau}.

\subsection{Comparison between entropic measures}\label{mmix}
Here we aim to find extremal values of $S_{p}$ (for $p\neq 2$) at
fixed $S_{L}$ in the general $N$-mode Gaussian instance, in order to
quantitatively compare the characterization of mixedness given by
the different entropic measures \cite{extremal}. For simplicity, in
calculations we will employ $\mu$ instead of $S_L$. In view of
Eqs.~\pref{pgau} and \pref{purgau}, the possible values taken by
$S_{p}$ for a given $\mu$ are determined by \bea
(p-1)S_{p}=1-\left(\prod_{i=1}^{N-1}g_{p}(s_i)\right)
g_{p}\left(\frac{1}{\mu \prod_{i=1}^{N-1}s_i}\right) \, ,\label{iter}\\
{\rm with}\quad 1 \le s_i \le\frac{1} {\mu\prod_{i\neq j}s_{j}}\, .
\label{domi} \eea The last constraint on the $N-1$ real auxiliary
parameters $s_i$ is a consequence of the uncertainty relation
\pref{sympheis}. We first focus on the instance $p<2$, in which the
function $S_{p}$ is concave with respect to any $s_{i}$, for any
value of the $s_i$'s. Therefore its minimum with respect to, say,
$s_{N-1}$ occurs at the boundaries of the domain, for $s_{N-1}$
saturating inequality \pref{domi}. Since $S_{p}$ takes the same
value at the two extrema and exploiting $g_{p}(1)=1$, one has \be
(p-1)\min_{s_{N-1}}S_{p}=1- \left(\prod_{i=1}^{N-2}g_{p}(s_i)\right)
g_{p}\left(\frac{1}{\mu \prod_{i=1}^{N-2}s_i}\right) \: . \ee
Iterating this procedure for all the $s_i$'s leads eventually to the
minimum value $S_{p\min}(\mu)$ of $S_p$ at given purity $\mu$, which
simply reads \be\label{spmin}
S_{p\min}(\mu)=\frac{1-g_p\left(\frac{1}{\mu}\right)}{p-1}\, ,\quad
p<2 \, . \ee For $p<2$, the mixedness of the states with minimal
generalized entropies at given purity is therefore concentrated in
one quadrature: the symplectic spectrum of such states is partially
degenerate, with $\nu_{1}=\ldots=\nu_{N-1}=1$ and $\nu_{N}=1/\mu$.
We have identified states of this form as being mixed states of
partial minimum uncertainty.
\par The maximum value $S_{p\max}(\mu)$ is achieved by states
satisfying the coupled transcendental equations \be
g_p\left(\frac{1}{\mu\prod s_i}\right) g'_{p}(s_j)= \frac{1}{\mu
s_{j}\prod s_i}\, g_p(s_j) g'_p\left(\frac{1}{\mu \prod
s_i}\right)\, , \ee where all the products $\prod$ run over the
index $i$ from $1$ to $N-1$, and \be
 g'_p(x)=\frac{-p\,2^p\left[(x+1)^{p-1}-(x-1)^{p-1}\right]}
 {\left[(x+1)^p-(x-1)^p\right]^2}\, .
\ee It is promptly verified that the above two conditions are
fulfilled by states with a completely degenerate symplectic
spectrum: $\nu_{1}=\ldots=\nu_{N}={\mu}^{-1/N}$, yielding
\be\label{spmax} S_{p\max}(\mu)=\frac{1-
g_p\left(\mu^{-\frac{1}{N}}\right)^N}{p-1}\, ,\quad p<2 \, . \ee

\begin{figure}[tb!]
\includegraphics[width=9cm]{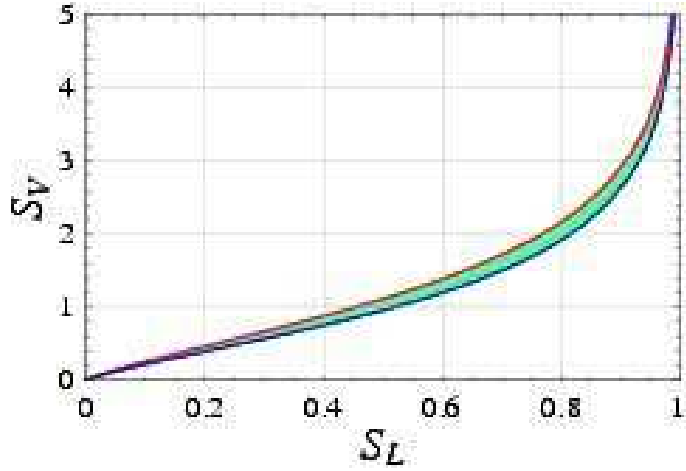}
\caption{Plot of the curves of maximal (red line) and minimal (blue
line) Von Neumann entropy at given linear entropy for two-mode
Gaussian states. All physical states lie in the shaded region
between the two curves.} \label{svvssl}
\end{figure}

The analysis that we carried out for $p<2$ can be straightforwardly
extended to the limit $p\rightarrow 1$, yielding the extremal values
of the Von Neumann entropy for given purity $\mu$ of $N$-mode
Gaussian states. Also in this case the states with maximal $S_{V}$
are those with a completely degenerate symplectic spectrum, while
the states with minimal $S_{V}$ are those with all the mixedness
concentrated in one quadrature. The extremal values $S_{V\min}(\mu)$
and $S_{V\max}(\mu)$ read \bea
S_{V\min}(\mu)&=&f\left(\frac{1}{\mu}\right) \; , \\
&&\nonumber\\
S_{V\max}(\mu)&=&N f\left(\mu^{-\frac1N}\right) \; . \eea The
behaviors of the Von Neumann and of the linear entropies for
two-mode Gaussian states are compared in Fig.~\ref{svvssl}.\par

 The instance $p>2$ can be treated in the same way, with the major
difference that the function $S_{p}$ of \eq{iter} is convex with
respect to any $s_i$ for any value of the $s_i$'s. As a consequence
we have an inversion of the previous expressions: for $p>2$, the
states with minimal $S_{p\min}(\mu)$ at given purity $\mu$ are those
with a fully distributed symplectic spectrum, with \be
S_{p\min}(\mu)=\frac{1- g_p\left(\mu^{-\frac{1}{N}}\right)^N}{p-1}\,
,\quad p>2 \, . \ee On the other hand, the states with maximal
$S_{p\max}$ at given purity $\mu$ are those with a spectrum of the
kind $\nu_{1}=\ldots=\nu_{N-1}=1$ and $\nu_{N}=1/\mu$. Therefore \be
S_{p\max}(\mu)=\frac{1-g_p\left(\frac{1}{\mu}\right)}{p-1}\, ,\quad
p>2 \, . \label{spglems} \ee
\par
The distance $|S_{p\max}-S_{p\min}|$ decreases with increasing $p$
\cite{extremal}. This is due to the fact that the quantity $S_{p}$
carries less information with increasing $p$, and the knowledge of
$\mu$ provides a more precise bound on the value of $S_{p}$.

\section{Standard forms of Gaussian covariance matrices}
\label{SecSFCM}

  We have seen that Gaussian states of $N$-mode CV systems are
special in that they are completely specified by the first and
second moments of the canonical bosonic operators. However, this
already reduced set of parameters (compared to a true
infinite-dimensional one needed to specify a generic non-Gaussian CV
state) contains many redundant degrees of freedom which have no
effect on the entanglement. A basic property of multipartite
entanglement is in fact its invariance under unitary operations
performed locally on the subsystems, \eq{E:Emixlocalinv}. To
describe entanglement efficiently, is thus natural to lighten
quantum systems of the unnecessary degrees of freedom adjustable by
local unitaries, and to classify states according to {\em standard
forms} representative of local-unitary equivalence classes
\cite{linden}. When applied to Gaussian states, the freedom arising
from the local invariance immediately rules out the vector of first
moments, as already mentioned. One is then left with the
$2N(2N+1)/2$ real parameters constituting the symmetric CM of the
second moments, \eq{CM}.

 In this Section  we study the action of local unitaries  on
a general CM of a multimode Gaussian state. We compute the minimal
number of parameters which completely characterize Gaussian states,
up to local unitaries. The set of such parameters will contain {\em
complete} information about any form of bipartite or multipartite
entanglement in the corresponding Gaussian states. We give
accordingly the standard form of the CM of a completely general
$N$-mode Gaussian state. We moreover focus on pure states, and on
(generally mixed) states with strong symmetry constraints, and for
both instances we investigate the further reduction of the minimal
degrees of freedom, arising due to the additional constraints on the
structure of the state. The analysis presented here will play a key
role in the investigation of bipartite and multipartite entanglement
of Gaussian states, as presented in the next Parts.

\subsection{Mixed states}\label{sform}

Here we discuss the standard forms of generic {\em mixed} $N$-mode
Gaussian states under local, single-mode symplectic operations,
following Ref.~\cite{sformato}. Let us express the CM $\sig$ as in
\eq{CM}, in terms of $2\times2$ sub-matrices $\sig_{jk}$, defined by
$$
\sig \equiv \left(\begin{array}{ccc}
\sig_{11} &\cdots &\sig_{1N}\\
\vdots & \ddots &  \vdots \\
\sig_{1N}^{\sf T} & \cdots  & \sig_{NN}
\end{array}\right) \; \
$$
each sub-matrix describing either the local CM of mode $j$
($\sig_{jj}$) or the correlations between the pair of modes $j$ and
$k$ ($\sig_{jk}$).

Let us recall the Euler decomposition (see Appendix \ref{SecEuler})
of a generic single-mode symplectic transformation
$S_1({\vartheta',\vartheta'',z})$, \be\label{eulersingle}
S_{1}({\vartheta',\vartheta'',z}) = \left(\begin{array}{cc}
\cos{\vartheta'} & \sin{\vartheta'} \\
-\sin{\vartheta'} & \cos{\vartheta'}
\end{array}\right)
\left(\begin{array}{cc}
z & 0 \\
0 & \frac1z
\end{array}\right)
\left(\begin{array}{cc}
\cos{\vartheta''} & \sin{\vartheta''} \\
-\sin{\vartheta''} & \cos{\vartheta''}
\end{array}\right) \,,
\ee
 into two single-mode rotations (``phase shifters'', with
reference to the ``optical phase'' in phase space) and one squeezing
operation. We will consider the reduction of a generic CM $\sig$
under local operations of the form $S_{l} \equiv \bigoplus_{j=1}^{N}
S_{1}(\vartheta'_{j},\vartheta''_j,z_j)$. The local symmetric blocks
$\sig_{jj}$ can all be diagonalized by the first rotations and then
symplectically diagonalized ({\em i.e.}, made proportional to the
identity) by the subsequent squeezings, such that $\sig_{jj}=a_j
\id_{2}$ (thus reducing the number of parameters in each diagonal
block to the local symplectic eigenvalue, determining the entropy of
the mode). The second series of local rotations can then be applied
to manipulate the non-local blocks, while leaving the local ones
unaffected (as they are proportional to the identity). Different
sets of $N$ entries in the non-diagonal sub-matrices can be thus set
to zero. For an even total number of modes, all the non-diagonal
blocks $\sig_{12}$, $\sig_{34}$,\ldots,$\sig_{(N-1)N}$ describing
the correlations between disjoint pairs of quadratures can be
diagonalized (leading to the singular-value diagonal form of each
block), with no conditions on all the other blocks. For an odd
number of modes, after the diagonalization of the blocks relating
disjoint quadratures, a further non-diagonal block involving the
last mode (say, $\sig_{1N}$) can be put in triangular form by a
rotation on the last mode.

Notice finally that the locally invariant degrees of freedom of a
generic Gaussian state of $N$ modes are $(2N+1)N-3N=2N^2-2N$, as
follows from the subtraction of the number of free parameters of the
local symplectics from the one of a generic state --- with an
obvious exception for $N=1$, for which the number of free parameters
is $1$, due to the rotational invariance of single-mode Williamson
forms (see the discussion about the vacuum state in Appendix
\ref{redu}).

\subsubsection{Standard form of two-mode Gaussian states} \label{Sec2M}
According to the above counting argument, an arbitrary (mixed)
Gaussian state of two modes can be described, up to local unitary
operations, by $4$ parameters. Let us briefly discuss this instance
explicitly, to acquaint the reader with the symplectic playground,
and since two-mode Gaussian states are the paradigmatic examples of
bipartite entangled states of CV systems.

 The expression of the
two-mode CM $\sig$ in terms of the three $2\times 2$ matrices
$\alp$, $\bet$, $\gr\gamma$, that will be useful in the following,
takes the form [see \eq{CM}]
\begin{equation}
\sig=\left(\begin{array}{cc}
{\alp}&{\gr\gamma}\\
{\gr\gamma}^{\sf T}&{\bet}
\end{array}\right)\, . \label{espre}
\end{equation}
For any two-mode CM ${\sig}$ there is a local symplectic operation
$S_{l}=S_{1}\oplus S_{2}$ which brings ${\sig}$ in the standard form
${\sig}_{sf}$ \cite{Simon00,Duan00}
\begin{equation}
S_{l}\T{\sig}S_{l}={\sig}_{sf} \equiv \left(\begin{array}{cccc}
a&0&c_{+}&0\\
0&a&0&c_{-}\\
c_{+}&0&b&0\\
0&c_{-}&0&b
\end{array}\right)\; . \label{stform}
\end{equation}
States whose standard form fulfills $a=b$ are said to be {\em
symmetric}.
%Let us recall that any pure state ($\mu = 1$) is symmetric and
%fulfills $c_{+}=-c_{-}=\sqrt{a^2-1}$.
The covariances $a$, $b$, $c_{+}$, and $c_{-}$ are determined by the
four local symplectic invariants ${\rm
Det}{\sig}=(ab-c_{+}^2)(ab-c_{-}^2)$, ${\rm Det}{\alp}=a^2$, ${\rm
Det}{\bet}=b^2$, ${\rm Det}{\gr\gamma}=c_{+}c_{-}$. Therefore, the
standard form corresponding to any CM is unique (up to a common sign
flip in $c_-$ and $c_{+}$).

Entanglement of two-mode Gaussian states is the topic of Chapter
\ref{Chap2M}.

\subsection{Pure states} \label{secpurecount}

The CM $\sig^p$ of a generic $N$-mode {\em pure} Gaussian state
satisfies the condition
\begin{equation}\label{osos}
- \Omega\ \sig^p\ \Omega\ \sig^p = \id\,. \end{equation} This
follows from the Williamson normal-mode decomposition of the CM,
$\sig^p = S \id S\T$, where $S$ is a symplectic transformation.
Namely, $$ -\Omega \sig^p \Omega \sig^p = -\Omega S S\T \Omega S S\T
= - \Omega S \Omega S\T = - \Omega  \Omega = \id\,.$$ The matrix
identity \eq{osos} provides a set of (not mutually independent)
polynomial equations that the elements of a generic CM have to
fulfill in order to represent a pure state. A detailed analysis of
the constraints imposed by \eq{osos}, as obtained in
Ref.~\cite{sformato},  is reported in Appendix \ref{ChapAppendixSF}.
Here, it suffices to say that by proper counting arguments the CM of
a pure $N$-mode Gaussian state is determined by $N^2+N$ parameters
in full generality. If one aims at evaluating entanglement, one can
then exploit the further freedom arising from local-unitary
invariance and reduce this minimal number of parameters to (see
Appendix \ref{redu})
\begin{equation}\label{npuri}
\#(\sig^p) = \left\{
  \begin{array}{ll}
    N(N-1)/2, & N \le 3\,; \\
    N(N-2), & N>3\,.
  \end{array}
\right.
\end{equation}
An important subset of pure $N$-mode Gaussian states is constituted
by those whose CM which can be locally put in a standard form with
zero direct correlations between position $\hat q_i$ and momentum
$\hat p_j$ operators, \ie with all diagonal submatrices in \eq{CM}.
This class encompasses basically all Gaussian states currently
produced in a multimode setting and employed in CV communication and
computation processes, and  in general all Gaussian states of this
form are ground states of some harmonic Hamiltonian with spring-like
interactions \cite{chain}. For these Gaussian states, which we will
refer to as {\em block-diagonal}
--- with respect to the vector of canonical operators reordered as
$(\hat q_1,\,\hat q_2,\,\ldots,\,\hat q_N,\,\hat p_1,\,\hat
p_2,\,\ldots,\,\hat p_N)$ --- we have proven that the minimal number
of local-unitary-invariant parameters reduces to $N(N-1)/2$ for any
$N$ \cite{generic}. As they deserve a special attention, their
structural properties will be investigated in detail in
Sec.~\ref{SecGeneric}, together with the closely related description
of their generic entanglement, and with the presentation of an
efficient scheme for their state engineering which involves exactly
$N(N-1)/2$ optical elements (single-mode squeezers and
beam-splitters) \cite{generic}. Notice from \eq{npuri} that not only
single-mode and two-mode states, but also {\em all} pure three-mode
Gaussian states are of this block-diagonal form: an explicit
standard form from them will be provided in Sec.~\ref{secpuri}, and
exploited to characterize their tripartite entanglement sharing as
in Ref.~\cite{3mpra}.

\subsubsection{Phase-space Schmidt decomposition}\label{SecSchmidtPS} In general, pure
Gaussian states of a bipartite CV system admit a physically
insightful decomposition at the CM level
\cite{holevo01,botero03,giedkeqic03}, which can be regarded as the
direct analogue of the Schmidt decomposition for pure
discrete-variable states (see Sec.~\ref{SecPure}). Let us recall
what happens in the finite-dimensional case. With respect to a
bipartition of a pure state $\ps_{A|B}$ into two subsystems $\s_A$
and $\s_B$, one can diagonalize (via an operation $U_A \otimes U_B$
which is local unitary according to the considered bipartition) the
two reduced density matrices $\ro_{A,B}$, to find that they have the
same rank and exactly the same nonzero eigenvalues $\set{\lambda_k}$
($k=1,\ldots,\min\{\dim\hh_A,\,\dim\hh_B\}$). The reduced state of
the higher-dimensional subsystem (say $\s_B$) will accomodate
$(\dim\hh_b-\dim\hh_A)$ additional $0$'s in its spectrum. The state
$\ps_{A|B}$ takes thus the Schmidt form of \eq{E:schmidt}.

Looking at the mapping provided by Table \ref{CVtabG}, one can guess
what happens for Gaussian states. Given a Gaussian CM $\sig_{A|B}$
of an arbitrary number $N$ of modes, where subsystem $\s_A$
comprises $N_A$ modes and subsystem $\s_B$, $N_B$ modes (with
$N_A+N_B=N$), then one can perform the Williamson decomposition
\eq{willia} in both reduced CMs (via a local symplectic operation
$S_A \oplus S_B$), to find that they have the same symplectic rank,
and the same non-unit symplectic eigenvalues $\set{\nu_k}$
($k=1,\ldots,\min\{N_A,\,N_B\}$). The reduced state of the
higher-dimensional subsystem (say $\s_B$) will accomodate
$(N_B-N_A)$ additional $1$'s in its symplectic spectrum. With
respect to an arbitrary $A|B$ bipartition, therefore, the CM
$\sig^p$ of any pure $N$-mode Gaussian state is locally equivalent
to the form $\sig^p_S = (S_A \oplus S_B) \sig^p (S_A \oplus S_B)\T$,
with
\begin{equation}\label{CMschmidt}
\sig^p_S = \left(
\begin{array}{c|c}
 \overbrace{\left.
\begin{array}{cccc}
 \gr C_1 & \diamond  & \diamond  & \diamond  \\
 \diamond  & \gr C_2 & \diamond  & \diamond  \\
 \diamond  & \diamond  & \ddots & \diamond  \\
 \diamond  & \diamond  & \diamond  & \gr C_{N_A}
\end{array}
\right.}^{N_A} & \overbrace{\left.
\begin{array}{ccccccc}
 \gr S_1 & \diamond  & \diamond  & \diamond  & \diamond  & \diamond  & \diamond  \\
 \diamond  & \gr S_2 & \diamond  & \diamond  & \diamond  & \diamond  & \diamond  \\
 \diamond  & \diamond  & \ddots & \diamond  & \diamond  & \diamond  & \diamond  \\
 \diamond  & \diamond  & \diamond  & \gr S_{N_A} & \diamond  & \diamond  & \diamond
\end{array}
\right.}^{N_B} \\ \hline
 \left.
\begin{array}{cccc}
 \gr S_1 & \diamond  & \diamond  & \diamond  \\
 \diamond  & \gr S_2 & \diamond  & \diamond  \\
 \diamond  & \diamond  & \ddots\,  & \diamond  \\
 \diamond  & \diamond  & \diamond  & \gr S_{N_A} \\
 \diamond  & \diamond  & \diamond  & \diamond  \\
 \diamond  & \diamond  & \diamond  & \diamond  \\
 \diamond  & \diamond  & \diamond  & \diamond
\end{array}
\right. & \left.
\begin{array}{ccccccc}
 \gr C_1 & \diamond  & \diamond  & \diamond  & \diamond  & \diamond  & \diamond  \\
 \diamond  & \gr C_2 & \diamond  & \diamond  & \diamond  & \diamond  & \diamond  \\
 \diamond  & \diamond  & \ddots & \diamond  & \diamond  & \diamond  & \diamond  \\
 \diamond  & \diamond  & \diamond  & \gr C_{N_A}\!\! & \diamond  & \diamond  & \diamond   \\
 \diamond  & \diamond  & \diamond  & \diamond  & \!\!\gr\id\!\! & \diamond  & \diamond \vspace*{-0.2cm} \\
 \diamond  & \diamond  & \diamond  & \diamond  & \diamond  & \!\!\!\ddots\!\!\! & \diamond  \\
 \diamond  & \diamond  & \diamond  & \diamond  & \diamond  & \diamond  &
\!\!\gr\id\!\!
\end{array}
\right.
\end{array}
\right).
\end{equation}
Here each element denotes a $2\times 2$ submatrix, in particular the
diamonds ($\diamond$) correspond to null matrices, $\gr\id$ to the
identity matrix, and
$$ \gr C_k = \left(
           \begin{array}{cc}
             \nu_k & 0 \\
             0 & \nu_k \\
           \end{array}
         \right)\,,\quad \gr S_k = \left(
           \begin{array}{cc}
             \sqrt {\nu_k^2-1} & 0 \\
             0 & -\sqrt {\nu_k^2-1} \\
           \end{array}
         \right)\,. $$
 The matrices $\gr C_k$ contain the symplectic
eigenvalues $\nu_k \neq 1$ of both reduced CMs. By expressing them
in terms of hyperbolic functions, $\nu_k = \cosh(2 r_k)$ and by
comparison with \eq{tms}, one finds that each two-mode CM
$$
\left(
  \begin{array}{cc}
    \gr C_k & \gr S_k \\
    \gr S_k & \gr C_k \\
  \end{array}
\right)\,,
$$
encoding correlations between a single mode from $\s_A$ and a single
mode from $\s_B$, is a two-mode squeezed state with squeezing $r_k$.
Therefore, the Schmidt form of a pure $N$-mode Gaussian state with
respect to a $N_A \times N_B$  bipartition (with $N=N_A+N_B$, $N_B
\ge N_A$) is that of a direct sum \cite{holevo01,giedkeqic03}
\begin{equation}\label{CVschmidt}
\sig^p_S = \bigoplus_{i=1}^{N_A} \sig^{sq}_{i,j}(r_i)
\bigoplus_{k=2N_A+1}^{N} \sig^0_k\,,
\end{equation}
where mode $i \in \s_A$, mode $j\equiv i+N_A \in \s_B$, and
$\sig^0_k = \id_2$ is the CM of the vacuum state of mode $k  \in
\s_B$. This corresponds, on the Hilbert space level, to the product
of two-mode squeezed states, tensor additional uncorrelated vacuum
modes in the higher-dimensional subsystem ($\s_B$ in our notation)
\cite{botero03}. The phase-space Schmidt decomposition is a very
useful tool both for the understanding of the structural features of
Gaussian states in the CM formalism, and for the evaluation of their
entanglement properties. Notice that the validity of such a
decomposition can be extended to mixed states with fully degenerate
symplectic spectrum, \ie Williamson normal form proportional to the
identity \cite{botero03,giedkeqic03}.

As a straightforward consequence of \eq{CVschmidt}, any pure
two-mode Gaussian state is equivalent, up to local unitary
operations, to a two-mode squeezed state of the form \eq{tms},
therefore the minimum number of local-unitary degrees of freedom for
pure Gaussian states with $N=2$ is just one (the squeezing degree),
in accordance with \eq{npuri}, and as explained by alternative
arguments in Appendix \ref{redu}. In other words, according to the
notation of \eq{stform}, any pure two-mode Gaussian state is
symmetric ($b=a$) and its standard form elements fulfill
$c_{\pm}=\pm \sqrt{a^2-1}$.

Notice also that the phase-space decomposition discussed here is
special to Gaussian states and is independent from the general
Schmidt decomposition at the Hilbert space level, \eq{E:schmidt},
which can be obtained for any pure state. For CV systems, it will
contain in principle infinite terms, as the local bases are
infinite-dimensional. To give an example, the two-mode squeezed
state, whose CM in its ``phase-space Schmidt decomposition'' is of
the form \eq{tms}, admits the following Hilbert-space Schmidt
decomposition \cite{barnett}
\begin{equation}\label{tmsN}
\ket{\psi^{sq}}_{i,j} =  \frac{1}{\cosh r} \sum_{n=0}^{\infty }\tanh
^{n}r\,\left| n\right\rangle_i\left| n\right\rangle_j\,,
\end{equation}
with local Schmidt bases given by the number states in the Fock
space of each mode.

We will now show that for (generally mixed) Gaussian states with
some local symmetry constraints, a similar phase-space reduction  is
available, such that multimode properties (like entanglement) can be
unitarily reduced to two-mode ones \cite{adescaling,unitarily}.

\subsection{Symmetric states} \label{SecSymm}

Very often in quantum information, and in particular in the theory
of entanglement, peculiar roles are played by {\em symmetric
states}, that is, states that are either invariant under a
particular group of transformations --- like Werner states of qudits
\cite{Werner89} --- or under permutation of two or more parties in a
multipartite system, like ground and thermal states of some
translationally invariant Hamiltonians ({\eg}of harmonic lattices)
 \cite{chain}.
Here we will introduce classes of Gaussian states invariant under
all permutation of the modes (fully symmetric states) or exhibiting
such permutation-invariance locally in each of the two subsystems
across a global bipartition of the modes (bisymmetric states). For
both we will provide a standard form based on the special properties
of their symplectic spectrum. We will limit ourself to a collection
of results, which will be useful for the computation and
exploitation of entanglement in the corresponding states. All the
proofs can be found in Ref.~\cite{unitarily}. Unless explicitly
stated, we are dealing with generally mixed states.

\subsubsection{Fully symmetric Gaussian states}

We shall say that a multimode Gaussian state $\varrho$ is  ``fully
symmetric'' if it is invariant under the exchange of any two modes.
In the following, we will consider the fully symmetric $M$-mode and
$N$-mode Gaussian states $\varrho_{\alp^M}$ and $\varrho_{\bet^N}$,
with CMs $\sig_{\alp^M}$ and $\sig_{\bet^N}$. Due to symmetry, we
have that the CM, \eq{CM}, of such states reduces to
\begin{equation}\label{fscm}
\sig_{\alp^M}={\left(%
 \begin{array}{cccc}
  \gr\alpha & \gr\varepsilon & \cdots & \gr\varepsilon \\
  \gr\varepsilon & \gr\alpha & \gr\varepsilon & \vdots \\
  \vdots & \gr\varepsilon & \ddots & \gr\varepsilon \\
  \gr\varepsilon & \cdots & \gr\varepsilon & \gr\alpha \\
\end{array}%
\right)}\,, \quad
\sig_{\bet^N}={\left(%
 \begin{array}{cccc}
  \gr\beta & \gr\zeta & \cdots & \gr\zeta \\
  \gr\zeta & \gr\beta & \gr\zeta & \vdots \\
  \vdots & \gr\zeta & \ddots & \gr\zeta \\
  \gr\zeta & \cdots & \gr\zeta & \gr\beta \\
\end{array}%
\right)}\,,
\end{equation}
where $\gr\alpha$, $\gr\varepsilon$, $\gr\beta$ and $\gr\zeta$ are
$2\times2$ real symmetric submatrices (the symmetry of
$\gr\varepsilon$ and $\gr\zeta$ stems again from the symmetry under
the exchange of any two modes).

\smallskip

\noindent {\rm Standard form.}--- Let $\sig_{\beta^N}$ be the CM of
a fully symmetric $N$-mode Gaussian state. The $2\times 2$ blocks
$\bet$ and $\gr\zeta$ of $\sig_{\beta^N}$, defined by \eq{fscm}, can
be brought by means of local, single-mode symplectic operations
$S\in \sy{2}^{\oplus N}$ into the form $\bet=\,{\rm diag}\,(b,b)$
and $\gr\zeta=\,{\rm diag}\,(z_1,z_2)$.

In other words, the standard form of fully symmetric $N$-mode states
is such that any reduced two-mode state is symmetric and in standard
form, see \eq{stform}.

\smallskip

\noindent {\rm Symplectic degeneracy.}--- The symplectic spectrum of
$\sig_{\beta^N}$ is $(N-1)$-times degenerate. The two symplectic
eigenvalues of $\sig_{\beta^N}$, $\nu_{\beta}^{-}$ and
$\nu_{\beta^N}^{+}$, read \be
\begin{split}
\nu_{\beta}^{-}&  =    \sqrt{(b-z_1)(b-z_2)} \; ,\\
\nu_{\beta^N}^{+}& =    \sqrt{(b+(N-1)z_1)(b+(N-1)z_2)} \; ,
\end{split}\label{fsspct}
\ee where $\nu_{\beta}^{-}$ is the $(N-1)$-times degenerate
eigenvalue.

\smallskip

Obviously, analogous results hold for the $M$-mode CM
$\sig_{\alpha^M}$ of \eq{fscm}, whose $2\times 2$ submatrices can be
brought to the form $\alp=\,{\rm diag}\,(a,a)$ and
$\gr\varepsilon=\,{\rm diag}\,(e_1,e_2)$ and whose $(M-1)$-times
degenerate symplectic spectrum reads \be
\begin{split}
\nu_{\alpha}^{-}&  =    \sqrt{(a-e_1)(a-e_2)} \; ,\\
\nu_{\alpha^M}^{+}& =    \sqrt{(a+(M-1)e_1)(a+(M-1)e_2)} \; .
\end{split}\label{fsspct2}
\ee

\subsubsection{Bisymmetric {\it M}\,$\times$\,{\it N} Gaussian states}

Let us now generalize this analysis to the $(M+N)$-mode Gaussian
states, whose CM $\sig$ result from a correlated combination of the
fully symmetric blocks $\sig_{\alp^M}$ and $\sig_{\bet^N}$, \be \sig
= \left(\begin{array}{cc}
\sig_{\alp^M} & \gr\Gamma\\
\gr\Gamma^{\sf T} & \sig_{\bet^N}
\end{array}\right) \; , \label{fulsim}
\ee where $\gr\Gamma$ is a $2M\times 2N$ real matrix formed by
identical $2\times 2$ blocks $\gr\gamma$. Clearly, $\gr\Gamma$ is
responsible of the correlations existing between the $M$-mode and
the $N$-mode parties. Once again, the identity of the submatrices
$\gr\gamma$ is a consequence of the local invariance under mode
exchange, internal to the $M$-mode and $N$-mode parties. States of
the form of \eq{fulsim} will be henceforth referred to as
``bisymmetric'' \cite{adescaling,unitarily}. A significant insight
into bisymmetric multimode Gaussian states can be gained by studying
the symplectic spectrum of $\sig$ and comparing it to the ones of
$\sig_{\alpha^M}$ and $\sig_{\beta^N}$.\smallskip

\noindent{\rm Symplectic degeneracy.}--- {The symplectic spectrum of
the CM $\sig$ \eq{fulsim} of a bisymmetric $(M+N)$-mode Gaussian
state includes two degenerate eigenvalues, with multiplicities $M-1$
and $N-1$. Such eigenvalues coincide, respectively, with the
degenerate eigenvalue $\nu_{\alpha}^{-}$ of the reduced CM
$\sig_{\alpha^M}$, and with the degenerate eigenvalue
$\nu_{\beta}^{-}$ of the reduced CM $\sig_{\beta^N}$.}

\smallskip

Equipped with these results, we are now in a position to show the
following central result \cite{unitarily}, which applies to all
(generally mixed) bisymmetric Gaussian states, and is  somehow
analogous to --- but independent of --- the phase-space Schmidt
decomposition of pure Gaussian states (and of mixed states with
fully degenerate symplectic spectrum).
\smallskip

\noindent{\rm Unitary localization of bisymmetric states.}--- { The
bisymmetric $(M+N)$-mode Gaussian state with CM $\sig$, \eq{fulsim}
can be brought, by means of a local unitary (symplectic) operation
with respect to the $M\times N$  bipartition with reduced CMs
$\sig_{\alpha^M}$ and $\sig_{\beta^N}$, to a tensor product of
$M+N-2$ single-mode uncorrelated states, and of a single two-mode
Gaussian state comprised of one mode from the $M$-mode block and one
mode from the $N$-mode block.}

For ease of the reader and sake of pictorial clarity, we can
demonstrate the mechanism of unitary reduction by explicitly writing
down the different forms of the CM $\sig$ at each step. The CM
$\sig$ of a bisymmetric $(M+N)$-mode Gaussian state reads,
 from \eq{fulsim}, \be \gr{\sigma}= \left(\begin{array}{cccc|cccc}
\gr{\alpha}&\gr{\varepsilon}&\ldots&\gr{\varepsilon}&
\gr{\gamma}&\cdots&\cdots&\gr{\gamma}\\
\gr{\varepsilon}&\ddots&\gr{\varepsilon}&\vdots&
\vdots&\ddots&&\vdots\\
\vdots&\gr{\varepsilon}&\ddots&\gr{\varepsilon}&
\vdots&&\ddots&\vdots\\
\gr{\varepsilon}&\cdots&\gr{\varepsilon}&\gr{\alpha}&
\gr{\gamma}&\cdots&\cdots&\gr{\gamma}\\ \hline
\overset{}{\gr{\gamma}^{\sf T}}&\cdots&\cdots&\gr{\gamma}^{\sf T}&
\gr{\beta}&\gr{\zeta}&\ldots&\gr{\zeta}\\
\vdots&\ddots&&\vdots&
\gr{\zeta}&\ddots&\gr{\zeta}&\vdots\\
\vdots&&\ddots&\vdots&
\vdots&\gr{\zeta}&\ddots&\gr{\zeta}\\
\gr{\gamma}^{\sf T}&\cdots&\cdots&\gr{\gamma}^{\sf T}&
\gr{\zeta}&\cdots&\gr{\zeta}&\gr{\beta}
\end{array}\right) \, . \label{totala}
\ee According to the previous results, by symplectically reducing
the block $\gr{\sigma}_{\beta^{N}}$ to its Williamson normal form,
the global CM $\sig$ is brought to the form
\[\gr{\sigma'}=
\left(\begin{array}{cccc|cccc}
\gr{\alpha}&\gr{\varepsilon}&\cdots&\gr{\varepsilon}&
\gr{\gamma'}&\diamond&\cdots&\diamond\\
\gr{\varepsilon}&\ddots&\gr{\varepsilon}&\vdots&
\vdots&\vdots&\ddots&\vdots\\
\vdots&\gr{\varepsilon}&\ddots&\gr{\varepsilon}&
\vdots&\vdots&\ddots&\vdots\\
\gr{\varepsilon}&\cdots&\gr{\varepsilon}&\gr{\alpha}&
\gr{\gamma'}&\diamond&\cdots&\diamond\\ \hline
\overset{}{\gr{\gamma'}^{\sf T}}&\cdots&\cdots&\gr{\gamma'}^{\sf T}&
\gr{\nu}_{\beta^N}^+&\diamond&\cdots&\diamond\\
\diamond&\cdots&\cdots&\diamond&
\diamond&\gr{\nu}_{\beta}^-&\diamond&\vdots\\
\vdots&\ddots&\ddots&\vdots&
\vdots&\diamond&\ddots&\diamond\\
\diamond&\cdots&\cdots&\diamond&
\diamond&\cdots&\diamond&\gr{\nu}_{\beta}^-
\end{array}\right) ,
\]
where the $2\times 2$ blocks $\gr{\nu}_{\beta^N}^{+}=
\nu_{\beta^n}^{+}{\mathbbm 1}_2$ and $\gr{\nu}_{\beta}^-=
\nu_{\beta}^{-}{\mathbbm 1}_2$ are the Williamson normal blocks
associated to the two symplectic eigenvalues of
$\gr{\sigma}_{\beta^{N}}$. The identity of the submatrices
$\gr{\gamma'}$ is due to the invariance under permutation of the
first $M$ modes, which are left unaffected. The subsequent
symplectic diagonalization of $\gr{\sigma}_{\alpha^{M}}$ puts the
global CM $\sig$ in the following form (notice that the first,
$(M+1)$-mode reduced CM is again a matrix of the same form of
$\sig$, with $N=1$), \be \gr{\sigma''}=
\left(\begin{array}{cccc|cccc}
\gr{\nu}_{\alpha}^-&\diamond&\cdots&\diamond&
\diamond&\diamond&\cdots&\diamond\\
\diamond&\ddots&\diamond&\vdots&
\vdots&\vdots&\ddots&\vdots\\
\vdots&\diamond&\gr{\nu}_{\alpha}^-&\diamond&
\diamond&\vdots&\ddots&\vdots\\
\diamond&\cdots&\diamond&\underset{}{\gr{\nu}_{\alpha^M}^+}&
\gr{\gamma''}&\diamond&\cdots&\diamond\\ \hline
\diamond&\cdots&\diamond&\overset{}{\gr{\gamma''}^{\sf T}}&
\gr{\nu}_{\beta^N}^+&\diamond&\cdots&\diamond\\
\diamond&\cdots&\cdots&\diamond&
\diamond&\gr{\nu}_{\beta}^-&\diamond&\vdots\\
\vdots&\ddots&\ddots&\vdots&
\vdots&\diamond&\ddots&\diamond\\
\diamond&\cdots&\cdots&\diamond&
\diamond&\cdots&\diamond&\gr{\nu}_{\beta}^-
\end{array}\right) , \label{final}
\ee with $\gr{\nu}_{\alpha^M}^{+}= {\nu}_{\alpha^M}^{+} {\mathbbm
1}_2$ and $\gr{\nu}_{\alpha}^{-}= {\nu}_{\alpha}^{-} {\mathbbm
1}_2$. \eq{final} shows explicitly that the state with CM
$\gr{\sigma''}$, obtained from the original state with CM $\sig$ by
exploiting local unitary operations, is the tensor product of
$M+N-2$ uncorrelated single-mode states and of a correlated two-mode
Gaussian state.

According to this reduction, one immediately has that the amount of
entanglement (quantum correlations) present in any bisymmetric
multimode Gaussian state can be localized (concentrated) in an
equivalent two-mode Gaussian state (\ie shared only by a single pair
of modes), via local unitary operations \cite{unitarily}. These
results and their consequences will be discussed in detail in
Chapter \ref{ChapUniloc}. Let us just note that fully symmetric
Gaussian states, \eq{fscm}, are special instances of bisymmetric
states with respect to any global bipartition of the modes.

 }

%\section{Experimental remarks}

\part{Bipartite~entanglement~of\ Gaussian~states}
{\vspace*{1cm}
\includegraphics[width=6cm]{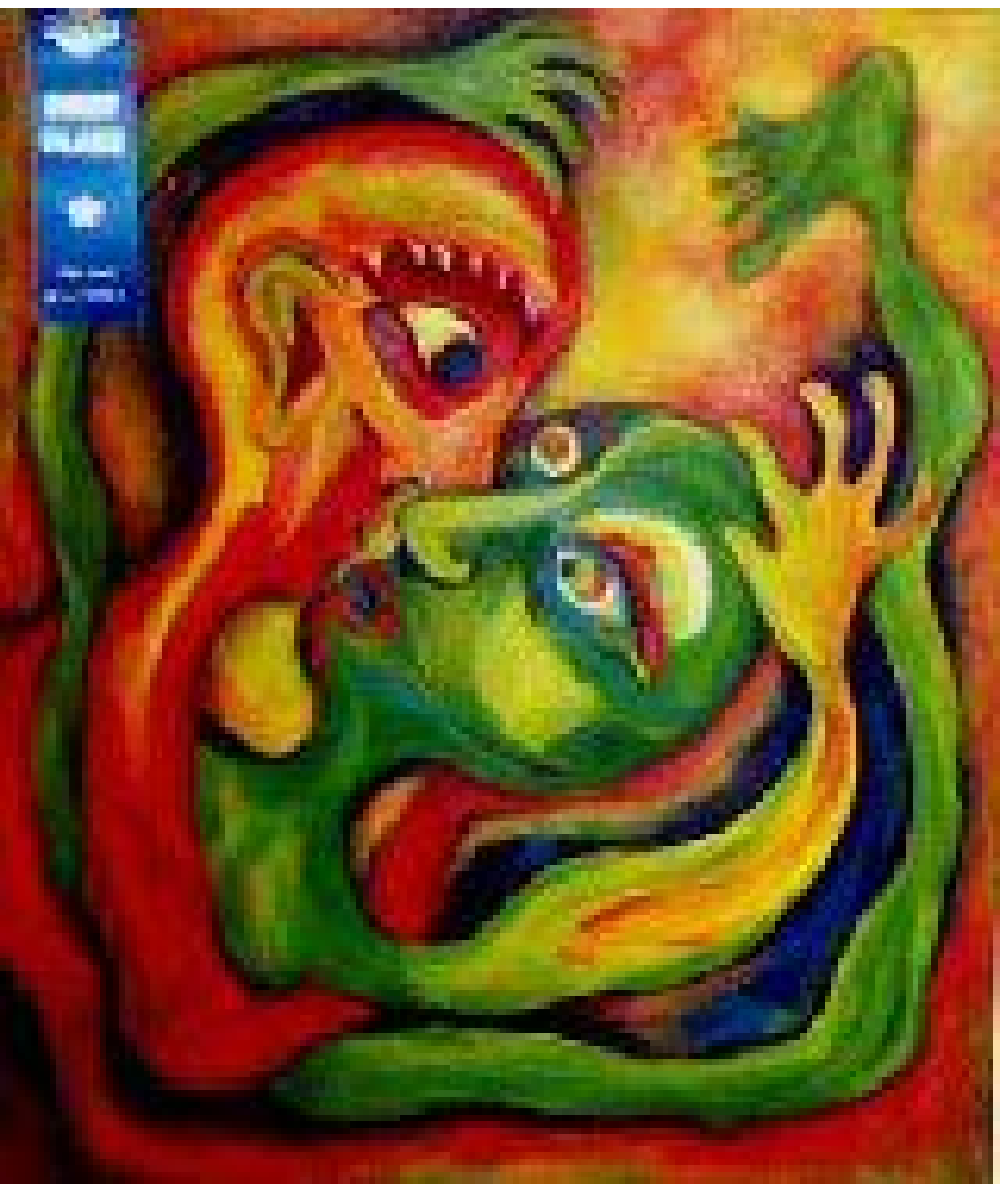} \\
\vspace*{0.6cm} {\rm \normalsize  {\em Entanglement.}  Pamela Ott,
2002.
\\ \vspace*{-0.4cm} \texttt{\footnotesize
http://www.hottr6.com/ott/}}} \label{PartBip}

\chapter{Characterizing entanglement of Gaussian states}
\label{ChapEntGauss}

{\sf

In this short Chapter we recall the main results on the
qualification and quantification of bipartite entanglement for
Gaussian states of CV systems. We will borrow some material from
\cite{adebook}.

\section{How to qualify bipartite Gaussian entanglement}

\subsection{Separability and distillability: PPT
criterion}\label{SecPPTG}

The positivity of the partially transposed state (Peres-Horodecki
PPT criterion \cite{Peres96,Horodecki96}, see Sec.~\ref{SecPPT}) is
necessary and sufficient for the separability of two-mode Gaussian
states \cite{Simon00} and, more generally, of all $(1+N)$-mode
Gaussian states under $1\times N$  bipartitions \cite{werewolf} and
--- as we are going to show --- of symmetric and bisymmetric $(M+N)$-mode Gaussian states (see
Sec.~\ref{SecSymm}) under $M\times N$ bipartitions \cite{unitarily}.
In general, the partial transposition $\varrho\PT{A}$ of a bipartite
quantum state $\varrho$ is defined as the result of the
transposition performed on only one of the two subsystems (say
$\s_A$) in some given basis. In phase space, the action of partial
transposition amounts to a mirror reflection of the momentum
operators of the modes comprising one subsystem \cite{Simon00}. The
CM $\sig_{A|B}$, where subsystem $\s_A$ groups $N_A$ modes, and
subsystem $\s_B$ is formed by $N_B$ modes, is then transformed into
a new matrix
\begin{equation}\label{cmpt}
\tilde\sig_{A|B} \equiv \gr\theta_{A|B}\ \sig_{A|B}\
\gr\theta_{A|B}\,,
\end{equation}
with \be\gr\theta_{A|B} = {\rm diag}
\{\underbrace{1,\,-1,\,1,\,-1,\,\ldots,\,1,-1}_{2N_A},\,\underbrace{1,\,1,\,1,\,1,\,\ldots,\,1,\,1}_{2N_B}\}\,.\label{thetatrans}\ee
Referring to the notation of \eq{CM}, the partially transposed
matrix $\tilde\sig_{A|B}$ differs from $\sig_{A|B}$ by a sign flip
in the determinants of the intermodal correlation matrices, $\det
\eps_{ij}$, with modes $i \in \s_A$ and modes $j \in s_B$.

The PPT criterion yields that a Gaussian state $\sig_{A|B}$ (with
$N_A=1$ and $N_B$ arbitrary) is {\em separable} if and only if the
partially transposed $\tilde\sig_{A|B}$ is a {\em bona fide} CM,
that is it satisfies the uncertainty principle \eq{bonfide},
\begin{equation}\label{bonfidept}
   \tilde\sig_{A|B} + i \Omega \ge 0\,.
\end{equation}
 This
reflects the positivity of the partially transposed density matrix
$\varrho\PT{A}$ associated to the state $\varrho$. For Gaussian
states with $N_A >1$ and not endowed with special symmetry
constraints, PPT condition is only necessary for separability, as
bound entangled Gaussian states, whose entanglement is
undistillable, have been proven to exist already in the instance
$N_A=N_B=2$ \cite{werewolf}.

We have demonstrated the existence of ``bisymmetric'' $(N_A +
N_B)$-mode Gaussian states for which PPT is again equivalent to
separability  \cite{unitarily}.  In view of the invariance of PPT
criterion under local unitary transformations (which can be
appreciated by the definition of partial transpose at the Hilbert
space level) and considering the results proved in
Sec.~\ref{SecSymm} on the unitary localization of bisymmetric
Gaussian states, see \eq{final}, it is immediate to verity that the
following property holds \cite{unitarily}.

\medskip

\begin{itemize}
\item[\ding{226}]
 \noindent{\rm\bf PPT criterion for bisymmetric multimode
Gaussian states.} {\it For generic $N_A \times N_B$ bipartitions,
the positivity of the partial transpose (PPT) is a necessary and
sufficient condition for the separability of bisymmetric
$(N_A+N_B)$-mode mixed Gaussian states of the form \eq{fulsim}. In
the case of fully symmetric mixed Gaussian states, \eq{fscm}, of an
arbitrary number of modes, PPT is equivalent to separability across
any global bipartition of the modes.}
\smallskip
\end{itemize}

This statement is a first important generalization to multimode
bipartitions of the equivalence between separability and PPT for
 $1 \times N$ bipartite Gaussian states \cite{werewolf}.
 In particular, it implies that no bisymmetric bound entangled
Gaussian states can exist \cite{werewolf,giedkeqic01} and all the
$N_A \times N_B$ multimode block entanglement of such states is
distillable. Moreover, it justifies the use of the negativity and
the logarithmic negativity as measures of entanglement for these
multimode Gaussian states, as will be done in Chapter
\ref{ChapUniloc}.

In general, the distillability problem for Gaussian states has been
also solved \cite{giedkeqic01}: the entanglement of any non-PPT
bipartite Gaussian state is distillable by LOCC. However, we remind
that this entanglement can be distilled only resorting to
non-Gaussian LOCC \cite{browne}, since distilling Gaussian states
with Gaussian operations is impossible \cite{nogo1,nogo2,nogo3}.

\subsubsection{Symplectic representation of PPT criterion}

The partially transposed matrix $\tilde\sig$ of any $N$-mode
Gaussian CM $\sig$ is still a positive and symmetric matrix. As
such, it admits a Williamson normal-mode decomposition
\cite{williamson36}, \eq{willia}, of the form
\begin{equation}
\tilde\sig=S\T \tilde{\gr\nu} S \; , \label{williapt}
\end{equation}
where $S\in Sp_{(2N,\mathbb{R})}$ and $\tilde{\gr\nu}$ is the CM
\begin{equation}
\tilde {\gr{\nu}}=\bigoplus_{k=1}^{N}\left(\begin{array}{cc}
\tilde \nu_k&0\\
0&\tilde\nu_k
\end{array}\right) \, , \label{thermapt}
\end{equation}

The $N$ quantities $\tilde\nu_{k}$'s are the symplectic eigenvalues
of the partially transposed CM $\tilde\sig$. While the symplectic
spectrum $\set{\nu_{k}}$ of $\sig$ encodes the structural and
informational properties of a Gaussian state, the partially
transposed spectrum $\set{\tilde\nu_{k}}$ encodes a complete
qualitative (and to some extent quantitative, see next Section)
characterization of entanglement in the state. Namely, the PPT
condition \pref{bonfidept}, \ie the uncertainty relation for
$\tilde\sig$, can be equivalently recast in terms of the parameters
$\tilde\nu_{k}$'s as
 \be
{\tilde\nu}_{k}\ge1 \; . \label{sympheispt} \ee

We can, without loss of generality, rearrange the modes of a
$N$-mode state such that the corresponding symplectic eigenvalues of
the partial transpose $\tilde\sig$ are sorted in ascending order
$$ \tilde\nu_-\equiv \tilde\nu_1 \le \tilde\nu_2 \le \ldots \le \tilde\nu_{N-1} \le \tilde\nu_N  \equiv \tilde \nu_+\,,$$
in analogy to what done in Sec.~\ref{SecSympHeis} for the spectrum
of $\sig$.  With this notation, PPT criterion across an arbitrary
bipartition reduces to $\tilde\nu_- \ge 1$ for all separable
Gaussian states. As soon as $\tilde\nu_- < 1$, the corresponding
Gaussian state $\sig$ is definitely entangled. The symplectic
characterization of physical versus PPT Gaussian states is
summarized in Table \ref{CVtabGE}.

\begin{table}[t!]   {\rm
\begin{tabular}{c||c|c}
%  \hline
   & Physicality & Separability \\
  \hline \hline
  density matrix & ${\ro \ge 0}$ & $\overset{}{\ro\PT{A}\ge 0}$  \\ \hline
  covariance matrix & $\sig + i \Omega \ge 0$ & $\overset{}{\tilde\sig + i \Omega \ge 0}$ \\ \hline
  symplectic spectrum & $\nu_k \ge 1$ & $\overset{}{\tilde\nu_k \ge 1}$  \\ \hline
 \end{tabular}}
 \caption{Schematic comparison between the existence conditions and
the separability conditions for Gaussian states, as expressed in
different representations. To be precise, the second column
qualifies the PPT condition, which is always implied by the
separability, and equivalent to it in general $1 \times N$ and
bisymmetric $M \times N$ Gaussian states.} \label{CVtabGE}
\end{table}

\subsection{Additional separability criteria}

Let us briefly mention alternative characterizations of separability
for Gaussian states.

For a general Gaussian state of any $N_A\times N_B$ bipartition, a
necessary and sufficient condition states that a CM $\sig$
corresponds to a separable state if and only if there exists a pair
of CMs $\sig_{A}$ and $\sig_{B}$, relative to the subsystems $\s_A$
and $\s_B$ respectively, such that the following inequality holds
\cite{werewolf}, $\sig \geq \sig_{A} \oplus \sig_{B}$. This
criterion is not very useful in practice. Alternatively, one can
introduce an operational criterion based on iterative applications
of a nonlinear map, that is independent of (and strictly stronger
than) the PPT condition, and completely qualifies separability  for
all bipartite Gaussian states  \cite{Giedke01}.

Note also that a comprehensive  characterization of linear and
nonlinear entanglement witnesses (see Sec.~\ref{SecWitn}) is
available for CV systems \cite{illuso}, as well as operational
criteria (useful in experimental settings) based on the violation of
inequalities involving combinations of variances of canonical
operators, for both two-mode \cite{Duan00} and multimode settings
\cite{vloock03}.

However, the range of results collected in this Dissertation deal
with classes of bipartite and multipartite Gaussian states in which
PPT holds as a necessary and sufficient condition for separability,
therefore it will be our preferred tool to check for the presence of
entanglement in the states under consideration.

\section{How to quantify bipartite Gaussian entanglement}

\subsection{Negativities} \label{secnega}

From a quantitative point of view, a family of entanglement measures
which are {\em computable} for general Gaussian states is provided
by the {\em negativities}. Both the negativity $\N$, defined by
\eq{E:N}, and the logarithmic negativity $E_\N$, \eq{E:EN}, are
entanglement monotones under LOCC
\cite{Zyczkowski98,VidalWerner02,EisertPHD,Plenio05}. Note that they
fail to be continuous in trace norm on infinite-dimensional Hilbert
spaces; however, this problem can be circumvented by restricting to
physical states with finite mean energy \cite{eisert02}. The
negativities provide a proper quantification of entanglement in
particular for arbitrary $1 \times N$ and bisymmetric $M \times N$
Gaussian states, directly quantifying the degree of violation of the
necessary and sufficient PPT criterion for separability,
\eq{sympheispt}.

Following \cite{VidalWerner02,SeralePHD} and
\cite{extremal,adescaling}, the negativity of a Gaussian state with
CM $\sig$ is given by \be \N(\sig)=\left\{\begin{array}{l}
\frac12\left(\prod_k {\tilde{\nu}_k}^{-1}-1\right) , \quad {\rm
for}\;
k: \tilde{\nu}_k<1   \; . \\
\\
0 \quad \,{\rm if}\; \tilde{\nu}_i\ge 1 \; \forall\, i \; .
\end{array}\right.\label{negagau}
\ee Here the set $\{\tilde{\nu}_k\}$ is constituted by the
symplectic eigenvalues of the partially transposed CM
$\tilde{\gr{\sigma}}$. Accordingly, the logarithmic negativity reads
\be E_\N(\sig)=\left\{\begin{array}{l} -\sum_k {\log{\tilde{\nu}_k}}
, \quad {\rm for}\;
k : \tilde{\nu}_k<1   \; . \\
\\
0 \quad \,{\rm if}\; \tilde{\nu}_i\ge 1 \; \forall\, i \; .
\end{array}\right.\label{lognegau}
\ee

For the interpretation and the computation of the negativities, a
little lemma by A. Serafini may be precious \cite{serafozziprl}. It
states that, in a  $(N_A +N_B)$-mode Gaussian state with CM
$\sig_{A|B}$, at most
\begin{equation}\label{littlelemm}
 N_{\min} \equiv \min\{N_A,\,N_B\}
\end{equation}
 symplectic eigenvalues $\tilde\nu_k$ of the partial transpose $\tilde\sig_{A|B}$ can
violate the PPT inequality~\pref{sympheispt}  with respect to a $N_A
\times N_B$ bipartition. Thanks to this result, in all $1 \times N$
Gaussian states and in all bisymmetric $M \times N$ Gaussian states
(whose symplectic spectra exhibit degeneracy, see
Sec.~\ref{SecSymm}), the entanglement  is not only qualified, but
also completely quantified (according to the negativities) in terms
of the smallest symplectic eigenvalue $\tilde\nu_-$ of the partially
transposed CM alone.\footnote{\sf Notice that such a result, in the
special instance of two-mode Gaussian states, had been originally
obtained in \cite{prl,extremal}, as detailed in the next Chapter.}
For $\tilde \nu_- \ge 1$ the state is separable, otherwise it is
entangled; the smaller $\tilde \nu_-$, the more entanglement is
encoded in the corresponding Gaussian state. In the limit of
vanishing partially transposed symplectic eigenvalue, $\tilde \nu_-
\rightarrow 0$, the negativities grow unboundedly.

\subsection{Gaussian convex-roof extended measures}\label{SecGEMS}

In this subsection, based on part of Ref.~\cite{ordering}, we
consider a family of entanglement measures exclusively defined for
Gaussian states of CV systems. The formalism of {\em Gaussian
entanglement measures} (Gaussian EMs) has been introduced in
Ref.~\cite{GEOF} where the ``Gaussian entanglement of formation''
has been defined and analyzed. Furthermore, the framework developed
in Ref.~\cite{GEOF} is general and enables to define generic
Gaussian EMs of bipartite entanglement by applying the Gaussian
convex roof, that is, the convex roof over pure Gaussian
decompositions only, to any {\em bona fide} measure of bipartite
entanglement defined for pure Gaussian states.

The original motivation for the introduction of Gaussian EMs stems
from the fact that the entanglement of formation
\cite{Bennett96pra}, defined by \eq{E:EF}, involves a nontrivial
minimization of the pure-state entropy of entanglement over convex
decompositions of bipartite mixed Gaussian states in ensemble of
pure states. These pure states may be, in principle, non-Gaussian
states of CV systems, thus rendering the analytical solution of the
optimization problem in \eq{E:EF} extremely difficult even in the
simplest instance of one mode per side. Nevertheless, in the special
subset of two-mode symmetric mixed Gaussian states [\ie with
$\det\gr\alpha=\det\gr\beta$ in \eq{espre}], the optimal convex
decomposition of \eq{E:EF} has been exactly determined, and it turns
out to be realized in terms of pure {\em Gaussian} states
\cite{giedke03}. Apart from that case (which will be discussed in
Sec.~\ref{SecEOFGauss}), the computation of the entanglement of
formation for nonsymmetric two-mode Gaussian states (and more
general Gaussian states) has not yet been solved, and it stands as
an open problem in the theory of entanglement \cite{QIProb}.
However, the task can be somehow simplified by restricting to
decompositions into pure Gaussian states only. The resulting
measure, known as Gaussian entanglement of formation  \cite{GEOF},
is an upper bound to the true entanglement of formation and
obviously coincides with it for symmetric two-mode Gaussian states.

In general, we can define a Gaussian EM $G_E$ as follows
\cite{ordering}. For any pure Gaussian state $\psi$ with CM
$\sig^p$, one has
\begin{equation}\label{Gaussian EMp}
G_E (\sig^p) \equiv E(\psi)\,,
\end{equation}
where $E$ can be {\em any} proper measure of entanglement of pure
states, defined as a monotonically increasing function of the
entropy of entanglement ({\ie}the Von Neumann entropy of the reduced
density matrix of one party).

For any mixed Gaussian state $\varrho$ with CM $\sig$, one has then
\cite{GEOF}
\begin{equation}\label{Gaussian EMm}
G_E (\sig) \equiv \inf_{\sig^p \le \sig} G_E(\sig^p)\,.
\end{equation}
If the function $E$ is taken to be exactly the entropy of
entanglement, \eq{E:E}, then the corresponding Gaussian EM is known
as {\em Gaussian entanglement of formation} \cite{GEOF}.

In general, the definition \eq{Gaussian EMm} involves an
optimization over all pure Gaussian states with CM $\sig^p$ smaller
than the CM $\sig$ of the mixed state whose entanglement one wishes
to compute. This corresponds to taking the Gaussian convex roof.
Despite being a simpler optimization problem than that appearing in
the definition \eq{E:EF} of the true entanglement of formation, the
Gaussian EMs cannot be expressed in a simple closed form, not even
in the simplest instance of (nonsymmetric) two-mode Gaussian states.
Advances on this issue were obtained in \cite{ordering}, and will be
presented in Sec.~\ref{secorder}.

 Before that let us recall, as an important side
remark, that any Gaussian EM is an entanglement monotone under
Gaussian LOCC. The proof given in Sec. IV of Ref.~\cite{GEOF} for
the Gaussian entanglement of formation, in fact, automatically
extends to every Gaussian EM constructed via the Gaussian convex
roof of any proper measure $E$ of pure-state entanglement.

}

\chapter{Two-mode entanglement} \label{Chap2M}

{\sf

This Chapter collects our theoretical results on the
characterization of the prototypical entangled states of CV systems,
\ie two-mode Gaussian states. Analytical quantification of the
negativities (and their relationship with global and marginal
entropic measures) \cite{prl,extremal,polacchi} and of the Gaussian
entanglement measures \cite{ordering} will be presented. An
experiment concerning the production and manipulation of two-mode
entanglement with linear optics \cite{francamentemeneinfischio} will
be described in Chapter \ref{Chap2MExp}. The present Chapter
represents, in our hope, a comprehensive reading for the basic
understanding of bipartite entanglement in CV systems.

\section{Symplectic parametrization of two-mode Gaussian states}
\label{SecSympParam}

To study entanglement and informational properties (like global and
marginal entropies) of two-mode Gaussian states, we can consider
without loss of generality states whose CM $\sig$ is in the
$\sy{2}\oplus\sy{2}$-invariant standard form, \eq{stform}
\cite{Simon00,Duan00}. Let us recall it here for the sake of
clarity,
\begin{equation}\label{stform2}
 {\sig}= \left(\begin{array}{cc}
{\alp}&{\gr\gamma}\\
{\gr\gamma}^{\sf T}&{\bet}
\end{array}\right) = \left(\begin{array}{cccc}
a&0&c_{+}&0\\
0&a&0&c_{-}\\
c_{+}&0&b&0\\
0&c_{-}&0&b
\end{array}\right)\,.
\end{equation}

For two-mode states, the uncertainty principle \ineq{bonfide} can be
recast as a constraint on the $Sp_{(4,{\mathbb R})}$ invariants
${\rm Det}\sig$ and $\Delta(\sig)={\rm Det}{\alp}+\,{\rm
Det}{\bet}+2 \,{\rm Det}{\gr\gamma}$ \cite{SymplecticInvariants},
\begin{equation}
\Delta(\sig)\le1+\,{\rm Det}\sig \label{sepcomp}\; .
\end{equation}

The symplectic eigenvalues of a two-mode Gaussian state will be
denoted as $\nu_{-}$ and $\nu_{+}$, with $\nu_{-}\le \nu_{+}$, with
the
 uncertainty relation \pref{sympheis} reducing to \be \label{symptwo} \nu_{-}\ge 1 \; .
\ee A simple expression for the $\nu_{\mp}$ can be found in terms of
the two $Sp_{(4,\mathbb{R})}$ invariants (invariants under global,
two-mode symplectic operations)
\cite{VidalWerner02,SymplecticInvariants,prl,extremal}
\begin{equation}
2{\nu}_{\mp}^2=\Delta(\sig)\mp\sqrt{\Delta^2(\sig) -4\,{\rm
Det}\,\sig} \, . \label{sympeig}
\end{equation}

According to \eq{stform2}, two-mode Gaussian states can be
classified in terms of  their four standard form covariances $a$,
$b$, $c_{+}$, and $c_{-}$. It is relevant to provide a
reparametrization of standard form states in terms of symplectic
invariants which admit a direct interpretation for generic Gaussian
states \cite{prl,extremal,polacchi}. Namely, the parameters of
\eq{stform2} can be determined in terms of the two local symplectic
invariants
\begin{equation}\label{mu12}
\mu_1 = (\det\gr\alpha)^{-1/2} = 1/a\,,\quad \mu_2 =
(\det\gr\beta)^{-1/2} = 1/b\,,
\end{equation}
which are the marginal purities of the reduced single-mode states,
and of the two global symplectic invariants
\begin{equation}\label{globinv}
\mu = (\det\gr\sigma)^{-1/2} =
[(ab-c_{+}^2)(ab-c_{-}^2)]^{-1/2}\,,\quad \Delta =
a^2+b^2+2c_+c_-\,,
\end{equation}
which are the global purity, \eq{purgau},  and the seralian,
\eq{seralian}, respectively. Eqs.~{\rm(\ref{mu12}--\ref{globinv})}
can be inverted to provide the following physical parametrization of
two-mode states in terms of the four independent parameters
$\mu_1,\,\mu_2,\,\mu$, and $\Delta$,
\begin{eqnarray}
  a  \,\,=\,\, \frac{1}{\mu_1}\,, \quad b &=& \frac{1}{\mu_2}\,, \quad
c_{\pm}\,\,=\,\,\frac{\sqrt{\mu_1 \mu_2}}4 \, \big( \epsilon_- \pm
\epsilon_+ \big)\,, \label{gabc} \\
{\rm with}\quad \epsilon_\mp &\equiv& \sqrt{\left[ \Delta -
\frac{(\mu_1 \mp \mu_2)^2}{\mu_1^2 \mu_2^2}\right]^2-\frac{4}{\mu^2}
}\, . \nonumber
\end{eqnarray}

The uncertainty principle \eq{sepcomp} and the existence of the
radicals appearing in \eq{gabc} impose the following constraints on
the four invariants in order to describe a physical state
\begin{eqnarray}
  0 &\le& \mu_{1,2} \,\,\le\,\, 1\,, \label{consmu12}\\
  \mu_1 \mu_2 &\le& \mu \,\,\le\,\,
  \frac{\mu_1 \mu_2}{\mu_1 \mu_2 + \abs{\mu_1-\mu_2}}\,, \label{consmu} \\
  \frac{2}{\mu} + \frac{(\mu_1 - \mu_2)^2}{\mu_1^2 \mu_2^2}
  &\le& \Delta  \,\,\le\,\,  \min \left\{ \frac{(\mu_1 + \mu_2)^2}{\mu_1^2 \mu_2^2}
  - \frac{2}{\mu} \; , \; 1+\frac{1}{\mu^2}\right\}  \, . \label{deltabnd}
\end{eqnarray}
The physical meaning of these constraints, and the role of the
extremal states [{\em i.e.~}states whose invariants saturate the
upper or lower bounds of Eqs.~{\rm(\ref{consmu}--\ref{deltabnd})}]
in relation to the entanglement, will be carefully investigated in
Sec.~\ref{secprl}.

\section{Entanglement and symplectic eigenvalues}
\label{secEnt2Sympl}

\subsection{Partial transposition and negativities}
\label{SecNega2M}

The PPT condition for separability, \eq{bonfidept} has obviously a
very simple form for two-mode Gaussian states. In terms of
symplectic invariants, partial transposition corresponds to flipping
the sign of ${\rm Det}\,\gr{\gamma}$,
\begin{equation}\label{GS:sig2PT}
\sig =\left(%
\begin{array}{cc}
  \gr\alpha & \gr\gamma \\
  \gr\gamma\T & \gr\beta \\
\end{array}%
\right)\quad \overset{\ro\,\rightarrow\,\ro\PT{i}}
{\overrightarrow{\quad\quad\quad}} \quad
\tilde{\sig} =\left(%
\begin{array}{cc}
  \gr\alpha & \tilde{\gr\gamma} \\
  \tilde{\gr\gamma}\T & \gr\beta \\
\end{array}%
\right)\,,
\end{equation}
with $\det\tilde{\gr\gamma} = - \det\gr\gamma$. For a standard form
CM, \eq{stform2}, this simply means $c_+ \rightarrow c_+$, $c_-
\rightarrow c_-$.
 Accordingly, the seralian $\Delta=\det\alp+\det\bet+2\,\det\gr\gamma$,
\eq{seralian}, is mapped under partial transposition into
\begin{equation}\label{deltatilde}
\begin{split} \tilde\Delta&=\det\alp+\det\bet+2\,\det\tilde{\gr\gamma}
=\det\alp+\det\bet-2\,\det\gr\gamma \\ &=\Delta-4\,{\rm
Det}\,\gr{\gamma} = -\Delta + 2/\mu_1^2 + 2/\mu_2^2\,. \end{split}
\end{equation}
From \eq{sympeig}, the symplectic eigenvalues of the partial
transpose $\tilde{\gr{\sigma}}$ of a two-mode CM $\sig$ are promptly
determined in terms of symplectic invariants
\cite{SymplecticInvariants,prl,extremal},
\begin{equation}
    2\tilde{\nu}_{\mp}^2 = \tilde{\Delta}\mp\sqrt{\tilde{\Delta}^2
-{\frac{4}{\mu^2}}}\,. \label{n1}
\end{equation}
The PPT criterion can be then recast as the following inequality
\begin{equation}
\tilde\Delta\le1+1/\mu^2 \label{sepcomppt}\,,
\end{equation}
equivalent to separability. In other words, it yields a state
$\gr{\sigma}$ separable if and only if
%\begin{equation}
$\tilde{\nu}_{-}\ge 1$.
%\label{lowest}
%\end{equation}
Accordingly, the logarithmic negativity \eq{lognegau} is a
decreasing function of $\tilde{\nu}_{-}$ only,
\begin{equation}\label{en}
E_{\N}=\max\{0,-\log\,\tilde{\nu}_{-}\}\,,
\end{equation}
as for the biggest symplectic eigenvalue of the partial transpose
one has $\tilde{\nu}_{+} > 1$ for all two-mode Gaussian states
\cite{prl,extremal}.

Note that from
Eqs.~{\rm(\ref{stform2},\ref{sepcomp},\ref{deltatilde},\ref{sepcomppt})}
the following necessary condition for two-mode entanglement follows
\cite{Simon00},
\begin{equation}\label{detgammanegative}
 \sig\hbox{ entangled}\quad \Rightarrow \quad \det{\gr\gamma} <0\,.
\end{equation}

\subsection{Entanglement of formation for symmetric states}
\label{SecEOFGauss}

The optimal convex decomposition involved in the definition
\eq{E:EF} of the entanglement of formation \cite{Bennett96pra}
(which, in principle, would run over ensembles of non-Gaussian pure
states), has been remarkably solved in the special instance of
two-mode symmetric mixed Gaussian states [\ie with
$\det\gr\alpha=\det\gr\beta$ in \eq{stform2}]   and turns out to be
Gaussian. Namely, the absolute minimum is realized within the set of
pure two-mode Gaussian states \cite{giedke03}, yielding \be E_F =
\max\left[ 0,h(\tilde{\nu}_{-}) \right] \; , \label{eofgau} \ee with
\begin{equation}\label{hentro}
h(x)=\frac{(1+x)^2}{4x}\log \left[\frac{(1+x)^2}{4x}\right]-
\frac{(1-x)^2}{4x}\log \left[\frac{(1-x)^2}{4x}\right].
\end{equation}
Such a quantity is, again, a monotonically decreasing function of
the smallest symplectic eigenvalue $\tilde{\nu}_{-}$ of the partial
transpose $\tilde\sig$ of a two-mode symmetric Gaussian CM $\sig$,
thus providing a quantification of the entanglement of symmetric
states {\em equivalent} to the one provided by the negativities.

As a consequence of this equivalence, it is tempting to conjecture
that there exists a unique quantification of entanglement for all
two-mode Gaussian states, embodied by the smallest symplectic
eigenvalue $\tilde \nu_-$ of the partially transposed CM, and that
the different measures simply provide trivial rescalings of the same
unique quantification. In particular, the {\em ordering} induced on
the set of entangled Gaussian state is uniquely defined for the
subset of symmetric two-mode states, and it is independent of the
chosen measure of entanglement. However, in Sec. \ref{secorder} we
will indeed show, within the general framework of Gaussian
entanglement measures (see Sec.~\ref{SecGEMS}), that different
families of entanglement monotones induce, in general, different
orderings on the set of nonsymmetric Gaussian states, as
demonstrated in \cite{ordering}.

Let us mention that, for nonsymmetric two-mode Gaussian states,
lower bounds on the entanglement of formation are available
\cite{rigesc04}.

\subsection{EPR correlations}\label{SecEPRcorrel} A
deeper insight on the relationship between correlations and the
smallest symplectic eigenvalue $\tilde{\nu}_{-}$ of the partial
transpose is provided by the following observation, which holds,
again, for symmetric two-mode Gaussian states only \cite{extremal}.

Let us define the EPR correlation $\xi$ \cite{giedke03,rigesc04} of
a CV two-mode quantum state as \be
\xi\equiv\frac{\delta_{\hat{q}_{1}-\hat{q}_{2}}+
\delta_{\hat{p}_{1}+\hat{p}_{2}}}{2} = \frac{{\rm
Tr}\,\gr{\sigma}}{2}-\sigma_{13} +\sigma_{24} \, , \ee where
$\delta_{\hat{o}}=\avr{\hat o^2}-\avr{\hat o}^2 $ for an operator
$\hat o$. If $\xi\ge 1$ then the state does not possess non-local
correlations \cite{Duan00}. The idealized EPR state \cite{EPR35}
(simultaneous eigenstate of the commuting observables $\hat
q_{1}-\hat q_{2}$ and $\hat p_{1}+\hat p_{2}$) has $\xi=0$. As for
standard form two-mode Gaussian states, \eq{stform2}, one has \bea
\delta_{\hat{q}_{1}-\hat{q}_{2}}&=&a+b-2c_{+}\; ,\\
\delta_{\hat{p}_{1}+\hat{p}_{2}}&=&a+b+2c_{-}\; ,\\
\xi&=&a+b-c_{+}+c_{-}\; . \eea Notice that $\xi$ is not by itself a
good measure of correlation because, as one can easily verify, it is
not invariant under local symplectic operations. In particular,
applying local squeezings with parameters $r_{i}=\log v_{i}$ and
local rotations with angles $\varphi_{i}$ to a standard form state,
we obtain \be \xi_{v_{i},\vartheta}=
\frac{a}{2}\left(v_{1}^{2}+\frac{1}{v_{1}^{2}}\right)+
\frac{b}{2}\left(v_{2}^{2}+\frac{1}{v_{2}^{2}}\right)
-\left(c_{+}v_{1}v_{2}-\frac{c_{-}}{v_{1}v_{2}}\right)\cos{\vartheta}
\, , \ee with $\vartheta=\varphi_{1}+\varphi_{2}$. Now, the quantity
\[
\bar\xi\equiv\min_{v_{i},\vartheta}\xi_{v_{i},\vartheta}
\]
has to be $Sp_{(2,\mathbb{R})}\oplus Sp_{(2,\mathbb{R})}$ invariant.
It corresponds to the maximal amount of EPR correlations which can
be distributed in a two-mode Gaussian state by means of local
operations. Minimization in terms of $\vartheta$ is immediate,
yielding $\bar \xi=\min_{v_{i}}\xi_{v_{i}}$, with \be \xi_{v_{i}}=
\frac{a}{2}\left(v_{1}^{2}+\frac{1}{v_{1}^{2}}\right)+
\frac{b}{2}\left(v_{2}^{2}+\frac{1}{v_{2}^{2}}\right)
-\left|c_{+}v_{1}v_{2}-\frac{c_{-}}{v_{1}v_{2}}\right| \, . \ee The
gradient of such a quantity is null if and only if \bea a
\left(v_{1}^{2}-\frac{1}{v_{1}^{2}}\right)-|c_{+}|v_{1}v_{2}-
\frac{|c_{-}|}{v_{1}v_{2}}&=&0 \, ,
\label{grad1} \\
b \left(v_{2}^{2}-\frac{1}{v_{2}^{2}}\right)-|c_{+}|v_{1}v_{2}-
\frac{|c_{-}|}{v_{1}v_{2}}&=&0 \, , \label{grad2} \eea where we
introduced the position $c_{+}c_{-}<0$, necessary to have
entanglement, see \eq{detgammanegative}. Eqs.~{\rm(\ref{grad1},
\ref{grad2})} can be combined to get \be
a\left(v_{1}^{2}-\frac{1}{v_{1}^{2}}\right)=
b\left(v_{2}^{2}-\frac{1}{v_{2}^{2}}\right) \; . \label{conda} \ee
Restricting to the symmetric ($a=b$) entangled ($\Rightarrow
c_{+}c_{-}<0$) case, \eq{conda} and the fact that $v_{i}>0$ imply
$v_{1}=v_{2}$. Under such a constraint, minimizing $\xi_{v_{i}}$
becomes a trivial matter and yields \be \bar\xi=2
\sqrt{(a-|c_{+}|)(a-|c_{-}|)}=2 \tilde{\nu}_{-} \; . \ee We thus see
that the smallest symplectic eigenvalue of the partially transposed
state is endowed with a direct physical interpretation: it
quantifies the greatest amount of EPR correlations which can be
created in a Gaussian state by means of local operations. \par As
can be easily verified by a numerical investigation, such a simple
interpretation is lost for nonsymmetric two-mode Gaussian states.
This fact properly exemplifies the difficulties of handling
optimization problems in nonsymmetric instances, encountered, {\em
e.g.}~in the computation of the entanglement of formation of such
states \cite{giedke03}. It also confirms that, in the special subset
of two-mode (mixed) symmetric Gaussian states, there is a unique
interpretation for entanglement and a unique ordering of entangled
states belonging to that subset, as previously remarked.

\section{Entanglement versus Entropic measures}\label{secEntvsMix}

Here we aim at a characterization of entanglement of two-mode
Gaussian states and in particular at unveiling its relationship with
the degrees of information associated with the global state of the
system, and with the reduced states of each of the two subsystems.

As extensively discussed in Chapter \ref{ChapEnt}, the concepts of
entanglement and  information encoded in a quantum state are closely
related. Specifically, for pure states bipartite entanglement is
equivalent to the lack of information (mixedness) of the reduced
state of each subsystem. For mixed states, each subsystem has its
own level of impurity, and moreover the global state is itself
characterized by a nonzero mixedness. Each of these properties can
be interpreted as information on the preparation of the respective
(marginal and global) states, as clarified in Sec.~\ref{ParInfo}.
Therefore, by properly accessing these degrees of information one is
intuitively expected to deduce, to some extent, the status of the
correlations between the subsystems.

The main question we are posing here is
\begin{quote}
 {\it What
can we say about the quantum correlations existing between the
subsystems of a quantum multipartite system in a mixed state, if we
know the degrees of information carried by the global and the
reduced states?}
\end{quote}
In this Section we provide an answer, which can be summarized as
\emph{``almost everything''},  in the context of two-mode Gaussian
states of CV systems. Based on our published work
\cite{prl,extremal,polacchi}, we will demonstrate, step by step, how
the entanglement --- specifically, measured by the logarithmic
negativity --- of two-mode Gaussian states can be accurately (both
qualitatively and quantitatively) characterized by the knowledge of
global and marginal degrees of information, quantified by the
purities, or by the generalized entropies of the global state and of
the reduced states of the two subsystems.

\subsection{Entanglement vs Information (I) -- Maximal negativities at fixed global purity}

The first step towards giving an answer to our original question is
to investigate the properties of extremally entangled states at a
given degree of global information. Let us mention that, for
two-qubit systems, the existence of maximally entangled states at
fixed mixedness (MEMS) was first found numerically by Ishizaka and
Hiroshima \cite{Ishizaka00}. The discovery of such states spurred
several theoretical works \cite{Verstraete01,Munro01}, aimed at
exploring the relations between different measures of entanglement
and mixedness \cite{Nemoto03} (strictly related to the questions of
the ordering induced by these different measures
\cite{VirmaniPlenioTheorem,VerstraeteJPA}, and of the volume of the
set of mixed entangled states \cite{Zyczkowski98,Zyczkowski99}).

Unfortunately, it is easy to show that a similar analysis in the CV
scenario is meaningless. Indeed, for any fixed, finite global purity
$\mu$ there exist infinitely many Gaussian states which are
infinitely entangled. As an example, we can consider the class of
(nonsymmetric) two-mode squeezed thermal states. Let $\hat
U_{1,2}(r)$, \eq{tmsU},  be the two mode squeezing operator with
real squeezing parameter $r\ge0$, and let
$\varrho^{_\otimes}_{\nu_i}$ be a tensor product of thermal states
with CM ${\gr\nu}_{\nu_{\mp}}= {\mathbbm 1}_{2}\nu_- \oplus
{\mathbbm 1}_{2}\nu_{+}$, where $\nu_{\mp}$ denotes, as usual, the
symplectic spectrum of the state. Then, a nonsymmetric two-mode
squeezed thermal state $\xi_{\nu_{i},r}$ is defined as
${\xi}_{\nu_{i},r}=\hat U(r)\varrho^{_\otimes}_{\nu_i}\hat
U^{\dag}(r)$, corresponding to a standard form CM with
\begin{eqnarray}
a&=&{\nu_-}\cosh^{2}r+{\nu_{+}}\sinh^{2}r \; ,\quad
b\,\,=\,\,{\nu_{-}}\sinh^{2}r+{\nu_{+}}\cosh^{2}r \; ,\label{2mst}\\
c_{\pm}&=&\pm\frac{\nu_-+\nu_+}{2}\sinh2r \;. \nonumber
\end{eqnarray}
Inserting Eqs.~\pref{2mst} into \eq{sepcomppt} yields the following
condition for a two-mode squeezed thermal state $\xi_{\nu_{i},r}$ to
be entangled \be \sinh^2(2r) >
\frac{(\nu_+^2-1)(\nu_-^2-1)}{(\nu_-+\nu_+)^2} \; . \label{2mstppt}
\ee

For simplicity we can consider the symmetric instance
($\nu_-=\nu_+=1/\sqrt\mu$) and compute the logarithmic negativity
\eq{en}, which takes the expression $$E_\N(r,\mu) = -(1/2)\log[{\rm
e}^{-4r}/\mu]\,.$$ Notice how the completely mixed state ($\mu
\rightarrow 0$) is always separable while, for any $\mu > 0$, we can
freely increase the squeezing $r$ to obtain Gaussian states with
arbitrarily large entanglement. For fixed squeezing, as naturally
expected, the entanglement decreases with decreasing degree of
purity of the state, analogously to what happens in
discrete-variable MEMS \cite{Nemoto03}.

It is in order to remark that the notion of Gaussian maximally
entangled mixed states acquires  significance if also the mean
energy is kept fixed \cite{ziman}, in which case the maximum
entanglement is indeed attained by (nonsymmetric) thermal squeezed
states. This is somehow expected given the result we are going to
demonstrate, namely that those states play the role of maximally
entangled Gaussian states at fixed global {\em and} local
mixednesses (GMEMS) \cite{prl,extremal}.

\subsection{Entanglement vs Information (II) -- Maximal negativities at fixed local
purities}\label{SecMEMMS}

The next step in the analysis is the unveiling of the relation
between the entanglement of a Gaussian state of CV systems and the
degrees of information related to the subsystems. Maximally
entangled states for given marginal mixednesses (MEMMS) had been
previously introduced and analyzed in detail in the context of qubit
systems \cite{AdessoPRA}. The MEMMS provide a suitable
generalization of pure states, in which the entanglement is
completely quantified by the marginal degrees of mixedness.

 For two-mode Gaussian states, it follows from the expression \eq{n1} of
$\tilde\nu_-$ that, for fixed marginal purities $\mu_{1,2}$ and
seralian $\Delta$, the logarithmic negativity is strictly increasing
with increasing $\mu$. By imposing the saturation of the upper bound
of \eq{consmu}, \be \label{mumax}
\mu=\mu^{\max}(\mu_{1,2})\equiv(\mu_1 \mu_2)/(\mu_1 \mu_2 +
\abs{\mu_1-\mu_2})\,,\ee we determine the most pure states for fixed
marginals; moreover, choosing $\mu=\mu^{\max}(\mu_{1,2})$
immediately implies that the upper and the lower bounds on $\Delta$
of \eq{deltabnd} coincide and $\Delta$ is uniquely determined in
terms of $\mu_{1,2}$, $\Delta = 1+1/\mu^{\max}$. This means that the
two-mode states with maximal purity for fixed marginals are indeed
the Gaussian maximally entangled states for fixed marginal
mixednesses (GMEMMS) \cite{extremal}. They can be seen as the CV
analogues of the MEMMS \cite{AdessoPRA}. The standard form of GMEMMS
can be determined by Eqs.~\pref{gabc}, yielding
\begin{equation}\label{sfgmemms}
c_{\pm} = \pm \sqrt{\frac{1}{\mu_1 \mu_2}-\frac{1}{\mu^{\max}}}
\end{equation}

\begin{figure}[t!]
\centering
 \subfigure[] {\includegraphics[width=6cm]{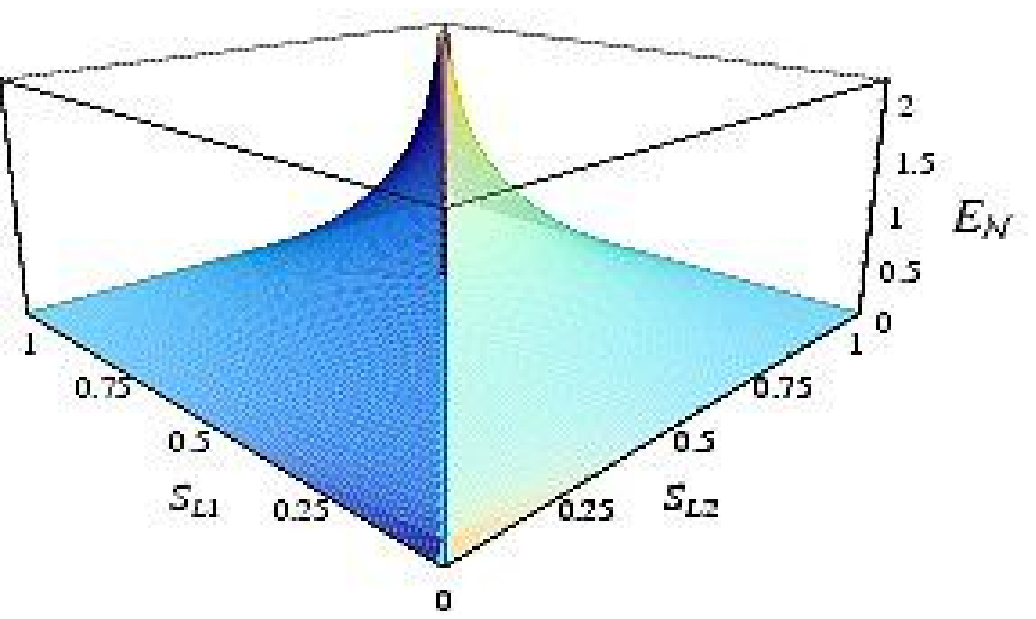}} \hspace{.2cm}
\subfigure[] {\includegraphics[width=6cm]{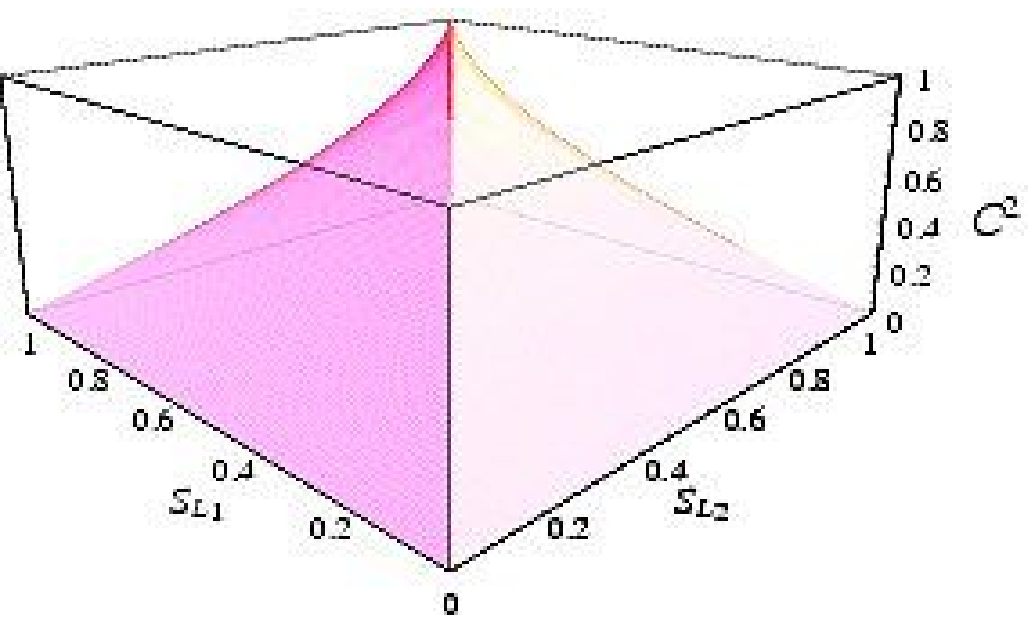}}
\caption{Plot of the maximal entanglement achievable by quantum
systems with given
 marginal linear entropies: {\rm (a)} logarithmic negativity of continuous
 variable GMEMMS, introduced in \cite{extremal},
which saturate the upper bound of inequality \pref{consmu};
 {\rm (b)} tangle of two-qubit MEMMS, introduced in \cite{AdessoPRA}.}
 \label{gmemms}
\end{figure}
In Fig.~\ref{gmemms} the logarithmic negativity of GMEMMS is plotted
as a function of the marginal linear entropies $S_{L{1,2}} \equiv 1-
\mu_{1,2}$, in comparison with the behavior of the tangle (an
entanglement monotone equivalent to the entanglement of formation
for two qubits \cite{Wootters98,CKW}, see Sec.~\ref{SecEnt2Q}) as a
function of $S_{L{1,2}}$ for discrete variable MEMMS. Notice, as a
common feature, how the maximal entanglement achievable by quantum
mixed states rapidly increases with increasing marginal mixednesses
(like in the pure-state instance) and decreases with increasing
difference of the marginals. This is natural, because the presence
of quantum correlations between the subsystems implies that they
should possess rather similar amounts of quantum information. Let us
finally mention that the ``minimally'' entangled states for fixed
marginals, which saturate the lower bound of \eq{consmu} ($\mu=\mu_1
\mu_2$), are just the tensor product states, {\em i.e.~}states
without any (quantum or classical) correlations between the
subsystems.\footnote{\sf Note that this is no longer true in
two-qubit systems. In that instance, there exist LPTP (``less pure
than product'') states, whose global purity is smaller than the
product of their two marginal purities, implying that they carry
 less information than the uncorrelated product states.
Surprisingly, they can even be entangled, meaning that they somehow
encode {\em negative} quantum correlations. The LPTP states of two
qubits have been discovered and characterized in \cite{AdessoPRA}.
Recently, the notion of negative quantum information has been
reinterpreted in a communication context \cite{qinfonegative}.}

\subsection{Entanglement vs Information (III) --
Maximal and minimal negativities at fixed global and local
purities}\label{secprl}

What we have shown so far, by simple analytical bounds, is a general
trend of increasing entanglement with increasing global purity, and
with decreasing marginal purities and difference between them. We
now wish to exploit the joint information about global and marginal
degrees of purity to achieve a significative characterization of
entanglement, both qualitatively and quantitatively. Let us first
investigate the role played by the seralian $\Delta$ in the
characterization of the properties of two-mode Gaussian states. To
this aim, we analyze the dependence of the eigenvalue
$\tilde{\nu}_-$ on $\Delta$, for fixed $\mu_{1,2}$ and $\mu$:
\begin{equation}
  \left. \frac{\partial\ \tilde{\nu}^2_{-}}{\partial\ \Delta}
  \right|_{\mu_1,\,\mu_2,\,\mu} \,
 = \; \frac12 \left(
\frac{\tilde{\Delta}}{\sqrt{\tilde{\Delta}^2 -{\frac{1}{4 \mu^2}}}}
-1 \right) \,
> 0 \, .
\label{derivata}\end{equation} The smallest symplectic eigenvalue of
the partially transposed state $\tilde\sig$ is strictly monotone in
$\Delta$. Therefore the entanglement of a generic Gaussian state
$\gr{\sigma}$ with given global purity $\mu$ and marginal purities
$\mu_{1,2}$, strictly increases with decreasing $\Delta$. The
seralian $\Delta$ is thus endowed with a direct physical
interpretation: given the global and the two marginal purities, it
exactly determines the amount of entanglement of the state.
Moreover, due to inequality \pref{deltabnd}, $\Delta$ is constrained
both by lower and upper bounds; therefore, both  {\em maximally}
and, remarkably,  \emph{minimally} entangled Gaussian states exist,
at fixed global and local degrees of purity. This fact admirably
elucidates the relation between quantum correlations and information
in two-mode Gaussian states \cite{prl,extremal,polacchi}, summarized
as follows.

\medskip

\begin{itemize}
\item[\ding{226}]
 \noindent{\rm\bf Entanglement at given degrees of information encoded in two-mode Gaussian states.}
{\it The entanglement, quantified by the negativities, of two-mode
(mixed) Gaussian states is tightly bound from above and from below
by functions of the global and the marginal purities, with only one
remaining degree of freedom related to the symplectic invariant
$\Delta$.}
\smallskip
\end{itemize}

\subsubsection{GMEMS and GLEMS: Extremally entangled states and purity-based separability criteria} \label{SecGmemsGlems}

We now aim to characterize extremal (maximally and minimally)
entangled Gaussian states for fixed global and marginal purities,
along the lines of \cite{prl,extremal}. As it is clear from
\eq{mu12}, the standard form of states with fixed marginal purities
always satisfies $a=1/\mu_1$, $b=1/\mu_2$. Therefore the complete
characterization of maximally and minimally entangled states is
achieved by specifying the expression of their  coefficients
$c_{\mp}$.

\smallskip

\noindent {\rm {\bf GMEMS}.}--- Let us first consider the states
saturating the lower bound in \eq{deltabnd}, which entails
\emph{maximal} entanglement. They are Gaussian maximally entangled
states for fixed global and local purities (GMEMS), admitting the
following standard form parametrization
\begin{equation} \label{gnsm}
c_{\pm}= \pm \sqrt{\frac{1}{\mu_1 \mu_2}-\frac{1}{\mu}} \; .
\end{equation}
It is easily seen that such states belong to the class of asymmetric
two-mode squeezed thermal states, \eq{2mst}, with squeezing
parameter and symplectic spectrum given by \be\label{2sqp}
\tanh2r=2(\mu_1\mu_2-\mu_1^2\mu_2^2/\mu)^{1/2}/(\mu_1+\mu_2)\, , \ee
\be
\nu_{\mp}^2=\frac{1}{\mu}+\frac{(\mu_1-\mu_2)^2}{2\mu_1^2\mu_2^2}\mp
\frac{|\mu_1-\mu_2|}{2\mu_1\mu_2}\sqrt{\frac{(\mu_1-\mu_2)^2}{\mu_1^2\mu_2^2}
+\frac{4}{\mu}}\, . \ee In particular, any GMEMS can be written as
an entangled two-mode squeezed thermal state [satisfying Ineq.~{\rm
(\ref{2mstppt})}]. This provides a characterization of two-mode
thermal squeezed states as maximally entangled states for given
global and marginal purities. We can restate this result as follows:
given an initial tensor product of (generally different) thermal
states, the unitary operation providing the maximal entanglement for
given values of the local purities $\mu_i$'s is given by a two-mode
squeezing, with squeezing parameter determined by \eq{2sqp}. Note
that the same states have also been proven to be maximally entangled
at fixed global purity and mean energy \cite{ziman}, as already
mentioned. Nonsymmetric two-mode thermal squeezed states turn out to
be {\em separable} in the range
\begin{equation}
\label{gnsmsep} \mu \le \frac{\mu_1 \mu_2}{\mu_1 + \mu_2 - \mu_1
\mu_2} \, .
\end{equation}
In such a {\em separable region} in the space of purities, no
entanglement can occur for states of the form \eq{gnsm}, while,
outside this region, they are properly GMEMS. As a consequence, we
obtain a sufficient entropic condition for separability: {\em all}
two-mode Gaussian states whose purities fall in the separable region
defined by inequality \pref{gnsmsep}, are separable.

\smallskip

\noindent {\rm {\bf GLEMS}.}--- We now consider the states that
saturate the upper bound in \eq{deltabnd}. They determine the class
of Gaussian least entangled states for given global and local
purities (GLEMS). Violation of inequality \pref{gnsmsep} implies
that
$$
1 + \frac{1}{\mu^2} \, \le \, \frac{(\mu_1 + \mu_2)^{2}}{\mu_1^{2}
\mu_2^{2}} - \frac{2}{\mu} \; .
$$
Therefore, outside the separable region, GLEMS fulfill
\begin{equation}
\label{glm} \D = 1+\frac{1}{\mu^2} \; .
\end{equation}
Considering the symplectic diagonalization of Gaussian states and
the definition of the seralian $\Delta = \nu_-^2 + \nu_+^2$,
\eq{seralian}, it immediately follows that the $Sp_{(4,\mathbb{R})}$
invariant condition \pref{glm} is fulfilled if and only if the
symplectic spectrum of the state takes the form $\nu_-=1$,
$\nu_+=1/\mu$. We thus find that GLEMS are characterized by a
peculiar spectrum, with all the mixedness concentrated in one
`decoupled' quadrature. Moreover, by comparing Eq.~\pref{glm} with
the uncertainty relation \pref{sympheis}, it follows that GLEMS are
the mixed Gaussian states of partial minimum uncertainty (see
Sec.~\ref{SecSympHeis}).  They are therefore the most ``classical''
mixed Gaussian states and, in a sense, this is compatible with their
property of having minimum entanglement at fixed purities. GLEMS are
determined by the standard form correlation coefficients
\begin{eqnarray}
c_{\pm} \!\!&=&\!\! \frac14 \sqrt{\mu_1 \mu_2
\left[-\frac{4}{\mu^2}+\left(1+\frac{1}{\mu^2}-\frac{(\mu_1-\mu_2)^2}{\mu_1^2
\mu_2^2}\right)^2\right]}\nonumber \\
\!\!&\pm&\!\! \frac{1}{4\mu}\sqrt{-4\mu_1\mu_2+\frac{\left[\left(
1+\mu^2\right)\mu_1^2\mu_2^2-\mu^2{\left(\mu_1+\mu_2\right)}^2\right]
^2}{\mu^2\mu_1^3\mu_2^3}} \; . \nonumber \\
\label{glems}
\end{eqnarray}
Quite remarkably, recalling the analysis presented in
Sec.~\ref{mmix}, it turns out that the GLEMS at fixed global and
marginal purities are also states of minimal global $p-$entropy for
$p<2$, and of maximal global $p-$entropy for $p>2$.
\par According to the PPT criterion, GLEMS are
separable only if $$\mu \le \mu_1 \mu_2 / \sqrt{\mu_1^2 + \mu_2^2 -
\mu_1^2 \mu_2^2}\,.$$ Therefore, in the range
\begin{equation}
\label{gnslsep} \frac{\mu_1 \mu_2}{\mu_1 + \mu_2 - \mu_1 \mu_2} <
\mu \le \frac{\mu_1 \mu_2}{\sqrt{\mu_1^2 + \mu_2^2 - \mu_1^2
\mu_2^2}}
\end{equation}
both separable and entangled states two-mode Gaussian states can be
found. Instead, the region \be \mu>\frac{\mu_1 \mu_2}{\sqrt{\mu_1^2
+ \mu_2^2 - \mu_1^2 \mu_2^2}} \label{sufent} \ee can only accomodate
\emph{entangled} states. The very narrow region defined by
inequality \pref{gnslsep} is thus the only {\em region of
coexistence} of both entangled and separable Gaussian two-mode mixed
states, compatible with a given triple of purities. We mention that
the sufficient condition for entanglement \pref{sufent}, first
obtained in Ref.~\cite{prl}, has been independently rederived in
Ref.~\cite{fiurasek04}.\par

\smallskip

Let us also recall that for Gaussian states whose purities saturate
the rightmost inequality in \eq{consmu}, GMEMS and GLEMS coincide
and we have a unique class of entangled states depending only on the
marginal purities $\mu_{1,2}$: they are the Gaussian maximally
entangled states for fixed marginals (GMEMMS), introduced in
Sec.~\ref{SecMEMMS}.

\smallskip

All the previous necessary and/or sufficient conditions for
entanglement --- which constitute the strongest entropic criteria
 for separability \cite{NielsKempe01} to date in the case of Gaussian
states --- are collected in Table~\ref{table1} and allow a graphical
display of the behavior of the entanglement of mixed Gaussian states
as shown in Fig.~\ref{fig2D}. These relations classify the
properties of separability of all two-mode Gaussian states according
to their degree of global and marginal purities.

\begin{table}[b]
  \begin{tabular*}{0.8\textwidth}{@{\extracolsep{\fill}} c c }
 \hline
\hline
% \hline \hline
\vspace*{-.2cm}\\
  {\rm \large Degrees of purity} &
{\rm \large Entanglement properties} \vspace*{0.2cm} \\  \hline \vspace*{-.1cm}\\
  $\mu<\mu_1 \mu_2$ & {\rm unphysical region} \vspace*{0.3cm}\\
  $\mu_1 \mu_2 \; \le \; \mu \; \le \; \frac{\mu_1 \mu_2}{\mu_1 + \mu_2 - \mu_1 \mu_2}$ &
  {\rm \emph{separable} states}
 \vspace*{0.3cm} \\
  $\frac{\mu_1 \mu_2}{\mu_1 + \mu_2 - \mu_1 \mu_2} < \mu \le
  \frac{\mu_1 \mu_2}{\sqrt{\mu_1^2 + \mu_2^2 - \mu_1^2 \mu_2^2}}$ &
  {\rm \emph{coexistence} region}
\vspace*{0.3cm}\\
  $\frac{\mu_1 \mu_2}{\sqrt{\mu_1^2 + \mu_2^2 - \mu_1^2 \mu_2^2}} <
  \mu \le \frac{\mu_1 \mu_2}{\mu_1 \mu_2 + \abs{\mu_1-\mu_2}}$ &
  {\rm \emph{entangled} states}
  \vspace*{0.3cm}\\
  $\mu > \frac{\mu_1 \mu_2}{\mu_1 \mu_2 + \abs{\mu_1-\mu_2}}$ &
  {\rm unphysical region}
  \vspace*{0.3cm}\\ \hline
\end{tabular*}
\caption{Classification of two-mode Gaussian states and of their
properties of separability according to their degrees of global
purity $\mu$ and of marginal purities $\mu_1$ and $\mu_2$.}
\label{table1}
 \end{table}

\begin{figure}[t!]
\centering
  \includegraphics[width=12cm]{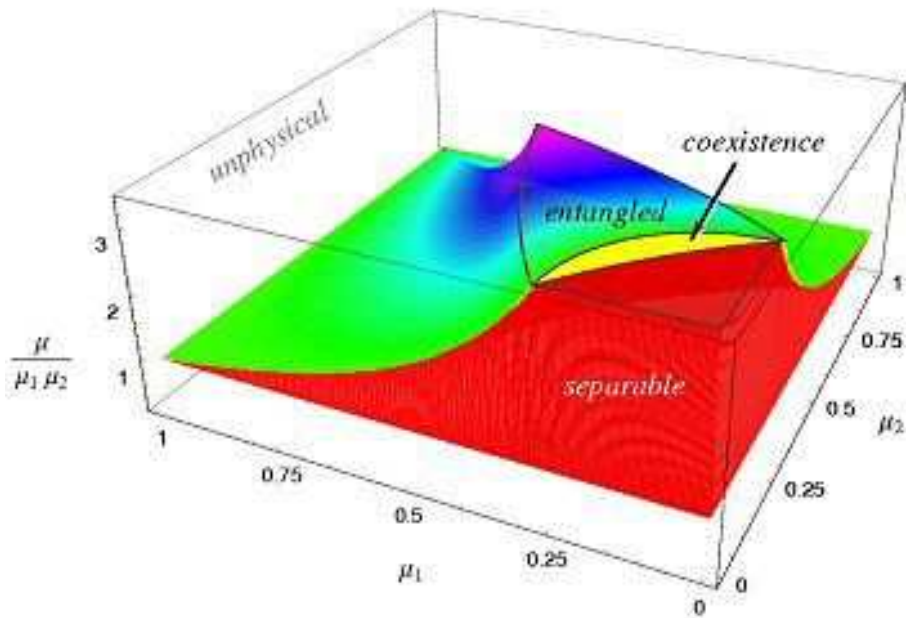}\\
  \caption{
  Summary of entanglement properties of two-mode (nonsymmetric)
  Gaussian states in the space of marginal purities $\mu_{1,2}$
  ($x$- and $y$-axes) and global purity $\mu$. In fact, on the $z$-axis
  we plot the ratio $\mu/\mu_1\mu_2$ to gain a better graphical
  distinction between the various regions. In this space, all physical
  states lay between the horizontal plane $z=1$ representing product states,
  and the upper limiting surface representing GMEMMS. Separable and entangled
  states are well separated except for a narrow region of coexistence (depicted
  in yellow). Separable states fill the region depicted in red,
  while in the region containing only entangled states we have depicted
  the average logarithmic negativity \eq{average}, growing from green to
  magenta.
  The mathematical relations defining the boundaries between all these regions
  are collected in Table~\ref{table1}.
  The three-dimensional envelope is cut at $z=3.5$.}
  \label{fig2D}
\end{figure}

\subsection{Entanglement vs Information (IV) --
Maximal and minimal negativities at fixed global and local
generalized entropies}\label{genent}

Here we introduce a more general characterization of the
entanglement of generic two-mode Gaussian states, by exploiting the
generalized $p-$entropies, defined by \eq{pgen} and computed for
Gaussian states in \eq{pgau}, as measures of global and marginal
mixedness. For ease of comparison we will carry out this analysis
along the same lines followed before, by studying the explicit
behavior of the global invariant $\D$, directly related to the
logarithmic negativity $E_\N$ at fixed global and marginal purities.
This study will clarify the relation between $\D$ and the
generalized entropies $S_p$ and the ensuing consequences for the
entanglement of Gaussian states.

We begin by observing that the standard form CM $\gr\sigma$ of a
generic two-mode Gaussian state can be parametrized by the following
quantities: the two marginals $\mu_{1,2}$ (or any other marginal
$S_{p_{1,2}}$ because all the local, single-mode entropies are
equivalent for any value of the integer $p$), the global $p-$entropy
$S_p$ (for some chosen value of the integer $p$), and the global
symplectic invariant $\D$. On the other hand,
Eqs.~{\rm(\ref{pgen},\ref{pgau},\ref{sympeig})} provide an explicit
expression for any $S_p$ as a function of $\mu$ and $\D$. Such an
expression can be exploited to study the behavior of $\D$ as a
function of the global purity $\mu$, at fixed marginals and global
$S_p$ (from now on we will omit the explicit reference to fixed
marginals). One has \bea \nonumber \left.\frac{\partial
\mu}{\partial \D}\right\vert_{S_p}=-\frac2{R^2} \left.
\frac{\partial R}{\partial \D}\right\vert_{S_p}=\frac2{R^2}
\frac{\left. \partial S_p / \partial \D \right\vert_{R}}{\left.
\partial
S_p / \partial R \right\vert_{\D}}\\
=\frac2{R^2} \frac{N_p(\D,R)}{D_p(\D,R)}\label{dubbino} \; , \eea
where we have defined the inverse participation ratio \be
R\equiv\frac2{\mu} \; , \ee and the remaining quantities $N_p$ and
$D_p$ read
\begin{eqnarray}
N_p(\D,R) &=& \left[
(R+2+2\sqrt{\D+R})^{p-1} -(R+2-2\sqrt{\D+R})^{p-1} \right] \sqrt{\D-R} \nonumber\\
   &-& \left[
(R-2+2\sqrt{\D-R})^{p-1}-(R-2-2\sqrt{\D-R})^{p-1} \right]\sqrt{\D+R} \; , \nonumber\\
&& \nonumber \\
D_p(\D,R) &=& \left[(\sqrt{\D+R}+1)
(R+2+2\sqrt{\D+R})^{p-1}\right.\nonumber\\
&+&\left.(\sqrt{\D+R}-1)(R+2-2\sqrt{\D+R})^{p-1}
\right] \sqrt{\D-R}\nonumber\\
&-& \left[(\sqrt{\D-R}+1)
(R-2-2\sqrt{\D-R})^{p-1}\right.\nonumber\\
&+&\left.(\sqrt{\D-R}-1)(R-2+2\sqrt{\D-R})^{p-1} \right] \sqrt{\D+R}
\label{npdp} \; .
\end{eqnarray}
Now, it is easily shown that the ratio $N_p(\D,R)/D_p(\D,R)$ is
increasing with increasing $p$ and has a zero at $p=2$ for any
$\D,R$; in particular, its absolute minimum $(-1)$ is reached in the
limit $(\D\rightarrow2,\,R\rightarrow2,\,p\rightarrow1)$. Thus the
derivative \eq{dubbino} is negative for $p<2$, null for $p=2$ (in
this case $\D$ and $S_2=1-\mu$ are of course regarded as independent
variables) and positive for $p>2$. This implies that, for given
marginals, keeping fixed any global $S_p$ for $p<2$ the minimum
(maximum) value of $\Delta$ corresponds to the maximum (minimum)
value of the global purity $\mu$. Instead, by keeping fixed any
global $S_p$ for $p>2$ the minimum of $\Delta$ is always attained at
the minimum of the global purity  $\mu$. In other words, for fixed
marginal entropies and global $S_V$, the quantity $\D$ decreases
with increasing global purity, while for fixed marginal properties
and global $S_p\ (p>2)$,  $\D$ increases with increasing $\mu$.

This observation allows to determine rather straightforwardly the
states with extremal $\D$. They are extremally entangled states
because, for fixed global and marginal entropies, the logarithmic
negativity of a state is determined only by the one remaining
independent global symplectic invariant, represented by $\D$ in our
choice of parametrization.
%In fact, looking \emph{e.g.} for a state with the
%minimal $\mu$ (or the maximal $S_L$) at fixed $S_V=S_V^\ast$ means
%finding that state such as $S_{V min}(S_L)=S_V^\ast$, \emph{i.e.}
%the state with the minimal $S_V$ for fixed $\mu$ (or $S_L$).
If, for the moment being, we neglect the fixed local purities, then
the states with maximal $\D$ are the states with minimal (maximal)
$\mu$ for a given global $S_p$ with $p<2\ (p>2)$ (see
Sec.~\ref{mmix} and Fig.~\ref{svvssl}). As found in Sec.~\ref{mmix},
such states are minimum-uncertainty two-mode states with mixedness
concentrated in one quadrature. We have shown in
Sec.~\ref{SecGmemsGlems} that they correspond to Gaussian least
entangled mixed states (GLEMS) whose standard form is given by
\eq{glems}. As can be seen from \eq{glems}, these states are
consistent with any legitimate physical value of the local
invariants $\mu_{1,2}$. We therefore conclude that {\em all}
Gaussian states with {\em maximal} $\D$ for any fixed triple of
values of global and marginal entropies are GLEMS. \par Viceversa
one can show that {\em all} Gaussian states with {\em minimal} $\D$
for any fixed triple of values of global and marginal entropies are
Gaussian maximally entangled mixed states (GMEMS). This fact is
immediately evident in the symmetric case because the extremal
surface in the  $S_p$ vs.~$S_L$ diagrams is always represented by
symmetric two-mode squeezed thermal states (symmetric GMEMS). These
states are characterized by a degenerate symplectic spectrum and
encompass only equal choices of the local invariants: $\mu_1=\mu_2$.
In the nonsymmetric case, the given values of the local entropies
are different, and the extremal value of $\D$ is further constrained
by inequality \pref{deltabnd} \be
\D-R\ge\frac{(\mu_1-\mu_2)^2}{\mu_1^2\mu_2^2}\; . \label{56furba}
\ee
%We
%want to show that the minimal $\D$ at fixed $S_p$ and (different)
%marginals is attained by those states which saturate the bound
%\eq{56furba}, \emph{i.e.} by nonsymmetric GMEMS. To do this, we
%should study the behavior of the function $\D-R$ (depending only
%on global symplectic invariants) with varying $\D$, at fixed
%marginals and global $S_p$.
From \eq{dubbino} it follows that
\begin{equation}\label{minihorror}
\left.\frac{\partial (\D-R)}{\partial
\D}\right\vert_{S_p,\mu_{1,2}}=1+\frac{N_p(\D,R)}{D_p(\D,R)} \ge
0\,,
\end{equation}
because $N_p(\D,R)/D_p(\D,R)>-1$. Thus, $\D-R$ is an increasing
function of $\D$ at fixed $\mu_{1,2}$ and $S_p$, and the minimal
$\D$ corresponds to the minimum of $\D-R$, which occurs if
inequality \pref{56furba} is saturated. Therefore, also in the
nonsymmetric case, the two-mode Gaussian states with minimal $\D$ at
fixed global and marginal entropies are GMEMS.

Summing up, we have shown that the two special classes of  GMEMS and
GLEMS, introduced in Sec.~\ref{secprl} for fixed global and marginal
linear entropies, are always \emph{extremally} entangled two-mode
Gaussian states, whatever triple of generalized global and marginal
entropic measures one chooses to fix. Maximally and minimally
entangled states of CV systems are thus very robust with respect to
the choice of different measures of mixedness. This is at striking
variance with the case of discrete variable systems, where it has
been shown that fixing different measures of mixedness yields
different classes of maximally entangled states \cite{Nemoto03}.

\subsubsection{Inversion of extremally entangled states}
We will now show that the characterization provided by the
generalized entropies leads to some remarkable new insight on the
behavior of the entanglement of CV systems. The crucial observation
is that for a generic $p$, the smallest symplectic eigenvalue of the
partially transposed CM, at fixed global and marginal $p-$entropies,
is {\em not} in general a monotone function of $\D$, so that the
connection between extremal $\D$ and extremal entanglement turns out
to be, in some cases, inverted. In particular, while for $p<2$ the
GMEMS and GLEMS surfaces tend to be more separated as $p$ decreases,
for $p>2$ the two classes of extremally entangled states get closer
with increasing $p$ and, within a particular range of global and
marginal entropies, they exchange their role. GMEMS ({\em
i.e.~}states with minimal $\D$) become minimally entangled states
and GLEMS ({\em i.e.~}states with maximal $\D$) become maximally
entangled states. This inversion always occurs for all $p>2$.

To understand this interesting behavior, let us study the dependence
of the symplectic eigenvalue $\tilde{\nu}_-$ on the global invariant
$\D$ at fixed marginals and at fixed $S_p$ for a generic $p$. Using
Maxwell's relations, we can write
\begin{equation}
\label{maxwell}
\kappa_p\equiv\left.\frac{\partial(2\tilde{\nu}_-^2)}{\partial
\D}\right\vert_{S_p}=\left.\frac{\partial(2\tilde{\nu}_-^2)}{\partial
\D}\right\vert_{R}-\left.\frac{\partial(2\tilde{\nu}_-^2)}{\partial
R}\right\vert_{\D} \cdot \frac{\left. \partial S_p / \partial \D
\right\vert_{R}}{\left. \partial S_p / \partial R
\right\vert_{\D}}\,.
\end{equation}
Clearly, for $\kappa_p>0$ GMEMS and GLEMS retain their usual
interpretation, whereas for $\kappa_p<0$ they exchange their role.
On the \emph{node} $\kappa_p=0$ GMEMS and GLEMS share the same
entanglement, \emph{i.e.}~the entanglement of all Gaussian states at
$\kappa_p=0$ is fully determined by the global and marginal
$p-$entropies alone, and does not depend any more on $\D$. Such
nodes also exist in the case $p \le 2$ in two limiting instances: in
the special case of GMEMMS (states with maximal global purity at
fixed marginals) and in the limit of zero marginal purities. We will
now show that, besides these two asymptotic behaviors, a nontrivial
node appears for all $p>2$, implying that on the two sides of the
node GMEMS and GLEMS indeed exhibit opposite behaviors. Because of
\eq{dubbino}, $\kappa_p$ can be written in the following form
\begin{equation}
\label{horror} \kappa_p=\kappa_2- \frac{R}{\sqrt{\tilde{\D}^2-R^2}}
\, \frac{N_p(\D,R)}{D_p(\D,R)} \; ,
\end{equation}
with $N_p$ and $D_p$ defined by \eq{npdp} and
\begin{eqnarray*}
\tilde{\D} &=& -\D+\frac2{\mu_1^2} + \frac2{\mu_2^2} \; , \\
\kappa_2 &=& -1+\frac{\tilde{\Delta}}{\sqrt{\tilde{\D}^2-R^2}} \; ,
\end{eqnarray*}

\begin{figure}[tb!]
\subfigure[] {\includegraphics[width=5cm]{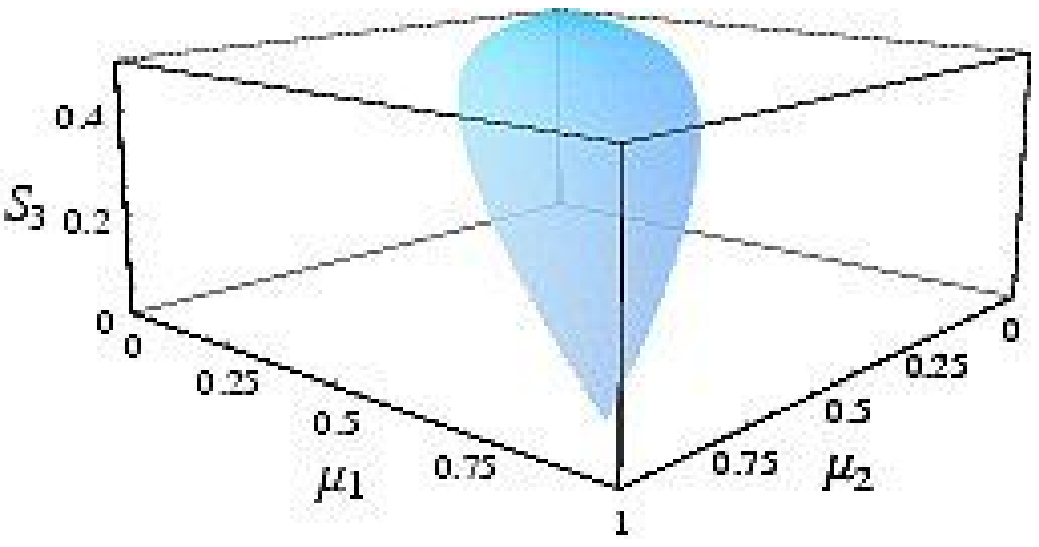}}
\hspace*{0.5cm} \subfigure[]
{\includegraphics[width=5cm]{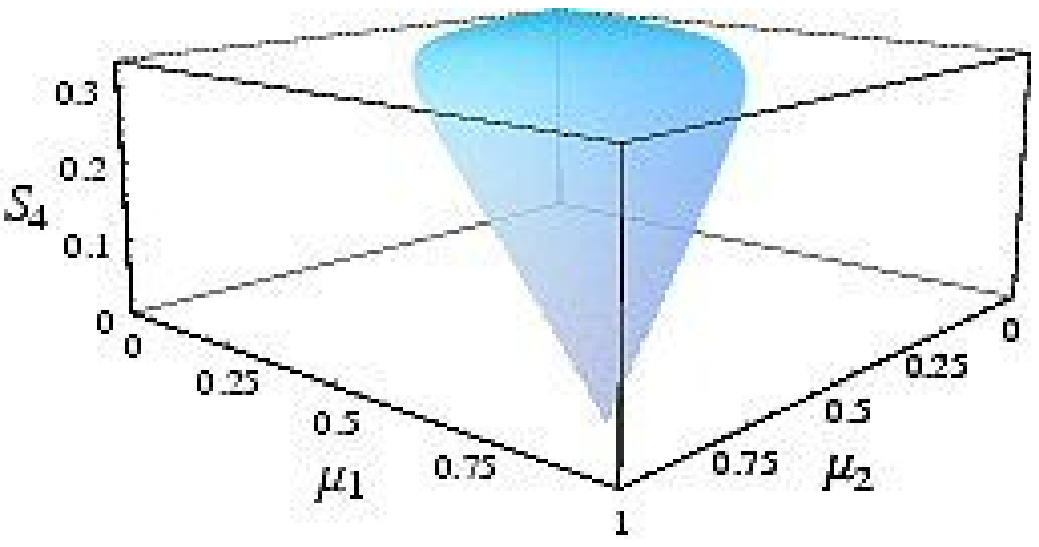}} \caption{Plot of the nodal
surface which solves the equation $\kappa_p=0$ with $\kappa_p$
defined by \eq{horror}, for {\rm (a)} $p=3$ and {\rm (b)} $p=4$. The
entanglement of Gaussian states that lie on the leaf--shaped
surfaces is fully quantified in terms of the marginal purities and
the global generalized entropy {\rm (a)} $S_3$ or {\rm (b)} $S_4$.
The equations of the surfaces in the space
$\mathcal{E}_p\equiv\{\mu_1,\mu_2,S_p\}$ are given by Eqs.~{\rm
(\ref{spleaf}--\ref{spleaf4})}.} \label{figleaf}
\end{figure}

The quantity $\kappa_p$ in \eq{horror} is a function of
$p,\,R,\,\D,\,$ and of the marginals; since we are looking for the
node (where the entanglement is independent of $\D$), we can
investigate the existence of a nontrivial solution to the equation
$\kappa_p=0$ fixing any value of $\D$. Let us choose $\D=1+R^2/4$
that saturates the   uncertainty relation and is satisfied by GLEMS.
With this position, \eq{horror} becomes
\begin{eqnarray}
\label{horrorl}
\kappa_p(\mu_1,\mu_2,R)&=&\kappa_2(\mu_1,\mu_2,R)\nonumber\\
&-&\frac{R}{\sqrt{\left(
\frac2{\mu_1^2}+\frac2{\mu_2^2}-\frac{R^2}4-1\right)^2-R^2}}\,
f_p(R) \, .
\nonumber \\
\end{eqnarray}
\noindent The existence of the node depends then on the behavior of
the function
\begin{equation}
\label{fp} f_p(R) \equiv \frac{2 \left[ (R+2)^{p-2}-(R-2)^{p-2}
\right]} {(R+4)(R+2)^{p-2}-(R-4)(R-2)^{p-2}} \; .
\end{equation}
In fact, as we have already pointed out, $\kappa_2$ is always
positive, while the function $f_p(R)$ is an increasing function of
$p$ and, in particular, it is negative for $p<2$, null for $p=2$ and
positive for $p>2$, reaching its asymptote $2/(R+4)$ in the limit $p
\rightarrow \infty$. This entails that, for $p\le2$, $\kappa_p$ is
always positive, which in turn implies that GMEMS and GLEMS are
respectively maximally and minimally entangled two-mode states with
fixed marginal and global $p-$entropies in the range $p\le2$
(including both Von Neumann and linear entropies). On the other
hand, for any $p>2$ one node can be found solving the equation
$\kappa_p(\mu_{1},\mu_{2},2/\mu)=0$. The solutions to this equation
can be found analytically for low $p$ and numerically for any $p$.
They form a continuum in the space $\{\mu_1,\mu_2,\mu\}$ which can
be expressed as a surface of general equation
$\mu=\mu^\kappa_p(\mu_1,\mu_2)$. Since the fixed variable is $S_p$
and not $\mu$ it is convenient to rewrite the equation of this
surface in the space $\mathcal{E}_p\equiv\{\mu_1,\mu_2,S_p\}$,
keeping in mind the relation~\pref{spglems}, holding for GLEMS,
between $\mu$ and $S_p$. In this way the nodal surface
$(\kappa_p=0)$ can be written in the form
\begin{equation}
\label{spleaf} S_p=S_p^\kappa(\mu_1,\mu_2) \equiv \frac{1 - g_p
\left[ (\mu_p^\kappa(\mu_1,\mu_2))^{-1}\right]}{p-1} \; .
\end{equation}
The entanglement of all Gaussian states whose entropies lie on the
surface $S_p^\kappa(\mu_1,\mu_2)$ is \emph{completely} determined by
the knowledge of $\mu_1$, $\mu_2$ and $S_p$. The explicit expression
of the function $\mu^\kappa_p(\mu_1,\mu_2)$ depends on $p$ but,
being the global purity of physical states, is constrained by the
inequality
$$
\mu_1 \mu_2 \, \le \, \mu^\kappa_p (\mu_1,\mu_2) \, \le \,
\frac{\mu_1 \mu_2}{\mu_1 \mu_2 + \abs{\mu_1-\mu_2}} \; .
$$
The nodal surface of \eq{spleaf} constitutes a `leaf', with base at
the point $\mu^\kappa_p(0,0)=0$ and tip at the point
$\mu^\kappa_p(\sqrt3/2,\sqrt3/2)=1$, for any $p>2$; such a leaf
becomes larger and flatter with increasing $p$ (see
Fig.~\ref{figleaf}).

For $p>2$, the function $f_p(R)$ defined by \eq{fp} is negative but
decreasing with increasing $R$, that is with decreasing $\mu$. This
means that, in the space of entropies $\mathcal{E}_p$, above the
leaf $(S_p>S_p^\kappa)$ GMEMS (GLEMS) are still maximally
(minimally) entangled states for fixed global and marginal
generalized entropies, while below the leaf they are
\emph{inverted}. Notice also that for $\mu_{1,2}>\sqrt3/2$ no node
and so no inversion can occur for any $p$. Each point on the
leaf--shaped surface of \eq{spleaf} corresponds to an entire class
of infinitely many two-mode Gaussian states (including GMEMS and
GLEMS) with the same marginals and the same global
$S_p=S_p^\kappa(\mu_1,\mu_2)$, which are {\em all equally
entangled}, since their logarithmic negativity is completely
determined by $\mu_1,\mu_2$ and $S_p$. For the sake of clarity we
provide the explicit expressions of $\mu_p^\kappa(\mu_1,\mu_2)$, as
plotted in Fig.~\ref{figleaf} for the cases {\rm (a)} $p=3$, and
{\rm (b)} $p=4$,
\begin{eqnarray}
  \mu_3^\kappa(\mu_1,\mu_2) &=&
  \left(\frac{6}{\frac{3}{\mu_1^2}+\frac{3}{\mu_1^2}-2}
  \right)^{\frac12},
  \label{spleaf3}\\
&& \nonumber \\
\mu_4^\kappa(\mu_1,\mu_2) &=& \sqrt{3}\,\mu_1 \mu_2\,\,\,
  \bigg/\,\,\,  \bigg(\mu_1^2+\mu_2^2-2\mu_1^2\mu_2^2+ \nonumber\\
  & & \sqrt{\left(\mu_1^2+\mu_2^2\right)\left(\mu_1^2+\mu_2^2-\mu_1^2 \mu_2^2\right)
  +\mu_1^4 \mu_2^4} \bigg)^{\frac12}.\nonumber \\
\label{spleaf4}
\end{eqnarray}

\begin{figure}[t!]
%\centering
 \hspace{1mm}
 \subfigure[\label{fig2D1}]
{\includegraphics[width=5.7cm]{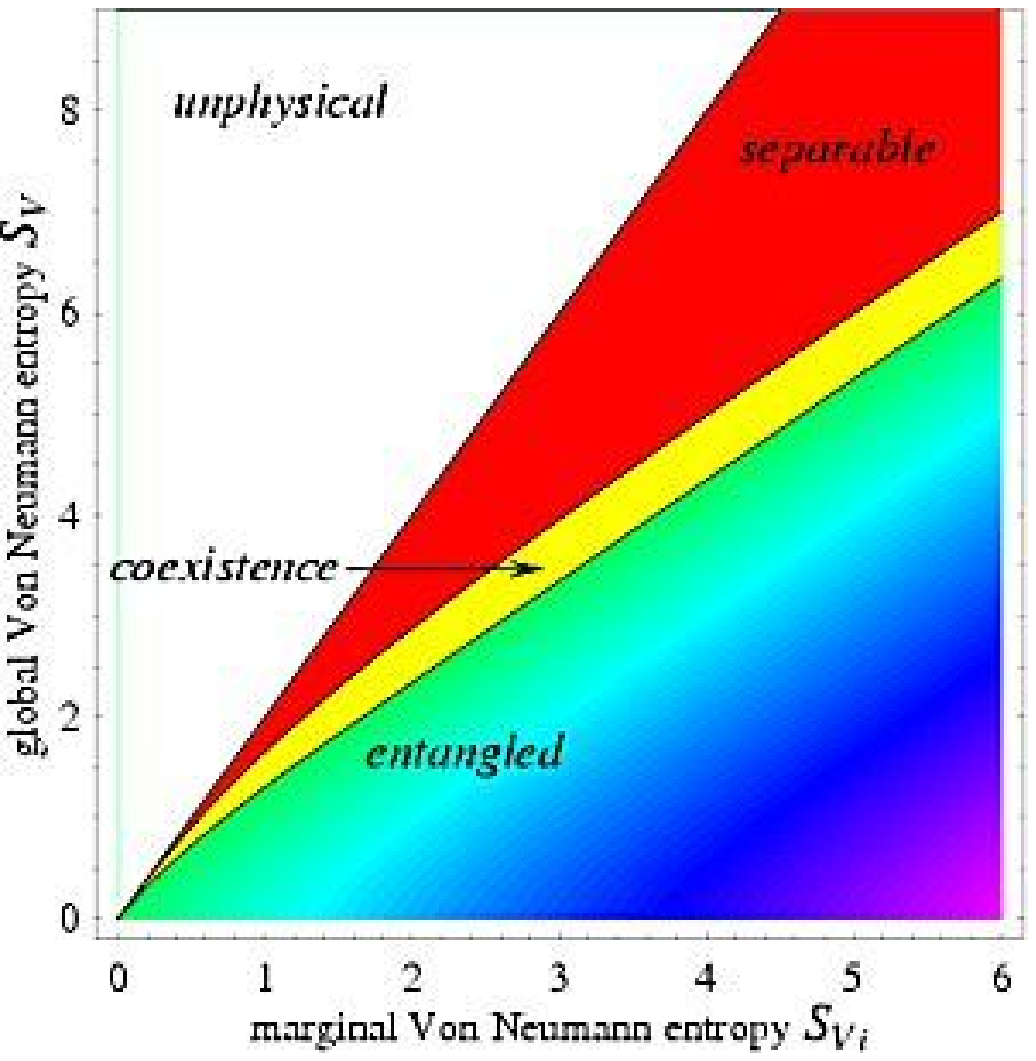}}
 \hspace{4mm}
\subfigure[\label{fig2D2}]
{\includegraphics[width=5.9cm]{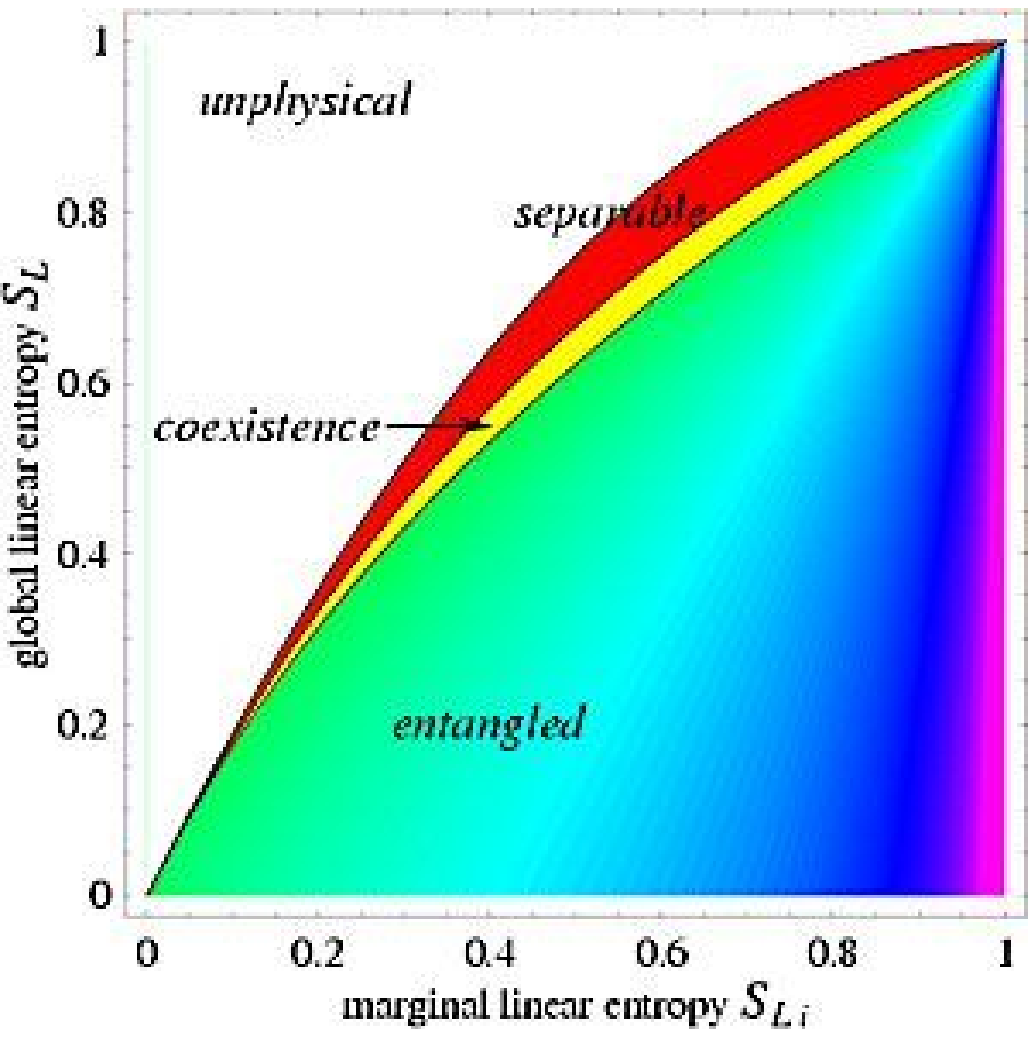}} \\
\subfigure[\label{fig2D3}]
{\includegraphics[width=6cm]{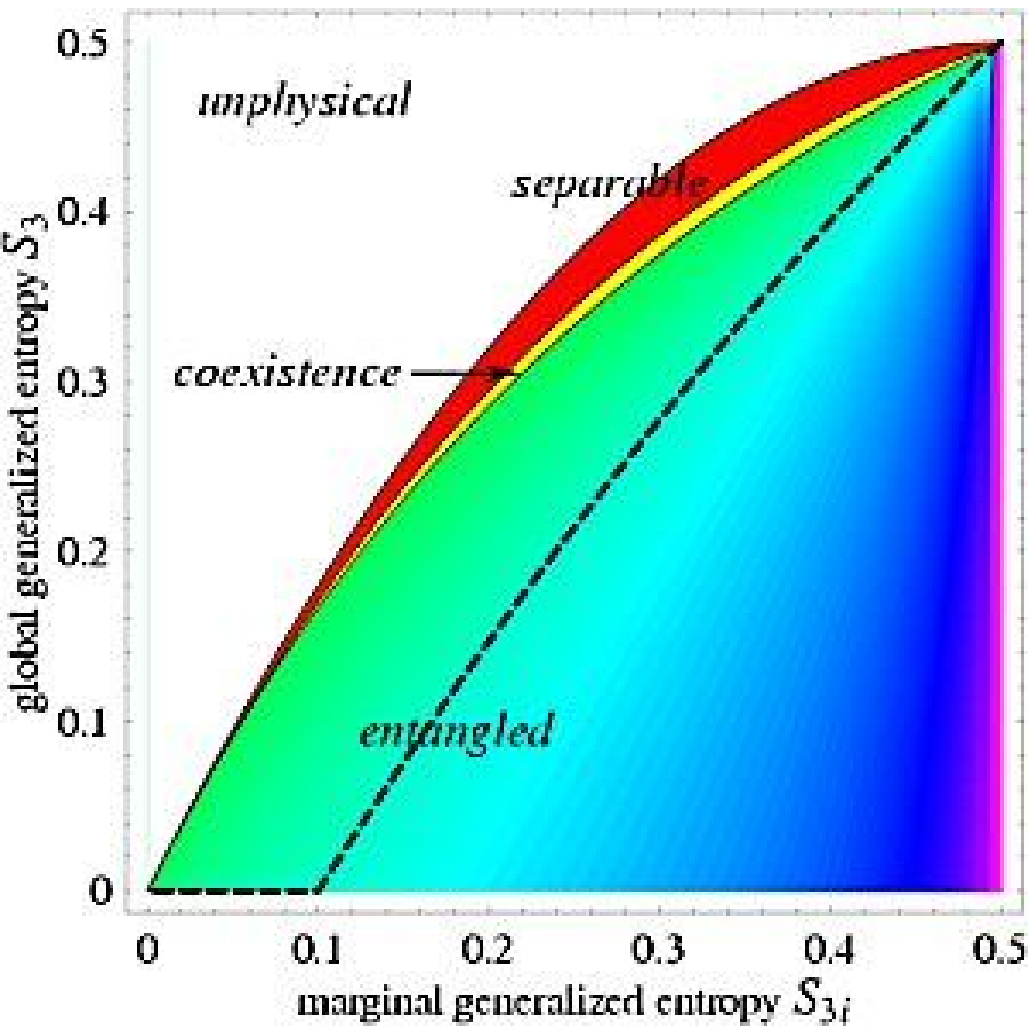}}
 \hspace{2mm}
\subfigure[\label{fig2D4}]
{\includegraphics[width=6.1cm]{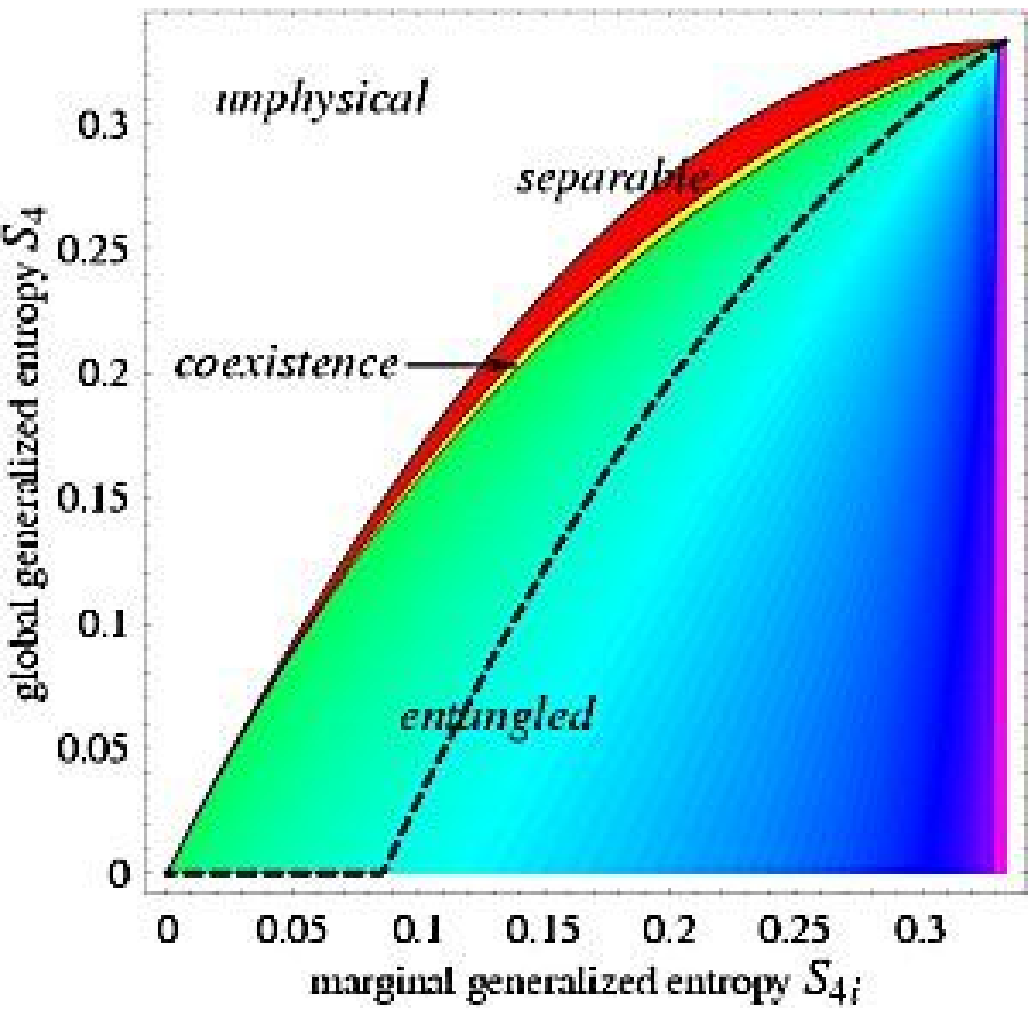}}
  \caption{Summary of the entanglement properties for symmetric Gaussian
  states at fixed global and marginal generalized $p-$entropies, for
  {\rm (a)} $p=1$ (Von Neumann entropies), {\rm (b)} $p=2$ (linear entropies), {\rm (c)} $p=3$,
  and {\rm (d)} $p=4$.  All states in the red region are separable.
In the entangled region, the average logarithmic
  negativity $\bar{E}_{\N}(S_{p_i},S_p)$ \eq{average} is depicted, growing from green to
  magenta. For $p>2$ an additional dashed curve is plotted;
  it represents the nodal line of inversion. Along it the entanglement is fully
  determined by the knowledge of the global and marginal generalized entropies
  $S_{p_i},S_p$, and GMEMS and GLEMS are equally entangled. On the left side of the nodal line GMEMS
  (GLEMS) are maximally (minimally) entangled Gaussian states at fixed $S_{p_i},S_p$.
  On the right side of the nodal line they are inverted: GMEMS (GLEMS) are minimally
  (maximally) entangled states. Also notice how the yellow region of
  coexistence (accommodating both separable and entangled states)
becomes narrower with increasing $p$. The equations
  of all boundary curves can be found in \eq{entsump}.}
\label{fig2Dm}
\end{figure}

\subsubsection{Classifying entangled states with generalized entropic
measures}

Apart from the relevant `inversion' feature shown by $p-$entropies
for $p>2$, the possibility of an accurate characterization of CV
entanglement based on global and marginal entropic measures still
holds in the general case for any $p$. In particular, the set of all
Gaussian states can be again divided, in the space of global and
marginal $S_p$'s, into three main areas: separable, entangled and
coexistence region. It can be thus very interesting to investigate
how the different entropic measures chosen to quantify the degree of
global mixedness (all marginal measures are equivalent) behave in
classifying the separability properties of Gaussian states.
Fig.~\ref{fig2Dm} provides a numerical comparison of the different
characterizations of entanglement obtained by the use of different
$p-$entropies, with $p$ ranging from $1$ to $4$, for symmetric
Gaussian states ($S_{p_1}=S_{p_2}\equiv S_{p_i}$). The last
restriction has been imposed just for ease of graphical display. The
following considerations, based on the exact numerical solutions of
the transcendental conditions, will take into account nonsymmetric
states as well.

The mathematical relations expressing the boundaries between the
different regions in Fig.~\ref{fig2Dm} are easily obtained for any
$p$ by starting from the relations holding for $p=2$ (see
Table~\ref{table1}) and by evaluating the corresponding
$S_p(\mu_{1,2})$ for each $\mu(\mu_{1,2})$. For any physical
symmetric state such a calculation yields
\begin{eqnarray}
 0  \le   (p-1) S_p &\!\!<\!\!&
1-g_p\left(\frac{\sqrt{2-\mu_i^2}}{\mu_i}\right)\nonumber\\
  &\Rightarrow& \textrm{entangled,} \nonumber\\
&& \nonumber \\
  1-g_p\left(\frac{\sqrt{2-\mu_i^2}}{\mu_i}\right)
\le  (p-1) S_p &\!\!<\!\!&
1-g_p^2\left(\sqrt{\frac{2-\mu_i}{\mu_i}}\right)\nonumber\\
&\Rightarrow& \textrm{coexistence,}  \label{entsump}\\
&& \nonumber \\
1-g_p^2\left(\sqrt{\frac{2-\mu_i}{\mu_i}}\right)  \le  (p-1)
S_p &\!\! \le \!\!& 1-g_p^2\left(\frac1{\mu_i^2}\right)\nonumber\\
  &\Rightarrow& \textrm{separable.} \nonumber
\end{eqnarray}
Equations~\pref{entsump} were obtained exploiting the
multiplicativity of $p-$norms on product states and using \eq{spmin}
for the lower boundary of the coexistence region (which represents
GLEMS becoming entangled) and \eq{spmax} for the upper one (which
expresses GMEMS becoming separable). Let us mention also that the
relation between any local entropic measure $S_{p_i}$ and the local
purity $\mu_i$ is obtained directly from \eq{pgau} and reads
\begin{equation}\label{spimui}
S_{p_i}=\frac{1-g_p(1/\mu_i)}{p-1}\,.
\end{equation}

We notice {\em prima facie} that, with increasing $p$, the
entanglement is more sharply qualified in terms of the global and
marginal $p-$entropies. In fact the region of coexistence between
separable and entangled states becomes narrower with higher $p$.
Thus, somehow paradoxically, with increasing $p$ the entropy $S_p$
provides less information about a quantum state, but at the same
time it yields a more accurate characterization and quantification
of its entanglement. In the limit $p \rightarrow \infty$ all the
physical states collapse to one point at the origin of the axes in
the space of generalized entropies, due to the fact that the measure
$S_\infty$ is identically zero.
%We will now show that the right compromise for a satisfactory
%characterization, both from a theoretical and an experimental
%point of view, of mixedness and entanglement in a Gaussian state,
%is represented by the instance $p=2$ and so by the purities of the
%state.
\begin{figure}[t!]
\subfigure[\label{fig3D1}] {\includegraphics[width=6cm]{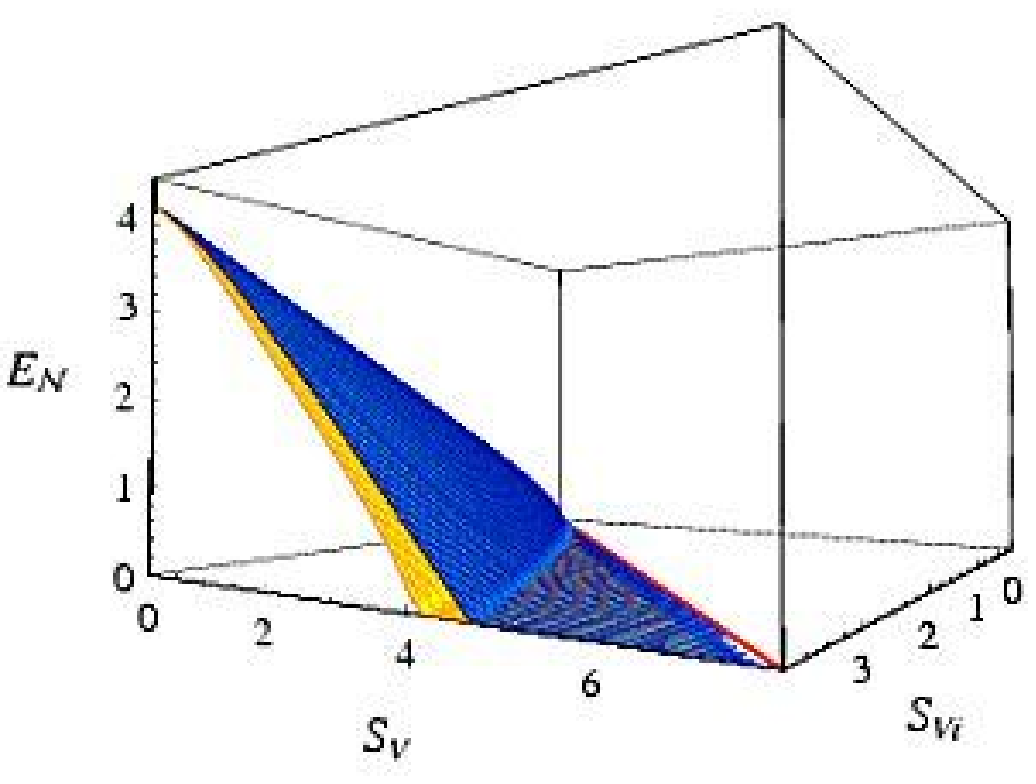}}
\hspace{2mm} \subfigure[\label{fig3D2}]
{\includegraphics[width=6cm]{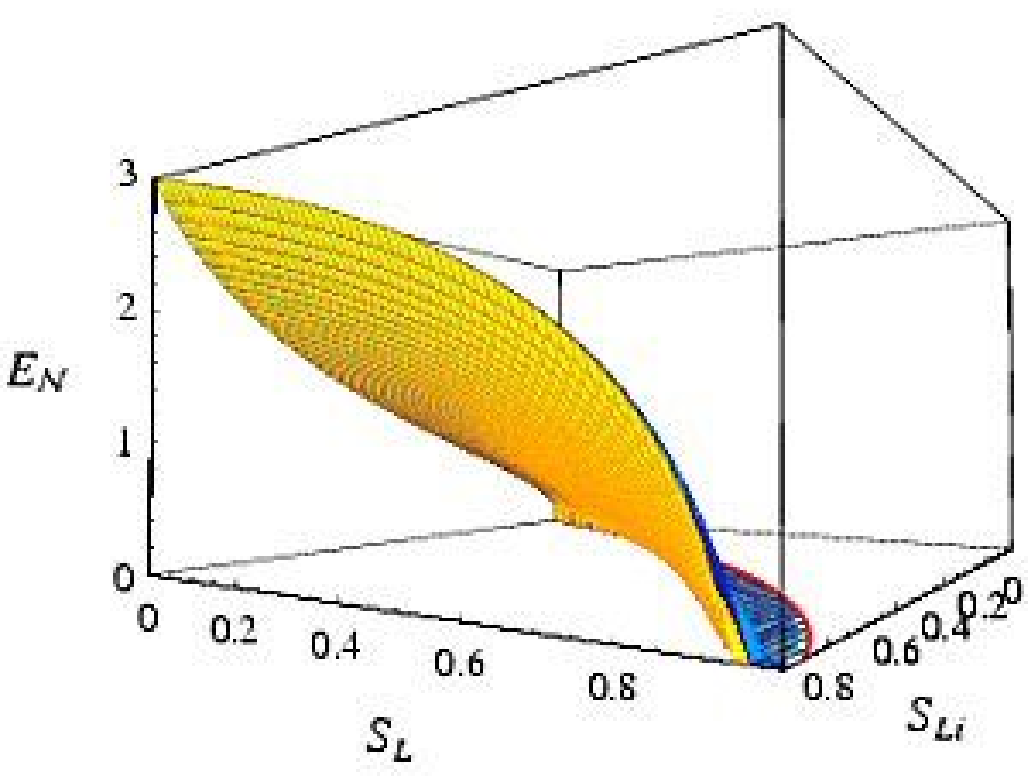}}
\\ \subfigure[\label{fig3D3}]
{\includegraphics[width=6cm]{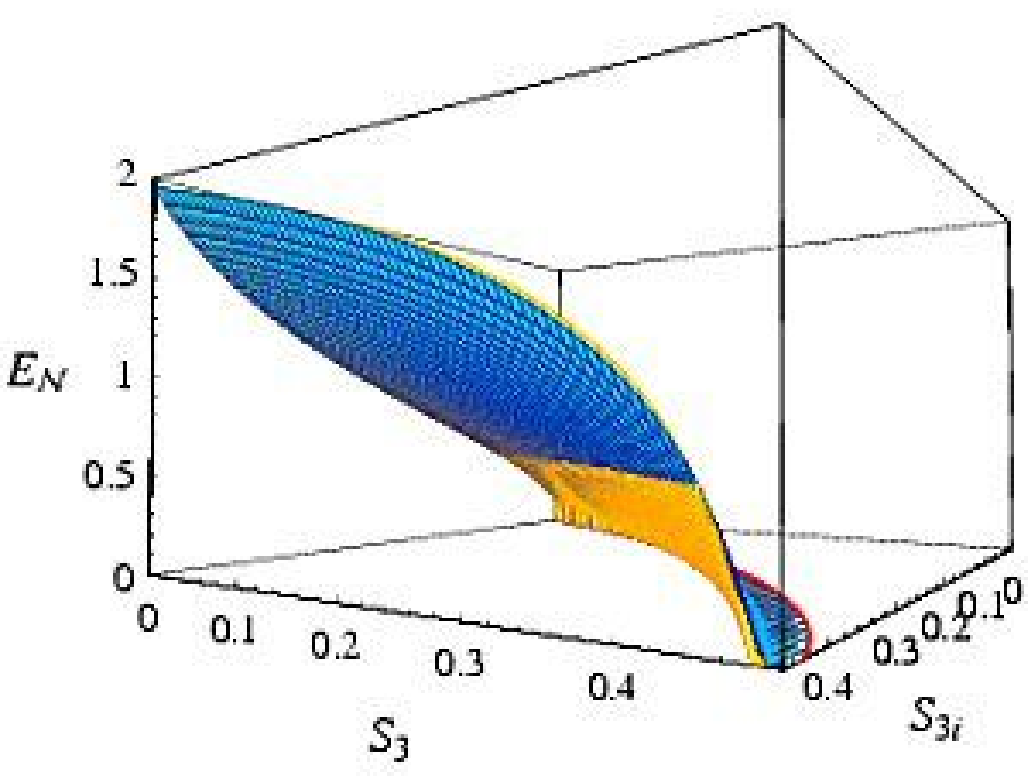}} \hspace{2mm}
\subfigure[\label{fig3D4}] {\includegraphics[width=6cm]{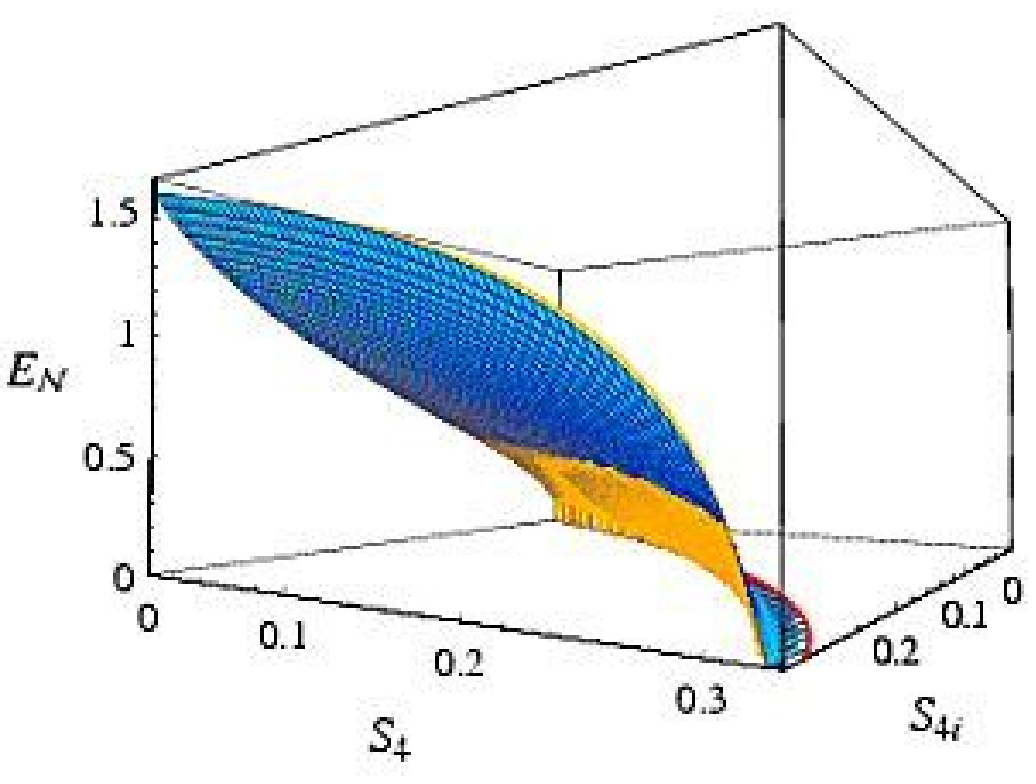}}
  \caption{Upper and lower bounds on the logarithmic
  negativity of symmetric Gaussian states
  as functions of the global and marginal generalized $p-$entropies, for
  {\rm (a)} $p=1$ (Von Neumann entropies), {\rm (b)} $p=2$ (linear entropies), {\rm (c)} $p=3$,
  and {\rm (d)} $p=4$. The blue (yellow) surface represents GMEMS
  (GLEMS). Notice that for $p>2$ GMEMS and GLEMS surfaces intersect
along the inversion line (meaning they are equally entangled along
that line), and beyond it they interchange their role. The equations
of the inversion lines are obtained from Eqs.~{\rm
(\ref{spleaf}--\ref{spleaf4})}, with the position
$S_{p_1}=S_{p_2}\equiv S_{p_i}$.}
  \label{fig3Dm}
\end{figure}

\section{Quantifying entanglement via purity measures: the average logarithmic negativity}\label{aln}

We have extensively shown that knowledge of the global and marginal
generalized $p-$entropies accurately characterizes the entanglement
of Gaussian states, providing strong sufficient and/or necessary
conditions. The present analysis naturally leads us to propose an
actual \emph{quantification} of entanglement, based exclusively on
marginal and global entropic measures, according to the approach
introduced in Refs.~\cite{prl,extremal}.

Outside the separable region, we can formally define the maximal
entanglement $E_{{\N}max}(S_{p_{1,2}},S_p)$ as the logarithmic
negativity attained by GMEMS (or GLEMS, below the inversion nodal
surface for $p>2$, see Fig.~\ref{figleaf}). In a similar way, in the
entangled region GLEMS (or GMEMS, below the inversion nodal surface
for $p>2$) achieve the minimal logarithmic negativity
$E_{{\N}min}(S_{p_{1,2}},S_p)$. The explicit analytical expressions
of these quantities are unavailable for any $p\neq2$ due to the
transcendence of the conditions relating $S_p$ to the symplectic
eigenvalues.

The surfaces of maximal and minimal entanglement in the space of the
global and local $S_p$ are plotted in Fig.~\ref{fig3Dm} for
symmetric states. In the plane $S_p=0$ the upper and lower bounds
correctly coincide, since for pure states the entanglement is
completely quantified by the marginal entropy. For mixed states this
is not the case but, as the plot shows, knowledge of the global and
marginal entropies strictly bounds the entanglement both from above
and from below. For $p>2$, we notice how GMEMS and GLEMS exchange
their role beyond a specific curve in the space of $S_p$'s. The
equation of this nodal curve is obtained from the general
leaf--shaped nodal surfaces of Eqs.~{\rm
(\ref{spleaf}--\ref{spleaf4})}, by imposing the symmetry constraint
($S_{p_1}=S_{p_2}\equiv S_{p_i}$). We notice again how the $S_p$'s
with higher $p$ provide a better characterization of the
entanglement, even quantitatively. In fact, the gap between the two
extremally entangled surfaces in the $S_p$'s space becomes smaller
with higher $p$. Of course the gap is exactly zero all along the
nodal line of inversion for $p>2$.

\begin{figure}[t!]
\subfigure[\label{figerror1}]
{\includegraphics[width=5.6cm]{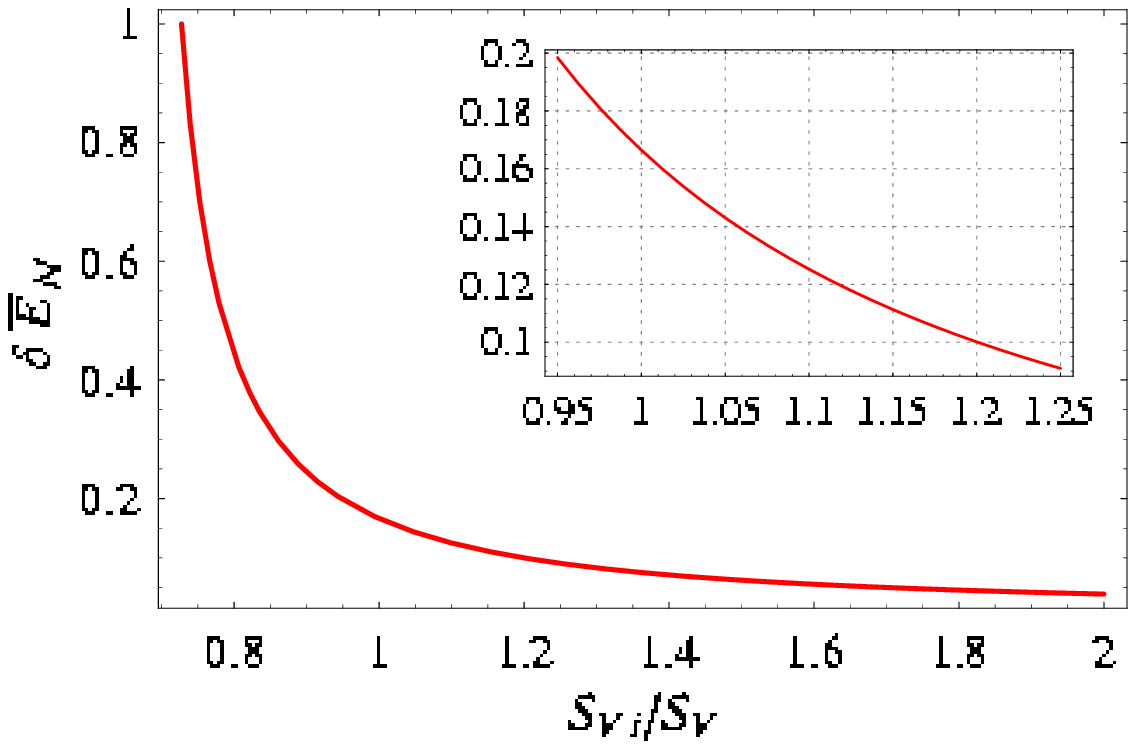}} \hspace{5mm}
\subfigure[\label{figerror2}]
{\includegraphics[width=5.6cm]{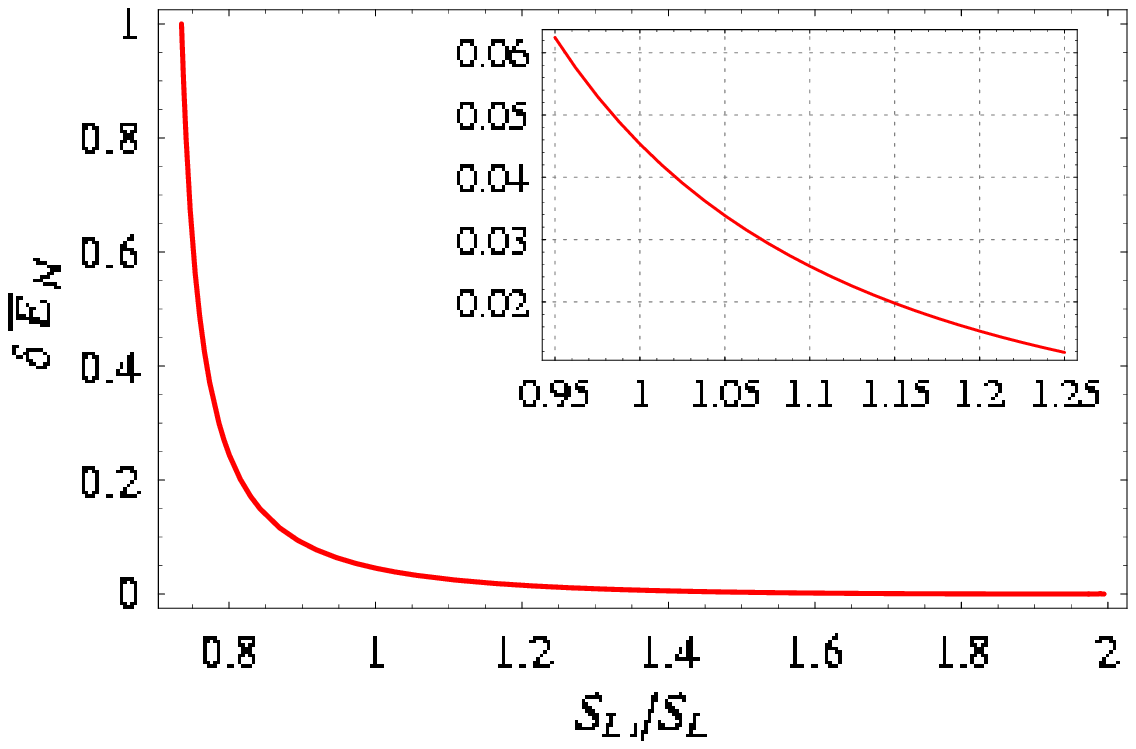}}\\
\subfigure[\label{figerror3}]
{\includegraphics[width=5.6cm]{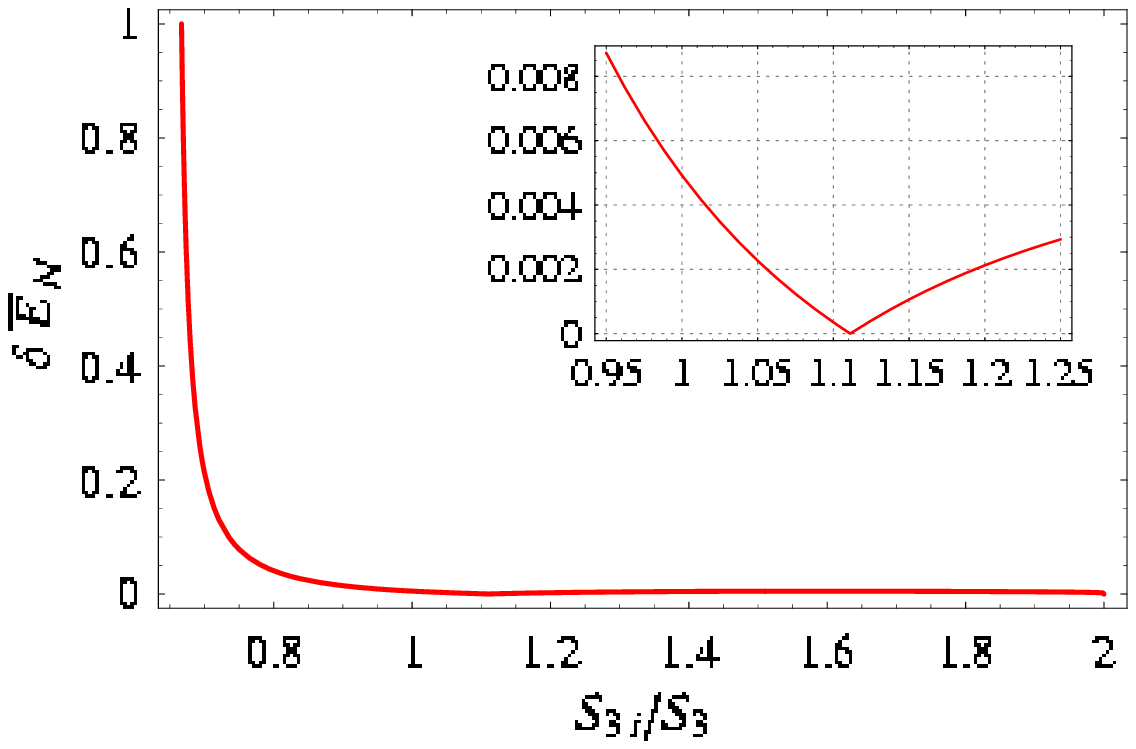}} \hspace{5mm}
\subfigure[\label{figerror4}]
{\includegraphics[width=5.6cm]{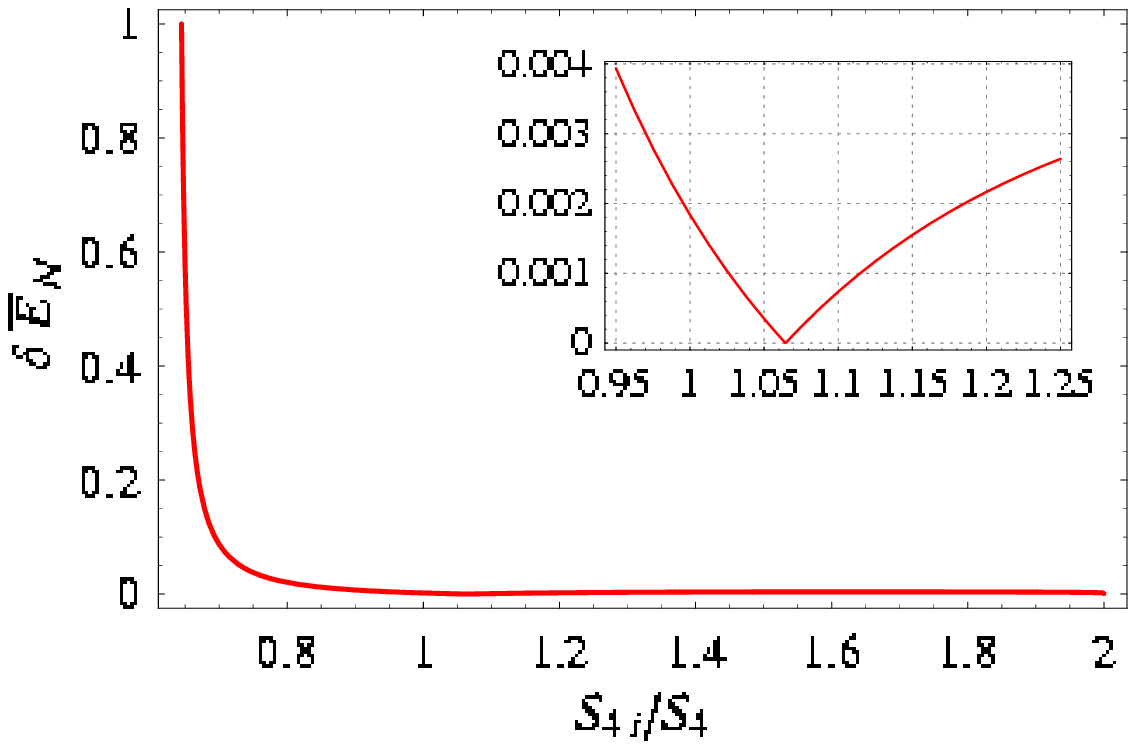}} \caption{The relative
error $\delta \bar{E}_{\N}$ Eq.~\pref{deltaen} on the average
logarithmic negativity as a function of the ratio $S_{p_i} /S_p$,
for {\rm (a)} $p=1$, {\rm (b)} $p=2$, {\rm (c)} $p=3$, {\rm (d)}
$p=4$, plotted at {\rm (a)} $S_V=1$, {\rm (b)} $S_L=1/2$, {\rm (c)}
$S_3=1/4$, {\rm (d)} $S_4=1/6$. Notice how, in general, the error
decays exponentially, and in particular faster with increasing $p$.
For $p>2$, notice how the error reaches zero on the inversion node
(see the insets), then grows and reaches a local maximum before
going back to zero asymptotically.}
  \label{figerrorm}
\end{figure}

Let us thus introduce a particularly convenient quantitative
estimate of the entanglement based only on the knowledge of the
global and marginal entropies. Let us define the {\em average
logarithmic negativity} $\bar{E}_{\N}$ as \be \bar{E}_{\N}
(S_{p_{1,2}},S_p) \equiv
\frac{E_{{\N}max}(S_{p_{1,2}},S_p)+E_{{\N}min}(S_{p_{1,2}},S_p)}{2}
\; . \label{average} \ee We will now show that this quantity, fully
determined by the global and marginal entropies, provides a reliable
quantification of entanglement (logarithmic negativity) for two-mode
Gaussian states. To this aim, we define the relative error $\delta
\bar{E}_{\N}$ on $\bar{E}_{\N}$ as
\begin{equation}
\label{deltaen} \delta \bar{E}_{\N} (S_{p_{1,2}},S_p) \equiv
\frac{E_{\N max}(S_{p_{1,2}},S_p) -E_{\N
min}(S_{p_{1,2}},S_p)}{E_{\N max}(S_{p_{1,2}},S_p) +E_{\N
min}(S_{p_{1,2}},S_p)}\,.
\end{equation}
As Fig.~\ref{figerrorm} shows, this error decreases
\emph{exponentially} both with decreasing global entropy and
increasing marginal entropies, that is with increasing entanglement.
In general the relative error $\delta \bar{E}_{\N}$ is `small' for
sufficiently entangled states; we will present more precise
numerical considerations in the subcase $p=2$. Notice that the
decaying rate of the relative error is faster with increasing $p$:
the average logarithmic negativity turns out to be a better estimate
of entanglement with increasing $p$. For $p>2$, $\delta
\bar{E}_{\N}$ is exactly zero on the inversion node, then it becomes
finite again and, after reaching a local maximum, it goes
asymptotically to zero (see the insets of Fig.~\ref{figerrorm}).

All the above considerations, obtained by an exact numerical
analysis, show that the average logarithmic negativity
$\bar{E}_{\N}$ at fixed global and marginal $p-$entropies is a very
good estimate of entanglement in CV systems, whose reliability
improves with increasing entanglement and, surprisingly, with
increasing order $p$ of the entropic measures.

\subsection{Direct estimate of two-mode entanglement}
\label{secFiura}

In the present general framework, a peculiar role is played by the
case $p=2$, {\em i.e.} by the linear entropy $S_L$ (or,
equivalently, the purity $\mu$). The previous general analysis on
the whole range of generalized entropies $S_p$, has remarkably
stressed the privileged theoretical role of the instance $p=2$,
which discriminates between the region in which extremally entangled
states are unambiguously characterized and the region in which they
can exchange their roles. Moreover, the graphical analysis shows
that, in the region where no inversion takes place $(p \le 2)$,
fixing the global $S_2=1-\mu$ yields the most stringent constraints
on the logarithmic negativity of the states (see Figs.~\ref{fig2Dm},
\ref{fig3Dm}, \ref{figerrorm}). Notice that such constraints,
involving no transcendental functions for $p=2$, can be easily
handled analytically. A crucial experimental consideration
strengthens these theoretical and practical reasons to privilege the
role of $S_2$. In fact, $S_2$ can indeed, assuming some prior
knowledge about the state (essentially, its Gaussian character), be
measured through conceivable direct methods, in particular by means
of single-photon detection schemes \cite{fiurasek04} (of which
preliminary experimental verifications are available
\cite{wenger04}) or of the upcoming quantum network architectures
\cite{Ekert02,Filip02,Oi-Aberg}.  Very recently, a scheme to locally
measure all symplectic invariants (and hence the entanglement) of
two-mode Gaussian states has been proposed, based on number and
purity measurements \cite{rigolinew}. Notice that no complete
homodyne reconstruction \cite{homotomo} of the CM is needed in all
those schemes.

As already anticipated, for $p=2$ we can provide analytical
expressions for the extremal entanglement in the space of global and
marginal purities \cite{prl} \bea E_{{\N}max}(\mu_{1,2},\mu)
\!\!&\!\!=\!\!& -\frac{\log \left[-\frac{1}{\mu}
  + \left(\frac{\mu_1+\mu_2}{2\mu_1^2 \mu_2^2} \right)
  \left(\mu_1+\mu_2 -
  \sqrt{(\mu_1+\mu_2)^2-\frac{4 \mu_1^2 \mu_2^2}{\mu}} \right)
\right]}{2}, \nonumber\\ && \label{enmax}\\
\nonumber \\E_{{\N}min}(\mu_{1,2},\mu) \!\!&\!\!=\!\!& - \frac{\log
\left[\frac{1}{\mu_1^2}+\frac{1}{\mu_2^2}-\frac{1}{2\mu^2} -
  \frac12 - \sqrt{\left( \frac{1}{\mu_1^2}+\frac{1}{\mu_2^2}- \frac{1}{2\mu^2}
- \frac12 \right)^2 - \frac{1}{\mu^2}} \; \right]}{2}.\nonumber \\
&& \label{enmin} \eea Consequently, both the average logarithmic
negativity $\delta \bar{E}_{\N}$, defined in \eq{average}, and the
relative error $\delta \bar{E}_{\N}$, given by \eq{deltaen}, can be
easily evaluated in terms of the purities. The relative error is
plotted in Fig.~\ref{figerrorm}{\rm (b)} for symmetric states as a
function of the ratio $S_{L_i} / S_L$. Notice, as already pointed
out in the general instance of arbitrary $p$, how the error decays
exponentially. In particular, it falls below $5\%$ in the range
$S_L<S_{L_i}\ (\mu
> \mu_i)$, which excludes at most very weakly entangled states
(states with $E_{\N} \lesssim 1$).\footnote{\sf It is
straightforward to verify that, in the instance of two-mode squeezed
thermal (symmetric) states, such a condition corresponds to
$\cosh(2r)\gtrsim \mu^{1/4}$. This constraint can be easily
satisfied with the present experimental technology: even for the
quite unfavorable case $\mu=0.5$ the squeezing parameter needed is
just $r\simeq 0.3$.} Let us remark that the accuracy of estimating
entanglement by the average logarithmic negativity proves even
better in the nonsymmetric case $\mu_1 \neq \mu_2$, essentially
because the maximal allowed entanglement decreases with the
difference between the marginals, as shown in Fig.~\ref{gmemms}{\rm
(a)}.

The above analysis proves that the average logarithmic negativity
$\bar{E}_{\N}$ is a reliable estimate of the logarithmic negativity
$E_{\N}$, improving as the entanglement increases
\cite{prl,extremal}. This allows for an accurate quantification of
CV entanglement by knowledge of the global and marginal purities. As
we already mentioned, the latter quantities may be in turn amenable
to direct experimental determination by exploiting recent
single-photon-detection proposals \cite{fiurasek04} or in general
interferometric quantum-network setups. Let us stress, even though
quite obvious, that the estimate becomes indeed an {\em exact}
quantification in the two crucial instances of GMEMS (nonsymmetric
thermal squeezed states) and GLEMS (mixed states of partial minimum
uncertainty), whose logarithmic negativity is completely determined
as a function of the three purities alone, see
Eqs.~{\rm(\ref{enmax},\,\ref{enmin})}.

\section{Gaussian entanglement measures versus Negativities}
\label{secorder}

In this Section, based on Ref.~\cite{ordering}, we add a further
piece of knowledge on the quantification of entanglement in two-mode
Gaussian states.  We compute the Gaussian entanglement of formation
and, in general, the family of Gaussian entanglement measures (see
Sec.~\ref{SecGEMS}), for two special classes of two-mode Gaussian
states, namely the states of extremal, maximal and minimal,
negativities at fixed global and local purities
 (GMEMS and GLEMS, introduced in
Sec.~\ref{secprl} \cite{prl,extremal}). We find  that the two
families of entanglement measures (negativities and Gaussian
entanglement measures) are not equivalent for nonsymmetric two-mode
states. Remarkably, they may induce a completely different {\em
ordering} on the set of entangled two-mode Gaussian states: a
nonsymmetric state $\varrho_A$ can be more entangled than another
state $\varrho_B$, with respect to negativities, and less entangled
than the same state $\varrho_B$, with respect to Gaussian
entanglement measures. However, the inequivalence between the two
families of measures is somehow bounded: we show that, at fixed
negativities, the Gaussian entanglement measures are rigorously
bounded from below. Moreover, we provide strong evidence hinting
that they should be bounded from above as well.

\subsection{Geometric framework for two-mode Gaussian entanglement measures}

The problem of evaluating Gaussian entanglement measures (Gaussian
EMs) for a generic two-mode Gaussian state has been solved in
Ref.~\cite{GEOF}. However, the explicit result contains so
``cumbersome'' expressions (involving the solutions of a
fourth-order algebraic equation), that they were judged of no
particular insight to be reported explicitly in  Ref.~\cite{GEOF}.

We recall here the computation procedure \cite{ordering} that we
will need in the following. For any two-mode Gaussian state with CM
$\sig \equiv \sig_{sf}$ in standard form \eq{stform2}, a generic
Gaussian EM $G_E$ is given by the entanglement $E$ of the least
entangled pure state with CM $\sig^p \le \sig$, see \eq{Gaussian
EMm}. Denoting by $\gamma_q$ (respectively $\gamma_p$) the $2 \times
2$ submatrix obtained from $\sig$ by canceling the even (resp. odd)
rows and columns, we have
\begin{equation}\label{cpcq}
\gamma_q = \left(
\begin{array}{ll}
  a & c_+ \\
  c_+ & b \\
\end{array}
\right)\,,\quad \gamma_p = \left(
\begin{array}{ll}
  a & c_- \\
  c_- & b \\
\end{array}
\right)\,.
\end{equation}
All the covariances relative to the ``position'' operators of the
two modes are grouped in $\gamma_q$, and analogously for the
``momentum'' operators in $\gamma_p$. The total CM can then be
written as a direct sum $\sig = \gamma_q \oplus \gamma_p$.
Similarly, the CM of a generic pure two-mode Gaussian state in
block-diagonal form  (it has been proven that the CM of the optimal
pure state has to be with all diagonal $2 \times 2$ submatrices as
well \cite{GEOF}) can be written as $\sig^p = \gamma_q^p \oplus
\gamma_p^p$, where the global purity of the state imposes
$(\gamma_p^p)^{-1} = \gamma_q^p \equiv \Gamma$ (see Appendix
\ref{redu}). The pure states involved in the definition of the
Gaussian EM must thus fulfill the condition
\begin{equation}\label{rim}
\gamma_p^{-1} \le \Gamma \le \gamma_q\,.
\end{equation}

This problem is endowed with a nice geometric description
\cite{GEOF}. Writing the matrix $\Gamma$ in the basis constituted by
the identity matrix and the three Pauli matrices,
\begin{equation}\label{Gamma} \Gamma = \left(
\begin{array}{cc}
  x_0 + x_3 & x_1 \\
  x_1 & x_0 - x_3 \\
\end{array}
\right)\,,
\end{equation}
the expansion coefficients $(x_0,x_1,x_3)$ play the role of
space-time coordinates in a three-dimensional Minkowski space. In
this picture, for example, the rightmost inequality in \eq{rim} is
satisfied by matrices $\Gamma$ lying on a cone, which is equivalent
to the (backwards) light cone of $\gamma_q$ in the Minkowski space;
and similarly for the leftmost inequality. Indeed, one can show
that, for the optimal pure state $\sig^p_{opt}$ realizing the
minimum in \eq{Gaussian EMm}, the two inequalities in \eq{rim} have
to be simultaneously saturated \cite{GEOF}. From a geometrical point
of view, the optimal $\Gamma$ has then to be found on the rim of the
intersection of the forward and the backward cones of
$\gamma_p^{-1}$ and $\gamma_q$, respectively. This is an ellipse,
and one is left with the task of minimizing the entanglement $E$ of
$\sig^p = \Gamma \oplus \Gamma^{-1}$ [see \eq{Gaussian EMp}] for
$\Gamma$ lying on this ellipse.\footnote{\sf The geometric picture
describing the optimal two-mode state which enters in the
determination of the Gaussian EMs is introduced in \cite{GEOF}. A
more detailed discussion, including the explicit expression of the
Lorentz boost needed to move into the plane of the ellipse, can be
found in \cite{noteole}.}

At this point, let us pause to briefly recall that any pure two-mode
Gaussian state $\sig^p$ is locally equivalent to a two-mode squeezed
state with squeezing parameter $r$, described by the CM of \eq{tms}.
The following statements are then equivalent: (i) $E$ is a
monotonically increasing function of the entropy of entanglement;
(ii) $E$ is a monotonically increasing function of the single-mode
determinant $m^2\equiv\det\gr\alpha \equiv \det\gr\beta$ [see
\eq{espre}]; (iii) $E$ is a monotonically decreasing function of the
local purity $\mu_i\equiv\mu_1\equiv\mu_2$ [see \eq{purgau}]; (iv)
$E$ is a monotonically decreasing function of the smallest
symplectic eigenvalue $\tilde\nu_-^p$ of the partially transposed CM
$\tilde\sig^p$; (v) $E$ is a monotonically increasing function of
the squeezing parameter $r$. This chain of equivalences is
immediately proven by simply recalling that a pure state is
completely specified by its single-mode marginals, and that for a
single-mode Gaussian state there is a unique symplectic invariant
(the determinant), so that all conceivable entropic quantities are
monotonically increasing functions of this invariant, as shown in
Sec.~\ref{SecEntroG} \cite{extremal}. In particular, statement (ii)
allows us to minimize directly the single-mode determinant over the
ellipse,
\begin{equation}\label{mdef}
m^2 = 1 + \frac{x_1}{\det \Gamma}\,,
\end{equation}
with $\Gamma$ given by \eq{Gamma}.

To simplify the calculations, one can move to the plane of the
ellipse with a Lorentz boost which preserves the relations between
all the cones; one can then choose the transformation so that the
ellipse degenerates into a circle (with fixed radius), and introduce
polar coordinates on this circle. The calculation of the Gaussian EM
for any two-mode Gaussian state is thus finally reduced to the
minimization of $m^2$ from \eq{mdef}, at given standard form
covariances of $\sig$, as a function of the polar angle $\theta$ on
the circle \cite{noteole}. This technique had been applied to the
computation of the Gaussian entanglement of formation by minimizing
\eq{mdef} {\em numerically} \cite{GEOF} (see also \cite{oleposter}).
In addition to that, as already mentioned, the Gaussian entanglement
of formation has been exactly computed for symmetric states, and it
has been proven that in this case the Gaussian entanglement of
formation is the true entanglement of formation \cite{giedke03}.

Here we are going to present new analytical calculations, first
obtained in \cite{ordering}, of the Gaussian EMs for two relevant
classes of nonsymmetric two-mode Gaussian states: the states of {\em
extremal} negativities at fixed global and local purities
\cite{prl,extremal}, introduced in Sec.~\ref{secprl}. We begin by
writing the general expression of the single-mode determinant
\eq{mdef} in terms of the standard form covariances of a generic
two-mode state, \eq{stform2}, and of the polar angle $\theta$. After
some tedious but straightforward algebra, one finds \cite{ordering}
\begin{eqnarray}\label{mfunc}
%\begin{minipage}
&\!\!\!&\!\!\!\!\!\! \hspace*{-0.3cm} m^2_\theta (a,b,c_+, c_-)\ = \\
&\!\!\!&\!\!\!\!\!\! 1 + \left\{\left[c_+(ab-c_-^2)-c_-+\cos \theta
\sqrt{\left[a -
b(ab-c_-^2)\right]\left[b-a(ab-c_-^2)\right]}\right]^2\right\}
\nonumber \\
&\!\!\!&\!\!\!\!\!\! \times  \Bigg\{
2\left(ab-c_-^2\right)\left(a^2+b^2+2c_+c_- \right)  \nonumber \\
&\!\!\!&\!\!\!\!\!\! -\ \frac{\cos \theta\left[2abc_-^3+\left(a^2+
b^2\right)c_+c_-^2+\left(\left(1-2b^2\right)a^2+
b^2\right)c_--ab\left(a^2+b^2- 2\right)c_+\right]}{\sqrt{\left[a -
b(ab-c_-^2)\right]\left[b-a(ab-c_-^2)\right]}} \nonumber \\
&\!\!\!&\!\!\!\!\!\! +\ \sin \theta\left(a^2-
b^2\right)\sqrt{1-\frac{\left[c_+(ab-c_-^2)+c_-\right]^2}{\left[a -
b(ab-c_-^2)\right]\left[b-a(ab-c_-^2)\right]}} \, \Bigg\}^{-1}\,,
\nonumber
\end{eqnarray}
 where we have assumed $c_+ \ge |c_-|$ without any loss of
generality. This implies that, for any entangled state, $c_+ > 0$
and $c_- < 0$, see \eq{detgammanegative}. The Gaussian EM, defined
in terms of the function $E$ on pure states [see \eq{Gaussian EMp}],
coincides then for  a generic two-mode Gaussian state with the
entanglement $E$ computed on the pure state with $m^2=m^2_{opt}$,
where $m^2_{opt} \equiv \min_\theta (m^2_\theta)$. Accordingly, the
symplectic eigenvalue $\tilde \nu_-$ of the partial transpose of the
corresponding optimal pure-state CM $\sig^{p}_{opt}$, realizing the
infimum in \eq{Gaussian EMm}, reads [see \eq{n1}]
\begin{equation}\label{nutopt}
\tilde\nu_{- opt}^{p} \equiv \tilde \nu_-(\sig^{p}_{opt}) = m_{opt}
- \sqrt{m^2_{opt}-1}\,.
\end{equation}
As an example, for the Gaussian entanglement of formation
\cite{GEOF} one has \be \label{geofm} G_{E_F}(\sig) =
h\left(\tilde\nu_{- opt}^{p}\right)\,,\ee with $h(x)$ defined by
\eq{hentro}.

Finding the minimum of \eq{mfunc} analytically for a generic state
is a difficult task. By numerical investigations, we have found that
the equation $\partial_\theta m^2_\theta = 0$ can have from one to
four physical solutions (in a period) corresponding to extremal
points, and the global minimum can be attained in any of them
depending on the parameters of the CM $\sig$ under inspection.
However, a closed solution can be found for two important classes of
nonsymmetric two-mode Gaussian states, GMEMS and GLEMS (see
Sec.~\ref{secprl}), as we will now show.

\subsection{Gaussian entanglement measures for extremal states}
\label{SecGEMextra}

We have shown in Sec.~\ref{secprl} that, at fixed global purity of a
two-mode Gaussian state $\sig$, and at fixed local purities of each
of the two reduced single-mode states, the smallest symplectic
eigenvalue $\tilde \nu_-$ of the partial transpose of the CM $\sig$
(which qualifies its separability by the PPT criterion, and
quantifies its entanglement in terms of the negativities) is
strictly bounded from above and from below. This entails the
existence of two disjoint classes of extremal states, namely the
states of maximum negativity for fixed global and local purities
(GMEMS), and the states of minimum negativity for fixed global and
local purities (GLEMS) \cite{prl,extremal}.

Recalling these results, it is useful to reparametrize the standard
form covariances \eq{stform2} of a general {\em entangled} two-mode
Gaussian states, whose purities satisfy \ineq{sufent} (see also
Table \ref{table1}), as follows,
\begin{eqnarray}
  a &=& s + d\,,\qquad b\,\,=\,\,s-d\,,\label{asd} \\
  c_{\pm} &=& \frac{1}{4\sqrt{s^2 -
            d^2}} \left\{\sqrt{\left[4 d^2 + \frac{1}{2} \left(g^2 +
                          1\right) (\lambda - 1) - \left(2 d^2 +
                          g\right) (\lambda + 1)\right]^2 -
            4 g^2} \right. \nonumber \\
            & \pm & \left.\sqrt{\left[4 s^2 + \frac{1}{2} \left(g^2 +
                          1\right) (\lambda - 1) - \left(2 d^2 +
                          g\right) (\lambda + 1)\right]^2 - 4
                          g^2}\right\} \label{cpm}\,,
\end{eqnarray}
where the two local purities are regulated by the parameters $s$ and
$d$, being $\mu_1 = (s+d)^{-1},\,\mu_2 = (s-d)^{-1}$, and the global
purity is $\mu = g^{-1}$. The coefficient $\lambda$ embodies the
only remaining degree of freedom (related to $\Delta$) needed for
the complete determination of the negativities, once the three
purities have been fixed. It ranges from the minimum $\lambda = -1$
(corresponding to the GLEMS) to the maximum $\lambda = +1$
(corresponding to the GMEMS). Therefore, as it varies, $\lambda$
encompasses all possible entangled two-mode Gaussian states
compatible with a given set of assigned values of the purities (\ie
those states which fill the entangled region in Table \ref{table1}).
The constraints that the parameters $s,\,d,\,g$ must obey for
\eq{stform} to denote a proper CM of a physical state are, from
Eqs.~{\rm(\ref{consmu12}--\ref{deltabnd})}: $s \ge 1$, $|d| \le
s-1$, and
\begin{equation}\label{gmbound}
g \ge 2 |d| + 1\,,
\end{equation}

If the global purity is large enough so that \ineq{gmbound} is
saturated, GMEMS and GLEMS coincide, the CM becomes independent of
$\lambda$, and the two classes of extremal states coalesce into a
unique class, completely determined by the marginals $s$ and $d$. We
have denoted these states as GMEMMS in Sec.~\ref{SecMEMMS}, that is,
Gaussian two-mode states of maximal negativity at fixed local
purities \cite{extremal}. Their CM, from \eq{gmemms}, is simply
characterized by $c_{\pm} = \pm \sqrt{s^2 -(d+1)^2}$, where we have
assumed without any loss of generality that $d \ge 0$ (corresponding
to choose, for instance, mode $1$ as the more mixed one: $\mu_1 \le
\mu_2$).

In general (see Table \ref{table1}), a GMEMS ($\lambda = +1$) is
entangled for \be\label{gmement} g < 2s-1\,, \ee while a GLEMS
($\lambda = -1$) is entangled for a smaller $g$, namely
\be\label{glement} g < \sqrt{2(s^2 + d^2) -1}\,. \ee

%We are now equipped with the necessary tools to compute Gaussian EMs
%for the two extremal classes of nonsymmetric two-mode Gaussian
%states, the GLEMS and the GMEMS.

\subsubsection{Gaussian entanglement of minimum-negativity states
(GLEMS)}

We want to find the optimal pure state $\sig^p_{opt}$ entering in
the definition \eq{Gaussian EMm} of the Gaussian EM. To do this, we
have to minimize the single-mode determinant of $\sig^p_{opt}$,
given by \eq{mfunc}, over the angle $\theta$. It turns out that, for
a generic GLEMS, the coefficient of $\sin \theta$ in the last line
of \eq{mfunc} vanishes, and the expression of the single-mode
determinant reduces to the simplified form
\begin{equation}\label{mglems}
{m^2_\theta}^{_{\rm  GLEMS}} = 1 + \frac{[A \cos \theta + B]^2}{2(a
b - c_-^2)[(g^2 - 1) \cos \theta  + g^2 + 1]}\,,
\end{equation}
with $A = c_+(a b - c_-^2) + c_-,\, B = c_+(a b - c_-^2) - c_-,$ and
$a,b,c_\pm$ are the covariances of GLEMS, obtained from Eqs.~{\rm
(\ref{asd},\ref{cpm})} setting $\lambda = -1$.

The only relevant solutions (excluding the unphysical and the
trivial ones) of the equation $\partial_\theta m^2_\theta = 0$ are
$\theta = \pi$ and $$\theta = \pm \theta^\ast \equiv \arccos
\left[\frac{3+g^2}{1-g^2} - \frac{2c_-}{c_+(a b - c_-^2) + c_-}
\right]\,.$$ Studying the second derivative $\partial^2_\theta
m^2_\theta$ for $\theta = \pi$ one finds immediately that, for
\begin{equation}\label{gminmax} g \ge
\sqrt{-\frac{2c_+(a b - c_-^2) + c_-}{c_-}}
\end{equation}
(remember that $c_- \le 0$), the solution $\theta = \pi$ is a
minimum. In this range of parameters, the other solution $\theta =
\theta^\ast$ is unphysical (in fact $|\cos \theta^\ast| \ge 1$), so
$m^2_{\theta=\pi}$ is the global minimum. When, instead,
\ineq{gminmax} is violated, $m^2_\theta$ has a local maximum for
$\theta = \pi$ and two minima appear at $\theta = \pm \theta^\ast$.
The global minimum is attained in any of the two, given that, for
GLEMS, $m^2_\theta$ is invariant under reflection with respect to
the axis $\theta = \pi$. Collecting, substituting, and simplifying
the obtained expressions, we arrive at the final result for the
optimal $m^2$:
\begin{eqnarray}
%\\ %\nonumber
\label{m2glems} {m^2}_{opt}^{_{\rm GLEMS}}=\left\{
\begin{array}{l}
    1, \\ \qquad g \ge \sqrt{2(s^2 + d^2)-1} \qquad {\hbox{[separable state]}}\; ;
\\ \\
\\
    \frac{16 s^2 d^2}{(g^2-1)^2}, \\ \quad \sqrt{\frac{\left(4 s^2 + 1\right) d^2 + s^2 +
            4 s \sqrt{\left(s^2 + 1\right) d^2 + s^2} |d|}{d^2 + s^2}}
\le g < \sqrt{2(s^2 + d^2)-1}\; ; \\ \\
 \\
    \frac{-g^4 +
      2 \left(2 d^2 + 2 s^2 +
            1\right) g^2 - \left(4 d^2 - 1\right) \left(4 s^2 -
            1\right) - \sqrt{\delta}}{8 g^2}, \\ \qquad 2 |d| + 1 \le g < \sqrt{\frac{\left(4 s^2 + 1\right) d^2 + s^2 +
            4 s |d| \sqrt{\left(s^2 + 1\right) d^2 + s^2} }{d^2 + s^2}}\,.
\\
\end{array}
\right.
\end{eqnarray}
Here $\delta \equiv (2 d - g - 1) (2 d - g + 1) (2 d + g - 1) (2 d +
                g + 1) (g - 2 s - 1) (g - 2 s + 1) (g + 2 s - 1) (g + 2 s +
                1)$.

Immediate inspection crucially reveals that ${m^2}_{opt}^{_{\rm
GLEMS}}$ is {\em not} in general a function of the symplectic
eigenvalue $\tilde\nu_-$ alone. Therefore,  the Gaussian EMs, and in
particular, the Gaussian entanglement of formation, are not
equivalent to the negativities for GLEMS. Further remarks will be
given in the following, when the Gaussian EMs of GLEMS and GMEMS
will be compared and their relationship with the negativities will
be elucidated.

\subsubsection{Gaussian entanglement of maximum-negativity states
(GMEMS)} \label{SecGEMGLEM}

The minimization of $m^2_\theta$ from \eq{mfunc} can be carried out
in a  simpler way in the case of GMEMS, whose covariances can be
retrieved from \eq{cpm} setting $\lambda = 1$. First of all, one can
notice that, when expressed as a function of the Minkowski
coordinates $(x_0,x_1,x_3)$, corresponding to the submatrix $\Gamma$
[\eq{Gamma}] of the pure state $\sig^p = \Gamma \oplus \Gamma^{-1}$
entering in the optimization problem \eq{Gaussian EMm}, the
single-mode determinant $m^2$ of $\sig^p$ is globally minimized for
$x_3=0$. In fact, from \eq{mdef}, $m^2$ is minimal, with respect to
$x_3$, when $\det\Gamma = x_0^2 - x_1^2 - x_3^2$ is maximal. Next,
one can show that for GMEMS there always exists a matrix $\Gamma$,
with $x_3=0$, which is a simultaneous solution of the two matrix
equations obtained by imposing the saturation of the two sides of
inequality \pref{rim}. As a consequence of the above discussion,
this matrix would denote the optimal pure state $\sig^p_{opt}$. By
solving the system of equations $\det(\gamma_q - \Gamma) =
\det(\Gamma - \gamma_p^{-1}) = 0$, where the matrices involved are
explicitly defined combining \eq{cpcq} and \eq{cpm} with $\lambda =
1$, one finds the following two solutions for the coordinates $x_0$
and $x_1$:
\begin{equation}
\begin{split}
\!\!x_0^{\pm} &= \frac{(g + 1) s \pm \sqrt{\left[(g - 1)^2 - 4
d^2\right] \left(-d^2 + s^2 - g\right)}}{2 \left(d^2 + g\right)}\,,\\
\!\!x_1^{\pm} &= \frac{(g + 1) \sqrt{-d^2 + s^2 - g}\pm s \sqrt{(g -
1)^2 - 4 d^2}}{2 \left(d^2 + g\right)}\,.
\end{split}
\end{equation}
The corresponding pure state $\sig^{p\pm} = \Gamma^{\pm} \oplus
{\Gamma^{\pm}}^{-1}$ turns out to be, in both cases, a two-mode
squeezed state described by a CM of the form \eq{tms}, with $\cosh
(2r) = x_0^\pm$. Because the single-mode determinant $m^2 = \cosh^2
(2r)$ for these states, the optimal $m^2$ for GMEMS is simply equal
to $(x_0^-)^2$. Summarizing,
\begin{equation}\label{m2gmems}
{m^2}_{opt}^{_{\rm GMEMS}}=\left\{
\begin{array}{ll}
    1, \qquad \qquad g \ge 2s-1 \quad {\hbox{[separable state]}}\; ; \\
& \\
    \frac{\left\{(g + 1) s - \sqrt{\left[(g - 1)^2 - 4 d^2\right] \left(-d^2 +
                      s^2 - g\right)}\right\}^2}{4 \left(d^2 + g\right)^2},\\
\qquad \qquad \ \ \; 2 |d| + 1 \le g  < 2s-1\; . \\
\end{array}
\right.
\end{equation}
Once again, also for the class of GMEMS the Gaussian EMs are not
simple functions of the symplectic eigenvalue $\tilde\nu_-$ alone.
Consequently, they provide a quantification of CV entanglement of
GMEMS inequivalent to the one determined by the negativities.
Furthermore, we will now show how these results raise the problem of
the ordering of two-mode Gaussian states according to their degree
of entanglement, as quantified by different families of entanglement
measures \cite{ordering}.

\subsection{Entanglement-induced ordering of two-mode Gaussian states}

%%%%%%%%%%%%%%%%%%%%%%%%%%%%%%%%%%%%%%%%%%%%%%%%%%
\begin{figure}[t!]
\includegraphics[width=8.5cm]{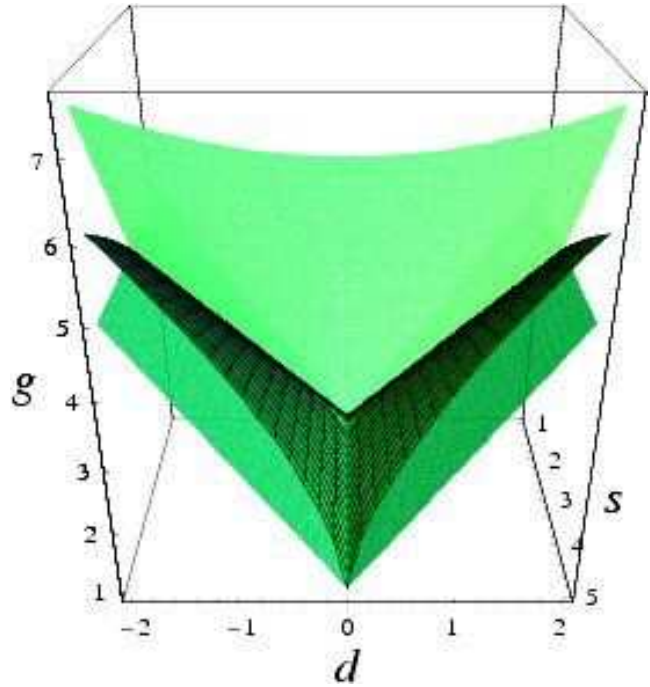}
\caption{Comparison between the ordering induced by Gaussian EMs on
the classes of states with extremal (maximal and minimal)
negativities. This {\em extremal ordering} of the set of entangled
two-mode Gaussian states is studied in the space of the CM's
parameters $\{s,d,g\}$, related to the global and local purities by
the relations $\mu_1 = (s+d)^{-1}$, $\mu_2 = (s-d)^{-1}$ and $\mu =
g^{-1}$.  The intermediate, meshed surface is constituted by those
global and local mixednesses such that the Gaussian EMs give equal
values for the corresponding GMEMS (states of maximal negativities)
and GLEMS (states of minimal negativities). Below this surface, the
extremal ordering is inverted (GMEMS have less Gaussian EM than
GLEMS). Above it, the extremal ordering is preserved (GMEMS have
more Gaussian EM than GLEMS). However, it must be noted that this
does not exclude that the individual orderings induced by the
negativities and by the Gaussian EMs on a pair of non-extremal
states may still be inverted in this region. Above the uppermost,
lighter surface, GLEMS are separable states, so that the extremal
ordering is trivially preserved. Below the lowermost, darker
surface, no physical two-mode Gaussian states can exist. }
\label{figorder3d}
\end{figure}
%%%%%%%%%%%%%%%%%%%%%%%%%%%%%%%%%%%%%%%%%%%%%%%%%%

We have more than once remarked that, in the context of CV systems,
when one restricts to symmetric, two-mode Gaussian states (which
include all pure states) the known computable measures of
entanglement all correctly induce {\em the same} ordering on the set
of entangled states \cite{ordering}. We will now show that, indeed,
this nice feature is not preserved moving to mixed, nonsymmetric
two-mode Gaussian states. We aim at comparing Gaussian EMs and
negativities on the two extremal classes of two-mode Gaussian states
\cite{extremal}, introducing thus the concept of {\em extremal
ordering}. At fixed global and local purities, the negativity of
GMEMS (which is the maximal one) is obviously always greater than
the negativity of GLEMS (which is the minimal one). If for the same
values of purities the Gaussian EMs of GMEMS are larger than those
of GLEMS, we will say that the extremal ordering is preserved.
Otherwise, the extremal ordering is inverted. In this latter case,
which is clearly the most intriguing, the states of minimal
negativities are more entangled, with respect to Gaussian EMs, than
the states of maximal negativities, and the inequivalence of the
orderings, induced by the two different families of entanglement
measures, becomes manifest.

The problem can be easily stated. By comparing $m_{opt}^{_{\rm
GLEMS}}$ from \eq{m2glems} and $m_{opt}^{_{\rm GMEMS}}$ from
\eq{m2gmems}, one has that in the range of global and local
purities, or, equivalently, of parameters $\{s,d,g\}$, such that
\begin{equation}\label{preserved}
m_{opt}^{_{\rm GMEMS}} \ge m_{opt}^{_{\rm GLEMS}}\; ,
\end{equation}
the extremal ordering is preserved. When \ineq{preserved} is
violated, the extremal ordering is inverted. The boundary between
the two regions, which can be found imposing the equality
$m_{opt}^{_{\rm GMEMS}} = m_{opt}^{_{\rm GLEMS}}$, yields the range
of global and local purities such that the corresponding GMEMS and
GLEMS, despite having different negativities, have equal Gaussian
EMs. This boundary surface can be found numerically, and the result
is shown in the 3D plot of Fig.~\ref{figorder3d}.

%%%%%%%%%%%%%%%%%%%%%%%%%%%%%%%%%%%%%%%%%%%%%%%%%%
\begin{figure}[t!]
\includegraphics[width=7.7cm]{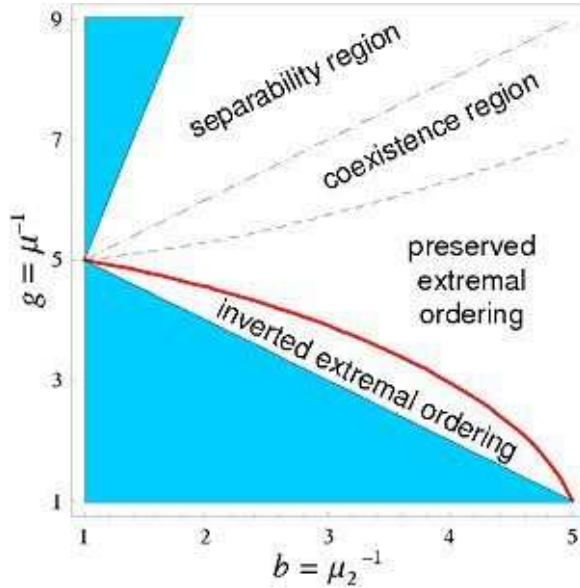}
\caption{Summary of entanglement properties of two-mode Gaussian
states, in the projected space of the local mixedness $b =
\mu_2^{-1}$ of mode $2$, and of the global mixedness $g = \mu^{-1}$,
while the local mixedness of mode $1$ is kept fixed at a reference
value $a = \mu_1^{-1} = 5$. Below the thick curve, obtained imposing
the equality in \ineq{preserved}, the Gaussian EMs yield GLEMS more
entangled than GMEMS, at fixed purities: the extremal ordering is
thus inverted. Above the thick curve, the extremal ordering is
preserved. In the coexistence region (see Fig.~\ref{fig2D} and Table
\ref{table1}), GMEMS are entangled while GLEMS are separable. The
boundaries of this region are given by \eq{glement} (dashed line)
and \eq{gmement} (dash-dotted line). In the separability region,
GMEMS are separable too, so all two-mode Gaussian states whose
purities lie in that region are not entangled. The shaded regions
cannot contain any physical two-mode Gaussian state. }
\label{figorder2d}
\end{figure}
%%%%%%%%%%%%%%%%%%%%%%%%%%%%%%%%%%%%%%%%%%%%%%%%%%

 One can see, as a crucial result, that a region where the extremal
ordering is inverted does indeed exist. The Gaussian EMs and the
negativities are thus definitely {\em not} equivalent for the
quantification of entanglement in nonsymmetric two-mode Gaussian
states. The interpretation of this result is quite puzzling. On the
one hand, one could think that the ordering induced by the
negativities is a natural one, due to the fact that such measures of
entanglement are directly inspired by the necessary and sufficient
PPT criterion for separability. Thus, one would expect that the
ordering induced by the negativities should be preserved by any {\em
bona fide} measure of entanglement, especially if one considers that
the extremal states, GLEMS and GMEMS, have a clear physical
interpretation. Therefore, as the Gaussian entanglement of formation
is an upper bound to the true entanglement of formation, one could
be tempted to take this result as an evidence that the latter is
globally minimized on non-Gaussian decomposition, at least for
GLEMS. However, this is only a qualitative/speculative argument:
proving or disproving that the Gaussian entanglement of formation is
the true one for any two-mode Gaussian state is still an open
question under lively debate \cite{QIProb}.

On the other hand, one could take the simplest discrete-variable
instance, constituted by a two-qubit system, as a test-case for
comparison. There, although for pure states the negativity coincides
with the concurrence, an entanglement monotone equivalent to the
entanglement of formation for all states of two qubits
\cite{Wootters97} (see Sec.~\ref{SecEnt2Q}),  the two measures cease
to be equivalent for mixed states, and the orderings they induce on
the set of entangled states can be different \cite{VerstraeteJPA}.
This analogy seems to support again (see Sec.~\ref{SecOrderDiscuss})
the stand that, in the arena of mixed states, a unique measure of
entanglement is a {\em chimera} and cannot really be pursued, due to
the different operative meanings and physical processes (in the
cases when it has been possible to identify them) that are
associated to each definition: one could think, for instance, of the
operative difference existing between the definitions of distillable
entanglement and entanglement cost (see Sec.~\ref{SecEntMeas}). In
other words, from this point of view, each inequivalent measure of
entanglement introduced for mixed states should capture physically
distinct aspects of quantum correlations existing in these states.
Then, joining this kind of outlook, one could hope that the Gaussian
EMs might still be considered as proper measures of CV entanglement,
adapted to a different context than negativities. This point of view
will be proven especially correct when constructing Gaussian EMs to
investigate entanglement sharing in multipartite Gaussian states, as
discussed in Part \ref{PartMulti}.

 Whatever be the case, we have
shown that two different families of measures of CV entanglement can
induce different orderings on the set of two-mode entangled states.
This is more clearly illustrated in Fig.~\ref{figorder2d}, where we
keep fixed one of the local mixednesses and we classify, in the
space of the other local mixedness and of the global mixedness, the
different regions related to entanglement and extremal ordering of
two-mode Gaussian states, completing diagrams like Fig.~\ref{fig2D}
and Fig.~\ref{fig2Dm}{\rm (b)}, previously introduced to describe
separability in the space of purities.

\subsection{Comparison between Gaussian entanglement measures and
negativities} \label{SecGEMvsNEGcomparison}

We wish to give now a more direct comparison of the two families of
entanglement measures for two-mode Gaussian states \cite{ordering}.
In particular, we are interested in finding the maximum and minimum
values of one of the two measures, if the other is kept fixed. A
very similar analysis has been performed by Verstraete {\em et al.}
\cite{VerstraeteJPA}, in their comparative analysis of the
negativity and the concurrence for states of two-qubit systems.

Here it is useful to perform the comparison directly between the
symplectic eigenvalue $\tilde\nu_-(\sig)$ of the partially
transposed CM $\tilde \sig$ of a generic two-mode Gaussian state
with CM $\sig$, and the symplectic eigenvalue
$\tilde\nu_-(\sig^p_{opt})$ of the partially transposed CM $\tilde
\sig^{p}_{opt}$ of the optimal pure state with CM $\sig^p_{opt}$,
which minimizes \eq{Gaussian EMm}. In fact, the negativities are all
monotonically decreasing functions of $\tilde\nu_-(\sig)$, while the
Gaussian EMs are all monotonically decreasing functions of
$\tilde\nu_-(\sig^p_{opt})$.

To start with, let us recall once more that for pure states and for
mixed symmetric states (in the set of two-mode Gaussian states), the
two quantities coincide \cite{giedke03}. For nonsymmetric states,
one can immediately prove the following bound
\begin{equation}\label{ubound}
\tilde\nu_-(\sig^p_{opt}) \le \tilde\nu_-(\sig)\,.
\end{equation}
In fact, from \eq{Gaussian EMm}, $\sig^p_{opt} \le \sig$
\cite{GEOF}. For positive matrices, $A \ge B$ implies $a_k \ge b_k$,
where the $a_k$'s (resp. $b_k$'s) denote the ordered symplectic
eigenvalues of $A$ (resp. $B$) \cite{giedkeqic03}. Because the
ordering $A \ge B$ is preserved under partial transposition,
\ineq{ubound} holds true. This fact induces a characterization of
symmetric states, which saturate \ineq{ubound}, as the two-mode
Gaussian states with {\em minimal} Gaussian EMs at fixed
negativities.

\begin{figure}[t!]
\includegraphics[width=10cm]{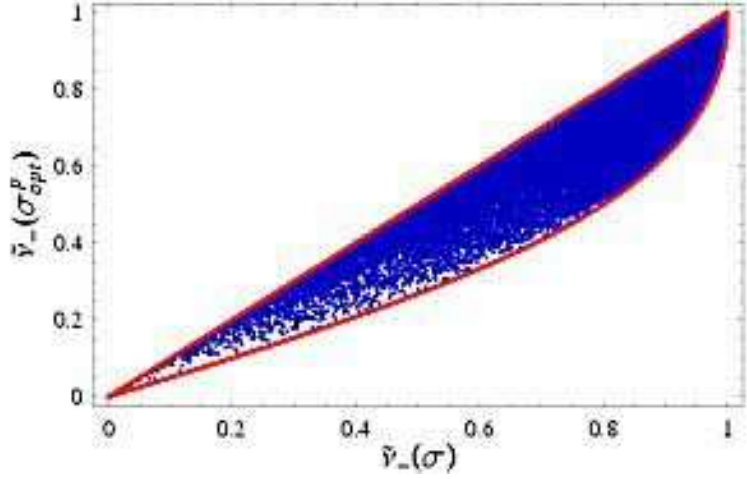}
\caption{Comparison between Gaussian EMs and negativities for
two-mode Gaussian states. On the horizontal axis we plot the
symplectic eigenvalue $\tilde\nu_-(\sig)$ of the partially
transposed CM $\tilde \sig$ of a generic two-mode Gaussian state
with CM $\sig$. On the vertical axis we plot the symplectic
eigenvalue $\tilde\nu_-(\sig^p_{opt})$ of the partially transposed
CM $\tilde \sig^{p}_{opt}$ of the optimal pure state with CM
$\sig^p_{opt}$, which minimizes \eq{Gaussian EMm}. The negativities
are all monotonically decreasing functions of $\tilde\nu_-(\sig)$,
while the Gaussian EMs are all monotonically decreasing functions of
$\tilde\nu_-(\sig^p_{opt})$. The equation of the two boundary curves
are obtained from the saturation of \ineq{ubound} (upper bound) and
\ineq{lbound} (lower bound), respectively. The dots represent 50~000
randomly generated CMs of two-mode Gaussian states. Of up to $1$
million random CMs, none has been found to lie below the lower
solid-line curve, enforcing the conjecture that it be an absolute
boundary for all two-mode Gaussian states.} \label{figfrank}
\end{figure}

It is then natural to raise the question whether an upper bound on
the Gaussian EMs at fixed negativities exists as well. It seems hard
to address this question directly, as one lacks a closed expression
for the Gaussian EMs of generic states. But we can promptly give
partial answers if we restrict to the classes of GLEMS and of GMEMS,
for which the Gaussian EMs have been explicitly computed in
Sec.~\ref{SecGEMextra}.

Let us begin with the GLEMS. We can compute the squared symplectic
eigenvalue $$\tilde\nu_-^2(\sig^{_{\rm GLEMS}}) =
 {\left[4 (s^2 +d^2) - g^2-1 -
 \sqrt{\left(4 (s^2 +d^2) - g^2-1\right)^2 - 4 g^2}\right]/2}\,.$$ Next,
 we can reparametrize the CM (obtained by \eq{cpm} with $\lambda =
 -1$) to make $\tilde\nu_-$ appear explicitly, namely
$g = \sqrt{\tilde\nu_-^2
[4(s^2+d^2)-1-\tilde\nu_-^2]/(1+\tilde\nu_-^2)}$. At this point, one
can study the piecewise function ${m^2}_{opt}^{_{\rm GLEMS}}$ from
\eq{m2glems}, and find out that it is a convex function of $d$ in
the whole space of parameters corresponding to entangled states.
Hence, ${m^2}_{opt}^{_{\rm GLEMS}}$, and thus the Gaussian EM, is
maximized at the boundary $|d| = (2 \tilde\nu_- s - \tilde\nu_-^2 -
1)/2$, resulting from the saturation of \ineq{gmbound}. The states
maximizing Gaussian EMs at fixed negativities, if we restrict to the
class of GLEMS, have then to be found in the subclass of GMEMMS
(states of maximal negativity for fixed marginals \cite{extremal},
defined by \eq{gmemms} in Sec.~\ref{SecMEMMS}), depending on the
parameter $s$ and on the eigenvalue $\tilde\nu_-$ itself, which
completely determines the negativity. For these states,
\begin{equation}\label{mgm}
    m^{_{\rm GMEMMS}}_{opt} (s, \tilde \nu_-) =
    \frac{2s}{1-\tilde\nu_-^2 + 2 \tilde\nu_- s}\,.
\end{equation}
The further optimization over $s$ is straightforward because
$m^{_{\rm GMEMMS}}_{opt}$ is an increasing function of $s$, so its
global maximum is attained for $s \rightarrow \infty$. In this
limit, one has simply
\begin{equation}\label{mmax}
    m^{_{\rm GMEMMS}}_{\max} (\tilde \nu_-) =\frac{1}{\tilde \nu_-}\,.
\end{equation}
From \eq{nutopt}, one thus finds that for all GLEMS the following
bound holds
\begin{equation}\label{lbound}
\tilde\nu_-(\sig^p_{opt}) \ge
\frac{1}{\tilde\nu_-(\sig)}\Big(1-\sqrt{1-\tilde\nu_-^2(\sig)}\Big)\,.
\end{equation}

One can of course perform a similar analysis for GMEMS. But, after
analogous reasonings and computations, what one finds is exactly the
same result. This is not so surprising, keeping in mind that GMEMS,
GLEMS and all two-mode Gaussian states with generic $s$ and $d$ but
with global mixedness $g$ saturating \ineq{gmbound}, collapse into
the same family of two-mode Gaussian states, the GMEMMS, completely
determined by the local single-mode properties (they can be viewed
as a generalization of the pure two-mode states: the symmetric
GMEMMS are in fact pure). Hence, the bound of \ineq{lbound},
limiting the Gaussian EMs from above at fixed negativities, must
hold for all GMEMS as well.

At this point, it is tempting to conjecture that \ineq{lbound} holds
for all two-mode Gaussian states. Unfortunately, the lack of a
closed, simple expression for the Gaussian EM of a generic state
makes the proof of this conjecture impossible, at the present time.
However, one can show, by analytical power-series expansions of
\eq{mfunc}, truncated to the leading order in the infinitesimal
increments, that, for any infinitesimal variation of the parameters
of a generic CM around the limiting values characterizing GMEMMS,
the Gaussian EMs of the resulting states lie always below the
boundary imposed by the corresponding GMEMMS with the same
$\tilde\nu_-$. In this sense, the GMEMMS are, at least, a {\em
local} maximum for the Gaussian EM versus negativity problem.
Furthermore, extensive numerical investigations of up to a million
CMs of randomly generated two-mode Gaussian states, provide
confirmatory evidence that GMEMMS attain indeed the {\em global}
maximum (see Fig.~\ref{figfrank}). We can thus quite confidently
conjecture, however, at the moment, without a complete formal proof
of the statement, that GMEMMS, in the limit of infinite average
local mixedness ($s \rightarrow \infty$), are the states of maximal
Gaussian EMs at fixed negativities, among {\em all} two-mode
Gaussian states.

\begin{figure}[tb]
\includegraphics[width=10.5cm]{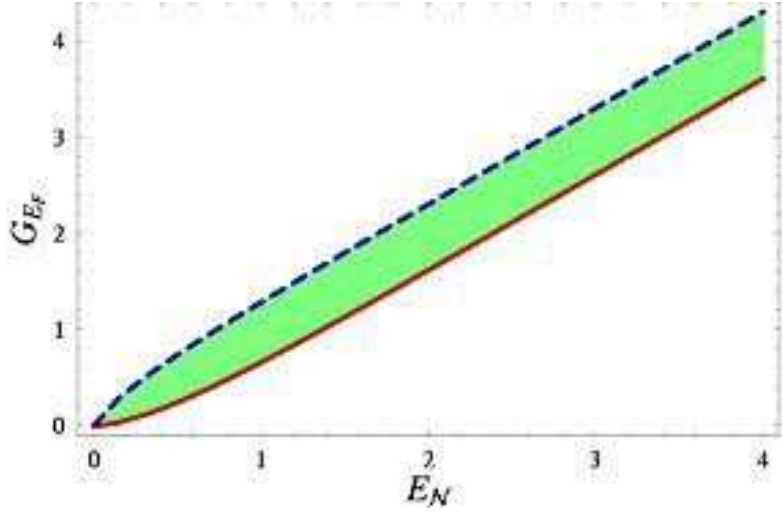}
\caption{Comparison between the Gaussian entanglement of formation
$G_{E_F}$ and the logarithmic negativity $E_\N$ for two-mode
Gaussian states. Symmetric states accomodate on the lower boundary
(solid line), determined by the saturation of \ineq{eflower}. GMEMMS
with infinite, average local mixedness, lie on the dashed line,
whose defining equation is obtained from the saturation of
\ineq{efupper}. All GMEMS and GLEMS lie below the dashed line. The
latter is conjectured, with strong numerical support, to be the
upper boundary for the Gaussian entanglement of formation of all
two-mode Gaussian states, at fixed negativity. } \label{geofvslone}
\end{figure}

A direct comparison between the two prototypical representatives of
the two families of entanglement measures, respectively the Gaussian
entanglement of formation $G_{E_F}$ and the logarithmic negativity
$E_\N$, is plotted in Fig.~\ref{geofvslone}. For any fixed value of
$E_\N$, \ineq{ubound} provides in fact a rigorous lower bound on
$G_{E_F}$, namely
\begin{equation}\label{eflower}
G_{E_F} \ge h[\exp(-E_\N)]\,,
\end{equation}
while \ineq{lbound} provides the conjectured lower bound
\begin{equation}\label{efupper}
G_{E_F} \le
h\left[\exp(E_\N)\left(1-\sqrt{1-\exp(-2E_\N)}\right)\right]\,,
\end{equation}
where we exploited Eqs.~{\rm(\ref{en},\ref{geofm})}, and $h[x]$ is
given by \eq{hentro}.

The existence of lower and upper bounds on the Gaussian EMs at fixed
negativities (the latter strictly proven only for extremal states),
limits to some extent the inequivalence arising between the two
families of entanglement measures, for nonsymmetric two-mode
Gaussian states.

We have thus demonstrated the following.

\medskip

\begin{itemize}
\item[\ding{226}]
 \noindent{\rm\bf Ordering two-mode Gaussian states with entanglement measures.}
{\it The Gaussian entanglement measures and the negativities induce
inequivalent orderings on the set of entangled, nonsymmetric,
two-mode Gaussian states. This inequivalence is however constrained:
at fixed negativities, the Gaussian measures of entanglement are
bounded from below (the states which saturate this bound are simply
symmetric two-mode states); moreover, we provided some strong
evidence suggesting that they are as well bounded from above.}
\smallskip
\end{itemize}

\section{Summary and further remarks}%\label{concl}

Summarizing, in this Chapter we focused on the simplest conceivable
states of a bipartite CV system: two-mode Gaussian states. We have
shown that, even in this simple instance, the theory of quantum
entanglement hides several subtleties and reveals some surprising
aspects.

Following Refs.~\cite{prl,extremal,polacchi}, we have pointed out
the existence of both maximally and minimally entangled two-mode
Gaussian states at fixed local and global generalized $p-$entropies.
The analytical properties of such states have been studied in detail
for any value of $p$. Remarkably, for $p\le 2$, minimally entangled
states are minimum-uncertainty states, saturating \ineq{bonfide},
while maximally entangled states are nonsymmetric two-mode squeezed
thermal states. Interestingly, for $p>2$ and in specific ranges of
the values of the entropic measures, the role of such states is
reversed. In particular, for such quantifications of the global and
local entropies, two-mode squeezed thermal states, often referred to
as CV analog of maximally entangled states, turn out to be minimally
entangled.  Moreover, we have introduced the notion of ``average
logarithmic negativity'' for given global and marginal generalized
$p$-entropies, showing that it provides a reliable estimate of CV
entanglement in a wide range of physical parameters.

Our analysis also clarifies the reasons why the linear entropy is a
`privileged' measure of mixedness in continuous variable systems. It
is naturally normalized between $0$ and $1$, it offers an accurate
qualification and quantification of entanglement of any mixed state
while giving significative information about the state itself and,
crucially, is the only entropic measure which could be directly
measured in the near future by schemes involving only single-photon
detections or the technology of quantum networks, without requiring
a full homodyne reconstruction of the state.

We have furthermore studied, following Ref.~\cite{ordering}, the
relations existing between different computable measures of
entanglement, showing how the negativities (including the standard
logarithmic negativity) and the Gaussian convex-roof extended
measures (Gaussian EMs, including the Gaussian entanglement of
formation) are inequivalent entanglement quantifiers for
nonsymmetric two-mode Gaussian states. We have computed Gaussian EMs
explicitly for the two classes of two-mode Gaussian states having
extremal (maximal and minimal) negativities at fixed purities. We
have highlighted how, in a certain range of values of the global and
local purities, the ordering on the set of entangled states, as
induced by the Gaussian EMs, is inverted with respect to that
induced by the negativities. The question whether a certain Gaussian
state is more entangled than another, thus, has no definite answer,
not even when only extremal states are considered, as the answer
comes to depend on the measure of entanglement one chooses. Extended
comments on the possible meanings and consequences of the existence
of inequivalent orderings of entangled states have been provided.
Furthermore, we have proven the existence of a lower bound holding
for the Gaussian EMs at fixed negativities, and that this bound is
saturated by two-mode symmetric Gaussian states. Finally, we have
provided some strong numerical evidence, and partial analytical
proofs restricted to extremal states, that an upper bound on the
Gaussian EMs at fixed negativities exists as well, and is saturated
by states of maximal negativity for given marginals, in the limit of
infinite average local mixedness.

We believe that our results will raise renewed interest in the
problem of the quantification of entanglement in CV systems, which
seemed fairly well understood in the special instance of two-mode
Gaussian states. Moreover, we hope that the present Chapter may
constitute a first step toward the solution of more general (open)
problems concerning the entanglement of Gaussian states
\cite{QIProb}, such as the computation of the entanglement of
formation for generic two-mode Gaussian states, and the proof (or
disproof) of its identity with the Gaussian entanglement of
formation in a larger class of Gaussian states beyond the symmetric
instance.

We are now going to show, in Chapter \ref{ChapUniloc}, how some of
the results here derived for two-mode states, can be extended for
the investigation of bipartite entanglement in multimode Gaussian
states endowed with peculiar symmetric structures. Last but not the
least, the results collected in the present Chapter might prove
useful as well in the task of quantifying multipartite entanglement
of Gaussian states. For instance, we should mention here that any
two-mode reduction of a pure three-mode Gaussian state is a GLEMS,
as we will show in Chapter \ref{Chap3M} (this straightforwardly
follows from the phase-space Schmidt decomposition discussed in
Sec.~\ref{SecSchmidtPS}). Therefore, one has then available the
tools and can apply them to investigate the sharing structure of
multipartite CV entanglement of three-mode, and, more generally,
multimode Gaussian states, as we will do in Chapter
\ref{ChapMonoGauss}.

Let us moreover mention that the experimental production and
manipulation of two-mode Gaussian entanglement will be discussed in
Chapter \ref{Chap2MExp}.

}

\chapter{Multimode entanglement under symmetry} \label{ChapUniloc}

{\sf

 In quantum information and computation science, it is of
particular relevance to provide theoretical methods to determine the
entanglement of systems susceptible to encompass many parties. Such
an interest does not stem only from pure intellectual curiosity, but
also from practical needs in the implementations of realistic
information protocols. This is especially true as soon as one needs
to encode two-party information in a multipartite structure in order
to minimize possible errors and decoherence effects
\cite{chuaniels,heiss}. The study of the structure of multipartite
entanglement poses many formidable challenges, concerning both its
qualification and quantification, and so far little progress has
been achieved for multi-qubit systems and in general for multi-party
systems in finite-dimensional Hilbert spaces. However, the situation
looks somehow more promising in the arena of CV systems, where some
aspects of genuine multipartite entanglement can be, to begin with,
qualitatively understood studying the entanglement of multimode
bipartitions.

In the present Chapter, based on Refs.~\cite{adescaling,unitarily}
we analyze in detail the entanglement properties of multimode
Gaussian states endowed with particular symmetry constraints under
mode permutations. Their usefulness arises in contexts like quantum
error correction \cite{BraunsteinERR}, where some redundancy is
required for a fault-tolerant encoding of information. Bisymmetric
and, as a special case, fully symmetric Gaussian states have been
introduced in Sec.~\ref{SecSymm}. An analysis of the symplectic
spectra of $(M + N)$-mode Gaussian states has revealed that, with
respect to the bipartition across which they exhibit the local
permutation invariance (any bipartition is valid for fully symmetric
states), local symplectic diagonalizations of the $M$-mode and the
$N$-mode blocks result in a complete reduction of the multimode
state to an equivalent two-mode state, tensor $M+N-2$ uncorrelated
thermal single-mode states. The equivalent two-mode state encodes
all the information of the original bisymmetric multimode state for
what concerns entropy and entanglement. As a consequence, the
validity of the PPT criterion as a necessary and sufficient
condition for separability has been extended to bisymmetric Gaussian
states in Sec.~\ref{SecPPTG}.

Here, equipped with the powerful theoretical tools for the analysis
of two-mode entanglement in Gaussian states, demonstrated in the
previous Chapter, we perform a close analysis of the multimode
entanglement in symmetric and bisymmetric Gaussian states. In
particular, we will investigate how the block entanglement scales
with the number of modes, hinting at the presence of genuine
multipartite entanglement arising among all the modes as their total
number increases, at a given squeezing degree. Motivated by this
analysis, in the next Part of this Dissertation we will face
full-force the problem of quantifying the crucial and hideous
property of genuine multipartite CV entanglement in Gaussian states.

 The central observation of the present Chapter is embodied by following result \cite{adescaling,unitarily},
straightforwardly deducible from the discussions in
Sec.~\ref{SecSymm} and Sec.~\ref{SecPPTG}.

\medskip

\begin{itemize}
\item[\ding{226}]
 \noindent{\rm\bf Unitarily localizable entanglement of bisymmetric Gaussian states.}
{\it The bipartite entanglement of bisymmetric $(M + N)$-mode
Gaussian states under $M \times N$ partitions is ``unitarily
localizable'', namely, through local unitary (reversible)
operations, it can be completely concentrated onto a single pair of
modes, each of them belonging respectively to the $M$-mode and to
the $N$-mode blocks.}
\smallskip
\end{itemize}

Hence the multimode block entanglement ({\ie}the entanglement
between blocks of modes) of bisymmetric (generally mixed) Gaussian
states can be determined as a two-mode entanglement. The
entanglement will be quantified by the logarithmic negativity in the
general instance because the PPT criterion holds,
 but we will also show some explicit nontrivial cases in which the
entanglement of formation, \eq{E:EF}, between $M$-mode and $N$-mode
parties can be exactly computed.

We remark that our notion of ``localizable entanglement'' is
different from that introduced by Verstraete, Popp, and Cirac for
spin systems \cite{localiz}. There, it was defined as the maximal
entanglement concentrable on two chosen spins through local {\em
measurements} on all the other spins.\footnote{\sf This
(non-unitarily) localizable entanglement will be also computed for
(mixed) symmetric Gaussian states of an arbitrary number of modes,
and it will be shown to be in direct quantitative connection with
the optimal fidelity of multiparty teleportation networks
\cite{telepoppate} (see Sec.~\ref{SecTelepoppy} and
Fig.~\ref{figlocaliz}).} Here, the local operations that concentrate
all the multimode entanglement on two modes are {\em unitary} and
involve the two chosen modes as well, as parts of the respective
blocks.

\section{Bipartite block entanglement of bisymmetric Gaussian states}\label{block}

In Sec.~\ref{SecSymm}, the study of the multimode CM $\sig$ of
\eq{totala} has been reduced to a two-mode problem by means of local
unitary operations. This finding allows for an exhaustive analysis
of the bipartite entanglement between the $M$- and $N$-mode blocks
of a multimode bisymmetric Gaussian state, resorting to the powerful
results available for two-mode Gaussian states (see Chapter
\ref{Chap2M}). For any multimode Gaussian state with CM $\sig$, let
us define the associated {\em equivalent} two-mode Gaussian state
$\varrho_{eq}$, with CM $\sig_{eq}$ given by \be \sig_{eq}=
\left(\begin{array}{cc}
\gr{\nu}_{\alpha^M}^{+}&\gr{\gamma''}\\
\gr{\gamma''}^{\sf T}& \gr{\nu}_{\beta^N}^{+}
\end{array}\right) \; ,
\ee where the $2\times 2$ blocks have been implicitly defined in the
CM,  \eq{final}. As already mentioned, the entanglement of the
bisymmetric state with CM $\sig$, originally shared among all the
$M+N$ modes, can be {\em completely} concentrated by local unitary
(symplectic) operations on a single pair of modes in the state with
CM $\sig_{eq}$. Such an entanglement is, in this sense, {\em
localizable}.

 We now move on to describe some consequences of this
result. A first qualificative remark has been explored in
Sec.~\ref{SecPPTG}. Namely, PPT criterion turns out to be
automatically necessary and sufficient for separability of
$(M+N)$-mode Gaussian states under $M \times N$ bipartitions
\cite{unitarily}. For a more quantitative investigation, the
following symplectic analysis, which completes that of
Sec.~\ref{SecSymm}, will be precious.

\subsection{Symplectic properties of symmetric states}

As a preliminary analysis, it is useful to provide a symplectic
parametrization for the standard form coefficients of any two-mode
reduced state of a fully symmetric $N$-mode CM $\sig_{\beta^N}$,
\eq{fscm}. Following the discussion in Sec.~\ref{Sec2M}, the
coefficients $b$, $z_1$, $z_2$ of the standard form are determined
by the local, single-mode invariant
$\det{\bet}\equiv\mu_{\beta}^{-2}$, and by the symplectic invariants
$\det{\sig_{\beta^2}}\equiv \mu_{\beta^2}^{-2}$ and
$\D_{\beta^2}\equiv \D(\sig_{\beta^2})$. Here $\mu_{\beta}$
($\mu_{\beta^2}$) is the marginal purity of the single-mode
(two-mode) reduced states, while $\D_{\beta^2}$ is the remaining
seralian invariant, \eq{seralian}, of the two-mode reduced states.
According to  Sec.~\ref{SecSympParam}, this parametrization is
provided, in the present instance, by the following equations \be
b=\frac{1}{\mu_{\beta}} \, , \quad z_1=\frac{\mu_{\beta}}{4}
(\epsilon_- - \epsilon_+)\,,\quad z_2=\frac{\mu_{\beta}}{4}
(\epsilon_- + \epsilon_+)\,,\label{parasim} \ee
\[
{\rm with}\quad
\epsilon_{-}=\sqrt{\D_{\beta^2}^2-\frac{4}{\mu_{\beta^2}^2}}\,,
\]
\[
{\rm and}\quad\epsilon_{+} =
\sqrt{\left(\D_{\beta^2}-\frac{4}{\mu_{\beta}^2}\right)^2-\frac{4}{\mu_{\beta^2}^2}}
\, .
\]
This parametrization has a straightforward interpretation, because
$\mu_{\beta}$ and $\mu_{\beta^2}$ quantify the local mixednesses and
$\D_{\beta^2}$ regulates the entanglement of the two-mode blocks at
fixed global and local purities \cite{prl} (see Sec.~\ref{secprl}).

Moreover, we can connect the symplectic spectrum of
$\sig_{\beta^N}$, given by \eq{fsspct}, to the known symplectic
invariants. The $(N-1)$-times degenerate eigenvalue $\nu_{\beta}^-$
is independent of $N$, while $\nu_{\beta^N}^{+}$ can be simply
expressed as a function of the single-mode purity $\mu_\beta$ and
the symplectic spectrum of the two-mode block with eigenvalues
$\nu_{\beta}^{-}$ and $\nu_{\beta^2}^{+}$:
\begin{equation}\label{fsnupiun}
(\nu_{\beta^N}^{+})^2 = -\frac{N(N-2)}{\mu_\beta^2}+\frac{(N-1)}
2\left( N({\nu^{+}_{\beta^2}})^2+(N-2)(\nu^{-}_{\beta})^2 \right)\,.
\end{equation}
In turn, the two-mode symplectic eigenvalues are determined by the
two-mode invariants by the relation \be
%\begin{split}
2(\nu^{\mp}_{\beta})^2 = \Delta_{\beta^2}\mp\sqrt{\Delta_{\beta^2}^2-4/\mu_{\beta^2}^2} \; .%\\
%   2\nu^{+2}_{\beta^2} = &
 % \Delta_{\beta^2}+\sqrt{\Delta_{\beta^2}^2-4/\mu_{\beta^2}^2} \, .
\label{symp2}
%\end{split}
\ee The global purity \eq{purgau} of a fully symmetric multimode
Gaussian state is
\begin{equation}\label{fsmu}
\mu_{\beta^N}\equiv\left(\det\sig_{\beta^N}\right)^{-1/2}=
\left((\nu_\beta^-)^{N-1} \nu_{\beta^{N}}^+\right)^{-1}\,,
\end{equation}
and, through \eq{fsnupiun}, can be fully determined in terms of the
one- and two-mode parameters alone. Analogous reasonings and
expressions hold of course for the fully symmetric $M$-mode block
with CM $\sig_{\alpha^M}$ given by \eq{fscm}.

\subsection{Evaluation of block entanglement in terms of symplectic invariants}
\label{SecBlockSymp}

We can now efficiently discuss the quantification of the multimode
block entanglement of bisymmetric Gaussian states. Exploiting our
results on the symplectic characterization of two-mode Gaussian
states \cite{prl,extremal} (see Sec.~\ref{SecSympParam}) we can
select the relevant quantities that, by determining the correlation
properties of the two-mode Gaussian state with CM $\sig_{eq}$, also
determine the entanglement and correlations of the multimode
Gaussian state with CM $\sig$. These quantities are, clearly, the
equivalent marginal purities $\mu_{\alpha eq}$ and $\mu_{\beta eq}$,
the global purity $\mu_{eq}$ and the equivalent two-mode invariant
$\D_{eq}$.

Let us remind that, by exploiting
Eqs.~{\rm(\ref{fsspct},\ref{fsspct2},\ref{parasim})}, the symplectic
spectra of the CMs $\gr{\sigma}_{\alpha^{m}}$ and $\sig_{\beta^{n}}$
may be recovered by means of the local two-mode invariants
$\mu_{\beta}$, $\mu_{\alpha}$, $\mu_{\beta^2}$, $\mu_{\alpha^2}$,
$\D_{\beta^2}$ and $\D_{\alpha^2}$. The quantities $\mu_{\alpha eq}$
and $\mu_{\beta eq}$ are easily determined in terms of local
invariants alone: \be \mu_{\alpha eq}=1/\nu_{\alpha^m}^{+} \; \quad
\mu_{\beta eq}=1/\nu_{\beta^n}^{+} \; . \label{localone} \ee On the
other hand, the determination of $\mu_{eq}$ and $\D_{eq}$ require
the additional knowledge of two global symplectic invariants of the
CM $\sig$; this should be expected, because they are susceptible of
quantifying the correlations between the two parties. The natural
choices for the global invariants are the global purity
$\mu=1/\sqrt{\det{\sig}}$ and the invariant $\D$, given by \bea
%\begin{split}
\D&=&M\det{\alp}+M(M-1)\det{\gr{\varepsilon}}+
N\det{\bet}\nonumber\\
&&+N(N-1)\det{\gr{\zeta}}+2 M N\det{\gr{\gamma}}\nonumber\, .
%\end{split}
\eea One has \bea
\mu_{eq}&=&(\nu_{\alpha}^{-})^{M-1}(\nu_{\beta}^{-})^{N-1}\mu \; ,\label{mueq}\\
\D_{eq}&=& \D-(M-1)(\nu_{\alpha}^{-})^{2}-(N-1)(\nu_{\beta}^{-})^{2}
\; . \eea

The entanglement between the $M$-mode and the $N$-mode subsystems,
quantified by the logarithmic negativity \eq{lognegau} can thus be
easily determined, as it is the case for two-mode states. In
particular, the smallest symplectic eigenvalue
${\tilde{\nu}_{-\,eq}}$ of the matrix $\tilde{\sig}_{eq}$, derived
from $\sig_{eq}$ by partial transposition, fully quantifies the
entanglement between the $M$-mode and $N$-mode partitions. Recalling
the results of Sec.~\ref{SecNega2M},  the quantity
${\tilde{\nu}_{-\,eq}}$ reads \bea 2{\tilde{\nu}_{-\,eq}}^2
&=&\tilde{\D}_{eq}-\sqrt{\tilde{\D}_{eq}^2-
\frac{4}{\mu_{eq}^2}}\, ,\label{numenotildeeqpene}\\
{\rm with} \quad \tilde{\D}_{eq}&=& \frac{2}{\mu_{\alpha
eq}^2}+\frac{2}{\mu_{\beta eq}^2}- \D_{eq} \, . \nonumber \eea The
logarithmic negativity $E_{\N}^{\alpha^M|\beta^N}$ measuring the
bipartite entanglement between the $M$-mode and $N$-mode subsystems
is then \be E_{\N}^{\alpha^M|\beta^N}=
\max\left[-\log{\tilde{\nu}_{-\,eq}},0\right] \; . \ee

In the case $\nu_{\alpha^M}^{+}=\nu_{\beta^N}^{+}$, corresponding to
the following condition \be (a+(M-1)e_1)(a+(M-1)e_2) =
(b+(N-1)z_1)(b+(N-1)z_2) \; , \label{possi} \ee on the standard form
covariances \eq{fscm}, the equivalent two-mode state is symmetric
and we can determine also the entanglement of formation, using
\eq{eofgau}. Let us note that the possibility of exactly determining
the entanglement of formation of a multimode Gaussian state under a
$M\times N$  bipartition is a rather remarkable consequence, even
under the symmetry constraints obeyed by the CM $\sig$. Another
relevant fact to point out is that, since both the logarithmic
negativity and the entanglement of formation are decreasing
functions of the quantity ${\tilde{\nu}_{-\,eq}}$, the two measures
induce the same entanglement hierarchy on such a subset of
``equivalently symmetric'' states ({\em i.e.}~states whose
equivalent two-mode CM $\sig_{eq}$ is symmetric).

From \eq{mueq} it follows that, if the $(M+N)$-mode bisymmetric
state is pure ($\mu=\nu_{\alpha^M}^-=\nu_{\beta^N}^-=1$), then the
equivalent two-mode state is pure as well ($\mu_{eq}=1$) and, up to
local symplectic operations, it is a two-mode squeezed vacuum.
Therefore {\em any pure bisymmetric multimode Gaussian state is
equivalent, under local unitary (symplectic) operations, to  a
tensor product of a single two-mode squeezed pure state and of
$M+N-2$ uncorrelated vacua}. This refines the somehow similar
phase-space Schmidt reduction holding for arbitrary pure bipartite
Gaussian states \cite{botero03,giedkeqic03}, discussed in
Sec.~\ref{SecSchmidtPS}.

More generally, if both the reduced $M$-mode and $N$-mode CMs
$\sig_{\alpha^M}$ and $\sig_{\beta^N}$ of a bisymmetric, mixed
multimode Gaussian state $\sig$ of the form \eq{fulsim} correspond
to Gaussian mixed states of partial minimum uncertainty (see
Sec.~\ref{SecSympHeis}), {\em i.e.}~if
$\nu_{\alpha^M}^-=\nu_{\beta^N}^-=1$, then \eq{mueq} implies
$\mu_{eq}=\mu$. Therefore, the equivalent two-mode state has not
only  the same entanglement, but also the same degree of mixedness
of the original multimode state. In all other cases of bisymmetric
multimode states one has that \be\label{superpurification} \mu_{eq}
\ge \mu \ee and the process of localization thus produces a two-mode
state with higher purity than the original multimode state. In this
specific sense, we see that {\em the process of unitary localization
implies a process of purification as well}.

\subsection{Unitary localization as a reversible multimode/two-mode entanglement
switch}\label{secswitch}

It is important to observe that the unitarily localizable
entanglement (when computable) is always stronger than the
localizable entanglement in the sense of \cite{localiz}. In fact, if
we consider a generic bisymmetric multimode state of a $M \times N$
bipartition, with each of the two target modes owned respectively by
one of the two parties (blocks), then the ensemble of optimal local
measurements on the remaining (``assisting'') $M+N-2$ modes belongs
to the set of  LOCC with respect to the considered bipartition. By
definition the entanglement cannot increase under LOCC, which
implies that the localized entanglement (in the sense of
\cite{localiz}) is always less or equal than the original $M\times
N$ block entanglement. On the contrary, {\em all} of the same
$M\times N$ original bipartite entanglement can be unitarily
localized onto the two target modes.

This is a key point, as such local unitary transformations are {\em
reversible} by definition.
 Therefore,  by only
using passive and active linear optics elements such as
beam-splitters, phase shifters and squeezers, one can in principle
implement a reversible machine ({\em entanglement switch}) that,
from mixed, bisymmetric multimode states with strong quantum
correlations between all the modes (and consequently between the
$M$-mode and the $N$-mode partial blocks) but weak couplewise
entanglement, is able to extract a highly pure, highly entangled
two-mode state (with no entanglement lost, all the $M \times N$
entanglement can be localized). If needed, the same machine would be
able, starting from a two-mode squeezed state and a collection of
uncorrelated thermal or squeezed states, to distribute the two-mode
entanglement between all modes, converting the two-mode into
multimode, multipartite quantum correlations, again with no loss of
entanglement. The bipartite or multipartite entanglement can then be
used on demand, the first for instance in a CV quantum teleportation
protocol \cite{Braunstein98}, the latter \eg to enable teleportation
networks \cite{network} or to perform multimode entanglement
swapping \cite{Multientswap}. We remark, once more, that such an
entanglement switch is endowed with maximum ($100 \%$) efficiency,
as no entanglement is lost in the conversions. This fact may have a
remarkable impact in the context of quantum repeaters \cite{briegel}
for communications with continuous variables.

\subsubsection{The case of the basset hound}

\begin{figure}[t!]
\includegraphics[width=7cm]{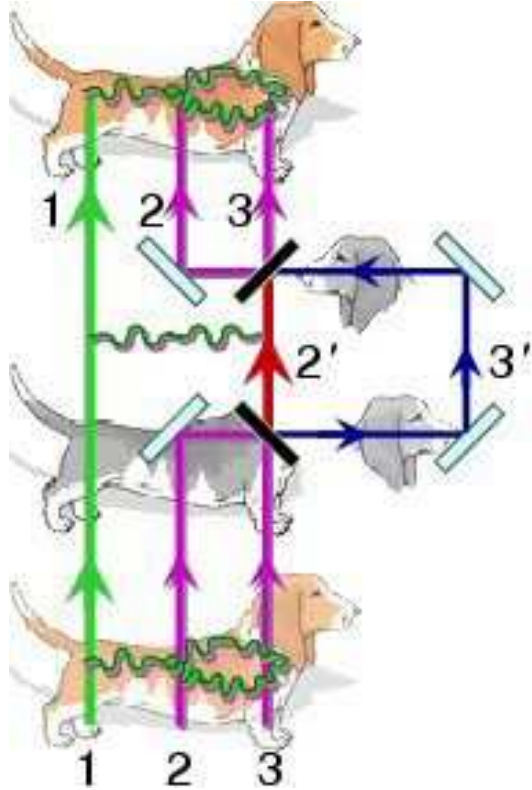}%
\caption{{\sc ``If You Cut The Head Of A Basset Hound, It Will Grow
Again''} (by F. Illuminati, 2001; see also \cite{SeralePHD}, Chapter
1). Graphical depiction of the process of unitary localization
(concentration) and delocalization (distribution) of entanglement in
three-mode bisymmetric Gaussian states \cite{unitarily} (or ``basset
hound'' states), described in the text. Initially, mode $1$ is
entangled (entanglement is depicted as a waving string) with both
modes $2$ and $3$. It exists a local (with respect to the $1|(23)$
bipartition) symplectic operation, realized {\eg}via a beam-splitter
(denoted by a black thick dash), such that all the entanglement is
concentrated between mode $1$ and the transformed mode $2'$, while
the other transformed mode $3'$ decouples from the rest of the
system ({\em unitary localization}). Therefore, the head of the
basset hound (mode $3'$) has been cut off. However, being realized
through a symplectic operation ({\ie}unitary on the density matrix),
the process is reversible: operating on modes $2'$ and $3'$ with the
inverse symplectic transformation, yields the original modes $2$ and
$3$ entangled again with mode $1$, without any loss of quantum
correlations ({\em unitary delocalization}): the head of the basset
hound is back again.} \label{figbasset}
\end{figure}

To give an example, we can consider a bisymmetric $1\times 2$
three-mode Gaussian state,\footnote{\sf The bipartite and genuinely
tripartite entanglement structure of three-mode Gaussian states will
be extensively investigated in Chapter \ref{Chap3M}, based on
Ref.~\cite{3mpra}. The bisymmetric three-mode Gaussian states will
be also reconsidered as efficient resources for $1 \rightarrow 2$
telecloning of coherent states in Sec.~\ref{sectlc}, based on
Ref.~\cite{3mj}.} where the CM of the last two modes (constituting
subsystem $\s_B$) is assumed in standard form, \eq{stform}.  Because
of the symmetry, the local symplectic transformation responsible for
entanglement concentration in this simple case is the identity on
the first mode (constituting subsystem $\s_A$) and just a 50:50
beam-splitter transformation $B_{2,3}(1/2)$, \eq{bbs}, on the last
two modes \cite{passive} (see also Sec.~\ref{SecPassive}). The
entire procedure of unitary localization and delocalization of
entanglement \cite{unitarily} is depicted in Fig.~\ref{figbasset}.
Interestingly, it may be referred to as ``cut-off and regrowth of
the head of a basset hound'', where in our example the basset hound
pictorially represents a bisymmetric three-mode state. However, the
breed of the dog reflects the fact that the unitary localizability
is a property that extends to all $1 \times N$ \cite{adescaling} and
$M \times N$ \cite{unitarily} bisymmetric Gaussian states (in which
case, the basset hound's body would be longer and longer with
increasing $N$). We can therefore address bisymmetric Gaussian
states as {\em basset hound states}, if desired.

In this canine analogy, let us take the freedom to remark that fully
symmetric states of the form \eq{fscm}, as a special case, are of
course bisymmetric under any bipartition of the modes;  this, in
brief, means that any conceivable multimode, bipartite  entanglement
is locally equivalent to the minimal two-mode, bipartite
entanglement (consequences of this will be deeply investigated in
the following). Pictorially, remaining in the context of three-mode
Gaussian states, this special type of basset hound state resembles a
{\em Cerberus} state, in which any one of the three heads can be cut
and can be reversibly regrown.

\section{Quantification and scaling of entanglement in fully symmetric
states} \label{SecScal}

In this Section we will explicitly compute the block entanglement
({\em i.e.~}the entanglement between different blocks of modes) for
some instances of multimode Gaussian states. We will study its
scaling behavior as a function of the number of modes and explore in
deeper detail the localizability of the multimode entanglement. We
focus our attention on fully symmetric $L$-mode Gaussian states (the
number of modes is denoted by $L$ in general to avoid confusion),
endowed with complete permutation invariance under mode exchange,
and described by a $2L \times 2L$ CM $\sig_{\beta^{L}}$ given by
\eq{fscm}. These states are trivially bisymmetric under any
bipartition of the modes, so that their block entanglement is always
localizable by means of local symplectic operations. Let us recall
that concerning the covariances in normal forms of fully symmetric
states (see Sec.~\ref{SecSymm}), {\em pure} $L$-mode states are
characterized by $\nu_{\beta}^{-} = \nu_{\beta^L}^{+}=1$ in
\eq{fsspct}, which yields
\begin{equation}\label{fspure}\begin{split}
  z_1 &= \frac{(L - 2) (b^2 -1)  + \sqrt{\left(b^2 -
                  1\right) \left[L \left(\left(b^2 - 1\right) L + 4\right) -
                4\right]}}{2 b (L - 1)}\,, \\
  z_2 &= \frac{(L - 2) (b^2 -1)
 - \sqrt{\left(b^2 -
                  1\right) \left[L \left(\left(b^2 - 1\right) L + 4\right) -
                4\right]} }{2 b (L - 1)}\,.\end{split}
\end{equation}
Pure fully symmetric Gaussian states are generated as the outputs of
the application of a sequence of $L-1$ beam-splitters to $L$
single-mode squeezed inputs \cite{network,vloock03}.
%In the limit
%of infinite squeezing, these states reduce to the simultaneous
%eigenstates of total momentum and all relative positions $\hat x_i -
%\hat x_j$ of a $L$-mode radiation field,  which define the proper
%GHZ states of CV systems \cite{network}.
The CM $\sig^{p}_{\beta^{L}}$ of this class of pure states, for a
given number of modes, depends only on the parameter $b\equiv
1/\mu_\beta \ge 1$, which is an increasing function of the
single-mode squeezing needed to prepare the state. Correlations
between the modes are induced according to the expression
\pref{fspure} for the covariances $z_{i}$. We will study their
multipartite entanglement sharing in Chapter \ref{ChapMonoGauss},
and their usefulness for teleportation networks in
Sec.\ref{SecTelepoppy}.

In general, exploiting our previous analysis, we can compute the
entanglement between a block of $K$ modes and the remaining $L-K$
modes, both for pure states (in this case the block entanglement is
simply equivalent to the Von Neumann entropy of each of the reduced
blocks) and, remarkably, also for mixed fully symmetric states under
any bipartition of the modes.

\subsection{1\,$\times$\,{\em N} entanglement} \label{Sec1N}

Based on Ref.~\cite{adescaling}, we begin by assigning a single mode
to subsystem $\s_A$, and an arbitrary number $N$ of modes to
subsystem $\s_B$, forming a CV system globally prepared in a fully
symmetric $(1+N)$-mode Gaussian state of  modes.

\subsubsection{Block entanglement hierarchy and signatures of genuine multipartite entanglement}

\begin{figure}[t!]
\includegraphics[width=11cm]{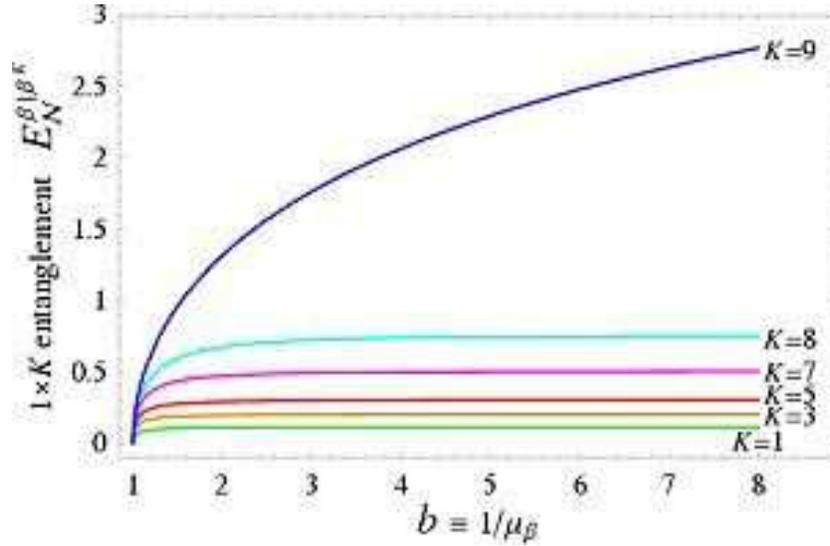}
\caption{Entanglement hierarchy for $(1+N)$-mode fully symmetric
pure Gaussian states ($N=9$).} \label{fighz}
\end{figure}

We consider pure fully symmetric states with CM
$\sig^p_{\beta^{1+N}}$, obtained by inserting \eq{fspure} into
\eq{fscm} with $L \equiv (1+N)$. Exploiting our previous analysis,
we can compute the entanglement between a single mode with reduced
CM $\sig^\beta$ and any $K$-mode partition of the remaining modes
($1 \le K \le N$), by determining the equivalent two-mode CM
$\sig_{eq}^{\beta\vert\beta^K}$.
 We remark that, for every $K$,
the $1 \times K$ entanglement is always equivalent to a $1 \times 1$
entanglement, so that the quantum correlations between the different
partitions of $\sig$ can be directly compared to each other: it is
thus possible to establish a multimode entanglement hierarchy
without any problem of ordering.

The $1\times K$ entanglement quantified by the logarithmic
negativity $E_\N^{\beta\vert\beta^K}$ is determined by the smallest
symplectic eigenvalue $\tilde{\nu}_{- eq}^{(K,N)}$ of the partially
transposed CM $\tilde{\sig}_{eq}^{\beta\vert\beta^K}$. For any
nonzero squeezing ({\em i.e.} $b>1$) one has that $\tilde{\nu}_{-
eq}^{(K,N)}<1$, meaning that the state exhibits genuine multipartite
entanglement: each mode is entangled with any other $K$-mode block,
as first remarked in Ref.~\cite{vloock03}. Further, the genuine
multipartite nature of the entanglement can be precisely quantified
by observing that $$E_\N^{\beta\vert\beta^K} \ge
E_\N^{\beta\vert\beta^{K-1}}\,,$$ as shown in Fig.~\ref{fighz}.

The $1\times 1$ entanglement between two modes is weaker than the
$1\times 2$ one between a mode and other two modes, which is in turn
weaker than the $1\times K$ one, and so on with increasing $K$ in
this typical cascade structure. From an operational point of view, a
signature of {\em genuine multipartite entanglement} is revealed by
the fact that performing {\em e.g.}~a local measurement on a single
mode will affect {\em all} the other $N$ modes. This means that the
quantum correlations contained in the state with CM
$\sig^{p}_{\beta^{1+N}}$ can be fully recovered only when
considering the $1 \times N$ partition.

In particular, the pure-state $1 \times N$ logarithmic negativity
is, as expected, independent of $N$, being a simple monotonic
function of the entropy of entanglement $E_V$, \eq{E:E} (defined as
the Von Neumann entropy of the reduced single-mode state with CM
$\sig_\beta$). It is worth noting that, in the limit of infinite
squeezing ($b \rightarrow \infty$), only the $1\times N$
entanglement diverges while all the other $1\times K$ quantum
correlations remain finite (see Fig.~\ref{fighz}). Namely, \be
\label{logneg1k}
E_\N^{\beta\vert\beta^K}\!\!\big(\!\sig^{p}_{\beta^{1+N}}\!\big)\overset{b\rightarrow\infty}{\longrightarrow}
-\frac12 \log\left[\frac{1- 4K}{N(K+1)-K(K-3)}\right]\,, \ee
 which cannot exceed
$\log\sqrt5 \simeq 0.8$ for any $N$ and for any $K<N$.

\subsubsection{Entanglement scaling with the number of modes}

\begin{figure}[t!]
 \includegraphics[width=10cm]{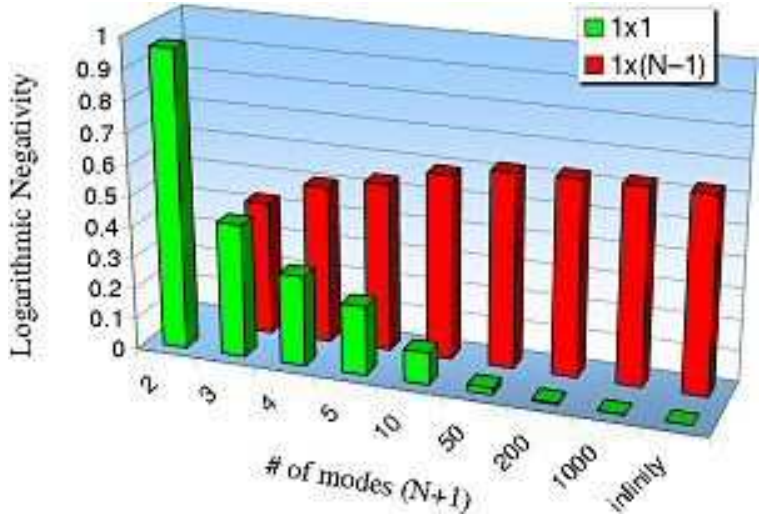}
 \caption{Scaling as a function of $N$ of the $1 \times 1$ entanglement (green bars)
and of the $1 \times (N-1)$ entanglement (red bars)
  for a $(1+N)$-mode pure fully symmetric Gaussian
state, at fixed squeezing ($b=1.5$).}
 \label{figscal}
\end{figure}

At fixed squeezing (\ie fixed local properties, $b \equiv
1/\mu_\beta$), the {\em scaling} with $N$ of the $1 \times (N-1)$
entanglement compared to the $1\times 1$ entanglement is shown in
Fig.~\ref{figscal} (we recall that the $1\times N$ entanglement is
independent on $N$). Notice how, with increasing number of modes,
the multimode entanglement increases to the detriment of the
two-mode one. The latter is indeed being {\em distributed} among all
the modes: this feature will be properly quantified within the
framework of CV entanglement sharing in Chapter \ref{ChapMonoGauss}
\cite{contangle}.

We remark that such a scaling feature occurs in any Gaussian state,
either fully symmetric or bisymmetric (think, for instance, to a
single-mode squeezed state coupled with a $N$-mode symmetric thermal
squeezed state), pure or mixed.  The simplest example of a mixed
state in which our analysis reveals the presence of genuine
multipartite entanglement is obtained from $\sig^{p}_{\beta^{1+N}}$
by tracing out some of the modes. Fig.~\ref{figscal} can then also
be seen as a demonstration of the scaling in such a $N$-mode mixed
state, where the $1 \times (N-1)$ entanglement is the strongest one.
Thus, with increasing $N$, the global mixedness can limit but not
destroy the distribution of entanglement in multiparty form among
all the modes.

\subsection{{\em M}\,$\times$\,{\em N} entanglement}

Based on Ref.~\cite{unitarily}, we can now consider a generic
$2N$-mode fully symmetric mixed state with CM
$\sig_{\beta^{2N}}^{p\backslash Q}$, see \eq{fscm}, obtained from a
pure fully symmetric $(2N+Q)$-mode state by tracing out $Q$ modes.

\subsubsection{Block entanglement hierarchy and optimal localizable entanglement}

For any $Q$, for any dimension $K$ of the block ($K \leq N$), and
for any nonzero squeezing ({\em i.e.~}for $b>1$) one has that
$\tilde{\nu}_K<1$, meaning that the state exhibits genuine
multipartite entanglement, generalizing the $1 \times N$ case
described before: each $K$-mode party is entangled with the
remaining $(2N-K)$-mode block. Furthermore, the genuine multipartite
nature of the entanglement can be precisely unveiled by observing
that, again, $E_\N^{\beta^K\vert\beta^{2N-K}}$ is an increasing
function of the integer $K \le N$, as shown in Fig.~\ref{fiscalb}.
Moreover, we note that the multimode entanglement of mixed states
remains finite also in the limit of infinite squeezing, while the
multimode entanglement of pure states diverges with respect to any
bipartition, as shown in Fig.~\ref{fiscalb}.

\begin{figure}[t!]
\includegraphics[width=11cm]{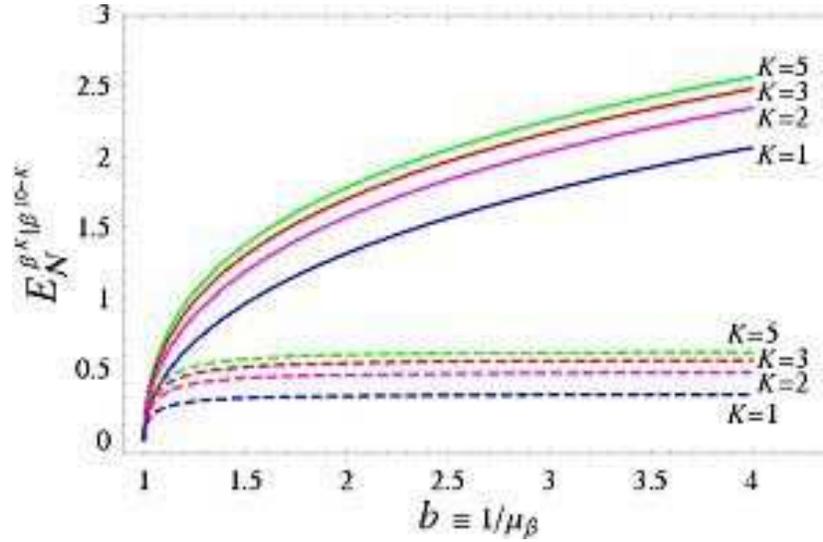}
\caption{Hierarchy of block entanglements of fully symmetric
$2N$-mode Gaussian states of $K \times (2N-K)$ bipartitions
($2N=10$) as a function of the single-mode squeezing $b$. The block
entanglements are depicted both for pure states (solid lines) and
for mixed states obtained from fully symmetric $(2N+4)$-mode pure
Gaussian states by tracing out $4$ modes (dashed lines).}
\label{fiscalb}
\end{figure}

In fully symmetric Gaussian states, the block entanglement is
unitarily localizable with respect to any $K \times (2N-K)$
bipartition. Since in this instance {\em all} the entanglement can
be concentrated on a single pair of modes, after the partition has
been decided, no strategy could grant a better yield than the local
symplectic operations bringing the reduced CMs in Williamson form
(because of the monotonicity of the entanglement under general
LOCC). However, the amount of block entanglement, which is the
amount of concentrated two-mode entanglement after unitary
localization has taken place, actually depends on the choice of a
particular $K \times (2N-K)$ bipartition, giving rise to a hierarchy
of localizable entanglements.

Let us suppose that a given Gaussian multimode state (say, for
simplicity, a fully symmetric state) is available and its
entanglement is meant to serve as a resource for a given protocol.
Let us further suppose that the protocol is optimally implemented if
the entanglement is concentrated between only two modes of the
global systems, as it is the case, {\em e.g.}, in a CV teleportation
protocol between two single-mode parties \cite{Braunstein98}. Which
choice of the bipartition between the modes allows for the best
entanglement concentration by a succession of local unitary
operations? In this framework, it turns out that assigning $K=1$
mode at one party and all the remaining modes to the other, as
discussed in Sec.~\ref{Sec1N}, constitutes the {\em worst}
localization strategy \cite{unitarily}. Conversely, for an even
number of modes the best option for localization is an equal $K=N$
splitting of the $2N$ modes between the two parties. The logarithmic
negativity $E_\N^{\beta^N\vert\beta^{N}}$, concentrated into two
modes by local operations, represents the optimal localizable
entanglement (OLE) of the state $\sig_{\beta^{2N}}$, where
``optimal'' refers to the choice of the bipartition. Clearly, the
OLE of a state with $2N+1$ modes is given by
$E_\N^{\beta^{N+1}\vert\beta^{N}}$. These results may be applied to
arbitrary, pure or mixed, fully symmetric Gaussian states.

\subsubsection{Entanglement scaling with the number of modes}

%%%%%%%%%%%%%%%%%%%%%%%%%%%%%%%%%%%%%%%%%%%%%%%%%%
\begin{figure}[t!]
 \subfigure[] {\includegraphics[width=9.3cm]{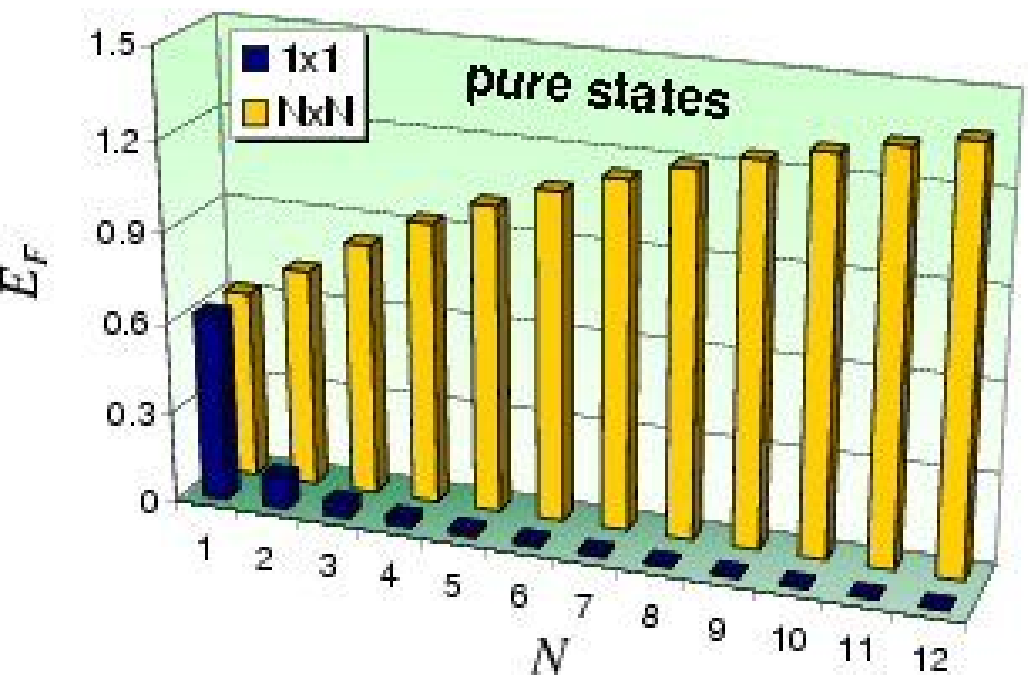}} %\hspace{.2cm}
\subfigure[] {\includegraphics[width=9.3cm]{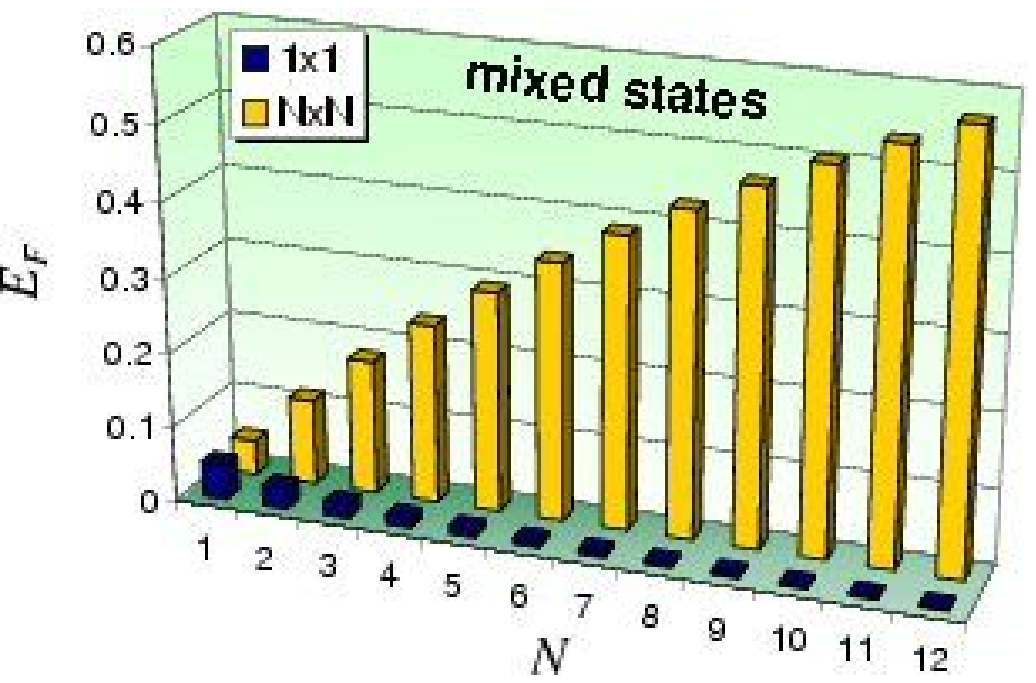}}
\caption{Scaling, with half the number of modes, of the entanglement
of formation in two families of fully symmetric $2N$-mode Gaussian
states. Plot {\rm(a)} depicts pure states, while mixed states
{\rm(b)}  are obtained from $(2N+4)$-mode pure states by tracing out
$4$ modes. For each class of states, two sets of data are plotted,
one referring to $N \times N$ entanglement (yellow bars), and the
other to $1 \times 1$ entanglement (blue bars). Notice how the $N
\times N$ entanglement, equal to the optimal localizable
entanglement (OLE) and estimator of genuine multipartite quantum
correlations among all the $2N$ modes, increases at the detriment of
the bipartite $1\times1$ entanglement between any pair of modes. The
single-mode squeezing parameter is fixed at $b=1.5$.} \label{fisnef}
\end{figure}
%%%%%%%%%%%%%%%%%%%%%%%%%%%%%%%%%%%%%%%%%%%%%%%%%%

We now turn to the study of the scaling behavior with $N$ of the OLE
of $2N$-mode states, to understand how the number of local
cooperating parties can improve the maximal entanglement that can be
shared between two parties. For generic (mixed) fully symmetric
$2N$-mode states of $N \times N$ bipartitions, the OLE can be
quantified also by the entanglement of formation $E_F$, \eq{eofgau},
as the equivalent two-mode state is symmetric, see
Sec.~\ref{SecBlockSymp}. It is then useful to compare, as a function
of $N$, the $1 \times 1$ entanglement of formation between a pair of
modes (all pairs are equivalent due to the global symmetry of the
state) before the localization, and the $N \times N$ entanglement of
formation, which is equal to the optimal entanglement concentrated
in a specific pair of modes after performing the local unitary
operations. The results of this study are shown in
Fig.~\ref{fisnef}. The two quantities are plotted at fixed squeezing
$b$ as a function of $N$ both for a pure $2N$-mode state with CM
$\sig_{\beta^{2N}}^p$ and a mixed $2n$-mode state with CM
$\sig_{\beta^{2N}}^{p\backslash4}$. As the number of modes
increases, any pair of single modes becomes steadily less entangled,
but the total multimode entanglement of the state grows and, as a
consequence, the OLE increases with $N$. In the limit $N \rightarrow
\infty$, the $N \times N$ entanglement diverges while the $1\times1$
one vanishes. This exactly holds both for pure {\em and} mixed
states, although the global degree of mixedness produces the typical
behavior that tends to reduce the total entanglement of the state.

\subsection{Discussion}

 We have shown that bisymmetric (pure or mixed) multimode Gaussian states,
whose structural properties are introduced in Sec.~\ref{SecSymm},
can be reduced by local symplectic operations to the tensor product
of a correlated two-mode Gaussian state and of uncorrelated thermal
states (the latter being obviously irrelevant as far as the
correlation properties of the multimode Gaussian state are
concerned). As a consequence, {\em all} the entanglement of
bisymmetric multimode Gaussian states of arbitrary $M \times N$
bipartitions is {\em unitarily localizable} in a single (arbitrary)
pair of modes shared by the two parties. Such a useful reduction to
two-mode Gaussian states is somehow similar to the one holding for
states with fully degenerate symplectic spectra
\cite{botero03,giedkeqic03}, encompassing the relevant instance of
pure states, for which all the symplectic eigenvalues are equal to
$1$ (see Sec.~\ref{SecSchmidtPS}). The present result allows to
extend the PPT criterion as a necessary and sufficient condition for
separability to all bisymmetric multimode Gaussian states of
arbitrary $M \times N$ bipartitions (as shown in
Sec.~\ref{SecPPTG}), and to quantify their entanglement
\cite{adescaling,unitarily}.

Notice that, in the general bisymmetric instance addressed in this
Chapter, the possibility of performing a two-mode reduction is
crucially partition-dependent. However, as we have explicitly shown,
in the case of fully symmetric states all the possible bipartitions
can be analyzed and compared, yielding remarkable insight into the
structure of the multimode block entanglement of Gaussian states.
This leads finally to the determination of the maximum, or optimal
localizable entanglement  that can be unitarily concentrated on a
single pair of modes.

It is important to notice that the multipartite entanglement in the
considered class of multimode Gaussian states can be produced and
detected \cite{network,vloock03}, and also, by virtue of the present
analysis, reversibly localized by all-optical means. Moreover, the
multipartite entanglement allows for a reliable ({\em i.e.}~with
fidelity ${\CMcal F}>{\CMcal F}_{cl}$, where ${\CMcal F}_{cl}=1/2$
is the classical threshold, see Chapter \ref{ChapCommun}) quantum
teleportation between any two parties with the assistance of the
remaining others \cite{network}. The connection between entanglement
in the symmetric Gaussian resource states and optimal
teleportation-network fidelity has been clarified in
\cite{telepoppate}, and will be discussed in
Sec.~\ref{SecTelepoppy}.

More generally, the present Chapter has the important role of
bridging between the two central parts of this Dissertation, the one
dealing with bipartite entanglement on one hand, and the one dealing
with multipartite entanglement on the other hand. We have
characterized entanglement in multimode Gaussian states by reducing
it to a two-mode problem. By comparing the equivalent two-mode
entanglements in the different bipartitions we have unambiguously
shown that genuine multipartite entanglement is present in the
studied Gaussian states. It is now time to analyze in more detail
the sharing phenomenon responsible for the distribution of
entanglement from a bipartite, two-mode form, to a genuine
multipartite manifestation in $N$-mode Gaussian states, under and
beyond symmetry constraints.

}

\part{Multipartite entanglement of Gaussian states}{\vspace*{1cm}
\includegraphics[width=10cm]{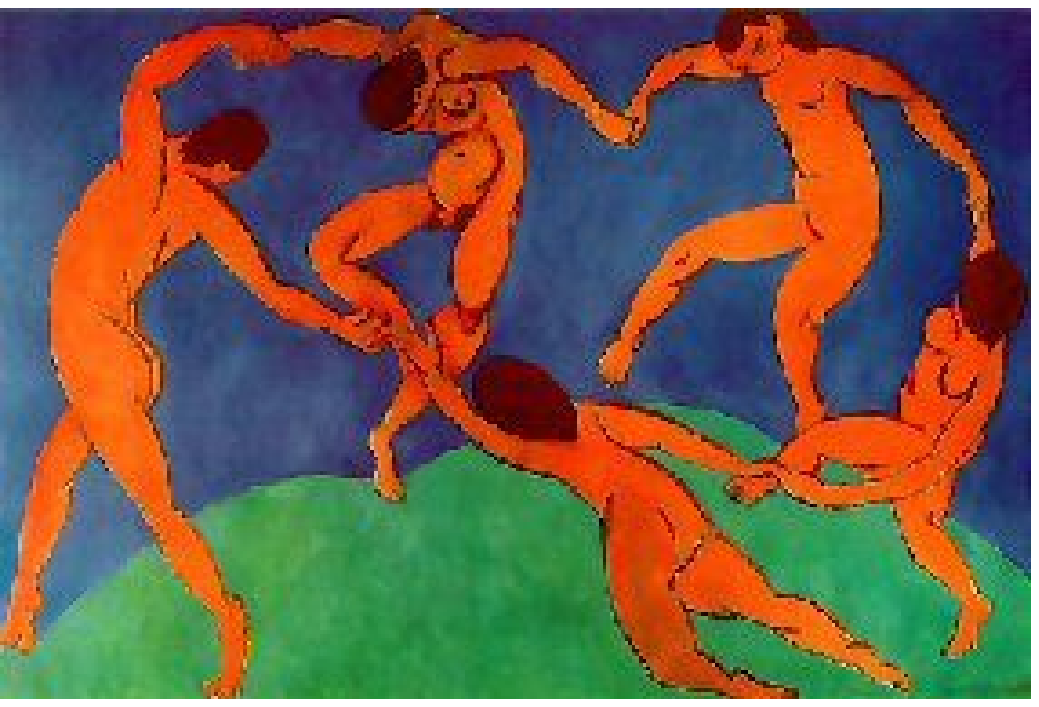} \\
\vspace*{0.6cm} {\normalsize \rm {\em Dance (II).} Henri Matisse,
1910.
\\ \vspace*{-0.4cm} \textrm{\footnotesize The Hermitage Museum,
St.~Petersburg}}}\label{PartMulti}
\chapter{Gaussian entanglement sharing}\label{ChapMonoGauss}

{\sf

One of the main challenges in fundamental quantum theory, as well as
in quantum information and computation sciences, lies in the
characterization and quantification of bipartite entanglement for
mixed states, and in the definition and interpretation of
multipartite entanglement both for pure states and in the presence
of mixedness \cite{chuaniels,heiss}. More intriguingly, a
quantitative, physically significant, characterization of the
entanglement of states shared by many parties can be attempted: this
approach, introduced in a seminal paper by Coffman, Kundu and
Wootters (CKW) \cite{CKW}, has lead to the discovery of so-called
``monogamy inequalities'' [see \eq{ckw3}], constraining the maximal
entanglement distributed among different internal partitions of a
multiparty system. Such inequalities are uprising as one of the
fundamental guidelines on which proper multipartite entanglement
measures have to be built \cite{pisa}.

While important progresses have been gained on these issues in the
context of qubit systems (as reviewed in Sec.~\ref{SecPisa}), a less
satisfactory understanding had been achieved until recent times on
higher-dimensional systems, associated to Hilbert spaces with an
increasingly complex structure. However, and quite remarkably, in
infinite-dimensional Hilbert spaces of CV systems, important
progresses have been obtained in the understanding of the
(bipartite) entanglement properties of the fundamental class of
Gaussian states, as it clearly emerges, we hope, from the previous
Parts of this Dissertation.

Building on these insights, we have performed {\em the first}
analysis of multipartite entanglement sharing in a CV scenario. This
has resulted, in particular, in the first (and unique to date)
mathematically and physically {\em bona fide} measure of genuine
tripartite entanglement for arbitrary three-mode Gaussian states
\cite{contangle,3mpra}, in a proof of the monogamy inequality on
distributed entanglement for all Gaussian states \cite{hiroshima},
and in the demonstration of the {\em promiscuous} sharing structure
of multipartite entanglement in Gaussian states \cite{contangle},
which arises in three-mode symmetric states \cite{3mpra,3mj} and can
be unlimited in states of more than three modes \cite{unlim}.

These and related results are the subject of the present Part of
this Dissertation.

\smallskip

We begin in this Chapter by introducing our novel entanglement
monotones ({\em contangle}, {\em Gaussian contangle} and {\em
Gaussian tangle}) apt to quantify distributed Gaussian entanglement,
thus generalizing to the CV setting the {\em tangle} \cite{CKW}
defined for systems of two qubits by \eq{tangle}.

Motivated by the analysis of the block entanglement hierarchy and
its scaling structure in fully symmetric Gaussian states (see
Sec.~\ref{SecScal}) we will proceed by establishing a monogamy
constraint on the entanglement distribution in such states. We will
then lift the symmetry requirements and prove that CV entanglement,
once properly quantified, is monogamous for {\em all} Gaussian
states \cite{hiroshima}. This is arguably the most relevant result
of this Chapter, and one of the milestones of this Dissertation.

The paradigmatic instance of tripartite CV entanglement, embodied by
three-mode Gaussian states, will be treated independently and in
full detail in the next Chapter. In that case, let us anticipate
that from the monogamy inequality a measure of genuine tripartite
entanglement emerges naturally ({\em residual Gaussian contangle}),
and we will prove it to be a full entanglement monotone under
Gaussian LOCC. Equipped with such a powerful tool to quantify
tripartite entanglement, we will proceed to investigate the
entanglement sharing structure in three-mode Gaussian states,
unveiling the original feature named {\em promiscuity}: it
essentially means that bipartite and multipartite entanglement can
enhance each other in Gaussian states and be simultaneously
maximized without violating the monogamy inequality on entanglement
sharing. In Chapter \ref{ChapUnlim}, the promiscuous sharing
structure of distributed CV entanglement will be shown to arise to
an unlimited extent in Gaussian states of at least four modes.

\section{Distributed entanglement in multipartite continuous variable systems}\label{seccontangle}

Our primary aim, as in Ref.~\cite{contangle}, is to analyze the
distribution of entanglement between different (partitions of) modes
in Gaussian states of CV systems. The reader is referred to
Sec.~\ref{SecPisa} for a detailed, introductory discussion on the
subject of entanglement sharing.

\subsection{The need for a new continuous-variable entanglement monotone}
 In Ref.~\cite{CKW} Coffman, Kundu
and Wootters (CKW) proved for system of three qubits, and
conjectured for $N$ qubits (this conjecture has now been proven by
Osborne and Verstraete \cite{osborne}), that the bipartite
entanglement $E$ (properly quantified) between, say, qubit A and the
remaining two-qubits partition (BC) is never smaller than the sum of
the A$|$B and A$|$C bipartite entanglements in the reduced states:
\begin{equation}
\label{CKWater} E^{A|(BC)} \ge E^{A|B} + E^{A|C} \; .
\end{equation}
This statement quantifies the so-called {\em monogamy} of quantum
entanglement \cite{monogamy}, in opposition to the classical
correlations, which are not constrained and can be freely shared.

One would expect a similar inequality to hold for three-mode
Gaussian states, namely
\begin{equation}\label{CKWine}
E^{i|(jk)}- E^{i|j} - E^{i|k} \ge 0 \; ,
\end{equation}
where $E$ is a proper measure of bipartite CV entanglement and the
indexes $\{i,j,k\}$ label the three modes. However, the
demonstration of such a property is plagued by subtle difficulties.

Let us for instance consider the simplest conceivable instance of a
pure three-mode Gaussian state completely invariant under mode
permutations. These pure Gaussian states are named fully symmetric
(see Sec.~\ref{SecSymm}), and their standard form CM [obtained by
inserting \eq{fspure} with $L=3$ into \eq{fscm}] is only
parametrized by the local mixedness $b=(1/\mu_{\beta}) \ge 1$, an
increasing function of the single-mode squeezing $r_{loc}$, with $b
\rightarrow 1^{+}$ when $r_{loc} \rightarrow 0^{+}$. For these
states, it is not difficult to show that the inequality
\pref{CKWine} can be violated for small values of the local
squeezing factor, using either the logarithmic negativity $E_\N$ or
the entanglement of formation $E_F$ (which is computable in this
case via \eq{eofgau}, because the two-mode reduced mixed states of a
pure symmetric three-mode Gaussian states are again symmetric) to
quantify the bipartite entanglement. This fact implies that none of
these two measures is the proper candidate for approaching the task
of quantifying entanglement sharing in CV systems. This situation is
reminiscent of the case of qubit systems, for which the CKW
inequality holds using the tangle $\tau$, defined in \eq{tangle} as
the square of the concurrence \cite{Wootters97}, but can fail if one
chooses equivalent measures of bipartite entanglement such as the
concurrence itself or the entanglement of formation \cite{CKW}.

It is then necessary to define a proper measure of CV entanglement
that specifically quantifies entanglement sharing according to a
monogamy inequality of the form \pref{CKWine}. A first important
hint toward this goal comes by observing that, when dealing with
$1\times N$ partitions of fully symmetric multimode pure Gaussian
states together with their $1 \times 1$ reduced partitions, the
desired measure should be a monotonically decreasing function $f$ of
the smallest symplectic eigenvalue $\tilde \nu_-$ of the
corresponding partially transposed CM $\tilde{\sig}$. This
requirement stems from the fact that $\tilde \nu_-$ is the only
eigenvalue that can be smaller than $1$, as shown in
Sec.~\ref{SecPPTG} and Sec.~\ref{SecBlockSymp}, violating the PPT
criterion with respect to the selected bipartition. Moreover, for a
pure symmetric three-mode Gaussian state, it is necessary to require
that the bipartite entanglements $E^{i|(jk)}$ and $E^{i|j}=E^{i|k}$
be respectively functions $f(\tilde \nu_-^{i|(jk)})$ and $f(\tilde
\nu_-^{i|j})$ of the associated smallest symplectic eigenvalues
$\tilde \nu_-^{i|(jk)}$ and $\tilde \nu_-^{i|j}$, in such a way that
they become infinitesimal of the same order in the limit of
vanishing local squeezing, together with their first derivatives:
\begin{equation}
\frac{f(\tilde \nu_-^{i|(jk)})}{2f(\tilde \nu_-^{i|j})} \cong
\frac{f'(\tilde \nu_-^{i|(jk)})}{2f'(\tilde \nu_-^{i|j})}
\rightarrow 1 \; \; \; \; \; \mbox{for} \; \; \; b \rightarrow 1^{+}
\; , \label{requir}
\end{equation}
where the prime denotes differentiation with respect to the
single-mode mixedness $b$. The violation of the sharing inequality
\pref{CKWine} exhibited by the logarithmic negativity can be in fact
traced back to the divergence of its first derivative in the limit
of vanishing squeezing. The above condition formalizes the physical
requirement that in a {\em symmetric} state the quantum correlations
should appear smoothly and be distributed uniformly among all the
three modes. One can then see that the unknown function $f$
exhibiting the desired property is simply the squared logarithmic
negativity\footnote{\sf Notice that an infinite number of functions
satisfying \eq{requir} can be obtained by expanding $f(\tilde
\nu_-)$ around $\tilde \nu_- = 1$ at any even order. However, they
are all monotonic convex functions of $f$. If the inequality
\pref{CKWine} holds for $f$, it will hold as well for any
monotonically increasing, convex function of $f$, such as the
logarithmic negativity raised to any even power $k \ge 2$, but not
for $k=1$. We will exploit this ``gauge freedom'' in the following,
to define an equivalent entanglement monotone in terms of squared
negativity \cite{hiroshima}.}
\begin{equation}\label{funcsq}
f(\tilde \nu_-)=[-\log \tilde \nu_-]^2\,.
\end{equation}
We remind again that for (fully symmetric) $(1+N)$-mode pure
Gaussian states, the partially transposed CM with respect to any $1
\times N$ bipartition, or with respect to any reduced $1 \times 1$
bipartition, has only one symplectic eigenvalue that can drop below
$1$; hence the simple form of the logarithmic negativity (and,
equivalently, of its square) in \eq{funcsq}.

\subsection{Squared negativities as continuous-variable tangles}

 Equipped with this finding,
one can give a formal definition of a bipartite entanglement
monotone $E_\tau$ that, as we will soon show, can be regarded as a
CV analogue of the tangle. Note that the context here is completely
general and we are not assuming that we are dealing with Gaussian
states only. For a generic pure state $\ket{\psi}$ of a $(1+N)$-mode
CV system, we define our measure as the square of the logarithmic
negativity [the latter defined by \eq{E:EN}]:
\begin{equation}
\label{etaupure} E_\tau (\psi) \equiv \log^2 \| \tilde \ro \|_1 \; ,
\quad \ro = \ketbra\psi\psi \; .
\end{equation}
This is a proper measure of bipartite entanglement, being a convex,
increasing function of the logarithmic negativity $E_\N$, which is
equivalent to the entropy of entanglement \eq{E:E} for arbitrary
pure states in any dimension.  Def.~\pref{etaupure} is naturally
extended to generic mixed states $\rho$ of $(N+1)$-mode CV systems
through the convex-roof formalism. Namely, we can introduce the
quantity
\begin{equation}\label{etaumix}
E_\tau(\rho) \equiv \inf_{\{p_i,\psi_i\}} \sum_i p_i
E_\tau(\psi_i)\; ,
\end{equation}
where the infimum is taken over all convex decompositions of $\rho$
in terms of pure states $\{\ket{\psi_i}\}$; if the index $i$ is
continuous, the sum in \eq{etaumix} is replaced by an integral, and
the probabilities $\{p_i\}$ by a probability distribution
$\pi(\psi)$. Let us now recall that, for two qubits, the tangle can
be defined as the convex roof of the squared negativity
\cite{crnega} (the latter being equal to the concurrence
\cite{Wootters97} for pure two-qubit states \cite{VerstraeteJPA}, as
mentioned in Sec.~\ref{SecEnt2Q}). Here, \eq{etaumix} states that
the convex roof of the squared logarithmic negativity properly
defines the continuous-variable tangle, or, in short, the {\em
contangle} $E_\tau(\rho)$, in which the logarithm takes into account
for the infinite dimensionality of the underlying Hilbert space.

On the other hand, by recalling the equivalence of negativity and
concurrence for pure states of qubits, the {\em tangle} itself can
be defined for CV systems as the convex-roof extension of the
squared negativity. Let us recall that the negativity $\N$, \eq{E:N}
of a quantum state $\ro$ is a convex function of the logarithmic
negativity $E_\N$, \eq{E:EN}, being
\begin{equation}\label{NvsEN}
\N(\ro) = \frac{\exp[E_\N(\ro)-1]}{2}\,.
\end{equation}

These definitions are sensible and applicable to a generic (in
principle non-Gaussian) state of a CV system.

\subsubsection{Gaussian contangle and Gaussian
tangle}\label{SecTau}

From now on, we will restrict our attention to Gaussian states.

\smallskip

\noindent {\rm {\bf Gaussian contangle}.}--- For any pure multimode
Gaussian state $\ket\psi$, with CM $\sig^p$, of $N+1$ modes assigned
in a generic $1 \times N$ bipartition, explicit evaluation gives
immediately that $E_\tau (\psi) \equiv E_\tau (\sig^{p})$ takes the
form
\begin{equation}
\label{piupurezzapertutti} E_\tau (\sig^{p}) = \log^2 \left(1/\mu_1
- \sqrt{1/\mu_1^2-1}\right) \; ,
\end{equation}
where $\mu_1 = 1/\sqrt{\det\sig_1}$ is the local purity of the
reduced state of mode $1$ with CM $\sig_1$.

For any multimode, mixed Gaussian states with CM $\sig$, we will
then denote the contangle by $E_\tau(\sig)$, in analogy with the
notation used for the contangle $E_\tau(\sig^{p})$ of pure Gaussian
states in \eq{piupurezzapertutti}. Any multimode mixed Gaussian
state with CM $\sig$, admits at least one decomposition in terms of
pure Gaussian states $\sig^p$ only. The infimum of the average
contangle, taken over all pure Gaussian state decompositions,
defines then the {\em Gaussian contangle} $G_\tau$,
\begin{equation}
G_\tau(\sig) \equiv \inf_{\{\pi(d\sig^p ), \sig^{p} \}} \int \pi
(d\sig^p) E_\tau (\sig^p) \; . \label{GaCoRo}
\end{equation}
It follows from the convex roof construction that the Gaussian
contangle $G_\tau(\sig)$ is an upper bound to the true contangle
$E_\tau(\sig)$ (as the latter can be in principle minimized over a
non-Gaussian decomposition),
\begin{equation}
E_\tau(\sig) \leq G_\tau(\sig) \; , \label{UpperCut}
\end{equation}
It can be shown that $G_\tau(\sig)$ is a bipartite entanglement
monotone under Gaussian LOCC: in fact, the Gaussian contangle
belongs to the general family of Gaussian entanglement measures,
whose properties as studied in Ref.~\cite{ordering} have been
presented in Sec.~\ref{SecGEMS}. Therefore, for Gaussian states, the
Gaussian contangle, similarly to the Gaussian entanglement of
formation \cite{GEOF}, takes the simple form
\begin{equation}
G_\tau (\sig) = \inf_{\sig^p \le \sig} E_\tau(\sig^p) \; ,
\label{simple}
\end{equation}
where the infimum runs over all pure Gaussian states with CM $\sig^p
\le \sig$. Let us remark that, if $\sig$ denotes a mixed symmetric
two-mode Gaussian state, then the Gaussian decomposition is the
optimal one \cite{giedke03} (see Sec.~\ref{SecEOFGauss}), and the
optimal pure-state CM $\sig^p$ minimizing $G_\tau(\sig)$ is
characterized by having $\tilde \nu_-({\sig}^p) = \tilde
\nu_-({\sig})$ \cite{GEOF} (see Sec.~\ref{SecGEMvsNEGcomparison}).
The fact that the smallest symplectic eigenvalue is the same for
both partially transposed CMs entails that for two-mode symmetric
(mixed) Gaussian states
\begin{equation} \label{etausym2} E_\tau(\sig) = G_\tau(\sig) =
[\max\{0,-\log \tilde \nu_-(\sig)\}]^2\,.\end{equation} We thus
consistently retrieve for the Gaussian contangle (or, equivalently,
the contangle, as they coincide in this specific case),  the
expression previously found for the mixed symmetric reductions of
fully symmetric three-mode pure states, \eq{funcsq}.

To our aims, it is useful here to provide a compact, operative
definition of the Gaussian contangle for $1 \times N$ bipartite
Gaussian states, based on the evaluation of Gaussian entanglement
measures in Sec.~\ref{SecGEMS}. If $\sig_{i\vert j} $ is the CM of a
(generally mixed) bipartite Gaussian state where subsystem $\s_i$
comprises one mode only, then the Gaussian contangle ${G_\tau} $ can
be computed as
\begin{equation}
\label{tau} {G_\tau} (\sig_{i\vert j} )\equiv {G_\tau} (\sig_{i\vert
j}^{opt} )=g[m_{i\vert j}^2 ],\;\;\;g[x]={\rm arcsinh}^2[\sqrt
{x-1}].
\end{equation}
Here $\sig_{i\vert j}^{opt} $ corresponds to a pure Gaussian state,
and $m_{i\vert j} \equiv m(\sig _{i\vert j}^{opt} )=\sqrt {\det
\sig_i^{opt} } =\sqrt {\det \sig_j^{opt}}$, with $\sig_{i(j)}^{opt}$
being the reduced CM of subsystem $\s_i$ $(\s_j)$ obtained by
tracing over the degrees of freedom of subsystem $\s_j$ ($\s_i$).
The CM $\sig_{i\vert j}^{opt} $ denotes the pure bipartite Gaussian
state which minimizes $m(\sig_{i\vert j}^p )$ among all pure-state
CMs $\sig_{i\vert j}^p $ such that $\sig_{i\vert j}^p \le
\sig_{i\vert j}$. If $\sig_{i\vert j}$ is already a pure state, then
$\sig_{i\vert j}^{opt} \equiv \sig_{i\vert j}$, while for a mixed
Gaussian state \eq{tau} is mathematically equivalent to constructing
the Gaussian convex roof. For a separable state $m(\sig_{i\vert
j}^{opt})=1$. Here we have implicitly used the fact that the
smallest symplectic eigenvalue $\tilde\nu_-$ of the partial
transpose of a pure $1 \times N$ Gaussian state
 $\sig_{i\vert j}^p$ is equal to $\tilde\nu_-=\sqrt{\det\sig_i}
-\sqrt{\det\sig_i-1}$, as follows by recalling that the $1 \times N$
entanglement is equivalent to a $1 \times 1$ entanglement by virtue
of the phase-space Schmidt decomposition (see
Sec.~\ref{SecSchmidtPS}) and by exploiting \eq{n1} with $\Delta=2$,
$\mu=1$ and $\mu_1=\mu_2 \equiv 1/\sqrt{\det\sig_i}$.

The Gaussian contangle ${G_\tau}$, like all members of the Gaussian
entanglement measures family (see Sec.~\ref{SecGEMS}) is completely
equivalent to the Gaussian entanglement of formation \cite{GEOF},
which quantifies the cost of creating a given mixed Gaussian state
out of an ensemble of pure, entangled Gaussian states.

\smallskip

\noindent {\rm {\bf Gaussian tangle}.}--- Analogously, for a $1
\times N$ bipartition associated to a pure Gaussian state
$\ro_{A|B}^{p} = \ps_{A|B}\bra{\psi}$ with $\s_A=\s_{1}$ (a
subsystem of a single mode) and $\s_B=\s_{2}\ldots \s_{N}$, we
define the following quantity
\begin{equation} \label{eq:G-tangle_pure}
\tau _{G}(\ro _{A|B}^{p})=\CMcal{N}^{2}(\ro _{A|B}^{p}).
\end{equation}
Here, $\CMcal{N}(\ro)$ is the negativity, \eq{E:N}, of the Gaussian
state $\ro$. The functional $\tau _{G}$, like the negativity
$\CMcal{N}$, vanishes on separable states and does not increase
under LOCC, \ie, it is a proper measure of pure-state bipartite
entanglement. It can be naturally extended to mixed Gaussian states
$\ro_{A|B}$ via the convex roof construction
\begin{equation} \label{eq:G-tangle_mixed}
\tau _{G}(\ro _{A|B})= \inf_{\{p_{i},\ro
_{i}^{(p)}\}}\sum_{i}p_{i}\tau _{G}(\ro _{i}^{p}),
\end{equation}
where the infimum is taken over all convex decompositions of
$\ro_{A|B}$ in terms of pure {\em Gaussian} states $\ro _{i}^{p}$:
$\ro _{A|B}=\sum_{i}p_{i}\ro_{i}^{p}$. By virtue of the Gaussian
convex roof construction,  the Gaussian entanglement measure
$\tau_{G}$ \eq{eq:G-tangle_mixed} is an entanglement monotone under
Gaussian LOCC  (see Sec.~\ref{SecGEMS}). Henceforth, given an
arbitrary $N$-mode Gaussian state $\ro _{\s_{1}|\s_{2}\ldots
\s_{N}}$, we refer to $\tau _{G}$, \eq{eq:G-tangle_mixed},  as the
{\em Gaussian tangle} \cite{hiroshima}. Obviously, in terms of CMs,
the analogous of the definition \pref{simple} is valid for the
Gaussian tangle as well, yielding it computable like the contangle
in \eq{tau}. Namely, exploiting \eq{negagau}, one finds
\begin{equation}
\label{Gtau} {\tau_G} (\sig_{i\vert j} )\equiv {\tau_G}
(\sig_{i\vert j}^{opt} )=w[m_{i\vert j}^2 ],\;\;\;w[x]=\frac{1}{4}
\left(\sqrt{x - 1} + \sqrt{x} - 1\right)^2.
\end{equation}
Refer to the discussion immediately after \eq{tau} for the
definition of the quantities involved in \eq{Gtau}.

\medskip

We will now proceed to investigate the entanglement sharing of
Gaussian states and to establish monogamy constraints on its
distribution. We remark that, being the (squared) negativity a
monotonic and convex function of the (squared) logarithmic
negativity, see \eq{NvsEN}, the validity of any monogamy constraint
on distributed Gaussian entanglement using as an entanglement
measure the ``Gaussian tangle'', is {\em implied} by the proof of
the corresponding monogamy inequality obtained using the ``Gaussian
contangle''. For this reasons, when possible, we will always employ
as a preferred choice the primitive entanglement monotone,
represented by the (Gaussian) contangle \cite{contangle} (which
could be generally referred to as a `logarithmic' tangle in quantum
systems of arbitrary dimension).

\section{Monogamy of distributed entanglement in {\em N}-mode Gaussian states}

We are now in the position to prove a collection of results
concerning the monogamy of distributed Gaussian entanglement in
multimode Gaussian states.

\subsection{General monogamy constraints and residual entanglement}
In the broadest setting we want to investigate whether a monogamy
inequality like \ineq{CKWine} holds in the general case of Gaussian
states with an arbitrary number $N$ of modes. Considering a Gaussian
state distributed among $N$ parties (each owning a single mode), the
monogamy constraint on distributed entanglement can be written as
\begin{equation}
\label{ckwine} E^{\s_i \vert (\s_1 \ldots \s_{i-1} \s_{i+1} \ldots
\s_N )} \ge \sum\limits_{j\ne i}^N {E^{\s_i \vert \s_j } }
\end{equation}
where the global system is multipartitioned in subsystems $\s_k$
($k=1,{\ldots},N$), each owned by a respective party, and $E$ is a
proper measure of bipartite entanglement. The corresponding general
monogamy inequality, see \eq{ckwn}, is known to hold for qubit
systems \cite{osborne}.

The left-hand side of inequality \pref{ckwine} quantifies the
bipartite entanglement between a probe subsystem $\s_i $ and the
remaining subsystems taken as a whole. The right-hand side
quantifies the total bipartite entanglement between $\s_i$ and each
one of the other subsystems $\s_{j\ne i}$ in the respective reduced
states. The non-negative difference between these two entanglements,
minimized over all choices of the probe subsystem, is referred to as
the \emph{residual multipartite entanglement}. It quantifies the
purely quantum correlations that are not encoded in pairwise form,
so it includes all manifestations of genuine $K$-partite
entanglement, involving $K$ subsystems (modes) at a time, with
$2<K\le N$.  The study of entanglement sharing and monogamy
constraints thus offers a natural framework to interpret and
quantify entanglement in multipartite quantum systems \cite{pisa}
(see Sec.~\ref{SecPisa}).

To summarize the results we are going to present, it is now known
that the (Gaussian) contangle --- and the Gaussian tangle, as an
implication --- is monogamous in fully symmetric Gaussian states of
$N$ modes \cite{contangle}. In general, {\em we have proven the
Gaussian tangle to be monogamous in  all, pure or mixed, Gaussian
states of an arbitrary number of modes} \cite{hiroshima}. A full
analytical proof of the monogamy inequality for the contangle in all
Gaussian states beyond the symmetry, is currently lacking; however,
numerical evidence obtained for randomly generated nonsymmetric
$4$-mode Gaussian states strongly supports the conjecture that the
monogamy inequality be true for all multimode Gaussian states, using
also the (Gaussian) contangle as a measure of bipartite entanglement
\cite{contangle}. Remarkably, for all (generally nonsymmetric)
three-mode Gaussian states the (Gaussian) contangle has been proven
to be monogamous, leading in particular to a proper measure of
tripartite entanglement in terms of the residual contangle: the
analysis of distributed entanglement in the special instance of
three-mode Gaussian states, with all the resulting implications, is
postponed to the next Chapter.

\subsection{Monogamy inequality for fully symmetric states}
\label{SecMonoFulSym}

The analysis of Sec.~\ref{SecScal} has revealed that in fully
permutation-invariant Gaussian states, by comparing the bipartite
block entanglement in the various bipartitions of the modes (which
is always unitarily localizable into a two-mode one
\cite{adescaling,unitarily}), the presence of genuine multipartite
entanglement is revealed. In general, with increasing number of
modes, we have evidenced by scaling arguments how the individual
pairwise entanglement between any two modes is redistributed into a
multipartite entanglement among all the modes.

How does this entanglement distribution mechanism work? We show
here, based on a simple computation first appeared in
\cite{contangle}, that it obeys the fundamental monogamy law.

We will employ the Gaussian contangle $G_\tau$, \eq{tau}, as a
measure of bipartite entanglement. Due  to the convex roof extension
involved in the definition of $G_\tau$, \eq{GaCoRo}, it will suffice
to prove monogamy for pure fully symmetric Gaussian states, for
which the Gaussian contangle $G_\tau$ coincides with the true
contangle $E_\tau$ in every bipartition thanks to the symmetry of
the considered states (see Sec.~\ref{SecTau}). Such a proof will
also imply the corresponding monogamy property for the (Gaussian)
tangle, \eq{Gtau}.

For a pure, fully symmetric Gaussian states of $N+1$ modes, we will
then prove the statement
\begin{equation}\label{monoN}
E_\tau^{i|(j_1 , \ldots , j_N)} - \sum_{l=1}^{N} E_\tau^{i|j_l} \geq
0 \; ,
\end{equation}
by induction. Referring to the notation of
Eqs.~{\rm(\ref{fscm},\,\ref{fspure})} with $L \equiv N$, for any $N$
and for $b> 1$ (for $b=1$ one has a product state), the $1 \times N$
contangle
\begin{equation}
E_\tau^{i|(j_1 , \ldots , j_N)} = \log^2(b-\sqrt{b^2-1})
\end{equation}
is independent of $N$, while the total two-mode contangle
\begin{eqnarray}
N E_\tau^{i|j_l} & = & \frac{N}{4} \log^2 \bigg\{\frac{1}{N}  \Big[ b^2 (N+1) - 1  \nonumber \\
& - &  \sqrt{(b^2-1)(b^2(N+1)^2-(N-1)^2)} \; \; \Big] \bigg\},
\end{eqnarray}
is a monotonically decreasing function of the integer $N$ at fixed
$b$.        \\  Because the sharing inequality trivially holds for
$N=1$, it is inductively proven for any $N$. \hfill $\blacksquare$

Entanglement --- specifically measured by any CV extension of the
tangle as introduced in Sec.~\ref{SecTau} --- in all (pure and
mixed) fully symmetric Gaussian states, defined in
Sec.~\ref{SecSymm}, is indeed {\em monogamous}.

We can now study the difference between left-hand and right-hand
sides in \eq{monoN}, quantifying the residual entanglement not
stored in bipartite correlations between any two modes. As apparent
from the above proof, this residual entanglement, for any squeezing
$b$, strictly grows with $N$. This proves quantitatively that (as
qualitatively clear from the analysis of Sec.~\ref{SecScal}) with
increasing number of modes $N$, the entanglement is gradually less
encoded in pairwise bipartite form, being conversely more and more
increasingly retrieved in multipartite (specifically, three-partite,
four-partite, $\ldots$, $N$-partite) quantum correlations among all
the single modes. Crucially, we have just proven that this
redistribution is always such that a general monogamy law on the
shared CV entanglement is satisfied.

The multipartite entanglement in (generally mixed) fully symmetric
Gaussian states will be operationally interpreted in terms of
optimal success of CV $N$-party teleportation networks in
Sec.~\ref{SecTelepoppy}.

\subsection{Monogamy inequality for {\em all} Gaussian
states}\label{SecHiro}

Following Ref.~\cite{hiroshima}, we state here the crucial result
which definitely solves the qualitative problem of entanglement
sharing in Gaussian states. Namely, we prove that the monogamy
inequality {\em does} hold for all Gaussian states of multimode CV
systems with an arbitrary number $N$ of modes and parties
$\s_{1},\ldots,\s_{N}$, thus generalizing the results of the
previous subsection.

As a measure of bipartite entanglement, we employ the Gaussian
tangle $\tau_G$ defined via the square of negativity,
Eqs.~{\rm(\ref{eq:G-tangle_pure}, \ref{eq:G-tangle_mixed})}, in
direct analogy with the case of $N$-qubit systems \cite{osborne}.
Our proof is based on the symplectic analysis of CMs (see Chapter
\ref{ChapGauss}) and on the properties of Gaussian entanglement
measures (see Sec.~\ref{SecGEMS}). The monogamy constraint has
important implications on the structural characterization of
entanglement sharing in CV systems
\cite{contangle,3mpra,3mj,hiroshima},  in the context of
entanglement frustration in harmonic lattices \cite{frusta}, and for
practical applications such as secure key distribution and
communication networks with continuous variables (see Part
\ref{Part4}).

Given an arbitrary $N$-mode Gaussian state $\ro
_{\s_{1}|\s_{2}\ldots \s_{N}}$, we  now prove the general monogamy
inequality
\begin{equation} \label{eq:Gaussian_monogamy}
\tau _{G}(\ro _{\s_{1}|\s_{2}\ldots \s_{N}}) \, \geq \,
\sum_{l=2}^{N}\tau _{G}(\ro _{\s_{1}|\s_{l}})\, ,
\end{equation}
where we have in general renamed the modes so that the probe
subsystem in \eq{ckwine} is $\s_1$, for mere convenience.

To this end, we can assume without loss of generality that the
reduced two-mode states $\ro _{\s_{1}|\s_{l}}=\mathrm{Tr}
_{\s_{2}\ldots \s_{l-1}\s_{l+1}\ldots \s_{N}}\ro
_{\s_{1}|\s_{2}\dots \s_{N}}$ of subsystems $(\s_{1}\s_{l})$
$(l=2,\ldots ,N)$ are all entangled. In fact, if for instance $\ro
_{\s_{1}|\s_{2}}$ is separable, then $\tau _{G}(\ro
_{\s_{1}|\s_{3}\ldots \s_{N}})\leq \tau _{G}(\ro
_{\s_{1}|\s_{2}\ldots \s_{N}})$ because the partial trace over the
subsystem $\s_{2}$ is a local Gaussian operation that does not
increase the Gaussian entanglement. Furthermore, by the convex roof
construction of the Gaussian tangle, it is sufficient to prove the
monogamy inequality for any {\em pure} Gaussian state $\ro
_{\s_{1}|\s_{2}\ldots \s_{N}}^{p}$ (see also
Refs.~\cite{CKW,osborne}). Therefore, in the following we can always
assume that $\ro _{\s_{1}|\s_{2}\ldots \s_{N}} $ is a pure Gaussian
state for which the reduced states $\ro _{\s_{1}|\s_{l}}$
$(l=2,\ldots ,N)$ are all entangled.

We start by computing the left-hand side of
\eq{eq:Gaussian_monogamy}. Since $\ro _{\s_{1}|\s_{2}\ldots \s_{N}}$
is a $1\times (N-1)$ pure Gaussian state,  its CM  $\gr\sigma$ is
characterized by the condition \eq{osos},  which implies
\begin{equation}
\det \gr\alpha + \sum_{l=2}^{N} \det \gr\gamma_{l} = 1\,,
\label{speci}
\end{equation}
where $\gr\gamma_l$ is the  matrix encoding intermodal correlations
between mode $1$ and mode $l$ in the reduced state
$\ro_{\s_{1}|\s_{l}}$ $(l=2,\ldots ,N)$, described by a CM [see
\eq{stform2}]
\begin{equation} \label{eq:reduced_gamma}
\gr\sigma _{\s_{1}|\s_{l}}=\left(
\begin{array}{cccc}
\sigma _{1,1} & \sigma _{1,2} & \sigma _{1,2l-1} & \sigma _{1,2l} \\
\sigma _{2,1} & \sigma _{2,2} & \sigma _{2,2l-1} & \sigma _{2,2l} \\
\sigma _{2l-1,1} & \sigma _{2l-1,2} & \sigma _{2l-1,2l-1} & \sigma
_{2l-1,2l}
\\
\sigma _{2l,1} & \sigma _{2l,2} & \sigma _{2l,2l-1} & \sigma
_{2l,2l}
\end{array}
\right) =\left(
\begin{array}{cc}
\gr\alpha  & \gr\gamma _{l} \\
\gr\gamma _{l}\T & \gr\beta _{l}
\end{array}
\right).
\end{equation}
As $\ro _{\s_{1}|\s_{l}}$ is entangled, $\det \gr\gamma _{l}$ is
negative \cite{Simon00}, see \eq{detgammanegative}. It is useful to
introduce the auxiliary quantities
\begin{equation}\label{Upsilon}
\Upsilon _{l}=-4\det \gamma _{l}>0\,,
\end{equation}
such that one has $\det\gr\alpha = 1+ \sum_l \Upsilon_l/4\,.$

From \eq{Gtau}, the Gaussian tangle for the pure Gaussian state
$\ro_{\s_{1}|\s_{2}\ldots \s_{N}}$ is then written as
\begin{equation} \label{eq:monogamy1} \tau _{G}(\ro
_{\s_{1}|\s_{2}\ldots \s_{N}})=w(\det\gr\alpha) \equiv f\left(
\sum_{l=2}^{N}\Upsilon _{l}\right),
\end{equation}
\begin{equation} \label{eq:f_g} {\rm with}\quad
f(t)=(g^{-1}(t)-1/2)^{2}\,,\qquad g(t)=\sqrt{t+4}-\sqrt{t}\,.
\end{equation}
We observe that  $f(t)/t$ is an increasing function for $t>0$ and
$f(0)=0$ so  $f$ is a star-shaped function: $f(ct)\leq cf(t)$ for
$c\in [0,1]$ and $t\geq 0$.\footnote{\sf $f(t)$ is convex for $t\geq
0$, which also implies that $f$ is star-shaped.} Therefore, we have
$f(t)\leq \frac{t}{t+s}f(t+s)$ and $f(s)\leq \frac{s}{t+s}f(t+s)$
for $t,s\geq 0$ to obtain $f(t)+f(s)\leq f(t+s)$. That is, $f$ is
superadditive \cite{MO}. Hence,\footnote{\sf If we chose to quantify
entanglement in terms of the contangle (rather than of the Gaussian
tangle), defined for pure Gaussian states as the squared logarithmic
negativity \eq{etaupure},  we would have, instead of $f(t)$ in
Eq.~\pref{eq:f_g}, the quantity $\log^{2}[g(t)/2]$ which lacks the
star-shape property. It can be confirmed numerically that the
function $\log^{2}[g(t)/2]$ ($t\geq 0$) is not superadditive.
However this does not imply the failure of the $N$-mode monogamy
inequality for the contangle \cite{contangle}, which might be proven
with different techniques than those employed here.}
\begin{equation} \label{eq:monogamy2}
f\left( \sum_{l=2}^{N}\Upsilon _{l}\right) \geq
\sum_{l=2}^{N}f(\Upsilon _{l}).
\end{equation}
Each term in the right-hand side is well defined since $\Upsilon
_{l}>0$, \eq{Upsilon}.

We are now left to compute the right-hand side of
\eq{eq:Gaussian_monogamy}, {\em i.e.}~the bipartite entanglement in
the reduced (mixed) two-mode states $\ro_{\s_{1}|\s_{l}}$
$(l=2,\ldots ,N)$. We will show that the corresponding Gaussian
tangle is bounded from above by $f(\Upsilon _{l})$, which will
therefore prove the monogamy inequality via \eq{eq:monogamy2}. To
this aim, we recall that any bipartite and multipartite entanglement
in a Gaussian state is fully specified in terms of its CM, as the
displacement vector of first moments can be always set to zero by
local unitary operations, which preserve entanglement by definition.
It is thus convenient to express the Gaussian tangle directly in
terms of the CMs. Recalling the framework of Gaussian entanglement
measures (Sec.~\ref{SecGEMS}), the definition
\pref{eq:G-tangle_mixed} given in Sec.~\ref{SecTau} for the Gaussian
tangle of a mixed Gaussian state with CM $\sigma _{\s_{1}|\s_{l}}$
can be rewritten as
\begin{equation} \label{eq:G-tangle_simple}
\tau _{G}(\sigma _{\s_{1}|\s_{l}})=\inf_{\sigma
_{\s_{1}|\s_{l}}^{p}}\left\{ \tau _{G}(\sigma
_{\s_{1}|\s_{l}}^{p})|\sigma _{\s_{1}:\s_{l}}^{p}\leq \sigma
_{\s_{1}|\s_{l}} \right\} \, ,
\end{equation}
where the infimum is taken over all CMs $\sigma
_{\s_{1}|\s_{l}}^{p}$ of pure Gaussian states such that $\sigma
_{\s_{1}|\s_{l}}\geq \sigma _{\s_{1}|\s_{l}}^{p}$, see \eq{Gaussian
EMm}.

The quantities $\Upsilon _{l}$, \eq{Upsilon}, and $\tau _{G}(\sigma
_{\s_{1}:\s_{l}})$ for any $l$, as well as every single-mode reduced
determinant, are ${\sy{2}}^{\oplus N}$-invariants, \ie they are
preserved under local unitary (symplectic at the CM level)
operations, as mentioned in Sec.~\ref{SecSympl}. For each two-mode
partition described by \eq{eq:reduced_gamma}, we can exploit such
local-unitary freedom to put the CM $\sigma _{\s_{1}:\s_{l}}$ in
standard form, \eq{stform2},\footnote{\sf The reduced two-mode CMs
$\sig _{\s_{1}|\s_{l}}$ cannot be all brought simultaneously in
standard form, as clarified in Appendix \ref{redu}. However, our
argument runs as follows \cite{hiroshima}. We apply ${\sy{2}}\oplus
{\sy{2}}$ operations on subsystems $\s_1$ and $\s_2$ to bring $\sig
_{\s_{1}|\s_{2}}$ in standard form, evaluate an upper bound on the
Gaussian tangle in this representation, and derive an inequality
between local-unitary invariants, \eq{eq:bound}, that is therefore
not relying on the specific standard form in which explicit
calculations are performed. We then repeat such computation for the
remaining matrices
 $\sig_{\s_{1}|\s_{l}}$ with
$l=3\ldots N$. At each step, only a single two-mode CM is in
standard form while the other ones will be transformed back in a
form with (in general) non-diagonal intermodal blocks $\gr\gamma_{k
\neq l}$. However, the determinant of these blocks --- and so
$\Upsilon_k$, \eq{Upsilon} --- and the corresponding two-mode
entanglement in the CMs $\sig _{\s_{1}|\s_{k}}$ are preserved, so
the invariant condition \eq{eq:bound} holds simultaneously for all
$l=2\ldots N$.} with $\alpha = {\rm diag}\{a,\,a\}$, $\beta_l = {\rm
diag}\{b,\,b\}$, and $\gamma_l = {\rm diag}\{c_{+},\,c_{-}\}$, where
$c_{+}\geq \left| c_{-}\right| $ \cite{Simon00,Duan00}. The
uncertainty condition \ref{bonfide}  for $\sigma _{\s_{1}|\s_{l}}$
is thus equivalent to the following inequalities [see also
\eq{sepcomp}]
\begin{eqnarray}&
a \geq 1\,,\quad b \geq 1\,,\quad ab-c_{\pm }^{2} \geq 1 \, ;  \label{eq:positivity1} \\
&\!\!\!\!\! \det \sigma
_{\s_{1}|\s_{l}}+1=(ab-c_{+}^{2})(ab-c_{-}^{2})+1 \geq
a^{2}+b^{2}+2c_{+}c_{-}\,. \quad \label{eq:positivity2}
\end{eqnarray}
Furthermore, since the state $\ro _{\s_{1}|\s_{l}}$ is entangled, we
have \cite{Simon00}
\begin{equation} \label{eq:inseparability}
(ab-c_{+}^{2})(ab-c_{-}^{2})+1 \, < \, a^{2}+b^{2}-2c_{+}c_{-} \, .
\end{equation}
From Eqs.~\pref{eq:positivity2} and \pref{eq:inseparability}, it
follows that $c_{-}<0 $.

In Eq.~\pref{eq:G-tangle_simple}, $\tau _{G}(\sig
_{\s_{1}|\s_{l}}^{p})=f(4\det \gr\alpha ^{p}-4)$, which is an
increasing function of $\det \gr\alpha ^{p}$, where $\gr\alpha ^{p}$
is the first $2\times 2$ principal submatrix of $\sig
_{\s_{1}:\s_{l}}^{p}$, see \eq{eq:reduced_gamma}. The infimum of the
right-hand side of Eq.~\pref{eq:G-tangle_simple} is achieved by the
pure-state CM $\sig _{\s_{1}|\s_{l}}^{p}$ (with $\sig
_{\s_{1}|\s_{l}}^{p}\leq \sig _{\s_{1}|\s_{l}}$ and
$\sig_{\s_{1}|\s_{l}}^{p}+i\Omega\geq 0$) that minimizes $\det
\gr\alpha ^{p}$. The minimum value of $\det \gr\alpha ^{p}$ is given
by $$\min_{0\leq \theta <2\pi }m^2_\theta(a,b,c_+,c_-)\,,$$ with
$m^2_\theta(a,b,c_+,c_-)$ defined by \eq{mfunc} \cite{ordering}.

Namely, $m^2_\theta(a,b,c_+,c_-)=1+{h_{1}^{2}(\theta
)}/{h_{2}(\theta )}\,, $ with $h_{1}(\theta )=\xi _{-}+\sqrt{\eta
}\cos \theta $, and $h_{2}(\theta ) =
2(ab-c_{-}^{2})(a^{2}+b^{2}+2c_{+}c_{-}) -({\zeta }/{\sqrt{\eta
}})\cos \theta +(a^{2}-b^{2})\sqrt{1-{\xi _{+}^{2}}/{\eta}}\sin
\theta$. Here
\begin{eqnarray}
\label{eq:xi} \xi _{\pm }&=&c_{+}(ab-c_{-}^{2})\pm c_{-}\,, \\
\label{eq:eta} \eta &=&[a-b(ab-c_{-}^{2})][b-a(ab-c_{-}^{2})]\,, \\
\zeta&=&2abc_{-}^{3}+(a^{2}+b^{2})c_{+}c_{-}^{2} \nonumber
\\&+&[a^{2}+b^{2}-2a^{2}b^{2}]c_{-}-ab(a^{2}+b^{2}-2)c_{+}\,.
\label{eq:zeta}
\end{eqnarray}
Moreover, it is obvious that $m^2_\pi \geq \min_{0\leq \theta <2\pi
}m^2_\theta$ and therefore
\begin{equation} \label{eq:monogamy3}
\tau _{G}(\sig _{\s_{1}|\s_{l}})\leq f(4m^2_\pi-4)=f\left( 4\
\zeta_{1}^{2}/{\zeta _{2}}\right) \, ,
\end{equation}
where $\zeta _{1}=h_{1}(\pi )$ and $\zeta _{2}=h_{2}(\pi )$.
Finally, one can prove (see the Appendix of Ref.~\cite{hiroshima})
that
\begin{equation}\label{eq:appendix}
\Upsilon _{l}=-4\det \gamma_{l}=-4c_{+}c_{-}\geq 4\ {\zeta
_{1}^{2}}/{\zeta _{2}}\,,
\end{equation}
which, being $f(t)$ [\eq{eq:f_g}] an increasing function of $t$,
entails that $f(\Upsilon _{l})\geq f( 4\ \zeta_{1}^{2}/{\zeta
_{2}})$. Combining this with \eq{eq:monogamy3} leads to the crucial
${\sy{2}}^{\oplus N}$-invariant condition
\begin{equation} \label{eq:bound}
\tau _{G}(\sigma _{\s_{1}:\s_{l}}) \le f(\Upsilon _{l})\,,
\end{equation}
which holds in general for all $l=2\ldots N$ and does not rely on
the specific standard form of the reduced CMs $\sig
_{\s_{1}|\s_{l}}$.

Then, recalling
Eqs.~{\rm(\ref{eq:monogamy1},\ref{eq:monogamy2},\ref{eq:bound})},
Inequality~\pref{eq:Gaussian_monogamy} is established. This
completes the proof of the monogamy constraint on CV entanglement
sharing for pure Gaussian states of an arbitrary number of modes. As
already mentioned, the proof immediately extends to arbitrary mixed
Gaussian states by the convexity of the Gaussian tangle,
Eq.~\pref{eq:G-tangle_mixed}. \hfill $\blacksquare$

\smallskip

Summarizing, we have proven the following \cite{hiroshima}.

\medskip

\begin{itemize}
\item[\ding{226}]
 \noindent{\rm\bf Monogamy inequality for all Gaussian states.}
{\it The Gaussian tangle $\tau_G$,  an entanglement monotone under
Gaussian LOCC, is monogamous for all, pure and mixed, $N$-mode
Gaussian states distributed among $N$ parties, each owning a single
mode.}
\smallskip
\end{itemize}

\subsubsection{Implications and perspectives}

The consequences of our result are manifold. The monogamy
constraints on entanglement sharing are essential for the security
of CV quantum cryptographic schemes \cite{Cry1,Cry2}, because they
limit the information that might be extracted from the secret key by
a malicious eavesdropper. Monogamy is useful as well in
investigating the range of correlations in Gaussian valence bond
states of harmonic rings \cite{gvbs} (see Chapter \ref{ChapGVBS}),
and in understanding the entanglement frustration occurring in
ground states of many-body harmonic lattice systems \cite{frusta},
which, following our findings, may be now extended to arbitrary
states beyond symmetry constraints.

On the other hand, the investigation of the consequences of the
monogamy property on the structure of entanglement sharing in
generic Gaussian states (as we will show in the next Chapters),
reveals that there exist states that maximize both the pairwise
entanglement in any reduced two-mode partition, and the residual
distributed (multipartite) entanglement obtained as a difference
between the left-hand and the right-hand side in
Eq.~\pref{eq:Gaussian_monogamy}. The simultaneous monogamy and {\em
promiscuity} of CV entanglement (unparalleled in qubit systems) may
allow for novel, robust protocols for the processing and
transmission of quantum and classical information. The monogamy
inequality \pref{eq:Gaussian_monogamy} bounds the persistency of
entanglement  when one or more nodes in a CV communication network
sharing generic $N$-mode Gaussian resource states are traced out.

At a fundamental level, the proof of the monogamy property for all
Gaussian states paves the way to a proper quantification of genuine
multipartite entanglement in CV systems in terms of the residual
distributed entanglement. In this respect, the intriguing question
arises  whether a {\em stronger} monogamy constraint exists on the
distribution of entanglement in many-body systems, which imposes a
physical trade-off on the sharing of both bipartite and genuine
multipartite quantum correlations.

It would be important to understand whether the inequality
\pref{eq:Gaussian_monogamy} holds as well for discrete-variable
qudits ($2 < d < \infty$), interpolating between qubits and CV
systems (see Sec.~\ref{SecPisa}). If this were the case, the
(convex-roof extended) squared negativity, which coincides with the
tangle for arbitrary states of qubits and with the Gaussian tangle
for Gaussian states of CV systems, would qualify as a universal {\em
bona fide}, dimension-independent quantifier of entanglement sharing
in all multipartite quantum systems. In such context, a deeper
investigation into the analogy between Gaussian states with finite
squeezing and effective finite-dimensional systems (see
Sec.~\ref{SecUnlimDiscuss}), focused on the point of view of
entanglement sharing, may be worthy.

All of this research is in full progress.

 }

\chapter{Tripartite entanglement in three-mode Gaussian
states}\label{Chap3M}

{\sf

In this Chapter, based on Refs.~\cite{contangle,3mpra,3mj}, we
present a complete analysis of entanglement in three-mode Gaussian
states of CV systems. They constitute the simplest instance of
infinite-dimensional states exhibiting multipartite entanglement.

We construct standard forms which characterize the CM of pure and
mixed three-mode Gaussian states, up to local unitary operations. We
approach the quantification of multipartite entanglement by
providing an independent proof of the monogamy of entanglement
specialized to a tripartite Gaussian setting, where the quantum
correlations are measured by the (Gaussian) contangle (convex-roof
extended squared logarithmic negativity), defined in
Sec.~\ref{seccontangle}. We adopt the ``residual (Gaussian)
contangle'', emerging from the monogamy inequality, as measure of
genuine tripartite entanglement, and prove it to be monotonically
nonincreasing under Gaussian LOCC. It embodies therefore the first
{\em bona fide} measure of multipartite (specifically, tripartite)
entanglement in CV systems.

We analytically compute the residual contangle for arbitrary pure
three-mode Gaussian states. We analyze the distribution of quantum
correlations and show that pure, fully symmetric three-mode Gaussian
states (see Sec.~\ref{SecSymm} for the definition of fully symmetric
states in general) allow a {\em promiscuous} entanglement sharing,
having both maximum tripartite residual entanglement and maximum
couplewise entanglement between any pair of modes, for any given
degree of squeezing. These states are thus simultaneous CV analogs
of both the GHZ \cite{GHZ} and the $W$ states \cite{wstates} of
three qubits (defined in Sec.~\ref{sec3tangle}) and are hence
rebaptized CV ``GHZ/$W$'' states. The persistency of promiscuity
against thermalization and lack of symmetry is investigated.

We finally consider the action of decoherence on tripartite
entangled Gaussian states, studying the decay of the residual
contangle. The Gaussian GHZ/$W$ states are shown to be maximally
robust against decoherence effects.

\section{Three-mode Gaussian states}\label{Sec3M}

To begin with, let us set the notation and review the known results
about three-mode Gaussian states of CV systems. We will refer to the
three modes under exam as mode $1$, $2$ and $3$. The $2 \times 2$
submatrices that form the CM $\sig \equiv \sig_{123}$ of a
three-mode Gaussian state are defined according to \eq{CM}, whereas
the $4 \times 4$ CMs of the reduced two-mode Gaussian states of
modes $i$ and $j$ will be denoted by $\sig_{ij}$. Likewise, the
local (two-mode) seralian invariants $\Delta_{ij}$, \eq{seralian},
will be specified by the labels $i$ and $j$ of the modes they refer
to, while, to avoid any confusion, the three-mode (global) seralian
symplectic invariant will be denoted by $\Delta\equiv\Delta_{123}$.
Let us recall the uncertainty relation \eq{sepcomp} for two-mode
Gaussian states, \be \Delta_{ij} - \det{\sig_{ij}} \le 1 \; .
\label{uncedue} \ee

\subsection{Separability properties}\label{secbarbie}

As it is clear from the discussion of Sec.~\ref{SecPPTG},    a
complete {\em qualitative} characterization of the entanglement of
three-mode Gaussian state is possible because the PPT criterion is
necessary and sufficient for their separability under {\em any},
partial or global (\ie $1\times 1$ or $1\times 2$), bipartition of
the modes. This has lead to an exhaustive classification of
three-mode Gaussian states in five distinct separability classes
\cite{kraus}. These classes take into account the fact that the
modes $1$, $2$ and $3$ allow for three distinct global bipartitions:
\begin{itemize}
\item{{\it Class 1}\!~: states not separable under all the three possible
bipartitions $i \times (jk)$  of the modes (fully inseparable
states, possessing genuine multipartite entanglement).}
\item{{\it Class 2}\!~: states separable under only one of the three possible
bipartitions (one-mode biseparable states).}
\item{{\it Class 3}\!~: states separable under only two of the three possible
bipartitions (two-mode biseparable states).}
\item{{\it Class 4}\!~: states separable under all the three possible bipartitions,
but impossible to write as a convex sum of tripartite products of
pure one-mode states (three-mode biseparable states).}
\item{{\it Class 5}\!~: states that are separable under all the three possible bipartitions,
and can be written as a convex sum of tripartite products of pure
one-mode states (fully separable states).}
\end{itemize}
Notice that Classes 4 and 5 cannot be distinguished by partial
transposition of any of the three modes (which is positive for both
classes). States in Class 4 stand therefore as nontrivial examples
of tripartite entangled states of CV systems with positive partial
transpose \cite{kraus}. It is well known that entangled states with
positive partial transpose possess {\em bound entanglement}, that
is, entanglement that cannot be distilled by means of LOCC.

\subsection{Pure states: standard form and local entropic triangle inequality}\label{secpuri}

We begin by focusing on {\em pure} three-mode Gaussian states, for
which one has \be \det{\sig} = 1 \; , \quad \Delta=3 \, .
\label{purinv} \ee The purity constraint requires the local entropic
measures of any $1\times 2$-mode bipartitions to be equal: \be
\det{\sig_{ij}}=\det{\sig_{k}} \; , \label{pur} \ee with $i$, $j$
and $k$ different from each other. This general, well known property
of the bipartitions of pure states may be easily proven resorting to
the Schmidt decomposition (see Sec.~\ref{SecSchmidtPS}).

A first consequence of Eqs.~\pref{purinv} and \pref{pur} is rather
remarkable. Combining such equations one easily obtains
$$
(\Delta_{12}-\det{\sig_{12}}) + (\Delta_{13}-\det{\sig_{13}}) +
(\Delta_{23}-\det{\sig_{23}}) = 3 \; ,$$ which, together with
Inequality \pref{uncedue}, implies \be \Delta_{ij} = \det{\sig_{ij}}
+ 1 \; , \quad \forall \, i,j: \; i\neq j \, . \label{glems3m} \ee
The last equation shows that any reduced two-mode state of a pure
three-mode Gaussian state saturates the partial uncertainty relation
\eq{uncedue}. The states endowed with such a partial minimal
uncertainty (namely, with their smallest symplectic eigenvalue equal
to $1$) are states of minimal negativity for given global and local
purities, alias GLEMS (Gaussian least entangled mixed states)
\cite{prl,extremal}, introduced in Sec.~\ref{SecGmemsGlems}. In
general, by invoking the phase-space Schmidt decomposition (see
Sec.~\ref{SecSchmidtPS}) \cite{holevo01,botero03,giedkeqic03}, it
immediately follows that any $(N-1)$-mode reduced state of a
$N$-mode pure Gaussian state is a mixed state of partial minimum
uncertainty (a sort of generalized GLEMS), with $N-2$ symplectic
eigenvalues fixed to $1$ and only one, in general, greater than $1$
 --- shortly, with symplectic rank $\aleph=1$, see
Sec.~\ref{SecSympHeis} --- thus saturating \eq{sympheis}. This
argument is resumed in Appendix \ref{condpuri}.

In fact, our simple proof, straightforwardly derived in terms of
symplectic invariants, provides some further insight into the
structure of CMs characterizing three-mode Gaussian states. What
matters to our aims, is that the standard form CM of Gaussian states
is completely determined by their global and local invariants, as
discussed in Sec.~\ref{SecSFCM}. Therefore, because of \eq{pur}, the
entanglement between any pair of modes embedded in a three-mode pure
Gaussian state is fully determined by the local invariants
$\det{\sig_{l}}$, for $l=1,2,3$, whatever proper measure we choose
to quantify it. Furthermore, the entanglement of a $\sig_{i|({jk})}$
bipartition of a pure three-mode state is determined by the entropy
of one of the reduced states, \ie, once again, by the quantity
$\det{\sig_{i}}$. Thus, {\em the three local symplectic invariants
$\det{\sig_{1}}$, $\det{\sig_{2}}$ and $\det{\sig_{3}}$ fully
determine the entanglement of any bipartition of a pure three-mode
Gaussian state}. We will show that they suffice to determine as well
the genuine tripartite entanglement encoded in the state
\cite{3mpra}.

For ease of notation, in the following we will denote by $a_l$ the
local single-mode symplectic eigenvalue associated to mode $l$ with
CM $\sig_l$, \be \label{al} a_l\equiv \sqrt{\det{\sig_l}} \; . \ee
\eq{purgau} shows that the quantities $a_l$ are simply related to
the purities of the reduced single-mode states, \ie the local
purities $\mu_l$, by the relation \be \mu_l = \frac{1}{a_{l}} \; .
\ee Since the set $\{a_l\}$, $l=1,2,3$, fully determines the
entanglement of any of the $1\times2$ and $1\times1$ bipartitions of
the state, it is important to determine the range of the allowed
values for such quantities. This is required in order to  provide a
complete quantitative characterization of the entanglement of
three-mode pure Gaussian states. To this aim, let us focus on the
reduced two-mode CM $\sig_{12}$ and let us bring it (by local
unitaries) in standard form \cite{Duan00,Simon00}, so that \eq{CM}
is recast in the form
\begin{eqnarray}
\label{2sform} \sig_{l} & = & {\rm diag}\{a_{l},\,a_{l}\}\,,\quad
l=1,2 \, ; \nonumber \\
\eps_{12} & = & {\rm diag}\{c_{12},\,d_{12}\} \; ,
\end{eqnarray}
where $c_{12}$ and $d_{12}$ are the intermodal covariances, and, as
we will show below, can be evaluated independently in pure
three-mode Gaussian states. Notice that no generality is lost in
assuming a standard form CM, because the entanglement properties of
any bipartition of the system are invariant under local
(single-mode) symplectic operations. Now, Eqs.~\pref{pur} and
\pref{purinv} may be recast as follows \bea
a_3^2 &=& a_1^2 + a_2^2 +2 c_{12} d_{12} -1 \; , \\
a_3^2 &=& (a_1 a_2 -c_{12}^2)(a_1 a_2 -d_{12}^2) \; , \eea showing
that we may eliminate one of the two covariances to find the
expression of the remaining one only in terms of the three local
mixednesses (inverse purities) $a_l$, \eq{al}. Defining the quantity
$\kappa$ as \be \kappa\equiv c_{12} d_{12} =
\frac{1+a_3^2-a_1^2-a_2^2}{2} \; , \ee leads to the following
condition on the covariance $c_{12}$, \be c_{12}^4 - \frac{1}{a_1
a_2}\left[(\kappa-1)^2+a_1^2 a_2^2-a_1^2-a_2^2\right] c_{12}^2 +
\kappa^2 = 0 \; . \label{eqe1} \ee Such a second-order algebraic
equation for $c_{12}^2$ admits a positive solution if and only if
its discriminant $\delta$ is positive, \be \delta \ge 0 \; .
\label{deltino} \ee After some algebra, one finds \bea
\delta&=&(a_1+a_2+a_3+1)(a_1+a_2+a_3-1) \nonumber\\
&\times& (a_1+a_2-a_3+1)(a_1-a_2+a_3+1) \nonumber \\
&\times&(-a_1+a_2+a_3+1)(a_1+a_2-a_3-1) \nonumber \\
&\times&(a_1-a_2+a_3-1)(-a_1+a_2+a_3-1) \, . \eea Aside from the
existence of a real covariance $c_{12}$, the further condition of
positivity of $\sig_{12}$ has to be fulfilled for a state to be
physical. This amounts to impose the inequality $a_1 a_2 -c_{12}^2
\ge 0$, which can be explicitly written, after solving \eq{eqe1}, as
$$
4\left[2a_1^2 a_2^2 -
\left((\kappa-1)^2+a_1^2a_2^2-a_1^2-a_2^2\right)\right] \ge
\sqrt{\delta} \; .
$$
This inequality is trivially satisfied when squared on both sides;
therefore it reduces to \be 2a_1^2a_2^2 -
\left((\kappa-1)^2+a_1^2a_2^2-a_1^2-a_2^2\right)\ge 0 \; .
\label{posi} \ee

\begin{figure}[t!]
\includegraphics[width=12cm]{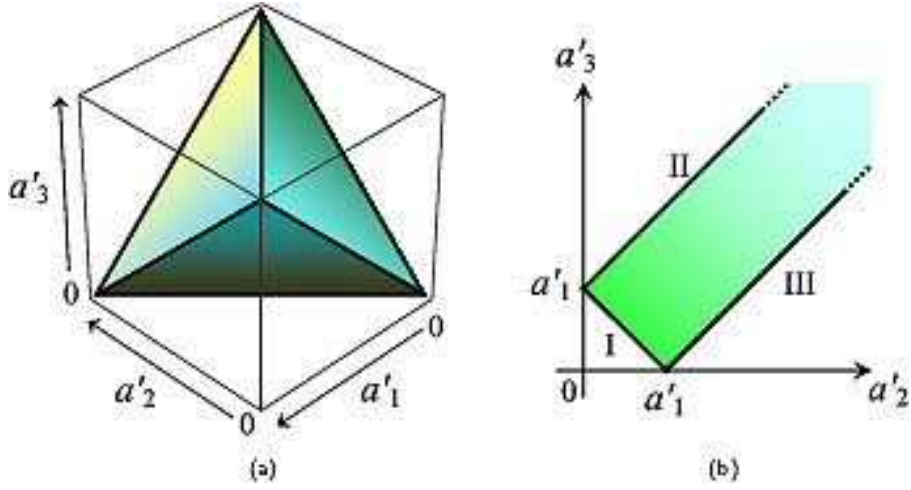}%
\caption{Range of the entropic quantities $a'_l = \mu_l^{-1}-1$  for
pure three-mode Gaussian states. The three parameters $a'_l$, with
$l=1,2,3$, have to vary inside the pyramid represented in plot
\textrm{(a)} or, equivalently, for fixed values of one of them, say
$a'_1$, inside the shaded slice represented in plot \textrm{(b)}, in
order to determine the CM of a physical state, \eq{cm3tutta}. The
expression of the boundary surfaces/curves come from the saturation
of the triangular inequality \pref{triangleprim} for all possible
mode permutations. In particular, for the projected two-dimensional
plot \textrm{(b)}, the equations of the three boundaries are:
I.~$a'_3=a'_1-a'_2$; II.~$a'_3=a'_1+a'_2$;
III.~$a'_3=a'_2-a'_1$.}\label{figangle}
\end{figure}

Notice that conditions \pref{deltino} and \pref{posi}, although
derived by assuming a specific bipartition of the three modes, are
independent on the choice of the modes that enter in the considered
bipartition, because they are invariant under all possible
permutations of the modes. Defining the parameters
\begin{equation}
\label{aprimi} a'_l \equiv a_l-1 \; ,
\end{equation}
the uncertainty principle \eq{sympheis} for single-mode states
reduces to \be a'_l \ge 0 \; \; \; \; \; \forall\, l=1,2,3 \; . \ee
This fact allows to greatly simplify the existence conditions
\pref{deltino} and \pref{posi}, which can be combined into the
following triangular inequality \be \label{triangleprim} |a'_i-a'_j|
\, \le \, a'_k \, \le \, a'_i+a'_j \; . \ee Inequality
\pref{triangleprim} is a condition invariant under all possible
permutations of the mode indexes $\{i,j,k\}$, and, together with the
positivity of each $a'_l$, fully characterizes the local symplectic
eigenvalues of the CM of three-mode pure Gaussian states. It
therefore provides a complete characterization of the entanglement
in such states. All standard forms of pure three-mode Gaussian
states and in particular, remarkably, all the possible values of the
negativities (Sec.~\ref{secnega}) and/or of the Gaussian
entanglement measures (Sec.~\ref{SecGEMS}) between {\em any} pair of
subsystems, can be determined by letting $a'_1$, $a'_2$ and $a'_3$
vary in their range of allowed values, as summarized in
Fig.~\ref{figangle}.

Let us remark that \eq{triangleprim} qualifies itself as an entropic
inequality, as the quantities $\{a'_j\}$ are closely related to the
purities and to the Von Neumann entropies of the single-mode reduced
states. In particular the Von Neumann entropies $S_{Vj}$ of the
reduced states are given by $S_{Vj}=f(a'_j+1)=f(a_j)$, where the
increasing convex entropic function $f(x)$ has been defined in
\eq{entfunc}. Now, Inequality \pref{triangleprim} is strikingly
analogous to the well known triangle (Araki-Lieb) and subadditivity
inequalities \cite{ArakiLieb70,Wehrl78} for the Von Neumann entropy,
which hold  for general systems [see \eq{QM:SVtriangle}], and in our
case read \be |f(a_i)-f(a_j)| \le f(a_k) \le f(a_i) + f(a_j) \; .
\label{arakilieb} \ee However, as the different convexity properties
of the  involved functions suggest, Inequalities \pref{triangleprim}
and \pref{arakilieb} are not equivalent. Actually, as can be shown
by exploiting the properties of the function $f(x)$, the
Inequalities \pref{triangleprim} imply the Inequalities
\pref{arakilieb} for both the leftmost and the rightmost parts. On
the other hand, there exist values of the local symplectic
eigenvalues $\{a_l\}$ for which Inequalities \pref{arakilieb} are
satisfied but \pref{triangleprim} are violated. Therefore, the
conditions imposed by \eq{triangleprim} on the local invariants, are
strictly {\em stronger} than the generally holding inequalities for
the Von Neumann entropy applied to pure quantum states.

We recall that the form of the CM of any Gaussian state can be
simplified through local (unitary) symplectic operations, that
therefore do not affect the entanglement or mixedness properties of
the state, belonging to ${Sp_{(2,\R)}}^{\oplus N}$. Such reductions
of the CMs are called ``standard forms'', as introduced in
Sec.~\ref{SecSFCM}. For the sake of clarity, let us write the
explicit standard form CM of a generic {\em pure} three-mode
Gaussian state \cite{3mpra},
\begin{equation}
\label{cm3tutta} \sig^p_{sf}=\left(
\begin{array}{cccccc}
 a_1 & 0 & e_{12}^+ & 0 & e_{13}^+ & 0 \\
 0 & a_1 & 0 & e_{12}^- & 0 & e_{13}^- \\
 e_{12}^+ & 0 & a_2 & 0 & e_{23}^+ & 0 \\
 0 & e_{12}^- & 0 & a_2 & 0 & e_{23}^- \\
 e_{13}^+ & 0 & e_{23}^+ & 0 & a_3 & 0 \\
 0 & e_{13}^- & 0 & e_{23}^- & 0 & a_3
\end{array}
\right)\; ,
\end{equation}
with
\begin{equation}\label{eij}
\begin{split}
e_{ij}^{\pm} \equiv \frac{1}{{4 \sqrt{a_i a_j}}} &\Bigg\{
\sqrt{\left[\left(a_i-a_j\right)^2-\left(a_k-1\right)^2\right]
\left[\left(a_i-a_j\right)^2-\left(a_k+1\right)^2\right]} \\ &\pm
\sqrt{\left[\left(a_i+a_j\right)^2-\left(a_k-1\right)^2\right]
\left[\left(a_i+a_j\right)^2-\left(a_k+1\right)^2\right]}\Bigg\} \,
. \end{split}
\end{equation}
By direct comparison with \eq{glems}, it is immediate to verify that
each two-mode reduced CM $\sig_{ij}$ denotes a standard form GLEMS
with local purities $\mu_i=a_i^{-1}$ and $\mu_j=a_j^{-1}$, and
global purity $\mu_{ij}\equiv \mu_k = a_k^{-1}$. Notice also that
the standard form of any pure three-mode Gaussian state,
\eq{cm3tutta}, admits all $2 \times 2$ subblocks of the CM
simultaneously in diagonal form; this is no longer possible for
completely general pure Gaussian states of $N \ge 4$ modes, as
clarified in Appendix \ref{redu}. However, pure Gaussian states
which, for an arbitrary number of modes, are reducible to such a
``block-diagonal'' standard form, are endowed with peculiar
entanglement properties \cite{generic}, which will be investigated
in Chapter \ref{ChapGeneric}.

Let us  stress that, although useful in actual calculations, the use
of CMs in standard form does not entail any loss of generality,
because all the results derived in the present Chapter for $N=3$ do
not depend on the choice of the specific form of the CMs, but only
on invariant quantities, such as the global and local symplectic
invariants.

A first qualitative result which immediately follows from our study
\cite{3mpra}, is that, regarding the classification of
Sec.~\ref{secbarbie} \cite{kraus}, pure three-mode Gaussian states
may belong either to Class 5, in which case they reduce to the
global three-mode vacuum, or to Class 2, reducing to the
uncorrelated product of a single-mode vacuum and of a two-mode
squeezed state, or to Class 1 (fully inseparable state). No two-mode
or three-mode biseparable pure three-mode Gaussian states are
allowed.

\subsection{Mixed states}
For the sake of completeness, let us briefly report that the most
general standard form $\sig_{sf}$ associated to the CM of any
(generally mixed) three-mode Gaussian state can be written as
\begin{equation}\label{tutta}
\sig_{sf}=\left(
\begin{array}{cccccc}
 a_1 & 0 & f_1 & 0 & f_3 & f_5 \\
 0 & a_1 & 0 & f_2 & 0 & f_4 \\
 f_1 & 0 & a_2 & 0 & f_6 & f_8 \\
 0 & f_2 & 0 & a_2 & f_9 & f_7 \\
 f_3 & 0 & f_6 & f_9 & a_3 & 0 \\
 f_5 & f_4  & f_8 & f_7 & 0 & a_3
\end{array}
\right)\, ,
\end{equation}
where the 12 parameters $\{a_k\}$ (inverse of the local purities)
and $\{f_k\}$ (the covariances describing correlations between the
modes) are only constrained by the uncertainty relations
\eq{bonfide}. The possibility of this useful reduction (the general
case of $N$-mode Gaussian mixed states has been discussed in
Sec.~\ref{sform}) can be easily proven along the same lines as the
two-mode standard form reduction \cite{Duan00}: by means of three
local symplectic operations one can bring the three blocks $\sig_1$,
$\sig_2$ and $\sig_3$ in Williamson form, thus making them
insensitive to further local rotations (which are symplectic
operations); exploiting such rotations on mode $1$ and $2$ one can
then diagonalize the block $\gr{\varepsilon}_{12}$ as allowed by its
singular value decomposition; finally, one local rotation on mode
$3$ is left, by which one can cancel one entry of the block
$\gr{\varepsilon}_{13}$. Indeed, the resulting number of free
parameters could have been inferred by subtracting the number of
parameters of an element of $Sp_{(2,\R)}\oplus Sp_{(2,\R)}\oplus
Sp_{(2,\R)}$ (which is $9$, as $Sp_{(2,\R)}$ has $3$ independent
generators) from the 21 entries of a generic $6\times 6$ symmetric
matrix.

\section{Distributed entanglement and genuine tripartite quantum correlations}

In this Section we approach in a systematic way the question of
distributing quantum correlations among three parties globally
prepared in a (pure or mixed) three-mode Gaussian state, and we deal
with the related problem of quantifying genuine tripartite
entanglement in such a state.

\subsection{Monogamy of the Gaussian contangle for all three-mode Gaussian states}
\label{secmono}

In Sec.~\ref{SecHiro}, we have established the monogamy of
distributed entanglement, \eq{ckwine}, for all Gaussian states of an
arbitrary number of modes, employing the Gaussian tangle, \eq{Gtau},
defined in terms of squared negativity, as a measure of bipartite
entanglement. We have also mentioned that, when possible, it is more
appropriate to adopt as a bipartite entanglement monotone the
(Gaussian) contangle, \eq{tau}, defined in terms of squared
logarithmic negativity. The (Gaussian) contangle is indeed the
primitive measure, whose monogamy implies by convexity the monogamy
of the Gaussian tangle. As the convex rescaling induced by the
mapping from the Gaussian contangle to the Gaussian tangle becomes
relevant once the bipartite entanglements in the different
bipartitions have to be compared to induce a proper tripartite
entanglement quantification, we will always commit ourselves to the
(Gaussian) contangle in the quantification of entanglement for
three-mode Gaussian states.

Henceforth, we now provide the detailed proof, which we originally
derived in Ref.~\cite{contangle}, that all three-mode Gaussian
states satisfy the CKW monogamy inequality \pref{CKWine}, using the
(Gaussian) contangle \eq{tau} to quantify bipartite entanglement.
Chronologically, this is actually the first monogamy proof ever
obtained in a CV scenario. The intermediate steps of the proof will
be then useful for the subsequent computation of the residual
genuine tripartite entanglement, as we will show in
Sec.~\ref{secresid}.

We start by considering pure three-mode Gaussian states, whose
standard form CM $\sig^p$ is given by \eq{cm3tutta}. As discussed in
Sec.~\ref{secpuri}, all the properties of bipartite entanglement in
pure three-mode Gaussian states are completely determined by the
three local purities. Reminding that the mixednesses $a_l \equiv
1/\mu_l$ have to vary constrained by the triangle inequality
\pref{triangleprim}, in order for $\sig^p$ to represent a physical
state, one has
\begin{equation}
\label{triangle} |a_j - a_k| + 1 \le a_i \le a_j + a_k - 1 \; .
\end{equation}
For ease of notation let us rename the mode indices so that
$\{i,j,k\} \equiv \{1,2,3\}$ in \ineq{CKWine}. Without any loss of
generality, we can assume $a_1 > 1$. In fact, if $a_1=1$ the first
mode is not correlated with the other two and all the terms in
\ineq{CKWine} are trivially zero. Moreover, we can restrict the
discussion to the case of both the reduced two-mode states
$\sig_{12}$ and $\sig_{13}$ being entangled. In fact, if {\em
e.g.}~$\sig_{13}$ denotes a separable state, then $E_\tau^{1|2} \le
E_\tau^{1|(23)}$ because tracing out mode $3$ is a LOCC, and thus
the sharing inequality is automatically satisfied. We will now prove
\ineq{CKWine} in general by using the Gaussian contangle $G_\tau$
[see \eq{GaCoRo}], as this will immediately imply the inequality for
the true contangle $E_\tau$ [see \eq{etaumix}] as well. In fact,
$G_\tau^{1|(23)}(\sig^p) = E_\tau^{1|(23)}(\sig^p)$, but
$G_\tau^{1|l}(\sig) \ge E_\tau^{1|l}(\sig),\; l=2,3$.

Let us proceed by keeping $a_1$ fixed. From \eq{piupurezzapertutti},
it follows that the entanglement between mode $1$ and the remaining
modes, $E_\tau^{1|(23)} = \arcsinh^2\sqrt{a_1^2-1}$, is constant. We
must now prove that the maximum value of the sum of the $1|2$ and
$1|3$ bipartite entanglements can never exceed $E_\tau^{1|(23)}$, at
fixed local mixedness $a_1$. Namely,
\begin{equation}
\label{maxQ} \max_{s,d} Q \, \le \, \arcsinh^2\sqrt{a^{2} - 1} \; ,
\end{equation}
where $a \equiv a_1$ (from now on we drop the subscript ``1''), and
we have defined
\begin{equation}
\label{QQ} Q \, \equiv \, G_\tau^{1|2}(\sig^p) +
G_\tau^{1|3}(\sig^p) \; .
\end{equation}
The maximum in \eq{maxQ} is taken with respect to the ``center of
mass'' and ``relative'' variables $s$ and $d$ that replace the local
mixednesses $a_2$ and $a_3$ according to
\begin{eqnarray}
s &=& \frac{a_2+a_3}2 \; , \label{s1} \\
& & \nonumber \\
d &=& \frac{a_2-a_3}2 \; . \label{d1}
\end{eqnarray}
The two parameters $s$ and $d$ are constrained to vary in the region
\begin{equation}
\label{sdangle} s \, \ge \, \frac{a+1}{2} \; , \; \; \qquad\abs{d}
\, \le \, \frac{a^2-1}{4s} \; .
\end{equation}
\ineq{sdangle} combines the triangle inequality \pref{triangle} with
the condition of inseparability for the states of the reduced
bipartitions $1|2$ and $1|3$, \eq{glement}.

We have used the fact that, as stated in Sec.~\ref{secpuri}, each
$\sig_{1l}$, $l=2,3$, is a state of partial minimum uncertainty
(GLEMS, see Sec.~\ref{SecGmemsGlems}). For this class of states the
Gaussian measures of entanglement, including $G_\tau$, have been
computed explicitly in Sec.~\ref{SecGEMGLEM} \cite{ordering},
yielding
\begin{equation}
\label{Qglems} Q = \arcsinh^2
\Big[\sqrt{m^2(a,s,d)-1}\Big]+\arcsinh^2
\Big[\sqrt{m^2(a,s,-d)-1}\Big] \; ,
\end{equation}
where $m = m_-$ if $D \le 0$, and $m = m_+$ otherwise (one has
$m_+=m_-$ for $D=0$). Here:
\begin{eqnarray}
\label{unsacco}
m_- & = & \frac{|k_-|}{(s-d)^2-1} \; , \nonumber \\
& & \nonumber \\
m_+ & = & \frac{\sqrt{2\,\left[2 a^2 (1+2 s^2 + 2 d^2) - (4 s^2 -
1)(4 d^2 - 1) -a^4 -
\sqrt{\delta}\right]}}{4(s-d)}\; , \nonumber \\
& & \nonumber \\
D & = & 2 (s - d) - \sqrt{2\left[k_-^2 + 2 k_++|k_-| (k_-^2 + 8
k_+)^{1/2}\right]/k_+} \; , \nonumber \\
& & \nonumber \\
k_\pm & = & a^2 \pm (s+d)^2 \; ,
\end{eqnarray}
and the quantity $\delta = (a - 2 d - 1) (a - 2 d + 1) (a + 2 d - 1)
(a + 2 d + 1) (a - 2 s - 1) (a - 2 s + 1) (a + 2 s - 1) (a + 2 s +
1)$ is the same as in \eq{deltino}. Note (we omitted the explicit
dependence for brevity) that each quantity in \eq{unsacco} is a
function of $(a,s,d)$. Therefore, to evaluate the second term in
\eq{Qglems} each $d$ in \eq{unsacco} must be replaced by $-d$.

Studying the derivative of $m_\mp$ with respect to $s$, it is
analytically proven that, in the whole range of parameters
$\{a,s,d\}$ defined by \ineq{sdangle}, both $m_{-}$ and $m_{+}$ are
monotonically decreasing functions of $s$. The quantity $Q$ is then
maximized over $s$ for the limiting value
\begin{equation}
\label{sopt} s \, = \, s^{\min} \, \equiv \, \frac{a + 1}2 \; .
\end{equation}
This value of $s$ corresponds to three-mode pure Gaussian states in
which the state of the reduced bipartition $2|3$ is always
separable, as one should expect because the bipartite entanglement
is maximally concentrated in the states of the $1|2$ and $1|3$
reduced bipartitions. With the position \eq{sopt}, the quantity $D$
defined in \eq{unsacco} can be easily shown to be always negative.
Therefore, for both reduced CMs $\sig_{12}$ and $\sig_{13}$, the
Gaussian contangle is defined in terms of $m_{-}$. The latter, in
turn, acquires the simple form
\begin{equation}
\label{msmin} m_-(a,s^{\min},d) \, = \, \frac{1 + 3a + 2d}{3 + a - 2
d} \; .
\end{equation}
Consequently, the quantity $Q$ turns out to be an even and convex
function of $d$, and this fact entails that it is globally maximized
at the boundary
\begin{equation}
\label{dopto} |d| \, = \, d^{\max} \, \equiv \, \frac{a-1}2 \; .
\end{equation}
We finally have that
\begin{eqnarray}
\label{Qmax} Q^{\max} & \equiv & Q \left[a,s=s^{\min},d=\pm
d^{\max}\right] \nonumber \\
& = & \arcsinh^2 \sqrt{a^2-1} \; ,
\end{eqnarray}
which implies that in this case the sharing inequality \pref{CKWine}
is exactly saturated and the genuine tripartite entanglement is
consequently zero. In fact this case yields states with $a_2=a_1$
and $a_3=1$ (if $d=d^{\max}$), or $a_3=a_1$ and $a_2=1$ (if
$d=-d^{\max}$), {\ie}tensor products of a two-mode squeezed state
and a single-mode uncorrelated vacuum. Being $Q^{\max}$ from
\eq{Qmax} the global maximum of $Q$, \ineq{maxQ} holds true and the
monogamy inequality \pref{CKWine} is thus proven for any pure
three-mode Gaussian state, choosing either the Gaussian contangle
$G_\tau$ or the true contangle $E_\tau$ as measures of bipartite
entanglement \cite{contangle}.

The proof immediately extends to all mixed three-mode Gaussian
states $\sig$, but only if the bipartite entanglement is measured by
$G_\tau(\sig)$.\footnote{\sf If $\sig$ is decomposed into pure
non-Gaussian states, it is not known at the present stage whether
the CKW monogamy inequality \eq{CKWine} is satisfied by each of
them.} Let $\{\pi(d\sig^p_{m}), \sig^{p}_{m}\}$ be the ensemble of
pure Gaussian states minimizing the Gaussian convex roof in
Eq.~\pref{GaCoRo}; then, we have
\begin{eqnarray}
\label{convext} G_\tau^{i|(jk)}(\sig) &=& \int \pi (d\sig^p_{m})
G_\tau^{i|(jk)}(\sig^p_{m}) \nonumber \\
& \ge & \int \pi (d\sig^p_{m})
[G_\tau^{i|j}(\sig^p_{m}) + G_\tau^{i|k}(\sig^p_{m})] \\
&\ge& G_\tau^{i|j}(\sig) + G_\tau^{i|k}(\sig) \; , \nonumber
\end{eqnarray}
where we exploited the fact that the Gaussian contangle is convex by
construction. This concludes the proof of the CKW monogamy
inequality~\pref{CKWine} for all three-mode Gaussian states. \hfill
$\blacksquare$

The above proof, as more than once remarked, implies the
corresponding monogamy proof for all three-mode Gaussian states by
using the Gaussian tangle \eq{Gtau} as a bipartite entanglement
monotone. Monogamy of the Gaussian tangle for {\em all} $N$-mode
Gaussian states has been established in Sec.~\ref{SecHiro}
\cite{hiroshima}.

\subsection{Residual contangle and genuine tripartite entanglement}
\label{secresid}

The sharing constraint leads naturally to the definition of the {\em
residual contangle} as a quantifier of genuine tripartite
entanglement in three-mode Gaussian states, much in the same way as
in systems of three qubits \cite{CKW} (see Sec.~\ref{sec3tangle}).
However, at variance with the three-qubit case (where the residual
tangle of pure states is invariant under qubit permutations), here
the residual contangle is partition-dependent according to the
choice of the probe mode, with the obvious exception of the fully
symmetric states. A {\em bona fide} quantification of tripartite
entanglement is then provided by the {\em minimum} residual
contangle \cite{contangle}
\begin{equation}
\label{etaumin} E_\tau^{i|j|k}\equiv\min_{(i,j,k)} \left[
E_\tau^{i|(jk)}-E_\tau^{i|j}-E_\tau^{i|k}\right] \; ,
\end{equation}
where the symbol $(i,j,k)$ denotes all the permutations of the three
mode indexes. This definition ensures that $E_\tau^{i|j|k}$ is
invariant under all permutations of the modes and is thus a genuine
three-way property of any three-mode Gaussian state. We can adopt an
analogous definition for the minimum residual Gaussian contangle
$G_\tau^{res}$, sometimes referred to as {\em arravogliament}
\cite{contangle,3mpra,3mj} (see Fig.~\ref{figomini} for a pictorial
representation):
\begin{equation}
\label{gtaures} G_\tau^{res} \equiv
G_\tau^{i|j|k}\equiv\min_{(i,j,k)} \left[
G_\tau^{i|(jk)}-G_\tau^{i|j}-G_\tau^{i|k}\right] \; .
\end{equation}

\begin{figure}[t!]
\includegraphics[width=8.5cm]{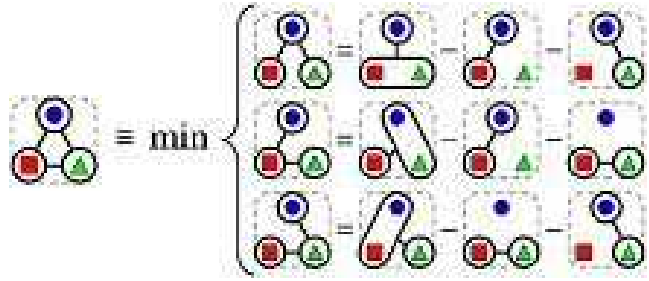}
\caption{Pictorial representation of \eq{gtaures}, defining the
residual Gaussian contangle $G_\tau^{res}$ of generic (nonsymmetric)
three-mode Gaussian states. $G_\tau^{res}$ quantifies the genuine
tripartite entanglement shared among mode $1$
(\textcolor[rgb]{0.00,0.00,1.00}{\ding{108}}), mode $2$
(\textcolor[rgb]{1.00,0.00,0.00}{\ding{110}}), and mode $3$
(\textcolor[rgb]{0.00,1.00,0.00}{\ding{115}}). The optimal
decomposition that realizes the minimum in \eq{gtaures} is always
the one for which the CM of the reduced state of the reference mode
has the smallest determinant.} \label{figomini}
\end{figure}

One can verify that
\begin{equation}
\label{refsat} (G_\tau^{i|(jk)} \, - \, G_\tau^{i|k}) \, - \,
(G_\tau^{j|(ik)} \, - \, G_\tau^{j|k}) \, \ge \, 0
\end{equation}
if and only if $a_i \ge a_j$, and therefore the absolute minimum in
\eq{etaumin} is attained by the decomposition realized with respect
to the reference mode $l$ of smallest local mixedness $a_l$, \ie for
the single-mode reduced state with CM of smallest determinant
(corresponding to the largest local purity  $\mu_{l}$).

\subsubsection{The residual Gaussian contangle is a Gaussian entanglement monotone}\label{secTauresMonotone}

A crucial requirement for the residual (Gaussian) contangle,
\eq{gtaures}, to be a proper measure of tripartite entanglement is
that it be nonincreasing under (Gaussian) LOCC.  The monotonicity of
the residual tangle was proven for three-qubit pure states in
Ref.~\cite{wstates}. In the CV setting we will now prove, based on
Ref.~\cite{contangle}, that for pure three-mode Gaussian states
$G_\tau^{res}$ is an entanglement monotone under tripartite Gaussian
LOCC, and that it is nonincreasing even under probabilistic
operations, which is a stronger property than being only monotone on
average.

We thus want to prove that
$$G_\tau^{res}(G_p(\sig^p)) \le G_\tau^{i|j|k}(\sig^p)\,,$$ where
$G_p$ is a pure Gaussian LOCC mapping pure Gaussian states
$\sig^{p}$ into pure Gaussian states \cite{giedkeqic03,nogo1}. Every
Gaussian LOCC protocol can be realized through a local operation on
one party only. Assume that the minimum in \eq{gtaures} is realized
for the probe mode $i$; the output of a pure Gaussian LOCC $G_p$
acting on mode $i$ yields a pure-state CM with $a_i' \le a_i$, while
$a_{j}$ and $a_{k}$ remain unchanged \cite{giedkeqic03}. Then, the
monotonicity of the residual Gaussian contangle $G_\tau^{res}$ under
Gaussian LOCC is equivalent to proving that $G_\tau^{res} =
G_\tau^{i|(jk)}-G_\tau^{i|j}-G_\tau^{i|k}$  is a monotonically
increasing function of $a_i$ for pure Gaussian states. One can
indeed show that the first derivative of $G_\tau^{res}$ with respect
to $a_i$, under the further constraint $a_i \le a_{j,k}$, is
globally minimized for $a_i=a_j=a_k \equiv a$, {\em i.e.}~for a
fully symmetric state. It is easy to verify that this minimum is
always positive for any $a >1$, because in fully symmetric states
the residual contangle is an increasing function of the local
mixedness $a$ (previously tagged as $b$, see
Sec.~\ref{SecMonoFulSym}). Hence the monotonicity of $G_\tau^{res}$,
\eq{gtaures}, under Gaussian LOCC for {\em all} pure three-mode
Gaussian states is finally proven. \hfill $\blacksquare$

Therefore, we have established the following \cite{contangle}.

\medskip

\begin{itemize}
\item[\ding{226}]
 \noindent{\rm\bf Monotonicity of the residual Gaussian contangle under Gaussian LOCC.}
{\it The residual Gaussian contangle $G_\tau^{res}$ is a proper and
computable measure of genuine multipartite  (specifically,
tripartite) entanglement in three-mode Gaussian states, being an
entanglement monotone under Gaussian LOCC.}
\smallskip
\end{itemize}

It is worth noting that the {\em minimum} in \eq{gtaures}, that at
first sight might appear a redundant (or artificial) requirement, is
physically meaningful and mathematically necessary. In fact, if one
chooses to fix a reference partition, or to take {\eg}the maximum
(and not the minimum) over all possible mode permutations in
\eq{gtaures}, the resulting ``measure'' is not monotone under
Gaussian LOCC and thus is definitely {\em not} a measure of
tripartite entanglement.

\subsection{Tripartite entanglement of pure three-mode Gaussian
states}\label{sec3purent}

We now work out in detail an explicit application, by describing the
complete procedure to determine the genuine tripartite entanglement
in a {\em pure} three-mode Gaussian state with a completely general
(not necessarily in standard form) CM $\sig^p$, as presented in
Ref.~\cite{3mpra}.

\begin{description}

\item[{\rm (i)} \it Determine the local purities] The
state is globally pure ($\det\sig^p = 1$). The only quantities
needed for the computation of the tripartite entanglement are
therefore the three local mixednesses $a_l$, defined by \eq{al}, of
the single-mode reduced states $\sig_l,\,l=1,2,3$ [see \eq{CM}].
Notice that the global CM $\sig^p$ needs not to be in the standard
form of \eq{cm3tutta}, as the single-mode determinants are local
symplectic invariants. From an experimental point of view, the
parameters $a_l$ can be extracted from the CM using the homodyne
tomographic reconstruction of the state \cite{homotomo}; or they can
be directly measured with the aid of single photon detectors
\cite{fiurasek04,wenger04}.

\item[{\rm (ii)} \it Find the minimum] From \eq{refsat}, the minimum in the definition
\pref{gtaures} of the residual Gaussian contangle $G_\tau^{res}$ is
attained in the partition where the bipartite entanglements are
decomposed choosing as probe mode $l$ the one in the single-mode
reduced state of smallest local mixedness $a_l \equiv a_{min}$.

\item[{\rm (iii)} \it Check range and compute] Given the mode
with smallest local mixedness $a_{min}$ (say, for instance, mode
$1$) and the parameters $s$ and $d$ defined in Eqs.~{\rm(\ref{s1},
\ref{d1})}, if $a_{min}=1$ then mode $1$ is uncorrelated from the
others: $G_\tau^{res}=0$. If, instead, $a_{min}>1$ then
\begin{equation}
\label{gtaurespur} G_\tau^{res} (\sig^p) =
\arcsinh^2\!\Big[\sqrt{a_{min}^2-1}\Big] - Q(a_{min},s,d) \; ,
\end{equation}
with $Q \equiv G_\tau^{1|2} + G_\tau^{1|3}$ defined by
Eqs.~{\rm(\ref{Qglems}, \ref{unsacco})}. Note that if
$d<-(a_{min}^2-1)/4s$ then $G_\tau^{1|2}=0$. Instead, if
$d>(a_{min}^2-1)/4s$ then $G_\tau^{1|3}=0$. Otherwise, all terms in
 \eq{gtaures} are nonvanishing.
\end{description}

\smallskip

\begin{figure}[t!]
\includegraphics[width=11cm]{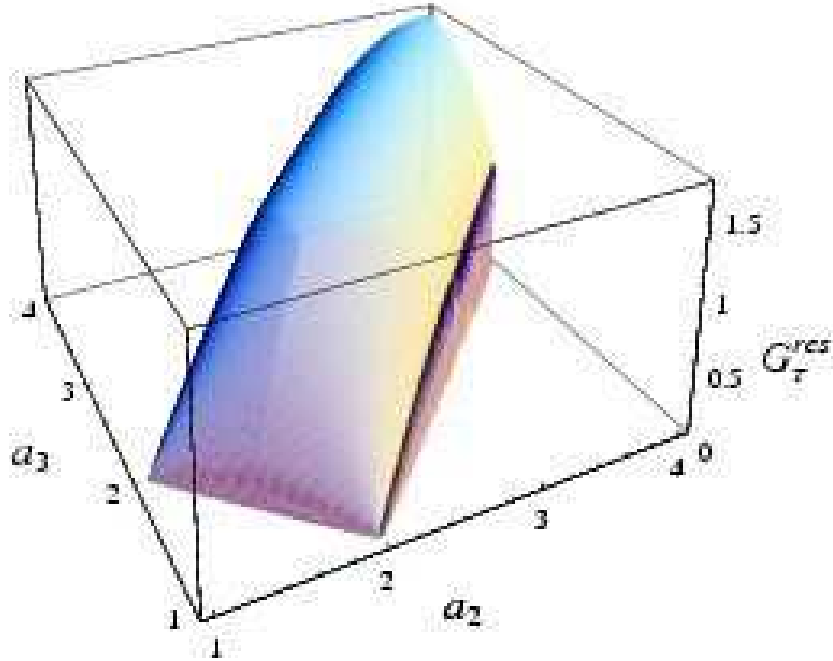}%
\caption{Three-dimensional plot of the residual Gaussian contangle
$G_\tau^{res}(\sig^p)$ in pure three-mode Gaussian states $\sig^p$,
determined by the three local mixednesses $a_l$, $l=1,2,3$. One of
the local mixednesses is kept fixed ($a_1=2$). The remaining ones
vary constrained by the triangle inequality \pref{triangle}, as
depicted in Fig.~\ref{figangle}{\rm (b)}. The explicit expression of
$G_\tau^{res}$ is given by \eq{gtaurespur}. See text for further
details.}\label{figsupposta}
\end{figure}

The residual Gaussian contangle \eq{gtaures} in generic pure
three-mode Gaussian states is plotted in Fig.~\ref{figsupposta} as a
function of $a_2$ and $a_3$, at constant $a_1=2$. For fixed $a_1$,
it is interesting to notice that $G_\tau^{res}$ is maximal for
$a_2=a_3$, {\ie}for bisymmetric states (see Fig.~\ref{figbasset}).
Notice also how the residual Gaussian contangle of these bisymmetric
pure states has a cusp for $a_1=a_2=a_3$. In fact, from \eq{refsat},
for $a_2=a_3 < a_1$ the minimum in \eq{gtaures} is attained
decomposing with respect to one of the two modes $2$ or $3$ (the
result is the same by symmetry), while for $a_2=a_3 > a_1$ mode $1$
becomes the probe mode.

\subsubsection{Residual contangle and distillability of mixed states}

For generic {\em mixed} three-mode  Gaussian states, a quite
cumbersome analytical expression for the $1|2$ and $1|3$ Gaussian
contangles may be written, which explicitly solves the minimization
over the angle $\theta$ in \eq{mfunc}. On the other hand, the
optimization appearing in the computation of the $1|(23)$ bipartite
Gaussian contangle [see \eq{tau}] has to be solved only numerically.
However, exploiting techniques like the unitary localization of
entanglement described in Chapter \ref{ChapUniloc}, and results like
that of \eq{etausym2}, closed expressions for the residual Gaussian
contangle can be found as well in relevant classes of mixed
three-mode Gaussian states endowed with some symmetry constraints.
Interesting examples of these states and the investigation of their
physical properties will be discussed in Sec.~\ref{secstructex}.

As an additional remark, let us recall that, although the
entanglement of Gaussian states is always distillable with respect
to $1\times N$ bipartitions \cite{werewolf} (see
Sec.~\ref{SecPPTG}), they can exhibit bound entanglement in $1
\times 1 \times  1$ tripartitions \cite{kraus}. In this case, the
residual Gaussian contangle cannot detect
 tripartite PPT entangled states. For example, the
residual Gaussian contangle in three-mode biseparable Gaussian
states (Class $4$ of Ref.~\cite{kraus}) is always zero, because
those bound entangled states are separable with respect to all
$1\times 2$ bipartitions of the modes. In this sense we can
correctly regard the residual Gaussian contangle as an estimator of
{\em distillable} tripartite entanglement, being strictly nonzero
only  on fully inseparable three-mode Gaussian states (Class 1 in
the classification of Sec.~\ref{secbarbie}).

\section{Sharing structure of tripartite entanglement: {\em promiscuous} Gaussian states}\label{secstructex}

We are now in the position to analyze the sharing structure of CV
entanglement in three-mode Gaussian states by taking the residual
Gaussian contangle as a measure of tripartite entanglement, in
analogy with the study done for three qubits \cite{wstates} using
the residual tangle \cite{CKW} (see Sec.~\ref{sec3tangle}).

The first task we face is that of identifying the three-mode
analogues of the two inequivalent classes of fully inseparable
three-qubit states, the GHZ state \cite{GHZ}, \eq{ghzstates}, and
the $W$ state \cite{wstates}, \eq{wstates}. These states are both
pure and fully symmetric, {\ie}invariant under the exchange of any
two qubits. On the one hand, the GHZ state possesses maximal
tripartite entanglement, quantified by the residual tangle
\cite{CKW,wstates}, with zero couplewise entanglement in any reduced
state of two qubits reductions. Therefore its entanglement is very
fragile against the loss of one or more subsystems. On the other
hand, the $W$ state contains the maximal two-party entanglement in
any reduced state of two qubits \cite{wstates} and is thus maximally
robust against decoherence, while its tripartite residual tangle
vanishes.

\subsection{CV finite-squeezing GHZ/{\em W} states}\label{secghzw}

To define the CV counterparts of the three-qubit states
$\ket{\psi_{\rm GHZ}}$ and $\ket{\psi_{W}}$, one must start from the
fully symmetric (generally mixed) three-mode CM $\sig_s$ of the form
$\sig_{\alp^3}$, \eq{fscm}. Surprisingly enough, in symmetric
three-mode Gaussian states, if one aims at maximizing, at given
single-mode mixedness $a\equiv\sqrt{\det\gr\alpha}$, either the
bipartite entanglement $G_\tau^{i|j}$ in any two-mode reduced state
({\em i.e.}~aiming at the CV $W$-like state), or the genuine
tripartite entanglement $G_\tau^{res}$ ({\em i.e.}~aiming at the CV
GHZ-like state), one finds {\em the same}, unique family of states.
They are exactly the {\em pure}, fully symmetric three-mode Gaussian
states (three-mode squeezed states) with CM $\sig^{p}_{s}$  of the
form $\sig_{\alp^3}$, \eq{fscm}, with $\gr\alpha=a \id_2$,
$\gr\varepsilon={\rm diag}\{e^+,\,e^-\}$ and
\begin{equation}
\label{epmfulsym} e^\pm = \frac{a^2-1 \pm \sqrt{\left(a^2 - 1\right)
\left(9 a^2 - 1\right)}}{4a} \; ,
\end{equation}
where we have used \eq{fspure} ensuring the global purity of the
state. In general, we have studied the entanglement scaling in fully
symmetric (pure) $N$-mode Gaussian states by means of the unitary
localization in Sec.~\ref{SecScal}. It is in order to mention that
these states were previously known in the literature as CV
``GHZ-type'' states \cite{network,vloock03}, as in the limit of
infinite squeezing ($a \rightarrow \infty$), they approach the
proper (unnormalizable) continuous-variable GHZ state $\int dx
\ket{x,x,x}$, a simultaneous eigenstate of total momentum
$\hat{p}_1+\hat{p}_2+\hat{p}_3$ and of all relative positions
$\hat{q}_i - \hat{q}_j$ ($i,j=1,2,3$), with zero eigenvalues
\cite{cvghz}.

For any finite squeezing (equivalently, any finite local mixedness
$a$), however, the above entanglement sharing study leads ourselves
to re-baptize these states as ``CV GHZ/$W$ states''
\cite{contangle,3mpra,3mj}, and denote their CM by $\sig_{s}^{_{{\rm
GHZ}/W}}$.

The residual Gaussian contangle of GHZ/$W$ states with finite
squeezing takes the simple form (see Sec.~\ref{SecMonoFulSym})
\begin{equation}
\label{gresghzw}
\begin{split}
G_\tau^{res}(\sig_{s}^{_{{\rm GHZ}/W}})&= \arcsinh^2\!\left[\sqrt{a^2 -1}\right] \\
&- \frac{1}{2} \log^2\!\left[\frac{3 a^2 - 1 -\sqrt{9 a^4 - 10 a^2 +
1}}{2}\right]\, .
\end{split}
\end{equation}
It is straightforward to see that $G_\tau^{res}(\sig_{s}^{_{{\rm
GHZ}/W}})$ is nonvanishing as soon as $a>1$. Therefore, the GHZ/$W$
states belong to the class of fully inseparable three-mode states
\cite{kraus,network,vanlokfortshit,vloock03} (Class 1, see
Sec.~\ref{secbarbie}). We finally recall that in a GHZ/$W$ state the
residual Gaussian contangle $G_\tau^{res}$ \eq{gtaures} coincides
with the true residual contangle $E_\tau^{1|2|3}$ \eq{etaumin}. This
property clearly holds because the Gaussian pure-state decomposition
is the optimal one in every bipartition, due to the fact that the
global three-mode state is pure and the reduced two-mode states are
symmetric (see Sec.~\ref{SecEOFGauss}).

\subsection{{\em T} states with zero reduced bipartite entanglement}\label{sectstates}

The peculiar nature of entanglement sharing in  CV GHZ/$W$ states is
further confirmed by the following observation. If one requires
maximization of the $1 \times 2$ bipartite Gaussian contangle
$G_\tau^{i|(jk)}$ under the constraint of separability of all the
reduced two-mode states (like it happens in the GHZ state of three
qubits), one finds a class of symmetric mixed states characterized
by being three-mode Gaussian states of partial minimum uncertainty
(see Sec.~\ref{SecSympHeis}). They are in fact characterized by
having their smallest symplectic eigenvalue equal to $1$, and
represent thus the three-mode generalization of two-mode symmetric
GLEMS (introduced in Sec.~\ref{SecGmemsGlems}).

We will name these states {\em $T$ states}, with $T$ standing for
{\em tripartite} entanglement only \cite{contangle,3mpra,3mj}. They
are described by a CM $\sig_s^T$ of the form \eq{fscm}, with
$\gr\alpha=a \id_2$, $\gr\varepsilon={\rm diag}\{e^+,\,e^-\}$ and
\begin{eqnarray}
\label{epmtstat} e^+  & = & \frac{a^2 -5 + \sqrt{9  a^2 \left(a^2 -
2\right) + 25}}{4a}
\; , \nonumber \\
e^- & = & \frac{5-9a^2 + \sqrt{9  a^2 \left(a^2 - 2\right) + 25}}{12
a} \; .
\end{eqnarray}
The $T$ states, like the GHZ/$W$ states, are determined only by the
local mixedness $a$, are fully separable for $a=1$, and fully
inseparable for $a>1$. The residual Gaussian contangle \eq{gtaures}
can be analytically computed for these mixed states as a function of
$a$. First of all one notices that, due to the complete symmetry of
the state, each mode can be chosen indifferently to be the reference
one in \eq{gtaures}. Being the $1 \times 1$ entanglements all zero
by construction, $G_\tau^{res} = G_\tau^{i|(jk)}$. The $1 \times 2$
bipartite Gaussian contangle can be in turn obtained exploiting the
unitary localization procedure (see Chapter \ref{ChapUniloc} and
Fig.~\ref{figbasset}). Let us choose mode $1$ as the probe mode and
combine modes $2$ and $3$ at a 50:50 beam-splitter, a local unitary
operation with respect to the bipartition $1|(23)$ that defines the
transformed modes $2'$ and $3'$. The CM $\sig_s^{T'}$ of the state
of modes $1$, $2'$, and $3'$ is then written in the following block
form:
\begin{equation}
\label{sigtprim} \sig_s^{T'} = \left(\begin{array}{ccc}
\sig_{1} & \eps_{12'} & {\bf 0} \\
\eps_{12'}^{\sf T} & \sig_{2'} & {\bf 0} \\
{\bf 0} & {\bf 0} & \sig_{3'} \\
\end{array}\right) \; ,
\end{equation}
where mode $3'$ is now disentangled from the others. Thus
\begin{equation}
\label{tlocato} G_\tau^{1|(23)}(\sig_s^T) =
G_\tau^{1|2'}(\sig_s^{T'}) \; .
\end{equation}
Moreover, the reduced CM $\sig_{12'}$ of modes $1$ and $2'$ defines
a nonsymmetric GLEMS, \eq{glems}, with
\begin{eqnarray*}
\det\sig_1 &=&a^2\,, \\
\det\sig_{2'} &=& \frac{1}{6} \left(3 a^2 +
\sqrt{9 \left(a^2 - 2\right) a^2 + 25} - 1\right)\,, \\
\det\sig_{12'} &=& \frac{1}{2} \left(3 a^2 - \sqrt{9 \left(a^2 -
2\right) a^2 + 25} + 3\right)\,,
\end{eqnarray*}
and we have shown that the Gaussian contangle (and the whole family
of Gaussian entanglement measures, Sec.~\ref{SecGEMS}) is computable
in two-mode GLEMS, via \eq{mglems}. After some algebra, one finds
the complete expression of $G_\tau^{res}$ for $T$ states:
\begin{eqnarray}
\label{grest} G_\tau^{res}(\sig_s^T) &=& \arcsinh^2 \Bigg\{ \bigg[25
R -9 a^4 + 3 R a^2
+ 6 a^2  -109 \nonumber \\
&-& \Big(81 a^8 - 432 a^6 + 954 a^4 -
1704 a^2 +2125 \nonumber \\
&-& \left(3 a^2 - 11\right) \left(3 a^2 - 7\right) \left(3 a^2 +
5\right)
R\Big)^{\frac12}\sqrt{2}\bigg]^{\frac12} \nonumber \\
&\times& \left[{18 \left(3 a^2 - R +
3\right)}\right]^{-\frac12}\Bigg\} \; ,
\end{eqnarray}
with $R \equiv \sqrt{9 a^2 (a^2 - 2) + 25}$.

What is remarkable about $T$ states is that their tripartite
Gaussian contangle, \eq{grest}, is strictly smaller than the one of
the GHZ/$W$ states, \eq{gresghzw}, for any fixed value of the local
mixedness $a$, that is, for any fixed value of the only parameter
(operationally related to the squeezing of each single mode) that
completely determines the CMs of both families of states up to local
unitary operations.\footnote{\sf Notice that this result cannot be
an artifact caused by restricting to pure Gaussian decompositions
only in the definition \eq{gtaures} of the residual Gaussian
contangle. In fact, for $T$ states the relation
$G_\tau^{res}(\sig_s^T) \ge E_\tau^{res}(\sig_s^T)$ holds due to the
symmetry of the reduced two-mode states, and to the fact that the
unitarily transformed state of modes $1$ and $2'$ is mixed and
nonsymmetric.} This hierarchical behavior of the residual Gaussian
contangle in the two classes of states is illustrated in
Fig.~\ref{figatua}. The crucial consequences of this result for the
structure of the entanglement trade-off in Gaussian states will be
discussed further in the next subsection.
\begin{figure}[t!]
\includegraphics[width=10.5cm]{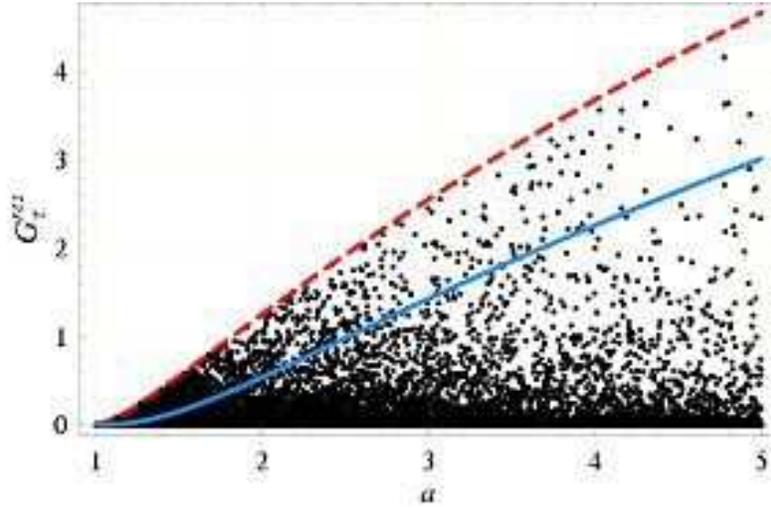}
\caption{Plot, as a function of the single-mode mixedness $a$, of
the tripartite residual Gaussian contangle $G_\tau^{res}$
\eq{gresghzw} in the CV GHZ/$W$ states (dashed red line); in the $T$
states \eq{grest} (solid blue line); and in 50 000 randomly
generated mixed symmetric three-mode Gaussian states of the form
\eq{fscm} (dots). The GHZ/$W$ states, that maximize any bipartite
entanglement, also achieve maximal genuine tripartite quantum
correlations, showing that CV entanglement distributes in a
promiscuous way in symmetric Gaussian states. Notice also how all
random mixed states have a nonnegative residual Gaussian contangle.
This confirms the results presented in Ref.~\cite{contangle}, and
discussed in detail and extended in Sec.~\ref{secmono}, on the
strict validity of the CKW monogamy inequality for CV entanglement
in three-mode Gaussian states.} \label{figatua}
\end{figure}

\subsection{Promiscuous continuous-variable entanglement
sharing} \label{secpromis}

The above results, pictorially illustrated in Fig.~\ref{figatua},
lead to the conclusion that in symmetric three-mode Gaussian states,
when there is no bipartite entanglement in the two-mode reduced
states (like in $T$ states) the genuine tripartite entanglement is
not enhanced, but frustrated. More than that, if there are maximal
quantum correlations in a three-party relation, like in GHZ/$W$
states, then the two-mode reduced states of any pair of modes are
maximally entangled mixed states.

These findings, unveiling a major difference between
discrete-variable (mainly qubits) and continuous-variable systems,
establish the {\em promiscuous} nature of CV entanglement sharing in
symmetric Gaussian states \cite{contangle}. Being associated with
degrees of freedom with continuous spectra, states of CV systems
need not saturate the CKW inequality to achieve maximum couplewise
correlations, as it was instead the case for $W$ states of qubits,
\eq{wstates}. In fact, the following holds.

\medskip

\begin{itemize}
\item[\ding{226}]
 \noindent{\rm\bf Promiscuous entanglement in continuous-variable GHZ/{\em W} three-mode Gaussian states.}
{\it  Without violating the monogamy constraint \ineq{CKWine}, pure
symmetric three-mode Gaussian states are maximally three-way
entangled and, at the same time, possess the maximum possible
entanglement between any pair of modes in the corresponding two-mode
reduced states. The two entanglements are mutually enhanced.}
\smallskip
\end{itemize}

 The notion of
``promiscuity'' basically means that bipartite and genuine
multipartite (in this case tripartite) entanglement are increasing
functions of each other, while typically in low-dimensional systems
like qubits only the opposite behavior is compatible with monogamy
(see Sec.~\ref{SecPisa}). The promiscuity of entanglement in
three-mode GHZ/$W$ states is, however, {\em partial}. Namely they
exhibit, with increasing squeezing, unlimited tripartite
entanglement (diverging in the limit $a \rightarrow \infty$) and
nonzero, accordingly increasing bipartite entanglement between any
two modes, which nevertheless stays finite even for infinite
squeezing. Precisely, from \eq{gresghzw}, it saturates to the value
\begin{equation}\label{gredmaxghzw}
G_\tau^{i|j}(\sig_s^{_{{\rm GHZ}/W}},\,a \rightarrow \infty) =
\frac{\log^2{3}}{4} \approx 0.3\,.
\end{equation}
We will show in the next Chapter that in CV systems with more than
three modes, entanglement can be distributed in an {\em infinitely}
promiscuous way.

More remarks are in order concerning the tripartite case. The
structure of entanglement in GHZ/$W$ states is such that, while
being maximally three-party entangled, they are also maximally
robust against the loss of one of the modes. This preselects GHZ/$W$
states also as optimal candidates for carrying quantum information
through a lossy channel, being intrinsically less sensitive to
decoherence effects.  In the next Section, we will exactly analyze
the effect of environmental decoherence on three-mode Gaussian
states and the sharing structure of noisy GHZ/$W$ states,
investigating the persistency of a promiscuous structure in the
presence of thermal noise. The usefulness of GHZ/$W$ states for CV
quantum communication will be analyzed in Sec.~\ref{SecTelepoppy}.

As an additional comment, let us mention that, quite naturally, not
all three-mode Gaussian states (in particular nonsymmetric states)
are expected to exhibit a promiscuous entanglement sharing. We will
provide in Sec.~\ref{secbas} an example of three-mode states with
not so strong symmetry constraints, where the entanglement sharing
structure is more traditional, \ie with bipartite and tripartite
quantum correlations being mutually competitors.

\section{Promiscuous entanglement versus noise and
asymmetry}\label{secstructexmix}

\subsection{Decoherence of three-mode states and decay of tripartite
entanglement} \label{decoherence}

Here we analyze, following Ref.~\cite{3mpra}, the action of
decoherence on tripartite entangled Gaussian states, studying the
decay of the residual contangle. The GHZ/$W$ states of
Sec.~\ref{secghzw} are shown to be maximally robust against
decoherence effects.

\subsubsection{Basics of decoherence theory for Gaussian
states}\label{Secdeco}

Among their many special features,  Gaussian states allow remarkably
for a straightforward, analytical treatment of decoherence,
accounting for the most common situations encountered in the
experimental practice (like fibre propagations or cavity decays) and
even for more general, `exotic' settings (like ``squeezed'' or
common reservoirs) \cite{serafozzijob05}. This agreeable feature,
together with the possibility --- extensively exploited in this
Dissertation
--- of exactly computing several interesting benchmarks for such
states, make Gaussian states a useful theoretical reference for
investigating the effect of decoherence on the information and
correlation content of quantum states.

In this Section, we will explicitly show how the decoherence of
three-mode Gaussian states may be exactly studied for any finite
temperature, focusing on the evolution of the residual Gaussian
contangle as a measure of tripartite correlations. The results here
obtained will be recovered in Sec.~\ref{qnoise}, and applied to the
study of the effect of decoherence on multiparty protocols of CV
quantum communication with the classes of states we are addressing,
thus completing the present analysis by investigating its precise
operational consequences.

In the most customary and relevant instances, the bath interacting
with a set of $N$ modes can be modeled by $N$ independent continua
of oscillators, coupled to the bath through a quadratic Hamiltonian
$H_{int}$ in the rotating wave approximation, reading \be
H_{int}=\sum_{i=1}^{N} \int
v_i(\omega)[a_i^{\dag}b_i(\omega)+a_ib_i^{\dag}(\omega)] \,{\rm
d}\omega \, , \label{coupling} \ee where $b_i(\omega)$ stands for
the annihilation operator of the $i$-th continuum's mode labeled by
the frequency $\omega$, whereas $v_i(\omega)$ represents the
coupling of such a mode to the mode $i$ of the system (assumed, for
simplicity, to be real). The state of the bath is assumed to be
stationary. Under the Born-Markov approximation,\footnote{\sf Let us
recall that such an approximation requires small couplings (so that
the effect of $H_{int}$ can be truncated to the first order in the
Dyson series) and no memory effects, in that the `future state' of
the system depends only on its `present state'.} the Hamiltonian
$H_{int}$ leads, upon partial tracing over the bath, to the
following master equation for the $N$ modes of the system (in
interaction picture) \cite{carmichael} \be \dot\varrho\;  = \;
\sum_{i=1}^{N} \frac{\gamma_i}{2}\Big(n_i \: L[a_i^{\dag}]\varrho
+(n_i+1)\:L[a_i]\varrho \Big) \, , \label{rhoev} \ee where the dot
stands for time-derivative, the Lindblad superoperators are defined
as\index{Lindblad superoperators} $L[\hat{o}]\varrho \equiv  2
\hat{o}\varrho \hat{o}^{\dag} - \hat{o}^{\dag} \hat{o}\varrho
-\varrho \hat{o}^{\dag} \hat{o}$, the couplings are $\gamma_i=2\pi
v_i^{2}(\omega_i)$, whereas the coefficients $n_i$ are defined in
terms of the correlation functions $\langle
b_i^{\dag}(\omega_i)b_i(\omega_i) \rangle = n_i$, where averages are
computed over the state of the bath and $\omega_i$ is the frequency
of mode $i$. Notice that $n_i$ is the number of thermal photons
present in the reservoir associated to mode $i$, related to the
temperature $T_i$ of the reservoir by the Bose statistics at null
chemical potential: \be n_i =
\frac{1}{{\exp}({\frac{\omega_i\hbar}{kT_{i}}})-1} \; . \label{bose}
\ee In the derivation, we have also assumed $\langle
b_i(\omega_i)b_i(\omega_i) \rangle = 0$, holding for a bath at
thermal equilibrium. We will henceforth refer to a ``homogeneous''
bath in the case $n_{i}=n$ and $\gamma_i=\gamma$ for all $i$.

Now, the master equation \pref{rhoev} admits a simple and physically
transparent representation as a diffusion equation for the
time-dependent characteristic function of the system $\chi(\xi,t)$
\cite{carmichael},  \be
\dot{\chi}(\xi,t)=-\sum_{i=1}^{N}\frac{\gamma_i}{2}\Bigg[ (x_i\;
p_i) {\partial{x_i}\choose \partial{p_i}} + (x_i\; p_i) \omega^{\sf
T}\gr{\sigma}_{i\infty}\omega {x_i\choose p_i} \Bigg] \chi(\xi,t) \,
, \label{fokpla} \ee   where $\xi\equiv(x_1,p_1,\ldots,x_N,p_N)$ is
a phase-space vector and $\sig_{i\infty}=\,{\rm
diag}\,(2n_i+1,2n_i+1)$ (for a homogeneous bath), while $\omega$ is
the symplectic form, \eq{symform}. The right hand side of the
previous equation contains a deterministic drift term, which has the
effect of damping the first moments to zero on a time scale of
$\gamma/2$, and a diffusion term with diffusion matrix
$\sig_{\infty}\equiv\oplus_{i=1}^{N}\sig_{i\infty}$. The essential
point here is that \eq{fokpla} preserves the Gaussian character of
the initial state, as can be straightforwardly checked for any
initial CM $\sig_0$ by inserting the Gaussian characteristic
function $\chi(\xi,t)$, \be \chi(\xi,t) = \,{\rm e}^{-\frac12
\xi^{\sf T}\Omega^{\sf T}\sig(t)\Omega\xi +i X^{\sf T}\Gamma_t\Omega
\xi}\;, \ee where $X$ are generic initial first moments,
$\sig(t)\equiv\Gamma_t^2\sig_0+(\id-\Gamma_t^2)\sig_{\infty}$, and
$\Gamma_{t}\equiv\oplus_{i}{\rm e}^{-\gamma_i t/2}\id_2$, into the
equation and verifying that it is indeed a solution. Notice that,
for a homogeneous bath, the diagonal matrices $\Gamma_t$ and
$\sig_{\infty}$ (providing a full characterization of the bath) are
both proportional to the identity. In order to keep track of the
decay of correlations of Gaussian states, we are interested in the
evolution of the initial CM $\sig_{0}$ under the action of the bath
which, recalling our previous Gaussian solution, is just described
by \be \sig (t) = \Gamma_{t}^2\sig_0+(\id-\Gamma_t^2)\sig_{\infty}
\; .\label{cmevo} \ee This simple equation describes the dissipative
evolution of the CM of any initial state under the action of a
thermal environment and, at zero temperature, under the action of
``pure losses'' (recovered in the instance $n_i=0$ for
$i=1,\ldots,N$). It yields a basic, significant  example of
`Gaussian channel', {\em i.e.}~of a map mapping Gaussian states into
Gaussian states under generally nonunitary evolutions. Exploiting
\eq{cmevo} and our previous findings, we can now study the exact
evolution of the tripartite entanglement of Gaussian states under
the decoherent action of losses and thermal noise. For simplicity,
we will mainly consider homogeneous baths.

\subsubsection{Robustness of tripartite entangled states}

As a first general remark let us notice that, in the case of a
zero-temperature bath ($n=0$), in which decoherence is entirely due
to losses, the bipartite entanglement between any different
partition decays in time but persists for an infinite time. This is
a general property of Gaussian entanglement \cite{serafozzijob05}
under any multimode bipartition. The same fact is also true for the
genuine tripartite entanglement, quantified by the residual Gaussian
contangle. If $n\neq0$, a finite time does exist for which
tripartite quantum correlations disappear. In general, the two-mode
entanglement between any given mode and any other of the remaining
two modes vanishes before than the three-mode bipartite entanglement
between such a mode and the other two --- not surprisingly, as the
former quantity is, at the beginning, bounded by the latter because
of the CKW monogamy inequality  \pref{CKWine}.

The main issue addressed in this analysis consists in inspecting the
robustness of different forms of genuine tripartite entanglement,
previously introduced in Sec.~\ref{secstructex}. Notice that an
analogous question has been addressed in the qubit scenario, by
comparing the action of decoherence on the residual tangle of the
inequivalent sets of GHZ and $W$ states: $W$ states, which are by
definition more robust under subsystem erasure, proved more robust
under decoherence as well \cite{carvalho04}. In our instance, the
symmetric GHZ/$W$ states constitute a promising candidate for the
role of most robust Gaussian tripartite entangled states, as somehow
expected. Evidence supporting this conjecture is shown in
Fig.~\ref{decofig1}, where the evolution in different baths of the
tripartite entanglement of GHZ/$W$ states, \eq{gresghzw}, is
compared to that of symmetric $T$ states, \eq{grest} (at the same
initial entanglement). No fully symmetric states with tripartite
entanglement more robust than GHZ/$W$ states were found by further
numerical inspection. Quite remarkably, the promiscuous sharing of
quantum correlations, proper to GHZ/W states, appears to better
preserve genuine multipartite entanglement against the action of
decoherence.

\begin{figure}[t!]
\centering{
\includegraphics[width=10cm]{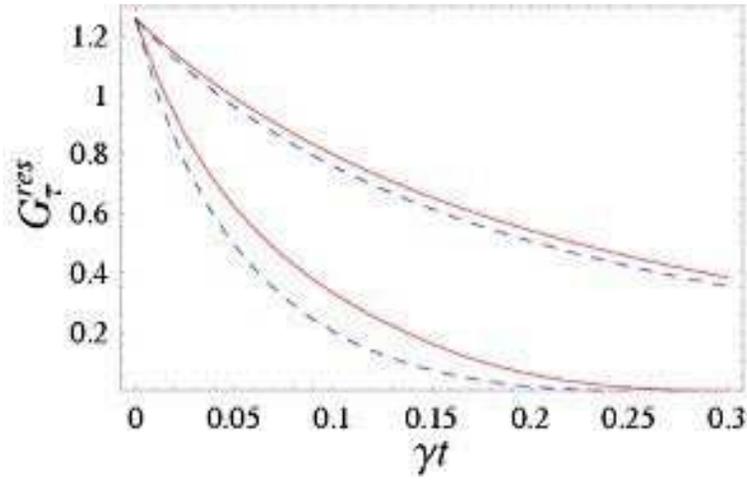}
\caption{Evolution of the residual Gaussian contangle
$G^{res}_{\tau}$ for GHZ/$W$ states with local mixedness $a=2$
(solid curves) and $T$ states with local mixedness $a=2.8014$
(dashed curves). Such states have equal initial residual contangle.
The uppermost curves refer to a homogeneous bath with $n=0$ (pure
losses), while the lowermost curves refer to a homogeneous bath with
$n=1$. As apparent, thermal photons are responsible for the
vanishing of entanglement at finite times.} \label{decofig1}}
\end{figure}

Notice also that, for a homogeneous bath and for all fully symmetric
and bisymmetric three-mode states, the decoherence of the global
{\em bipartite} entanglement of the state is the same as that of the
corresponding equivalent two-mode states (obtained through unitary
localization, see Fig.~\ref{figbasset}). Indeed, for any bisymmetric
state which can be localized by an orthogonal transformation (like a
beam-splitter), the unitary localization and the action of the
decoherent map of \eq{cmevo} commute, because
$\sig_{\infty}\propto\id$ is obviously preserved under orthogonal
transformations (note that the bisymmetry of the state is maintained
through the channel, due to the symmetry of the latter). In such
cases, the decoherence of the bipartite entanglement of the original
three-mode state (with genuine tripartite correlations) is exactly
equivalent to that of the corresponding initial two-mode state
obtained by unitary localization. This equivalence breaks down, even
for GHZ/$W$ states which can be localized through an (orthogonal)
beam-splitter transformation, for non homogeneous baths, {\em
i.e.~}if the thermal photon numbers $n_i$ related to different modes
are different --- which is the case for different temperatures $T_i$
or for different frequencies $\omega_i$, according to \eq{bose} ---
or if the couplings $\gamma_i$ are different. In this instance let
us remark that the unitary localization could provide a way to cope
with decoherence, limiting its hindering effect on entanglement. In
fact, let us suppose that a given amount of genuine tripartite
entanglement is stored in a symmetric (unitarily localizable)
three-mode state and is meant to be exploited, at some (later) time,
to implement tripartite protocols. During the period going from its
creation to its actual use such an entanglement decays under the
action of decoherence. Suppose the three modes involved in the
process do not decay with the same rate (different $\gamma_i$) or
under the same amount of thermal photons (different $n_i$), then the
obvious, optimal way to shield tripartite entanglement is
concentrating it, by unitary localization, in the two least
decoherent modes. The entanglement can then be redistributed among
the three modes by a reversal unitary operation, just before
employing the state. Of course, the concentration and distribution
of entanglement require a high degree of non-local control on two of
the three-modes, which would not always be allowed in realistic
operating conditions.

As a final remark, let us mention that the {\em bipartite}
entanglement of GHZ/$W$ states  under $1 \times 2$  bipartitions,
decays slightly faster (in homogeneous baths with equal number of
photons) than that of an initial pure two-mode squeezed state  with
the same initial entanglement. In this respect, the multimode
entanglement is more fragile than the two-mode one, as the Hilbert
space exposed to decoherence which contains it is larger.

\subsection{Entanglement distribution in noisy GHZ/{\em W} states}
\label{secnoisyghzw}

We consider here the noisy version of the GHZ/$W$ states previously
introduced (Sec.~\ref{secghzw}), which are a family of mixed
Gaussian fully symmetric states, also called three-mode squeezed
thermal states \cite{3mcinese}. They result in general from the
dissipative evolution of pure GHZ/$W$ states in proper Gaussian
noisy channels, as shown in Sec.~\ref{decoherence}.
 Let us mention
that various properties of noisy three-mode Gaussian states have
already been addressed, mainly regarding their effectiveness in the
implementation of CV protocols \cite{pirandolo,paris05}. Here, based
on Ref.~\cite{3mj}, we focus on the multipartite entanglement
properties of noisy states. This analysis will allow us to go beyond
the set of pure states, thus gaining deeper insight into the role
played by realistic quantum noise in the sharing and
characterization of tripartite entanglement.

Noisy GHZ/$W$ states are described by a CM $\sig_s^{th}$ of the form
\eq{fscm}, with $\gr\alpha=a \id_2$, $\gr\varepsilon={\rm
diag}\{e^+,\,e^-\}$ and
\begin{equation}\label{epmthermal}
e^\pm = \frac{a^2 - n^2 \pm \sqrt{\left(a^2 - n^2\right) \left(9 a^2
- n^2\right)}}{4 a}\,,
\end{equation}
where $a \ge n$ to ensure the physicality of the state. Noisy
GHZ/$W$ states have a completely degenerate symplectic spectrum
(their symplectic eigenvalues being all equal to $n$) and represent
thus, somehow, the three-mode generalization of two-mode squeezed
thermal states (also known as symmetric GMEMS, states of maximal
negativity at fixed purities, see Sec.~\ref{SecGmemsGlems}). The
state $\sig_s^{th}$ is completely determined by the local purity
$\mu_l=a^{-1}$ and by the global purity $\mu = n^{-3}$. Noisy
GHZ/$W$ states reduce to pure GHZ/$W$ states ({\ie}three-mode
squeezed {\em vacuum} states) for $n=1$.

For ease of notation, let us replace the parameter $a$ with the
effective ``squeezing degree'' $s$, defined by \be\label{noiseffs} s
= \frac{1}{2} \sqrt{\frac{3 \left(3 a^2+\sqrt{9 a^4-10 n^2
a^2+n^4}\right)}{n^2}-5}\,,\ee whose physical significance will
become clear once the optical state engineering of noisy GHZ/$W$
will be described in Sec.~\ref{SecEngiGHZWnoisy}.

\subsubsection{Separability properties} Depending on the defining parameters $s$ and $n$, noisy GHZ/$W$
states can belong to three different separability classes
\cite{kraus} according to the classification of Sec.~\ref{secbarbie}
(and not to only two classes like the previously considered
examples). Namely, as explicitly computed in Ref.~\cite{3mcinese},
we have in our notation
\begin{eqnarray}
\hspace*{-1cm}
  s > \frac{\sqrt{9 n^4 - 2 n^2 + 9 +
          3 \left(n^2 - 1\right) \sqrt{9 n^4 + 14 n^2 + 9}}}{4 n}
          &\Rightarrow& \mbox{\it Class 1;} \label{termic1} \\
\hspace*{-1cm}  n < s \le \frac{\sqrt{9 n^4 - 2 n^2 + 9 +
          3 \left(n^2 - 1\right) \sqrt{9 n^4 + 14 n^2 + 9}}}{4 n}
          &\Rightarrow& \mbox{\it Class 4;}  \label{termic4}\\
\hspace*{-1cm}  s \le n &\Rightarrow& \mbox{\it Class
5.}\label{termic5}
\end{eqnarray}
States which fulfill \ineq{termic1} are fully inseparable (Class 1,
encoding genuine tripartite entanglement), while states that violate
it have a positive partial transpose with respect to all
bipartitions. However, as already mentioned in Sec.~\ref{secbarbie},
the PPT property does not imply separability. In fact, in the range
defined by \ineq{termic4}, noisy GHZ/$W$ states are three-mode
biseparable (Class 4), that is they exhibit tripartite {\em bound
entanglement}. This can be verified by showing, using the methods of
Ref.~\cite{kraus}, that such states cannot be written as a convex
combination of separable states. Finally, noisy GHZ/$W$ states that
fulfill \ineq{termic5} are fully separable (Class 5), containing no
entanglement at all.

\begin{figure}[t!]
\includegraphics[width=9.5cm]{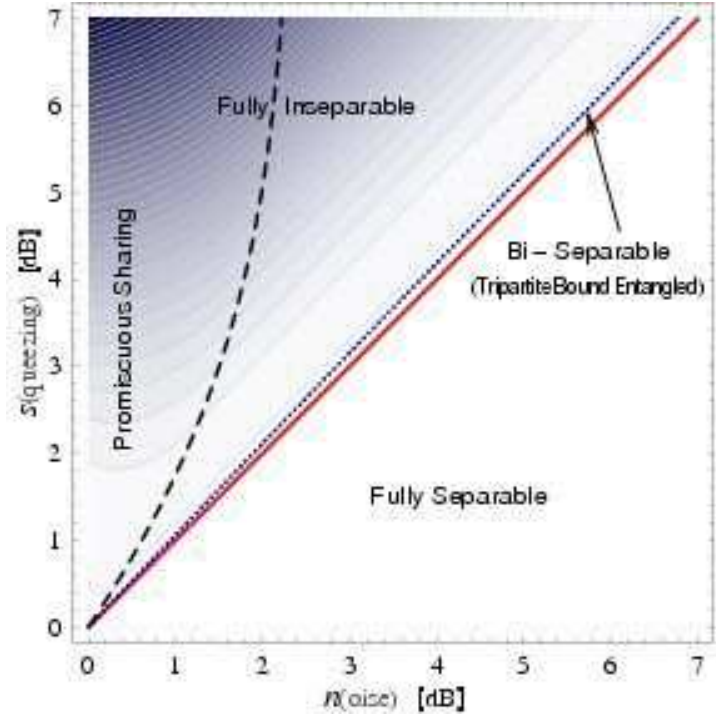}
\caption{Summary of separability and entanglement properties of
three-mode squeezed thermal states, or noisy GHZ/$W$ states, in the
space of the two parameters $n$ and $s$. The separability is
classified according to the scheme of Sec.~\ref{secbarbie}. Above
the dotted line the states are fully inseparable (Class 1); below
the solid line they are fully separable (Class 5). In the narrow
intermediate region, noisy GHZ/$W$ states are three-mode biseparable
(Class 4), {\ie}they exhibit tripartite bound entanglement. The
relations defining the boundaries for the different regions are
given in Eqs.~{\rm (\ref{termic1}--\ref{termic5})}. In the fully
inseparable region, the residual (Gaussian) contangle \eq{gtauresth}
is depicted as a contour plot, growing with increasing darkness from
$G_\tau^{res}=0$ (along the dotted line) to $G_\tau^{res}\approx
1.9$ (at $n=0$ dB, $s=7$ dB). On the left side of the dashed line,
whose expression is given by \eq{termprom}, not only genuine
tripartite entanglement is present, but also each reduced two-mode
bipartition is entangled. In this region, $G_\tau^{res}$ is strictly
larger than in the region where the two-mode reductions are
separable. This evidences the {\em promiscuous} sharing structure of
multipartite CV entanglement in symmetric, even mixed, three-mode
Gaussian states.} \label{figacunt}
\end{figure}

The tripartite residual Gaussian contangle \eq{gtaures}, which is
nonzero only in the fully inseparable region, can be explicitly
computed. In particular, the $1 \times 2$ Gaussian contangle
$G_\tau^{i|(jk)}$ is obtained following a similar strategy to that
employed for $T$ states (see Sec.~\ref{sectstates}). Namely, if one
performs a unitary localization on modes $2$ and $3$ that decouples
the transformed mode $3'$, one finds that the resulting equivalent
two-mode state of modes $1$ and $2'$ is symmetric. The bipartite
Gaussian contangle of the three-mode state follows then from
\eq{etausym2}. As for the two-mode Gaussian contangles
$G_\tau^{1|2}=G_\tau^{1|3}$, the same formula can be used, as the
reduced states are symmetric too. Finally one gets, in the range
defined by \ineq{termic1}, a tripartite entanglement given by
\cite{3mj}
\begin{eqnarray} \label{gtauresth}
  G_\tau^{res}(\sig_s^{th}) &=& \frac14 \log^2\!\left\{\frac{n^2 \left[4 s^4 + s^2 +4 -
          2 \left(s^2 - 1\right) \sqrt{4 s^4 + 10 s^2 + 4}\right]}{9 s^2} \right\}
  \nonumber \\
   &-& 2 \left[\max \left\{0, -\log \left(\frac{n \sqrt{s^2 +
                    2}}{\sqrt{3} s}\right)\right\}\right]^2\,,
\end{eqnarray}
and $G_\tau^{res}(\sig_s^{th})=0$ when \ineq{termic1} is violated.
For noisy GHZ/$W$ states, the residual Gaussian contangle
\eq{gtauresth} is still equal to the true one \eq{etaumin} (like in
the special instance of pure GHZ/$W$ states), thanks to the symmetry
of the two-mode reductions, and of the unitarily transformed state
of modes $1$ and $2'$.

\subsubsection{Sharing structure} The second term in \eq{gtauresth} embodies the sum of
the couplewise entanglement in the $1|2$ and $1|3$ reduced
bipartitions. Therefore, if its presence enhances the value of the
tripartite residual contangle (as compared to what happens if it
vanishes), then one can infer that entanglement sharing is
`promiscuous' in the (mixed) three-mode squeezed thermal Gaussian
states as well (`noisy GHZ/$W$' states). And this is exactly the
case, as shown in the contour plot of Fig.~\ref{figacunt}, where the
separability and entanglement properties of noisy GHZ/$W$ states are
summarized, as functions of the parameters $n$ and $s$ expressed in
decibels.\footnote{\label{notedb}\sf The noise expressed in decibels
(dB) is obtained from the covariance matrix elements via the formula
$N_{ij}(dB) = 10 \log_{10}(\sigma_{ij})$.} Explicitly, one finds
that for
\begin{equation}\label{nsoglia}
n \ge \sqrt{3}\,,
\end{equation}
the entanglement sharing can never be promiscuous, as the reduced
two-mode entanglement is zero for any (even arbitrarily large)
squeezing $s$. Otherwise, applying PPT criterion, one finds that for
sufficiently high squeezing bipartite entanglement is also present
in any two-mode reduction, namely
\begin{eqnarray}
  n<\sqrt{3}\,,\quad s > \frac{\sqrt{2} n}{\sqrt{3-n^2}}
          &\quad\Rightarrow\quad& \mbox{{\em promiscuous} sharing}\,. \label{termprom}
\end{eqnarray}

Evaluation of \eq{gtauresth}, as shown in Fig.~\ref{figacunt},
clearly demonstrates that the genuine tripartite entanglement
increases with increasing bipartite entanglement in any two-mode
reduction, unambiguously confirming that CV quantum correlations
distribute in a promiscuous way not only in pure
\cite{contangle,3mpra}, but also in {\em mixed} \cite{3mj} symmetric
three-mode Gaussian states. However, the global mixedness is prone
to affect this sharing structure, which is completely destroyed if,
as can be seen from \eq{nsoglia}, the global purity $\mu$ falls
below $1/(3\sqrt{3}) \approx 0.19245$. This purity threshold is
remarkably low: a really strong amount of global noise is necessary
to destroy the promiscuity of entanglement distribution.

\subsection{Basset hound states: a `traditional' sharing of entanglement} \label{secbas}

Let us finally consider an instance of tripartite entangled states
which are not fully symmetric, but merely bisymmetric pure Gaussian
states (in this specific case, invariant under the exchange of modes
$2$ and $3$, see the general definition in Sec.~\ref{SecSymm}).
Following the arguments summarized in Fig.~\ref{figbasset}, they
will be named {\em basset hound} states. Such states are denoted by
a CM $\sig_{B}^p$ of the form \eq{CM} for $N=3$, with
\begin{eqnarray}
  &   \sig_1 = a \id_2,\quad &
  \sig_2 = \sig_3 =  \left(\frac{a+1}{2}\right) \id_2,
  \label{bassigl} \\
  &   \eps_{23} = \left(\frac{a-1}{2}\right) \id_2,\quad &
  \eps_{12} = \eps_{13} = {\rm
  diag}\left\{\frac{\sqrt{a^2-1}}{\sqrt{2}},\,-\frac{\sqrt{a^2-1}}{\sqrt{2}}\right\}.
  \label{basseps}
\end{eqnarray}
They belong to a family of states introduced in
Ref.~\cite{telecloning} as resources for optimal CV telecloning
({\ie}cloning at distance, or equivalently teleportation to more
than one receiver) of single-mode coherent states. A more detailed
discussion of this protocol will be given in Sec.~\ref{sectlc}.

\subsubsection{Tripartite entanglement}
 From a qualitative point of view, basset hound states are fully
inseparable for $a>1$ and fully separable for $a=1$, as already
remarked in Ref.~\cite{telecloning}; moreover, the PPT criterion
(see Sec.~\ref{SecPPTG}) entails that the two-mode reduced state of
modes $2$ and $3$ is always separable. Following the guidelines of
Sec.~\ref{secresid}, the residual Gaussian contangle $G_\tau^{res}$
is easily computable. From \eq{refsat}, it follows that the minimum
in \eq{gtaures} is attained if one sets either mode $2$ or mode $3$
(indifferently, due to the bisymmetry) to be the probe mode. Let us
choose mode $3$; then we have
\begin{equation}\label{gtauresbh}
G_\tau^{res}(\sig_B^p) = G_\tau^{3|(12)}(\sig_B^p) -
G_\tau^{3|1}(\sig_B^p)\,,
\end{equation}
with
\begin{eqnarray}
G_\tau^{3|(12)}(\sig_B^p) &=& \arcsinh^2\left[\frac{1}{2}
\sqrt{(a - 1) (a + 3)}\right]\,, \label{g12bh}\\
G_\tau^{3|1}(\sig_B^p) &=& \arcsinh^2 \left[\sqrt{\frac{(3 a + 1)^2}
{(a + 3)^2} - 1}\right]\,. \label{g11bh}
\end{eqnarray}
The tripartite entanglement \eq{gtauresbh} is strictly smaller than
that of both GHZ/$W$ states and $T$ states, but it can still diverge
in the limit of infinite squeezing ($a \rightarrow \infty$) due to
the global purity of basset hound states. Instead, the bipartite
entanglement $G_\tau^{1|2}=G_\tau^{1|3}$ between mode $1$ and each
of the modes $2$ and $3$ in the corresponding two-mode reductions,
given by \eq{g11bh}, is strictly {\em larger} than the bipartite
entanglement in any two-mode reduction of GHZ/$W$ states. This does
not contradict the  previously given characterization of GHZ/$W$
states as maximally three-way and two-way entangled (maximally
promiscuous). In fact, GHZ/$W$ states have maximal couplewise
entanglement between {\em any} two-mode reduction, while in basset
hound states only two (out of three) two-mode reductions are
entangled, allowing this entanglement to be larger. This is the
reason why these states are well-suited for telecloning, as we will
detail in Sec.~\ref{telebass}. Nevertheless, this reduced bipartite
entanglement cannot increase arbitrarily in the limit of infinite
squeezing, because of the monogamy inequality \pref{CKWine}; in fact
it saturates to
\begin{equation}\label{gredmaxbass}
G_\tau^{1|l}(\sig_B^p,\,a \rightarrow \infty) = \log ^2\left[3 + 2
\sqrt{2}\right] \approx 3.1\,,
\end{equation}
which is about ten times the asymptotic value of the reduced
bipartite two-mode entanglement for GHZ/$W$ states,
\eq{gredmaxghzw}.

\subsubsection{Sharing structure} It is interesting to notice that entanglement
sharing in basset hound states is {\em not} promiscuous.  Tripartite
and bipartite entanglement coexist (the latter only in two of the
three possible two-mode reductions), but the presence of a strong
bipartite entanglement does not help the tripartite one to be
stronger (at fixed local mixedness $a$) than in other states, like
GHZ/$W$ states or even $T$ states (which are globally mixed and
moreover contain no reduced bipartite entanglement at all).

\subsection{The origin of tripartite entanglement promiscuity?}

The above analysis of the entanglement sharing structure in several
instances of three-mode Gaussian states (including the
non-fully-symmetric basset hound states, whose entanglement
structure is not promiscuous) delivers a clear hint that, in the
tripartite Gaussian setting, {\em `promiscuity'} is a peculiar
consequence not of the global purity (noisy GHZ/$W$ states remain
promiscuous for quite strong mixedness), but of the complete {\em
symmetry} under mode exchange. Beside frustrating the maximal
entanglement between pairs of modes \cite{frusta}, symmetry also
constrains the multipartite sharing of quantum correlations. In
basset hound states, the separability of the reduced state of modes
$2$ and $3$, prevents the three modes from having a strong genuine
tripartite entanglement among them all, despite the heavy quantum
correlations shared by the two couples of modes $1|2$ and $1|3$.

This argument will not hold anymore in the case of Gaussian states
with four and more modes, where relaxing the symmetry constraints
will allow for an enhancement of the distributed entanglement
promiscuity to an unlimited extent, as we will show in the next
Chapter.

%\vfill
%
%\begin{flushright}
%\includegraphics[width=3cm]{threesome.eps} \\\vspace*{.3cm}
% {\normalsize \rm {\em Creative threesome} \\
%{\small Michael Perez, 2004}
%\\  \texttt{\footnotesize
%http://www.michaelperez-artist.com}}
%\end{flushright}

}

%%%%%%%%%%%%%%%%%%%%%%%%%%%%%%%%%%%%

\chapter{Unlimited promiscuity of multipartite Gaussian
entanglement}\label{ChapUnlim}

{\sf

The structure of multipartite entanglement of Gaussian states, as
explored up to now,  opens interesting perspectives which are
driving us towards the last part of this Dissertation, namely the
one concerning production and applications of multiparty Gaussian
entangled resources. This Chapter, based on Ref.~\cite{unlim},
provides an additional, exceptional motivation to select CV systems,
and specifically Gaussian states, as ideal candidates for physical
realizations of current and perhaps revolutionary quantum
information and communication implementations. The findings
described here  are also of  importance from a fundamental point of
view, for the quantification and primarily the understanding  of
shared quantum correlations in systems with infinitely large state
space.

We have seen indeed in the previous Chapter that in the most basic
multipartite CV setting, namely that of three-mode Gaussian states,
a partial ``promiscuity'' of entanglement can be achieved.
Permutation-invariant states exist which are the simultaneous
analogs of GHZ and $W$ states of qubits, exhibiting unlimited
tripartite entanglement (with increasing squeezing) and nonzero,
accordingly increasing bipartite entanglement which nevertheless
stays finite even for infinite squeezing \cite{contangle}. We will
now show that in CV systems with more than three modes, entanglement
can be distributed in an {\em infinitely} promiscuous way.

\section{Continuous variables versus qubits}

From an operational perspective, qubits are the main logical units
for standard realizations of quantum information  protocols
\cite{chuaniels}. Also CV Gaussian entangled resources have  been
proven useful for all known implementations of quantum information
processing \cite{brareview}, including quantum computation
\cite{menicucci}, sometimes outperforming more traditional
qubit-based approaches as in the case of unconditional teleportation
\cite{Furusawa98}. It is therefore important to understand if
special features of entanglement appear in states of infinite
Hilbert spaces, which are unparalleled in the corresponding states
of qubits. Such findings may lead to new ways of manipulating
quantum information in the CV setting. For instance, there exist
infinitely many inequivalent classes of bipartite entangled pure CV
states, meaning that a substantially richer structure is available
for quantum correlations and it could be potentially exploited for
novel realizations of quantum information protocols \cite{slocc}.

Here, we address this motivation on a clearcut physical ground,
aiming in particular to show whether the unboundedness of the mean
energy characterizing CV states enables a qualitatively richer
structure for distributed quantum correlations.  We prove that
multimode Gaussian states exist, that can possess simultaneously
arbitrarily large pairwise bipartite entanglement between some pairs
of modes and arbitrarily large genuine multipartite entanglement
among all modes without violating the monogamy
inequality~\pref{ckwine} on entanglement sharing.
%These
%states asymptotically reach the form of two perfectly entangled EPR
%pairs  that can moreover be arbitrarily intercorrelated quantumly.
The class of states exhibiting such unconstrained simultaneous
distribution of quantum correlations are producible with standard
optical means (as we will detail in Sec.~\ref{Sec4Mengi}), the
achievable amount of entanglement being technologically limited only
by the attainable degree of squeezing. This unexpected feature of
entanglement sheds new light on the role of the fundamental laws of
quantum mechanics in curtailing the distribution of information. On
a more applicative ground, it serves as a prelude to implementations
of quantum information processing in the infinite-dimensional
scenario that {\em cannot} be achieved with qubit resources.

 To illustrate the existence of such phenomenon, we consider the
simplest nontrivial instance of a family of four-mode Gaussian
states, endowed with a partial symmetry under mode exchange.

% may play a leading
%role in the description of many-body systems, whose phase
%transitions are driven by strong correlations between the canonical
%variables of the quantum constituents, such as vibrational modes. On
%a more applicative ground, it

\section{Entanglement in partially symmetric four-mode Gaussian states}

\subsection{State definition}

We start with an uncorrelated state of four modes, each one
initially in the vacuum of the respective Fock space, whose
corresponding CM is the identity. We apply a two-mode squeezing
transformation $S_{2,3}(s)$, \eq{tmsS}, with squeezing $s$ to modes
2 and 3, then two further two-mode squeezing transformations
$S_{1,2}(a)$ and $S_{3,4}(a)$, with squeezing $a$, to the pairs of
modes $\{1,2\}$ and $\{3,4\}$. The two last transformations serve
the purpose of redistributing the original bipartite entanglement,
created between modes 2 and 3 by the first two-mode squeezing
operations, among all the four modes. For any value of the
parameters $s$ and $a$, the output is a pure four-mode Gaussian
state  with CM $\gr\sigma$,
\begin{equation}\label{s4}
\gr\sigma =S_{3,4}(a)S_{1,2}(a)S_{2,3}(s)S_{2,3}\T (s)S_{1,2}\T
(a)S_{3,4}\T (a)\,.
\end{equation}
Explicitly, $\sig$ is of the form \eq{CM} where
\begin{eqnarray*}
\sig_1 = \sig_4 &=& [\cosh ^2(a) + \cosh (2 s) \sinh ^2(a)] \id_2 \,,\\
\sig_2 = \sig_3 &=& [\cosh (2 s) \cosh ^2(a) + \sinh ^2(a)] \id_2 \,,\\
\eps_{1,2} = \eps_{3,4} &=& [\cosh ^2(s) \sinh (2 a)] Z_2 \,,\\
\eps_{1,3} = \eps_{2,4} &=& [\cosh (a) \sinh (a) \sinh (2 s)]
\id_2\,,\\
\eps_{1,4} &=& [\sinh ^2(a) \sinh (2 s)] Z_2\,,\\
\eps_{2,3} &=& [\cosh ^2(a) \sinh (2 s)] Z_2\,,
\end{eqnarray*}
 with $Z_2={{1\ \ \ 0}\choose {0 \ -1}}$.
It is immediate to see that a state of this form is  invariant under
the double exchange of modes $1\leftrightarrow 4$ and
$2\leftrightarrow 3$, as $S_{i,j} =S_{j,i} $ and operations on
disjoint pairs of modes commute. Therefore, there is only a partial
symmetry under mode permutations, not a full one like in the case of
the three-mode GHZ/$W$ states and in general the states of
Sec.~\ref{SecSymm}.

\subsection{Structure of bipartite entanglement}\label{SecUnlimBip}

Let us recall that in a pure four-mode Gaussian state and in its
reductions, bipartite entanglement is equivalent to  negativity of
the partially transposed CM, obtained by reversing time in the
subspace of any chosen single subsystem \cite{Simon00,werewolf} (PPT
criterion, see Sec.~\ref{SecPPTG}). This inseparability condition is
readily verified for the family of states in \eq{s4} yielding that,
for all nonzero values of the squeezings $s$ and $a$, $\gr\sigma$ is
entangled with respect to any global bipartition of the modes. This
follows from the global purity of the state, together with the
observation that the determinant of each reduced one- and two-mode
CM obtainable from \eq{s4} is strictly bigger than $1$ for any
nonzero squeezings. The state is thus said to be {\em fully
inseparable} \cite{heiss}, \ie it contains genuine four-partite
entanglement.

Following our previous studies on CV entanglement sharing (see
Chapters \ref{ChapMonoGauss} and \ref{Chap3M}) we choose to measure
bipartite entanglement by means of the Gaussian contangle
${G_\tau}$, an entanglement monotone under Gaussian LOCC, computable
according to \eq{tau}.

In the four-mode state with CM $\gr\sigma$, we can evaluate the
bipartite Gaussian contangle in closed form  for all pairwise
reduced (mixed) states of two modes $i$ and $j$, described by a CM
$\gr\sigma_{i\vert j} $. By applying again PPT criterion (see
Sec.~\ref{SecPPTG}), one finds that the two-mode states indexed by
the partitions $1\vert3$, $2\vert4$, and $1\vert4$, are separable.
For the remaining two-mode states the computation is possible thanks
to the results of Sec.~\ref{SecGEMextra}. Namely, the reduced state
of modes 2 and 3, $\sig_{23}$, belongs to the class of GMEMS
(defined in Sec.~\ref{SecGmemsGlems}); for it \eq{m2gmems}
yields\footnote{\sf We refer to the notation of \eq{tau} and write,
for each partition $i|j$, the corresponding parameter $m_{i|j}$
involved in the optimization problem which defines the Gaussian
contangle.}
\begin{equation}\label{unlim23}
m_{2\vert 3}=\left\{
  \begin{array}{ll}
    \frac{-1+2\cosh ^2(2a)\cosh ^2s+3\cosh (2s)-4\sinh
^2a\sinh (2s)}{4[\cosh ^2a+e^{2s}\sinh ^2a]}, & a<{\rm
arcsinh}[\sqrt{\tanh s}]\,; \\
    1, & \hbox{otherwise.}
  \end{array}
\right.
\end{equation}
On the other hand, the  states $\sig_{1|2}$ and $\sig_{3|4}$ are
GMEMMS (defined in Sec.~\ref{SecMEMMS}), \ie simultaneous GMEMS and
GLEMS, for which either \eq{m2glems} or \eq{m2gmems} give
\begin{equation}\label{unlim1234}
m_{1\vert 2} =m_{3\vert 4} =\cosh 2a\,.
\end{equation}
Accordingly, one can compute the pure-state entanglements between
one probe mode and the remaining three modes. In this case one has
simply $m_{i|(jkl)}= \det\sig_i$. One finds from \eq{s4},
\be\begin{split}m_{1\vert (234)} &=m_{4\vert (123)} =\cosh ^2a+\cosh
(2s)\sinh ^2a\,,\\m_{2\vert (134)} &=m_{3\vert (124)} =\sinh
^2a+\cosh (2s)\cosh ^2a\,.\end{split}\ee

\begin{figure}[t!]
\includegraphics[width=6cm]{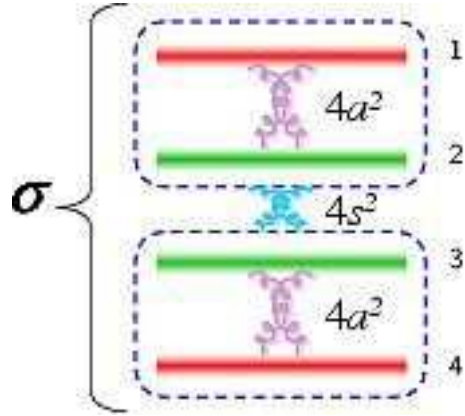} \caption{Bipartite
entanglement structure in the four-mode Gaussian states $\gr\sigma$
of \eq{s4}. The block of modes 1,2 shares with the block of modes
3,4 all the entanglement created originally between modes 2 and 3,
which is an increasing function of $s$ (blue springs). Moreover,
modes 1 and 2 internally share an entanglement arbitrarily
increasing as a function of $a$, and the same holds for modes 3 and
4 (pink springs). For $a$ approaching infinity, each of the two
pairs of modes 1,2 and 3,4 reproduces the entanglement content of an
ideal EPR state, while being the same pairs arbitrarily entangled
with each other according to value of $s$.} \label{figprep}
\end{figure}

Concerning the structure of bipartite entanglement, Eqs.~{\rm
(\ref{tau}, \ref{unlim1234})} imply that the Gaussian contangle in
the mixed two-mode states $\gr\sigma_{1\vert 2} $ and
$\gr\sigma_{3\vert 4} $ is $4a^2$, irrespective of the value of $s$.
This quantity is exactly equal to the degree of entanglement in a
pure two-mode squeezed state $S_{i,j} (a)S_{i,j}\T (a)$ of modes $i$
and $j$ generated with the same squeezing $a$. In fact, the two-mode
mixed state $\sigma_{1\vert 2} $ (and, equivalently, $\sigma_{3\vert
4}$) serves as a proper resource for CV teleportation
\cite{Braunstein98,Furusawa98}, realizing a perfect transfer (unit
fidelity\footnote{\sf The {\em fidelity} ${\CMcal F} \equiv
\bra{\psi^{in}} \varrho^{out}\ket{\psi^{in}}$ (``in'' and ``out''
being input and output state, respectively) quantifies the
teleportation success, as detailed in Chapter \ref{ChapCommun}. For
single-mode coherent input states and $\gr\sigma_{1|2}$ or
$\gr\sigma_{3|4}$ employed as entangled resources, ${\CMcal F}=
(1+{\rm e}^{-2a}\cosh ^2s)^{-1}$. It reaches unity for $a\gg 0$ even
in presence of high interpair entanglement ($s \gg 0$), provided
that $a$ approaches infinity much faster than $s$.}) in the limit of
infinite squeezing $a$.

The four-mode state $\gr\sigma$ reproduces therefore (in the regime
of very high $a$) the entanglement content of two EPR-like pairs
($\{1,2\}$ and $\{3,4\}$). Remarkably, there is an additional,
independent entanglement {\em between} the two pairs, given by
${G_\tau} (\sigma _{(12)\vert (34)} )=4s^2$ --- the original
entanglement in the two-mode squeezed state $S_{2,3} (s)S_{2,3}\T
(s)$ after the first construction step --- which can be itself
increased arbitrarily with increasing $s$. This peculiar
distribution of bipartite entanglement, pictorially displayed in
 Fig.~\ref{figprep}, is a first remarkable signature of an
unmatched freedom of entanglement sharing in multimode Gaussian
states as opposed for instance to states of the same number of
qubits, where a similar situation is {\em impossible}. Specifically,
if in a pure state of four qubits the first two approach unit
entanglement and the same holds for the last two, the only
compatible global state reduces necessarily to a product state of
the two singlets: no interpair entanglement is allowed by the
fundamental monogamy constraint \cite{CKW,osborne}

\subsection{Distributed entanglement and multipartite sharing
structure}

We can now move to a closer analysis of entanglement distribution
and genuine multipartite quantum correlations.

\subsubsection{Monogamy inequality}

A primary step is to verify whether the fundamental monogamy
inequality~\pref{ckwine} is satisfied on the four-mode state
$\gr\sigma$ distributed among the four parties (each one owning a
single mode).\footnote{\sf In Sec.~\ref{SecHiro} the general
monogamy inequality for $N$-mode Gaussian states has been
established by using the Gaussian tangle, \eq{Gtau}. No complete
proof is available to date for the monogamy of the (Gaussian)
contangle, \eq{tau}, beyond the tripartite case. Therefore, we need
to check its validity explicitly on the state under consideration.}
In fact, the problem reduces to proving that $$\min \{g[m_{1\vert
(234)}^2 ]-g[m_{1\vert 2}^2 ],\,g[m_{2\vert (134)}^2 ]-g[m_{1\vert
2}^2 ]-g[m_{2\vert 3}^2 ]\}$$ is nonnegative. The first quantity
always achieves the minimum yielding
\begin{eqnarray}
\label{taures} {G_\tau}^{res}(\gr\sigma ) &\!\!\equiv\!\!& {G_\tau}
(\gr\sigma _{1\vert (234)} )-{G_\tau} (\gr\sigma _{1\vert 2} )
\\ &\!\!=\!\!&{\rm arcsinh}^2\!\left\{ {\sqrt {[\cosh
^2a+\cosh (2s)\sinh ^2a]^2-1} } \right\}-4a^2. \nonumber
\end{eqnarray}
Since $\cosh (2s)>1$ for $s>0$, it follows that ${G_\tau}^{res}>0$
and \ineq{ckwine} is satisfied.

The entanglement in the global Gaussian state is therefore
distributed according to the laws of quantum mechanics, in such a
way that the residual Gaussian contangle ${G_\tau}^{res}$ quantifies
the multipartite entanglement not stored in couplewise form. Those
quantum correlations, however, can be either tripartite involving
three of the four modes, and/or genuinely four-partite among all of
them. We can now quantitatively estimate to what extent such
correlations are encoded in some tripartite form: as an
anticipation, we will find them negligible in the limit of high
squeezing $a$.

\subsubsection{Tripartite entanglement estimation}
 Let us first observe that in the
tripartitions 1$\vert $2$\vert $4 and 1$\vert $3$\vert $4 the
tripartite entanglement is zero, as mode 4 is not entangled with the
block of modes 1,2, and mode 1 is not entangled with the block of
modes 3,4 (the corresponding three-mode states are then said to be
biseparable \cite{kraus}, see Sec.~\ref{secbarbie}). The only
tripartite entanglement present, if any, is equal in content (due to
the symmetry of the state $\gr\sigma$) for the tripartitions 1$\vert
$2$\vert $3 and 2$\vert $3$\vert $4, and can be quantified via the
residual Gaussian contangle determined by the corresponding
three-mode monogamy inequality~\pref{ckwine}.

The residual Gaussian contangle of a Gaussian state $\sig$ of the
three modes $i$, $j$, and $k$  (which is an entanglement monotone
under tripartite Gaussian LOCC for pure states \cite{contangle}, see
Sec.~\ref{secTauresMonotone}), takes the form as in \eq{gtaures},
\begin{equation}\label{taures3}
{G_\tau}(\sig_{i|j|k})\equiv\min_{(i,j,k)} \left[
{G_\tau}(\sig_{i|(jk)})-{G_\tau}(\sig_{i|j})-{G_\tau}(\sig_{i|k})\right]\,,
\end{equation}
where the symbol $(i,j,k)$ denotes all the permutations of the three
mode indexes.

\begin{figure}[t!]
\includegraphics[width=11cm]{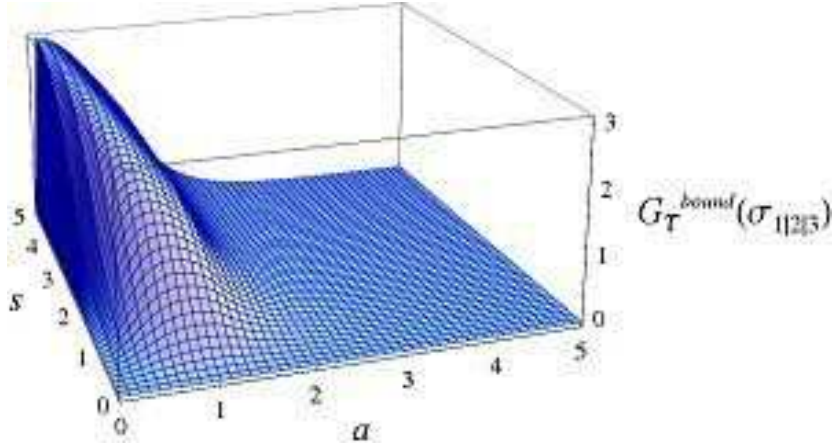} \caption{
Upper bound ${G_\tau} ^{bound}(\gr\sigma _{1\vert 2\vert 3} )$,
\eq{taubnd}, on the tripartite entanglement between modes 1, 2 and 3
(and equivalently 2, 3, and 4) of the four-mode Gaussian state
defined by \eq{s4}, plotted as a function of the squeezing
parameters $s$ and $a$. The plotted upper bound on the tripartite
entanglement among modes 1, 2, 3 (and equivalently 2, 3, 4)
asymptotically vanishes for $a$ going to infinity, while any other
form of tripartite entanglement among any three modes is always
zero. } \label{figtrip}
\end{figure}

We are interested here in computing the residual tripartite Gaussian
contangle, \eq{taures3}, shared among modes 1, 2 and 3 in the
reduced mixed three-mode state $\sig_{123}$ obtained from \eq{s4} by
tracing over the degrees of freedom of mode 4. To quantify such
tripartite entanglement exactly, it is necessary to compute the
mixed-state $1 \times 2$ Gaussian contangle between one mode and the
block of the two other modes. This requires solving the nontrivial
optimization problem of \eq{tau} over all possible pure three-mode
Gaussian states --- not necessarily in standard form, \eq{cm3tutta}.
However, from the definition itself, \eq{tau}, the bipartite
Gaussian contangle ${G_\tau} (\sig_{i\vert (jk)})$ (with $i,j,k$ a
permutation of $1, 2, 3$) is bounded from above by the corresponding
bipartite Gaussian contangle ${G_\tau} (\sig_{i\vert (jk)}^p )$ in
any pure, three-mode Gaussian state with CM $\sig_{i\vert (jk)}^p
\le \sig_{i\vert (jk)}$. As an ansatz we can choose pure three-mode
Gaussian states whose CM $\sig_{123}^p$ has the same matrix
structure of our mixed state $\sig_{123}$ (in particular, zero
correlations between position and momentum operators, and diagonal
subblocks proportional to the identity), and restrict the
optimization to such a class of states. This task is accomplished by
choosing a pure state given by the following CM
\begin{equation}\label{sigpbound}
\gr\gamma^p_{123}=S_{1,2}(a)S_{2,3}(t)S_{2,3}\T (t)S_{1,2}\T (a)\,,
\end{equation}
where the two-mode squeezing transformation $S_{i,j}$ is defined by
\eq{tmsS}, and
$$
t=\frac12{\rm arccosh}\left[\frac{1+{\rm sech}^2a\tanh ^2s}{1-{\rm
sech}^2a\tanh^2s}\right]\,. $$ We have then
\begin{equation}\label{taubndbip}
{G_\tau} (\sig_{i\vert (jk)} )\le g[(m_{i\vert (jk)}^{\gamma})^2]\,,
\end{equation}
where $m^\gamma$ is meant to determine entanglement in the state
$\gr\gamma^p$, \eq{sigpbound}, via \eq{tau}. The bipartite
entanglement properties of the state $\gr\gamma^p$  can be
determined analogously to what done in Sec.~\ref{SecUnlimBip}. We
find
\begin{eqnarray}
 m^\gamma_{3|(12)} &=& \frac{1+{\rm sech}^2a\tanh^2s}
{1-{\rm sech}^2a\tanh^2s}\,, \\
  m^\gamma_{1|(23)} &=& \cosh^2a+m^\gamma_{4|(1 2)} \sinh^2a\,, \\
   m^\gamma_{2|(13)} &=& \sinh^2a+m^\gamma_{4|(1 2)}
   \cosh^2a\,.
\end{eqnarray}
Eqs.~{\rm(\ref{taures3},\ref{taubndbip})} thus lead to an upper
bound on the genuine tripartite entanglement   between modes 1, 2
and 3 (and equivalently 2, 3, and 4),
\begin{equation}\label{taubnd}
{G_\tau}(\sig_{1|2|3}) \le {G_\tau} ^{bound}(\sig_{1|2|3}) \equiv
\min \{g[(m_{1\vert (23)}^{\gamma})^2]-g[m_{1\vert 2}^2
],\,g[(m_{3\vert (12)}^{\gamma})^2]-g[m_{2\vert 3}^2 ]\}\,,
\end{equation}
where the two-mode entanglements regulated by the quantities
$m_{i|j}$ without the superscript ``$\gamma$'' are referred to the
reductions of the mixed state $\sig_{123}$ and are listed in
Eqs.~{\rm(\ref{unlim23}, \ref{unlim1234})}. In \eq{taubnd} the
quantity $g[(m_{2\vert (1 3)}^{\gamma} )^2]-g[m_{1\vert 2}^2
]-g[m_{2\vert 3}^2 ]$ is not included in the minimization, being
always larger than the other two terms. Numerical investigations in
the space of all pure three-mode Gaussian states seem to confirm
that the upper bound of \eq{taubnd} is actually sharp (meaning that
the three-mode contangle is globally minimized on the state
$\gr\gamma^p$), but this statement can be left here as a conjecture
since it is not required for our subsequent analysis.

The upper bound ${G_\tau} ^{bound}(\gr\sigma _{1\vert 2\vert 3} )$
is always nonnegative (as an obvious consequence of monogamy, see
Sec.~\ref{secmono}), moreover it is decreasing with increasing
squeezing $a$, and vanishes in the limit $a\to \infty $, as shown in
Fig.~\ref{figtrip}. Therefore, in the regime of increasingly high
$a$, eventually approaching infinity, any form of tripartite
entanglement among any three modes in the state $\gr\sigma $ is
negligible (exactly vanishing in the limit). As a crucial
consequence, in that regime the residual entanglement ${G_\tau}
^{res}(\gr\sigma )$ determined by \eq{taures} is all stored in
four-mode quantum correlations and quantifies the {\em genuine}
four-partite entanglement.

\subsubsection{Genuine four-partite entanglement: promiscuous beyond
limits}

\begin{figure}[t!]
\includegraphics[width=11cm]{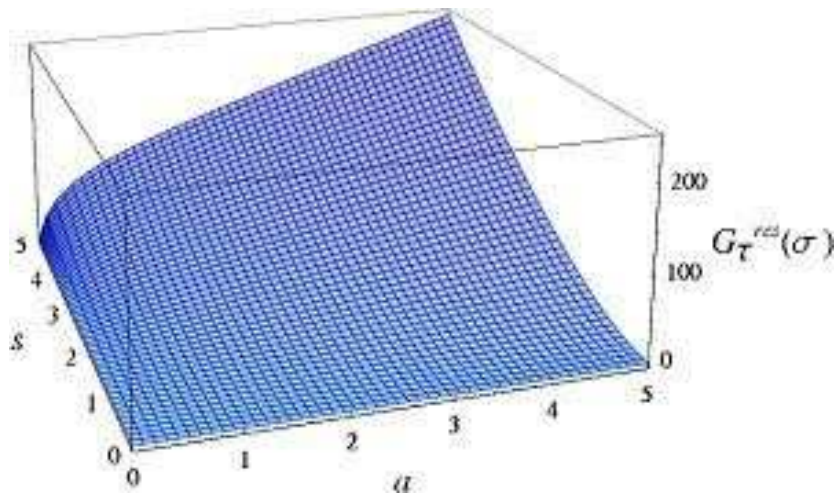} \caption{Residual
multipartite entanglement ${G_\tau} ^{res}(\gr\sigma )$ [see
\eq{taures}], which in the regime of large squeezing $a$ is
completely distributed in the form of genuine four-partite quantum
correlations. The four-partite entanglement is monotonically
increasing with increasing squeezing $a$, and diverges as $a$
approaches infinity.  The multimode Gaussian state $\sig$
constructed with an arbitrarily large degree of squeezing $a$,
consequently, exhibits a coexistence of unlimited multipartite and
pairwise bipartite entanglement in the form of EPR correlations. In
systems of many qubits, and even in Gaussian states of CV systems
with a number of modes smaller than four (see Chapter \ref{Chap3M}),
such an unlimited and unconstrained promiscuous distribution of
entanglement is strictly forbidden. } \label{figres}
\end{figure}

We finally observe that ${G_\tau} ^{res}(\gr\sigma )$, \eq{taures},
is an increasing function of $a$ for any value of $s$ (see
Fig.~\ref{figres}), and it {\em diverges} in the limit $a\to
\infty$. This proves that the class of pure four-mode Gaussian
states with CM $\gr\sigma$ given by \eq{s4} exhibits genuine
four-partite entanglement which grows unboundedly with increasing
squeezing $a$ and, {\em simultaneously}, possesses pairwise
bipartite entanglement in the mixed two-mode reduced states of modes
$\{1,2\}$ and $\{3,4\}$, that increases unboundedly as well with
increasing $a$.\footnote{\sf The notion of {\em unlimited}
entanglement has to be interpreted in the usual asymptotic sense.
Namely, given an arbitrarily large entanglement threshold, one can
always pick a state in the considered family with squeezing high
enough so that its entanglement exceeds the threshold.} Moreover, as
previously shown, the two pairs can themselves be arbitrarily
entangled with each other with increasing squeezing $s$.

By constructing a simple and feasible example we have shown that,
when the quantum correlations arise among degrees of freedom
spanning an infinite-dimensional space of states (characterized by
unbounded mean energy), an accordingly infinite freedom is tolerated
for quantum information to be doled out.  This happens with no
violation of the fundamental monogamy constraint that retains its
general validity in quantum mechanics. In the CV instance
demonstrated here, the only effect of monogamy is to bound the
divergence rates of the individual entanglement contributions as the
squeezing parameters are increased. Within the restricted Hilbert
space of four or more qubits, instead, an analogous entanglement
structure between the single qubits is strictly forbidden.

Quite naturally, the procedure presented here can be in principle
extended to investigate the increasingly richer structure of
entanglement sharing in $N$-mode ($N$ even) Gaussian states, via
additional pairs of redistributing two-mode squeezing operations.

In summary, the main result of this Chapter may be stated as follows
\cite{unlim}.

\medskip

\begin{itemize}
\item[\ding{226}]
 \noindent{\rm\bf Unlimited promiscuity of multipartite Gaussian entanglement.}
{\it The entanglement in $N$-mode Gaussian states ($N \ge 4$) can
distribute in such a way that it approaches infinity both as a
genuinely multipartite quantum correlation shared among all modes,
and as a bipartite, two-mode quantum correlation in different pairs
of modes, within the validity of the general monogamy constraints on
entanglement sharing.}
\smallskip
\end{itemize}

%\section{Practical consequences: an eye to the future}

\subsection{Discussion} \label{SecUnlimDiscuss}

The discovery of an unlimited promiscuous entanglement sharing,
while of inherent importance in view of novel implementations of CV
systems for multiparty quantum information protocols, opens
unexplored perspectives for the understanding and characterization
of entanglement in multiparticle systems.

Gaussian states with finite squeezing (finite mean energy) are
somehow analogous to discrete systems with an effective dimension
related to the squeezing degree \cite{brareview}. As the promiscuous
entanglement sharing arises in Gaussian states by asymptotically
increasing the squeezing to infinity, it is natural to expect that
dimension-dependent families of states will exhibit an entanglement
structure gradually more promiscuous with increasing Hilbert space
dimension towards the CV limit. A proper investigation  into the
huge moat of qudits (with dimension $2 < d < \infty$) is therefore
the next step to pursue, in order to aim at developing the complete
picture of entanglement sharing in many-body systems, which is
currently lacking (see Sec.~\ref{SecPisa}).  Here \cite{unlim}, we
have initiated this program by establishing a sharp discrepancy
between the two {\em extrema} in the ladder of Hilbert space
dimensions: namely, entanglement of CV systems in the limit of
infinite mean energy has been proven infinitely more shareable than
that of individual qubits.

Once a more comprehensive understanding will be available of the
distributed entanglement structure in high-dimensional and CV
systems (also beyond the special class of Gaussian states), the task
of devising new protocols to translate such potential into
full-power practical implementations for what concerns encoding,
processing and distribution of shared quantum information, can be
addressed as well.

We will briefly discuss the optical generation and exploitation of
promiscuous entanglement in four-mode Gaussian states in
Sec.~\ref{Sec4Mengi}. }

%
%\vfill
%
%\begin{flushright}
%\includegraphics[width=4cm]{perezfire.eps} \\\vspace*{.3cm}
% {\normalsize \rm {\em Fire} \\
%{\small Michael Perez, 2003}
%\\  \texttt{\footnotesize
%http://www.michaelperez-artist.com}}
%\end{flushright}
%

%%%%%%%%%%%%%%%%%%%%%%%%%%%%%%%%%%%%%%%%%%%%%%

\part{Quantum state engineering of entangled Gaussian
states}{\vspace*{1cm}
\includegraphics[width=9.5cm]{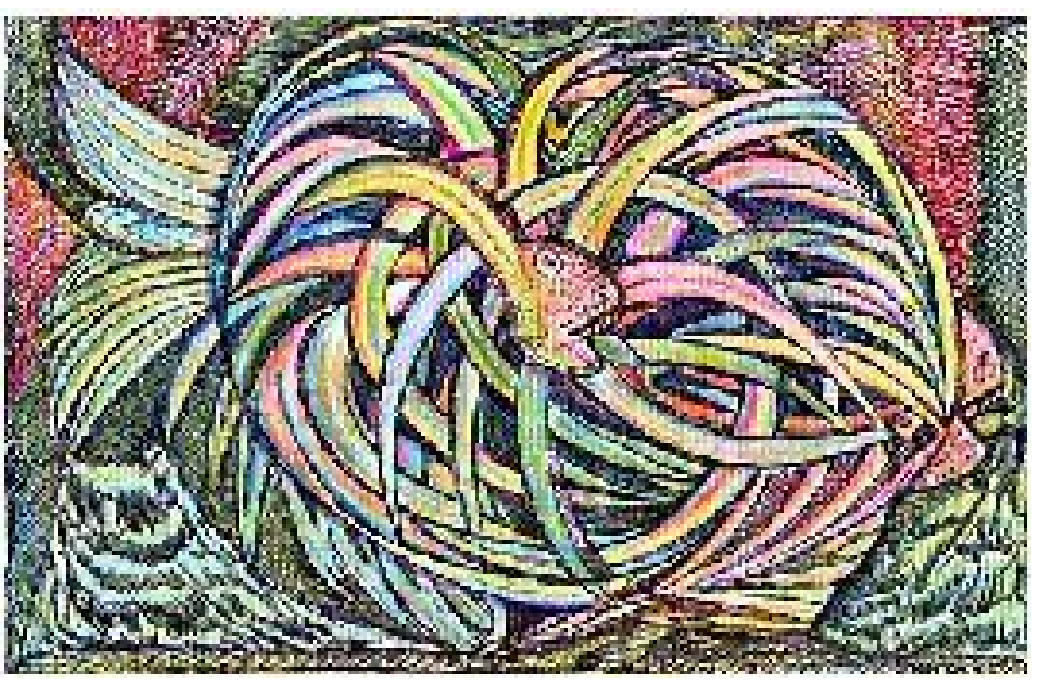} \\
\vspace*{0.6cm} {\normalsize \rm {\em The Entangle Fishes.} Louis
Monza, 1970.
\\ \vspace*{-0.4cm} \texttt{\footnotesize http://www.americaohyes.com/pages/monza.htm}}}
 \label{PartEngi}

\chapter{Two-mode Gaussian states in the lab}\label{Chap2MExp}

{\sf

One of the strength points of the CV quantum information science
with Gaussian states, alongwith the mathematical structure which
enables an accurate description of their informational properties
(see Chapter \ref{ChapGauss}), has surely to be traced back to the
astonishing progress obtained on the experimental side for what
concerns preparation, processing and characterization  of entangled
Gaussian resources, and their successful implementation for the most
diverse communication and computation tasks.
 We have already
stressed, for instance, that one of the main byproducts of our study
on bipartite entanglement versus purity, presented in
Sec.~\ref{secEntvsMix}, is that of having provided a direct,
reliable way to estimate entanglement of arbitrary unknown two-mode
Gaussian states in terms of experimentally accessible measurements
of purity \cite{prl} (see Sec.~\ref{secFiura}).

This Chapter mainly originates from our collaboration  to an
experiment which illustrates the state-of-the-art in the engineering
and processing of two-mode Gaussian states, via an original optical
set-up based on a type-II optical parametric oscillator (OPO) with
adjustable mode coupling \cite{francamentemeneinfischio}.
Experimental results allow a direct verification of many theoretical
predictions and provide a sharp insight into the general properties
of two-mode Gaussian states, elucidated in Chapter \ref{Chap2M}, and
the manipulation of the entanglement resource. We will discuss this
experiment in Sec.~\ref{secFrancesi}.

As a disclaimer, we remark  that the main focus of this Dissertation
is of a theoretical nature, as our primary aim has been up to now to
develop strong mathematical tools to define and characterize
entanglement of Gaussian states. Therefore, many experimental
details, largely available elsewhere (see, as a guide,
Refs.~\cite{brareview,SeralePHD,sculzub,paris01,LauratPHD}) will be
surely lacking here. However, and thanks to the close contact with
the ``reality'' of experiments achieved during the preparation of
Ref.~\cite{francamentemeneinfischio}, we have in parallel devoted a
special attention towards the practical production of CV
entanglement on one side, and its interpretation in connection with
operational settings on the other.

These two aspects of our work are respectively treated in this, and
in the next Part of this Dissertation.

Let us first briefly comment on the latter, namely the investigation
of the usefulness of entangled Gaussian states for the most common
implementations of CV quantum information and communication
protocols \cite{brareview}. This side of our research  enriches the
mathematical analysis and clarifies the physical understanding of
our results: an example is provided by the full equivalence between
(bipartite and multipartite) entanglement and optimal success of
(two-party and multi-party) CV quantum teleportation experiments
with (two-mode and multimode) symmetric, generally mixed, Gaussian
resources \cite{telepoppate}, which will be established in Chapter
\ref{ChapCommun}.

Before turning to the operational interpretation of entanglement, we
judge of interest to discuss, at this point of the Dissertation, the
issue of providing efficient recipes to engineer, in the lab, the
various classes of Gaussian states that we have singled out in the
previous Parts for their remarkable entanglement properties. These
optimal production schemes (among which we mention the one for all
pure Gaussian states exhibiting generic entanglement \cite{generic},
presented in Chapter \ref{ChapGeneric}) are of inherent usefulness
to experimentalists who need to prepare entangled Gaussian states
with minimal resources. Unless explicitly stated, we will always
consider as preferred realistic setting for Gaussian state
engineering that of {\em quantum optics} \cite{fabio}.

In this Chapter, we thus begin by first completing the analysis of
Chapter \ref{Chap2M} on the two special classes of two-mode
 ``extremally'' entangled Gaussian states that have arisen both in the negativity
versus purity analysis, and in the comparison between Gaussian
entanglement measures and negativities, namely GMEMS and GLEMS
\cite{extremal}. We discuss how both families of Gaussian states can
be obtained experimentally, adding concreteness to the plethora of
results previously presented on their entanglement characterization.
After that, we move more deeply into the description of the
experiment concerning production and optimization of entanglement in
two-mode Gaussian states by optical means, reported in
\cite{francamentemeneinfischio}.

State engineering of Gaussian states of more than two modes will be
addressed in the next two Chapters.

\section{Schemes to realize extremally entangled states in
experimental settings}\label{exp}

We discuss here how to obtain the two-mode states introduced in
Sec.~\ref{SecGmemsGlems} in a practical setting.

\begin{figure}[t]
%\begin{center}
\includegraphics[width=5.5cm]{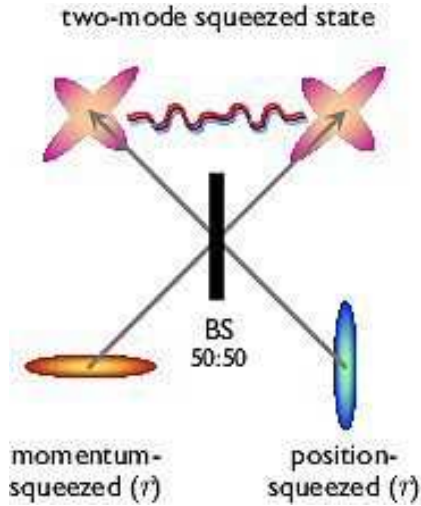}
\caption{Optical generation of two-mode squeezed states (twin-beams)
by superimposing two single-mode beams, independently squeezed of
the same amount $r$ in orthogonal quadratures, at a 50:50
beam-splitter. The two operations (individual squeezings plus
beam-splitter), taken together, correspond to acting with the
twin-beam transformation \eq{twin} on two vacuum beams.}
\label{figtms}
\end{figure}

\subsection{GMEMS state engineering} As we have seen, GMEMS are
two-mode squeezed thermal states, whose general CM is described by
Eqs.~\pref{stform} and \pref{2mst}. A realistic instance giving rise
to such states is provided by the dissipative evolution of an
initially pure two-mode squeezed vacuum with CM \eq{tms}. The latter
may be created \emph{e.g.} by mixing two independently squeezed
beams (one in momentum and one in position, with equal squeezing
parameter $r$) at a 50:50 beam-splitter $B_{1,2}(1/2)$, \eq{bbs}, as
shown in Fig.~\ref{figtms}.\footnote{\sf See also
Sec.~\ref{SecSympl}. A more detailed discussion concerning the
production of two-mode squeezed states is deferred to
Sec.~\ref{secFrancesi}.}

Let us denote by $\gr{\sigma}_r$ the CM of a two mode squeezed
vacuum with squeezing parameter $r$, \eq{tms}, derived from
Eqs.~\pref{2mst} with $\nu_{\mp}=1$. The interaction of this initial
state with a thermal noise results in the following dynamical map
describing the time evolution of the CM $\gr{\sigma}(t)$
\cite{serafinipra04} \be \gr{\sigma}(t)=\,{\rm e}^{-\Gamma
t}\gr{\sigma}_{r} + (1-\,{\rm e}^{-\Gamma t})\gr{\sigma}_{n_1,n_2}
\; ,\label{realgmems} \ee where $\Gamma$ is the coupling to the
noisy reservoir (equal to the inverse of the damping time) and
$\gr{\sigma}_{n_1,n_2}=\oplus_{i=1}^{2} n_i {\mathbbm 1}_2$ is the
CM of the thermal noise (see also Sec.~\ref{Secdeco}). The average
number of thermal photons $n_i$ is given by \eq{temperature},
\[
n_i=\frac{1}{\exp \left( \hbar \omega_{i}/k_{B}T \right) - 1}\,
\]
in terms of the frequencies of the modes $\omega_i$ and of the
temperature of the reservoir $T$. It can be easily verified that the
CM \eq{realgmems} defines a two-mode thermal squeezed state,
generally nonsymmetric (for $n_1 \neq n_2$). However, notice that
the entanglement of such a state cannot persist indefinitely,
because after a given time inequality \pref{2mstppt} will be
violated and the state will evolve into a disentangled two-mode
squeezed thermal state. We also notice that the relevant instance of
pure loss ($n_1=n_2=0$) allows the realization of symmetric GMEMS.

\subsection{GLEMS state engineering}

\begin{figure}[t]
%\begin{center}
\includegraphics[width=8cm]{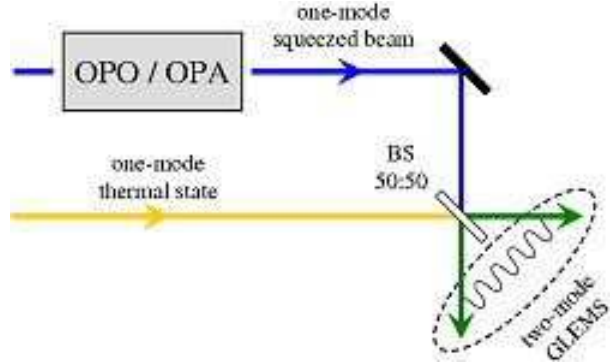}
\caption{Possible scheme for the generation of Gaussian least
entangled mixed states (GLEMS), introduced in
Sec.~\ref{SecGmemsGlems}. A single-mode squeezed state (obtained,
for example, by an optical parametric oscillator or amplifier)
interferes with a thermal state through a 50:50 beam-splitter. The
resulting two-mode state is a minimally entangled mixed Gaussian
state at given global and marginal purities. }
%\end{center}
\label{glemsetup}
\end{figure}

Concerning the experimental characterization of minimally entangled
Gaussian states (GLEMS), defined by \eq{glems}, one can envisage
several explicit experimental settings for their realization. For
instance, let us consider (see Fig.~\ref{glemsetup}) a beam-splitter
with transmittivity $\tau=1/2$,  corresponding to a two-mode
rotation of angle $\pi/4$ in phase space, \eq{bbs}.

Suppose that a single-mode squeezed state, with CM
$\gr{\sigma}_{1r}=\,{\rm diag}\,(\,{\rm e}^{2r},\,{\rm e}^{-2r})$
(like, {\em e.g.}, the result of a degenerate parametric down
conversion in a nonlinear crystal), enters in the first input of the
beam-splitter. Let the other input be an incoherent thermal state
produced from a source at equilibrium at a temperature $T$. The
purity $\mu$ of such a state can be easily computed in terms of the
temperature $T$ and of the frequency of the thermal mode $\omega$,
\be \mu= \frac{\exp \left( \hbar \omega / k_{B} T \right) - 1}{\exp
\left( \hbar \omega / k_{B} T \right) + 1} \; . \ee The state at the
output of the beam-splitter will be a correlated two-mode Gaussian
state with CM $\gr{\sigma}_{out}$ that reads
\[
\gr{\sigma}_{out}=\frac{1}{2}\left(
\begin{array}{cccc}
n+k&0&n-k&0\\
0&n+k^{-1}&0&n-k^{-1}\\
n-k&0&n+k&0\\
0&n-k^{-1}&0&n+k^{-1}
\end{array}
\right) \; ,
\]
with $k=\,{\rm e}^{2r}$ and $n=\mu^{-1}$. By immediate inspection,
the symplectic spectrum of this CM is $\nu_-=1$ and $\nu_+=1/\mu$.
Therefore the output state is always a state with extremal
generalized entropy at a given purity (see Sec.~\ref{SecEntroG}).
Moreover, the state is entangled if \be \cosh(2r) >
\frac{\mu^2+1}{2\mu} = \frac{\exp \left( 2\hbar \omega / k_{B} T
\right) + 1}{\exp \left( 2\hbar \omega / k_{B} T \right) - 1} \, .
\ee Tuning the experimental parameters to meet the above condition,
indeed makes the output state of the beam-splitter a symmetric
GLEMS. It is interesting to observe that nonsymmetric GLEMS can be
produced as well by choosing a beam-splitter with transmittivity
different from $1/2$.

\section{Experimental production and manipulation of two-mode
entanglement}\label{secFrancesi}

Experimentally, CV entanglement can be obtained directly by type-II
parametric interaction deamplifying either the vacuum fluctuations
as was demonstrated in the seminal experiment by Ou \emph{et al.}
\cite{ou92} (or in recent experiments \cite{bf,saopaulo}) or the
fluctuations of a weak injected beam \cite{eprtaiyuan}. It can also
be obtained indirectly by mixing on a beam-splitter two independent
squeezed beams, as shown in Fig.~\ref{figtms}. The required
squeezing can be produced by Kerr effects --- using optical fibers
\cite{erlangen} or cold atoms in an optical cavity \cite{josse} ---
or by type-I parametric interaction in a cavity
\cite{wu86,australie}. Single-pass type-I interaction in a
non-collinear configuration can also generate directly entangled
beams as demonstrated recently by Wenger \emph{et al.} in the pulsed
regime \cite{wengerEPR}. All these methods produce a symmetric
entangled state enabling dense coding, the teleportation of coherent
\cite{Furusawa98,australie,furunew} or squeezed states
\cite{teleportationfurusawa} or entanglement swapping
\cite{entswap,swappingtaiyuan,furunew}.\footnote{\sf
Continuous-variable ``macroscopic'' entanglement can also be induced
between two micromechanical oscillators via entanglement swapping,
exploiting the quantum effects of radiation pressure
\cite{pirlandomacro}.} These experiments generate an entangled
two-mode Gaussian state with a CM in the so-called `standard form'
\cite{Duan00,Simon00}, \eq{stform2}, without having to apply any
linear unitary transformations such as beam-splitting or
phase-shifts to improve its entanglement in order to exploit it
optimally in quantum information protocols.

However, it has been recently shown \cite{eprparis} that, when a
birefringent plate is inserted inside the cavity of a type-II
optical parametric oscillator, \emph{i.e.}~when mode coupling is
added, the generated two-mode state remains symmetric but
entanglement is not observed on orthogonal quadratures: the state
produced is not in the standard form. The entanglement of the two
emitted modes in this configuration is not optimal: it is indeed
possible by passive non-local operations to select modes that are
more entangled. The original system of
Ref.~\cite{francamentemeneinfischio} provides thus a good insight
into the quantification and manipulation of the entanglement
resources of two-mode Gaussian states. In particular, as just
anticipated, it allows to confirm experimentally the theoretical
predictions on the entangling capacity of passive optical elements
and on the selection of the optimally entangled bosonic modes
\cite{passive}.

We start by recalling such theoretical predictions.

\subsection{Entangling power of passive optical elements on
symmetric Gaussian states} \label{SecPassive}

Let us now briefly present some additional results on the
entanglement qualification of symmetric two-mode Gaussian states,
which will be the subject of the experimental investigations
presented in the following. We remark that symmetric Gaussian
states, which carry the highest possible entanglement among all
thermal squeezed states [see Fig.~\ref{gmemms}{\rm (a)}], are the
resources that enable CV teleportation of an unknown coherent state
\cite{Braunstein98,Furusawa98} with a fidelity arbitrarily close to
1 even in the presence of noise (mixedness), provided that the state
is squeezed enough (ideally, a unit fidelity requires infinite
squeezing). Actually, the fidelity of such an experiment, if the
squeezed thermal states employed as shared resource are optimally
produced, turns out to be itself a measure of entanglement and
provides a direct, operative quantification of the entanglement of
formation present in the resource \cite{telepoppate}, as presented
in Chapter \ref{SecTelepoppy}.

It is immediately apparent that, because $a=b$ in \eq{stform2}, the
partially transposed CM in standard form $\tilde{\sig}$ (obtained by
flipping the sign of $c_-$) is diagonalized by the orthogonal and
symplectic beam-splitter transformation (with $50\%$ transmittivity)
$B(1/2)$, \eq{bbs},  resulting in a diagonal matrix with entries
$a\mp |c_\mp|$. The symplectic eigenvalues of such a matrix are then
easily retrieved by applying local squeezings. In particular, the
smallest symplectic eigenvalue $\tilde{\nu}_{-}$ (which completely
determines entanglement of symmetric two-mode states, with respect
to any known measure, see Sec.~\ref{SecEOFGauss}) is simply given by
\be \tilde{\nu}_{-}=\sqrt{(a-|c_{+}|)(a-|c_{-}|)} \; .
\label{symeig} \ee Note that also the original standard form CM
$\sig$ with $a=b$ could be diagonalized ({\em not symplectically},
since the four diagonal entries are generally all different) by the
same beam-splitter transformation $B(1/2)$, with the same orthogonal
eigenvalues $a\mp |c_\mp|$. It is immediate to verify that
$\tilde{\nu}_-$ is just given by the geometric average between the
two smallest of such orthogonal eigenvalues of $\sig$. The two
quadratures resulting from the previous beam-splitter transformation
select orthogonal directions in phase space with respect to the
original ones, so they will be referred to as `orthogonal'
quadratures. Notice that, in the experimental practice, this allows
for the determination of the entanglement through the measurement of
diagonal entries (noise variances) of the CM after the application
of a balanced beam-splitter, which embodies the transformation
$B(1/2)$.

To explore further consequences of this fact, let us briefly recall
some theoretical results on the generation of entanglement under
passive (energy-preserving) transformations, which will be applied
in the following. As shown in Ref.~\cite{passive}, the minimum value
for $\tilde{\nu}_{-}$ ({\ie}the maximal entanglement) attainable by
passive transformations is determined by \be
\tilde{\nu}_{-}^{2}=\lambda_1\lambda_2 \; , \label{mineig} \ee where
$\lambda_1$ and $\lambda_2$ are the two smallest eigenvalues of
$\sig$. Therefore, the entanglement of symmetric states {\em in
standard form}, \eq{stform2}, cannot be increased through energy
preserving operations, like beam-splitters and phase shifters, as
the symplectic eigenvalue $\tilde{\nu}_{-}$ given by \eq{symeig}
already complies with the optimal condition \pref{mineig}. On the
other hand, as it will be discussed in detail in the following, the
insertion of a birefringent plate in a type-II optical parametric
oscillator results in states symmetric but not in standard form. In
such a case the entanglement can be optimized by the action of a
(passive) phase shifter, as we will explicitly show, through both
theoretical proof (Sec.~\ref{sec:modecoupling}) and experimental
demonstration (Sec.~\ref{sec:nonsf}).

%%%%%%%%%%%%%%%%%%%%%%%%%%%%%%%%%%%%%%%%%%%%%%%%%
\subsection{Effects of mode coupling on the entanglement
generation} \label{sec:modecoupling}

As mentioned in the introductory Sec.~\ref{secFrancesi}, CV
entanglement is very often produced by mixing two squeezed modes on
a beam-splitter. In the general case, the squeezed quadratures have
an arbitrary phase difference. We denote by $\theta+\pi/2$ the phase
difference between the two squeezed quadratures. The CM of the
squeezed modes is then
\begin{equation}
 \sig_{A_{+}\,\!A_{-}} = \left(
\begin{array}{cc|cc}
a & 0 & 0 & 0 \\
0 & 1/a & 0 & 0 \\
\hline 0 & 0 & b & c\\
0 & 0 & c & b'
\end{array} \right), \label{eq:sqnondiag}
\end{equation}
while the CM of the two modes after the beam-splitter is
\begin{equation}
\sig_{A_{1}\,\!A_{2}} = B(1/2)\T  \sig_{A_{+}\,\!A_{-}} B(1/2)=
\left(
\begin{array}{cc|cc}
n_1 & k' & k & k' \\
k' & n_2 & k' & -k \\
\hline k & k' & n_1 & k'\\
k' & -k & k' & n_2
\end{array} \right)\,,\label{eq:nonsform}
\end{equation}
where
\begin{eqnarray*}
  b\!\!&=&\!\! \frac{\cos^2 \theta}a + a \sin^2 \theta\,,\quad
  b'= a \cos^2 \theta + \frac{\sin^2 \theta}a\,,\quad
  c = \left(a - \frac1a\right) \sin\theta \cos\theta\,,\\
  n_1 \!\!&=&\!\! \frac{\cos^2\theta + a^2 (\sin^2\theta+1)}{2a}\,,\quad
  n_2 = \frac{a^2\cos^2\theta +
\sin^2\theta+1}{2a}\,, \\
  k\!\!&=&\!\! \left(\frac{1-a^2}{2a}\right) \cos^2\theta\,,\quad
  k' = \left(\frac{a^2-1}{2a}\right) \sin\theta \cos\theta\,.\\
 \end{eqnarray*}
%Let us first note that expression \ref{eq:nonsform} can be brought
%back to the expression given on Fig.~5 of \cite{eprparis} via local
%unitary operations which do not modify the entanglement.

The CM of the squeezed ($A_\pm$) modes gives a good insight into the
properties of the two-mode state. One can see that the intermodal
blocks are zero, meaning that the two modes are uncorrelated.
Consequently, they are the two most squeezed modes of the system (no
further passive operation can select more squeezed quadratures). But
one can also note that the two diagonal blocks are not diagonal
simultaneously. This corresponds to the tilt angle of the squeezed
quadrature. In order to maximize the entanglement, the two squeezed
quadratures have to be made orthogonal, which can be done by a
phase-shift of one mode relative to the other.

\begin{figure}[t!]
\centering{\includegraphics[width=.85\columnwidth]{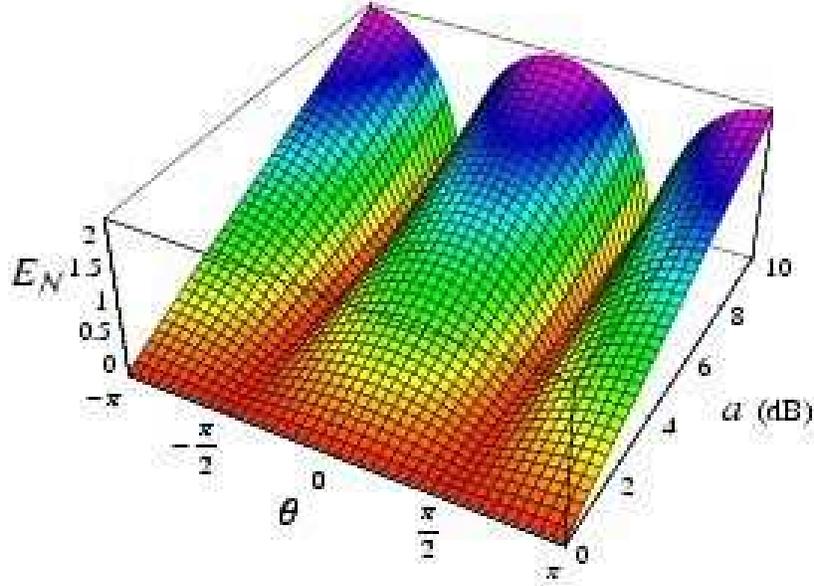}%
\caption{\label{loneper}Logarithmic negativity as a function of the
single-mode squeezing $a$ and the tilt angle $\theta$ between the
two non-orthogonal quadratures in presence of mode coupling.}}
\end{figure}

It is easy in fact to compute the logarithmic negativity quantifying
entanglement between the entangled modes $A_1$ and $A_2$, when the
two squeezed quadratures are rotated of $\pi/2 + \theta$. One has
from \eq{lognegau}, $E_\N (\sig_{A_{1}\,\!A_{2}}) = -(1/2) \log
\tilde \nu_{-}^{2}$, with
\begin{eqnarray}
\tilde\nu_{-}^2 &=& \left(\frac{1}{4a^2}\right)\Bigg\{2 \left(a^4 +
1\right) \cos ^2(\theta ) + 4 a^2 \sin ^2(\theta ) \nonumber \\
&-& \sqrt{2} \left(a^2 -1\right) \sqrt{\cos ^2(\theta) \left[a^4 + 6
a^2 + \left(a^2 - 1\right)^2 \cos (2 \theta ) + 1\right]}\Bigg\}\,.
\label{eq:nu}
\end{eqnarray}
The symplectic eigenvalue $\tilde\nu_{-}$ is obviously a periodic
function of $\theta$, and is globally minimized for $\theta =
k\,\pi$, with $k\in {\mathbbm Z}$. The entanglement, in other words,
is maximized for orthogonal modes in phase space, as already
predicted in Ref.~\cite{passive}. Notice that this results holds for
general nonsymmetric states, \ie also in the case when the two modes
$A_1$ and $A_2$ possess different individual squeezings. For
symmetric states, the logarithmic negativity is depicted as a
function of the single-mode squeezing $a$ and the tilt angle
$\theta$ in Fig.~\ref{loneper}.

In the experiment we will discuss below, the entanglement is
produced by a single device, a type-II OPO operated below threshold.
When no coupling is present in the optical cavity, the entangled
modes are along the neutral axis of the crystal while the squeezed
modes corresponds to the $\pm 45^\circ$ linear polarization basis.
However, it has been shown theoretically and experimentally
\cite{eprparis} that a coupling can be added via a birefringent
plate which modifies the quantum properties of this device: the most
squeezed quadratures are non-orthogonal with an angle depending on
the plate angle. When the plate angle increases, the squeezed
($A_-$) quadrature rotates of a tilt angle $\theta$ and the
correlations are degraded. The evolution is depicted in
Fig.~\ref{ellipse} through the noise ellipse of the superposition
modes.

\begin{figure}[t!]
\centering{\includegraphics[width=.8\columnwidth]{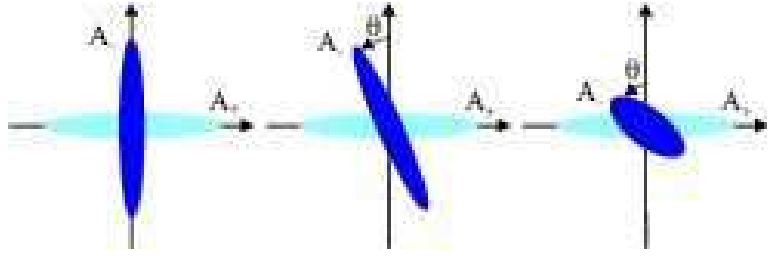}%
\caption{\label{ellipse}Fresnel representation of the noise ellipse
of the $\pm 45^{\circ}$ rotated modes when the coupling is
increased. The noise ellipse of the $-45^{\circ}$ mode rotates and
the noise reduction is degraded when the coupling increases while
the $+45^{\circ}$ rotated mode is not affected.}}
\end{figure}

\eq{eq:nu} shows that when coupling is present, it is necessary to
perform an operation on the two modes in order to optimize the
available entanglement. Before developing experimental measures of
entanglement and optimization of the available resource in our
system, let us detail our experimental setup.

\subsection{Experimental setup and homodyne measurement}
\label{sec:setup}

The experimental scheme is depicted in Fig.~\ref{setup} and relies
on a frequency-degenerate type-II OPO below threshold. The system is
equivalent to the one of the seminal experiment by Ou \emph{et al.}
\cite{ou92} but a $\lambda/4$ birefringent plate has been inserted
inside the optical cavity. When this plate is rotated, it results in
a linear coupling between the signal and idler modes which induces
above threshold a phase locking effect at exact frequency degeneracy
\cite{wong98,opophlockc}. This triply-resonant OPO is pumped below
threshold with a continuous frequency-doubled Nd$:$YAG laser. The
input flat mirror is directly coated on one face of the 10mm-long
KTP crystal. The reflectivities for the input coupler are 95\% for
the pump (532nm) and almost 100\% for the signal and idler beams
(1064nm). The output coupler (R=38mm) is highly reflecting for the
pump and its transmission is 5\% for the infrared. At exact triple
resonance, the oscillation threshold is less than 20 mW. The OPO is
actively locked on the pump resonance by the Pound-Drever-Hall
technique. The triple resonance is reached by adjustment of both the
crystal temperature and the frequency of the pump laser. Under these
conditions, the OPO can operate stably during more than one hour
without mode-hopping. The birefringent plate inserted inside the
cavity is exactly $\lambda/4$ at 1064 nm and almost $\lambda$ at the
532 nm pump wavelength. Very small rotations of this plate around
the cavity axis can be operated thanks to a piezoelectric actuator.

Measurements of the quantum properties of arbitrary quadratures of
light mode are generally made using homodyne detection
\cite{yuen83}. When an intense local oscillator is used, one obtains
a photocurrent which is proportional to the quantum noise of the
light in a quadrature defined by the phase-shift between the local
oscillator and the beam measured. This photocurrent can be either
sent to a spectrum analyzer which calculates the noise power
spectrum, or numerized for further treatments like tomographic
measurements of the Wigner function \cite{tomography} or selection
\cite{laurat03}. As mentioned above, one can also characterize the
entanglement by looking at linear combinations of the optical modes
as opposed to linear combinations of the photocurrents
\cite{Duan00,Simon00}. The two modes which form the entangled state
must be transformed via the beam-splitter transformation, that is
they are mixed on a 50/50 beam-splitter or a polarizing
beam-splitter, into two modes which will be both squeezed if the
original state is entangled.

Homodyne detection allows for a simple and direct measurement of the
$2\times 2$ diagonal blocks of the $4\times 4$ CM. In order to
measure the $2\times 2$ off-diagonal blocks, linear combinations of
the photocurrents can be used as we will show below. In order to
characterize two modes simultaneously, a single phase reference is
needed. To implement this, we have built a simultaneous double
homodyne detection (Fig.~\ref{setup}, in box). The difference
photocurrents are sent into two spectrum analyzers triggered by the
same signal (SA$_1$ and SA$_3$). Two birefringent plates (Q$_4$,
H$_3$) inserted in the local oscillator path are rotated in order to
compensate residual birefringence. A $\lambda/4$ (Q$_3$) plate can
be added on the beam exiting the OPO in order to transform the
in-phase detections into in-quadrature ones, making the
transformation $(\hat q_+, \hat p_+, \hat q_-, \hat p_-)
\ensuremath{\to} (\hat q_+, \hat p_+, \hat p_+, \hat q_+)$. In such
a configuration, two states of light with squeezing on orthogonal
quadratures give in-phase squeezing curves on the spectrum
analyzers. This has two goals: first, it simplifies the measurements
of the phase shift between the two homodyne detections, and
secondly, it is necessary for the measurement of the off-diagonal
blocks of the CM, as we will  now show.

\begin{figure}[t!]
\centering{\includegraphics[width=\columnwidth]{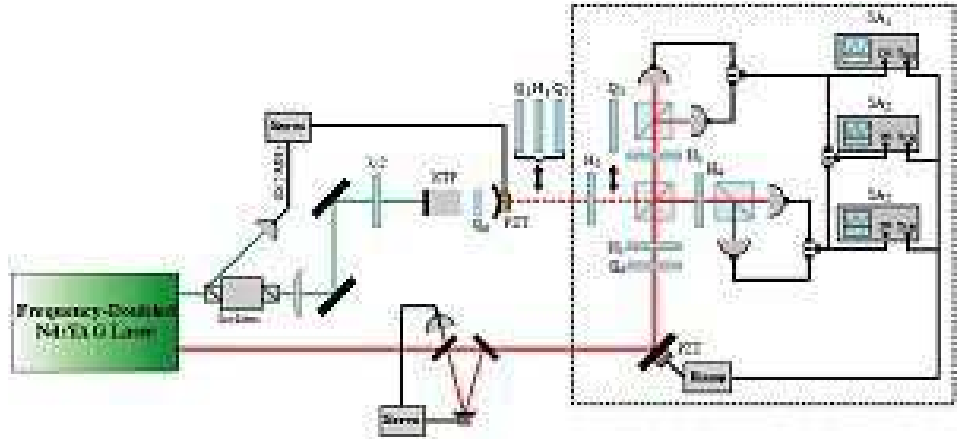}
\caption{Experimental setup. A continuous-wave frequency-doubled
Nd:YAG laser pumps below threshold a type II OPO with a $\lambda/4$
plate inserted inside the cavity (Q$_0$). The generated two-mode
vacuum state is characterized by two simultaneous homodyne
detections. The infrared output of the laser is used as local
oscillator after filtering by a high-finesse cavity. SA$_{1,2,3}$:
spectrum analyzers. Q$_{1,\ldots,4}$ and H$_{1,\ldots,5}$:
respectively quarter and half waveplates. PD Lock: FND-100
photodiode for locking of the OPO. PD Split: split two-element
InGaAs photodiode for tilt-locking of the filtering
cavity.}\label{setup}}
\end{figure}

Let us describe precisely the procedure used to extract the values
of the CM from the homodyne detection signals. These signals consist
in an arbitrary pair of spectrum analyzer traces which are
represented in Fig.~\ref{francesiextra}. The horizontal axis is the
local oscillator (LO) phase which is scanned as a function of the
time, while the vertical axis gives the noise power relative to the
shot noise expressed in decibels (dB) (for the definition of dB see
footnote {\rm \ref{notedb}} on page {\rm \pageref{notedb}}).

We make no assumption on the form of the CM which is written in the
general case
\[
\sig = \left(\begin{array}{c|c} \sig_+ & \gr\gamma_\pm\\ \hline
\overset{}{\gr\gamma\T_\pm} & \sig_-
\end{array}\right)= \left( \begin{array}{cc|cc} a & b & c & d \\
b & e & f & g
\\ \hline c & f & h & i \\ d & g & i &j \end{array}\right)\,.
\]
When the LO phase is chosen so that zero corresponds to the long
axis of the noise ellipse of the first mode, the CM is written in
the form
\[
\sig = \left( \begin{array}{cc|cc} a' & 0 & c' & d'\\ 0 & e' & f' &
g'\\ \hline c' & f' & h' & i' \\ d' & g' & i' &j'
\end{array}\right) = \left(\begin{array}{c|c} \sig_+' &
\gr\gamma_\pm^{\prime}
\\\hline \overset{}{\gr\gamma_\pm^{\prime {\sf T}}} & \sig_-'
\end{array}\right)\,,
\]
where $a'$ and $c'$ correspond respectively to the maximum and
minimum noise levels measured in a linear scale on the spectrum
analyzer for the first mode, which we will choose arbitrarily to be
$A_+$. One can also easily determine $h'$, $i'$ and $j'$ from the
spectrum analyzer traces for $A_-$: when the LO phase is chosen so
that zero corresponds to the long axis of the noise ellipse of the
second mode, the CM is written in the form
\[
\sig_-'' = \left(\begin{array}{cc} h'' & 0 \\ 0 & j''
\end{array}\right)
\]
and $\sig_-'$ can be easily deduced from $\sig_-''$ by applying a
rotation. The angle of this rotation is given by the phase shift
$\varphi$ between the two traces (see Fig.~\ref{francesiextra}).
This operation is performed numerically. We have now measured both
diagonal blocks.

\begin{figure}[t!]
\centering{\includegraphics[width=8cm]{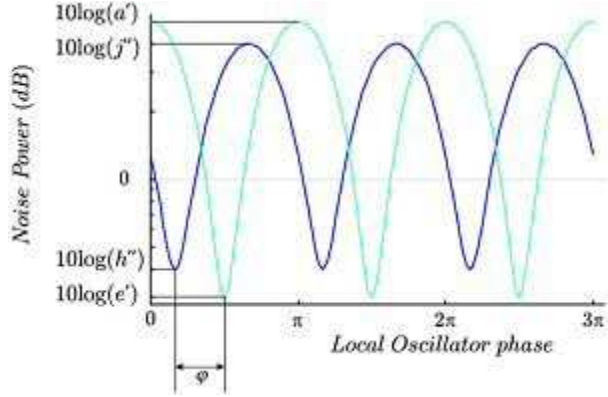}
\caption{Spectrum analyzer traces as a function of the local
oscillator phase.} \label{francesiextra}}
\end{figure}

In order to measure the non-diagonal blocks, one records on an
additional spectrum analyzer a third signal, the difference between
the two homodyne detection signals (these signals being themselves
the difference between their respective photodiodes photocurrents,
see Fig.~\ref{setup}). Let us consider the case where the waveplate
$Q_3$ is not present; for a given LO phase $\psi_1$, the homodyne
detections will give photocurrents which are proportional to the
amplitude noise for the $A_+$ beam, $\hat q_+$, and to the phase
noise for the $A_-$ beam, $\hat p_-$. The signal recorded on
spectrum analyzer SA$_2$ is, in this case, proportional to $\hat s =
\hat q_+ - \hat p_-$ whose variance is
\[
\langle \hat  i^2\rangle = \langle \hat  q_+^2\rangle + \langle \hat
p_-^2 \rangle - 2 \langle \hat  q_+ \hat  p_-\rangle = a' + j' - 2
d'.
\]
$a'$ and $j'$ being already known, it is easy to extract $d'$ from
this measurement. For a LO phase $\psi_1+\pi/2$, one will get using
a similar procedure $e'$, $h'$ and $f'$. Let us now add the wave
plate $Q_3$. For a LO phase $\psi_1$,  one will get $a'$, $h'$ and
$c'$ and for $\psi_1+\pi/2$ $e'$, $j'$ and $g'$ thus completing the
measurement of the CM.

\subsection{Experimental measures of entanglement by the
negativity} \label{sec:expmesent}

As all experimental measurements, the measurement of the CM  is
subject to noise. It is thus critical to evaluate the influence of
this noise on the entanglement. A quantitative analysis, relating
the errors on the measured CM entries (in the $A_\pm$ basis) to the
resulting error in the determination of the logarithmic negativity
(the latter quantifying entanglement between the corresponding $A_1$
and $A_2$ modes) has been carried out and is summarized in
Fig.~\ref{error} in absence of mode coupling. In general, the
determination of the logarithmic negativity is much more sensitive
to the errors on the diagonal $2\times 2$ blocks $\alpha$ and
$\beta$ (referring to the reduced states of each mode) of the CM
$\sig$ [see \eq{espre}] than to those on the off-diagonal ones
($\gamma$, and its transpose $\gamma\T$, whose expectations are
assumed to be null). Let us remark that the relative stability of
the logarithmic negativity with respect to the uncertainties on the
off-diagonal block adds to the reliability of our experimental
estimates of the entanglement. Notice also that, concerning the
diagonal blocks, the errors on diagonal (standard form) entries turn
out to affect the precision of the logarithmic negativity more than
the errors on off-diagonal (non standard form) entries. This
behavior is reversed in the off-diagonal block, for which errors on
the off-diagonal (non standard form) entries affect the uncertainty
on the entanglement more than errors on the diagonal (standard form)
entries.

\begin{figure}[t!]
\includegraphics[width=\columnwidth]{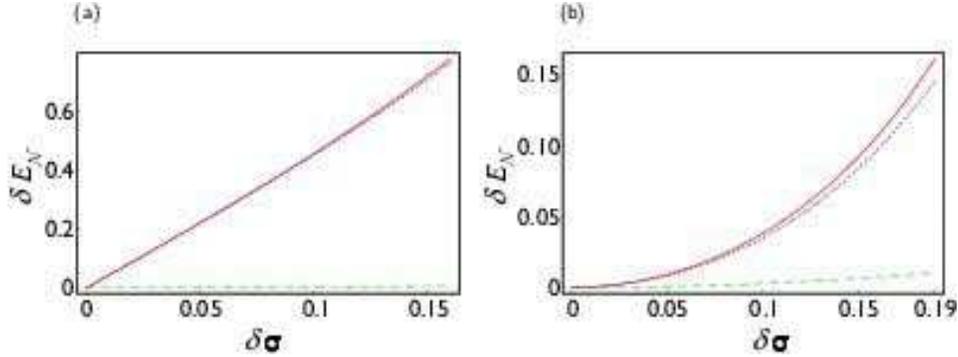}
\caption{Error $\delta E_{\CMcal N}$ on the logarithmic negativity
between modes $A_1$ and $A_2$, as a function of the error
$\delta\sig$ on the entries of the diagonal {\rm (a)} and
off-diagonal {\rm (b)} $2\times 2$ blocks of the measured CM $\sig$
in the $A_\pm$ basis, given by \eq{exem}. In plot {\rm (a)}: the
solid red curve refers to equal errors (of value $\delta\sig$) on
the eight entries of the diagonal blocks (standard form entries),
the dotted blue curve refers to equal errors on the four diagonal
entries of the diagonal blocks, while the dashed green curve refers
to equal errors on the off-diagonal entries of the diagonal blocks
(non standard form entries). At $\delta\sig\gtrsim0.16$ some of the
considered states get unphysical. In plot {\rm (b)}: the solid red
curve refers to equal errors on the four entries of the off-diagonal
block, the dotted blue curve refers to equal errors on the two
off-diagonal entries of the off-diagonal block (non standard form
entries), while the dashed green curve refers to equal errors on the
diagonal entries of the off-diagonal block (standard form entries).
At $\delta\sig\gtrsim0.19$ some of the considered states get
unphysical.}\label{error}
\end{figure}

Experimentally, we have measured the noise on the CM elements to be
at best on the order of a few percents of the measured values for
the diagonal blocks, corresponding to a fraction of a dB (see
footnote {\rm \ref{notedb}} on page {\rm \pageref{notedb}}). This is
the case for the diagonal blocks which are well-known since they are
directly related to the noise measurements of $A_+$ and $A_-$. The
situation is less favorable for the off-diagonal blocks: the
off-diagonal elements of these blocks show a large experimental
noise which, as shown on Fig.~\ref{error}{\rm(b)}, may lead in some
cases to unphysical CMs, yielding for instance a negative
determinant and complex values for the logarithmic negativity. In
the following, we will set these terms to zero in agreement with the
form of the CM of Eq.~\pref{eq:sqnondiag}.

Let us first give an example of entanglement determination from
measurements of CM elements, in the absence of mode coupling.
Without the plate, the squeezing of the two superposition modes is
expected on orthogonal quadratures: the ideal CM is then in the form
\eq{eq:sqnondiag} with $\theta=0$. Spectrum analyzer traces while
scanning the local oscillator phase are shown in Fig.~\ref{scan}:
the rotated modes are squeezed on orthogonal quadratures. The state
is produced directly in the standard form and the CM in the $A_\pm$
basis can be extracted from this measurement:
\begin{equation} \label{exem}
\sig (\rho=0) = \left(\begin{array}{cc|cc}
0.33 & 0 & (0) & (0)\\
0 & 7.94 & (0) & (0) \\ \hline
(0) & (0) & 7.94 & 0\\
(0) & (0) & 0 & 0.33
\end{array}\right)\,.
\end{equation}
The resulting smallest symplectic eigenvalue is the geometric
average of the two minimal diagonal elements, $\tilde{\nu}_- =
0.33$, yielding a logarithmic negativity $E_{\N} = - \log
(\tilde{\nu}_-) = 1.60$ between the modes $A_1$ and $A_2$.

\begin{figure}[t!]
\centering{\includegraphics[width=.85\columnwidth,clip=]{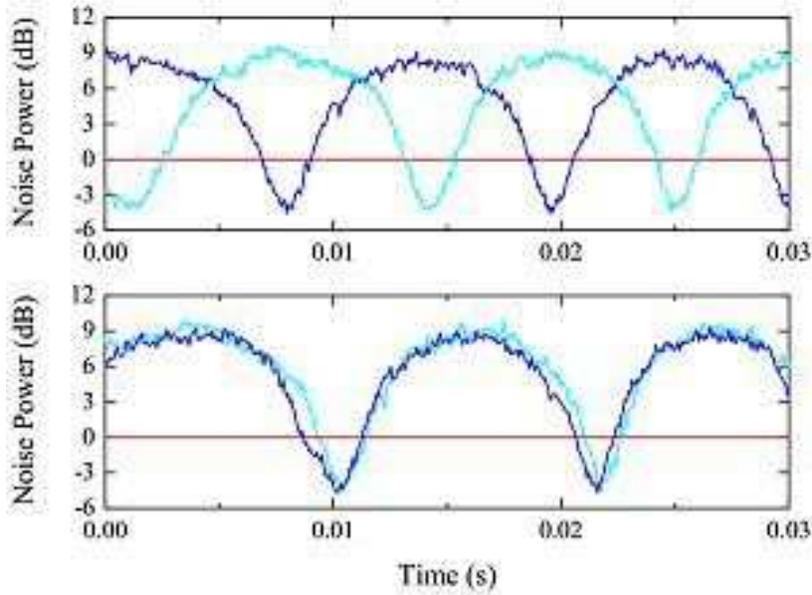}%
\caption{\label{scan}Normalized noise variances at 3.5 MHz of the
$\pm 45^{\circ}$ modes while scanning the local oscillator phase.
The first plot corresponds to in-phase homodyne detections and the
second one in-quadrature. Squeezing is well observed on orthogonal
quadratures. (RBW 100 kHz, VBW 1 kHz)}}
\end{figure}

\subsection{Experimental non standard form and optimization by
linear optics} \label{sec:nonsf}

As discussed previously, when the plate angle is increased, the
state produced is not anymore in the standard form but rather
similar to Eq.~\pref{eq:sqnondiag}. Fig.~\ref{corrige} gives the
normalized noise variances at 3.5~MHz of the rotated modes while
scanning the local oscillator phase for an angle of the plate of
$0.3^{\circ}$. The first plot shows that the squeezing is not
obtained on orthogonal quadratures. The CM takes the following form
in the `orthogonal quadratures' $A_{\mp}$:
\begin{equation}
\sig (\rho=0.3^{\circ}) = \left(\begin{array}{cc|cc}
 0.4 & 0 & (0) & (0)\\
0 & 12.59 & (0) & (0) \\ \hline
(0) & (0) & 9.54 & -5.28\\
(0) & (0) & -5.28 & 3.45
\end{array}\right) \equiv \alpha'\oplus \alpha''\, , \label{ortho}
\end{equation}
where $\alpha'$ and $\alpha''$ are $2\times 2$ submatrices, which
corresponds to the following symmetric non standard form CM on the
original quadratures $A_{1,2}$ [see \eq{eq:nonsform}]
\begin{equation}
\begin{split}
\sig' (\rho=0.3^{\circ}) &= \left(\begin{array}{cc|cc}
4.97 & -2.64 & 4.57 & -2.64\\
-2.64 & 8.02 & -2.64 & -4.57 \\ \hline
4.57 & -2.64 & 4.97 & -2.64\\
-2.64 & -4.57 & -2.64 & 8.02
\end{array}\right)\\  &= B\T(1/2)\ \sig (\rho=0.3^{\circ})\ B(1/2) \, .
\end{split}
\end{equation}
In this instance one finds for the partially transposed symplectic
eigenvalue $\tilde{\nu}_{-}\simeq 0.46$ (corresponding to a
logarithmic negativity between $A_1$ and $A_2$ much lower than the
previous value: $E_{\N}  = 1.13$), whereas the smallest eigenvalues
read $\lambda_1=\lambda_2 \simeq 0.40$. The entanglement can thus be
raised via passive operations to the optimal value $E_{\N}  = 1.32$,
corresponding to $\tilde{\nu}_{-}=\sqrt{\lambda_1\lambda_2}$.

\begin{figure}[t!]
\centering{\includegraphics[width=.85\columnwidth,clip=]{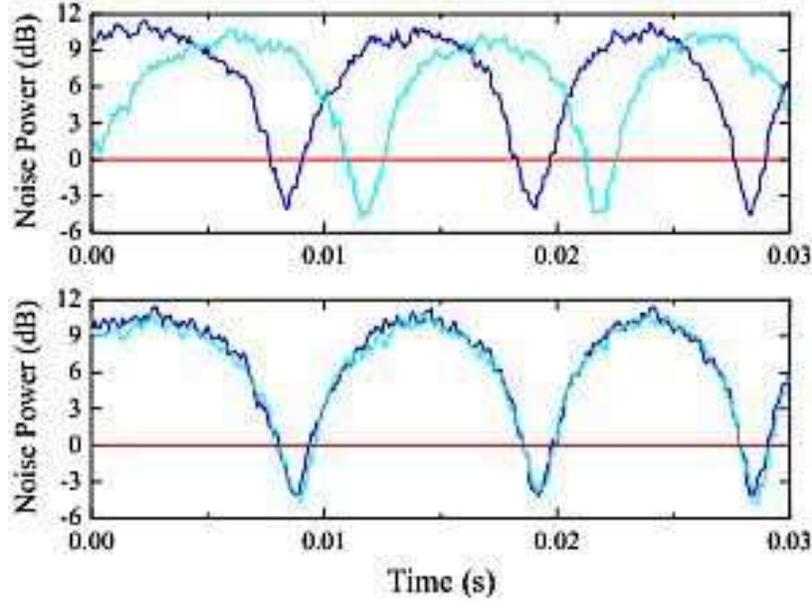}%
\caption{\label{corrige} Normalized noise variances at 3.5 MHz of
the rotated modes while scanning the local oscillator phase for an
angle of the plate of $0.3^{\circ}$, before and after the non-local
operation. The homodyne detections are in-quadrature. After this
operation, squeezing is observed on orthogonal quadratures.}}
\end{figure}

The passive transformation capable of optimizing the entanglement is
easily found, according to the theoretical analysis of
Sec.~\ref{sec:modecoupling}. If $O$ is the rotation diagonalizing
the $2\times 2$ symmetric matrix $\alpha''$ defined in \eq{ortho},
then the transformation $K\equiv B\T(1/2) ({\mathbbm 1} \oplus O)
B(1/2)$ turns the CM $\sig (\rho=0.3^{\circ})$ into $\bar{\sig}
(\rho=0.3^{\circ})$, which is diagonal in the orthogonal quadratures
$A_{\mp}$ and in a symmetric standard form in the quadratures
$A_{1,2}$. The entanglement of such a matrix is therefore optimal
under passive operations. The optimal symplectic operation $K$
consists in a `phase-shift' of the rotated modes $A_{1,2}$. In the
experimental practice, such an operation can be readily performed on
co-propagating, orthogonally polarized beams \cite{simon94}. The
minimal combination of waveplates needed can be shown to consist in
three waveplates: two $\lambda/4$ waveplates denoted Q and one
$\lambda/2$ waveplate denoted H. When using any combination of these
three plates, the state can be put back into standard form which
will maximize the entanglement.  Fig.~\ref{corrige} gives the
normalized noise variances before and after this operation. In
agreement with the theory, the CM is changed into:
\begin{equation}
\bar{\sig} (\rho=0.3^{\circ}) = \left(\begin{array}{cc|cc}
0.4 & 0 & (0) & (0)\\
0 & 12.59 & (0) & (0) \\ \hline
(0) & (0) & 12.59 & 0\\
(0) & (0) & 0 & 0.4
\end{array}\right) \, .
\end{equation}
giving the expected optimal logarithmic negativity $E_{\N} = 1.32$
between $A_1$ and $A_2$, larger than the value before the operation.
{\em No more entanglement} can be generated by passive operations on
this Gaussian state, which has been experimentally transformed into
the form which achieves the maximum possible bipartite quantum
correlations.

Let us remark again that this transformation is non-local in the
sense of the EPR argument \cite{EPR35}: it has to be performed
before spatially separating the entangled modes for a quantum
communication protocol for instance.

\subsection{Summary of the experiment}

We have given the flavor of the powerful tools underlying the
description of CV systems in quantum optics. These tools allow for a
nice pictorial view of two-mode Gaussian entangled states.
Specifically, we have experimentally achieved the following
\cite{francamentemeneinfischio}.

\medskip

\begin{itemize}
\item[\ding{226}]
 \noindent{\rm\bf Experimental production and manipulation of two-mode entanglement.}
{\it Continuous-variable entanglement in two-mode Gaussian states
has been produced experimentally with an original device, a type-II
optical parametric oscillator containing a birefringent plate; it
has been quantitatively measured by homodyne reconstruction of the
standard form covariance matrix, and optimized using purely passive
operations.}
\smallskip
\end{itemize}

We have also studied quantitatively the influence of the noise,
affecting the measurement of the elements of the CM, on the
entanglement, showing that the most significant covariances
(exhibiting the highest stability against noise) for an accurate
entanglement quantification are the diagonal terms of the diagonal
single-mode blocks, and the off-diagonal terms of the intermodal
off-diagonal block, the latter being the most difficult to measure
with high precision. Alternative methods have been devised to tackle
this problem \cite{fiurasek04,rigolinew} based on direct
measurements of global and local invariants of the CM \cite{prl}, as
introduced in Sec.~\ref{secFiura}. Such techniques have been
implemented in the case of pulsed beams \cite{wenger04} but no
experiment to date has been performed for continuous-wave beams.

}

%\chapter{Optical state engineering of Gaussian states}

\chapter{Tripartite and four-partite state engineering}
\label{Chap3M4Mengi}

{\sf

In this Chapter, based mainly on Ref.~\cite{3mj}, we provide the
reader with a systematic analysis of state engineering of the
several classes of three-mode Gaussian states characterized by
peculiar structural and/or entanglement properties, introduced in
Chapter \ref{Chap3M} (Secs.~\ref{secstructex} and
\ref{secstructexmix}). We will also briefly discuss the instance of
those four-mode Gaussian states exhibiting unlimited promiscuous
entanglement \cite{unlim}, introduced in Chapter \ref{ChapUnlim}.
For every family of Gaussian states, we will outline practical
schemes for their production with current optical technology.

General recipes to produce pure Gaussian states of an arbitrary
number of modes will be presented in the next Chapter.

\section{Optical production of three-mode Gaussian states}

\subsection{The ``allotment'' box for engineering arbitrary three-mode pure states}\label{secallot}
The structural properties of generic  {\em pure} three-mode Gaussian
states, and their standard form under local operations, have been
discussed in Sec.~\ref{secpuri}. The computation of the tripartite
entanglement, quantified by the residual Gaussian contangle of
\eq{gtaures},  for those states has been presented in full detail in
Sec.~\ref{sec3purent}. Here we investigate how to produce pure
Gaussian states of three modes  with optical means, allowing for any
possible entanglement structure.

A viable scheme to produce all pure three-mode Gaussian states, as
inspired by the Euler decomposition \cite{pramana} (see also
Appendix \ref{SecEuler}), would combine three independently squeezed
modes (with in principle all different squeezing factors) into any
conceivable combination of orthogonal (energy preserving) symplectic
operations (essentially, beam-splitters and phase-shifters, see
Sec.~\ref{SecSympl}). This procedure, that is obviously legitimate
and will surely be able to generate any pure state, is however not,
in general, the most economical one in terms of physical resources.
Moreover, this procedure is not particularly insightful because the
degrees of bipartite and tripartite entanglement of the resulting
output states are not, in general, easily related to the performed
operations.

Here, we want instead to give a precise recipe providing the exact
operations to achieve an arbitrary three-mode pure Gaussian state
with CM in the standard form of \eq{cm3tutta}. Therefore, we aim at
producing states with any given triple $\{a_1,\,a_2,\,a_3\}$ of
local mixednesses, and so with any desired `physical'
[{\ie}constrained by \ineq{triangleprim}] asymmetry among the three
modes and any needed amount of tripartite entanglement. Clearly,
such a recipe is {\em not} unique. We provide here one possible
scheme,\footnote{\sf An alternative scheme to produce pure
three-mode Gaussian states is provided in Sec.~\ref{SecEngiGeneric},
where the state engineering of generic pure $N$-mode Gaussian states
is discussed.} which may not be the cheapest one but possesses a
straightforward physical interpretation: the distribution, or {\em
allotment} of two-mode entanglement among three modes. Our scheme
will be optimal in that it relies exactly on 3 free parameters, the
same number as the degrees of freedom (the three local mixednesses)
characterizing any pure three-mode Gaussian state up to local
unitaries.

Explicitly, one starts with modes $1$ and $2$ in a two-mode squeezed
state, and mode $3$ in the vacuum state. In Heisenberg picture:
\begin{eqnarray}
  &\hat q_1 = \frac{1}{\sqrt{2}}\left(e^{r}\ \hat q_1^0+e^{-r}\ \hat q_2^0\right)\,,&\quad
  \hat p_1 =  \frac{1}{\sqrt{2}}\left(e^{-r}\ \hat p_1^0+e^{r}\ \hat p_2^0\right)\,, \label{mode1bas}  \\
    &\hat q_2 = \frac{1}{\sqrt{2}}\left(e^{r}\ \hat q_1^0-e^{-r}\ \hat q_2^0\right)\,,&\quad
  \hat p_2 =  \frac{1}{\sqrt{2}}\left(e^{-r}\ \hat p_1^0-e^{r}\ \hat p_2^0\right)\,, \label{mode2bas}  \\
  &\hat q_3 = \hat q_3^0\,,&\quad
  \hat p_3 =  \hat p_3^0\,, \label{mode3bas}
\end{eqnarray}
where the suffix ``0'' refers to the vacuum.  The reason why we
choose to have from the beginning a two-mode squeezed state, and not
one or more independently squeezed single modes, is that, as already
mentioned (see also Sec.~\ref{SecSympl}), two-mode squeezed states
can be obtained in the lab either directly in non-degenerate
parametric processes (as demonstrated in Sec.~\ref{secFrancesi}) or
indirectly by mixing two squeezed vacua at a beam-splitter (as
depicted in Fig.~\ref{figtms}). Depending on the experimental setup,
any means to encode two-mode squeezing will be therefore legitimate
to the aim of re-allot it among three modes, as we will now show.

The three initial modes are then sent in a sequence of three
beam-splitters [see \eq{bsplit}], which altogether realize what we
call {\em allotment} operator \cite{3mj} and denote by
$\hat{A}_{123}$, see Fig.~\ref{allocco}:
\begin{equation}\label{allot}
\hat{A}_{123} \equiv \hat{B}_{23} (\arccos\sqrt{2/3}) \cdot
\hat{B}_{12} (\arccos\sqrt{t}) \cdot \hat{B}_{13}
(\arccos\sqrt{s})\,.
\end{equation}

\begin{figure}[t!]
\centering{
\includegraphics[width=7.5cm]{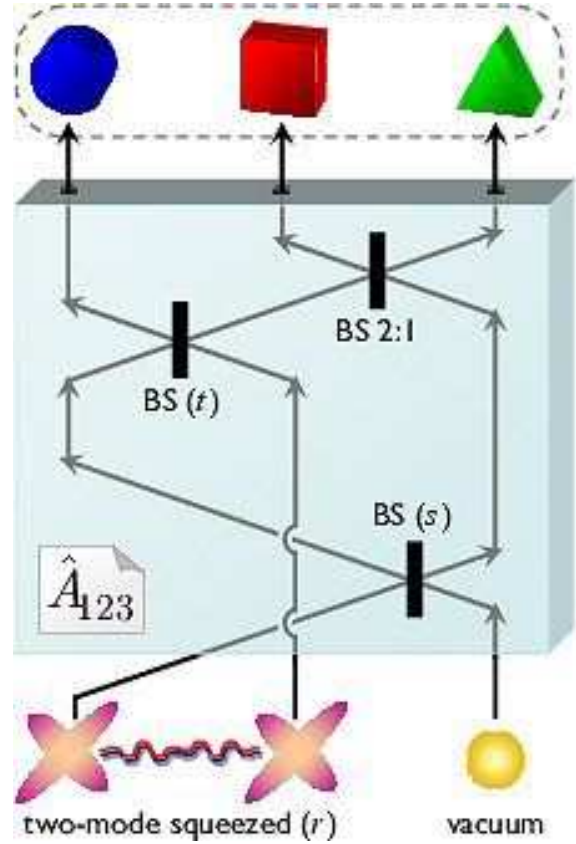}
\caption{Scheme to produce generic pure three-mode Gaussian states.
A two-mode squeezed state and a single-mode vacuum are combined by
the ``allotment'' operator $\hat{A}_{123}$, which is a sequence of
three beam-splitters, \eq{allot}. The output yields a generic pure
Gaussian state of modes $1$
(\textcolor[rgb]{0.00,0.00,1.00}{\ding{108}}), $2$
(\textcolor[rgb]{1.00,0.00,0.00}{\ding{110}}), and $3$
(\textcolor[rgb]{0.00,1.00,0.00}{\ding{115}}), whose CM depends on
the initial squeezing factor $m = \cosh(2r)$ and on two
beam-splitter transmittivities $s$ and $t$.} \label{allocco}}
\end{figure}

It is convenient in this instance to deal with the phase-space
representations of the states ({\ie}their CM) and of the operators
({\ie}the associated symplectic transformations, see
Sec.~\ref{SecSympl}). The three-mode input state is described by a
CM $\sig^p_{in}$ of the form \eq{CM} for $N=3$, with [see \eq{tms}]
\begin{eqnarray}
  &\sig_1=\sig_2=m\ \id_2\,,\quad&\sig_3=\id_2\,, \\
  &\eps_{12} = {\rm diag}
  \left\{\sqrt{m^2-1},\,-\sqrt{m^2-1}\right\}\,,\quad&\eps_{13}=\eps_{23}= {\bf
  0}\,,
\end{eqnarray}
and $m\equiv \cosh(2r)$. A beam-splitter with transmittivity $\tau$
corresponds to a rotation of $\theta = \arccos\sqrt{\tau}$ in phase
space, see \eq{bsplit}. In a three-mode system, the symplectic
transformation corresponding to $\hat B_{ij}(\theta)$ is a direct
sum of the matrix $B_{ij}(\tau)$, \eq{bbs}, acting on modes $i$ and
$j$, and of the identity $\id_2$ acting on the remaining mode $k$.

The output state after the allotment will be denoted by a CM
$\sig_{out}^p$ given by
\begin{equation}\label{aftall}
\sig_{out}^p = A_{123} \sig_{in}^p A_{123}^{\sf T}\,,
\end{equation}
where $A_{123}$ is the phase-space representation of the allotment
operator \eq{allot}, obtained from the matrix product of the three
beam-splitter transformations. The output state is clearly pure
because the allotment is a unitary operator (symplectic in phase
space). The elements of the CM $\sig_{out}^p$,  not reported here
for brevity, are functions of the three parameters
\begin{equation}\label{mst}
m \in [1,\,\infty),\,\quad s \in [0,\,1],\,\quad t \in [0,\,1]\,,
\end{equation}
being respectively related to the initial squeezing in the two-mode
squeezed state of modes $1$ and $2$, and two beam-splitter
transmittivities (the transmittivity of the third beam-splitter is
fixed). In fact, by letting these three parameters vary in their
respective domain, the presented procedure allows for the creation
of {\em arbitrary} three-mode pure Gaussian states (up to local
unitaries), with any possible triple of local mixednesses
$\{a_1,\,a_2,\,a_3\}$ ranging in the physical region defined by the
triangle inequality \pref{triangleprim}.

This can be shown as follows. Once identified $\sig_{out}^p$ with
the block form of \eq{CM} (for $N=3$), one can solve analytically
the equation $\det \sig_1 = a_1^2$ to find
\begin{equation}\label{mgen}
m(a_1,s,t)=\begin{array}{c} \frac{t \left[t (s - 1)^2 + s - 1\right]
+ \sqrt{a_1^2 (s t + t - 1)^2  +
            4 s (t - 1) t (2 t - 1) (2 s t - 1)}}{(s t + t -
            1)^2}\,.
            \end{array}
\end{equation}
Then, substituting \eq{mgen} in $\sig_{out}^p$ yields a
reparametrization of the output state in terms of $a_1$ (which is
given), $s$ and $t$. Now solve (numerically) the system of nonlinear
equations $\{\det\sig_2 = a_2^2,\,\det\sig_3 = a_3^2\}$ in the
variables $s$ and $t$. Finally, substitute back the obtained values
of the two transmittivities in \eq{mgen}, to have the desired triple
$\{m,\,s,\,t\}$ as functions of the local mixednesses
$\{a_1,\,a_2,\,a_3\}$ characterizing the target state.

\begin{figure}[t!]
\centering{
\includegraphics[width=8cm]{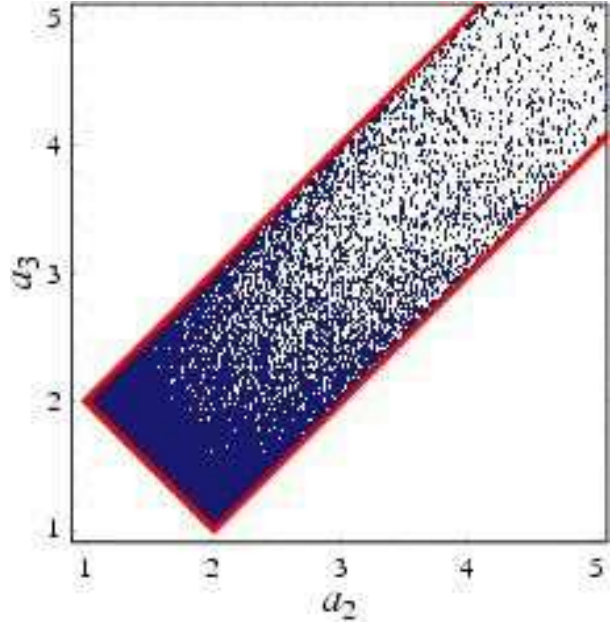}
\caption{Plot of 100 000 randomly generated pure three-mode Gaussian
states, described by their single-mode mixednesses $a_2$ and $a_3$,
at fixed $a_1=2$. The states are produced by simulated applications
of the allotment operator with random beam-splitter transmittivities
$s$ and $t$, and span the whole physical range of parameters allowed
by \ineq{triangleprim}. A comparison of this plot with
Fig.~\ref{figangle}{\rm(b)}  may be instructive. See text for
further details.} \label{pompilio}}
\end{figure}

We have therefore demonstrated the following \cite{3mj}.

\medskip

\begin{itemize}
\item[\ding{226}]
 \noindent{\rm\bf State engineering of pure three-mode Gaussian states.}
{\it  An arbitrary pure three-mode Gaussian state, with a CM locally
equivalent to the standard form of \eq{cm3tutta}, can be produced
with the current experimental technology by linear quantum optics,
employing the allotment box --- a passive redistribution of two-mode
entanglement among three modes --- with exactly tuned amounts of
input two-mode squeezing and beam-splitter properties, without any
free parameter left.}
\smallskip
\end{itemize}

 A pictorial test of this procedure is shown in
Fig.~\ref{pompilio}, where at a given local mixedness of mode $1$
($a_1=2$), several runs of the allotment operator have been
simulated with randomized beam-splitter transmittivities $s$ and
$t$. Starting from a two-mode squeezed input with $m$ given by
\eq{mgen}, tensor a vacuum, the resulting output states are plotted
in the space of $a_2$ and $a_3$. By comparing Fig.~\ref{pompilio}
with Fig.~\ref{figangle}{\rm(b)}, one clearly sees that the randomly
generated states distribute towards a complete fill of the physical
region emerging from the triangle inequality \pref{triangleprim},
thus confirming the generality of our scheme.

\subsection{Tripartite state engineering handbook and simplified schemes}

Having a generalization of the ``allotment'' for the production of
arbitrary {\em mixed} three-mode Gaussian states turns out to be a
quite involved task. However, for many classes of tripartite states
introduced in Chapter \ref{Chap3M}, efficient state engineering
schemes can be devised. Also in special instances of pure states,
depending in general on less than three parameters, cheaper recipes
than the general one in terms of the allotment box are available. We
will now complement the entanglement analysis of
Secs.~\ref{secstructex} and \ref{secstructexmix} with such practical
proposals, as presented in Ref.~\cite{3mj}.

\subsubsection{CV GHZ/{\it W} states}\label{SecEngiGHZW}

\begin{figure}[t!]
\centering{
\includegraphics[width=9.5cm]{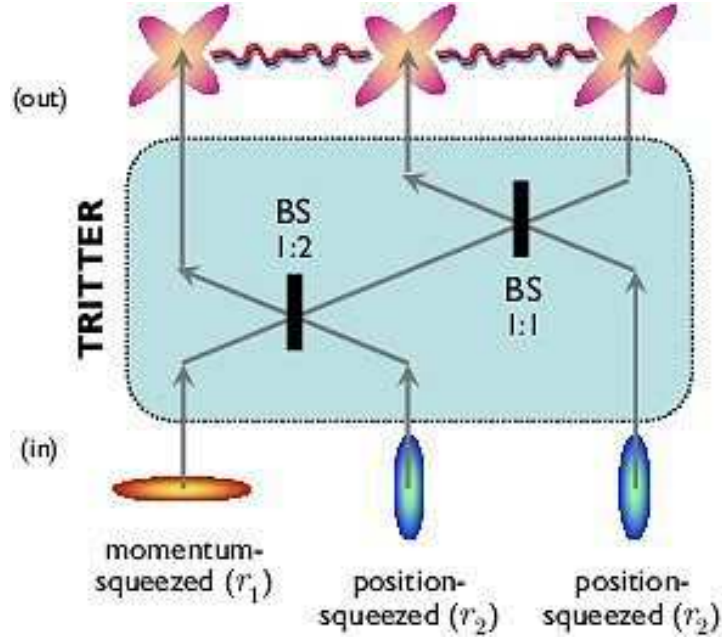}
\caption{Scheme to produce CV GHZ/$W$ states, as proposed in
Ref.~\cite{network} and implemented in Ref.~\cite{3mexp}. Three
independently squeezed beams, one in momentum and two in position,
are combined through a double beam-splitter (tritter). The output
yields a pure, symmetric, fully inseparable three-mode Gaussian
state, also known as CV GHZ/$W$ state.} \label{fighzw}}
\end{figure}

Several schemes have been proposed to produce what we call
finite-squeezing GHZ/$W$ states of continuous variables, \ie fully
symmetric pure three-mode Gaussian states with promiscuous
entanglement sharing (see Sec.~\ref{secghzw}). In particular, as
discussed by van Loock and Braunstein \cite{network}, these states
can be produced by mixing three squeezed beams through a double
beam-splitter, or {\em tritter} \cite{BraunsteinERR}. One starts
with mode $1$ squeezed in momentum, and modes $2$ and $3$ squeezed
in position. In Heisenberg picture:
\begin{eqnarray}
  &\hat q_1 = e^{r_1}\ \hat q_1^0\,,&\quad
  \hat p_1 =  e^{-r_1}\ \hat p_1^0\,, \label{momsq3}  \\
   &\hat q_{2,3} =  e^{-r_2}\ \hat q_{2,3}^0\,,&\quad
  \hat p_{2,3} =  e^{r_2}\ \hat p_{2,3}^0\,, \label{possq3}
\end{eqnarray}
where the suffix ``0'' refers to the vacuum. Then one combines the
three modes in a tritter
\begin{equation}\label{tritter}
\hat{B}_{123} \equiv \hat{B}_{23}(\pi/4)\cdot\hat{B}_{12}
(\arccos\sqrt{1/3})\,,
\end{equation}
where the action of an ideal (phase-free) beam-splitter operation
$\hat{B}_{ij}$ on a pair of modes $i$ and $j$ is defined by
\eq{bsplit}.

 The output of the tritter yields a CM of the form
\eq{fscm} with
\begin{eqnarray}
\alp = {\rm diag} \left\{
 \frac{1}{3} \left(e^{2 r_1} + 2 e^{-2 r_2}\right)\,,\quad
 \frac{1}{3} \left(e^{-2 r_1} + 2 e^{2 r_2}\right)\right\}\,, \label{sighzwa}\\
\eps = {\rm diag} \left\{
 \frac{1}{3} \left(e^{2 r_1} - e^{-2 r_2}\right)\,,\quad
\frac{1}{3} \left(e^{-2 r_1} - e^{2 r_2}\right)\right\}\,.
\label{sighzwe}
\end{eqnarray}
This resulting pure and fully symmetric three-mode Gaussian state,
obtained in general with differently squeezed inputs $r_1 \neq r_2$,
is locally equivalent to the state prepared with all initial
squeezings equal to the average $\bar r = (r_1 + r_2)/2$ (this will
be discussed in more detail in connection with teleportation
experiments, see Sec.~\ref{SecTelepoppy}).

The CM described by Eqs.~{\rm(\ref{sighzwa},\ref{sighzwe})}
represents a CV GHZ/$W$ state. It can be in fact transformed, by
local symplectic operations, into the standard form CM given by
\eq{epmfulsym}, with
\begin{equation}\label{aghzw}
a=\frac{1}{3} \sqrt{4 \cosh \left[2 \left(r_1 + r_2\right)\right] +
5}\,.
\end{equation}
The preparation scheme of CV GHZ/$W$ states is depicted in
Fig.~\ref{fighzw}. It has been experimentally implemented
\cite{3mexp}, and the fully inseparability of the produced states
has been verified through the violation of the separability
inequalities derived in Ref.~\cite{vloock03}. Very recently, the
production of strongly entangled GHZ/$W$ states has also been
demonstrated by using a novel optical parametric oscillator, based
on concurrent $\chi^{(2)}$ nonlinearities \cite{pfister}.

\subsubsection{Noisy GHZ/{\it W} states} \label{SecEngiGHZWnoisy}

Noisy GHZ/$W$ states, whose entanglement has been characterized in
Sec.~\ref{secnoisyghzw}, can be obtained as GHZ/$W$ states generated
from (Gaussian) thermal states: one starts with three single-mode
squeezed thermal states (with average photon number $\bar n
=[n-1]/2$) and combine them through a tritter \eq{tritter}, with the
same procedure described in Fig.~\ref{fighzw} for $n=1$. The initial
single, separable, modes are thus described by the following
operators in Heisenberg picture,
\begin{eqnarray}
  &\hat q_1 = \sqrt{n} e^{r}\ \hat q_1^0\,,&\quad
  \hat p_1 =  \sqrt{n} e^{-r}\ \hat p_1^0\,, \label{momthsq}  \\
   &\hat q_{2,3} =  \sqrt{n} e^{-r}\ \hat q_{2,3}^0\,,&\quad
  \hat p_{2,3} =  \sqrt{n} e^{r}\ \hat p_{2,3}^0\,. \label{posthsq}
\end{eqnarray}
Defining $s \equiv e^{2r}$, at the output of the tritter one obtains
a CM of the form \eq{fscm}, with
\begin{eqnarray}
\alp = {\rm diag} \left\{
 \frac{n(s^2+2)}{3s}\,,\quad
 \frac{n(2s^2+1)}{3s}\right\}\,, \label{sitha}\\
\eps = {\rm diag} \left\{
 \frac{n(s^2-1)}{3s}\,,\quad
-\frac{n(s^2-1)}{3s}\right\}\,. \label{sithe}
\end{eqnarray}
This resulting CM is locally equivalent to the standard form of
\eq{epmthermal}, with
\begin{equation}\label{athermal}
a = \frac{n \sqrt{2 s^4 + 5 s^2 + 2}}{3 s}\,.
\end{equation}
Here $s$ is the same as in \eq{noiseffs}, and was indeed defined
there by inverting \eq{athermal}.

Let us also mention again that noisy GHZ/$W$ states would also
result from the dissipative evolution of pure GHZ/$W$ states in
proper Gaussian noisy channels (see Sec.~\ref{decoherence}).

\subsubsection{{\it T} states}\label{SecEngiT}

The $T$ states have been introduced in Sec.~\ref{sectstates} to show
that in symmetric three-mode Gaussian states, imposing the absence
of reduced bipartite entanglement between any two modes results in a
frustration of the genuine tripartite entanglement. It may be useful
to know how to produce this novel class of mixed Gaussian states in
the lab.

The simplest way to engineer $T$ states is to reutilize the scheme
of Fig.~\ref{fighzw}, {\ie}basically the tritter, but with different
inputs. Namely,  one has mode $1$ squeezed again in momentum (with
squeezing parameter $r$), but this time modes $2$ and $3$ are in a
thermal state (with average photon number $\bar n = [n(r)-1]/2$,
depending on $r$). In Heisenberg picture:
\begin{eqnarray}
  &\hat q_1 = e^{r}\ \hat q_1^0\,,&\quad
  \hat p_1 =  e^{-r}\ \hat p_1^0\,, \label{momsqt}  \\
   &\hat q_{2,3} =  \sqrt{n(r)}\ \hat q_{2,3}^0\,,&\quad
  \hat p_{2,3} =  \sqrt{n(r)}\ \hat p_{2,3}^0\,, \label{thermt}
  \\ \mbox{with}\quad &n(r) = \sqrt{3 + e^{-4 r}} - e^{-2 r}
  \nonumber \,.&
\end{eqnarray}
Sending these three modes in a tritter \eq{tritter} one recovers, at
the output, a $T$ state whose CM is locally equivalent to the
standard form of \eq{epmtstat}, with
\begin{equation}\label{atstat}
a = \frac{1}{3} \sqrt{2 e^{-2 r} \sqrt{3 + e^{-4 r}} \left(-3 + e^{4
r}\right) + 6 e^{-4 r} + 11}\,.
\end{equation}

\subsubsection{Basset hound states}\label{secbasengi}

A scheme for producing the basset hound states of Sec.~\ref{secbas},
and in general the whole family of pure bisymmetric Gaussian states
known as ``multiuser quantum channels'' (due to their usefulness for
telecloning, as we will show in Sec.~\ref{sectlc}), is provided in
Ref.~\cite{telecloning}. In the case of three-mode basset hound
states of the form given by Eqs.~{\rm(\ref{bassigl},\ref{basseps})},
one can use a simplified version of the allotment introduced in
Sec.~\ref{secallot} for arbitrary pure states. One  starts with a
two-mode squeezed state (with squeezing parameter $r$) of modes $1$
and $2$, and mode $3$ in the single-mode vacuum,  like in
Eqs.~{\rm(\ref{mode1bas}--\ref{mode3bas})}.

 Then, one combines one half (mode 2) of the two-mode squeezed state  with
the vacuum mode 3 via a 50:50 beam-splitter, described in phase
space by $B_{23}(1/2)$, \eq{bbs}. The resulting three-mode state is
exactly a basset hound state described by
Eqs.~{\rm(\ref{bassigl},\ref{basseps})}, once one identifies
%\begin{equation}\label{acoshbas}
$a \equiv \cosh(2r)$.
%\end{equation}
In a realistic setting, dealing with noisy input modes, mixed
bisymmetric states can be obtained as well by the same procedure.

\section{How to produce and exploit unlimited promiscuous
entanglement?}\label{Sec4Mengi}

In Chapter \ref{ChapUnlim}, four-mode Gaussian states with an
unlimited promiscuous entanglement sharing have been introduced.
Their definition, \eq{s4}, involves three instances of a two-mode
squeezing transformation of the form \eq{tmsS}. From a practical
point of view, two-mode squeezing transformations are basic tools in
the domain of quantum optics \cite{barnett}, occurring \eg in
parametric down-conversions (see Sec.~\ref{secFrancesi} for more
technical details). Therefore, we can readily provide an all-optical
preparation scheme for our promiscuous states \cite{unlim}, as shown
in Fig.~\ref{figprep4}.

\begin{figure}[t!]
\includegraphics[width=8cm]{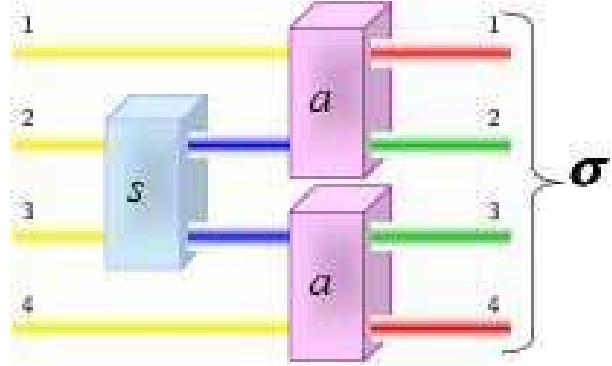} \caption{
Preparation of the four-mode Gaussian states $\gr\sigma$ of \eq{s4}.
Starting with four initially uncorrelated modes of light, all
residing in the respective vacuum states (yellow beams), one first
applies a two-mode squeezing transformation (light blue box), with
squeezing $s$, between the central modes 2 and 3, and then two
additional two-mode squeezing transformations (light pink boxes),
each one with equal squeezing $a$, acting on the pair of modes 1,2
and 3,4 respectively.  The resulting state is endowed with a
peculiar, yet insightful bipartite entanglement structure,
pictorially depicted in Fig.~\ref{figprep}.} \label{figprep4}
\end{figure}

It is interesting to observe that the amount of producible squeezing
in optical experiments is constantly improving \cite{furuapl}. Only
technological, no {\em a priori} limitations need to be overcome to
increase $a$ and/or $s$ to the point of engineering excellent
approximations to the demonstrated promiscuous entanglement
structure, elucidated in Chapter \ref{ChapUnlim}, in multimode
states of light and atoms (see also \cite{suexp}).

To make an explicit example, already with realistic squeezing
degrees like $s=1$ and $a=1.5$ (corresponding to $\sim$ 3 dB and 10
dB, respectively, where decibels are defined in  footnote {\rm
\ref{notedb}} on page {\rm \pageref{notedb}}), one has a bipartite
entanglement of
${G_\tau}(\gr\sigma_{1|2})={G_\tau}(\gr\sigma_{3|4})=9$ ebits
(corresponding to a Gaussian entanglement of formation \cite{GEOF},
see Sec.~\ref{SecGEMS}, of $\sim 3.3\ $ebits), coexisting with a
residual multipartite entanglement of
${G_\tau}^{res}(\gr\sigma)\simeq 5.5$ ebits, of which the tripartite
portion is at most ${G_\tau}^{bound}(\gr\sigma _{1\vert 2\vert 3})
\simeq 0.45$ ebits. This means that one can simultaneously extract
at least $3$ qubit singlets from each pair of modes $\{1,2\}$ and
$\{3,4\}$, and more than a single copy of genuinely four-qubit
entangled states (like cluster states \cite{cluster}). Albeit with
imperfect efficiency, this entanglement transfer can be realized by
means of Jaynes-Cummings interactions \cite{pater}, representing a
key step for a reliable physical interface between fields and qubits
in a distributed quantum information processing network (see also
Refs.~\cite{memorypolzik,telepolzik}).

}

%%%%%%%%%%%%%%%%%%%%%%%%%%%%%%%%%%%%%%%%%%%%%%%%%%%%%%%%%%%%%%%%%
\chapter{Efficient production of pure {\em N}-mode Gaussian
states}\label{ChapGeneric}
%\label{SecGeneric}

%\section{prl}

{\sf

Recently, a great insight into the role of entanglement in the
description of quantum systems has been gained through the quantum
information perspective, mostly focusing on the usefulness of
entanglement, rather than on its meaning. In these years, quantum
entanglement has turned from a paradoxical concept into a physical
resource allowing for the encoding, manipulation, processing and
distribution of information in ways forbidden by the laws of
classical physics. In this respect, we have evidenced how CV
entanglement between canonically conjugate observables of
infinite-dimensional systems, like harmonic oscillators, light modes
and atomic ensembles, has emerged as a versatile and powerful
resource. In particular, multimode Gaussian states have been proven
useful for a wide range of implementations in CV quantum information
processing \cite{brareview}, and advances in the characterization of
their bipartite and multipartite entanglement have been recently
recorded (see the previous Parts of this Dissertation).

In experiments, one typically aims at preparing pure states, with
the highest possible entanglement, even though unavoidable losses
and thermal noises will affect the purity of the engineered
resources, and hence the efficiency of the realized protocols (a
direct evidence is reported in Sec.~\ref{secFrancesi}). It is
therefore important to understand the structure of correlations in
pure Gaussian states of an arbitrary number of modes, and to provide
economical schemes to produce such states in the lab with minimal
elements, thus reducing the possibility of accumulating errors and
unwanted noise.

\section{Degrees of freedom of pure Gaussian states: practical
perspectives}\label{secdegreesoperational}

In the instance of two- and three-mode Gaussian states, efficient
schemes for their optical production with minimal resources have
been presented, respectively, in Chapters \ref{Chap2MExp} and
\ref{Chap3M4Mengi}. In the general case of pure $N$-mode Gaussian
states with $N>3$, we know from \eq{npuri} that the minimal number
of degrees of freedom characterizing all their structural and
informational properties (up to local unitaries) is $(N^2-2N)$ (see
also Sec.~\ref{secpurecount} and Appendix \ref{ChapAppendixSF}).

It would be desirable to associate the mathematically clear number
$(N^2-2N)$ with an operational, physical insight. In other words, it
would be useful for experimentalists (working, for instance, in
quantum optics) to be provided with a recipe to create pure $N$-mode
Gaussian states with completely general entanglement properties in
an economical way, using exactly $(N^2-2N)$ optical elements, such
as squeezers, beam-splitters and phase shifters. A transparent
approach to develop such a procedure consists in considering the
reverse of the phase space $1 \times (N-1)$ Schmidt decomposition,
as introduced in Sec.~\ref{SecSchmidtPS}. Namely, a completely
general (not accounting for the local invariances) state engineering
prescription for pure Gaussian states can be cast in two main steps:
(i) create a two-mode squeezed state of modes 1 and 2, which
corresponds to the multimode state in its Schmidt form; (ii) operate
with the most general $(N-1)$-mode symplectic transformation
$S^{-1}$ on the block of modes $\{2,3,\ldots,N\}$ (with modes
$i=3,\ldots,N$ initially in the vacuum state) to redistribute
entanglement among all modes. The operation $S^{-1}$ is the inverse
of the transformation $S$ which brings the reduced CM of modes
$\{2,3,\ldots,N\}$ in its Williamson diagonal form, see
Sec.~\ref{SecWilly}. It is also known that any such symplectic
transformation $S^{-1}$ (unitary on the Hilbert space) can be
decomposed in a network of optical elements \cite{reckzeil}. The
number of elements required to accomplish this network, however,
will in general greatly exceed the minimal number of parameters on
which the entanglement between any two sub-systems depends. Shifting
the local-unitary optimization from the final CM, back to the
engineering symplectic network, is in principle an extremely
involved and nontrivial task.

This problem has been solved in Ref.~\cite{generic} for a special
subclass of Gaussian states, which is of null measure but still of
central importance for practical implementations. It is constituted
by those pure $N$-mode Gaussian states which can be locally put in a
standard form with all diagonal $2\times 2$ submatrices in \eq{CM}
(\ie with null $\sig_{qp}$ in the notation of Appendix
\ref{ChapAppendixSF}). This class encompasses generalized GHZ-type
Gaussian states, useful for CV quantum teleportation networks
\cite{network} (see Sec.~\ref{SecTelepoppy}), Gaussian cluster
states \cite{zhang,clusterloo} employed in CV implementations of
one-way quantum computation \cite{menicucci}, and states of four or
more modes with an unlimited promiscuous entanglement sharing (see
Chapter \ref{ChapUnlim}). It also comprises (as proven in
Sec.~\ref{secpuri}) {\em all} three-mode pure Gaussian states, whose
usefulness for CV quantum communication purposes has been thoroughly
investigated in this Dissertation (see Chapter \ref{Chap3M}, Chapter
\ref{Chap3M4Mengi}, and Sec.~\ref{sectlc}). In the physics of
many-body systems, those states are quite ubiquitous as they are
ground states of harmonic Hamiltonians with spring-like interactions
\cite{chain}. As such, they admit an efficient ``valence bond''
description, as discussed in Chapter \ref{ChapGVBS}.

For these Gaussian states, which we will call here {\em
block-diagonal}
--- with respect to the  canonical operators reordered as $(\hat
q_1,\,\hat q_2,\,\ldots,\,\hat q_N,\,\hat p_1,\,\hat
p_2,\,\ldots,\,\hat p_N)$ ---  the minimal number of
local-unitarily-invariant parameters reduces to $N(N-1)/2$ for any
$N$.\footnote{\sf This number is easily derived from the general
framework developed in Appendix \ref{condpuri}: for $\sig_{qp}=0$,
Eqs.~\pref{first} and \pref{second} reduce to
$\sig_{q}=\sig_{p}^{-1}$. The only further condition to impose after
the local reduction is then ${\rm diag}(\sig_{q})= {\rm diag}
(\sig_{q}^{-1})$, which brings the number of free parameters of the
symmetric $\sig_{q}$ from $(N+1)N/2$ down to $N(N-1)/2$.}
Accordingly, one can show that an efficient scheme  can be devised
to produce block-diagonal pure Gaussian states, involving exactly
$N(N-1)/2$ optical elements which in this case are only constituted
by single-mode squeezers and beam-splitters, in a given sequence
\cite{generic}.

We will now detail the derivation of these results explicitly, as it
will lead to an important physical insight into the entanglement
structure (which we define ``generic'') of such block-diagonal
Gaussian states. The latter, we recall, are basically all the
resources currently produced and employed in optical realizations of
CV quantum information and communication processing.

\section[Generic entanglement and state engineering of block-diagonal pure states]
{Generic entanglement, standard form and state engineering of
block-diagonal pure Gaussian states}

\subsection{Generic entanglement of Gaussian
states}\label{SecGeneric}

In this Section, based on Ref.~\cite{generic} we address the
question of how many physical resources are really needed to
engineer and characterize entanglement in pure Gaussian states of an
arbitrary number of modes, up to local unitary operations. Let us
recall again (see Sec.~\ref{secpurecount}) that for states of $N \le
3$ modes, it has been shown that such a number of minimal degrees of
freedom scales as $N(N-1)/2$ . For a higher number of modes,
however, a richer structure is achievable by pure Gaussian states,
as from symplectic arguments like those of Appendix \ref{redu}  a
minimal number of parameters given by $N(N-2)$ can be inferred. A
random state of $N \ge 4$ modes, selected according to the uniform
distribution over pure Gaussian states, will be thus reducible to a
form characterized by such a number of independent quantities.

However, in practical realizations of CV quantum information one is
interested in states which, once prepared with efficient resources,
still achieve an almost complete structural variety in their
multipartite entanglement properties. Such states will be said to
possess {\em generic entanglement} \cite{genericP}, where generic
means practically equivalent to that of random states, but
engineered (and described) with a considerably smaller number of
degrees of freedom.

Precisely, we define as ``generic-entangled'' those Gaussian states
whose local entropies of entanglement in any single mode are
independent, and bipartite entanglements between any pair of modes
are unconstrained. Having a standard form for such $N$-mode Gaussian
states, may be in fact extremely helpful in understanding and
quantifying multipartite CV entanglement, in particular from the
theoretical point of view of entanglement sharing and monogamy
constraints (see Chapter \ref{ChapMonoGauss}), and from a more
pragmatical approach centered on using entanglement as a resource.

We show that, to achieve generic entanglement, for the global pure
$N$-mode Gaussian state it is enough to be described by a minimal
number of
 parameters (corresponding to the local-unitarily invariant
degrees of freedom) equal to $N(N-1)/2$ for any $N$, and thus much
smaller than the $2N(2N+1)/2$ of a completely general CM. Therefore,
generic entanglement appears in states which are highly {\em not}
`generic' in the sense usually attributed to the term, {\em
i.e.}~randomly picked. Crucially, we demonstrate that
``generic-entangled'' Gaussian states coincide with the above
defined ``block-diagonal'' Gaussian states, \ie with the resources
typically employed in experimental realizations of CV quantum
information \cite{brareview}. Accordingly, we provide an optimal and
practical scheme for their state engineering.

\subsection{Minimal number of parameters}\label{secprop1}

Adopting the above definition of generic entanglement, we prove now
the main
\begin{prop} \label{propgeneric}
A generic-entangled $N$-mode pure Gaussian state is described, up to
local symplectic (unitary) operations, by $N(N-1)/2$ independent
parameters.
\end{prop}

\smallskip

\noindent {\it Proof.} Let us start with a $N$-mode pure state,
described by a CM $\sig^p \equiv \sig$ as in \eq{CM}, with all
single-mode blocks $\sig_i$ ($i=1\ldots N$) in diagonal form: we can
always achieve this by local single-mode Williamson diagonalizations
in each of the $N$ modes. Let $\sig^{\backslash 1}\equiv
\sig_{2,\cdots,N}$ be the reduced CM of modes $(2,\dots,N)$. It can
be diagonalized by means of a symplectic $S_{2,\cdots,N}$, and
brought thus to its Williamson normal form, characterized by a
symplectic spectrum $\{a,1,\cdots,1\}$, where $a =
\sqrt{\det{\sig_1}}$. Transforming $\sig$ by $S=\id_1 \oplus
S_{2,\cdots,N}$, brings the CM into its Schmidt form, constituted by
a two-mode squeezed state between modes $1$ and $2$ (with squeezing
$a$), plus $N-2$ vacua \cite{holevo01,botero03,giedkeqic03} (see
Sec.~\ref{SecSchmidtPS}).

All $N$-mode pure Gaussian states are thus completely specified by
the symplectic $S_{2,\cdots,N}$, plus the parameter $a$.
Alternatively, the number of parameters of $\sig$ is also equal to
that characterizing an arbitrary mixed $(N-1)$-mode Gaussian CM,
with symplectic rank $\aleph=1$ ({\em i.e.}~with $N-2$ symplectic
eigenvalues equal to $1$, see Sec.~\ref{SecSympHeis}). This means
that, assigning the reduced state $\sig_{2,\cdots,N}$, we have
provided a complete description of $\sig$. In fact,  the parameter
$a$ is determined as the square root of the determinant of the CM
$\sig_{2,\cdots,N}$.

We are now left to compute the minimal set of parameters of an
arbitrary mixed state of $N-1$ modes, with symplectic rank
$\aleph=1$. While we know that for $N\ge4$ this number is equal to
$N(N-2)$ in general, we want to prove that for generic-entangled
Gaussian resource states this number reduces to
\begin{equation}
\label{numero} \Xi_N={N(N-1)}/{2}\,.
\end{equation}
We prove it by induction. For a pure state of one mode only, there
are no reduced ``zero-mode'' states, so the number is zero. For a
pure state of two modes, an arbitrary one-mode mixed CM with
$\aleph=1$ is completely determined by its own determinant, so the
number is one. This shows that our law for $\Xi_N$ holds true for
$N=1$ and $N=2$.

Let us now suppose that it holds for a generic $N$, {\em i.e.}~we
have that a mixed $(N-1)$-mode CM with $\aleph=1$ can be put in a
standard form specified by $N(N-1)/2$ parameters. Now let us check
what happens for a $(N+1)$-mode pure state, {\em i.e.}~for the
reduced $N$-mode mixed state with symplectic rank equal to $1$. A
general way (up to LUs) of constructing a $N$-mode CM with
$\aleph=1$ yielding generic entanglement is the following: (a) take
a generic-entangled $(N-1)$-mode CM with $\aleph=1$ in standard
form; (b) append an ancillary mode ($\sig_N$) initially in the
vacuum state (the mode cannot be thermal as $\aleph$ must be
preserved); (c) squeeze mode $N$ with an arbitrary $s$ (one has this
freedom because it is a local symplectic operation); (d) let mode
$N$ interact couplewise with all the other modes, via a chain of
beam-splitters, \eq{bbs},  with arbitrary transmittivities
$b_{i,N}$, with $i=1,\cdots,N-1$;\footnote{\sf Squeezings and
beam-splitters are basic entangling tools in CV systems, see
Sec.~\ref{SecSympl}. For $N \ge 4$, steps (c) and (d) should be
generalized to arbitrary one- and two-mode symplectic
transformations to achieve all possible Gaussian states, as
discussed in Sec.~\ref{secdegreesoperational}.} (e) if desired,
terminate with $N$ suitable single-mode squeezing operations (but
with all squeezings now {\em fixed} by the respective reduced CM's
elements) to symplectically diagonalize each single-mode CM.

With these steps one is able to construct a mixed state of $N$
modes, with the desired rank, and with  generic
(local-unitarily-invariant) properties for each single-mode
individual CM. We will show in the following that in the considered
states the pairwise quantum correlations between any two modes are
unconstrained. To conclude, let us observe that the constructed
generic-entangled state is specified by a number of parameters equal
to: $N(N-1)/2$ (the parameters of the starting $(N-1)$-mode mixed
state of the same form) plus $1$ (the initial squeezing of mode $N$)
plus $N-1$ (the two-mode beam-splitter interactions between mode $N$
and each of the others). Total: $(N+1)N/2 = \Xi_{N+1}$.  \hfill
$\blacksquare$

\subsection{Quantum state engineering}\label{SecEngiGeneric}

Following the ideas of the above proof, a physically insightful
scheme to produce generic-entangled  $N$-mode pure Gaussian states
can be readily presented (see Fig.~\ref{figuro}). It consists of
basically two main steps: (1) creation of the state in the $1 \times
(N-1)$ Schmidt decomposition; (2) addition of modes and entangling
operations between them. One starts with a chain of $N$ vacua.

First of all (step 1), the recipe is to squeeze mode $1$ of an
amount $s$, and mode 2 of an amount $1/s$ ({\em i.e.}~one squeezes
the first mode in one quadrature and the second, of the same amount,
in the orthogonal quadrature); then one lets the two modes interfere
at a $50:50$ beam-splitter. One has so created a two-mode squeezed
state between modes $1$ and $2$ (as in Fig.~\ref{figtms}), which
corresponds to the Schmidt form of $\sig$ with respect to the $1
\times (N-1)$ bipartition. The second step basically corresponds to
create the most general mixed state with $\aleph=1$, of modes
$2,\cdots,N$, out of its Williamson diagonal form. This task can be
obtained, as already sketched in the above proof, by letting each
additional mode interact step-by-step with all the previous ones.
Starting with mode $3$ (which was in the vacuum like all the
subsequent ones), one thus squeezes it (of an amount $r_3$) and
combines it with mode $2$ via a beam-splitter $B_{2,3}(b_{2,3})$,
\eq{bbs} (characterized by a transmittivity $b_{2,3}$). Then one
squeezes mode $4$ by $r_4$ and lets it interfere sequentially, via
beam-splitters, both with mode $2$ (with transmittivity $b_{2,4}$)
and with mode $3$ (with transmittivity $b_{3,4}$). This process can
be iterated for each other mode, as shown in Fig.~\ref{figuro},
until the last mode $N$ is squeezed ($r_N$) and entangled with the
previous ones via beam-splitters with respective transmittivities
$b_{i,N}$, $i=2,\cdots,N-1$. Step 2 describes the redistribution of
the two-mode entanglement created in step 1, among all modes. We
remark that mode $1$ becomes entangled with all the other modes as
well, even if it never comes to a direct interaction with each of
modes $3,\cdots,N$.

\begin{figure}[t!]
\centering{\includegraphics[width=5.5cm]{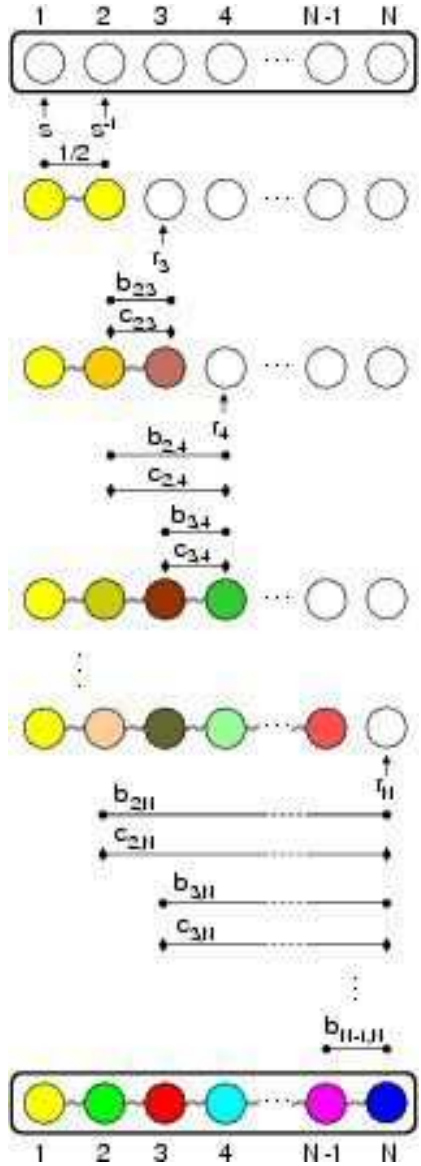}%
\caption{\label{figuro}How to create a generic-entangled  $N$-mode
pure Gaussian state. White balls are vacua, while each color depicts
a different single-mode determinant ({\em i.e.}~different degrees of
local mixedness). Vertical arrows denote single-mode squeezing
operations with squeezing parameters $r_j$, while horizontal
circle-ended lines denote beam-splitting operations, \eq{bbs}, with
transmittivities $b_{i,j}$, acting on modes $i$ and $j$. See text
for details.}}
\end{figure}

The presented prescription enables to create a generic form (up to
local unitaries) of multipartite entanglement among $N$ modes in a
pure Gaussian state, by means of active (squeezers) and passive
(beam-splitters) linear optical elements. What is relevant for
practical applications, is that the state engineering is implemented
with minimal resources. Namely, the process is characterized by  one
squeezing degree (step 1), plus $N-2$ individual squeezings for step
2, together with $\sum_{i=1}^{N-2} i = (N-1)(N-2)/2$ beam-splitter
transmittivities, which amount to a total of $N(N-1)/2 \equiv \Xi_N$
quantities. The optimally produced Gaussian states can be readily
implemented for $N$-party CV communication networks (see Chapter
\ref{ChapCommun}).

A remark is in order. From \eq{npuri}, it follows that the scheme of
Fig.~\ref{figuro}, in the special case $N=3$, allows for the
creation of {\em all} pure three-mode Gaussian states with CM in
standard form, \eq{cm3tutta}. In other words, up to local unitaries,
all pure three-mode Gaussian states exhibit generic entanglement as
defined above. Therefore, the state engineering recipe  presented
here represents an alternative to the allotment box of
Fig.~\ref{allocco}, introduced in Sec.~\ref{secallot}. In both
schemes, the input is a two-mode squeezed state of modes 1 and 2
(whose squeezing degree accounts for one of the three degrees of
freedom of pure three-mode Gaussian states, see Sec.~\ref{secpuri})
 and mode 3 in the vacuum state. The main difference is that in the
present scheme (Fig.~\ref{figuro}), an additional {\em active}
operation is implemented, as the third mode is squeezed in the first
place; a single beam-splitter (between modes 2 and 3) is then enough
to achieve a completely general entanglement structure  in the three
modes, up to local unitaries. On the other hand, the allotment of
Fig.~\ref{allocco}, as the name itself suggests, is realized by a
{\em passive} redistribution of entanglement only, as the third mode
is not squeezed but the three modes interfere with each other via
three beam-splitters (one of which has fixed transmittivity) and
this again yields a completely general entanglement freedom, up to
local unitaries. Depending on the experimental facilities, one can
thus choose either scheme when aiming to produce pure three-mode
Gaussian states.

\subsection{Standard forms: generic-entangled \ding{214}
block-diagonal} \label{secgenericsf}

The special subset of pure $N$-mode Gaussian states emerging from
our constructive proof exhibits a distinct property: all
correlations between ``position'' $\hat q_i$ and ``momentum'' $\hat
p_j$ operators are vanishing. Looking at \eq{CM}, this means that
any such generic-entangled pure Gaussian state can be put in a {\em
standard form} where all the $2\times2$ submatrices of its CM are
diagonal. The class of pure Gaussian states exhibiting generic
entanglement {\em coincides} thus with that formed by the
``block-diagonal'' states discussed in Secs.~\ref{secpurecount} and
\ref{secdegreesoperational}.

The diagonal subblocks $\sig_i$ can be additionally made
proportional to the identity by local Williamson diagonalizations in
the individual modes. This standard form for generic-entangled
$N$-mode Gaussian states, as already mentioned, can be achieved by
{\em all} pure Gaussian states for $N=2$ \cite{Duan00} and $N=3$
\cite{3mpra} (see Sec.~\ref{secpuri}); for $N\ge4$, pure Gaussian
states can exist whose number of independent parameters scales as
$N(N-2)$  and which cannot thus be brought in the $\hat q$-$\hat p$
block-diagonal form. Interestingly, all pure Gaussian states in our
considered block-diagonal standard form, are ground states of
quadratic Hamiltonians with spring-like interactions \cite{chain}.
Let us now investigate the physical meaning of the standard form.

Vanishing $\hat q$-$\hat p$ covariances imply that the $N$-mode CM
can be written as a direct sum (see also Appendix
\ref{ChapAppendixSF}) $\sig = \sig_q \oplus \sig_p$, when the
canonical operators are arranged as $(\hat q_1,\ldots,\hat
q_N,\,\hat p_1,\ldots,\hat p_N)$. Moreover, the global purity of
$\sig$ imposes $\sig_p = \sig_q^{-1}$. Named
 $(\sig_q)_{ij}=v_{q_{ij}}$ and
 $(\sig_p)_{hk}=v_{p_{hk}}$,
this means that each $v_{p_{hk}}$ is a  function of the
$\{v_{q_{ij}}\}$'s. The additional $N$ Williamson conditions
$v_{p_{ii}}=v_{q_{ii}}$ fix the diagonal elements of $\sig_q$. The
standard form is thus completely specified by the off-diagonal
elements of the symmetric $N\times N$ matrix $\sig_q$, which are, as
expected, $N(N-1)/2 \equiv \Xi_N$ from \eq{numero}.

 Proposition \rref{propgeneric} of Sec.~\ref{secprop1} acquires now
the following remarkable physical insight \cite{generic}.

\medskip

\begin{itemize}
\item[\ding{226}]
 \noindent{\rm\bf Generic entanglement of pure Gaussian states.}
{\it The structural properties of pure block-diagonal $N$-mode
Gaussian states, and in particular their bipartite and multipartite
entanglement, are completely specified (up to local unitaries) by
the `two-point correlations' $v_{q_{ij}} = \langle \hat q_i \hat q_j
\rangle$ between any pair of modes, which amount to $N(N-1)/2$
locally invariant degrees of freedom.}
\smallskip
\end{itemize}

For instance, the entropy of entanglement between one mode (say $i$)
and the remaining $N-1$ modes, which is monotonic in $\det \sig_i$
(see Sec.~\ref{SecEntroG}), is completely specified by assigning all
the pairwise correlations between mode $i$ and any other mode $j
\neq i$, as $\det \sig_i =1- \sum_{j \neq i} \det \eps_{ij}$ from
\eq{osos}. The rationale is that entanglement in such states is
basically reducible to a mode-to-mode one. This statement, strictly
speaking true only for the pure Gaussian states for which
Proposition \rref{propgeneric} holds, acquires a general validity in
the context of the modewise decomposition of arbitrary pure Gaussian
states \cite{holevo01,botero03,giedkeqic03}, as detailed in
Sec.~\ref{SecSchmidtPS}. We remark that such an insightful
correlation picture breaks down for mixed Gaussian states, where
also classical, statistical-like correlations arise.

\section{Economical state engineering of arbitrary pure Gaussian states?}

Borrowing the ideas leading to the state engineering of
block-diagonal pure Gaussian states \cite{generic}, see
Fig.~\ref{figuro}, we propose here a scheme \cite{sformato},
involving $(N^2-2N)$ independent optical elements, to produce more
general $N$-mode pure Gaussian states, encoding correlations between
positions and momentum operators as well. To this aim, we introduce
`counter-beam-splitter' transformations, named ``{\em
seraphiques}'', which, recovering the phase space ordering of
Sec.~\ref{secIntroCV}, act on two modes $j$ and $k$ as
\be\label{serafica}C_{j,k}(\tau) = \left(
\begin{array}{cccc}
 \sqrt{\tau} & 0 & 0 & \sqrt{1-\tau} \\
 0 & \sqrt{\tau} & - \sqrt{1-\tau} & 0 \\
 0 & \sqrt{1-\tau} & \sqrt{\tau} & 0 \\
 - \sqrt{1-\tau} & 0 & 0 & \sqrt{\tau}
\end{array}
\right)\,, \ee where the amplitude $\tau$ is related to an angle
$\theta$ in phase space by $\tau = \cos^2\theta$. Such operations
can be obtained from usual beam-splitters $B_{j,k}(\tau)$, \eq{bbs},
by applying a $\pi/2$ phase shifter $P_{k}$ on {\em only one} of the
two considered modes. $P_k$ is a local rotation mapping, in
Heisenberg picture, $\hat{q}_{k}\mapsto-\hat{p}_{k}$ and
$\hat{p}_{k}\mapsto\hat{q}_{k}$. In phase space, one has
$C_{j,k}(\tau)=P_{k}^{\sf T} B_{j,k}(\tau)P_{k}$. Notice that, even
though $C_{j,k}(\tau)$ is equal to the product of single-mode
operations and beam-splitters, this does not mean that such a
transformation is ``equivalent'' to a beam-splitter in terms of
state generation. In fact, the local operations do not commute with
the beam-splitters, so that a product of the kind
$B_{j,k}(\tau')C_{j,k}(\tau'')$ {\em cannot} be written as
$B_{j,k}(\tau)S_{l}$ for some local operation $S_{l}$ and $\tau$.

\begin{figure}[t!]
\centering{\includegraphics[width=5.5cm]{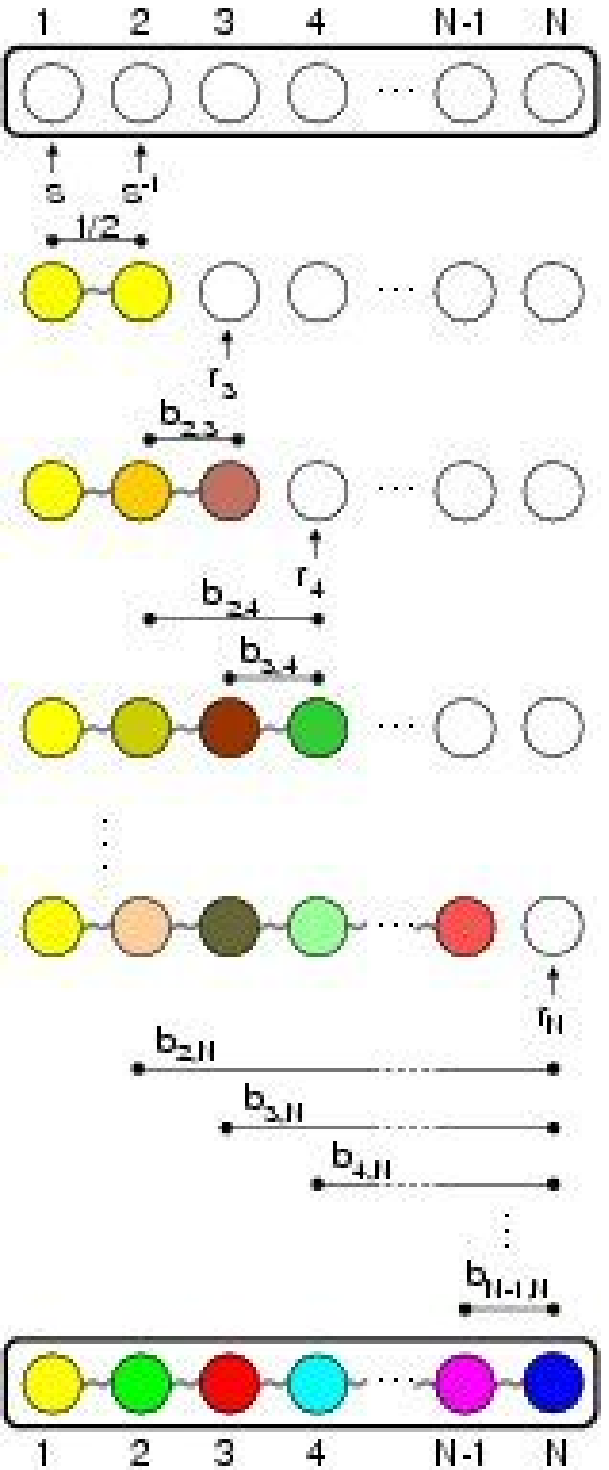}%
\caption{\label{schemino}Possible scheme to create a general
$N$-mode pure Gaussian state. White balls are vacua, while each
color depicts a different single-mode determinant ({\em
i.e.}~different degrees of local mixedness). Vertical arrows denote
single-mode squeezing operations with squeezing parameters $r_j$,
horizontal circle-ended lines denote beam-splitting operations,
\eq{bbs}, with transmittivity $b_{i,j}$ between modes $i$ and $j$,
and horizontal diamond-ended lines denote two-mode seraphiques,
\eq{serafica}, with amplitudes $c_{i,j}$. See text for details.}}
\end{figure}

The state engineering scheme runs along exactly the same lines as
the one for the block-diagonal states, Sec.~\ref{SecEngiGeneric},
the only modification being that for each pair of modes except the
last one ($N-1,N$), a beam-splitter transformation is followed by a
seraphique. In more detail (see Fig.~\ref{schemino}): first of all
(step 1), one squeezes mode $1$ of an amount $s$, and mode $2$ of an
amount $1/s$ ({\em i.e.}~one squeezes the first mode in one
quadrature and the second, of the same amount, in the orthogonal
quadrature); then one lets the two modes interfere at a $50:50$
beam-splitter. One has so created a two-mode squeezed state between
modes $1$ and $2$, which corresponds to the Schmidt form of the pure
Gaussian state with respect to the $1 \times (N-1)$ bipartition. The
second step basically corresponds to a re-distribution of the
initial two-mode entanglement among all modes; this task can be
obtained  by letting each additional $k$ mode ($k=3\ldots N$)
interact step-by-step with all the previous $l$ ones ($l=2\ldots
k-1$), via beam-splitters and seraphiques (which are in turn
combinations of beam-splitters and phase shifters). It is easy to
see that this scheme is implemented with minimal resources. Namely,
the state engineering process is characterized by one squeezing
degree (step 1), plus $N-2$ individual squeezings, together with
$\sum_{i=1}^{N-2} i = (N-1)(N-2)/2$ beam-splitter transmittivities,
and $[\sum_{i=1}^{N-2} i]-1 = N(N-3)/2$ seraphique amplitudes, which
amount to a total of $(N^2-2N)$ quantities, exactly the ones
parametrizing a generic pure Gaussian state of $N\ge3$ modes up to
local symplectic operations, \eq{npuri}.

While this scheme (Fig.~\ref{schemino}) is surely more general than
the one for block-diagonal states (Fig.~\ref{figuro}), as it enables
to efficiently create a broader class of pure Gaussian states for
$N>3$, we will leave it as an open question to check if it is
general enough to produce {\em all} pure $N$-mode Gaussian states up
to local unitaries. Verifying this analytically leads to pretty
intractable expressions already for $N=4$. Instead, it would be very
interesting to investigate if the average entanglement of the output
Gaussian states numerically obtained by a statistically significant
sample of applications of our scheme with random parameters, matches
the {\em typical} entanglement of pure Gaussian states under
``thermodynamical'' state-space measures as computable along the
lines of Ref.~\cite{typical}. This would prove the optimality and
generality of our scheme in an operational way, which is indeed more
useful for practical applications.

%%%%%%%%%%%%%%%%%%%%%%%%%%%%%%%%%%%%%%%%%%%%%%%%%%%%%%%%%%%
\section{Generic versus typical entanglement: are off-block-diagonal correlations
relevant?} \label{secgentyp}

The structural properties of pure $N$-mode Gaussian states under
local operations have been addressed in Sec.~\ref{secpurecount} and
completely characterized in Appendix \ref{ChapAppendixSF}
\cite{sformato}. Here \cite{generic}, block-diagonal states (\ie
with no correlations between position and momentum operators) have
been in particular proven to possess generic entanglement in the
sense of Sec.~\ref{SecGeneric}, and their standard form covariances
(determining any form of entanglement in such states) have been
physically understood in terms of two-body correlations. It is thus
quite natural to question if the $N(N-3)/2$ additional parameters
encoded in $\hat q$-$\hat p$ correlations for non-block-diagonal
pure states, have a definite impact or not on the bipartite and
multipartite entanglement.

 At present,
usual CV protocols are devised, even in multimode settings (see
Chapter \ref{ChapCommun}), to make use of states without any $\hat
q$-$\hat p$ correlations. In such cases, the economical (relying on
$N(N-1)/2$ parameters) ``block-diagonal state engineering'' scheme
detailed in Fig.~\ref{figuro} is clearly the optimal general
strategy for the production of entangled resources. However,
theoretical considerations strongly suggest that states with
$\sig_{qp}\neq0$ [adopting the notation of \eq{sig}] might have
remarkable potential for improved quantum-informational
applications. In fact, considering as just mentioned the
thermodynamical entanglement framework of Gaussian states
\cite{typical}, one can define natural averages either on the whole
set of pure Gaussian states, or restricting to states with
$\sig_{qp}=0$. Well, numerics unambiguously show \cite{sformato}
that the (thermodynamically-averaged) ``generic'' entanglement
(under any bipartition) of Gaussian states without $\hat q$-$\hat p$
correlations (like the ones considered in Sec.~\ref{SecGeneric}) is
systematically lower than the ``typical''  entanglement of
completely general pure Gaussian states, with this behavior getting
more and more manifest as the total number of modes $N$ increases
(clearly, according to Sec.~\ref{secpurecount}, this discrepancy
only arises for $N>3$). In a way, the full entanglement potential of
Gaussian states is diminished by the restriction to block-diagonal
states.

On the other hand,  the comparison between the average entanglement
generated in randomizing processes based on the engineering scheme
of Fig.~\ref{schemino}, and the block-diagonal one of
Fig.~\ref{figuro}, is under current investigation as well. If the
general scheme of Fig.~\ref{schemino}, based on beam-splitters and
seraphiques, turned out to be out-performing the simpler ones (like
the one of Fig.~\ref{figuro}, based on beam-splitters only) in terms
of entanglement generation --- as expected in view of the argument
above --- this would provide us with a formidable motivation to
explore novel CV protocols capable of adequately exploiting $\hat
q$-$\hat p$ correlated resources.

}

%%%%%%%%%%%%%%%%%%%%%%%%%%%%%%%%%%%%%%%%%%%%%%%%%%%%%%%%%%%%%%%%
\part{Operational interpretation and applications of Gaussian
entanglement}{\vspace*{1cm}
\includegraphics[width=\columnwidth]{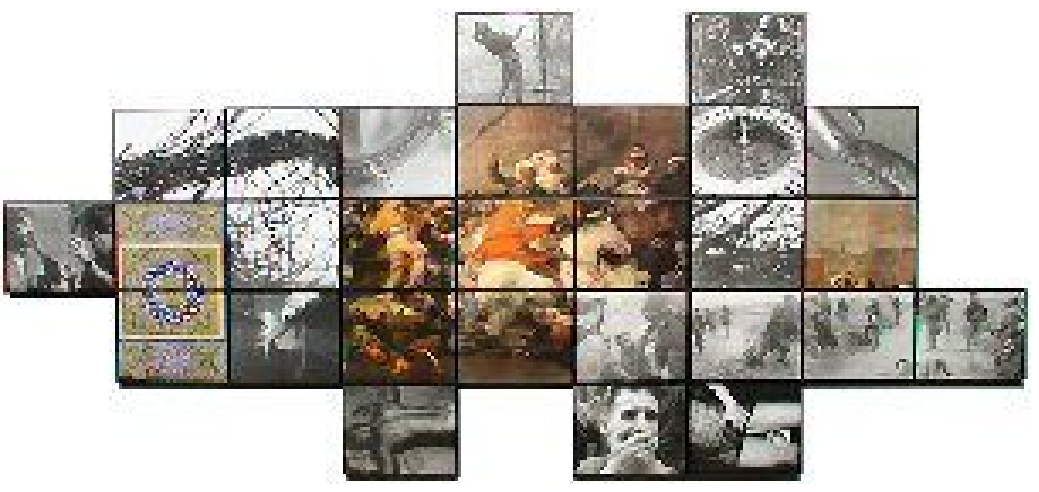} \\
\vspace*{0.6cm} {\normalsize \rm {\em Entanglement.} Anne Kesler
Shields, 2004.
\\ \vspace*{-0.4cm} \texttt{\footnotesize
http://annekeslershields.com/portfolio/port10.html}}}\label{Part4}
%%%%%%%%%%%%%%%%%%%%%%%%%%%%%%%%%%%%%%%%%%%%%%%%%%%%%%%%%%%%%%%%

\chapter{Multiparty quantum communication with Gaussian
resources}\label{ChapCommun}

{\sf

The field of quantum information with continuous variables is
flourishing with theoretical and experimental successes. It can be
considered mature for what concerns {\em two-party} information and
communication processing, exploiting in particular bipartite
entangled resources such as two-mode Gaussian states. We have
ourselves contributed some experimental results on the production
and manipulation of two-mode entanglement in Gaussian optical beams
(see Sec.~\ref{secFrancesi}). An in-depth {\em excursus} in this
multifaceted physical sector is beyond the scope of the present
Dissertation, and the interested reader may surely find  a
comprehensive and quite up-to-date review on the subject in
Ref.~\cite{brareview}.

In this Part, having laid the foundations for a proper
quantification of multipartite entanglement in CV system in Part
\ref{PartMulti}, and having dealt with the issue of efficiently
producing multipartite entangled resources in practical optical
settings in Part \ref{PartEngi},  we judge a wise choice to collect
our results concerning the usefulness and optimal exploitation of
Gaussian entanglement for processes involving more than two parties.
We will also include in this Part applications of our machinery to
study the structure and distribution of correlations in harmonic
lattices with an underlying valence bond structure (Chapter
\ref{ChapGVBS}), as well as to investigate entanglement sharing
between non-inertial observers in a relativistic setting (Chapter
\ref{ChapIvette}).

We begin in this Chapter by epitomizing  the capability of Gaussian
states for quantum communication and by providing their theoretical
entanglement characterization with a significant operative
background. To this aim we will focus on the transmission of quantum
states by means of classical communication and entangled Gaussian
resources shared by $N$ parties: specifically, CV teleportation
networks and telecloning.

\section{Quantum teleportation with continuous variables}

For two parties, the
 process of {\em quantum teleportation} using  entanglement and with the aid of
 classical communication was originally proposed for qubit systems \cite{Telep}, and
experimentally  implemented with polarization-entangled photons
\cite{zeilinger97,demartini98}.

The CV counterpart of discrete-variable teleportation, using
quadrature entanglement, is in principle imperfect due to the
impossibility of achieving infinite squeezing. Nevertheless, by
considering the finite EPR correlations between the quadratures of a
two-mode squeezed Gaussian state, \eq{tms}, a realistic scheme for
CV teleportation  was proposed \cite{vaidman,Braunstein98} and
experimentally implemented  to teleport coherent states with a
measured fidelity up to ${\CMcal F} = 0.70 \pm 0.02$
\cite{Furusawa98,australie,furunew}. Without using entanglement, by
purely classical communication, an average fidelity of
\begin{equation}\label{fcl}
{\CMcal F}_{cl}\equiv\frac{1}{2}
\end{equation}
is the best that can be achieved if the alphabet of input states
includes all coherent states with even weight
\cite{bfkjmo,hammerer}. Let us recall that the fidelity ${\CMcal
F}$, which is the figure of merit quantifying the success of a
teleportation experiment, is defined with respect to a pure state
$\ket{\psi^{in}}$ as
\begin{equation}\label{fidelity}
{\CMcal F} \equiv \bra{\psi^{in}} \ro^{out}\ket{\psi^{in}}\, .
\end{equation}
Here ``in'' and ``out'' denote the input and the output states (the
latter being generally mixed) of a teleportation process,
respectively. ${\CMcal F}$ reaches unity only for a perfect state
transfer, $\ro^{out} = \ket{\psi^{in}}\!\bra{\psi^{in}}$. To
accomplish teleportation with high fidelity, the sender (Alice) and
the receiver (Bob) must share an entangled state (resource). The
{\em sufficient} fidelity criterion \cite{bfkjmo} states that, if
teleportation is performed with ${\CMcal F} > {\CMcal F}_{cl}$, then
the two parties exploited an entangled state. The converse is
generally false, that is, quite surprisingly, some entangled
resources may in principle yield lower-than-classical fidelities.
This point will be discussed thoroughly in the following and the
solution to such a puzzling issue, obtained in \cite{telepoppate},
will be explained.

Let us briefly mention how to compute \eq{fidelity}, in terms of
CMs. Setting, as usual, all first moments to zero, the fidelity of
two-user teleportation of arbitrary single-mode Gaussian states
exploiting two-mode Gaussian resources can be computed directly from
the respective second moments \cite{fiuratele}. Being $\sig_{in}$
the CM of the unknown input state, and
\begin{equation}\label{sig2}
\sig_{ab} = \left(
              \begin{array}{cc}
                \sig_a & \eps_{ab} \\
                \eps_{ab}^{\sf T} & \sig_b \\
              \end{array}
            \right)\,,
\end{equation}
the CM of the shared two-mode resource, and defining the matrix
$\gr{\xi} = {\rm diag}\{-1\,,1\}$, the fidelity reads
\cite{fiuratele}
\begin{equation}\label{ficm}
{\CMcal F} = \frac2{\sqrt{\det \gr\Sigma}}\,,\qquad \gr\Sigma \equiv
2\, \sig_{in} + \gr{\xi} \sig_a \gr{\xi} + \sig_b + \gr{\xi}
\eps_{ab} + \eps_{ab}^{\sf T} \gr{\xi}\,.
\end{equation}
We will exploit this formula in the following.

To generalize the process of CV teleportation from two to three and
more users, one can consider two basic possible scenarios. On the
one hand, a network can be created where each user is able to
teleport states with better-than-classical efficiency (being the
same for all sender/receiver pairs) to any chosen receiver {\em with
the assistance of the other parties}. On the other hand, one of the
parties acts as the fixed sender, and distributes many approximate
copies (with in principle different cloning fidelities) to all the
others acting as remote receivers. These two protocols, respectively
referred to as ``teleportation network'' \cite{network} and
``telecloning'' \cite{telecloning}, will be described in the two
following sections, and the connections between their successful
implementation with multimode Gaussian resources and the amounts of
shared bipartite and multipartite entanglement, as obtained in
Refs.~\cite{telepoppate,3mj}, will be elucidated. We just mention
that several interesting variants to these basic schemes do exist,
(see {\em e.g.}, in a tripartite setting, the `cooperative
telecloning' of Ref.~\cite{pirgame}, where two receivers, instead of
two senders, are cooperating).

\section[Equivalence between entanglement and optimal teleportation fidelity]
{Equivalence between entanglement in symmetric Gaussian resource
states and optimal nonclassical teleportation fidelity}
\label{SecTelepoppy}

The original CV teleportation protocol \cite{Braunstein98} has been
generalized to a multi-user teleportation network requiring
multiparty entangled Gaussian states in Ref.~\cite{network}. The
tripartite instance of such a network has been recently
experimentally demonstrated by exploiting three-mode squeezed
states, yielding a maximal fidelity of ${\CMcal F} = 0.64 \pm 0.02$
\cite{naturusawa}.

Here, based on Ref.~\cite{telepoppate}, we investigate the relation
between the fidelity of a CV teleportation experiment and the
entanglement present in the shared
 resource Gaussian states. We find in particular that, while all the states belonging to the
same local-equivalence class (\ie convertible into each other by
local unitary operations) are undistinguishable from the point of
view of their entanglement properties, {\em they generally behave
differently when employed in quantum information and communication
processes}, for which the local properties such as the optical phase
reference get relevant. Hence we show that the optimal teleportation
fidelity, maximized over all local single-mode unitary operations
(at fixed amounts of noise and entanglement in the resource), is
{\em necessary and sufficient} for the presence of bipartite
(multipartite) entanglement in  two-mode (multimode) Gaussian states
employed as shared resources. Moreover, the optimal fidelity allows
for the quantitative definition of the {\em entanglement of
teleportation}, an operative estimator of bipartite (multipartite)
entanglement in CV systems. Remarkably, in the multi-user instance,
the optimal shared entanglement is exactly the ``localizable
entanglement'', originally introduced for spin systems
\cite{localiz} (not to be confused with the unitarily localizable
entanglement of bisymmetric Gaussian states, discussed in Chapter
\ref{ChapUniloc}), which thus acquires for Gaussian states a
suggestive operative meaning in terms of teleportation processes.
Moreover, let us recall that our previous study   on CV entanglement
sharing led to the definition of the residual Gaussian contangle,
\eq{gtaures}, as a tripartite entanglement monotone under Gaussian
LOCC for  three-mode Gaussian states \cite{contangle} (see
Sec.~\ref{secresid}). This measure too is here operationally
interpreted via the success of a three-party teleportation network.

Besides these fundamental theoretical results, our findings are of
important practical interest, as they answer the experimental need
for the best preparation recipe for entangled squeezed resources, in
order to implement CV teleportation (in the most general setting)
with the highest fidelity. It is indeed crucial in view of
experimental implementations, to provide optimal ways to engineer
quantum correlations, such that they are not wasted but optimally
exploited for the specifical task to be realized. We can see that
this was the {\em leitmotiv} of the previous Chapter as well.

We will now detail the results obtained in Ref.~\cite{telepoppate},
starting with the two-party teleportation instance, and then facing
with the general (and more interesting) $N$-party teleportation
network scenario. Notice that, by the defining structure itself of
the protocols under consideration, the employed resources will be
(both in the two-party and in the general $N$-party case) {\em fully
symmetric}, generally mixed Gaussian states (see
Sec.~\ref{SecSymm}). Therefore the equivalence between optimal
nonclassical fidelity and entanglement strictly holds only for fully
symmetric Gaussian resources. We will discuss this thoroughly in the
following, and show how this interesting connection actually is not
valid anymore for nonsymmetric, even two-mode resources.

\subsection{Optimal fidelity of two-party teleportation and bipartite
entanglement} \label{secpoppy2}

The two-user CV teleportation protocol \cite{Braunstein98} would
require, to achieve unit fidelity, the sharing of an ideal
(unnormalizable) EPR resource state \cite{EPR35}, \ie the
simultaneous eigenstate of relative position and total momentum of a
two-mode radiation field. An arbitrarily good approximation of the
EPR state, as we know, is represented by two-mode squeezed Gaussian
states of \eq{tms} with squeezing parameter $r \rightarrow \infty$.

As remarked in Sec.~\ref{secFrancesi}, a two-mode squeezed state can
be, in principle, produced by mixing a momentum-squeezed state and a
position-squeezed state, with squeezing parameters $r_1$ and $r_2$
respectively, through a 50:50 ideal (lossless) beam-splitter. In
practice, due to experimental imperfections and unavoidable thermal
noise the two initial squeezed states will be mixed. To perform a
realistic analysis, we must then consider two thermal squeezed
single-mode states,\footnote{\sf Any losses due to imperfect optical
elements and/or to the fibre or open-air propagation of the beams
can be embedded into the initial single-mode noise factors.}
described by the following quadrature operators in Heisenberg
picture
\begin{eqnarray}
  \hat q_1^{sq} \!&=&\! \sqrt{n_1} e^{r_1} \hat q_1^0\,,\quad
\  \hat p_1^{sq} = \sqrt{n_1} e^{-r_1} \hat p_1^0\,, \label{momsq}  \\
   \hat q_2^{sq} \!&=&\! \sqrt{n_2} e^{-r_2} \hat q_2^0\,,\quad
  \hat p_2^{sq} = \sqrt{n_2} e^{r_2} \hat p_2^0\,, \label{possq}
\end{eqnarray} where the suffix ``0'' refers to the
vacuum. The action of an ideal (phase-free) beam-splitter operation
$\hat{B}_{i,j}(\theta)$ on a pair of modes $i$ and $j$,
corresponding to a phase-space rotation of an angle $\theta$, is
defined by \eq{bsplit}.

\begin{figure}[t]
%\begin{center}
\includegraphics[width=5.5cm]{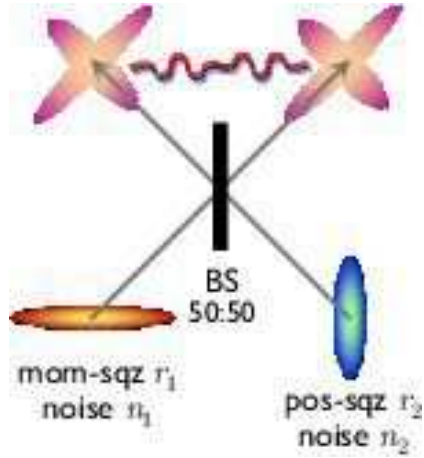}
\caption{Optical generation of two-mode symmetric mixed Gaussian
states,  by superimposing two independently squeezed single-mode
noisy beams at a 50:50 beam-splitter. The output states can be
employed as resources for CV teleportation of unknown coherent
states. For $r_1=r_2 \equiv r$ and $n_1=n_2=1$ (meaning absence of
noise), the output states reduce to those of Fig.~\ref{figtms}.}
\label{teleprep2}
\end{figure}

When applied to the two modes of
Eqs.~{\rm(\ref{momsq},\ref{possq})}, the beam-splitter entangling
operation ($\theta = \pi/4$) produces a symmetric mixed state
\cite{refbs}, depending on the squeezings $r_{1,2}$ and on the
thermal noises $n_{1,2}$, as depicted in Fig.~\ref{teleprep2}. The
CM $\sig$ of such a state reads
\begin{equation}\label{sigpoppy2}
\sig = \frac12 \left(
\begin{array}{llll}
 e^{r_1} n_1 + e^{-r_2} n_2 & 0 & e^{r_1} n_1 - e^{-r_2} n_2 & 0 \\
 0 & e^{-r_1} n_1 + e^{r_2} n_2 & 0 & e^{-r_1} n_1 - e^{r_2} n_2 \\
 e^{r_1} n_1 - e^{-r_2} n_2 & 0 & e^{r_1} n_1 + e^{-r_2} n_2 & 0 \\
 0 & e^{-r_1} n_1 - e^{r_2} n_2 & 0 & e^{-r_1} n_1 + e^{r_2} n_2
\end{array}
\right)\,.
\end{equation}

The noise can be difficult to control and reduce in the lab, but we
assume it is at least quantifiable (see Sec.~\ref{secFrancesi}).
Now, keeping $n_1$ and $n_2$ fixed, all states produced starting
with different $r_1$ and $r_2$, but with equal average
\begin{equation}\label{rmean}
\bar r \equiv \frac{r_1+r_2}{2}\,,
\end{equation}
are completely equivalent up to local unitary operations and
possess, by definition, the same entanglement. Let us recall that,
as we are dealing with symmetric two-mode Gaussian states, all
conceivable entanglement quantifications are equivalent, including
the computable entanglement of formation (see
Sec.~\ref{SecEOFGauss}), and are all decreasing functions of the
smallest symplectic eigenvalue $\tilde\nu_-$ of the partially
transposed CM, computable as in \eq{n1}.

 For the mixed two-mode
states considered here, we have
\begin{equation}\label{eta}
 \tilde\nu_- = \sqrt{n_1 n_2}e^{-(r_1+r_2)}\,.
\end{equation}
The entanglement thus depends both on the arithmetic mean of the
individual squeezings, and on the geometric mean of the individual
noises, which is related to the purity of the state $\mu = (n_1
n_2)^{-1}$. The teleportation success, instead, depends separately
on each of the four single-mode parameters. The fidelity ${\CMcal
F}$ \eq{fidelity} (averaged over the complex plane) for teleporting
an unknown single-mode coherent state can be computed by writing the
quadrature operators in  Heisenberg picture
\cite{network,vanlokfortshit},
\begin{equation}\label{fid}
{\CMcal F} \equiv \phi ^{-1/2},\:\: \phi = \left\{\left[\avr{(\hat
q_{tel})^2}+1\right]\left[\avr{(\hat
p_{tel})^2}+1\right]\right\}/4\,,
\end{equation}
where $\avr{(\hat q_{tel})^2}$ and $\avr{(\hat p_{tel})^2}$ are the
variances of the canonical operators $\hat q_{tel}$ and $\hat
p_{tel}$ which describe the teleported mode. For the utilized
states, we have
\begin{equation}\label{xptel}
\begin{split}
\hat q_{tel} &= \hat q^{in} -
\sqrt{2 n_2} e^{-r_2} \hat q_2^0\,, \\
\hat p_{tel} &= \hat p^{in} + \sqrt{2 n_1} e^{-r_1} \hat p_1^0\,,
\end{split}
\end{equation}
where the suffix ``in'' refers to the input coherent state to be
teleported. Recalling that, in our units (see
Sec.~\ref{secIntroCV}), $\avr{(\hat q_{i}^0)^2}=\avr{(\hat
p_{i}^0)^2}=\avr{(\hat q^{in})^2}=\avr{(\hat p^{in})^2}=1$, we can
 compute the fidelity from \eq{fid}, obtaining
$$\phi(r_{1,2},n_{1,2}) =
e^{-2(r_1+r_2)}(e^{2r_1}+n_1)(e^{2r_2}+n_2)\,.$$ It is convenient to
replace $r_1$ and $r_2$ by $\bar r$, \eq{rmean}, and
\begin{equation}\label{rdiff}
d \equiv \frac{r_1-r_2}{2}\,.
\end{equation}
One has then
\begin{equation}\label{fidd}
\phi(\bar r,d,n_{1,2}) = e^{-4 \bar r}(e^{2(\bar r +
d)}+n_1)(e^{2(\bar r - d)}+n_2)\,.\end{equation}

Maximizing the fidelity, \eq{fid}, for given entanglement and noises
of the Gaussian resource state (\ie for fixed $n_{1,2},\bar r$)
simply means finding the $d=d^{opt}$ which minimizes the quantity
$\phi$ of \eq{fidd}. Being $\phi$ a convex function of $d$, it
suffices to find the zero of $\partial \phi/\partial d$, which
yields an optimal $d \equiv d^{opt}$ given by
\begin{equation}
\label{dopt2} d^{opt} = \frac14 \log{\frac{n_1}{n_2}}\,.
\end{equation}
For equal noises ($n_1=n_2$), $d^{opt}=0$, indicating that the best
preparation of the entangled resource state needs two equally
squeezed single-mode states, in agreement with the results presented
in Ref.~\cite{bowen} for pure states. For different noises, however,
the optimal procedure involves two different squeezings, biased such
that $r_1 - r_2 = 2 d^{opt}$. Inserting $d^{opt}$ from \eq{dopt2},
in \eq{fidd}, we have the optimal fidelity
\begin{equation}\label{fiopt2}
{\CMcal F}^{opt} = \frac{1}{{1+\tilde\nu_-}}\,,
\end{equation} where $\tilde\nu_-$ is exactly the
smallest symplectic eigenvalue of the partial transpose $\tilde\sig$
of the CM $\sig$, \eq{sigpoppy2}, defined by \eq{eta}.

\eq{fiopt2} clearly shows that the optimal teleportation fidelity
depends only on the entanglement of the resource state, and vice
versa. In fact, the fidelity criterion becomes {\em necessary and
sufficient} for the presence of the entanglement, if ${\CMcal
F}^{opt}$ is considered: the optimal fidelity is classical  for
$\tilde\nu_- \ge 1$ (separable state) and it exceeds the classical
threshold for any entangled state. Moreover, ${\CMcal F}^{opt}$
provides a {\em quantitative} measure of entanglement completely
equivalent to the negativities and to the two-mode entanglement of
formation \cite{telepoppate}.  Namely, from
Eqs.~{\rm(\ref{eofgau},\ref{fiopt2})}, \be\label{efpoppy2} E_F =
\max \left\{0,\, h\left(\frac{1}{{\CMcal
F}^{opt}}-1\right)\right\}\,, \ee with $h(x)$ defined by
\eq{hentro}. In the limit of infinite squeezing ($\bar r \rightarrow
\infty$), ${\CMcal F}^{opt}$ reaches $1$ for any amount of finite
thermal noise.

On the other extreme, due to the convexity of $\phi$, the lowest
fidelity  is attained at one of the boundaries, $d = \pm \bar r$.
Explicitly, the worst teleportation success, corresponding to the
maximum waste of bipartite entanglement, is achieved by encoding
zero squeezing in the more mixed mode, \ie $r_1=0$ if $n_1 \ge n_2$,
and $r_2=0$ otherwise. For infinite squeezing, the worst fidelity
cannot exceed $1/\sqrt{\max\{n_1,n_2\}}$, easily falling below the
classical bound ${\CMcal F}^{cl}\equiv 1/2$ for strong enough noise.

\subsection{Optimal fidelity of {\em N}-party teleportation networks and multipartite
entanglement} \label{secpoppyN}

We now extend our analysis \cite{telepoppate} to a quantum
teleportation-network protocol, involving $N$ users who share a
genuine $N$-partite entangled Gaussian resource, fully symmetric
under permutations of the modes \cite{network} (see
Sec.~\ref{SecSymm} for the analysis of fully symmetric Gaussian
states).

Two parties are randomly chosen as sender (Alice) and receiver
(Bob), but this time, in order to accomplish teleportation of an
unknown coherent state, Bob needs the results of $N-2$ momentum
detections performed by the other cooperating parties. A
nonclassical teleportation fidelity (\ie ${\CMcal F} > {\CMcal
F}^{cl}\equiv1/2$) between {\em any} pair of parties is sufficient
for the presence of genuine $N$-partite entanglement in the shared
resource, while in general the converse is false [see {\em
e.g.}~Fig.~1 of Ref.~\cite{network}, reproduced in
Fig.~\ref{finopt}{\rm(b)}]. Our aim is to determine the optimal
multi-user teleportation fidelity, and to extract from it a
quantitative information on the multipartite entanglement in the
shared resources.

We begin with the state engineering of the shared $N$-partite
resource. Let us consider, generalizing Sec.~\ref{secpoppy2}, a
mixed momentum-squeezed state described by $r_1,n_1$ as in
\eq{momsq}, and $N-1$ position-squeezed states of the form
\eq{possq},
\begin{eqnarray}
  \hat q_1^{sq} \!&=&\! \sqrt{n_1} e^{r_1} \hat q_1^0\,,\quad
\  \hat p_1^{sq} = \sqrt{n_1} e^{-r_1} \hat p_1^0\,, \label{momsqNN}  \\
   \hat q_j^{sq}  \!&=&\! \sqrt{n_2} e^{-r_2} \hat q_j^0\,,\quad
  \hat p_j^{sq} = \sqrt{n_2} e^{r_2} \hat p_j^0\,, \label{possqNN}
\end{eqnarray}
with $j=2,\ldots,N$.
 We then combine the $N$ beams into an $N$-splitter,
which is a sequence of suitably tuned beam-splitters \cite{network}:
\begin{equation}\label{nsplit}
\hat{N}_{1\ldots N} \equiv \hat{B}_{N-1,N}(\pi/4)\hat{B}_{N-2,N-1}
(\cos^{-1}1/\sqrt{3})\cdot
\ldots\cdot\hat{B}_{1,2}(\cos^{-1}1/\sqrt{N})\,, \end{equation}
where the unitary beam-splitter operator $\hat B_{i,j}(\theta)$
acting on modes $i$ and $j$ is defined by \eq{bsplit}. \eq{nsplit}
represents the generalization to $N$ modes of the ``tritter'',
\eq{tritter}.

The resulting state (see Fig.~\ref{teleprepN}) is a completely
symmetric mixed Gaussian state of a $N$-mode CV system, with a CM
$\sig$ of the form \eq{fscm}, parametrized by $n_{1,2}$, $\bar r$
and $d$. Once again, all states with equal $\{n_{1,2},\bar r\}$
belong to the same iso-entangled class of equivalence, up to local
unitaries. For $\bar r \rightarrow \infty$ and for $n_{1,2}=1$ (pure
states), these states reproduce the (unnormalizable) CV generalized
GHZ state, $\int dx \ket{x,x,\ldots,x}$, an eigenstate with total
momentum $\sum_{i=1}^N \hat p_i = 0$ and all relative positions
$\hat q_i - \hat q_j = 0$ ($i,j=1,\ldots,N$).

\begin{figure}[t!]
%\begin{center}
\includegraphics[width=12cm]{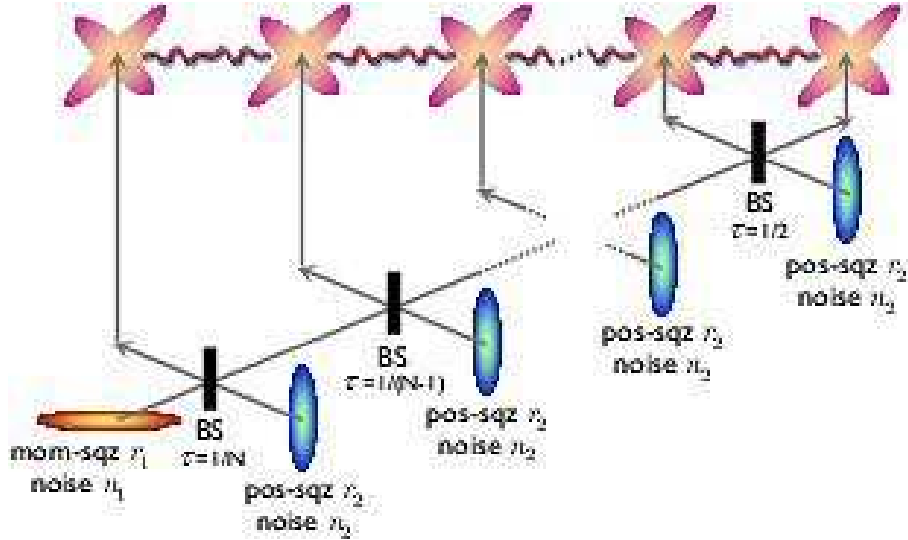}
\caption{Optical generation of $N$-mode fully symmetric mixed
Gaussian states, by combining $N$ independently squeezed single-mode
noisy beams (one squeezed in momentum, and $N-1$ in position) via a
cascade of $N-1$ beam-splitters with sequentially tuned
transmittivities.  The output states can be employed as shared
resources for CV teleportation networks among $N$ users. For pure
inputs ($n_i=1$) and $N=3$ we recover the scheme of
Fig.~\ref{fighzw}.} \label{teleprepN}
\end{figure}

Choosing randomly two modes, denoted by the indices $k$ and $l$, to
be respectively the sender and the receiver, the teleported mode is
described by the following quadrature operators (see Refs.
\cite{network,vanlokfortshit} for further details):
\begin{equation}\label{xpteln}
\begin{split}
\hat q_{tel} &= \hat q_{in} - \hat q_{rel}\,,\\
\hat p_{tel} &= \hat p_{in} + \hat p_{tot}\,, \end{split}
\end{equation}
with
\begin{equation}\label{xrelptotn}
\begin{split}
\hat q_{rel} &= \hat q_{k} - \hat q_{l}\,,\\
\hat p_{tot} &= \hat p_{k} + \hat p_{l} + g_N \sum_{j \ne k,l} {\hat
p_{j}}\,, \end{split}
\end{equation}
 where $g_N$ is an experimentally adjustable gain. To compute the
teleportation fidelity from \eq{fid}, we need the variances of the
operators $\hat q_{rel}$ and $\hat p_{tot}$ of \eq{xrelptotn}. From
the action of the $N$-splitter, \eq{nsplit}, we find
\begin{eqnarray}\label{avrel}
  \avr{(\hat q_{rel})^2} &=& 2 n_2 e^{-2 (\bar r - d)}\,, \nonumber\\
  \avr{(\hat p_{tot})^2} &=& \big\{ [2+ (N-2) g_N]^2 n_1 e^{-2(\bar r + d)}\\
  &+& 2 [g_N-1]^2(N-2) n_2 e^{2(\bar r - d)}  \big\}/4\,. \nonumber
 \end{eqnarray}

The optimal fidelity can  be now found in two straightforward steps:
1) minimizing $\avr{(\hat p_{tot})^2}$ with respect to $g_N$ (i.e.
finding the optimal gain $g_N^{opt}$); 2) minimizing the resulting
$\phi$ with respect to $d$ (\ie finding the optimal bias
$d_N^{opt}$). The results are
\begin{eqnarray}
  g_N^{opt} &=& 1- N/\left[(N-2)+2e^{4 \bar r} n_2/n_1\right]\,, \label{gopt}\\
  d_N^{opt} &=& \bar r + \frac14 \log\left[ \frac{N}{(N-2)+2e^{4 \bar r}
n_2/n_1}   \right]\,.  \label{dopt}
\end{eqnarray}
Inserting Eqs.~{\rm(\ref{avrel}--\ref{dopt})} in \eq{fid}, we find
the optimal teleportation-network fidelity, which can be put in the
following general form for $N$ modes
\begin{equation}\label{fidn}
{\CMcal F}_N^{opt} = \frac{1}{1+\tilde\nu_-^{(N)}}\,,\quad
\tilde\nu_-^{(N)} \equiv \sqrt{\frac{N n_1 n_2}{2e^{4 \bar r}
+(N-2)n_1/n_2}}\,.
 \end{equation} For $N=2$, $\tilde\nu_-^{(2)} \equiv \tilde\nu_-$ from
\eq{eta}, showing that the general multipartite protocol comprises
the standard bipartite one as a special case.

\begin{figure}[t!]
\includegraphics[width=8.3cm]{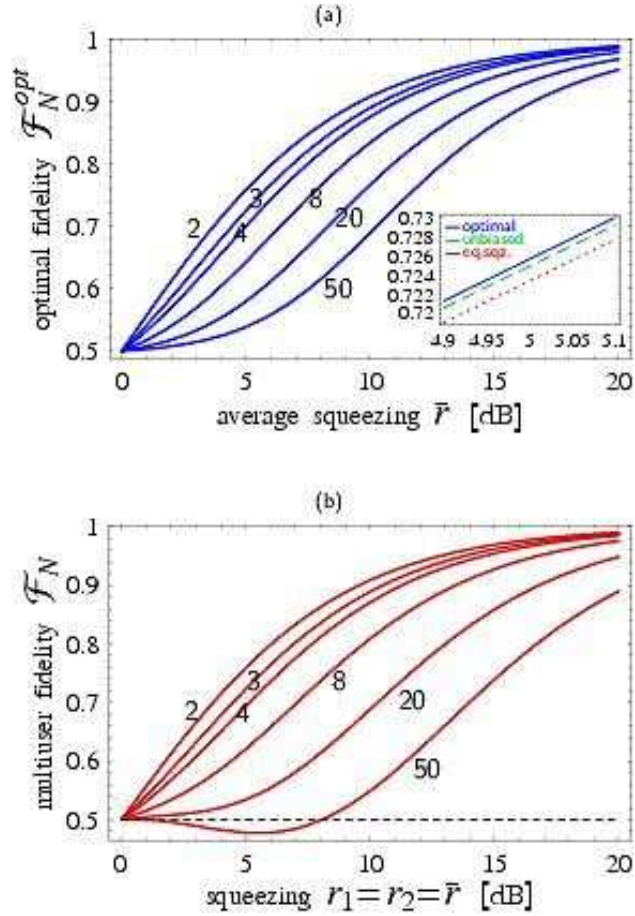}
\caption{{\rm (a)} Optimal fidelity for teleporting unknown coherent
states from any sender to any receiver chosen from $N$ (= 2, 3, 4,
8, 20, and 50) parties, sharing pure $N$-party entangled symmetric
Gaussian resources and with the aid of $N-2$ cooperating parties,
plotted as a function of the average squeezing used in the resource
production (expressed in decibels, for the definition of dB see
footnote {\rm \ref{notedb}} on page {\rm \pageref{notedb}}). The
optimal fidelity is nonclassical (${\CMcal F}^{opt}
> {\CMcal F}^{cl}\equiv 0.5$) for any $N$, if the initial squeezings are adjusted as in
\eq{dopt} \cite{telepoppate}. At fixed entanglement, states produced
with all equal squeezers yield lower-than-classical fidelities
(${\CMcal F}< {\CMcal F}^{cl}\equiv 0.5$) for $N \ge 30$, as shown
in {\rm (b)} (adapted from Fig.~1 of Ref.~\cite{network}). In the
inset of Plot {\rm(a)} we compare, for $N=3$ and a window of average
squeezing, the optimal fidelity (blue solid line), the fidelity
obtained with states having all unbiased quadratures \cite{bowen}
(green dashed line), and the fidelity obtained with equally squeezed
states \cite{network} (red dotted line). The three curves are close
to each other, but the optimal preparation yields always the highest
fidelity.} \label{finopt}
\end{figure}

By comparison with \eq{fiopt2}, we observe that, for any $N>2$, the
quantity $\tilde\nu_-^{(N)}$ plays the role of a ``generalized
symplectic eigenvalue'', whose physical meaning will be clear soon.
Before that, it is worth commenting on the form of the optimal
resources, focusing for simplicity on the pure-state setting
($n_{1,2}=1$). The optimal form of the shared $N$-mode symmetric
Gaussian states, for $N>2$, is neither unbiased in the $q_i$ and
$p_i$ quadratures (like the states discussed in Ref.~\cite{bowen}
for three modes), nor constructed by $N$ equal squeezers ($r_1=r_2=
\bar r$). This latter case, which has been implemented
experimentally for $N=3$ \cite{naturusawa}, is clearly not optimal,
yielding fidelities lower than $1/2$ for $N\ge30$ and $\bar r$
falling in a certain interval \cite{network} [see
Fig.~\ref{finopt}{\rm (b)}]. The explanation of this paradoxical
behavior, provided by the authors of Ref.~\cite{network}, is that
their teleportation scheme might not be optimal. Our analysis
\cite{telepoppate} shows instead that the problem does not lie in
the choice of the protocol, but rather in the form of the employed
states. If the shared $N$-mode resources are prepared by suitable
pre-processing --- or transformed by local unitary (symplectic on
the CM) operations --- into the optimal form of \eq{dopt}, the
teleportation fidelity is guaranteed to be nonclassical [see
Fig.~\ref{finopt}{\rm(a)}] as soon as $\bar r>0$ for any $N$, in
which case the considered class of pure states is genuinely
multiparty entangled (we have shown this unambiguously in
Sec.~\ref{SecScal}). Therefore, we can state the following
\cite{telepoppate}.

\medskip

\begin{itemize}
\item[\ding{226}]
 \noindent{\rm\bf Equivalence between entanglement and optimal fidelity of continuous variable teleportation.}
{\it A nonclassical optimal fidelity is  necessary and sufficient
for the presence of multipartite entanglement in any multimode
symmetric Gaussian state, shared as a resource for CV teleportation
networks.}
\smallskip
\end{itemize}

This {\em equivalence} silences the embarrassing question that
entanglement might not be an actual physical resource, as protocols
based on some entangled states might behave worse than their
classical counterparts in processing quantum information.

On the opposite side, the worst preparation scheme of the multimode
resource states, even retaining the optimal protocol
($g_N=g_N^{opt}$), is obtained setting $r_1=0$ if $n_1 > 2 n_2 e^{2
\bar r}/(N e^{2 \bar r} + 2 - N)$, and $r_2=0$ otherwise. For equal
noises ($n_1=n_2$), the case $r_1=0$ is always the worst one, with
asymptotic fidelities (in the limit $\bar r \rightarrow \infty$)
equal to $1/\sqrt{1+N n_{1,2}/2}$, and so rapidly dropping with $N$
at given noise.

\subsubsection{Entanglement of teleportation and localizable entanglement}

The meaning of $\tilde\nu_-^{(N)}$, \eq{fidn}, crucial for the
quantification of the multipartite entanglement, stems from the
following argument. The teleportation network \cite{network} is
realized in two steps: first, the $N-2$ cooperating parties perform
local measurements on their modes, then Alice and Bob exploit their
resulting highly entangled two-mode state to accomplish
teleportation. Stopping at the first stage, the protocol describes a
concentration, or {\em localization} of the original $N$-partite
entanglement, into a bipartite two-mode entanglement
\cite{network,vanlokfortshit}. The maximum entanglement that can be
concentrated on a pair of parties by locally measuring the others,
is known as the {\em localizable entanglement}\footnote{\sf This
localization procedure, based on measurements, is different from the
unitary localization which can be performed on bisymmetric Gaussian
states, as discussed in Chapter \ref{ChapUniloc}.}  of a multiparty
system \cite{localiz}, as depicted in Fig.~\ref{figlocaliz}.

Here, the localizable entanglement is the maximal entanglement
concentrable onto two modes, by unitary operations and nonunitary
momentum detections performed locally on the other $N-2$ modes. The
two-mode entanglement of the resulting state (described by a CM $\gr
\sigma^{loc}$) is quantified in general in terms of the symplectic
eigenvalue $\tilde\nu_-^{loc}$ of its partial transpose. Due to the
symmetry of both the original state and the teleportation protocol
(the gain is the same for every mode), the localized two-mode state
$\gr \sigma^{loc}$ will be symmetric too. We have shown in
Sec.~\ref{SecEPRcorrel} that, for two-mode symmetric Gaussian
states, the symplectic eigenvalue $\tilde\nu_-$ is related to the
EPR correlations by the expression \cite{extremal}
$$4 \tilde\nu_-
=\avr{(\hat q_1 - \hat q_2)^2} + \avr{(\hat p_1 + \hat p_2)^2}\,.$$
For the state $\gr\sigma^{loc}$, this means $4\tilde\nu_-^{loc} =
\avr{(\hat q_{rel})^2} + \avr{(\hat p_{tot})^2}$, where the
variances have been computed in \eq{avrel}. Minimizing
$\tilde\nu_-^{loc}$ with respect to $d$ means finding the optimal
set of local unitary operations (unaffecting multipartite
entanglement) to be applied to the original multimode mixed resource
described by $\{n_{1,2},\bar r, d\}$; minimizing then
$\tilde\nu_-^{loc}$ with respect to $g_N$ means finding the optimal
set of momentum detections to be performed on the transformed state
in order to localize the highest entanglement on a pair of modes.
From \eq{avrel}, the optimizations are readily solved and yield  the
same optimal $g_N^{opt}$ and $d_N^{opt}$ of
Eqs.~{\rm(\ref{gopt},\ref{dopt})}.

The resulting optimal two-mode state $\sig^{loc}$ contains a
localized entanglement which is {\em exactly} quantified by the
quantity $$\tilde\nu_-^{loc} \equiv \tilde\nu_-^{(N)}\,.$$ It is now
clear that $\tilde\nu_-^{(N)}$ of \eq{fidn} is a proper symplectic
eigenvalue, being the smallest one of the partial transpose
$\tilde\sig^{loc}$ of the optimal two-mode state $\sig^{loc}$ that
can be extracted from a $N$-party entangled resource by local
measurements on the remaining modes (see Fig.~\ref{figlocaliz}).
\eq{fidn} thus provides a bright connection between two {\em
operative} aspects of multipartite entanglement in CV systems:  the
maximal fidelity achievable in a multi-user teleportation network
\cite{network}, and the CV localizable entanglement \cite{localiz}.

\begin{figure}[t!]
%\begin{center}
\includegraphics[width=10cm]{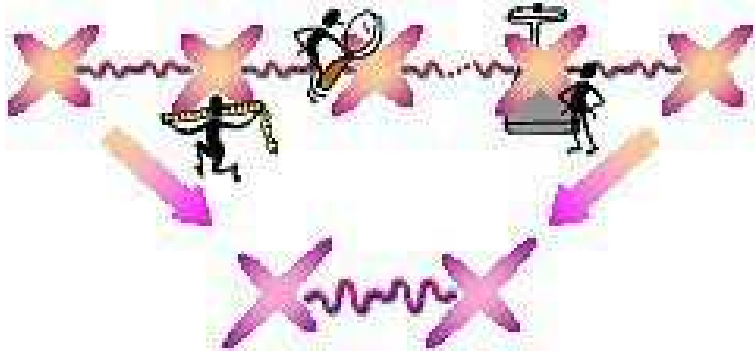}
\caption{Localizable entanglement in the sense of \cite{localiz}. By
optimal local measurements on $N-2$ subsystems in a $N$-party
system, a highly entangled two-mode state is (probabilistically, in
principle) obtained between the two non-measuring parties. For
Gaussian states and measurements the localization process in indeed
deterministic, as the entanglement properties of the resulting
states are independent of the measurement outcomes.}
\label{figlocaliz}
\end{figure}

This results yield quite naturally a direct operative way to
quantify multipartite entanglement in $N$-mode (mixed) symmetric
Gaussian states, in terms of the so-called {\em Entanglement of
Teleportation}, defined as the normalized optimal fidelity
\cite{telepoppate}
\begin{equation}\label{et}
E_T^{(N)} \equiv \max\left\{0,\frac{{\CMcal F}_N^{opt}-{\CMcal
F}_{cl}}{1-{\CMcal
F}_{cl}}\right\}=\max\left\{0,\frac{1-\tilde\nu_-^{(N)}}{1+\tilde\nu_-^{(N)}}\right\}\,,
\end{equation} and thus ranging from 0 (separable
states) to 1 (CV generalized GHZ state). The localizable
entanglement of formation $E_F^{loc}$ of $N$-mode symmetric Gaussian
states $\sig$ of the form \eq{fscm} is a monotonically increasing
function of $E_T^{(N)}$, namely: \be \label{efloc} E_F^{loc}(\sig) =
h\left[\frac{1-E_T^{(N)}}{1+E_T^{(N)}}\right]\,,\ee with $h(x)$
given by \eq{hentro}. For $N=2$ the state is already localized and
$E_F^{loc} \equiv E_F$, \eq{efpoppy2}

In the next subsection we will see how the entanglement of
teleportation relates, for three-mode states, to the residual
Gaussian contangle introduced in Sec.~\ref{secresid}.

\subsection{Operational interpretation of tripartite Gaussian
entanglement and how to experimentally investigate its sharing
structure}

\subsubsection{Entanglement of teleportation and residual contangle}

Let us focus, for the following discussion, on the case $N=3$,
{\ie}on three-mode states shared as resources for a three-party
teleportation network. This protocol is a basic, natural candidate
to operationally investigate the sharing structure of CV
entanglement in three-mode symmetric Gaussian states.

A first theoretical question that arises is to compare the
tripartite entanglement of teleportation \eq{et}, which possesses a
strong operational motivation, and the tripartite residual
(Gaussian) contangle \eq{gtaures} (defined in Sec.~\ref{secresid}),
which is endowed with a clear physical interpretation in the
framework of entanglement sharing and is built on solid mathematical
foundations (being an entanglement monotone under Gaussian LOCC, see
Sec.~\ref{secTauresMonotone}). Remarkably, in the case of pure
three-mode shared resources --- {\ie}CV GHZ/$W$ states, obtained by
setting $n_1=n_2=1$ in Eqs.~{\rm(\ref{momsqNN},\ref{possqNN})}, see
Sec.~\ref{secghzw} and Fig.~\ref{fighzw} --- {\em the two measures
are completely equivalent}, being monotonically increasing functions
of each other. Namely, from \eq{gresghzw},
\begin{equation}\label{etetau}
G_\tau^{res}(\sig_{s}^{_{{\rm GHZ}/W}}) = \log^2{\frac{2\sqrt2
E_T-(E_T+1)\sqrt{E_T^2+1}}{(E_T-1)\sqrt{E_T(E_T+4)+1}}}-\frac12
\log^2{\frac{E_T^2+1}{E_T(E_T+4)+1}}\,,
\end{equation}
where $E_T \equiv E_T^{(3)}$ in \eq{et}. Let us moreover recall that
$G_\tau^{res}$ coincides with the true residual contangle (globally
minimized in principle over all, including non-Gaussian,
decompositions), \eq{etaumin}, in these states (see
Sec.~\ref{secghzw}). The residual (Gaussian) contangle is thus
enriched of an interesting meaning as a {\em resource} enabling a
better-than-classical three-party teleportation experiment, while no
operational interpretations are presently known for the three-way
residual tangle quantifying tripartite entanglement sharing in qubit
systems \cite{CKW} (see Sec.~\ref{sec3tangle}).

We remark that in the tripartite instance, the optimal
teleportation-network fidelity of \eq{fidn} ($N=3$) achieves indeed
its {\em global} maximum over all possible Gaussian POVMs performed
on the shared resource, as can be confirmed with the methods of
Ref.~\cite{pirandolassisted}.

\subsubsection{The power of promiscuity in symmetric three-mode
resources}\label{secpromischeck}

The relationship between optimal teleportation fidelity and residual
(Gaussian) contangle, embodied by \eq{etetau},  entails that there
is a `unique' kind of three-party CV entanglement in pure {\em
symmetric} three-mode Gaussian states (alias CV finite-squeezing
GHZ/$W$ states, introduced in Sec.~\ref{secghzw}), which merges at
least three (usually inequivalent) properties: those of being
maximally genuinely tripartite entangled, maximally bipartite
entangled in any two-mode reduction, and `maximally efficient' (in
the sense of the optimal fidelity) for three-mode teleportation
networks. Recall that the first two properties, taken together,
label such entanglement as {\em promiscuous}, as discussed in
Sec.~\ref{secpromis}. These features add up to the property of
tripartite GHZ/$W$ Gaussian states of being maximally robust against
decoherence effects among all three-mode Gaussian states, as shown
in Sec.~\ref{decoherence}.

All this theoretical evidence strongly promotes GHZ/$W$ states,
experimentally realizable with current optical technology
\cite{3mexp,pfister} (see Sec.~\ref{SecEngiGHZW}), as paradigmatic
candidates for the encoding and transmission of CV quantum
information and in general for reliable CV quantum communication.
 Let us mention that, in particular, these tripartite entangled symmetric Gaussian states
have been successfully employed to demonstrate quantum secret
sharing \cite{secret}, controlled dense coding \cite{dense}, and the
above discussed teleportation network \cite{naturusawa}. Recently, a
theoretical solution for CV Byzantine agreement has been reported
\cite{sanpera}, based on the use of sufficiently entangled states
from the family of CV GHZ/$W$ states.

Building on our entanglement analysis, we can precisely enumerate
the peculiarities of those states which make them so appealing for
practical implementations \cite{3mj}.  Exploiting a strongly
entangled three-mode CV GHZ/$W$ state as a quantum channel affords
one with a number of simultaneous advantages:
\begin{enumerate}\label{checklist}
  \item[{(i)}] the ``guaranteed success'' ({\ie}with better-than-classical
figures of merit) of any known tripartite CV quantum information
protocol;
  \item[{(ii)}] the ``guaranteed success'' of any standard two-user CV
protocol, because a highly entangled two-mode channel is readily
available after a unitary (reversible) localization of entanglement
has been performed through a single beam-splitter (see
Fig.~\ref{figbasset});
  \item[{(iii)}] the ``guaranteed success'' (though with nonmaximal
efficiency) of any two-party quantum protocol through each two-mode
channel obtained discarding one of the three modes.
\end{enumerate}
Point (iii) ensures that, even when one mode is lost, the remaining
(mixed) two-mode resource can be still implemented for a two-party
protocol with better-than-classical success. It is realized with
 nonmaximal efficiency since, from
\eq{gredmaxghzw}, the reduced entanglement in any two-mode partition
remains finite even with infinite squeezing (this is the reason why
promiscuity of tripartite Gaussian entanglement is only partial,
compared to the four-partite case of Chapter \ref{ChapUnlim}).

We can now readily provide an explicit proposal to implement the
above checklist in terms of CV teleportation networks.

\subsubsection{Testing the promiscuous sharing of tripartite
entanglement}\label{sectest}

The results just elucidated pave the way towards an experimental
test for the promiscuous sharing of CV entanglement in symmetric
Gaussian states \cite{telepoppate,3mj}. To unveil this peculiar
feature, one should prepare a pure CV GHZ/$W$ state ---
corresponding to $n_1=n_2=1$ in
Eqs.~{\rm(\ref{momsqNN},\ref{possqNN})} --- according to
Fig.~\ref{fighzw}, in the optimal form given by \eq{dopt}. It is
worth remarking that, in the case of three modes, non-optimal forms
like that produced with equal single-mode squeezings $r_1=r_2$
\cite{3mexp,naturusawa} yield fidelities really close to the maximal
one [see the inset of Fig.~\ref{finopt}{\rm(a)}], and are thus
practically as good as the optimal states (if not even better,
taking into account that the states with $r_1=r_2$ are generally
easier to produce in practice, and so less sensitive to
imperfections).

To detect the presence of tripartite entanglement, one should be
able to implement the network in at least two different combinations
\cite{naturusawa}, so that the teleportation would be accomplished,
for instance, from mode $1$ to mode $2$ with the assistance of mode
$3$, and from mode $2$ to mode $3$ with the assistance of mode $1$.
To be complete (even if it is not strictly needed \cite{vloock03}),
one could also realize the transfer from mode $3$ to mode $1$ with
the assistance of mode $2$. Taking into account a realistic
asymmetry among the modes, the average experimental fidelity
${\CMcal F}_3^{opt}$ over the three possible situations would
provide a direct quantitative measure of tripartite entanglement,
through Eqs.~{\rm(\ref{fidn}, \ref{et}, \ref{etetau})}.

To demonstrate the promiscuous sharing, one would then need to
discard each one of the modes at a time, and perform standard
two-user teleportation between the remaining pair of parties. The
optimal fidelity for this two-user teleportation, which is achieved
exactly for $r_1=r_2$  [see \eq{dopt2}], is
\begin{equation}\label{f2red}
{\CMcal F}^{opt}_{2:red} = \frac{3}{3+\sqrt{3+6e^{-4 \bar r}}}\,.
\end{equation}
Again, one should implement the three possible configurations and
take the average fidelity as figure of merit. As anticipated in
Sec.~\ref{secpromischeck}, this fidelity cannot reach unity because
the entanglement in the shared mixed resource remains finite, and in
fact ${\CMcal F}^{opt}_{2:red}$ saturates to $3/(3+\sqrt{3}) \approx
0.634$ in the limit of infinite squeezing.

Finding simultaneously both ${\CMcal F}_3^{opt}$ and ${\CMcal
F}^{opt}_{2:red}$ above the classical threshold ${\CMcal F}^{cl}
\equiv 1/2$, \eq{fcl}, at fixed squeezing $\bar r$, would be a clear
experimental fingerprint of the promiscuous sharing of tripartite CV
entanglement. Theoretically, this is true for all $\bar r > 0$, as
shown in Fig.~\ref{figapoppa}. From an experimental point of view,
the tripartite teleportation network has been recently implemented,
and the genuine tripartite shared entanglement unambiguously
demonstrated by obtaining a nonclassical teleportation fidelity (up
to $0.64 \pm 0.02$) in all the three possible user configurations
\cite{naturusawa}. Nevertheless, a nonclassical fidelity ${\CMcal
F}_{2:red}$ in the teleportation exploiting any two-mode reduction
was not observed. This fact can be consistently explained by taking
into account experimental noise. In fact, even if the desired
resource states were pure GHZ/$W$ states, the unavoidable effects of
decoherence and imperfections resulted in the experimental
production of {\em mixed} states, namely of the noisy GHZ/$W$ states
discussed in Sec.~\ref{secnoisyghzw}. It is very likely that the
noise was too high compared with the pumped squeezing, so that the
actual produced states were still fully inseparable, but laid
outside the region of promiscuous sharing (see Fig.~\ref{figacunt}),
having no entanglement left in the two-mode reductions. However,
increasing the degree of initial squeezing, and/or reducing the
noise sources might be accomplished with the state-of-the-art
equipment employed in the experiments of Ref.~\cite{naturusawa} (see
also \cite{furunew}). The conditions required for a proper test (to
be followed by actual practical applications) of the promiscuous
sharing of CV entanglement in symmetric three-mode Gaussian states,
as detailed in Sec.~\ref{secpromischeck}, should be thus met
shortly. As a final remark, let us observe that repeating the same
experiment but employing $T$ states, introduced in
Sec.~\ref{sectstates}, as resources (engineerable as detailed in
Sec.~\ref{SecEngiT}), would be another interesting option. In fact,
in this case the expected optimal fidelity is strictly smaller than
in the case of GHZ/$W$ states, confirming the promiscuous structure
in which the reduced bipartite entanglement enhances the value of
the genuine tripartite one.

\begin{figure}[t!]
\centering{\includegraphics[width=9cm]{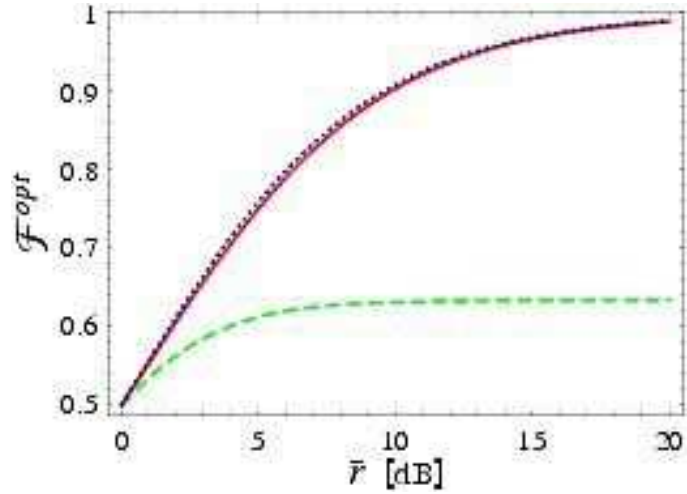}%
\caption{\label{figapoppa}  Expected success for an experimental
test of the promiscuous sharing of CV tripartite entanglement in
finite-squeezing GHZ/$W$ states. Referring to the check-list in
Sec.~\ref{secpromischeck}: the solid curve realizes point (i), being
the optimal fidelity ${\CMcal F}_3^{opt}$ of a three-party
teleportation network; the dotted curve realizes point (ii), being
the optimal fidelity ${\CMcal F}^{opt}_{2:uni}$ of two-party
teleportation exploiting the two-mode pure resource obtained from a
unitary localization applied on two of the modes; the dashed curve
realizes point (iii), being the optimal fidelity ${\CMcal
F}^{opt}_{2:red}$ of two-party teleportation exploiting the two-mode
mixed resource obtained discarding a mode. All of them lie above the
classical threshold ${\CMcal F}^{cl}\equiv 0.5$, providing a direct
evidence of the promiscuity of entanglement sharing in the employed
resources.}}
\end{figure}

With the same GHZ/$W$ shared resources (but also with all symmetric
and bisymmetric three-mode Gaussian states, including $T$ states,
noisy GHZ/$W$ states and basset hound states, all introduced in
Chapter \ref{Chap3M}), one may also test the power of the unitary
localization of entanglement by local symplectic operations
\cite{unitarily} (presented in Chapter \ref{ChapUniloc}), as opposed
to the nonunitary localization of entanglement by measurements
\cite{localiz} (described in Fig.~\ref{figlocaliz}), needed for the
teleportation network. Suppose that the three parties Alice, Bob and
Claire share a GHZ/$W$ state. If Bob and Claire are allowed to
cooperate (non-locally), they can combine their respective modes at
a 50:50 beam-splitter, as depicted in Fig.~\ref{figbasset}. The
result is an entangled state shared by Alice and Bob, while Claire
is left with an uncorrelated state. The optimal fidelity of standard
teleportation from Alice to Bob with the unitarily localized
resource, reads
\begin{equation}\label{f2uni}
{\CMcal F}^{opt}_{2:uni} = \left[\frac{1}{3} \left(\sqrt{4 \cosh (4
\bar r) + 5} - 2 \sqrt{\cosh (4 \bar r) - 1}\right) +
1\right]^{-1}\,.
\end{equation}
Notice that ${\CMcal F}_{2:uni}$ is  larger than ${\CMcal
F}_3^{opt}$, as shown in  Fig.~\ref{figapoppa}. This is true for any
number $N$ of modes, and the difference between the two fidelities
--- the optimal teleportation fidelity employing the  unitarily-localized two-mode resource,
minus the $N$-party optimal teleportation-network fidelity,
corresponding to a two-party teleportation with
nonunitarily-localized resources, \eq{fidn} --- at fixed squeezing
increases with $N$. This confirms that the unitarily localizable
entanglement of Chapter \ref{ChapUniloc} is strictly stronger than
the (measurement-based) localizable entanglement \cite{localiz} of
Fig.~\ref{figlocaliz}, as discussed in Sec.~\ref{secswitch}. This is
of course not surprising, as the unitary localization generally
requires a high degree of non-local control on the two subset of
modes, while the localizable entanglement of Ref.~\cite{localiz} is
defined in terms of LOCC alone.

\subsection{Degradation of teleportation efficiency under quantum noise}\label{qnoise}

In Sec.~\ref{decoherence} we have addressed the decay of
three-partite entanglement (as quantified by the residual Gaussian
contangle) of three-mode states in the presence of losses and
thermal noise. We aim now at relating such an `abstract' analysis to
precise operational statements, by investigating the decay of the
optimal teleportation fidelity, \eq{fidn} ($N=3$), of shared
three-mode resources subject to environmental decoherence. This
study will also provide further heuristic justification for the
residual Gaussian contangle, \eq{gtaures}, as a proper measure of
tripartite entanglement even for mixed (`decohered') Gaussian
states. Notice that the effect of decoherence occurring {\em during}
the creation of the state on the teleportation fidelity has been
already implicitly considered in
Eqs.~{\rm(\ref{momsqNN}--\ref{possqNN})}, by the noise terms $n_1$
and $n_2$. Here, we will instead focus on the decay of the
teleportation efficiency under decoherence affecting the resource
states {\em after} their distribution to the distant parties.

We will assume, realistically, a local decoherence ({\em i.e.}~with
no correlated noises) for the three modes, in thermal baths with
equal average photon number $n$. The evolving states maintain their
Gaussian character under such evolution (for a detailed description
of the master equation governing the system and of its Gaussian
solutions, refer to Sec.~\ref{decoherence}).

\begin{figure}[t!]
\centering{
\includegraphics[width=10cm]{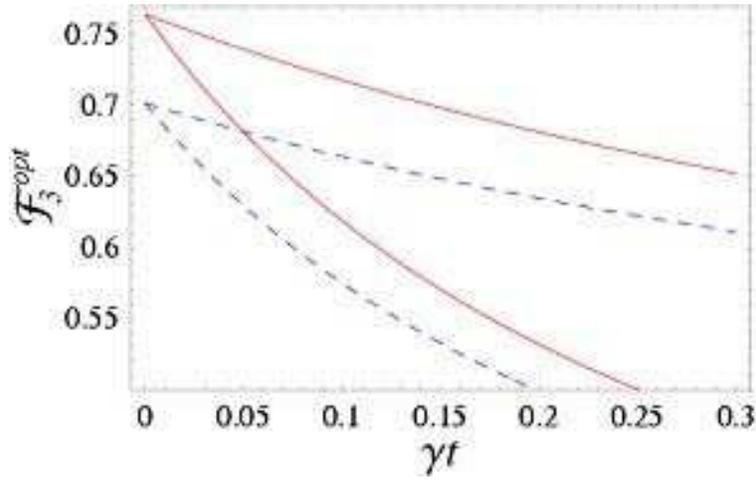}
\caption{Evolution of the optimal fidelity ${\CMcal F}^{opt}_{3}$
for GHZ/$W$ states with local mixedness $a=2$ (corresponding to
$\bar{r} \simeq 0.6842$) (solid lines) and $T$ states with local
mixedness $a=2.8014$. Such states have equal initial residual
Gaussian contangle. Uppermost curves refer to baths with $n=0$
(`pure losses'), while lowermost curves refer to baths with $n=1$.
$T$ states affording for the same initial fidelity as the considered
GHZ/$W$ state were also considered, and found to degrade faster than
the GHZ/$W$ state.} \label{teledeco}}
\end{figure}

As initial resources, we have considered both pure GHZ/$W$ states,
described in Sec.~\ref{secghzw}, and mixed $T$ states, described in
Sec.~\ref{sectstates}. The results, showing the exact evolution of
the fidelity ${\CMcal F}^{opt}_{3}$ (optimized over local unitaries)
of teleportation networks exploiting such initial states, are shown
in Fig.~\ref{teledeco}. GHZ/$W$ states, already introduced as
``optimal'' resources for teleportation networks, were also found to
allow for protocols most robust under decoherence. Notice how the
qualitative behavior of the curves of Fig.~\ref{teledeco} follow
that of Fig.~\ref{decofig1}, where the evolution of the residual
Gaussian contangle of the same states under the same conditions is
plotted. Also the vanishing of entanglement at finite times
(occurring only in the presence of thermal photons, {\em i.e.}~for
$n>0$) reciprocates the fall of the fidelity below the classical
threshold of $0.5$. The status of the residual Gaussian contangle as
a measure reflecting operational aspects of the states is thus
strengthened in this respect, even in the region of mixed states.
Notice, though, that Fig.~\ref{teledeco} also shows that  the
entanglement of teleportation is {\em not} in general quantitatively
equivalent (but for the pure-state case) to the residual Gaussian
contangle, as the initial GHZ/$W$ and $T$ states of
Fig.~\ref{teledeco} have the same initial residual Gaussian
contangle but grant manifestly different fidelities and, further,
the times at which the classical threshold is trespassed do not
exactly coincide with the times at which the residual contangle
vanishes.

This confirms the special role of pure fully symmetric GHZ/$W$
Gaussian states in tripartite CV quantum information, and the
``uniqueness'' of their entanglement under manifold interpretations
as discussed in Sec.~\ref{secpromischeck}, much on the same footage
of the ``uniqueness'' of entanglement in symmetric (mixed) two-mode
Gaussian states (see Sec.~\ref{SecEOFGauss})

\subsection{Entanglement and optimal fidelity for nonsymmetric
Gaussian resources?}

Throughout the whole Sec.~\ref{SecTelepoppy}, we have only dealt
with completely symmetric resource states, due to the invariance
requirements of the considered teleportation-network protocol. In
Ref.~\cite{telepoppate}, the question whether expressions like
\eq{fidn} and \eq{et}, connecting the optimal fidelity and the
entanglement of teleportation to the symplectic eigenvalue
$\tilde{\nu}_-^{(N)}$, were valid as well for nonsymmetric entangled
resources, was left open (see also Ref.~\cite{pirandolareview}). In
Sec.~\ref{sectlc}, devoted to telecloning, we will show with a
specific counterexample that this is {\em not} the case, not even in
the simplest case of $N=2$.

In this respect, let us  mention that the four-mode states of
Chapter \ref{ChapUnlim}, exhibiting an unlimited promiscuous
entanglement sharing, are not completely symmetric and as such they
are {\em not} suitable resources for efficient implementations of
four-partite teleportation networks. Therefore, alternative, maybe
novel communication and/or computation protocols are needed to
demonstrate in the lab
--- and take advantage of --- their unconstrained distribution of
entanglement in simultaneous bipartite and multipartite form. A
suggestion in terms of entanglement transfer from CV systems to
qubits was proposed in Sec.~\ref{Sec4Mengi}.

\section{1\,\ding{221}\,2 telecloning with bisymmetric and
nonsymmetric three-mode resources}\label{sectlc}

\subsection{Continuous variable ``cloning at a distance''}
Quantum {\em telecloning} \cite{telequb} among $N+1$ parties is
defined as a process in which one party (Alice) owns an unknown
quantum state, and wants to distribute her state, via teleportation,
to all the other $N$ remote parties. The no-cloning theorem
\cite{nocloning1,nocloning2} yields that the $N-1$ remote clones can
resemble the original input state only with a finite, nonmaximal
fidelity. In CV systems, $1 \rightarrow N$ telecloning of arbitrary
coherent states was proposed in Ref.~\cite{telecloning}, involving a
special class of $(N+1)$-mode multiparty entangled Gaussian states
(known as ``multiuser quantum channels'') shared as resources among
the $N+1$ users. The telecloning is then realized by a succession of
standard two-party teleportations between the sender Alice and each
of the $N$ remote receivers, exploiting each time the corresponding
reduced two-mode state shared by the selected pair of parties.

Depending on the symmetries of the shared resource, the telecloning
can be realized with equal fidelities for all receivers ({\em
symmetric} telecloning) or with unbalanced fidelities among the
different receivers ({\em asymmetric} telecloning). In particular,
in the first case, the needed resource must have complete invariance
under mode permutations in the $N$-mode block distributed among the
receivers: the resource state has to be thus a $1 \times N$
bisymmetric state \cite{adescaling,unitarily} (see
Sec.~\ref{SecSymm} and Chapter \ref{ChapUniloc}).

Here, based on Ref.~\cite{3mj}, we specialize on $1 \rightarrow 2$
telecloning, where Alice, Bob and Claire share a tripartite
entangled three-mode Gaussian state and Alice wants to teleport
arbitrary coherent states to Bob and Claire with certain fidelities.
As the process itself suggests, the crucial resource enabling
telecloning is not the genuine tripartite entanglement (needed
instead for a successful `multidirectional' teleportation network,
as shown in the previous Section), but the couplewise entanglement
between the pair of modes $1|2$ and $1|3$. We are assuming that
  the sender (Alice) owns mode $1$, while the receivers (Bob and
Claire) own modes $2$ and $3$.

\subsection{Symmetric telecloning}\label{telebass}

Let us first analyze the case of symmetric telecloning, occurring
when Alice aims at sending two copies of the original state with
equal fidelities to Bob and Claire. In this case it has been proven
\cite{cerfclon,iblisdir,telecloning} that Alice can teleport an
arbitrary coherent state to the two distant twins Bob and Claire
(employing a Gaussian cloning machine)
 with the maximal fidelity
\begin{equation}\label{f23}
{\CMcal F}_{\max}^{1 \rightarrow 2}=\frac{2}{3}\,.
\end{equation}

Very recently, unconditional symmetric $1 \rightarrow 2$ telecloning
of unknown coherent states has been demonstrated experimentally
\cite{exptelecloning}, with a fidelity for each clone of ${\CMcal F}
= 0.58 \pm 0.01$, surpassing the classical threshold of $0.5$,
\eq{fcl}.

The argument accompanying \eq{f23} inspired the introduction of the
`no-cloning threshold' for two-party teleportation \cite{grosgran},
basically stating that only a fidelity exceeding $2/3$ --- thus
greater than the previously introduced ``classical'' threshold of
$1/2$, \eq{fcl}, which implies the presence of entanglement ---
ensures the realization of actual two-party quantum teleportation of
a coherent state. In fact, if the fidelity falls in the range $1/2 <
{\CMcal F} < 2/3$, then Alice could have kept a better copy of the
input state for herself, or sent it to a `malicious' Claire. In this
latter case, the whole process would result into an asymmetric
telecloning, with a fidelity ${\CMcal F} > 2/3$ for the copy
received by Claire. It is worth remarking that two-party CV
teleportation beyond the no-cloning threshold has been recently
demonstrated experimentally, with a fidelity ${\CMcal F} = 0.70 \pm
0.02$ \cite{furunew}. Another important and surprising remark is
that the fidelity of $1 \rightarrow 2$ cloning of coherent states,
given by \eq{f23}, is {\em not} the optimal one. As recently shown
in Ref.~\cite{clingon}, using non-Gaussian operations as well, two
identical copies of an arbitrary coherent state can be obtained with
optimal single-clone fidelity ${\CMcal F} \approx 0.6826$.

In our setting, dealing with Gaussian states and Gaussian operations
only, \eq{f23} represents the maximum achievable success for
symmetric $1 \rightarrow 2$ telecloning of coherent states. As
previously anticipated,  the {\em basset hound states} $\sig_B^p$ of
Sec. \ref{secbas} are  the best suited resource states for symmetric
telecloning. Such states belong to the family of multiuser quantum
channels introduced in Ref.~\cite{telecloning}, and are $1 \times 2$
bisymmetric pure states (see Fig.~\ref{figbasset}), parametrized by
the single-mode mixedness $a$ of mode $1$, according to
Eqs.~{\rm(\ref{bassigl}, \ref{basseps})}.  In particular, it is
interesting to study how the single-clone telecloning fidelity
behaves compared with the actual amount of entanglement in the $1|l$
($l=2,3$) nonsymmetric two-mode reductions of $\sig_B^p$ states.

\begin{figure}[t!]
\centering{
\includegraphics[width=10cm]{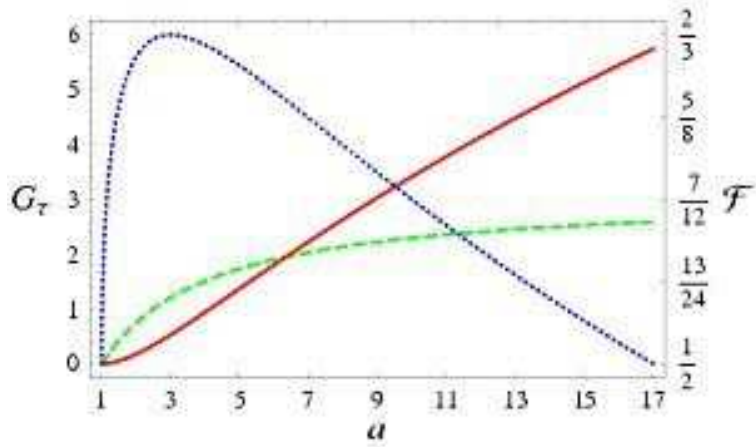}
\caption{Bipartite entanglement $G_\tau^{1|l}$ (dashed green line)
in $1|l$ ($l=2,3$) two-mode reductions of basset hound states, and
genuine tripartite entanglement $G_\tau^{res}$ (solid red line)
among the three modes, versus the local mixedness $a$ of mode $1$.
Entanglements are quantified by the Gaussian contangle (see Chapter
\ref{ChapMonoGauss}). The fidelity ${\CMcal F}^{1 \rightarrow
2}_{sym}$ of symmetric $1 \rightarrow 2$ telecloning employing
basset hound resource states is plotted as well (dotted blue line,
scaled on the right axis), reaching its optimal value of $2/3$ for
$a=3$.} \label{fitette}}
\end{figure}

The fidelity for teleporting a single-mode input Gaussian state
$\sig_{in}$ via a two-mode Gaussian entangled resource $\sig_{ab}$
is given by \eq{ficm}. In our case, $\sig_{in}= \id_2$ because Alice
is teleporting coherent states, while the resource $\sig_{ab}$ is
obtained by discarding either the third ($a=1$, $b=2$) or the second
($a=1$, $b=3$) mode from the CM $\sig_B^p$ of basset hound states.
From Eqs.~{\rm(\ref{bassigl}, \ref{basseps}, \ref{ficm})}, the
single-clone fidelity for symmetric $1 \rightarrow 2$ telecloning
exploiting basset hound states is:
\begin{equation}\label{fitsym}
{\CMcal F}^{1 \rightarrow 2}_{sym} = \frac{4}{3 a - 2 \sqrt{2}
\sqrt{a^2 - 1} + 5}\,.
\end{equation}
Notice, remembering that each of modes $2$ and $3$ contains an
average number of photons $\bar n = (a-1)/2$, that \eq{fitsym} is
the same as Eq.~{\rm(19)} of Ref.~\cite{paris}, where a production
scheme for three-mode Gaussian states by interlinked nonlinear
interactions in $\chi^{(2)}$ media is presented, and the usefulness
of the produced resources for $1 \rightarrow 2$ telecloning is
discussed as well.

 The basset hound states realize an optimal
symmetric cloning machine, {\ie}the fidelity of both clones
saturates \eq{f23}, for the finite value $a=3$. Surprisingly, with
increasing $a > 3$, the fidelity \eq{fitsym} starts decreasing, even
if the two-mode entanglements \eq{g11bh} in the reduced
(nonsymmetric) bipartitions of modes $1|2$ and $1|3$, as well as the
genuine tripartite entanglement \eq{gtauresbh}, increase with
increasing $a$. As shown in Fig.~\ref{fitette}, the telecloning
fidelity is not a monotonic function of the employed bipartite
entanglement. Rather, it roughly follows the difference
$G_\tau^{1|l}-G_\tau^{res}$, being maximized where the bipartite
entanglement is stronger than the tripartite one.  This fact
heuristically confirms that in basset hound states bipartite and
tripartite entanglements are competitors, meaning that the CV
entanglement sharing in these states is not promiscuous, as
described in Sec.~\ref{secbas}.

\subsubsection{Entanglement and teleportation fidelity are
inequivalent for nonsymmetric resources}

The example of basset hound states represents a clear hint that the
teleportation fidelity with generic two-mode (pure or mixed)
nonsymmetric resources is {\em not} monotone with the entanglement.
Even if an hypothetical optimization of the fidelity over the local
unitary operations could be performed (on the guidelines of
 Sec.~\ref{SecTelepoppy} \cite{telepoppate}), it would entail a fidelity growing up to $2/3$
and then staying constant while entanglement increases, which means
that no direct estimation of the entanglement can be extracted from
the nonsymmetric teleportation fidelity, at variance with the
symmetric case. To exhibit a quantitative argument, suppose that
\eq{fidn} (with $N=2$) held for nonsymmetric resources as well.
Applying it to the $1|l$ ($l=2,3$) two-mode reduced resources
obtained from basset hound states, would imply an ``optimal''
fidelity reaching $3/4$ in the limit $a \rightarrow \infty$. But
this value is impossible to achieve, even considering non-Gaussian
cloning machines \cite{clingon}: thus, the simple relation between
teleportation fidelity and entanglement, formalized by \eq{fidn},
{\em fails} to hold for nonsymmetric resources, even in the basic
two-mode instance \cite{3mj}.

This somewhat controversial result can be to some extent interpreted
as follows. For symmetric Gaussian states, there exists a `unique
type' of bipartite CV entanglement. In fact, measures such as the
logarithmic negativity (quantifying the violation of the
mathematical PPT criterion), the entanglement of formation (related
to the entanglement cost, and thus quantifying how expensive is the
process of creating a mixed entangled state through LOCC), and the
degree of EPR correlation (quantifying the correlations between the
entangled degrees of freedom) are {\em all} completely equivalent
for such states, being monotonic functions of only the smallest
symplectic eigenvalue $\tilde{\nu}_-$ of the partially transposed CM
(see Sec.~\ref{secEnt2Sympl}). As we have seen, this equivalence
extends also to the efficiency of two-user quantum teleportation,
quantified by the fidelity optimized over local unitaries (see
Sec.~\ref{secpoppy2}). For nonsymmetric states, the chain of
equivalences breaks down. In hindsight, this could have been somehow
expected, as there exist several inequivalent but legitimate
measures of entanglement, each of them capturing distinct aspects of
the quantum correlations (see \eg the discussion in
Sec.~\ref{SecOrderDiscuss}).

In the specific instance of nonsymmetric two-mode Gaussian states,
we have shown that the negativity is neither equivalent  to the
(Gaussian) entanglement of formation (the two measures may induce
inverted orderings on this subset of entangled states, see
Sec.~\ref{secorder}) \cite{ordering}, nor to the EPR correlation
(see Sec.~\ref{SecEPRcorrel}) \cite{extremal}. It is thus justified
that a process like teleportation emphasizes a distinct aspect of
the entanglement encoded in nonsymmetric resources. Notice also that
the richer and more complex entanglement structure of nonsymmetric
states, as compared to that of symmetric states, reflects a crucial
operational difference in the respective (asymmetric and symmetric)
teleportation protocols. While in the symmetric protocols the choice
of sender and receiver obviously does not affect the fidelity, this
is no longer the case in the asymmetric instance: this physical
asymmetry between sender and receiver properly exemplifies the more
complex nature of the two-mode asymmetric entanglement.

\subsection{Asymmetric telecloning}

\begin{figure}[t!]
\centering{\includegraphics[width=9cm]{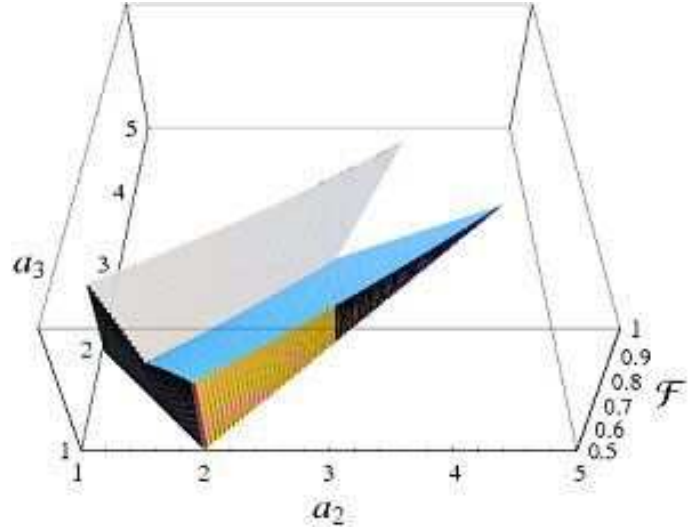}%
\caption{\label{fiaclo}Fidelities for asymmetric telecloning with
three-mode pure Gaussian resources, at a fixed $a_1=2$, as functions
of $a_2$ and $a_3$, varying in the allowed range of parameters
constrained by \ineq{triangleprim} (see also Fig.~\ref{figangle}).
The darker surface on the right-hand side of the diagonal $a_2=a_3$
(along which the two surfaces intersect) is the fidelity of Bob's
clone, ${\CMcal F}_{asym:2}^{1 \rightarrow 2}$, while the lighter,
`mirror-reflected' surface on the left-hand side of the diagonal is
the fidelity of Claire's clone, ${\CMcal F}_{asym:3}^{1 \rightarrow
2}$. Only nonclassical fidelities ({\ie}${\CMcal F} > 1/2$) are
shown.}}
\end{figure}

We focus now on the {\em asymmetric} telecloning of coherent states,
through generic pure three-mode Gaussian states shared as resources
among the three parties. Considering states in standard form,
\eq{cm3tutta} (see Sec.~\ref{secpuri}), parametrized by the local
single-mode mixednesses $a_i$ of modes $i=1,2,3$, the fidelity
${\CMcal F}_{asym:2}^{1 \rightarrow 2}$ of Bob's clone (employing
the $1|2$ two-mode reduced resource) can be computed from \eq{ficm}
and reads
\begin{eqnarray}\label{fiasym2}
 {\CMcal F}_{asym:2}^{1 \rightarrow 2} &=& 2\
\Bigg\{-2 a_3^2 + 2 a_1 a_2 + 4 \left(a_1 + a_2\right) +
          3 \left(a_1^2 +
                a_2^2\right) \\
   &-& \left(a_1 + a_2 +
                  2\right) \sqrt{\frac{[(a_1 + a_2 - a_3)^2 -
                          1] [(a_1 + a_2 + a_3)^2 -
                        1]}{a_1 a_2}} + 2
                        \Bigg\}^{-\frac{1}{2}}\,,
                        \nonumber
\end{eqnarray}
Similarly, the fidelity ${\CMcal F}_{asym:3}^{1 \rightarrow 2}$ of
Claire's clone can be obtained from \eq{fiasym2} by exchanging the
roles of ``$2$'' and ``$3$''.

It is of great interest to explore the space of parameters
$\{a_1,\,a_2,\,a_3\}$ in order to find out which three-mode states
allow for an asymmetric telecloning with the fidelity of one clone
above the symmetric threshold of $2/3$, while keeping the fidelity
of the other clone above the classical threshold of $1/2$. Let us
keep $a_1$ fixed. With increasing difference between $a_2$ and
$a_3$, one of the two telecloning fidelities increases at the
detriment of the other, while with increasing sum $a_2+a_3$ both
fidelities decrease to fall eventually below the classical
threshold, as shown in Fig.~\ref{fiaclo}. The asymmetric telecloning
is thus {\em optimal} when the sum of the two local mixednesses of
modes $2$ and $3$ saturates its lower bound. From
\ineq{triangleprim}, the optimal resources must have
\begin{equation}\label{asybest}
a_3 = a_1 - a_2 + 1\,,
\end{equation}
A suitable parametrization of these states is obtained setting $a_1
\equiv a$ and
\begin{equation}\label{basy}
a_2 = 1+(a-1)t\,,\qquad 0 \le t \le 1\,.
\end{equation}
For $t < 1/2$ the fidelity of Bob's clone is smaller than that of
Claire's one, ${\CMcal F}_{asym:2}^{1 \rightarrow 2} <  {\CMcal
F}_{asym:3}^{1 \rightarrow 2}$, while for $t > 1/2$ the situation is
reversed. In all the subsequent discussion, notice that Bob and
Claire swap their roles if $t$ is replaced by $1-t$. For $t=1/2$,
the asymmetric resources reduce to the bisymmetric basset hound
states useful for symmetric telecloning. The optimal telecloning
fidelities then read
\begin{equation}\label{fiazz12}
\begin{array}{c} {\CMcal F}_{asym:2}^{opt:1 \rightarrow 2} =\frac{2}{\sqrt{(a + 3)^2 +
(a - 1)^2 t^2 + 2 (a - 1) (3 a + 5) t-4 \sqrt{\left(a^2 - 1\right)
t} [a + (a - 1) t + 3]}}\,,
  \end{array}
\end{equation}
and similarly for ${\CMcal F}_{asym:3}^{opt:1 \rightarrow 2}$
replacing $t$ by $1-t$. The two optimal fidelities are plotted in
Fig.~\ref{fiapeto}.

\begin{figure}[t!]
\centering{\includegraphics[width=9cm]{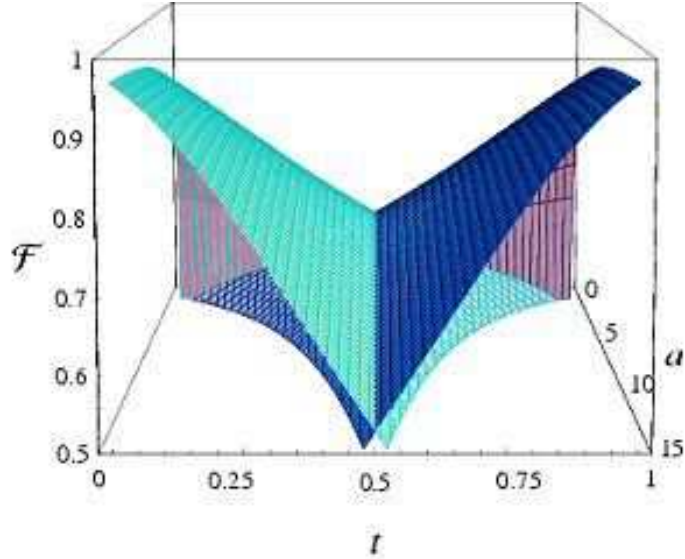}%
\caption{\label{fiapeto}Optimal fidelities for asymmetric
telecloning with three-mode pure Gaussian resources, as functions of
the single-mode mixedness $a$ of mode $1$, and of the parameter $t$
determining the local mixednesses of the other modes, through
Eqs.~{\rm(\ref{asybest}, \ref{basy})}. The darker, rightmost surface
is the optimal fidelity of Bob's clone, ${\CMcal F}_{asym:2}^{opt:1
\rightarrow 2}$, while the lighter, leftmost surface is the optimal
fidelity of Claire's clone, ${\CMcal F}_{asym:3}^{opt:1 \rightarrow
2}$. Along the intersection line $t=1/2$ the telecloning is
symmetric. Only nonclassical fidelities ({\ie}${\CMcal F}
> 1/2$) are shown.}}
\end{figure}

With these pure nonsymmetric resources, further optimizations can be
performed depending on the needed task. For instance, one may need
to implement telecloning with the highest possible fidelity of one
clone, while keeping the other nonclassical. This problem is of
straightforward solution, and yields optimal asymmetric resources
with
\begin{equation}\label{atas}
a=\frac{7}{2}\,,\quad t=\frac{4}{5}\,.
\end{equation}
In this case the fidelity of Claire's clone saturates the classical
threshold, ${\CMcal F}_{asym:3}^{opt:1 \rightarrow 2}=1/2$, while
the fidelity of Bob's clone reaches ${\CMcal F}_{asym:3}^{opt:1
\rightarrow 2}=4/5$, which is the maximum allowed value for this
setting \cite{fiuraclon}. Also, choosing $t=1/5$, Bob's fidelity
gets classical and Claire's fidelity is maximal.

In general, a telecloning with ${\CMcal F}_{asym:2}^{opt:1
\rightarrow 2} \ge 2/3$ and ${\CMcal F}_{asym:3}^{opt:1 \rightarrow
2} \ge 1/2$ is possible only in the window
\begin{equation}\label{awindow}
 1.26 \approx 2 \sqrt{2} \left[2 - \sqrt{1 + \sqrt{2}}\right]  \le a
\le 2 \sqrt{2} \left[2 + \sqrt{1 + \sqrt{2}}\right] \approx 10.05\,
\end{equation}
and, for each $a$ falling in the region defined by \ineq{awindow},
in the specific range
\begin{equation}\label{trange}
\frac{a - 2 \sqrt{a + 1} + 2}{a - 1} \le t \le \frac{2
\left(\sqrt{2} \sqrt{a + 1} - 2\right)}{a - 1}\,.
\end{equation}
For instance, for $a=3$, the optimal asymmetric telecloning (with
Bob's fidelity above no-cloning and Claire's fidelity above
classical bound) is possible in the whole range $1/2 \le t \le
2\sqrt{2}-1$, where the boundary $t=1/2$ denotes the basset hound
state realizing optimal symmetric telecloning (see
Fig.~\ref{fitette}). The sum $${\CMcal S}^{opt:1 \rightarrow
2}={\CMcal F}_{asym:2}^{opt:1 \rightarrow 2} + {\CMcal
F}_{asym:3}^{opt:1 \rightarrow 2}$$ can be maximized as well, and
the optimization is realized by values of $a$ falling in the range
$2.36 \lesssim a \le 3$, depending on $t$. The absolute maximum of
${\CMcal S}^{opt:1 \rightarrow 2}$ is reached, as expected, in the
fully symmetric instance $t=1/2$, $a=3$,
and yields ${\CMcal S}^{opt:1 \rightarrow 2}_{\max} = %2 \times 2/3 =
4/3$.

We finally recall that optimal three-mode Gaussian resources, can be
produced by implementing the allotment operator (see
Sec.~\ref{secallot}) \cite{3mj}, and employed to perform all-optical
symmetric and asymmetric telecloning machines
\cite{telecloning,fiuraclon}.

}

%section{Tre-modi parte 2 e Telepoppate}
\chapter{Entanglement in Gaussian valence bond
states}\label{ChapGVBS}

{\sf

The description of
 many-body systems and the understanding of multiparticle
 entanglement are among the hardest challenges of quantum physics.
 The two issues are entwined: recently, the basic tools
 of quantum information theory have found useful applications
 in condensed matter physics. In particular, the formalism of {\em valence bond states}
\cite{aklt} and more generally that of the so-called ``matrix
product representations'' \cite{maria},
 have led to an efficient simulation
 of many-body spin Hamiltonians \cite{vidal2} and to a deeper understanding of quantum phase transitions \cite{wolfito}.

On the wave of the growing interest which is being witnessed in the
theoretical and experimental applications of CV systems to quantum
information and communication processing \cite{brareview},  the
extension of the valence bond framework  to Gaussian states of CV
systems has been recently introduced \cite{GMPS}. In this Chapter,
based on Refs.~\cite{gvbs,minsk} we adopt a novel point of view,
aimed to comprehend the correlation picture of the considered
many-body systems from the physical structure of the underlying
valence bond framework. In the case of harmonic lattices, we
demonstrate that the quantum correlation length (the maximum
distance between pairwise entangled sites) of translationally
invariant Gaussian valence bond states is determined by the amount
of entanglement encoded in a smaller structure, the `building
block', which is a Gaussian state isomorphic to the valence bond
projector at each site. This connection provides a series of
necessary and sufficient conditions for bipartite entanglement of
distant pair of modes in Gaussian valence bond states depending on
the parameters of the building block, as explicitly shown for a
six-mode harmonic ring.

For any size of the ring we show remarkably that, when single
ancillary bonds connect neighboring sites, an infinite entanglement
in the building block leads to {\em fully symmetric}
(permutation-invariant, see Sec.~\ref{SecSymm}) Gaussian valence
bond states where each individual mode is equally entangled with any
other, independently of the distance. As the block entropy of these
states can diverge for any bipartition of the ring (see
Sec.~\ref{SecScal}), our results unveil a basic difference with
finite-dimensional valence bond states of spin chains, whose
entanglement is limited by the bond dimensionality \cite{vidal} and
is typically short-ranged \cite{kore}.

We finally focus on the experimental realization of Gaussian valence
bond states by means of quantum optics, provide a scheme for their
state engineering, and discuss the applications of such resources in
the context of CV telecloning (see Sec.~\ref{sectlc}) on multimode
harmonic rings.

\section{Gaussian valence bond states}
\label{SecGVBS}

Let us introduce the basic definitions and notations for Gaussian
valence bond states (GVBS), as adopted in Ref.~\cite{gvbs}. The
so-called matrix product Gaussian states introduced in
Ref.~\cite{GMPS} are $N$-mode states obtained by taking a fixed
number, $M$, of infinitely entangled ancillary bonds (EPR pairs)
shared by adjacent sites, and applying an arbitrary $2M \rightarrow
1$ Gaussian operation on each site $i=1,\ldots,N$. Such a
construction, more properly definable as a ``valence bond'' picture
for Gaussian states, can be better understood by resorting to the
Jamiolkowski isomorphism between quantum operations and quantum
states \cite{nogo2,nogo3}. In this framework, one starts with a
chain of $N$ Gaussian states of $2M+1$ modes (the {\em building
blocks}). The global Gaussian state of the chain is described by a
CM $\gr\Gamma = \bigoplus_{i=1}^N \gr\gamma^{[i]}$. As the interest
in GVBS lies mainly in their connections with ground states of
Hamiltonians invariant under translation \cite{GMPS}, we can focus
on pure ($\det \gr\gamma^{[i]} = 1$), translationally invariant
($\gr\gamma^{[i]} \equiv \gr\gamma \, \forall i$) GVBS. Moreover, in
this Chapter we consider single-bonded GVBS, \ie with $M=1$. This is
also physically motivated in view of experimental implementations of
GVBS, as more than one EPR bond would result in a building block
with five or more correlated modes, which appears technologically
demanding. However, our analysis can be easily generalized to
multiple bonds ($M >1$), and to mixed Gaussian states as well.

\begin{figure}[t!]
\includegraphics[width=9cm]{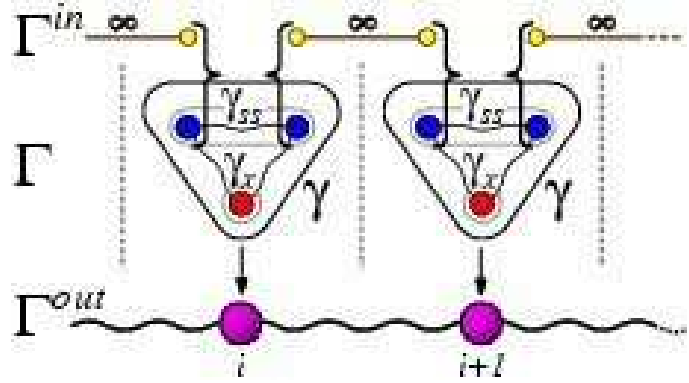} \caption{Gaussian valence bond states.
$\gr\Gamma^{in}$ is the state of $N$ EPR bonds and $\gam$ is the
three-mode building block. After the EPR measurements (depicted as
curly brackets), the chain of modes $\gr\gamma_x$ collapses into a
Gaussian valence bond state with global state $\gr\Gamma^{out}$.}
\label{fiocco}
\end{figure}

Under the considered  prescriptions, the building block $\gr\gamma$
is a pure Gaussian state of three modes (see Sec.~\ref{secpuri} for
an extended discussion on the structural properties of pure
three-mode Gaussian states). As we aim to construct a
translationally invariant GVBS, it is convenient to consider a
$\gr\gamma$ whose first two modes, which will be combined with two
identical halves of consecutive EPR bonds (see Fig.~\ref{fiocco}),
have the same reduced CM. This yields a pure, three-mode Gaussian
building block with the property of being a $2 \times 1$ {\em
bisymmetric} state (see Fig.~\ref{figbasset}), that is with a CM
invariant under permutation of the first two modes. This choice of
the building block is further justified by the fact that, among all
pure three-mode Gaussian states, bisymmetric states maximize the
genuine tripartite entanglement (see Fig.~\ref{figsupposta}): no
entanglement is thus wasted in the projection process.

The $6 \times 6$ CM $\gr\gamma$ of the building block can be written
as follows in terms of $2\times2$ submatrices (see
Sec.~\ref{SecSymm}),
\begin{equation}\label{bblock}
\gr\gamma = \left(
           \begin{array}{ccc}
             \gam_{s} & \eps_{ss} & \eps_{sx} \\
             \eps_{ss}\T & \gam_{s} & \eps_{sx} \\
             \eps_{sx}\T & \eps_{sx}\T & \gam_{x} \\
           \end{array}
         \right)\!.
\end{equation}
The $4\times 4$ CM of the first two modes (each of them having
reduced CM $\gr\gamma_s$) will be denoted by $\gr\gamma_{ss}$, and
will be regarded as the {\em input} port of the building block. On
the other hand, the CM $\gr\gamma_x$ of mode $3$ will play the role
of the {\em output} port. The intermodal correlations are encoded in
the off-diagonal $\eps$ matrices. Without loss of generality, we can
assume $\gam$ to be, up to local unitary operations, in the standard
form of \eq{cm3tutta}, with
\begin{eqnarray}
  &&\gam_s = {\rm diag}\{s,\,s\}\,,\quad \gam_x = {\rm diag}\{x,\,x\}\,, \label{bsform}\\
    &&\eps_{ss} = {\rm diag}\{t_+,\,t_-\}\,,\quad \eps_{sx} = {\rm diag}\{u_+,\,u_-\}\,; \nonumber \\
 && t_{\pm}=\frac{1}{4s}\left[x^2-1 \pm \sqrt{16 s^4 - 8 (x^2 + 1) s^2 + (x^2 -
1)^2}\right]\,, \nonumber \\
  &&u_\pm = \frac{1}{4} \sqrt{\frac{x^2 - 1}{s x}} \left[\sqrt{(x - 2
s)^2 - 1} \pm \sqrt{(x + 2 s)^2 - 1}\right]\,. \nonumber
\end{eqnarray}

The valence bond construction works as follows (see
Fig.~\ref{fiocco}). The global CM $\gr\Gamma = \bigoplus_{i=1}^N
\gr\gamma$ acts as the projector from the  state $\gr\Gamma^{in}$ of
the $N$ ancillary EPR pairs, to the final $N$-mode GVBS
$\gr\Gamma^{out}$. This is realized by collapsing the  state
$\gr\Gamma^{in}$, transposed in phase space, with the `input port'
$\gr\Gamma_{ss}=\bigoplus_i \gr\gamma_{ss}$ of $\gr\Gamma$, so that
the `output port' $\gr\Gamma_{x} = \bigoplus_i \gr\gamma_x$ turns
into the desired $\gr\Gamma^{out}$. Here collapsing means that, at
each site, the two two-mode states, each constituted by one mode
($1$ or $2$)  of $\gr\gamma_{ss}$ and one half of the EPR bond
between site $i$ and its neighbor ($i-1$ or $i+1$, respectively),
undergo an ``EPR measurement'' \ie are projected onto the infinitely
entangled EPR state \cite{nogo2,nogo3,GMPS}. An EPR pair between
modes $i$ and $j$ can be described as a two-mode squeezed state
$\sig_{i,j}(r) \equiv \sig^{sq}_{i,j}(r)$, \eq{tms}, in the limit of
infinite squeezing ($r \rightarrow \infty$). The input state is then
$$\gr\Gamma^{in} = \lim_{r \rightarrow \infty} \bigoplus_{i}^{N}
\gr\sigma_{i,i+1}(r)\,,$$ where we have set periodic boundary
conditions so that $N+1 = 1$ in labeling the sites. The projection
corresponds mathematically to taking a Schur complement (see
Refs.~\cite{GMPS,nogo2,nogo3} for details), yielding an output pure
GVBS of $N$ modes on a ring with a CM
\begin{equation}\label{cmout}
\gr\Gamma^{out} = \gr\Gamma_{x} - \gr\Gamma_{sx}\T (\gr\Gamma_{ss} +
\gr\theta \gr\Gamma^{in} \gr\theta)^{-1} \gr\Gamma_{sx}\,,
\end{equation}
where $\gr\Gamma_{sx}=\bigoplus^N \gr\gamma_{sx}$, and $\gr\theta
=\bigoplus^N {\rm diag}\{1,\,-1,\,1,\,-1\}$ represents transposition
in phase space \cite{Simon00} ($\hat q_i \rightarrow \hat q_i,\,\hat
p_i \rightarrow - \hat p_i$), see \eq{thetatrans}.

Within the building block picture, the valence bond construction can
be {\em in toto} understood as a multiple CV entanglement swapping
\cite{entswap}, as shown in Fig.~\ref{vbswap}: the GVBS is created
as the entanglement in the bonds is swapped to the chain of output
modes via CV teleportation \cite{Braunstein98} (see Chapter
\ref{ChapCommun}) through the input port of the building blocks. It
is thus clear that at a given initialization of the output port (\ie
at fixed $x$), changing the properties of the input port (\ie
varying $s$), which corresponds to implementing different Gaussian
projections from the ancillary space to the physical one, will
affect the structure and entanglement properties of the target GVBS.
This link is explored in the following Section.

\begin{figure}[t!]
\includegraphics[width=\columnwidth]{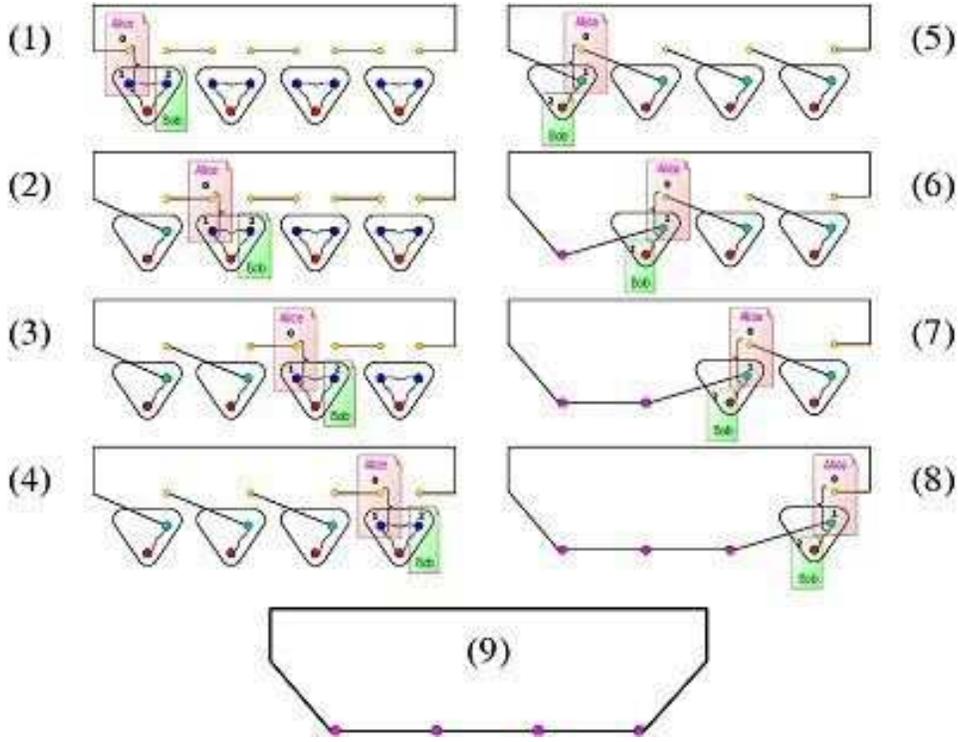} \caption{How a Gaussian valence bond state is created via continuous-variable
entanglement swapping. At each step, Alice attempts to teleport her
mode 0 (half of an EPR bond, depicted in yellow) to Bob, exploiting
as an entangled resource two of the three modes of the building
block (denoted at each step by 1 and 2). The curly bracket denotes
homodyne detection, which together with classical communication and
conditional displacement at Bob's side achieves teleportation. The
state will be approximately recovered in mode 2, owned by Bob. Since
mode 0, at each step, is entangled with the respective half of an
EPR bond, the process swaps entanglement from the ancillary chain of
the EPR bonds to the modes in the building block. The picture has to
be followed column-wise. For ease of clarity, we depict the process
as constituted by two sequences: in the first sequence [frames (1)
to (4)] modes 1 and 2 are the two input modes of the building block
(depicted in blue); in the second sequence [frames (5) to (8)] modes
1 and 2 are respectively an input and an output mode of the building
block. As a result of the multiple entanglement swapping [frame (9)]
the chain of the output modes (depicted in red), initially in a
product state, is transformed into a translationally invariant
Gaussian valence bond state, possessing in general multipartite
entanglement among all the modes (depicted in magenta).}
\label{vbswap}
\end{figure}

We note here that the Gaussian states generally constructed
according to the above
 procedure are ground states of harmonic Hamiltonians (a
property of all GVBS \cite{GMPS}). This follows as no mutual
correlations are ever created between the operators $\hat q_i$ and
$\hat p_j$ for any $i,j=1,\ldots,N$, due to the fact that  both EPR
bonds and building blocks are chosen from the beginning in standard
form. The final CM \eq{cmout} thus takes the form
\begin{equation} \label{circ}
\gr\Gamma^{out} = C^{-1} \oplus C\,, \end{equation} where $C$ is a
circulant $N \times N$ matrix \cite{bathia} and the phase space
operators are assumed here to be ordered as $(\hat q_1,\,\hat
q_2,\,\ldots,\,\hat q_N,\,\hat p_1,\,\hat p_2,\,\ldots,\,\hat p_N)$.
It can be shown that
 a CM of the form \eq{circ} corresponds to the ground state of the quadratic
Hamiltonian $$\hat H = \frac12\big(\sum_{i} \hat p_i^2 + \sum_{i,j}
\hat q_i V_{ij} \hat q_j \big)\,,$$ with the potential matrix given
by $V=C^2$ \cite{chain}. The GVBS we are going to investigate,
therefore, belong exactly to the class of block-diagonal pure
$N$-mode Gaussian states which, in Sec.~\ref{SecGeneric}, have been
shown to achieve ``generic entanglement''. We will now interpret the
entanglement and in general the distribution of correlations in GVBS
in terms of the structural and entanglement properties of the
building block $\gam$.

\subsection{Properties of the building block} In the Jamiolkowski
picture of Gaussian operations \cite{GMPS,nogo2,nogo3}, different
valence bond projectors correspond to differently entangled Gaussian
building blocks. Let us
 recall some results on the characterization of bipartite entanglement
from Part~\ref{PartBip} of this Dissertation.

According to the PPT criterion, a Gaussian state is separable (with
respect to a $1 \times N$ bipartition) if and only if the partially
transposed CM satisfies the uncertainty principle, see
Sec.~\ref{SecPPTG}. As a measure of entanglement, for two-mode {\em
symmetric} Gaussian states $\gr\gamma_{i,j}$ we can adopt either the
logarithmic negativity $E_\N$, \eq{lognegau}, or the entanglement of
formation $E_F$, computable in this case \cite{giedke03} via the
formula \eq{eofgau}. Both measures are equivalent being
monotonically decreasing functions of the positive parameter
$\tilde\nu_{i,j}$, which is the smallest symplectic eigenvalue of
the partial transpose $\tilde{\gr\gamma}_{i,j}$ of
$\gr\gamma_{i,j}$. For a two-mode state, $\tilde\nu_{i,j}$ can be
computed from the symplectic invariants of the state \cite{extremal}
(see Sec.~\ref{SecNega2M}) , and the PPT criterion \eq{sympheispt}
simply yields $\gr\gamma_{i,j}$ entangled as soon as
$\tilde\nu_{i,j}<1$, while infinite entanglement (accompanied by
infinite energy in the state) is reached for $\tilde\nu_{i,j}
\rightarrow 0^+$.

%\subsection{Properties of the building block}

We are interested in studying the quantum correlations of GVBS of
the form \eq{cmout}, and in relating them to the entanglement
properties of the building block $\gr\gamma$, \eq{bblock}. The
building block is a pure three-mode Gaussian state. As discussed in
Sec.~\ref{secpuri}, its standard form covariances \eq{bsform} have
to vary constrained to the triangle inequality \pref{triangleprim}
for $\gr\gamma$ to describe a physical state \cite{3mpra}. This
results in the following constraints on the parameters $x$ and $s$,
\begin{equation}\label{xs}
x \ge 1\,,\quad s\ge s_{\min}\equiv\frac{x+1}{2}\,.
\end{equation}

\begin{figure}[t!]
 \subfigure[] {\includegraphics[width=6cm]{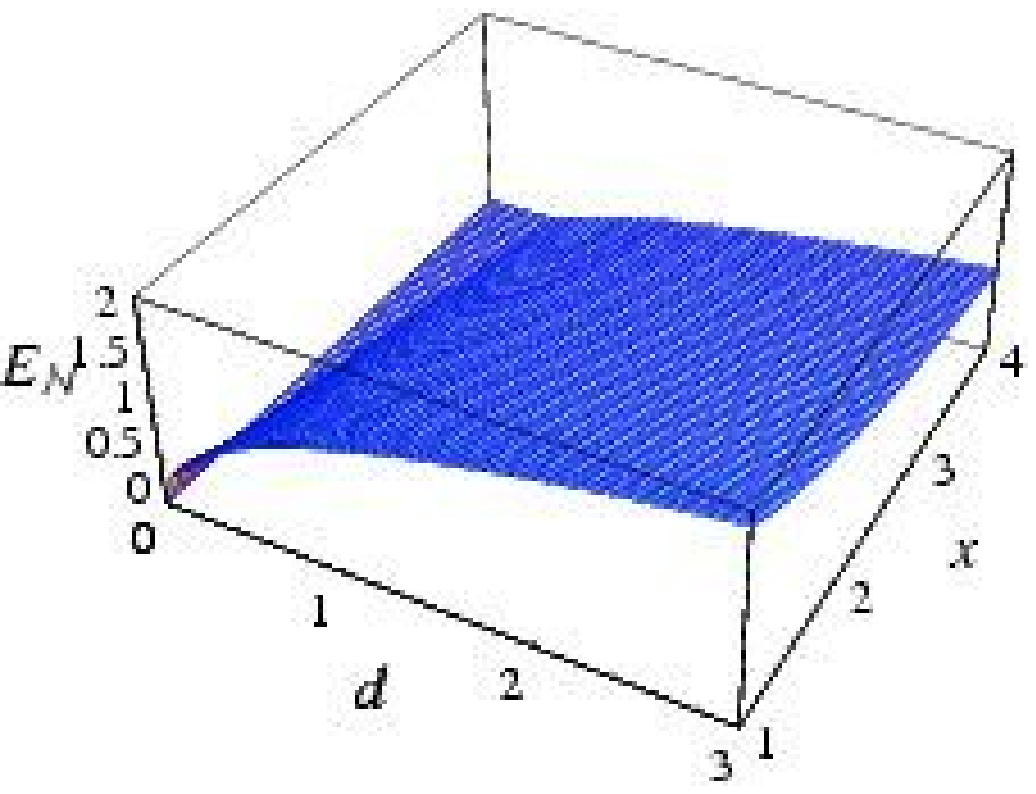}} \hspace{.3cm}
\subfigure[] {\includegraphics[width=6cm]{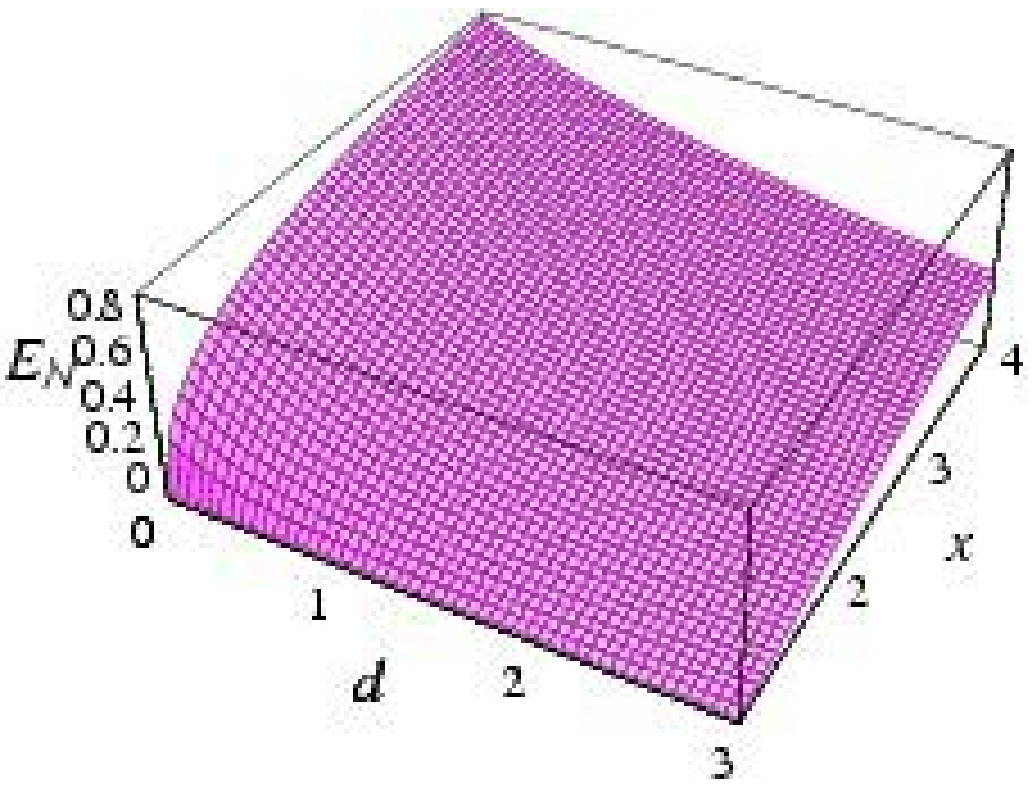}} %\\ %\caption{Plot
\subfigure[] {\includegraphics[width=6cm]{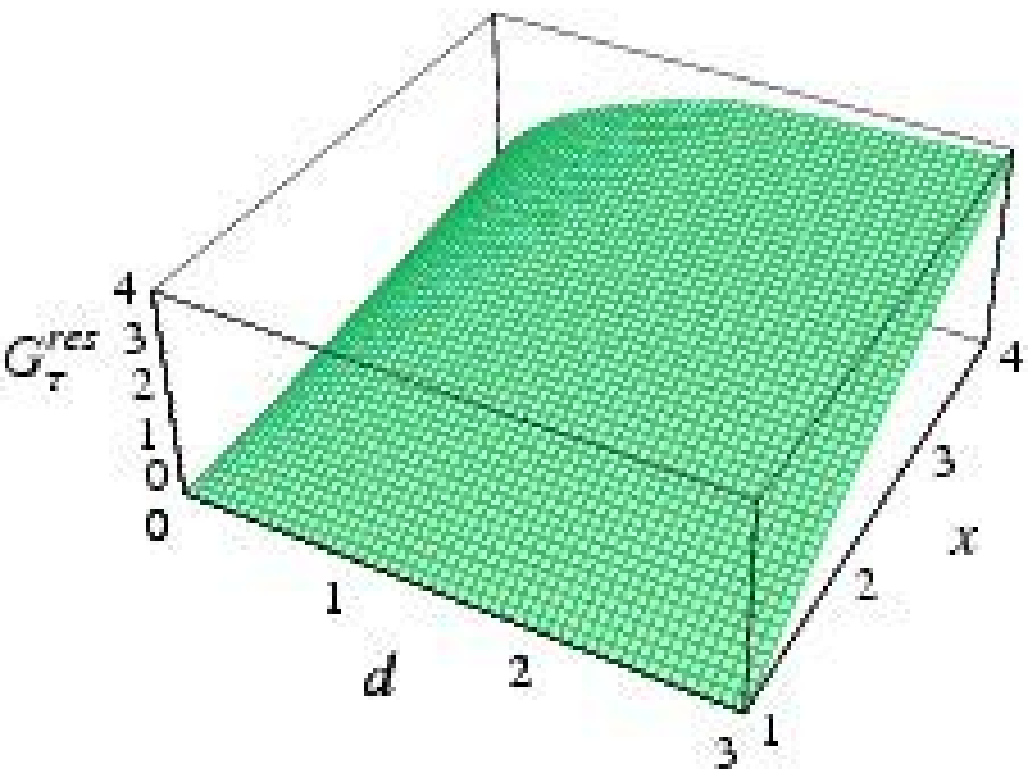}}
\caption{Entanglement properties of the three-mode building block
$\gr\gamma$, \eq{bblock}, of the Gaussian valence bond construction,
as functions of the standard form covariances $x$ and $d \equiv
s-s_{\min}$. {\rm (a)} Bipartite entanglement, as quantified by the
logarithmic negativity, between the first two input-port modes 1 and
2; {\rm (b)} Bipartite entanglement, as quantified by the
logarithmic negativity, between each of the first two modes and the
output-port mode 3; {\rm (c)} Genuine tripartite entanglement, as
quantified by the residual Gaussian contangle, among all the three
modes.}
 \label{figentbblock}
\end{figure}

Let us keep the  parameter $x$ fixed: this corresponds to assigning
the CM of mode 3 (output port). Straightforward applications of the
PPT separability conditions, and consequent calculations of the
logarithmic negativity \eq{lognegau}, reveal that the entanglement
between the first two modes in the CM $\gr\gamma_{ss}$ (input port)
is monotonically increasing as a function of $s$, ranging from the
case $s=s_{\min}$ when $\gr\gamma_{ss}$ is separable to the limit $s
\rightarrow \infty$ when the block $\gr\gamma_{ss}$ is infinitely
entangled. Accordingly, the entanglement between each of the first
two modes $\gam_s$ of $\gr\gamma$ and the third one $\gam_x$
decreases with $s$. One can also show that the genuine tripartite
entanglement in the building block, as quantified by the residual
Gaussian contangle \eq{gtaures} (see Sec.~\ref{sec3purent}),
increases both as a function of $x$ and with increasing difference
\begin{equation}\label{dsmin}
d \equiv s-s_{\min}\,.
\end{equation}
The bipartite and tripartite entanglement properties of the building
block are summarized in Fig.~\ref{figentbblock}.

\section{Entanglement distribution in Gaussian valence bond states}
\label{secmarie}

The main question we raise is how the initial entanglement in the
building block $\gr\gamma$ gets distributed in the GVBS
$\gr\Gamma^{out}$. The answer will be that the more entanglement we
prepare in the input port $\gr\gamma_{ss}$, the longer the range of
the quantum correlations in the output GVBS will be \cite{gvbs}. We
start from the case of minimum $s$.

\subsection{Short-range correlations}\label{secmarieshort}

Let us consider a building block $\gr\gamma$ with
$s=s_{\min}\equiv(x+1)/2$.
%This three-mode Gaussian
%state can be produced experimentally and is useful, for instance, to
%perform $1 \rightarrow 2$ telecloning of unknown coherent states
%\cite{telecloning}.
It is straightforward to evaluate, as a function of $x$, the GVBS in
\eq{cmout} for an arbitrary number of modes (we omit the CM here, as
no particular insight can be drawn from the explicit expressions of
the covariances). By repeatedly applying the PPT criterion, one can
analytically check that each reduced two-mode block
$\gr\gamma^{out}_{i,j}$ is separable for $|i-j|>1$, which means that
the output GVBS $\gr\Gamma^{out}$ exhibits bipartite entanglement
only between nearest neighbor modes, for any value of $x>1$ (for
$x=1$ we obtain a product state).

While this certainly entails that $\gr\Gamma^{out}$ is genuinely
multiparty entangled, due to the translational invariance, it is
interesting to observe that, without feeding entanglement in the
input port $\gr\gamma_{ss}$ of the original building block, the
range of quantum correlations in the output GVBS is  minimum. The
pairwise entanglement between nearest neighbors will naturally
decrease with increasing number of modes, being frustrated by the
overall symmetry and by the intrinsic limitations on entanglement
sharing (the so-called {\em monogamy} constraints \cite{contangle},
see Chapter \ref{ChapMonoGauss}). We can study the asymptotic
scaling of this entanglement in the limit  $x \rightarrow \infty$.
One finds that the corresponding partially-transposed symplectic
eigenvalue $\tilde\nu_{i,i+1}$ is equal to $(N-2)/N$ for even $N$,
and $[(N-2)/N]^{1/2}$ for odd $N$: neighboring sites are thus
considerably
 more entangled if the ring size is even-numbered. Such frustration effect on entanglement
 in odd-sized rings, already devised in a similar context in Ref.~\cite{frusta}, is quite puzzling.
 An explanation may follow from counting arguments
 applied to the number of parameters (which are related to the degree of
 pairwise entanglement) characterizing a block-diagonal pure state on
 harmonic lattices (see Sec.~\ref{SecGeneric}), as we will now show.

\subsubsection{Valence bond representability and entanglement
frustration} \label{secgvbsgeneric}

Let us make a brief digression. It is conjectured that {\em all}
pure $N$-mode Gaussian states can be described as GVBS \cite{GMPS}.
Here we provide a lower bound on the number $M$ of ancillary bonds
required to accomplish this task, as a function of $N$. We restrict
to ground states of harmonic chains with spring-like interactions,
\ie to the block-diagonal Gaussian states of Sec.~\ref{SecGeneric},
which have been proven to rely on $N(N-1)/2$ parameters
\cite{generic}, and which are GVBS with a CM of the form \eq{circ}.

With a simple counting argument using \eq{numero}, the total number
of parameters of the initial chain $\Gamma$ of building blocks
should be at least equal to that of the target state, {\em
i.e.}~$$N(2M+1)(2M)/2 \ge N(N-1)/2\,,$$ which means $M \ge {\rm
IntPart} [(\sqrt{4N-3}-1)/4]$. This implies, for instance, that to
describe general pure states with at least $N>7$ modes, a single EPR
bond per site is no more enough (even though the simplest case of
$M=1$ yields interesting families of $N$-mode GVBS for any $N$, as
shown in the following).

The minimum $M$ scales as $N^{1/2}$, diverging in the field limit $N
\rightarrow \infty$. As infinitely many bonds would be necessary
(and maybe not even sufficient) to describe general infinite
harmonic chains, the valence bond formalism is probably not helpful
to prove or disprove statements related to the entropic area scaling
law \cite{area} for critical bosons, which in general do not fall in
special subclasses of finite-bonded GVBS.\footnote{\sf Recently,
analytical progress on the area law issue (complementing the known
results for the noncritical bosonic case \cite{area}) has been
obtained for the continuum limit of the real scalar Klein-Gordon
massless field \cite{areanew}. It is known \cite{GMPS} that the
ground state of such critical model does not admit a GVBS
representation with a finite number $M$ of ancillary EPR bonds.}

The valence bond picture however effectively captures the
entanglement distribution in translationally invariant $N$-mode
harmonic rings \cite{gvbs}, as we are demonstrating in this Chapter.
In this case the GVBS building blocks are equal at all sites,
$\gr\gamma^{[i]} \equiv \gr\gamma \ \forall i$, while the number of
parameters \eq{numero} of the target state reduces, see
Sec.~\ref{secgenericsf}, to the number of independent pairwise
correlations (only functions of the distance between the two sites),
which by our counting argument is $\Theta_N \equiv (N - N\!\mod
2)/2$. The corresponding threshold for a GVBS representation becomes
$M \ge {\rm IntPart} [(\sqrt{8\Theta_N+1}-1)/4]$. As $\Theta_N$ is
bigger for even $N$, so it is the resulting threshold, which means
that in general a higher number of EPR bonds is needed, and so more
entanglement is inputed in the GVBS projectors and gets distributed
in the target $N$-mode Gaussian state, as opposed to the case of an
odd $N$. This finally clarifies why nearest-neighbor entanglement in
ground states of pure translationally invariant $N$-mode harmonic
rings (which belong to the class of states characterized by
Proposition \rref{propgeneric} of Sec.~\ref{secprop1}) is frustrated
for odd $N$ \cite{frusta}.

\subsection{Medium-range correlations}\label{secmariemedium}

\begin{figure}[t!]
\includegraphics[width=9cm]{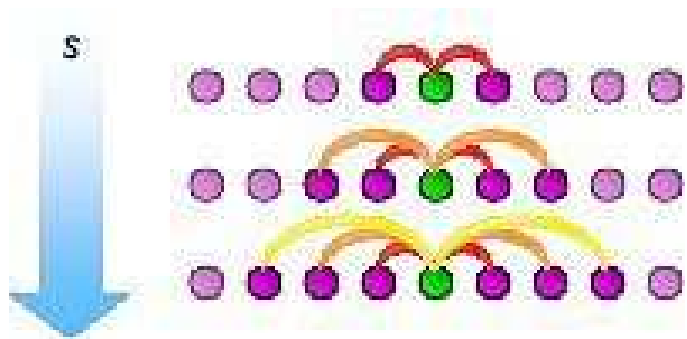}
\caption{Pictorial representation of the entanglement between a
probe (green) mode and its neighbor (magenta) modes on an harmonic
ring with an underlying valence bond structure. As soon as the
parameter $s$ (encoding entanglement in the input port of the
valence bond building block) is increased, pairwise entanglement
between the probe mode and its farther and farther neighbors
gradually appears in the corresponding output Gaussian valence bond
states. By translational invariance, each mode exhibits the same
entanglement structure with its respective neighbors. In the limit
$s \rightarrow \infty$, every single mode becomes equally entangled
with every other single mode on the ring, independently of their
relative distance: the Gaussian valence bond state is in this case
fully symmetric.} \label{figvbs}
\end{figure}

Back to the main track, the connection between input entanglement in
the building block and output correlation length in the destination
GVBS, can be investigated in detail considering a general building
block $\gr\gamma$ with $s > s_{\min}$. The GVBS CM in \eq{cmout} can
still be worked out analytically for a low number of modes, and
numerically for higher $N$. Let us keep the parameter $x$ fixed; we
find that with increasing $s$ the correlations extend smoothly to
distant modes. A series of thresholds $s_k$ can be found such that
for $s > s_k$, two given modes $i$ and $j$ with $|i-j| \le k$ are
entangled. While trivially $s_1(x) = s_{\min}$ for any $N$ (notice
that nearest neighbors are entangled also for $s=s_1$), the
entanglement boundaries for $k>1$ are in general different functions
of $x$, depending on the number of modes. We observe however a
 certain regularity in the process:  $s_k(x,N)$ always increases with the
 integer $k$.
 These considerations follow from analytic calculations on up to
ten-modes GVBS, and we can infer them to hold true for higher $N$ as
well, given the overall scaling structure of the GVBS construction
process. This entails the following remarkable result \cite{gvbs},
pictorially summarized in Fig.~\ref{figvbs}.

\medskip

\begin{itemize}
\item[\ding{226}]
 \noindent{\rm\bf Entanglement distribution in Gaussian valence bond states.} {\it
The maximum range of bipartite entanglement  between two modes, \ie
the maximum distribution of multipartite entanglement, in a GVBS on
a translationally invariant harmonic ring, is monotonically related
to the amount of entanglement in the reduced two-mode input port of
the building block.}
\smallskip
\end{itemize}

Moreover, no complete transfer of entanglement to more distant modes
occurs:  closer sites remain still entangled even when correlations
between farther pairs arise. This feature will be precisely
understood in the limit $s \rightarrow \infty$.

\subsubsection{Example: a six-mode harmonic ring}\label{secmarie6m}

\begin{figure}[t!]
\includegraphics[width=12.5cm]{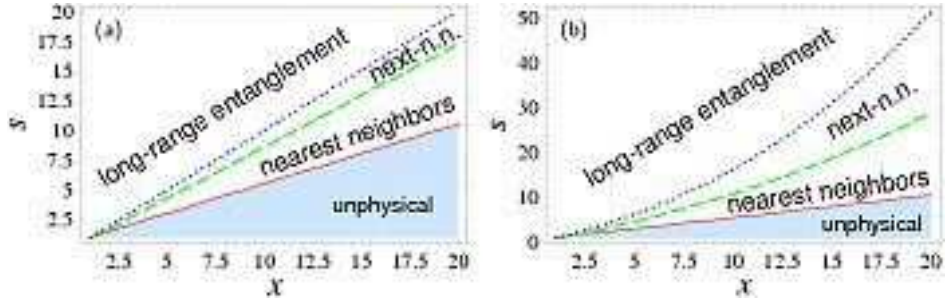}
\caption{Entanglement distribution for a six-mode GVBS constructed
from {\rm(a)} infinitely entangled EPR bonds and {\rm(b)} finitely
entangled bonds given by two-mode squeezed states of the form
\eq{tms} with $r=1.1$. The entanglement thresholds $s_k$ with $k=1$
(solid red line), $k=2$ (dashed green line) and $k=3$ (dotted blue
line) are depicted as functions of the parameter $x$ of the building
block. For $s > s_k$, all pairs of sites $i$ and $j$ with $|i-j| \le
k$ are entangled (see text for further details).} \label{thresh}
\end{figure}

To clearly demonstrate the intriguing connection described above,
let us consider the example of a GVBS with $N=6$ modes. In  a
six-site translationally invariant ring, each mode can be correlated
with another being at most $3$ sites away ($k=1,2,3$). From a
generic building block \eq{bblock}, the $12 \times 12$ CM \eq{cmout}
can be analytically computed as a function of $s$ and $x$. We can
construct the reduced CMs $\gr\gamma^{out}_{i,i+k}$ of two modes
with distance $k$, and evaluate for each $k$ the respective
symplectic eigenvalue $\tilde\nu_{i,i+k}$ of the corresponding
partial transpose. The entanglement condition $s > s_k$ will
correspond to the inequality $\tilde\nu_{i,i+k} < 1$. With this
conditions one finds that $s_2(x)$ is the only acceptable solution
to the equation: $72 s^8 - 12 (x^2 + 1) s^6 + (-34 x^4 + 28 x^2 -
34) s^4 + (x^6 - 5 x^4 - 5 x^2 + 1) s^2 + (x^2 - 1)^2 (x^4 - 6 x^2 +
1)=0$, while for the next-next-nearest neighbors threshold one has
simply $s_3(x)=x$. This enables us to classify the entanglement
distribution and, more specifically, to observe the interaction
scale in the GVBS $\gr\Gamma^{out}$: Fig.~\ref{thresh}{\rm(a)}
clearly shows how, by increasing initial entanglement in
$\gr\gamma_{ss}$, one can gradually switch on quantum correlations
between more and more distant sites.

We can also study entanglement quantitatively. Fig.~\ref{efidi}
shows the entanglement of formation $E_F$ of
$\gr\gamma^{out}_{i,i+k}$ for $k=1,2,3$ (being computable in such
symmetric two-mode reductions, see Sec.~\ref{SecEOFGauss}), as a
function of the standard form covariances $x$ and $d$, \eq{dsmin},
of the building block. For any $(x,d)$ the entanglement is a
decreasing function of the integer $k$, \ie quite naturally it is
always stronger for closer sites. However, in the limit of high $d$
(or, equivalently, high $s$), the three surfaces become close to
each other. We want now to deal exactly with this limit, for a
generic number of modes.

\begin{figure}[t!]
\includegraphics[width=12.5cm]{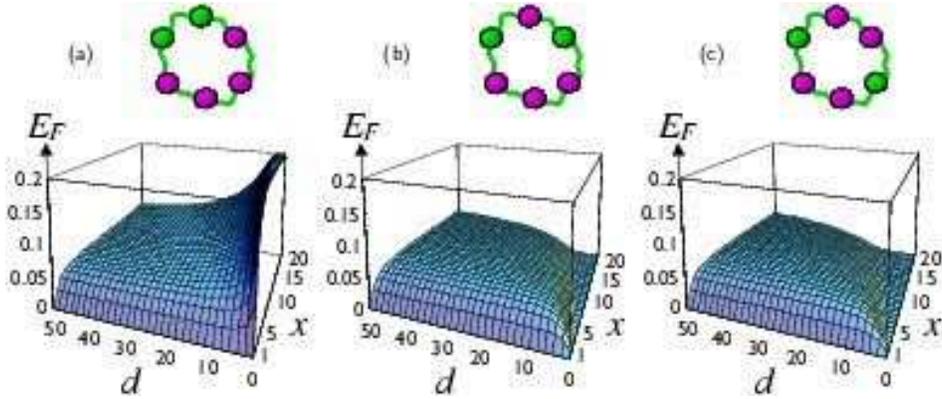}
\caption{Entanglement of formation between two sites $i$ and $j$ in
a six-mode GVBS, with $|i-j|$ equal to: {\rm(a)} $1$, {\rm(b)} $2$,
and {\rm(c)} $3$, as a function of the parameters $x$ and $d \equiv
s-s_{\min}$ determining the building block. For each plot, the modes
whose entanglement is displayed are schematically depicted as well
(green balls).} \label{efidi}
\end{figure}

\subsection{Long-range correlations}\label{secmarielong}

The most interesting feature is perhaps obtained when infinite
entanglement is fed in the input port of the building block ($s
\rightarrow \infty$). In this limit, the expressions greatly
simplify and we obtain a $N$-mode GVBS $\gr\Gamma^{out}$ of the form
\eq{circ}, where $C$ and $C^{-1}$ are completely degenerate
circulant matrices, with \begin{eqnarray*}(C^{-1})_{i,i}=a_q = [(N -
1) + x^2]/(N x)\,,&\quad& (C^{-1})_{i,j\ne i}=c_q =(x^2-1)/(N
x)\,;\\
(C)_{i,i}=a_p = [1 + (N - 1)x^2]/(N x)\,,&\quad& (C^{-1})_{i,j\ne
i}=c_p =-c_q\,.\end{eqnarray*} For any $N$, thus, each individual
mode is {\em equally entangled} with any other, no matter how
distant they are.

The asymptotic limit of our analysis  shows then that an infinitely
entangled input port of the building block results in GVBS with {\em
maximum} entanglement range. These $N$-mode Gaussian states are
well-known as ``fully symmetric'' (permutation-invariant) Gaussian
states, introduced in Sec.~\ref{SecSymm}. The CM $\gr\Gamma^{out}$
of these GVBS can indeed be put, by local symplectic (unitary)
operations, in a standard form parametrized by the single-mode
purity $$\mu_{loc} = 1/\sqrt{\det\gr\gam_k} = (a_q a_p)^{-1/2}$$ as
in \eq{fscm}. Remarkably, in the limit $\mu_{loc} \rightarrow 0$
(i.e.~$x \rightarrow \infty$), the entropy of any $K$-sized ($K<N$)
sub-block of the ring, quantifying entanglement between $K$ modes
and the remaining $N-K$, is {\em infinite}, as shown in
Sec.~\ref{SecScal} (see Fig.~\ref{fiscalb}).

This observation unveils a striking difference between
finite-dimensional and infinite-dimensional valence bond states, as
the former are by construction slightly entangled for a low
dimensionality of the bonds \cite{vidal}, and their entanglement is
short-ranged \cite{kore}. We have just shown instead that pure,
fully symmetric, $N$-mode Gaussian states are exactly GVBS with
minimum bond cardinality ($M=1$): yet, their entanglement can
diverge across any global splitting of the modes, and their pairwise
quantum correlations have maximum range. How this feature connects
with the potential validity of an area law  for all critical bosonic
systems \cite{area,areanew}, as already remarked, is currently an
open question.

\subsubsection{Permutation-invariance and promiscuity from the valence bond construction}

Within the valence bond framework, we also understand the  peculiar
``promiscuous'' entanglement sharing  of fully symmetric $N$-mode
Gaussian states (evidenced for $N=3$ in Sec.~\ref{secpromis}): being
them built by a symmetric distribution of infinite pairwise
entanglement among multiple modes, they achieve maximum genuine
multiparty entanglement while keeping the strongest possible
bipartite one in any pair. Let us also note that in the field limit
($N \rightarrow \infty$) each single pair of
 modes is in a separable state, as they have to
mediate a genuine multipartite entanglement distributed among {\em
all} the  infinite modes under a monogamy constraint (see
Sec.~\ref{SecMonoFulSym}).

Keeping Fig.~\ref{vbswap} in mind, we can conclude that having the
two input modes initially entangled in the building blocks,
increases the efficiency of the entanglement-swapping mechanism,
inducing  correlations between distant modes on the GVBS chain,
which enable to store and distribute joint information. In the
asymptotic limit of an infinitely entangled input port of the
building block, the entanglement range in the target GVBS states is
engineered to be maximum, and communication between any two modes,
independently of their distance, is enabled nonclassically.

In the next Sections, based on Ref.~\cite{minsk}, we investigate the
possibility of producing GVBS with linear optics, and discuss with a
specific example the usefulness of such resource states for
multiparty CV quantum communication protocols such as $1 \rightarrow
(N-1)$ telecloning of unknown coherent states \cite{telecloning}
(see Sec.~\ref{sectlc}).

\section{Optical implementation of Gaussian valence bond states}
\label{secopt}

The power of describing the production of GVBS in terms of physical
states, the building blocks, rather than in terms of arbitrary
non-unitary Gaussian maps, lies not only in the immediacy of the
analytical treatment. From a practical point of view, the recipe of
Fig.~\ref{fiocco} can be directly implemented to produce GVBS
experimentally in the domain of quantum optics. We first note that
the EPR measurements are realized by the standard toolbox of a
beam-splitter plus homodyne detection \cite{nogo3}, as demonstrated
in several CV teleportation experiments \cite{Furusawa98}.

\begin{figure}[t!]
\includegraphics[width=8.5cm]{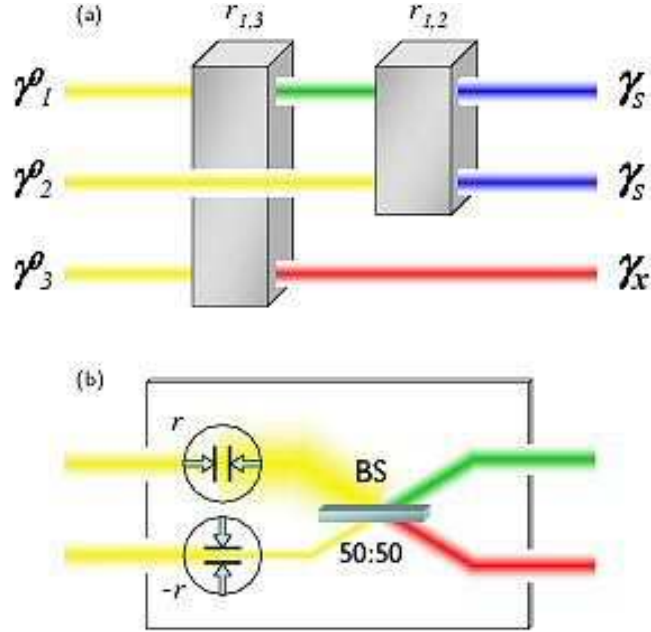} \caption{Optical production of bisymmetric three-mode
Gaussian states, used as building blocks for the valence bond
construction. {\rm(a)} Three initial vacuum modes are entangled
through two sequential twin-beam boxes, the first (parametrized by a
squeezing degree $r_{1,3}$) acting on modes $1$ and $3$, and the
second (parametrized by a squeezing degree $r_{1,2}$) acting on the
transformed mode $1$ and mode $2$. The output is a pure three-mode
Gaussian state whose CM is equivalent, up to local unitary
operations, to the standard form given in \eq{bblock}. {\rm(b)}
Detail of the entangling twin-beam transformation. One input mode is
squeezed in a quadrature, say momentum,  of a degree $r$ (this
transformation is denoted by stretching arrows
$\rightarrow\!\!\!|\,|\!\!\!\leftarrow$); the other input mode is
squeezed in the orthogonal quadrature, say position, of the same
amount (this anti-squeezing transformation is denoted by the
corresponding rotated symbol). Then the two squeezed modes are
combined at a 50:50 beam-splitter. If the input modes are both in
the vacuum state, the output is a pure two-mode squeezed Gaussian
state ({\em twin-beam} state), with entanglement proportional to the
degree of squeezing $r$, as in Fig.~\ref{figtms}.} \label{bbopt}
\end{figure}

The next ingredient to produce a $N$-mode GVBS is constituted by $N$
copies of the building block $\gam$. We provide here an easy scheme,
following Sec.~\ref{secbasengi} (see also Ref.~\cite{telecloning}),
to realize bisymmetric three-mode Gaussian states of the form
\eq{bblock}. As shown in Fig.~\ref{bbopt}, one can start from three
vacuum modes and first apply a two-mode squeezing operation
(twin-beam box) to modes $1$ and $3$, characterized by a squeezing
$r_{1,3}$, then apply another twin-beam operation to modes $1$ and
$2$, parametrized by $r_{1,2}$. The symplectic operation
$T_{i,j}(r_{i,j})$ describing the twin-beam transformation (two-mode
squeezing plus balanced beam-splitter) acting on modes $i$ and $j$
is given by \eq{twin} and pictorially represented in
Fig.~\ref{bbopt}{\rm (b)}. The output of this optical network is a
pure, bisymmetric, three-mode Gaussian state with a CM $$\gam_B =
T_{1,2}(r_{1,2})T_{1,3}(r_{1,3})T_{1,3}\T(r_{1,3})T_{1,2}\T(r_{1,2})\,,$$
of the form \eq{bblock}, with
\begin{eqnarray}
  \gam_s &=& {\rm diag}\left\{\frac{1}{2} e^{-2 r_{1,2}} \left(e^{4 r_{1,2}} \cosh \left(2 r_{1,3}
  \right) + 1\right),\, \frac{1}{2} e^{-2 r_{1,2}} \left(\cosh \left(2
r_{1,3}\right) \ + e^{4 r_{1,2}}\right)\right\}\!, \nonumber \\
 \gam_x &=& {\rm diag} \left\{\cosh \left(2 r_{1,3}\right),\, \cosh\left(2 r_{1,3}\right)\right\}\!, \nonumber \\
    \eps_{ss} &=& {\rm diag} \left\{\frac{1}{2} e^{-2 r_{1,2}} \left(e^{4 r_{1,2}} \cosh \left(2 r_{1,3}
    \right) - 1\right),\, \frac{1}{2} e^{-2 r_{1,2}} \left(\cosh \left(2
r_{1,3}\right) \
- e^{4 r_{1,2}}\right)\right\}\!,   \nonumber \\
 \eps_{sx} &=& {\rm diag} \left\{\sqrt{2} e^{r_{1,2}} \cosh
\left(r_{1,3}\right) \sinh \left(r_{1,3}\right),\, -\sqrt{2}
e^{-r_{1,2}} \cosh \left(r_{1,3}\right) \sinh
\left(r_{1,3}\right)\right\}\!.
 \nonumber \\ \label{bsformopt}
\end{eqnarray}
By means of  {\em local} symplectic operations (unitary on the
Hilbert space), like additional single-mode squeezings, the CM
$\gam_B$ can be brought in the standard form of \eq{bsform}, from
which one has \be\begin{split}r_{1,3}&=\arccos\left(\frac{\sqrt{x +
1}}{\sqrt{2}}\right)\,, \\r_{1,2}&=\arccos\sqrt{\frac{\sqrt{-x^3 + 2
x^2 + 4 s^2 x - x}}{4 x} + \frac{1}{2}}\,.\end{split}\ee For a given
$r_{1,3}$ (\ie at fixed $x$), the quantity $r_{1,2}$ is a
monotonically increasing function of the standard form covariance
$s$, so this squeezing parameter which enters in the production of
the building block (see Fig.~\ref{bbopt}) directly regulates the
entanglement distribution in the target GVBS, as discussed in
Sec.~\ref{secmarie}.

The only unfeasible part of the scheme seems constituted by the
ancillary EPR pairs. But are {\em infinitely} entangled bonds truly
necessary? In Ref.~\cite{gvbs} we have considered the possibility of
using a $\gr\Gamma^{in}$ given by the direct sum of two-mode
squeezed states of \eq{tms}, but with finite $r$. Repeating the
analysis of Sec.~\ref{secmarie} to investigate the entanglement
properties of the resulting GVBS with finitely entangled bonds, it
is found that, at fixed $(x,s)$, the entanglement in the various
partitions is degraded as $r$ decreases, as somehow expected.

Crucially, this does not affect the connection between input
entanglement and output correlation length. Numerical investigations
show that, while the thresholds $s_k$ for the onset of entanglement
between distant pairs are quantitatively modified  ---  namely, a
bigger $s$ is required at a given $x$ to compensate the less
entangled bonds
--- the overall structure stays untouched. As an example,
Fig.~\ref{thresh}{\rm(b)} depicts the entanglement distribution  in
six-mode GVBS obtained from finitely entangled bonds with $r=1.1$,
corresponding to $\approx 6.6$ dB of squeezing (an achievable value
\cite{furuapl}).

This ensures that the possibility of engineering the entanglement
structure in GVBS via the properties of the building block is robust
against imperfect resources, definitely meaning that the presented
scheme is feasible. Alternatively, one could from the beginning
observe that the triples consisting of two projective measurements
and one EPR pair can be replaced by a single projection onto the EPR
state, applied at each site $i$ between the input mode $2$ of the
building block and the consecutive input mode $1$ of the building
block of site $i+1$ \cite{GMPS}. The output of all the homodyne
measurements would conditionally realize the target GVBS.

\section{Telecloning with Gaussian valence bond
resources}\label{secclon}

The protocol of CV quantum  telecloning  among multiple parties
\cite{telecloning} has been described in Sec.~\ref{sectlc}. We can
now consider the general setting of asymmetric $1\rightarrow N-1$
telecloning on harmonic rings, where $N$ parties share a $N$-mode
GVBS as an entangled resource, and one of them plays the role of
Alice (the sender) distributing imperfect copies of unknown coherent
states to all the $N-1$ receivers \cite{minsk}. For any $N$, the
fidelity can be easily computed from the reduced two-mode CMs via
\eq{ficm} and will depend, for translationally invariant states,
only on the relative distance between the two considered modes.

\begin{figure}[t!]
\includegraphics[width=12.5cm]{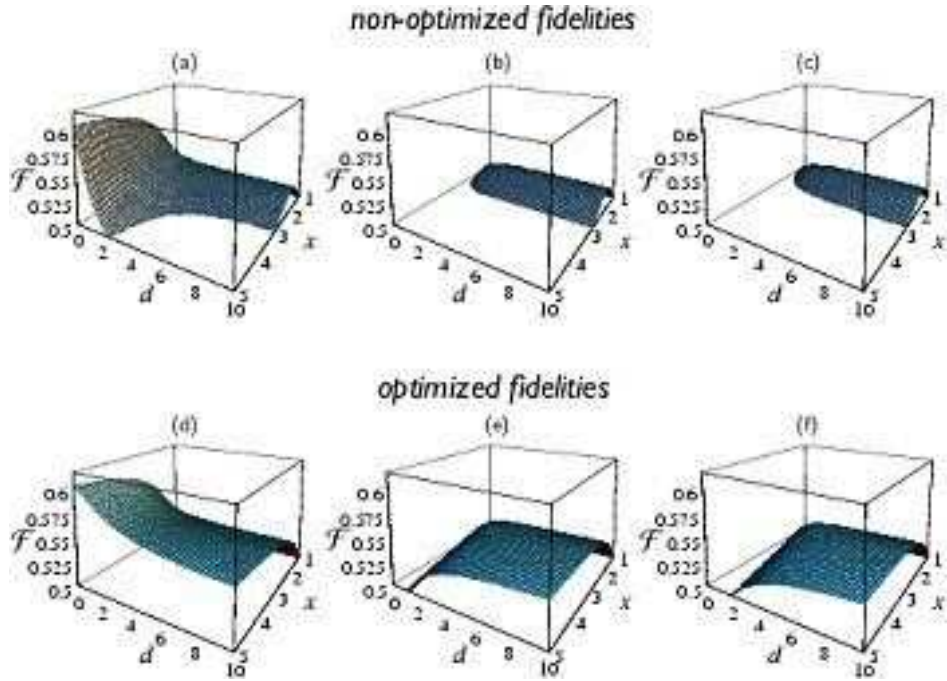}
\caption{$1\rightarrow 5$ quantum telecloning of unknown coherent
states exploiting a six-mode translationally invariant Gaussian
valence bond state as a shared resource. Alice owns mode $i$.
Fidelities $\CMcal F$ for distributing clones to modes $j$ such as
$k=|i-j|$ are plotted for  $k=1$ [{\rm(a)},{\rm(d)}];  $k=2$
[{\rm(b)},{\rm(e)}]; and
 $k=3$ [{\rm(c)},{\rm(f)}], as functions of the local invariants $s$ and $x$
 of the building block. In the first row [{\rm(a)}--{\rm(c)}]
 the fidelities are achieved exploiting the non-optimized
 Gaussian valence bond resource in standard form.
In the second row [{\rm(d)}--{\rm(f)}] fidelities optimized over
local unitary operations on the resource (see
Sec.~\ref{SecTelepoppy}) are displayed, which are equivalent to the
entanglement in the corresponding reduced two-mode states (see
Fig.~\ref{efidi}). Only nonclassical values of the fidelities
(${\CMcal F} > 0.5$) are shown.} \label{televbs}
\end{figure}

We focus here on the practical example of a GVBS on a
translationally invariant harmonic ring, with $N=6$ modes. In
 Sec.~\ref{secmarie6m}, the entanglement distribution in
a six-site GVBS has been studied, finding in particular that, by
increasing initial entanglement in $\gr\gamma_{ss}$, one can
gradually switch on pairwise quantum correlations between more and
more distant sites. Accordingly, it is interesting to test whether
this entanglement is useful to achieve nonclassical telecloning
towards distant receivers [\ie with fidelity ${\CMcal F} > {\CMcal
F}^{cl} \equiv 1/2$, see \eq{fcl}]. In this specific instance, Alice
will send two identical (approximate) clones to her nearest
neighbors, two other identical clones (with in principle different
fidelity than the previous case) to her next-nearest neighbors, and
one final clone to the most distant site. The fidelities for the
three transmissions can be computed from \eq{ficm} and are plotted
in Fig.~\ref{televbs}{\rm(a)}. For $s=s_{\min}$, obviously, only the
two nearest neighbor clones can be teleported with nonclassical
fidelity, as the reduced states of more distant pairs are separable.
With increasing $s$ also the state transfer to more distant sites is
enabled with nonclassical efficiency, but not in the whole region of
the space of parameters $s$ and $x$ in which the corresponding
two-mode resources are entangled [see, as a comparison,
Fig.~\ref{thresh}{\rm(a)}].

As elucidated in Sec.~\ref{SecTelepoppy}, one can optimize the
telecloning fidelity considering resources prepared in a different
way but whose CM can be brought by local unitary operations
(single-mode symplectic transformations) in the standard form of
\eq{cmout}. We know that for symmetric resources --- in this case
the two-mode reduced Gaussian states relative to the sender and each
receiver at a time --- the optimal teleportation fidelity obtained
in this way is equivalent to the shared entanglement
\cite{telepoppate}. For GVBS resources, this local-unitary freedom
can be transferred to the preparation of the building block. A more
general $\gam$ locally equivalent to the standard form given in
\eq{bsform}, can be realized by complementing the presented state
engineering scheme for the three-mode building block as in
\eq{bsformopt} [see Fig.~\ref{bbopt}{\rm(a)}], with additional
single-mode rotations and squeezing transformations aimed at
increasing the output fidelity in the target GVBS states, while
keeping both the entanglement in the building block and consequently
the entanglement in the final GVBS unchanged by definition.

The optimal telecloning fidelity, obtained in this way exploiting
the results of Sec.~\ref{SecTelepoppy}, is plotted in
Fig.~\ref{televbs}{\rm(b)} for the three bipartite teleportations
between modes $i$ and $j$ with $k=|i-j|=1,\,2,\,3$. In this case,
one immediately recovers a non-classical fidelity as soon as the
separability condition $s\le s_k$ is violated in the corresponding
resources. Moreover, the optimal telecloning fidelity at a given $k$
is itself a quantitative measure of the entanglement in the reduced
two-mode resource, being equal to [see \eq{fiopt2}]
\begin{equation}\label{fiopt2g}
{\CMcal F}^{opt}_k = 1/({1+{\tilde\nu_{i,i+k}}})\,,
\end{equation}
where $\tilde\nu_{i,i+k}$ is the smallest symplectic eigenvalue of
the partially transposed CM in the corresponding bipartition. The
optimal fidelity is thus completely equivalent to the entanglement
of formation \eq{eofgau} and to the logarithmic negativity
\eq{lognegau}: the optimal fidelity surfaces of
Fig.~\ref{televbs}{\rm(b)} and  the corresponding entanglement
surfaces of Fig.~\ref{efidi} exhibit indeed the same, monotonic,
behavior.

In the limit $s \rightarrow \infty$, as discussed in
Sec.~\ref{secmarielong}, the GVBS become fully permutation-invariant
for any $N$. Consequently, the (optimized and non-optimized)
telecloning fidelity for distributing coherent states is equal for
any pair of sender-receiver parties. These resources are thus useful
for $1 \rightarrow N-1$ symmetric telecloning. However, due to the
monogamy constraints on distribution of CV entanglement
\cite{contangle} (see Sec.~\ref{SecMonoFulSym}), this two-party
fidelity will decrease with increasing $N$, vanishing in the limit
$N \rightarrow \infty$ where the resources become completely
separable. In this respect, it is worth pointing out that the fully
symmetric GVBS resources (obtained from an infinitely entangled
building block) are more useful for teleportation networks as
thoroughly discussed in Sec.~\ref{secpoppyN}. In this case, more
economical state engineering procedures are available than the
impractical (requiring infinite entanglement) valence bond
construction, as exemplified by Fig.~\ref{teleprepN}.

\section{Discussion}

The valence bond picture is a valuable framework to study the
structure of correlations in quantum states of harmonic lattices. In
fact, the motivation for such a formalism is quite different from
that underlying the finite-dimensional case, where valence
bond/matrix product states are useful to efficiently approximate
ground states of $N$-body systems --- generally described by a
number of parameters exponential in $N$ --- with polynomial
resources \cite{maria}. In CV systems, the key feature of GVBS lies
in the understanding of their entanglement distribution as governed
by the properties of simpler structures. We have in fact shown that
the range of pairwise quantum correlations in translationally
invariant $N$-mode GVBS is determined by the entanglement in the
input port of the building block \cite{gvbs}. To the best of our
knowledge, such an interesting connection had not been pointed out
in traditional discrete-variable matrix product states, and further
investigation in this direction, still resorting to the Jamiolkowski
isomorphism, may be worthy.

Our analysis has also experimental implications giving a robust
recipe to engineer correlations in many-body Gaussian states from
feasible operations on the building blocks. We have provided a
simple scheme to produce bisymmetric three-mode building blocks with
linear optics, and discussed the subsequent implementation of the
valence bond construction. We have also investigated the usefulness
of such GVBS as resources for nonclassical communication, like
telecloning of unknown coherent states to distant receivers on a
harmonic ring \cite{minsk}.

It would be interesting to employ the valence bond picture to
describe quantum computation with CV cluster states
\cite{menicucci}, and to devise efficient protocols for its optical
implementation \cite{clusterloo}.

 }

\chapter{Gaussian entanglement sharing in non-inertial
frames}\label{ChapIvette}

{\sf

%%%%%%%%%%%%%%%%%%%%%%%%%%%%%%%%%%%%%%%%%%%%%%%%%%%%%%%%%%%%%%%%%%%%%%%%%%%%%%%%%%%%%%%%%%%%%%
%\section{Quantum information in a relativistic world}

In the study of most quantum information tasks such as teleportation
and quantum cryptography, non-relativistic observers share entangled
resources to perform their experiments \cite{chuaniels}. Apart from
a few studies \cite{relativistic,alice,dirac,ball,alsing,ahn}, most
works on quantum information assume a world without gravity where
space-time is flat. But the world is relativistic and any serious
theoretical study must take this into account. It is therefore of
fundamental interest to revise quantum information protocols in
relativistic settings \cite{peresreview}. It has been shown that
relativistic effects on quantum resources are not only
quantitatively important but also induce novel, qualitative features
\cite{alice,dirac,ball,ahn}. For example, it has been shown that the
dynamics of space-time can generate entanglement \cite{ball}. This,
in principle, would have a consequence in any entanglement-based
protocol performed in curved space-time. Relativistic effects have
also been found to be relevant in a flat spacetime where the
entanglement measured by observers in relative acceleration is
observer-dependent since it is degraded by the Unruh effect
\cite{alice,dirac,ahn}. In the infinite acceleration limit, the
entanglement vanishes for bosons \cite{alice,ahn} and reaches a
non-vanishing minimum for fermions \cite{dirac}. This degradation on
entanglement results in the loss of fidelity of teleportation
protocols which involve observers in relative acceleration
\cite{alsing}.

Understanding entanglement in a relativistic framework is not only
of interest to quantum information. Relativistic entanglement plays
an important role in black hole entropy \cite{bombelli,wilz,area}
and in the apparent loss of information in black holes \cite{loss},
one of the most challenging problems in theoretical physics at the
moment \cite{hawking1,hawking2}. Understanding the entanglement
between modes of a field close to the horizon of a black hole might
help to understand some of the key questions in black hole
thermodynamics and their relation to information.

In this Chapter, based on Ref.~\cite{ivette}, we interpret the loss
of bipartite entanglement between two modes of a scalar field in
non-inertial frames as an effect of entanglement redistribution. We
consider two observers, each with a detector sensitive to a single
mode. The observers make measurements on the field and look for
correlations to determine the degree to which the field modes are
entangled. Suppose that the observers are inertial and that the two
field modes measured are entangled to a given degree. The state will
appear less entangled if the observers move with uniform
acceleration in a non-inertial frame \cite{alice}. This is because
each inertial mode becomes a two-mode squeezed state in non-inertial
coordinates \cite{davies,unruh}. Therefore, the two-mode entangled
state in the inertial frame becomes a three-mode state when one
observer is in uniform acceleration and a four-mode state if both
observers are accelerated. The observers moving with uniform
acceleration have access only to one of the non-inertial modes.
Therefore, when measuring the state (which involves tracing over the
unaccessible modes) the observers find that some of the correlations
are lost. This effect, from the quantum information perspective, was
first studied for bosonic scalar fields \cite{alice} (considering
one inertial observer and the other one undergoing uniform
acceleration) and later for a fermionic Dirac field \cite{dirac}.
Although entanglement is in both cases degraded as a function of the
acceleration, there are important differences in the results. For
example, in the infinite acceleration limit, the entanglement
reaches a non-vanishing minimum value for fermions, while it
completely disappears in the bosonic case. The loss of entanglement
was explained in the fermionic case in the light of the entanglement
sharing framework (see Sec.~\ref{SecPisa}) as an effect of the
redistribution of entanglement among all, accessible and
unaccessible, modes. Although the loss of entanglement was first
studied for scalar fields (considering an inertial entangled state
which is maximally entangled in a two-qubit space, $\ket{\psi} \sim
\ket{00}+\ket{11}$), entanglement sharing was not analyzed in that
instance, due to the difficulty of computing entanglement in such a
hybrid qubit--continuous-variable system. Fortunately, as we have
thoroughly demonstrated in the previous Parts of this dissertation,
the theory of CV entanglement has been in recent times developed,
allowing for the exact, quantitative study of bipartite entanglement
and its distribution in the special class of Gaussian states.

Here, we consider a free scalar field which is, from an inertial
perspective, in a two-mode squeezed state. The choice of the state
is motivated by different observations. We have seen that it is the
paradigmatic entangled state of a CV system, approximating to an
arbitrarily good extent the EPR pair \cite{EPR35}, and as member of
the Gaussian family it admits an exact description of classical and
quantum correlations. Since the Unruh transformations are Gaussian
themselves, it is possible to characterize analytically the
redistribution of correlations (see Chapter \ref{ChapMonoGauss}) due
to relativistic effects. Furthermore, the two-mode squeezed state
plays a special role in quantum field theory. It is possible to
define particle states (necessary in any entanglement discussion)
when the spacetime has at least two asymptotically flat regions
\cite{dewitt,ball}. In this case, the most general particle states
correspond to multi-mode squeezed states in which all field modes
are in a pair-wise squeezed entangled state. The state we consider
in our entanglement discussion is the simplest multi-mode squeezed
state possible in which all modes are in the vacuum except for two
entangled modes.

A first investigation on the degradation of CV entanglement in a
two-mode squeezed state due to the Unruh effect has been recently
reported \cite{ahn}. The entanglement loss, quantified by the
logarithmic negativity \cite{VidalWerner02} [see \eq{lognegau}], was
analyzed when one of the observers is accelerated and found to
decrease more drastically when the entanglement in the inertial
frame is stronger and to vanish in the infinite acceleration limit.

We perform an extensive study of both quantum (entanglement) and
classical correlations of the two-mode squeezed state in
non-inertial frames. Our work aims at a conclusive understanding and
characterization of the relativistic effects on shared correlations
detected by observers in uniform acceleration. Therefore, we
evaluate not only the bipartite entanglement as degraded by the
Unruh thermalization, but remarkably, the multipartite entanglement
which arises among all modes in Rindler coordinates. Our analysis is
possible thanks to the analytical results on entanglement sharing
and the quantification of multipartite entanglement in Gaussian
states, presented in Part \ref{PartMulti} of this Dissertation. This
analysis relays on the (Gaussian) {\em contangle} \cite{contangle},
\eq{tau}, introduced in Sec.~\ref{SecTau}. The Gaussian contangle
for mixed states is not fully equivalent to the negativity (the
former belonging to the Gaussian entanglement measures, see
Sec.~\ref{secorder}). Therefore, in the case of a single accelerated
observer, our results will evidence significant differences with the
results presented in Ref.~\cite{ahn}. The main novel result we find
in this case, is that in the infinite acceleration limit, all the
bipartite entanglement in the inertial frame is exactly
redistributed into genuine tripartite correlations in the
non-inertial frame. We also analyze total correlations, finding that
the classical correlations are invariant under acceleration when one
observer is non-inertial.

Furthermore, we present an original analysis of the Unruh effect on
CV entanglement when both observers undergo uniform acceleration.
This analysis yields a series of significant new results. First, the
bipartite entanglement measured by observers in non-inertial frames
may vanish completely at finite acceleration even when the state
contains an infinite amount of entanglement in the inertial frame.
Second, the acceleration induces a redistribution of entanglement,
such that the modes in the non-inertial frame share genuine
four-partite entanglement. This entanglement increases unboundedly
with the acceleration, easily surpassing the original inertial
bipartite entanglement (the parametric four-mode state one obtains
is exactly the same as that discussed in Chapter \ref{ChapUnlim}).
Third, classical correlations are also degraded as function of the
acceleration. The degradation is of at most one unit with respect to
the case of a single non-inertial observer. Moreover, we study the
dependence of the bipartite entanglement on the frequency of the
modes detected by the non-inertial observers, finding that with
increasing acceleration the range of entangled frequencies gets
narrower and narrower, becoming empty in the limit of infinite
acceleration.

Our results are on one hand an interesting application of the
Gaussian quantum-information machinery, developed in this
Dissertation (and commonly confined to quantum optics or
light--matter interfaces, as we have seen in the previous Chapters)
to a relativistic setting. On the other hand, they provide a deeper
understanding of the characterization of the inherent relativistic
effects on the distribution of information. This may lead to a
better understanding of the behavior of information in presence of a
black hole \cite{myletter}.

\section{Entanglement in non-inertial frames: the Unruh
effect}\label{SecUnruh}

To study entanglement from the point of view of parties in relative
acceleration is necessary to consider that field quantization in
different coordinates is inequivalent. While an inertial observer
concludes that the field is in the vacuum state, an observer in
relative acceleration detects a thermal distribution of particles
proportional to his/her acceleration. This is known as the Unruh
effect \cite{davies,unruh} and it has important consequences on the
entanglement between (bosonic and/or fermionic) field modes and its
distribution properties \cite{alice,dirac}. We will unveil such
consequences in the case of bosonic scalar fields and a two-mode
squeezed state shared by two observers in an inertial perspective.
Let us first discuss how the Unruh effect arises.

Consider an observer moving in the $(t,z)$ plane ($c=1$) with
constant acceleration $a$. Rindler coordinates $(\tau,\zeta)$ are
appropriate for describing the viewpoint of an uniformly accelerated
observer. Two different sets of Rindler coordinates, which differ
from each other by an overall change in sign, are necessary for
covering Minkowski space. These sets of coordinates define two
Rindler regions ($I$ and $II$) that are causally disconnected from
each other:
\begin{eqnarray*}\label{Rindler_coords}
at &=&  e^{a\zeta}\sinh(a\tau), \quad a z = e^{a\zeta}\cosh(a\tau);
\\  a t &=& -e^{a\zeta}\sinh(a\tau), \quad a z =
-e^{a\zeta}\cosh(a\tau).
\end{eqnarray*} A particle
undergoing eternal uniform acceleration remains constrained to
either Rindler region $I$ or $II$ and has no access to the opposite
region, since these two regions are causally disconnected.

Now consider a free quantum scalar field in a flat background. The
quantization of a scalar field in the Minkowski coordinates is not
equivalent to its quantization in Rindler coordinates. However, the
Minkowski vacuum state can be expressed in terms of a product of
two-mode squeezed states of the Rindler vacuum \cite{BWallsMilburn}
\begin{equation} \label{eq:vacuum}
\left| 0\right\rangle_{\rho_M} = \frac{1}{\cosh r}
\sum_{n=0}^{\infty }\tanh ^{n}r\,\left| n\right\rangle
_{\rho_I}\!\left| n\right\rangle _{\rho_{II}}=U(r)\left|
n\right\rangle _{\rho_I}\!\left| n\right\rangle _{\rho_{II}},
\end{equation}
where
\begin{equation}\label{accparam}
\cosh r
=\left(1-e^{-\frac{2\pi|\omega_\rho|}{a}}\right)^{-\frac12}\,,
\end{equation}
and $U(r)$ is the two-mode squeezing operator introduced in
Eq.~\pref{tmsU}. Each Minkowski mode of frequency $|\omega_\rho|$
has a Rindler mode expansion given by Eq.~\pref{eq:vacuum}. The
relation between higher energy states can be found using
Eq.~\pref{eq:vacuum} and the Bogoliubov transformation between the
creation and annihilation operators, $\hat a_{\rho}=\cosh{r}\hat
b_{\rho_I}-\sinh{r}\hat b^{\dagger}_{\rho_{II}}$, where $\hat
a_{\rho}$ is the annihilation operator in Minkowski space for mode
$\rho$ and $\hat b_{\rho_{I}}$ and $\hat b_{\rho_{II}}$ are the
annihilation operators for the same mode in the two Rindler regions
\cite{davies,unruh}. A Rindler observer moving in region $I$ needs
to trace over the modes in region $II$ since he has no access to the
information in this causally disconnected region. Therefore, while a
Minkowski observer concludes that the field mode $\rho$ is in the
vacuum $|0\rangle_{\rho_M}$, an accelerated observer constrained to
region $I$, detects the state
\begin{eqnarray}
\ketbra{0}{0}_{\rho_M}\rightarrow
{1\over\cosh^2r}\sum_{n=0}^\infty\tanh^{2n}r\ketbra{n}{n}_{\rho_I}
\end{eqnarray}
which is a thermal state with temperature $T=\frac{a}{2\pi k_B}$
where $k_B$ is Boltzmann's constant.

\section{Distributed Gaussian entanglement due to one accelerated
observer}\label{secOne} We consider two inertial observers with
detectors sensitive to modes $\alpha$ and $\rho$, respectively. All
field modes are in the vacuum state except for modes $\alpha$ and
$\rho$ which are originally in a  two-mode squeezed state with
squeezing parameter $s$, as in \cite{ahn}. This Gaussian state,
which is the simplest multi-mode squeezed state (of central
importance in quantum field theory \cite{dewitt}), clearly allows
for the exact quantification of entanglement in all partitions of
the system in inertial and non-inertial frames.

From an inertial perspective, we can describe the two-mode squeezed
state via its CM [see \eq{tms}]
\begin{equation}\label{inAR}
\sig^{p}_{AR}(s)=S_{\alpha_M,\rho_M}(s)\id_{4}S_{\alpha_M,\rho_M}^T(s)\,,
\end{equation}
where $\id_{4}$ is the CM of the vacuum
$\ket{0}_{\alpha_M}\!\otimes\!\ket{0}_{\rho_M}$. If the observer
(Rob) who detects mode $\rho$ is in uniform acceleration, the state
corresponding to this mode must be described in Rindler coordinates,
so that the Minkowski vacuum is given by $\ket{0}_{\rho_M}=\hat
U_{\rho_I,\rho_{II}}(r)
\left(\ket{0}_{\rho_I}\!\otimes\!\ket{0}_{\rho_{II}}\right)$, with
$\hat U(r)$ given by \eq{tmsU}. Namely, the acceleration of Rob
induces a further two-mode squeezing transformation, with squeezing
$r$ proportional to Rob's acceleration [see \eq{accparam}]. As a
consequence of this transformation, the original two-mode
entanglement in the state \eq{inAR} shared by Alice (always
inertial) and Rob from an inertial perspective, becomes distributed
among Alice, the accelerated Rob moving in Rindler region $I$, and a
virtual anti-Rob ($\bar R$) theoretically able to detect the mode
$\rho_{II}$ in the complimentary Rindler region $II$.

 Our aim is to
investigate the distribution of entanglement induced by the purely
relativistic effect of Rob's acceleration. It is clear that the
three-mode state shared by Alice, Rob and anti-Rob is obtained from
the vacuum by the application of Gaussian unitary operations only,
therefore, it is a pure Gaussian state. Its CM, according to the
above description, is (see also \cite{ahn})
\begin{eqnarray}\label{in3}
\sig_{AR \bar R}(r,s)&=&[\id_{\alpha_M} \oplus
S_{\rho_I,\rho_{II}}(r)]\cdot[S_{\alpha_M,\rho_I}(s) \oplus
\id_{\rho_{II}}]\nonumber \\
&\ \ \cdot&\id_{6}\cdot[S_{\alpha_M,\rho_I}^T(s) \oplus
\id_{\rho_{II}}][\id_{\alpha_M} \oplus S_{\rho_I,\rho_{II}}^T(r)]\,,
\nonumber \\
\end{eqnarray}
where the symplectic transformations $S$ are given by \eq{tmsS}, and
 $\id_{6}$ is the CM of the vacuum
$\ket{0}_{\alpha_M}\!\otimes\!\ket{0}_{\rho_I}\!\otimes\!\ket{0}_{\rho_{II}}$.
Explicitly,
\begin{equation}\label{sig3}
\sig_{AR \bar
R}={\left(%
 \begin{array}{ccc}
  \sig_A& \eps_{AR} &  \eps_{A \bar R} \\
  \eps^T_{AR} & \sig_R & \eps_{R \bar R}\\
  \eps^T_{A \bar R}  & \eps^T_{R \bar R} & \sig_{\bar R} \\
\end{array}%
\right)}\,,
\end{equation}
where:
\begin{eqnarray*}
&& \sig_A = \cosh (2 s) \id_2\,, \\ && \sig_R = [\cosh (2 s) \cosh
^2(r) + \sinh ^2(r)] \id_2\,, \\ && \sig_{\bar R} = [\cosh ^2(r) +
\cosh (2 s) \sinh ^2(r)] \id_2\,, \\ &&\eps_{AR}=[\cosh ^2(r) +
\cosh (2 s) \sinh ^2(r)] Z_2\,, \\ &&\eps_{A \bar R} = [\sinh (r)
\sinh (2 s)] \id_2\,, \\ && \eps_{R \bar R} = [\cosh ^2(s) \sinh (2
r)] Z_2\,,
\end{eqnarray*}
 with $Z_2={{1\ \ \ 0}\choose {0 \ -1}}$.
We remind the reader to Chapter \ref{Chap3M} for an introduction to
the structural and informational properties of three-mode Gaussian
states.

As pointed out in Ref.~\cite{alice} the infinite acceleration limit
($r \rightarrow \infty$) can be interpreted as Alice and Rob moving
with their detectors close to the horizon of a black hole. While
Alice falls into the black hole Rob escapes the fall by accelerating
away from it with uniform acceleration $a$.

\subsection{Bipartite entanglement}
We now turn to an analysis of the entanglement between the different
observers. As already mentioned, we adopt the Gaussian contangle
\cite{contangle} (see Sec.~\ref{SecTau}) as entanglement quantifier.
Hence we refer to the notation of \eq{tau} and write, for each
partition $i|j$, the corresponding parameter $m_{i|j}$ involved in
the optimization problem which defines the Gaussian contangle for
bipartite Gaussian states.

The Gaussian contangle $G_\tau(\sig^p_{A|R})$, which quantifies the
bipartite entanglement shared by two Minkowski observers, is equal
to $4s^2$, as can be straightforwardly found by inserting
$m^p_{A|R}=\cosh(2s)$ in \eq{tau}.

Let us now compute the bipartite entanglement in the various $1
\times 1$ and $1 \times 2$ partitions of the state $\sig_{AR \bar
R}$. The $1 \times 2$ (Gaussian) contangles are immediately obtained
from the determinants of the reduced single-mode states of the
globally pure state $\sig_{AR \bar R}$, \eq{sig3}, yielding
\begin{eqnarray}\label{m3_12}
  m_{A|(R \bar R)} &=& \sqrt{\det\sig_A} = \cosh(2s)\,, \\
  m_{R|(A \bar R)} &=& \sqrt{\det\sig_R} = \cosh (2 s) \cosh
^2(r) + \sinh ^2(r)\,, \nonumber \\
  m_{\bar R|(AR)} &=& \sqrt{\det\sig_{\bar R}} = \cosh ^2(r) +
\cosh (2 s) \sinh ^2(r)\nonumber\,.
\end{eqnarray}
For any nonzero value of the two squeezing parameters $s$ and $r$
(\ie entanglement in the inertial frame and Rob's acceleration,
respectively), each single party is in an entangled state with the
block of the remaining two parties, with respect to all possible
global splitting of the modes. This classifies the state $\sig_{AR
\bar R}$ as {\em fully inseparable} according to the scheme of
Sec.~\ref{secbarbie} \cite{kraus}: it contains therefore genuine
tripartite entanglement, which will be precisely quantified in the
next subsection, thanks to the results of Sec.~\ref{sec3purent}.

Notice also that $m_{A|(R \bar R)}=m^p_{A|R}$, \ie all the inertial
entanglement is shared, from a non-inertial perspective, between
Alice and the group \{Rob, anti-Rob\}, as expected since the
coordinate transformation $S_{\rho_I,\rho_{II}}(r)$ is a local
unitary operation with respect to the considered bipartition, which
preserves entanglement by definition. In the following, we will
always assume $s \neq 0$ to rule out trivial circumstances.

Interestingly, as already pointed out in Ref.~\cite{ahn}, Alice
shares no direct entanglement with anti-Rob, because the reduced
state $\sig_{A|\bar R}$ is separable by inspection, being $\det
\eps_{A \bar R} \ge 0$. Actually, we notice that anti-Rob shares the
{\em minimum} possible bipartite entanglement with the group
constituted by Alice and Rob. This follows by recalling that, in any
pure three-mode Gaussian state $\sig_{123}$, the local single-mode
determinants have to satisfy the triangle inequality \pref{triangle}
\cite{3mpra}, which reads
$$
\left|m_1-m_2\right|+1 \le m_3 \le m_1+m_2-1\,, $$ with $m_i \equiv
\sqrt{\det{\sig_i}}$. In our case, identifying mode 1 with Alice,
mode 2 with Rob, and mode 3 with anti-Rob, \eq{m3_12} shows that the
state $\sig_{AR \bar R}$ saturates the leftmost side of the triangle
inequality~\pref{triangle}, $$m_{\bar R|(AR)}=m_{R|(A \bar
R)}-m_{A|(R \bar R)}+1\,.$$ In other words, the mixedness of
anti-Rob's mode, which is directly related to his entanglement with
the other two parties, is the smallest possible one. The values of
the entanglement parameters $m_{i|(jk)}$ from \eq{m3_12} are plotted
in Fig.~\ref{mmshow} as a function of the acceleration $r$, for a
fixed degree of initial squeezing $s$.

\begin{figure}[t!]
\centering{\includegraphics[width=9cm]{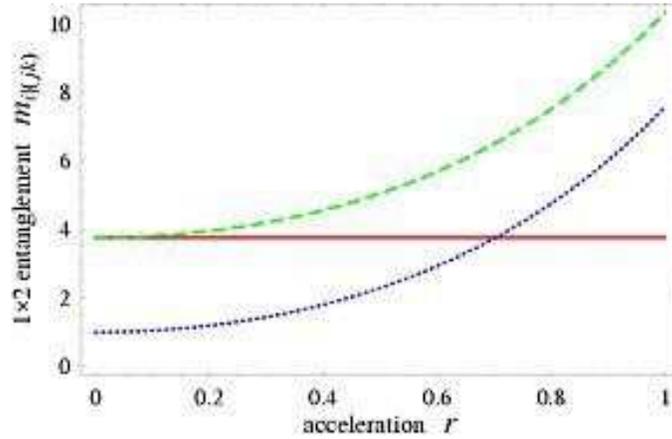}%
\caption{\label{mmshow} Plot, as a function of the acceleration
parameter $r$, of the bipartite entanglement between one observer
and the group of the other two, as expressed by the single-mode
determinants $m_{i(jk)}$ defined in \eq{m3_12}. The inertial
entanglement is kept fixed at $s=1$. The solid red line represents
$m_{A|(R \bar R)}$, the dashed green line corresponds to $m_{R|(A
\bar R)}$, while the dotted blue line depicts $m_{\bar R|(AR)}$.}}
\end{figure}

On the other hand, the PPT criterion (see Sec.~\ref{SecPPTG}) states
that the reduced two-mode states $\sig_{A|R}$ and $\sig_{R|\bar R}$
are both entangled. To compute the Gaussian contangle in those
partitions, we first observe that all the two-mode reductions of
$\sig_{AR \bar R}$ belong to the special class of GMEMMS
\cite{extremal}, mixed Gaussian states of maximal negativities at
given marginal mixednesses, introduced in Sec.~\ref{SecMEMMS} [see
Fig.~\ref{gmemms}{\rm(a)}] This is a curious coincidence because,
when considering entanglement of Dirac fields in non-inertial frames
\cite{dirac}, and describing the effective three-qubit states shared
by the three observers, also in that case all two-qubit reduced
states belong to the corresponding family of MEMMS \cite{AdessoPRA},
mixed two-qubit states of maximal entanglement at fixed marginal
mixednesses [see Fig.~\ref{gmemms}{\rm(b)}]. Back to the CV case,
this observation is useful as we know that for two-mode GMEMMS the
Gaussian entanglement measures, including the Gaussian contangle,
are computable in closed form \cite{ordering}, as detailed in
Sec.~\ref{SecGEMextra}. GMEMMS are indeed simultaneous GMEMS and
GLEMS (see Sec.~\ref{SecGmemsGlems}), therefore either \eq{m2glems}
or \eq{m2gmems} can be used to evaluate their Gaussian contangle. We
have then
\begin{eqnarray}
m_{A|R} &=& \frac{2 \sinh ^2(r) + (\cosh (2 r) +
              3) \cosh (2 s)}{2 \cosh (2 s) \sinh ^2(r) + \cosh (2 r) +
              3}\,, \label{m3_ar} \\
              m_{R|\bar R} &=& \cosh(2r) \label{m3_rr}\,.
\end{eqnarray}

\begin{figure}[t!]
\centering{
 \subfigure[] {\includegraphics[height=5cm]{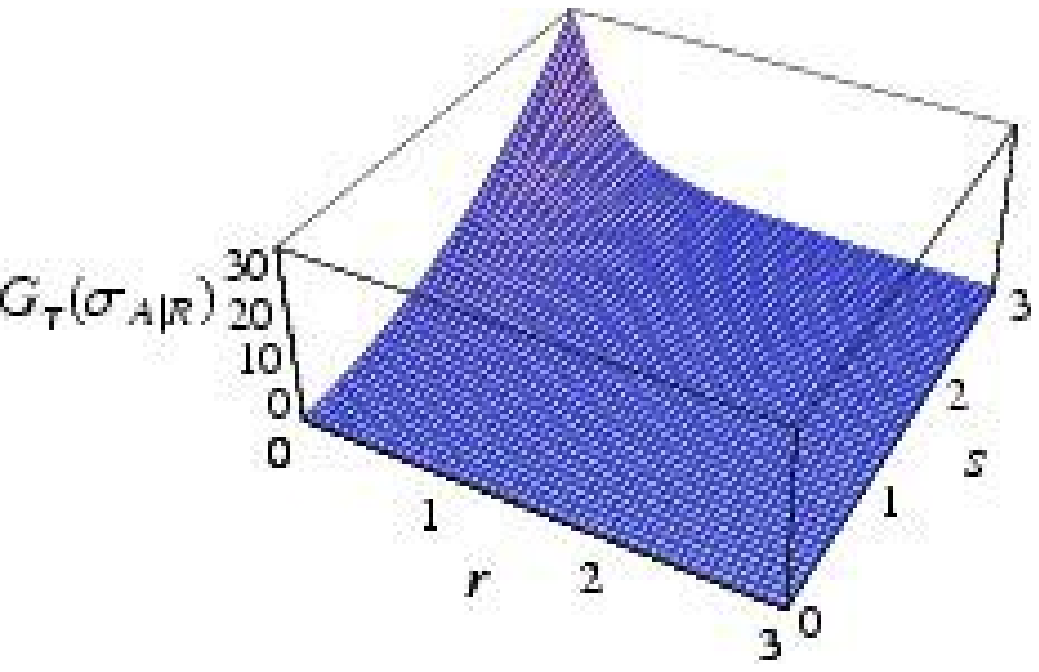}} \\
\subfigure[] {\includegraphics[height=5cm]{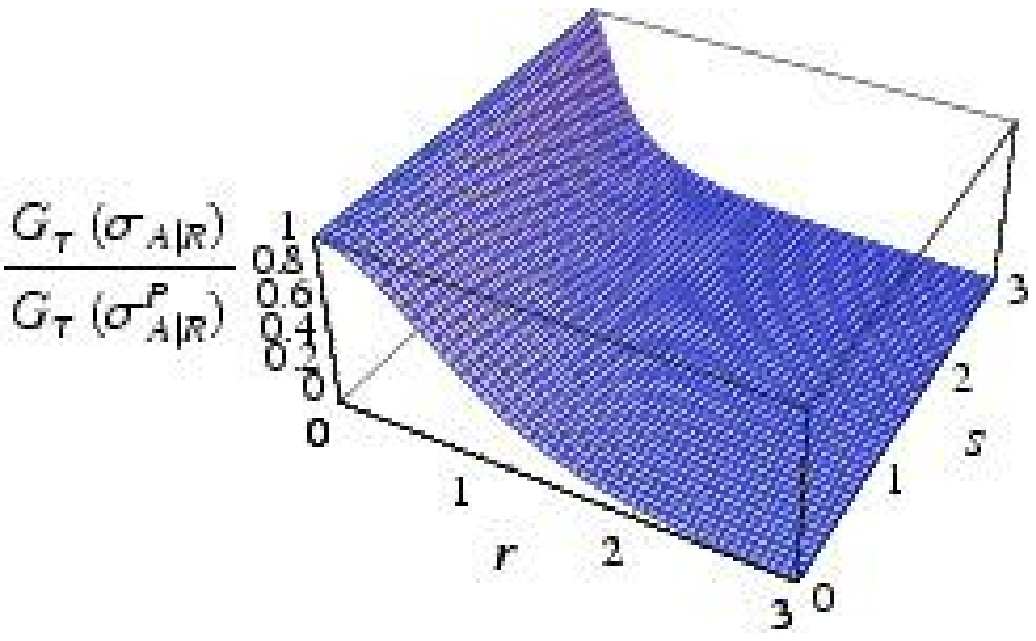}}
 \caption{\label{fi3decay} Bipartite entanglement between
Alice and the non-inertial observer Rob, who moves with uniform
acceleration parametrized by the effective squeezing $r$. From an
inertial perspective, the two observers share a two-mode squeezed
state with squeezing degree $s$. Plot {\rm(a)} depicts the Gaussian
contangle $G_\tau(\sig_{A|R})$, given by Eqs.~{\rm(\ref{tau},
\ref{m3_ar})}, as a function of $r$ and $s$. In plot {\rm(b)} the
same quantity is normalized to the original contangle as seen by
inertial observers, $G_\tau(\sig^p_{A|R})=4s^2$. Notice in {\rm(a)}
how the bipartite Gaussian contangle is an increasing function of
the entanglement, $s$, while it decreases with increasing Rob's
acceleration, $r$, vanishing in the limit $r \rightarrow \infty$.
This decay is faster for higher $s$, as clearly visible in
{\rm(b)}.}}
\end{figure}

Let us first comment on the quantum correlations created between the
two Rindler regions $I$ and $II$, given by \eq{m3_rr}. Note that the
entanglement in the mixed state $\sig_{R \bar R}$ is exactly equal,
in content, to that of a pure two-mode squeezed state with squeezing
$r$, irrespective of the inertial Alice-Rob entanglement quantified
by $s$. This provides a clearcut interpretation of the Unruh
mechanism, in which the acceleration alone is responsible of the
creation of entanglement between the accessible degrees of freedom
belonging to Rob, and the unaccessible ones belonging to the virtual
anti-Rob. By comparison with Ref.~\cite{ahn}, we remark that if the
logarithmic negativity is used as an entanglement measure, this
insightful picture is no longer true, as in that case the
entanglement between Rob and anti-Rob depends on $s$ as well. While
this is not surprising given the aforementioned inequivalence
between negativities and Gaussian entanglement measures in
quantifying quantum correlation of nonsymmetric mixed Gaussian
states \cite{ordering} (see Sec.~\ref{secorder}), it gives an
indication that the negativity is probably not the best quantifier
to capture the transformation of quantum information due to
relativistic effects.

The proper quantification of Gaussian entanglement, shows that the
bipartite quantum correlations are regulated by two competing
squeezing degrees. On one hand, the resource parameter $s$ regulates
the entanglement
 $G_\tau(\sig^p_{A|R})=4s^2$ measured by inertial
observers. On the other hand, the acceleration parameter $r$
regulates the uprising entanglement $G_\tau(\sig_{R|\bar R})=4s^2$
between the non-inertial Rob and his {\em alter ego} anti-Rob. The
latter entanglement, obviously, increases to the detriment of the
Alice-Rob entanglement $G_\tau(\sig_{A|R})=g[m^2_{A|R}]$ perceived
by the accelerating observer. \eq{m3_ar} shows in fact that
$G_\tau(\sig_{A|R})$ is increasing with $s$ and decreasing with $r$,
as pictorially depicted in Fig.~\ref{fi3decay}. Interestingly,
 the rate at which this bipartite entanglement decays with $r$,
$|\partial G_\tau(\sig_{A|R})/
\partial r|$, increases with $s$: for higher $s$ Alice and Rob share
more entanglement (in the inertial frame which corresponds to
$r=0$), but it drops faster when the acceleration ($r$) comes into
play. The same behavior is observed for the negativity \cite{ahn}.
For any inertial entanglement $s$, no quantum correlations are left
in the infinite acceleration limit ($r \rightarrow \infty$), when
the state $\sig_{A|R}$ becomes asymptotically separable.

It is instructive to compare these results to the analysis of
entanglement when the field (for $r=0$) is in a two-qubit Bell state
$\sqrt{\frac12}\left(\ket{0}_{\alpha_M}\ket{0}_{\rho_M} +
\ket{1}_{\alpha_M}\ket{1}_{\rho_M}\right)$, where $\ket{1}$ stands
for the single-boson Fock state \cite{alice}. When one observer is
non-inertial, the state belongs to a three-partite Hilbert space
with dimension $2\times\infty\times\infty$. The free entanglement in
the state is degraded with the acceleration and vanishes in the
infinite acceleration limit. Fig.~\ref{comparob} plots the
entanglement between Alice and the non-inertial Rob in such a
qubit-CV setting \cite{alice}, compared with the fully CV scenario
considered here \cite{ivette}. When the field in the inertial frame
is in a two-mode squeezed Gaussian state with $s
> 1/2$, the entanglement is always stronger than the
entanglement in the Bell-state case. We also observe that, even for
$s< 1/2$, the decay of entanglement with acceleration is slower for
the Gaussian state. The exploitation of all the infinitely-many
degrees of freedom available in the Hilbert space, therefore,
results in an improved robustness of the entanglement against the
thermalization induced by the Unruh effect.

\begin{figure}[t!]
\centering{\includegraphics[width=9cm]{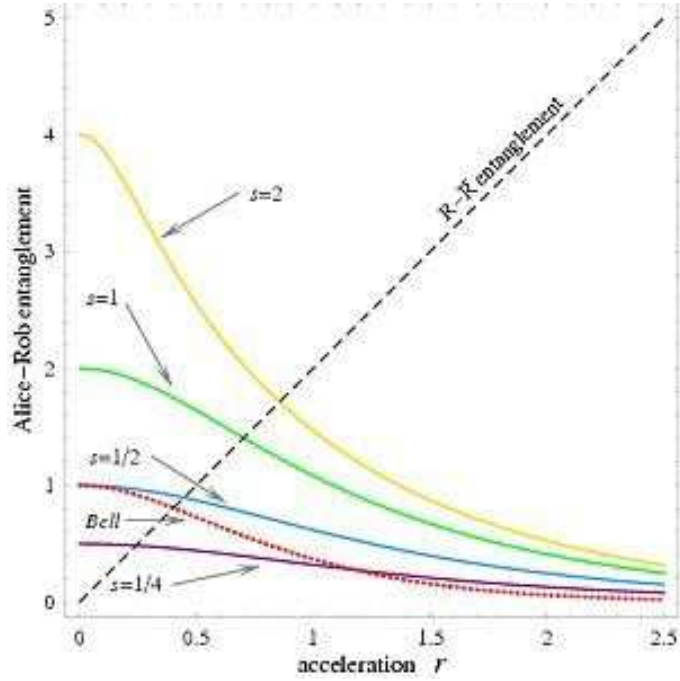}%
\caption{\label{comparob} Bipartite entanglement between Alice and
the non-inertial Rob moving with uniform acceleration parametrized
by $r$. The dotted red curve depicts the evolution of the
logarithmic negativity between Alice and Rob in the instance of them
initially sharing a two-qubit Bell state, as computed in
Ref.~\cite{alice}. The other solid curves correspond to
$\sqrt{G_\tau(\sig_{A|R})}$  (the square root of the Gaussian
contangle is taken to provide a fair dimensional comparison) as
computed in Ref.~\cite{ivette} [see \eq{m3_ar}], in the instance of
Alice and Rob initially sharing a two-mode squeezed state, with
different squeezing parameters $s=0.25,\,0.5,\,1,\,2$ (referring to
the purple, blue, green and gold curve, respectively). As a further
comparison, the entanglement between Rob and anti-Rob, given by
$\sqrt{G_\tau(\sig_{R|\bar R})}=2r$ [see \eq{m3_rr}] independently
of $s$, is plotted as well (dashed black diagonal line).}}
\end{figure}

In this context, we can pose the question of how much entanglement,
at most, can Alice and the non-inertial Rob hope to maintain, given
that Rob is moving with a finite, known acceleration $r$. Assuming
that from an inertial perspective the state is a perfect EPR state,
we find
\begin{equation}\label{limtau}
\lim_{s \rightarrow \infty} m_{A|R} = 1+2/\sinh^2(r)\,,
\end{equation}
meaning that the maximum entanglement left by the Unruh
thermalization, out of an initial unlimited entanglement, approaches
asymptotically
\begin{equation}\label{taumax}
{G_\tau}_r^{\max}(\sig_{A|R}) = {\rm
arcsinh}^2\left[\frac{2\cosh(r)}{\sinh^2(r)}\right]\,.
\end{equation}
Only for zero acceleration, $r=0$, this maximum entanglement
diverges. For any nonzero acceleration, the quantity
${G_\tau}_r^{\max}(\sig_{A|R})$ is finite and rapidly decays with
$r$. This provides an upper bound to the effective quantum
correlations and thus, the efficiency of any conceivable quantum
information protocol that Alice and the non-inertial Rob may wish to
implement. For example, if Rob travels with a modest acceleration
given by $r=0.5$, no more than 8 ebits of entanglement are left
between Alice and Rob, even if they shared an infinitely entangled
state in the Minkowski frame.  This apparent `loss' of quantum
information will be precisely understood in the next subsection,
where we will show that the initial bipartite entanglement does not
disappear, but is redistributed into tripartite correlations among
Alice, Rob and anti-Rob.

\subsection{Tripartite entanglement}

 We have introduced a proper measure of genuine tripartite entanglement for all three-mode Gaussian
 states in Chapter \ref{Chap3M} \cite{contangle,3mpra}, see \eq{gtaures}. This measure, known as
  the ``residual Gaussian contangle'', emerges from the monogamy
inequality~\pref{ckwine} and is an entanglement monotone under
tripartite Gaussian LOCC for pure states, as proven in
Sec.~\ref{secTauresMonotone}. The residual Gaussian contangle
$G_\tau(\sig_{i|j|k})$ of a generic three-mode ($i$, $j$, and $k$)
pure Gaussian state $\sig$ has been computed in
Sec.~\ref{sec3purent}.

\begin{figure}[t!]
\centering{\includegraphics[width=11cm]{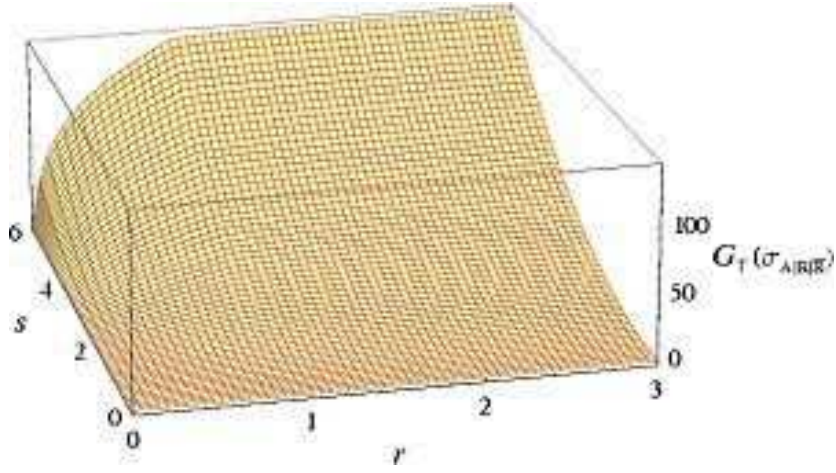}%
\caption{\label{tricont} Genuine tripartite entanglement, as
quantified by the residual Gaussian contangle \eq{tau3unr}, among
the inertial Alice, Rob in Rindler region $I$, and anti-Rob in
Rindler region $II$, plotted as a function of the initial squeezing
$s$ and of Rob's acceleration $r$. The tripartite entanglement
increases with $r$, and for ${r \rightarrow \infty}$ it approaches
the original entanglement content $4s^2$ shared by Alice and Rob in
the Minkowski modes.}}
\end{figure}

We can promptly apply such definition to compute the shared
tripartite entanglement in the state $\sig_{AR \bar R}$. From
\eq{m3_12}, we find that $m_{\bar R|(AR)}<m_{ A|(R \bar R)}$ for
$r<r^\ast$, with
$$r^\ast = {\rm arccosh}\sqrt{\tanh ^2(s) + 1}\,,$$
 while $m_{R|(A \bar R)}$ is always bigger than the other two
quantities. Therefore, by using Eqs.~{\rm(\ref{tau}, \ref{gtaures},
\ref{m3_12}, \ref{m3_ar}, \ref{m3_rr})} together with
$G_\tau(\sig_{A|\bar R})=0$, we find that the residual Gaussian
contangle is given by
\begin{eqnarray}\label{tau3unr}
G_\tau(\sig_{A|R|\bar R}) &=& \left\{
                            \begin{array}{ll}
g[m^2_{\bar R|(AR)}]-g[m^2_{R|\bar R}], & \ r<r^\ast; \\
g[m^2_{A|(R \bar R)}]-g[m^2_{A|R}], & \hbox{otherwise.}
                            \end{array}
                          \right.   \\
&=&  \left\{
                            \begin{array}{l}
                            - 4 r^2+{\rm arcsinh}^2 \sqrt{\left[\cosh ^2(r) + \cosh (2 s) \sinh ^2(r)\right]^2 \
- 1}, \\  \qquad r<r^\ast; \\
 4 s^2 - {\rm arcsinh}^2 \sqrt{\frac{\left[2 \sinh ^2(r) + (\cosh (2 r) +
                                        3) \cosh (2 s)\right]^2}{\left[2 \
\cosh (2 s) \sinh ^2(r) + \cosh (2 r) + 3\right]^2} - 1}, \\
\qquad \hbox{otherwise.} \nonumber
                            \end{array}
                          \right.
\end{eqnarray}

The tripartite entanglement is plotted in Fig.~\ref{tricont} as a
function of $r$ and $s$. Very remarkably, for any initial squeezing
$s$ it increases with increasing acceleration $r$. In the limit of
infinite acceleration, the bipartite entanglement between Alice and
Rob vanishes so we have that
\begin{equation}\label{tau3asi}
\lim_{r \rightarrow \infty} G_\tau(\sig_{A|R|\bar R}) =
G_\tau(\sig_{A|(R \bar R)}) = G_\tau(\sig^p_{A|R})=4s^2\,.
\end{equation}
Precisely, {\em the genuine tripartite entanglement tends
asymptotically to the two-mode squeezed entanglement measured by
Alice and Rob in the inertial frame}.

We have now all the elements necessary to fully understand the Unruh
effect on CV entanglement of bosonic particles, when a single
observer is accelerated. The acceleration of Rob, produces basically
the following effects:
\begin{itemize}
  \item a bipartite entanglement is created {\em ex novo} between the two Rindler
regions in the non-inertial frame. This entanglement is only
function of the acceleration and increases unboundedly with it.
  \item the bipartite entanglement measured by two inertial observers is
redistributed into a genuine tripartite entanglement shared by
Alice, Rob and anti-Rob. Therefore, as a consequence of the monogamy
of entanglement, the entanglement between Alice and Rob is degraded
and eventually disappears for infinite acceleration.
\end{itemize}
In fact, bipartite entanglement is never created between the modes
measured by Alice and anti-Rob. This is very different to the
distribution of entanglement of Dirac fields in non-inertial frames
\cite{dirac}, where the fermionic statistics does not allow the
creation of maximal entanglement between the two Rindler regions,
implying that  the entanglement between Alice and Rob is never fully
degraded; as a result of the monogamy constraints on entanglement
sharing, the mode measured by Alice becomes entangled with the mode
measured by anti-Rob and the entanglement in the resulting
three-qubit system is distributed in couplewise correlations, and a
genuine tripartite entanglement is never created in that case
\cite{dirac}.

In the next Section, we will show how in the bosonic case the
picture radically changes when both observers undergo uniform
acceleration, in which case the relativistic effects are even more
surprising.

\subsection{Mutual information}

It is interesting to compute the total (classical and quantum)
correlations between Alice and the non-inertial Rob, encoded in the
reduced (mixed) two-mode state $\sig_{A|R}$ of \eq{sig3}, using the
mutual information $I(\sig_{A|R})$, \eq{migau}. The symplectic
spectrum of such state is constituted by $\nu_-(\sig_{A|R})=1$ and
$\nu_+(\sig_{A|R})=\sqrt{\det\sig_{\bar R}}$: since it belongs to
the class of GMEMMS, it is in particular a mixed state of partial
minimum uncertainty (GLEMS), which saturates \ineq{bonfide} (see
Sec.~\ref{SecGmemsGlems}). Therefore, the mutual information reads
\begin{equation}\label{mi3}
I(\sig_{A|R}) = f(\sqrt{\det\sig_A})+f(\sqrt{\det\sig_R}) -
f(\sqrt{\det\sig_{\bar R}})\,,
\end{equation}
with $f(x)$ given by \eq{entfunc}

\begin{figure}[t!]
\centering{
 \subfigure[] {\includegraphics[width=6cm]{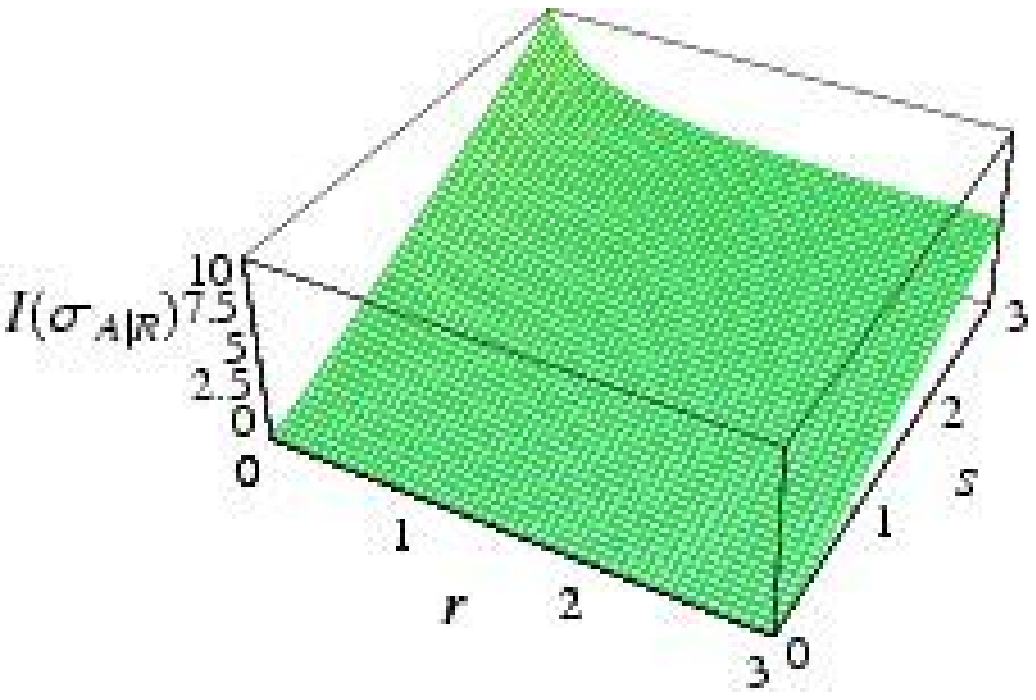}}
 \hspace*{0.2cm}
\subfigure[] {\includegraphics[width=6cm]{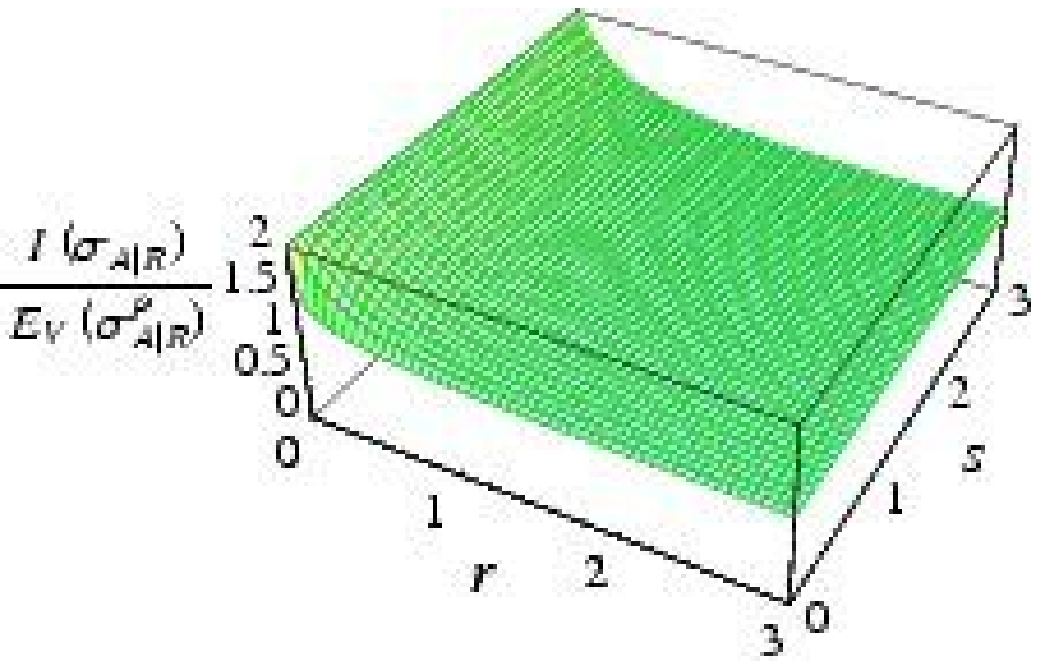}}
\caption{\label{mi3decay}Total correlations between Alice and the
non-inertial observer Rob, moving with acceleration given by the
effective squeezing parameter $r$. In an inertial frame the two
observers shared a two-mode squeezed state with squeezing degree
$s$. Plot {\rm(a)} depicts the evolution of the mutual information
$I(\sig_{A|R})$, given by \eq{mi3}, as a function of $r$ and $s$. In
plot {\rm(b)} the same quantity is normalized to the  entropy of
entanglement as perceived by inertial observers,
$E_V(\sig^p_{A|R})$, \eq{ev3}. Notice in {\rm(a)} how the mutual
information is an increasing function of the initial shared
entanglement, $s$; at variance with the entanglement (see
Fig.~\ref{fi3decay}), it saturates to a nonzero value in the limit
of infinite acceleration. From plot {\rm(b)}, one clearly sees that
this asymptotic value is exactly equal to the inertial entropy of
entanglement.}}
\end{figure}

\noindent Explicitly:

\smallskip \noindent $I(\sig_{A|R}) \\ \hspace*{0.5cm}= \log [\cosh
^2(s) \sinh ^2(r)] \sinh
^2(r) \cosh ^2(s) + \log [\cosh ^2(s)] \cosh ^2(s) \\
 \hspace*{0.5cm}+ \log [\cosh ^2(r) \cosh ^2(s)] \cosh ^2(r) \cosh
^2(s) - \log [\sinh ^2(s)] \sinh ^2(s) \\ \hspace*{0.5cm}-
\frac{1}{2} \log \{\frac{1}{2} [\cosh (2 s) \cosh ^2(r) + \sinh
^2(r) - 1]\} [\cosh (2 s) \cosh ^2(r) + \sinh ^2(r) -
        1] \\ \hspace*{0.5cm}- \frac{1}{2} \log \{\frac{1}{2} [\cosh ^2(r) + \cosh (2 s) \sinh ^2(r) +
              1]\} [\cosh ^2(r) + \cosh (2 s) \sinh ^2(r) +
        1]$.

\medskip

The mutual information of \eq{mi3} is plotted in
Fig.~\ref{mi3decay}{\rm(a)} as a function of the squeezing degrees
$s$ (corresponding to the entanglement in the inertial frame) and
$r$ (reflecting Rob's acceleration). It is interesting to compare
the mutual information with the original two-mode squeezed
entanglement measured between the inertial observers. In this case,
it is more appropriate to quantify the entanglement in terms of the
entropy of entanglement, $E_V(\sig_{A|R}^p$), defined as the Von
Neumann entropy of each reduced single-mode CM [see \eq{E:E}],
$E_V(\sig_{A|R}^p) \equiv S_V(\sig_A^p)\equiv S_V(\sig_B^p)$.
Namely,
\begin{equation}\label{ev3}
E_V(\sig_{A|R}^p)= f(\cosh 2s)\,,
\end{equation}
with $f(x)$ given by \eq{entfunc}. In the inertial frame ($r=0$),
the observers share a pure state, $\sig_{A|R}\equiv\sig_{A|R}^p$ and
the mutual information is equal to twice the entropy of entanglement
of \eq{ev3}, meaning that the two parties are correlated both
quantumly and classically to the same degree. When Rob is under
acceleration ($r \neq 0$), the entanglement shared with Alice is
degraded by the Unruh effect (see Fig.~\ref{fi3decay}), but the
classical correlations are left untouched. In the limit $r
\rightarrow \infty$, all entanglement is destroyed and the remaining
mutual information $I(\sig_{A|R})$, quantifying classical
correlations only, saturates to $E_V(\sig_{A|R}^p)$ from \eq{ev3}.
For any $s>0$ the mutual information of \eq{mi3}, once normalized by
such entropy of entanglement [see Fig.~\ref{mi3decay}{\rm(b)}],
ranges between $2$ ($1$ normalized unit of entanglement plus $1$
normalized unit of classical correlations) at $r=0$, and $1$ (all
classical correlations and zero entanglement) at $r \rightarrow
\infty$. The same behavior is found for classical correlations in
the case of Alice and Rob sharing a bosonic two-qubit Bell state in
an inertial perspective \cite{alice}.

\section{Distributed Gaussian entanglement due to both accelerated
observers} \label{secTwo}

A natural question arises whether the mechanism of degradation or,
to be precise, distribution of entanglement due to the Unruh effect
is qualitatively modified according to the number of accelerated
observers, or it only depends on the establishment of some relative
acceleration between the observers. One might guess that when both
observers travel with constant acceleration, basically the same
features as unveiled above for the case of a single non-inertial
observer will manifest, with a merely quantitative rescaling of the
relevant figures of merit (such as bipartite entanglement decay
rate). Indeed, we will now show that this is {\em not} the case
\cite{ivette}.

We consider here two non-inertial observers, with different names
for ease of clarity and to avoid confusion with the previous
picture. Leo and Nadia both travel with uniform accelerations $a_L$
and $a_N$, respectively. They have single-mode detectors sensitive
to modes $\lambda$ and $\nu$, respectively. We consider, that in the
inertial frame all the field modes are in the vacuum except for
modes $\lambda$ and $\nu$ which are in the pure two-mode squeezed
state $\sig^{p}_{LN}(s)$ of the form \eq{tms}, with squeezing
parameter $s$ as before. Due to their acceleration, two horizons are
created so the entanglement is redistributed among four parties:
Leo, anti-Leo (living respectively in Rindler region $I$ and $II$ of
Leo's horizon), Nadia, anti-Nadia (living respectively in Rindler
region $I$ and $II$ of Nadia's horizon). These four (some real and
some virtual) parties will detect modes $\lambda_{I}$,
$\lambda_{II}$, $\nu_I$, $\nu_{II}$, respectively. By the same
argument of Sec.~\ref{secOne}, the four observers will share a pure,
four-mode Gaussian state, with CM given by
\begin{equation}\label{in4}
\begin{split}
\sig_{\bar L L N \bar N}(s,l,n) &= S_{\lambda_I,\lambda_{II}}(l)
S_{\nu_I,\nu_{II}}(n) S_{\lambda_I,\nu_I}(s)\cdot \id_{8} \\ &\
\cdot\ S_{\lambda_I,\nu_I}^T(s) S_{\nu_I,\nu_{II}}^T(n)
S_{\lambda_I,\lambda_{II}}^T(l)\,, \end{split}
\end{equation}
where the symplectic transformations $S$ are given by \eq{tmsS},
 $\id_{8}$ is the CM of the vacuum
$\ket{0}_{\lambda_{II}}\!\otimes\!\ket{0}_{\lambda_{I}}\!\otimes\!\ket{0}_{\nu_I}\!\otimes\!\ket{0}_{\nu_{II}}$,
while $l$ and $n$ are the squeezing parameters associated with the
respective  accelerations $a_L$ and $a_N$ of Leo and Nadia [see
\eq{accparam}]. Explicitly,
\begin{equation}\label{sig4}
\sig_{\bar L L N \bar N}={\left(%
 \begin{array}{cccc}
  \sig_{\bar L} & \eps_{\bar L L} &  \eps_{\bar L N}  & \eps_{\bar L \bar N}\\
  \eps^T_{\bar L L} & \sig_L & \eps_{L N} & \eps_{L \bar N}\\
  \eps^T_{\bar L N}  & \eps^T_{L N} & \sig_{N} & \eps_{N \bar N} \\
  \eps^T_{\bar L \bar N} & \eps^T_{L \bar n} & \eps^T_{N \bar N} &
  \sig_{\bar N}
\end{array}%
\right)}\,,
\end{equation}
where:
\begin{eqnarray*}
\sig_{\bar X} &=& [\cosh^2(x) + \cosh (2 s) \sinh^2(x)] \id_2\,, \\
 \sig_{X} &=& [\cosh^2(x)\cosh(2s)+\sinh^2(x)] \id_2\,, \\
\eps_{\bar X X}=\eps_{X \bar X} &=& [\cosh^2(s) \sinh (2 x)]
Z_2\,,\\
\eps_{\bar X Y}=\eps_{Y \bar X} &=& [\cosh (y) \sinh (2 s) \sinh
(x)] \id_2 \,, \\
\eps_{\bar X \bar Y}&=&[\sinh (2 s) \sinh (x) \sinh (y)] Z_2\,,\\
\eps_{X Y} &=& [\cosh (x) \cosh (y) \sinh (2 s)] Z_2\,,
\end{eqnarray*}
 with $Z_2={{1\ \ \ 0}\choose {0 \ -1}}$;  $X,Y=\{L,N\}$ with $X \ne Y$,  and
 accordingly for the lower-case symbols $x,y=\{l,n\}$.

The infinite acceleration limit ($l,n \rightarrow \infty$) can now
be interpreted as Leo and Nadia both escaping the fall into a black
hole by accelerating away from it with acceleration $a_L$ and $a_N$,
respectively. Their entanglement will be degraded since part of the
information is lost through the horizon into the black hole
\cite{myletter}. Their acceleration makes part of the information
unavailable to them. We will show that this loss involves both
quantum and classical information.

\subsection{Bipartite entanglement}
We first recall that the original pure-state contangle
$G_\tau(\sig^p_{L|N})=4s^2$ detected by two inertial observers is
preserved under the form of bipartite four-mode entanglement
$G_\tau(\sig_{(\bar L L)|(N \bar N)})$ between the two horizons, as
the two Rindler change of coordinates amount to local unitary
operations with respect to the $(\bar L L)|(N \bar N)$ bipartition.
The computation of the bipartite Gaussian contangle in the various
$1 \times 1$ partitions of the state $\sig_{\bar L L N \bar N}$ is
still possible in closed form thanks to the results of
Sec.~\ref{SecGEMextra} \cite{ordering}. From
Eqs.~{\rm(\ref{m2glems}, \ref{m2gmems}, \ref{tau}, \ref{sig4})}, we
find
\begin{eqnarray}
 m_{L|\bar N} &=& m_{N|\bar L}\,=\,m_{\bar L|\bar
N}\,=\,1\,,   \label{m411sep} \\
 m_{L|\bar L} &=& \cosh(2l)\,,\quad m_{N|\bar N}\,=\,\cosh(2n)\,, \label{m4xbarx}\\
  m_{L|N} &=& \left\{
                \begin{array}{l}
                  1\,, \\  \qquad \tanh (s)\leq \sinh (l) \sinh (n)\,;
                  \\ \\
                  \frac{2 \cosh (2 l) \cosh (2 n) \cosh ^2(s) + 3 \cosh (2 s) -
      4 \sinh (l) \sinh (n) \sinh (2 s) -
      1}{2 \left[(\cosh (2 l) + \cosh (2 n)) \cosh ^2(s) - 2 \sinh ^2(s) +
          2 \sinh (l) \sinh (n) \sinh (2 s)\right]}, \\
          \qquad \hbox{otherwise}\,.
                \end{array}
              \right. \label{m4ln}
\end{eqnarray}

Let us first comment on the similarities with the setting of an
inertial Alice and a non-inertial Rob, discussed in the previous
Section. In the present case of two accelerated observers,
\eq{m411sep} entails (we remind that $m=1$ means separability) that
the mode detected by Leo (Nadia) never gets entangled with the mode
detected by anti-Nadia (anti-Leo). Naturally, there is no bipartite
entanglement generated between the modes detected by the two virtual
observers $\bar L$ and $\bar N$. Another similarity found in
\eq{m4xbarx}, is that the reduced two-mode state $\sig_{X \bar X}$
assigned to each observer $X=\{L,N\}$ and her/his respective virtual
counterpart $\bar X$, is exactly of the same form as $\sig_{R \bar
R}$, and therefore we find again that a bipartite Gaussian contangle
is created {\em ex novo} between each observer and her/his {\em
alter ego}, which is a function of the corresponding acceleration
$x=\{l,n\}$ only. The two entanglements corresponding to each
horizon are mutually independent, and for each the $X|\bar X$
entanglement content is again the same as that of a pure, two-mode
squeezed state created with squeezing parameter $x$.

The only entanglement which is physically accessible to the
non-inertial observers is encoded in the two modes $\lambda_I$ and
$\nu_I$ corresponding to Rindler regions $I$ of Leo and Nadia. These
two modes are left in the state $\sig_{LN}$, which is not a GMEMMS
(like the state $\sig_{AR}$ in Sec.~\ref{secOne}) but a nonsymmetric
thermal squeezed state (GMEMS \cite{extremal}), for which the
Gaussian entanglement measures are available as well [see
\eq{m2gmems}]. The Gaussian contangle of such state is in fact given
by \eq{m4ln}. Here we find a first significant qualitative
difference with the case of a single accelerated observer: a state
entangled from an inertial perspective can become disentangled for
two non-inertial observers, both traveling with {\em finite}
acceleration. \eq{m4ln} shows that there is a trade-off between the
amount of entanglement ($s$) measured from an inertial perspective,
and the accelerations of both parties ($l$ and $n$). If the
observers are highly accelerated --- namely, if $\sinh (l) \sinh
(n)$ exceeds $\tanh (s)$ --- the entanglement in the state
$\sig_{LN}$ vanishes, or better said, becomes physically
unaccessible to the non-inertial observers. Even in the ideal case,
where the state contains infinite entanglement (corresponding to $s
\rightarrow \infty$) in the inertial frame, the entanglement {\em
completely} vanishes in the non-inertial frame if $\sinh (l) \sinh
(n) \ge 1$.  Conversely, for any nonzero, arbitrarily small
accelerations $l$ and $n$, there is a threshold on the entanglement
$s$ such that, if the entanglement is smaller than the threshold, it
vanishes when observed in the non-inertial frames. With only one
horizon, instead (Sec.~\ref{secOne}), any infinitesimal entanglement
will survive for arbitrarily large acceleration, vanishing only in
the infinite acceleration limit. The present feature is also at
variance with the Dirac case, where entanglement never vanishes for
two non-inertial observers \cite{dirac}.

To provide a better comparison between the two settings, let us
address the following question. Can the entanglement degradation
observed by Leo and Nadia (with acceleration parameters $l$ and $n$
respectively) be observed by an inertial Alice and a non-inertial
Rob traveling with some effective acceleration $r^{eff}$? We will
look for a value of $r^{eff}$ such that the reduced state
$\sig_{AR}$ of the three-mode state in \eq{sig3} is as entangled as
the reduced state $\sig_{LN}$ of the four-mode state in \eq{sig4}.
The problem can be straightforwardly solved by equating the
corresponding Gaussian contangles \eq{m3_ar} and \eq{m4ln}, to
obtain
\begin{equation}\label{reff}
r^{eff}=\left\{
  \begin{array}{ll}
    {\rm arccosh}\left[\frac{\cosh (l) \cosh (n) \sinh (s)}{\sinh (s) - \cosh (s) \sinh \
(l) \sinh (n)}\right], & \\ & \hspace*{-3cm} \tanh (s)> \sinh (l)
\sinh (n)\,; \\ &  \\
    \infty, & \hspace*{-.5cm}\hbox{otherwise.}
  \end{array}
\right.
\end{equation}
Clearly, for very high accelerations $l$ and $n$ (or, equivalently,
very small inertial entanglement $s$) the information loss due to
the formation of the second horizon is only matched by an infinite
effective acceleration in the case of a single horizon. In the
regime in which entanglement does not decay completely, the
effective acceleration of Rob in the equivalent
single--non-inertial--observer setting is a function of the inertial
entanglement $s$, as well as of the accelerations of Leo and Nadia.

\subsubsection{Entanglement between different frequency modes}

The condition on the acceleration parameters $l$ and $n$ for which
the entanglement of the maximally entangled state ($s \rightarrow
\infty$) vanishes, from \eq{m4ln}, corresponds to the following
condition
$$e^{\pi\Omega_{L}}+e^{\pi\Omega_{N}}-e^{\pi(\Omega_{L}+\Omega_{N})}\geq
0\,, $$ where $\Omega_{L}=2\lambda/(a_{L})$ and
$\Omega_{N}=2\nu/(a_{N})$. Here we recall that $a_{L,N}$ are the
proper accelerations of the two non-inertial observers and
$\lambda,\nu$ the frequencies of the respective modes, see
\eq{accparam}. We assume now that Leo and Nadia have the same
acceleration $a_{L}=a_{N}\equiv \bar a$ and that they carry with
them single-frequency detectors which can be tuned, in principle, to
any frequency. We ask the question of, given their acceleration,
which frequency modes would they find entangled. This provides a
different, more operationally-oriented view on the setting of this
Chapter and in general on the effect of the Unruh thermalization on
the distribution of CV correlations.

\begin{figure}[t!]
\centering{
 \subfigure[] {\includegraphics[width=6cm]{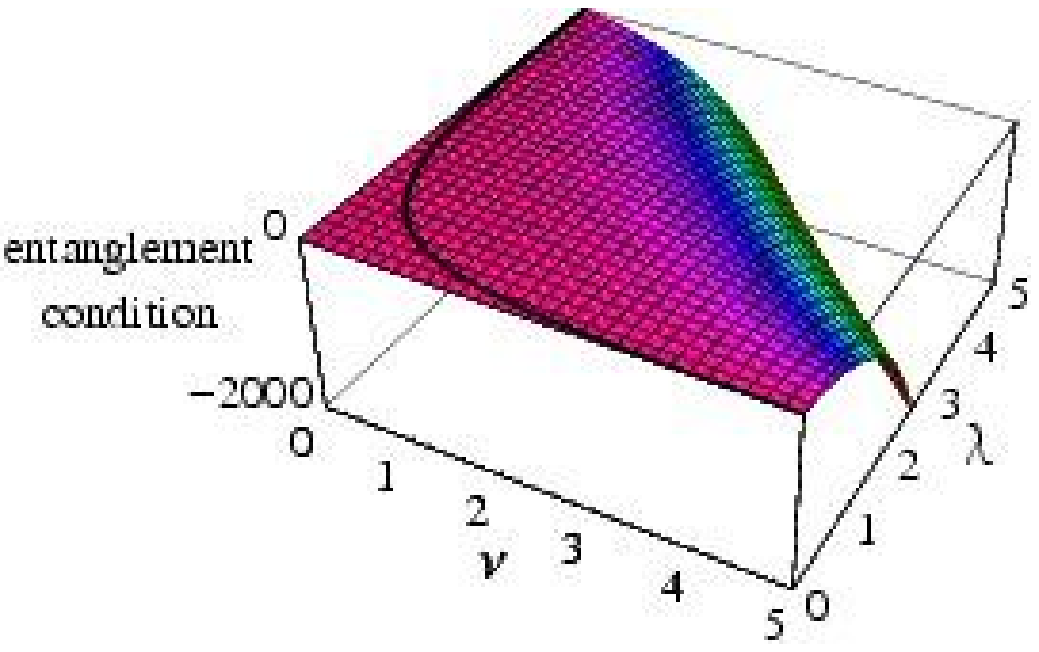}} \hspace{.2cm}
\subfigure[] {\includegraphics[width=6cm]{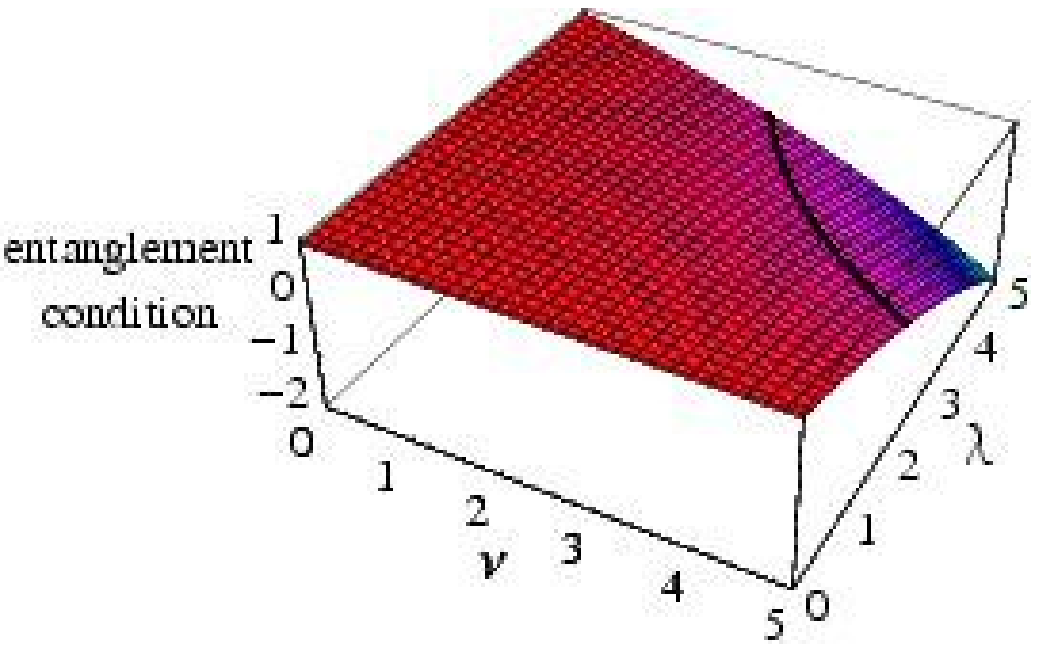}}
\caption{\label{condition} Entanglement condition,
\ineq{freqcondom}, for different frequency modes assuming that Leo
and Nadia have the same acceleration {\rm(a)} $\bar a=2\pi$ and
{\rm(b)} $\bar a=10\pi$. Entanglement is only present in the
frequency range where the plotted surfaces assume negative values,
and vanishes for frequencies where the plots become positive; the
threshold [saturation of \ineq{freqcondom}] is highlighted with a
black line. Only modes whose frequencies are sufficiently high
exhibit bipartite entanglement. For higher accelerations of the
observers, the range of entangled frequency modes gets narrower, and
in the infinite acceleration limit the bipartite entanglement
between all frequency modes vanish.}}
\end{figure}

Our results immediately show that in this context the entanglement
{\em vanishes} between field modes such that
\begin{equation}
\label{freqcondom} e^{\frac{2\pi}{\bar a}\lambda}+e^{\frac{2\pi
}{\bar a}\mu}-e^{\frac{2\pi}{\bar a}(\lambda+\nu)}\geq 0\,.
\end{equation}
This means that if the field is in a two-mode squeezed state with
frequencies satisfying  \eq{freqcondom}, Leo and Nadia would detect
no entanglement in the field. We have thus a practical  condition to
determine which modes would be entangled from Leo and Nadia's
non-inertial perspective, depending on their frequency.

In Fig.~\ref{condition} we plot the condition~\pref{freqcondom} on
entanglement for different frequency modes. The modes become
disentangled when the graph takes positive values. We see that only
modes with the highest frequencies exhibit bipartite entanglement
for a given acceleration $\bar a$ of the observers. The larger the
acceleration the less modes remain entangled, as expected. In the
limit of infinite acceleration, $\lambda/(a_{L}),\nu/(a_{N}) \gg 0$,
the set of entangled modes becomes empty.

Considering once more equally accelerated observers,
$a_{L}=a_{N}\equiv \bar a$ with finite $\bar a$, it is
straightforward to compute the Gaussian contangle of the modes that
do remain entangled, in the case of a maximally entangled state in
the inertial frame. From \eq{m4ln}, we have
\begin{equation}\label{entlims}
 m_{L|N}(s\rightarrow\infty)=\frac{\cosh (2 l) \cosh (2 n)-4
   \sinh (l) \sinh (n)+3}{2
   [\sinh (l)+\sinh (n)]^2}\,.
\end{equation}
In Fig.~\ref{entfrq} we plot the entanglement between the modes,
\eq{entlims}, as a function of their frequency $\lambda$ and $\nu$
[using \eq{accparam}] when Leo and Nadia travel with the same
acceleration $\bar a= 2\pi$. We see that, consistently with the
previous analysis, at fixed acceleration, the entanglement is larger
for higher frequencies.  In the infinite acceleration limit, as
already remarked, entanglement vanishes for all frequency modes.

\begin{figure}[t!]
\centering{\includegraphics[width=8.5cm]{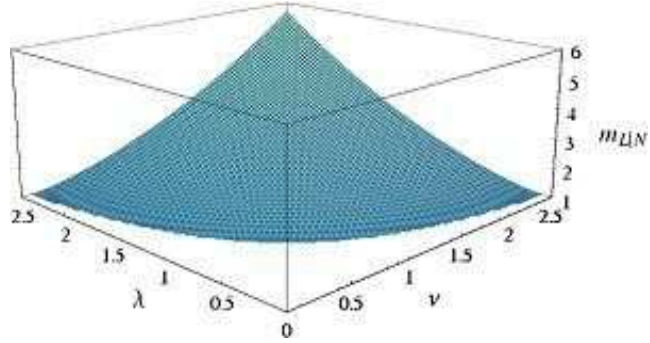}%
\caption{\label{entfrq} Entanglement between different frequency
modes assuming that Leo and Nadia have the same acceleration $\bar
a=2\pi$.  }} \end{figure}

\subsubsection{Equal acceleration parameters}

\begin{figure}[t!]
\centering{
 \subfigure[] {\includegraphics[height=5cm]{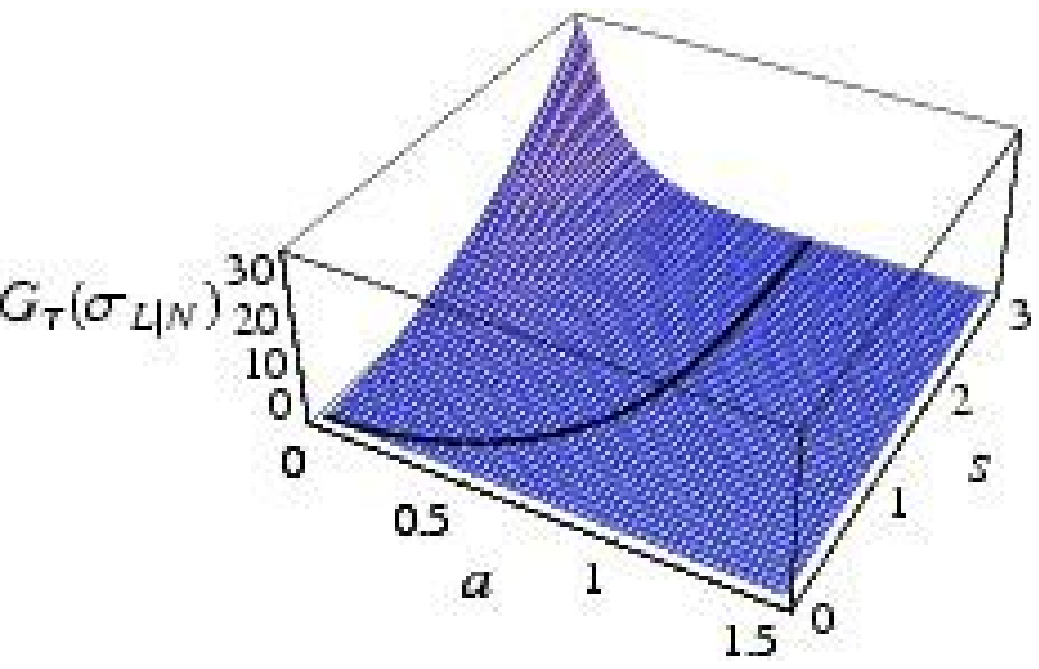}} \\
\subfigure[] {\includegraphics[height=5cm]{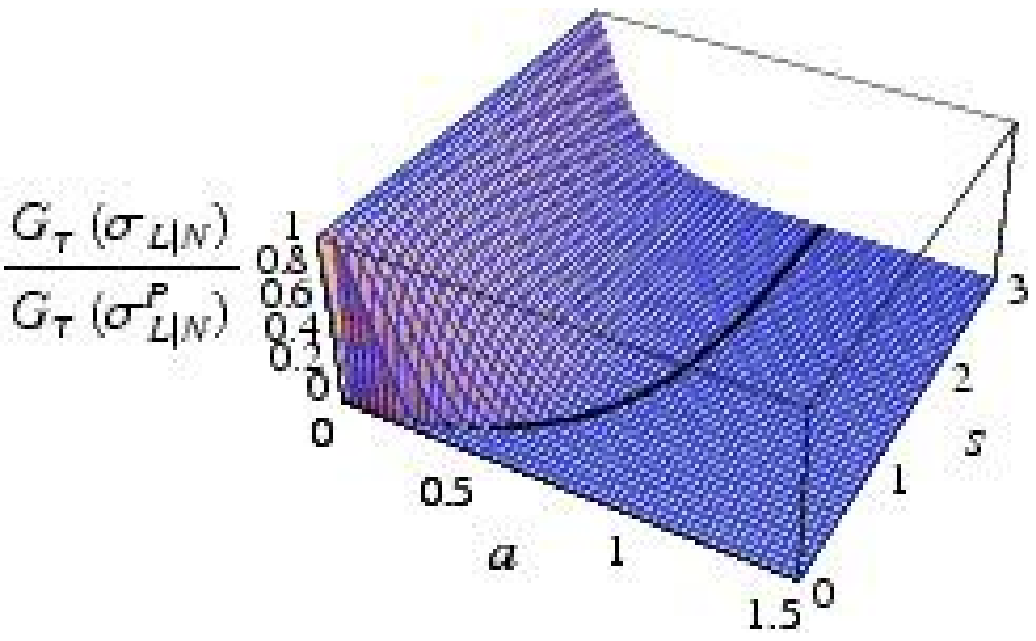}}
\caption{\label{fi4decay} Bipartite entanglement between the two
non-inertial observers Leo and Nadia, both traveling with uniform
acceleration given by the effective squeezing parameter $a$. From an
inertial perspective the two observers share a two-mode squeezed
state with squeezing degree $s$. Plot {\rm(a)} depicts the Gaussian
contangle $G_\tau(\sig_{L|N})$, given by Eqs.~{\rm(\ref{tau},
\ref{m4lnaa})}, as a function of $a$ and $s$. In plot {\rm(b)} the
same quantity is normalized to the contangle in the Minkowski frame,
$G_\tau(\sig^p_{L|N})=4s^2$. Notice in {\rm(a)} how the bipartite
Gaussian contangle is an increasing function of the inertial
entanglement, $s$, while it decreases with increasing acceleration,
$a$. This decay is faster for higher $s$, as clearly visible in
{\rm(b)}. At variance with the case of only one accelerated observer
(Fig.~\ref{fi3decay}), in this case the bipartite entanglement can
be completely destroyed at finite acceleration. The black line
depicts the threshold acceleration $a^\ast(s)$, \eq{ast}, such that
for $a \ge a^\ast(s)$ the bipartite entanglement shared by the two
non-inertial observers is exactly zero.}}
\end{figure}

We return to consider detectors sensitive to a single mode frequency
and, for simplicity, we restrict our attention to the case where Leo
and Nadia's trajectories have the same acceleration parameter
\begin{equation}\label{equa}
l=n\equiv a\,.
\end{equation}
This means that $\lambda/a_{L}=\nu/a_{N}$. While the following
results do not rely on this assumption, it is particularly useful in
order to provide a pictorial representation of entanglement in the
four-mode state $\sig_{\bar L L N \bar N}$, which is now
parametrized only by the two competing squeezing degrees, the
inertial quantum correlations ($s$) and the acceleration parameter
of both observers ($a$). In this case, the acceleration parameter
$a^\ast$ for which the entanglement between the modes detected by
Leo and Nadia vanishes, is
\begin{equation}\label{ast}
a^\ast(s) = {\rm arcsinh} \left[\sqrt{\tanh(s)}\right]\,,
\end{equation}
where we used \eq{m4ln}. The Gaussian contangle in the state
$\sig_{LN}$ is therefore given by
\begin{equation}\label{m4lnaa}
m_{L|N} = \left\{
                \begin{array}{l}
                  1\,, \hspace*{0.6cm} a \ge a^\ast(s)\,; \\ \\
                  \frac{2 \cosh ^2(2 a) \cosh ^2(s) + 3 \cosh (2 s) -
      4 \sinh ^2(a) \sinh (2 s) -
      1}{4 \left[\cosh ^2(a) + e^{2 s} \sinh ^2(a)\right]}
      ,  \\ \hspace*{1cm}
          \hbox{otherwise}\:\!,
                \end{array}
              \right.
\end{equation}
which we plot in Fig.~\ref{fi4decay}. The entanglement increases
with $s$ and decreases with $a$ with a stronger rate of decay
 for increasing $s$. The main difference with Fig.~\ref{fi3decay}
 is that entanglement here completely vanishes at
finite acceleration. Even for infinite entanglement in the inertial
frame, entanglement vanishes at $a \ge {\rm arcsinh} (1) \approx
0.8814$.

%\textbf{Let us explore this point further. Note that for the
%acceleration parameter to be equal, it is necessary that
%$\lambda/a_{L}=\nu/a_{N}$. Since it makes no sense to consider
%entanglement for $\lambda=\nu$, we must have $a_{L}\neq a_{N}$.
%Therefore $\nu=\alpha\lambda$ where $\alpha$ is a constant of
%proportionality. We then must fix $a_{L}=\alpha a_{N}$. Given the
%modes $\lambda$ and $\nu$, Nadia's acceleration is determined for a
%given acceleration of Leo. Lets then consider as the acceleration
%parameter $a=\lambda/a_{L}=K/A$ where $K$ is a frequency and $A$ an
%acceleration. There is a natural frequency of the system defined as
%$\omega=A/c$. The entanglement between modes $\lambda$ and $\nu$
%completely vanishes when the natural frequency of the system is
%close to the frequency $K$, exactly $\omega=(\log{2}/\pi)K$. If the
%acceleration $A$ is larger than that limit, the modes will be always
%disentangled. The Unruh temperature corresponding to this is
%$T_{U}=\hbar K/2\log{2}k_{B}$.}. \textbf{One can also consider that
%the accelerations of Nadia and Leo are fixed. Then only certain
%frequency modes $\nu=\frac{a_{N}}{a_{L}}\lambda$ would remain
%entangled: those modes whose for which $\lambda$ is smaller than
%$(\log{2}a_{L}/c\pi)$.}

\subsection{Residual multipartite entanglement} \label{secTwoMulti}
It is straightforward to show that the four-mode state $\sig_{\bar L
L N \bar N}$ of \eq{sig4} is fully inseparable, which means that it
contains multipartite entanglement shared among all the four parties
involved. This follows from the observation that the determinant of
each reduced one- and two-mode CM obtainable from $\sig_{\bar L L N
\bar N}$ is strictly bigger than $1$ for any nonzero squeezings.
This in addition to the global purity of the state means that there
is entanglement across all global bipartitions of the four modes. We
now aim to provide a quantitative characterization of such
multipartite entanglement. This analysis in the general case $l \neq
n$ is performed in Ref.~\cite{myletter}.

Here, following Ref.~\cite{ivette}, we focus once more for ease of
simplicity on the case  of two observers with equal acceleration
parameters $l=n \equiv a$. The state under consideration is obtained
from \eq{sig4} via the prescription \eq{equa}, and it turns out to
be exactly the four-mode state described in Chapter \ref{ChapUnlim}
in an optical setting, \eq{s4}. The entanglement properties of this
four-mode pure Gaussian state have been therefore already
investigated in detail in Chapter \ref{ChapUnlim} \cite{unlim},
where we showed in particular that the entanglement sharing
structure in such state is infinitely {\em promiscuous}. The state
admits the coexistence of an unlimited, genuine four-partite
entanglement, together with an accordingly unlimited bipartite
entanglement in the reduced two-mode states of two pair of parties,
here referred to as \{Leo, anti-Leo\}, and \{Nadia, anti-Nadia\}.
Both four-partite and bipartite correlations increase with $a$. We
will now recall the main results of the study of multipartite
entanglement in this four-partite Gaussian state, with the
particular aim of showing the effects of the relativistic
acceleration on the distribution of quantum information.

We have that the residual Gaussian contangle [see \eq{taures}],
\begin{eqnarray}\label{tau4}
G_\tau^{res}(\sig_{\bar L L N \bar N})&\equiv& G_\tau (\sig_{\bar L
| (L
N \bar N)})-G_\tau (\sig_{\bar L|L})\\
&=&{\rm arcsinh}^2\left\{ {\sqrt {[\cosh ^2a+\cosh (2s)\sinh
^2a]^2-1} } \right\}-4a^2\,, \nonumber
\end{eqnarray}
quantifies precisely the multipartite correlations that cannot be
stored in bipartite form. Those quantum correlations, however, can
be either tripartite involving three of the four modes, and/or
genuinely four-partite among all of them.

The tripartite portion (only present in equal content in the
tripartitions $\bar L | L | N$ and $L |N|\bar N$) can be estimated
as in Fig.~\ref{figtrip}, and specifically it decays to zero in the
limit of high acceleration.
 Therefore, in the regime of increasingly
high $a$, eventually approaching infinity, any form of tripartite
entanglement among any three modes in the state $\sig_{\bar L L N
\bar N}$ is negligible (exactly vanishing in the limit of infinite
acceleration).

It follows that, exactly like in Chapter \ref{ChapUnlim}, in the
regime of high acceleration $a$, the residual entanglement
$G_\tau^{res}$ determined by \eq{tau4} is stored entirely in the
form of four-partite quantum correlations. Therefore, the residual
entanglement in this case is a good measure of \textit{genuine}
four-partite entanglement among the four Rindler spacetime modes. It
is now straightforward to see that $G_\tau^{res}(\sig_{\bar L L N
\bar N})$ is itself an {\em increasing} function of $a$ for any
value of $s$ (see Fig.~\ref{figres}), and it \textit{diverges} in
the limit $a\to \infty $.

The four-mode state \eq{tau4} obtained with an arbitrarily large
acceleration $a$, consequently, exhibits a coexistence of unlimited
genuine four-partite entanglement, and pairwise bipartite
entanglement in the reduced two-mode states $\sig_{L|\bar L}$ and
$\sig_{N | \bar N}$. This peculiar distribution of CV entanglement
 in the considered Gaussian state has been
defined as {\em infinitely promiscuous} in Chapter \ref{ChapUnlim},
where its consequences are discussed in a practical optical setting
\cite{unlim}. It is interesting to note that in the relativistic
analysis we present here \cite{ivette}, the genuine four-partite
entanglement increases unboundedly with the observers' acceleration.
This is in fact in strong contrast with the case of an inertial
observer and an accelerating one (Sec.~\ref{secOne}), where we find
that, in the infinite acceleration limit, the genuine tripartite
entanglement saturates at $4s^2$ (\ie the original entanglement
encoded between the two inertial observers).

In the scenario considered in this Section, the acceleration of Leo
and Nadia creates {\em ex novo} entanglement (function of the
acceleration) between the respective Rindler regions of both
observers independently. The information loss at the double horizon
is such that even an infinite entangled state in the inertial frame
contains no quantum correlations when detected by two observers
traveling at finite acceleration. If one considers even higher
acceleration of the observers, it is basically the entanglement
between the Rindler regions which is redistributed into genuine
four-partite form. The tripartite correlations tend to vanish as a
consequence of the thermalization which destroys the inertial
bipartite entanglement. The multipartite entanglement, obviously,
increases infinitely with acceleration because the entanglement
between the Rindler regions increases without bound with
acceleration.  It is remarkable that such promiscuous distribution
of entanglement can occur without violating the fundamental monogamy
constraints on entanglement sharing (see Chapter
\ref{ChapMonoGauss}).

To give a simple example, suppose the bipartite entanglement in the
inertial frame is given by $4s^2 = 16$ for $s = 2$. If both
observers travel with an effective acceleration parameter $a=7$, the
four-partite entanglement [given by \eq{tau4}] among all Rindler
modes is $81.2$ ebits, more than $5$ times the inertial bipartite
entanglement. At the same time, a bipartite entanglement of $4a^2 =
196$ is generated between region $I$ and region $II$ of each
observer.

A final {\em caveat} needs to be stated. The above results show that
unbounded entanglement is created by merely the observers' motion.
This requires of course an unlimited energy needed to fuel their
spaceships. Unfortunately, in this setting such entanglement is
mostly unaccessible, as both Leo and Nadia are confined in their
respective Rindler region $I$. The only entangled resource that can
be used is the degraded two-mode thermal squeezed state of modes
$\lambda_I$ and $\nu_I$, whose entanglement soon vanishes for
sufficiently high, finite acceleration.

Let us remark, once more, that in the practical setting of quantum
optics the same four-mode Gaussian states of light beams can be
instead accessed and manipulated, as shown in Chapter
\ref{ChapUnlim}. The role of the acceleration on the detection of
the field is played in that case by the effects of a nonlinear
crystal through the mechanism of parametric down-conversion.  In
such a non-relativistic context, the different types of entanglement
can be readily used as a resource for bipartite and/or multipartite
transmission and processing of CV quantum information \cite{unlim}.

\subsection{Mutual information}

It is very interesting to evaluate the mutual information
$I(\sig_{L|N}$) between the states measured by Leo and Nadia, both
moving with acceleration parameter $a$.

In this case the symplectic spectrum of the reduced (mixed) two-mode
CM   $\sig_{L|N}$ of \eq{sig4} is degenerate as it belongs to the
family of GMEMS (see Sec.~\ref{SecGmemsGlems}), yielding
$\nu_-(\sig_{L|N})=\nu_+(\sig_{L|N})=\left({\det\sig_{L|N}}\right)^{\frac14}$.
From \eq{migau}, the mutual information then reads
\begin{equation}\label{mi4}
I(\sig_{L|N}) = f(\sqrt{\det\sig_{L}})+f(\sqrt{\det\sig_{N}}) - 2
f\left[\left({\det\sig_{L|N}}\right)^{\frac14}\right]\,,
\end{equation}
with $f(x)$ defined by \eq{entfunc}.

Explicitly:

\smallskip

 \noindent $I(\sig_{L|N})= 2 \cosh ^2(a)\cosh ^2(s)
\log [\cosh ^2(a) \cosh ^2(s)]  - [\cosh (2 s) \cosh ^2(a) + \sinh
^2(a) - 1]
      \log \{\frac{1}{2} [\cosh (2 s)  \cosh ^2(a) + \sinh ^2(a) -
                  1]\}  + \frac12 \{[2 \cosh (2 s) \sinh ^2(2 a) + \cosh (4 a) + 3]^{\frac12} -
            2\} \log \{[2 \cosh (2 s) \sinh ^2(2 a) + \cosh (4 a) + 3]^{\frac12} -
            2\} - \frac12 \{[2 \cosh (2 s) \sinh ^2(2 a) + \cosh (4 a) + 3]^{\frac12}
                  +
            2\} \log \{[2 \cosh (2 s) \sinh ^2(2 a) + \cosh (4 a) + 3]^{\frac12}
            +
            2\}+\log(16)$.

\medskip

We plot the mutual information both directly, and normalized to the
inertial entropy of entanglement, which is equal to \eq{ev3},
\begin{equation}\label{ev4}
E_V(\sig_{L|N}^p)= f(\cosh 2s)\,,
\end{equation}
with $f(x)$ given by \eq{entfunc}. We immediately notice another
novel effect. Not only the entanglement is completely destroyed at
finite acceleration, but also classical correlations are degraded,
see Fig.~\ref{mi4decay}{\rm(b)}. This is very different to the case
of a single non-inertial observer where classical correlations
remain invariant.

\begin{figure}[t!]
\centering{
 \subfigure[] {\includegraphics[width=6cm]{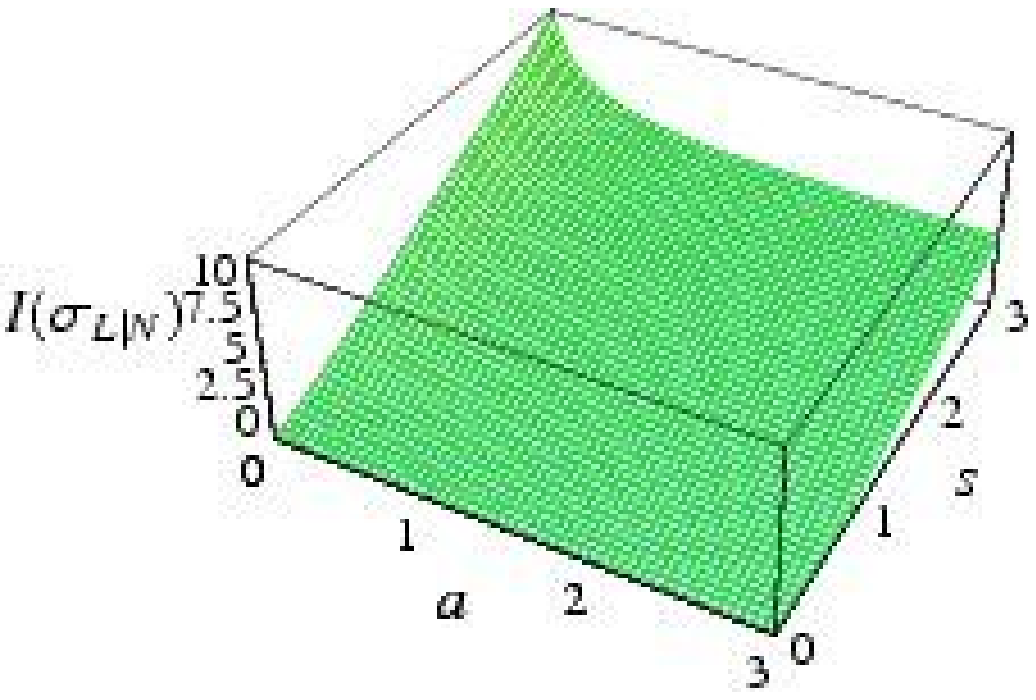}} \hspace{.2cm}
\subfigure[] {\includegraphics[width=6cm]{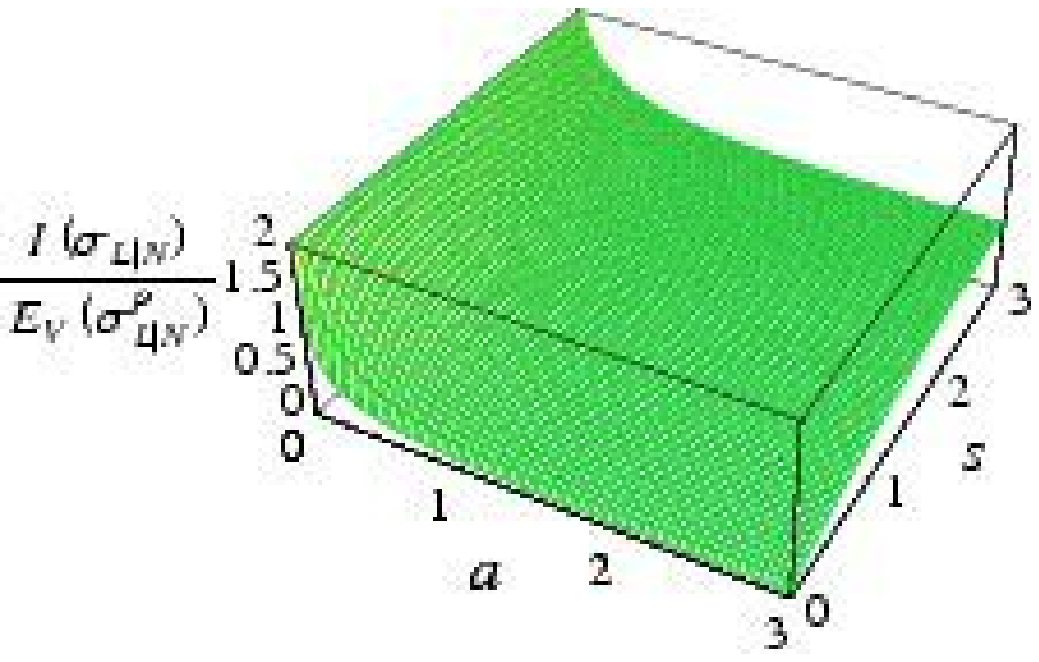}}
\caption{\label{mi4decay} Total correlations between the two
non-inertial observers Leo and Nadia, traveling with equal, uniform
acceleration given by the effective squeezing parameter $a$. In the
inertial frame, the modes are in a two-mode squeezed state with
squeezing degree $s$. Plot {\rm(a)} shows the mutual information
$I(\sig_{L|N})$, given by \eq{mi4}, as a function of $a$ and $s$. In
plot {\rm(b)} the same quantity is normalized to the entropy of
entanglement perceived by inertial observers, $E_V(\sig^p_{L|N})$,
\eq{ev4}. Notice in {\rm(a)} how the mutual information is an
increasing function of the squeezing parameter $s$ and saturates to
a nonzero value in the limit of infinite acceleration; in contrast,
the entanglement vanishes at finite acceleration (see
Fig.~\ref{fi4decay}). Plot {\rm(b)}, shows that this asymptotic
value is smaller than the entropy of entanglement in the Minkowski
frame (which is equal to the classical correlations detected by the
inertial observers). Therefore, classical correlations are also
degraded when both observers are accelerated, in contrast to the
case where only one observer is in uniform acceleration (see
Fig.~\ref{fi4decay}).}}
\end{figure}

The asymptotic state detected by Leo and Nadia, in the infinite
acceleration limit ($a \rightarrow \infty$), contains indeed some
residual classical correlations (whose amount is an increasing
function of the squeezing $s$). But these correlations are {\em
always} smaller than the classical correlations in the inertial
frame given by \eq{ev4}. Classical correlations are robust against
the effects of the double acceleration only when the classical
correlations in the inertial frame are infinite (corresponding to
infinite shared entanglement in the inertial frame, $s \rightarrow
\infty$). The entanglement, however, is always fragile, since we
have seen that it is completely destroyed at a finite, relatively
small acceleration parameter $a$.

Another intriguing fact is that, comparing
Figs.~\ref{mi3decay}{\rm(a)} and~\ref{mi4decay}{\rm(a)}, one sees
that in both cases (either one or two non-inertial observers) the
mutual information between the two ``real'' observers is a function
of the acceleration parameter and of the initial squeezing. In the
case of both accelerated observers, however, the mutual information
is always smaller, as we have just discussed. We can  study the
difference between them, once we set for ease of comparison equal
acceleration parameters, $r=a$, where $r$ regulates Rob's
acceleration when Alice is inertial, and $a$ is related to the
acceleration of both Leo and Nadia in the present situation:
\begin{equation}\label{defect}
D(a,s) = \left.I(\sig_{A|R})\right|_{r=a} - I(\sig_{L|N})\,.
\end{equation}
The quantity $D(a,s)$ is plotted in Fig.~\ref{difetto}:
surprisingly, it is strictly bounded. It increases both with $s$ and
$a$, but in the asymptotic limit of infinite inertial shared
entanglement, $D(a,s \rightarrow \infty)$ saturates exactly to $1$
(as it can be checked analytically) for any $a > 0$. We remark that
both mutual informations $I(\sig_{A|R})$ and $I(\sig_{L|N})$ diverge
in this limit: yet their difference is finite and equal to one.
Clearly, the small deficit of the mutual information seen when both
observers are accelerated, is detected as loss of classical
correlations, as plotted in Fig.~\ref{mi4decay}{\rm(b)}.
Mysteriously, the Unruh thermalization affects classical
correlations when both observers are accelerated: however, it
degrades at most one absolute unit of classical correlations. This
means that in the case when both Leo and Nadia escape the fall into
a black hole, not only their entanglement is degraded but {\em there
is also a loss of classical information}.

\begin{figure}[t!]
\centering{\includegraphics[width=8.5cm]{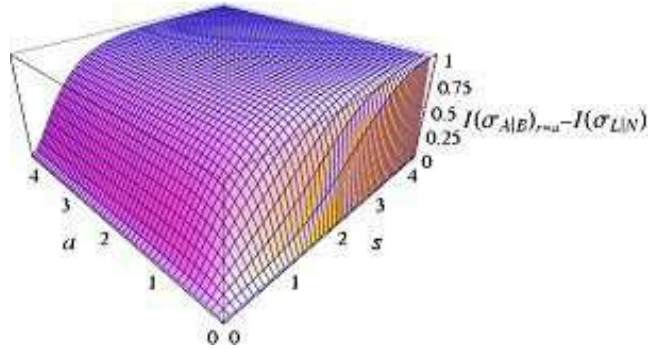}%
\caption{\label{difetto} Plot, as a function of the acceleration
parameter $a$ and the squeezing parameter $s$, of the difference
between the mutual information shared by the inertial Alice and the
non-inertial Rob, and the mutual information shared by the uniformly
accelerating Leo and Nadia, as given by \eq{defect}.}}
\end{figure}

\section{Discussion and outlook}\label{SecConcl}

In this Chapter, based on Ref.~\cite{ivette}, we presented a
thorough study of classical and quantum correlations between modes
of a scalar field measured by observers in relative acceleration. By
considering the state of the field in the inertial frame in the
simplest multi-mode squeezed state possible (the two-mode case) we
were able to investigate in detail the entanglement in all
partitions of the system. We considered two observers carrying
single mode detectors and discussed the correlations on their
measurements when both observers are in uniform acceleration and
when only one of them is non-inertial. We find that in both settings
entanglement is degraded with acceleration and we explain this
degradation as an effect of re-distribution of the entanglement
measured in an inertial frame.

Our main results can be summarized as follows. When one of the
observers is non-inertial the entanglement lost between the modes
measured by him and the inertial observer is re-distributed in
tripartite correlations. No entanglement is generated between the
modes measured by the inertial observer and the modes in the
causally disconnected region $II$. This shows that indeed the
behavior for bosonic fields is very different to the Dirac case
where the entanglement lost in the non-inertial frame is
re-distributed not into tripartite correlations but into bipartite
correlations between the mode measured by the inertial observer and
the mode in region $II$. The analysis of the mutual information
shows that in this case classical correlations are conserved
independently of the acceleration. The situation changes drastically
by considering that both observers are non-inertial. In this case
the entanglement lost between two non-inertial observers is
re-distributed into mainly four-partite correlations although some
tripartite correlations exist for finite acceleration. The
surprising result here (though expected in the framework of
distributed entanglement, as the additional fourth mode comes into
play) is that entanglement vanishes completely at a finite
acceleration. This is also strikingly different to the results in
the Dirac case where entanglement remains positive for all
accelerations (as a direct consequence of the restricted Hilbert
space in that instance). Another surprising result in this case is
that we find that classical correlations are no longer invariant to
acceleration but are also degraded to some extent. We analyzed the
entanglement between the modes of the field detected by two
non-inertial observers as a function of the frequencies of their
modes, and found that for a fixed acceleration high frequency modes
remain entangled while lower frequency modes disentangle. In the
limit of infinitely accelerated observers, the field modes are in a
separable state for any pair of frequencies.

The take-home message of this Chapter is the following.

\medskip

\begin{itemize}
\item[\ding{226}]
 \noindent{\rm\bf Continuous variable entanglement in non-inertial reference frames.}
 {\it The
degradation of entanglement due to the Unruh effect is analytically
studied for two parties sharing a two-mode squeezed state in an
inertial frame, in the cases of either one or both observers
undergoing uniform acceleration. For two non-inertial observers
moving with finite acceleration, the entanglement vanishes between
the lowest frequency modes. The loss of entanglement is precisely
explained as a redistribution of the inertial entanglement into
multipartite quantum correlations among accessible and unaccessible
modes from a non-inertial perspective. Classical correlations are
also lost for two accelerated observers but conserved if one of the
observers remains inertial.}
\smallskip
\end{itemize}

The tools developed in this Chapter can be used to
 investigate the problem of information loss in  black
 holes \cite{myletter}. There is a correspondence \cite{unruh} between the
 Rindler-Minkowski spacetime and the Schwarzschild-Kruskal spacetime, that
 allows us to study the loss (and re-distribution) of quantum and
 classical correlations for observers outside the black hole,
 extending and re-interpreting the results presented in
 Sec.~\ref{secTwo}. In that case the degradation of
 correlations can be understood as essentially being due to the
 Hawking effect \cite{hawking1,hawking2}.

The next step concerns the study of classical and quantum
correlations in the most general particle states definable in a
spacetime with at least two asymptotically flat regions, represented
by multi-mode squeezed states which involve all modes being
pair-wise entangled (like in the phase-space Schmidt decomposition,
see Sec.~\ref{SecSchmidtPS}). The study of entanglement in this
state, from a relativistic perspective, will provide a deeper
understanding of quantum information in quantum field theory in
curved spacetime \cite{dewitt}.

}

\part{Closing remarks}
{\vspace*{1cm}
\includegraphics[width=6.5cm]{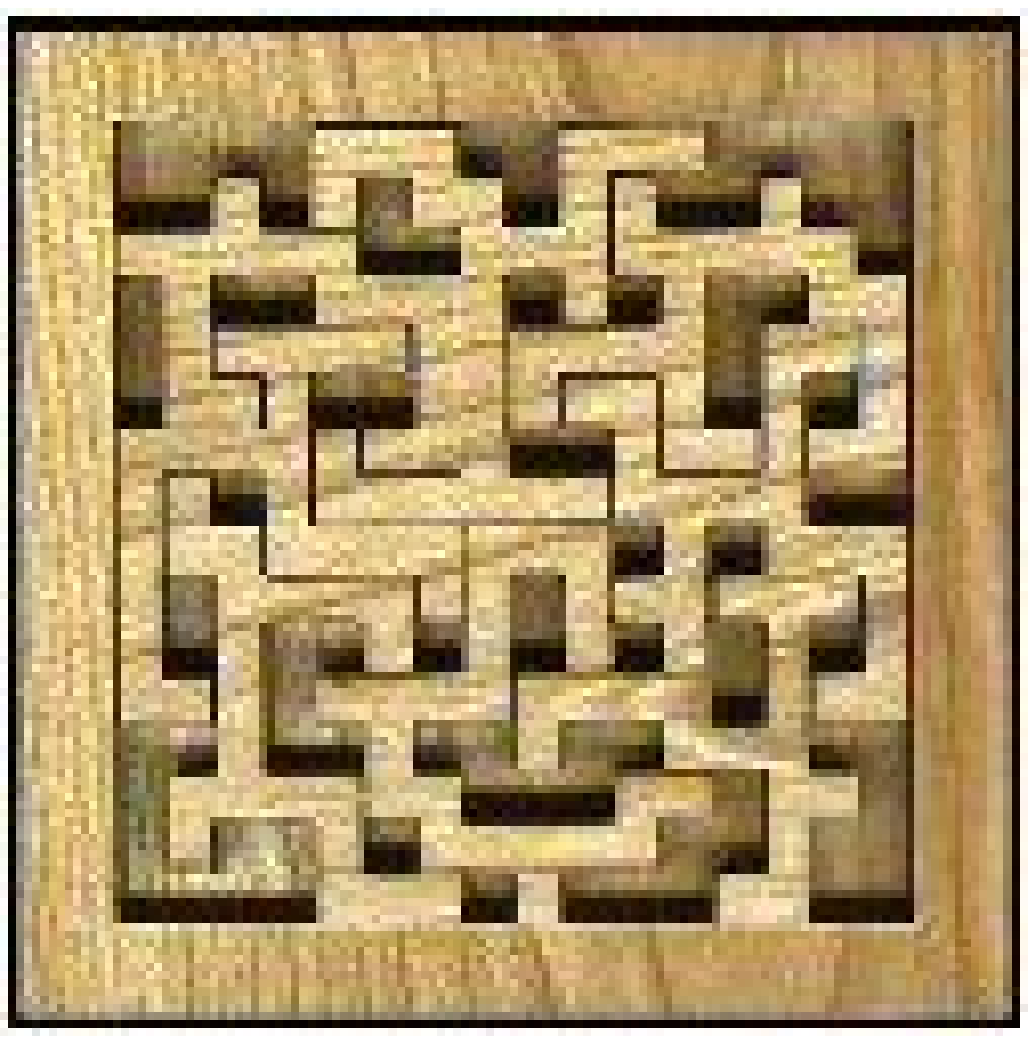} \\
\vspace*{0.6cm} {\rm \normalsize  {\em Entanglement Puzzle.} Tom
Jolly, 2005.
\\ \vspace*{-0.4cm} \texttt{\footnotesize http://www.abstractstrategy.com/2-entanglement.html}}}
\label{PartConcl}

\chapter*{Conclusion and Outlook}

{\sf

\subsection*{Entanglement of Gaussian states: what next?}

The centrality of Gaussian states in CV quantum information is
motivated not only by their peculiar structural properties which
make their description amenable of an analytical analysis, but also
by the ability to produce, manipulate and detect such states with
remarkable accuracy in realistic, experimental settings.

The scope of this Dissertation has been almost entirely theoretical.
We provided important advances for what concerns the structural and
informational characterization of bipartite entanglement, and the
definition and quantification of multipartite entanglement in
Gaussian states. State engineering prescriptions and several
applications to diverse fields (quantum communication, quantum
optics, many-body physics, relativity) were discussed as well. We
are not going here to list again the individual and numerous results
obtained in all those contexts --- retrievable in
Refs.~{\rm[\href{#cite.prl}{GA2}---\href{#cite.ivette}{GA20}]} and
in the previous Parts of this Dissertation --- to avoid unnecessary
repetitions with the front matter. We will try instead to frame our
results into a broader perspective, with the aim of providing an as
self-contained as possible outlook of the current directions of the
CV quantum information research, with and beyond Gaussian states.

 For reasons of space and
time, we cannot discuss in sufficient detail all the additional
proposals and experimental demonstrations concerning on one hand the
state engineering of two-, three- and in general $N$-mode Gaussian
states, and on the other hand the use of such states as resources
for the realization of quantum information protocols, which were not
covered by the present Dissertation. Excellent review papers are
already available for what concerns both the optical state
engineering of multiphoton quantum states of discrete and CV systems
\cite{fabio}, and the implementations of quantum information and
communication with continuous variables
\cite{vanlokfortshit,brareview}.

Let us just mention that, from a practical point of view, Gaussian
resources have been widely used to implement paradigmatic protocols
of CV quantum information, such as two-party and multiparty
teleportation
\cite{Braunstein98,Furusawa98,network,naturusawa,pirandolareview}
(see Chapter \ref{ChapCommun}), and quantum key distribution
\cite{Cry1,Cry2,gran}; they have been proposed for achieving one-way
quantum computation with CV generalizations of cluster states
\cite{menicucci}, and in the multiparty setting they have been
proven useful to solve Byzantine agreement \cite{sanpera}. Gaussian
states are currently considered key resources to realize
light-matter interfaced quantum communication networks. It has been
experimentally demonstrated how a coherent state of light can be
stored onto an atomic memory \cite{memorypolzik}, and teleported to
a distant atomic ensemble via a hybrid light-matter two-mode
entangled Gaussian resource \cite{telepolzik}.

Gaussian states also play a prominent role in many-body physics,
being ground and thermal states of harmonic lattice Hamiltonians
\cite{chain}. Entanglement entropy scaling in these systems has been
shown to follow an area law \cite{area,areanew}.  Thermodynamical
concepts have also been applied to the characterization of Gaussian
entanglement: recently, a ``microcanonical'' measure over the second
moments of pure Gaussian states under an energy constraint has been
introduced \cite{typical} (see also Sec.~\ref{secgentyp}), and
employed to investigate the statistical properties of the bipartite
entanglement in such states. Under that measure, the distribution of
entanglement concentrates around a finite value at the
thermodynamical limit and, in general, the typical entanglement of
Gaussian states with maximal energy $E$ is {\em not} close to the
maximum allowed by $E$.

A rather recent field of research concerns the investigation of
Gaussian states in a relativistic setting \cite{ahn,ivette}, as we
have seen in Chapter \ref{ChapIvette}. Within the general framework
of relativistic quantum information \cite{peresreview}, such studies
are of relevance to understand the phenomenon of information loss
through a black hole horizon \cite{myletter}, and more generally to
gain some knowledge on the structure of the curved spacetime
\cite{dewitt,ball}.

In a non-relativistic framework, the investigation of the structure
of entanglement in hybrid CV-qubit systems is not only of conceptual
importance, but it is relevant for applications as well.  From the
monogamy point of view (see Chapter \ref{ChapMonoGauss}), some
interesting hints come from a recent study of the ground-state
entanglement in highly connected systems made of harmonic
oscillators and spin-$1/2$ systems \cite{acin}. On a more practical
ground, we should at least mention a proposal for a quantum optical
implementation of hybrid quantum computation, where qubit degrees of
freedom for computation are combined with Gaussian modes for
communication \cite{ibridovanloock}, and a suggested scheme of
hybrid quantum repeaters for long-distance distribution of quantum
entanglement based on dispersive interactions between coherent light
with large average photon number and single, far-detuned atoms or
semiconductor impurities in optical cavities
\cite{ripetitoreibrido}. A hybrid CV memory realized by indirect
interactions between different Gaussian modes, mediated by qubits,
has been recently shown to have very appealing features compared to
pure-qubit quantum registers \cite{paternostro}.

It seems fitting to conclude this overview by commenting on the
intriguing possibility of observing CV (Gaussian) entanglement at
the interface between microscopic and macroscopic scales. In this
context, it is encouraging that the existence of optomechanical
entanglement between a macroscopic movable mirror and a cavity field
has been theoretically demonstrated and predicted to be quite robust
in realistic experimental situations, up to temperatures well in
reach of current cryogenic technologies \cite{vitali}.

\subsection*{Entanglement of non-Gaussian states: a new arena}

The infinite-dimensional quantum world, however, is not confined to
Gaussian states. In fact, some recent results demonstrate that
basically the current state-of-the-art in the theoretical
understanding and experimental control of CV entanglement is
strongly pushing towards the boundaries of the oasis of Gaussian
states and Gaussian operations. For instance, the entanglement of
Gaussian states cannot be increased (distilled) by resorting to
Gaussian operations only \cite{nogo1,nogo2,nogo3}. Similarly, for
universal one-way quantum computation using Gaussian cluster states,
a single-mode non-Gaussian measurement is required \cite{menicucci}.

There is indeed a fundamental motivation for investigating
entanglement in non-Gaussian states, as the {\em extremality} of
Gaussian states imposes that they are the minimally entangled states
among all states of CV systems with given second moments
\cite{extra}. Experimentally, it has been recently demonstrated
\cite{nongaussexp} that a two-mode squeezed Gaussian state can be
``de-Gaussified'' by coherent subtraction of a single photon,
resulting in a mixed non-Gaussian state whose non-local properties
and entanglement degree are enhanced (enabling a better efficiency
for teleporting coherent states \cite{kitagawa}). Theoretically, the
characterization of even bipartite entanglement (let alone
multipartite) in non-Gaussian states stands as a formidable task.

One immediate observation is that any (non-Gaussian) multimode state
with a CM corresponding to an entangled Gaussian state is itself
entangled too \cite{vanlokfortshit,extra}. Therefore, most of the
results presented in this Dissertation may serve to detect
entanglement in a broader class of states of infinite-dimensional
Hilbert spaces. They are, however, all sufficient conditions on
entanglement based on the second moments only of the canonical
operators. As such, for arbitrary non-Gaussian states, they are in
general very inefficient
--- meaning that most entangled non-Gaussian states fail to be
detected by these criteria. The description of non-Gaussian states
requires indeed (an infinite set of) high order statistical moments:
as an obvious consequence, also an inseparability criterion for
these states should involve high order correlations. Recently, some
separability criteria based on hierarchies of conditions involving
higher moments of the canonical operators have been introduced to
provide a sharper detection of inseparability in generic
non-Gaussian states \cite{agarwalbiswas,nhakim,hilzub,mirapiani}.

In particular, Shchukin and Vogel \cite{shukvog} introduced an
elegant and unifying approach to separability based on the PPT
requirement, that is constructed in the form of an infinite series
of inequalities, and includes as special cases all the above cited
results (including the conditions on second moments
\cite{Duan00,Simon00} qualifying separability in Gaussian states,
see Sec.~\ref{SecPPTG}), thus demonstrating the important role of
PPT in building a strong criterion for the detection of
entanglement. The conditions by Shchukin and Vogel can be applied to
distinguish between the several separability classes in a
multipartite CV system \cite{shukvogmulti}. To this aim,
entanglement witnesses are useful as well \cite{illuso}.

The efficiency of some of the above-mentioned inseparability
criteria based on higher order moments, for detecting bipartite
entanglement in the non-Gaussian family of squeezed number states of
two-mode radiation fields, has been recently evaluated
\cite{fabiotest}. We mention a further interesting approach to
non-Gaussian entanglement reported by McHugh {\em et al.}
\cite{mchugh}, who showed that entanglement of multiphoton squeezed
states is completely characterized by observing that with respect to
a new set of modes, those non-Gaussian states actually assume
Gaussian character.

\subsection*{Future perspectives}

Many open issues and unanswered intriguing questions naturally arise
when peeping out of the parental house of Gaussian states. There is
always the risk of being trapped in the infinite mathematical
complexity of the CV Hilbert space losing the compass which points
towards the physics under investigation. However, the brief hints
summarized above concerning the study of non-Gaussian entanglement
in its actual infancy, seem to suggest at least two things. On one
hand, that it is worth taking the risk, as the possibilities offered
by non-Gaussian states may be really intriguing;  on the other hand,
that wise footpaths in the CV labyrinth may be traced and followed
back and forth, leading to physically insightful, novel results on
both fundamental and practical grounds, obtainable with a finite,
accountable complexity rise compared to the Gaussian case.

The most exciting challenge for me is to enter this huge, largely
unexplored treasure island aiming to draw a map first of its
underworld (foundations), and then of its colorful surface
(applications). The structure and distribution of entanglement in
non-Gaussian states have to be understood, qualified and quantified
properly, at least in restricted families of states, in order to
single out their usefulness for quantum information (and not only)
implementations. For instance, we have learned (see Part
\ref{PartMulti}) how the monogamy constraint imposes a natural
hierarchical structure on multipartite entanglement of Gaussian
states. In this context, the promiscuity of some classes of Gaussian
states was established, opening new frontiers for the implementation
of such resources for multiparty communication purposes. Inspired by
these results, and bearing in mind that Gaussian states are extremal
in the sense of possessing minimal entanglement compared to the
non-Gaussian cousins, it appears as an exciting perspective to look
for exotic states in the CV arena with an enhancedly promiscuous
sharing structure of quantum correlations, with a monogamy of
entanglement stretched to its limits, and so with exceptional
predispositions for the transfer of quantum information.

These considerations, along with the few other examples mentioned
above, should suffice to convince the reader that CV entanglement of
Gaussian and non-Gaussian states, together with its applications in
fundamental quantum mechanics, quantum ``multimedia'', and several
other areas of physics, is a very active and lively field of
research, where more progress and new fascinating developments may
be forecast in the near future. The results of this Dissertation,
while of inherent fundamental interest for quantum information
theory, are thus expected to play --- either directly or as premises
for new advances --- an increasingly important role in the practical
characterization of the physical processes which underlie these
multifaceted, sometimes stunningly revolutionary situations.

}

\appendix
\chapter{Standard forms of pure  Gaussian states under local
operations}\label{ChapAppendixSF}

{\sf In this Appendix, based on Ref.~\cite{sformato}, we study the
action of local unitary operations on a general CM of a
 pure $N$-mode Gaussian state and compute the minimal number of
parameters which completely characterize pure Gaussian states up to
local unitaries. }

\section{Euler decomposition of symplectic
operations}\label{SecEuler}

{\sf
 Central to our analysis will be the following general
decomposition of a symplectic transformation $S$ (referred to as the
``Euler'' or ``Bloch-Messiah'' decomposition
\cite{pramana,braunsqueezirreducibile}):
\begin{equation}
S = O' Z O ,\label{euler1}
\end{equation}
where $O, O' \in K(N)= Sp_{(2N,\R)}\cap SO(2N)$ are orthogonal
symplectic transformations, while
$$
Z=\oplus_{j=1}^{N}\left(\begin{array}{cc}
z_j & 0 \\
0 & \frac{1}{z_j}
\end{array}\right) \; ,
$$
with $z_{j}\ge 1$ $\forall$ $j$. The set of such $Z$'s forms a
non-compact subgroup of $Sp_{(2N,\R)}$ comprised of local
(single-mode) squeezing operations (borrowing the terminology of
quantum optics, where such transformations arise in degenerate
parametric down-conversion processes). Moreover, let us also mention
that the compact subgroup $K(N)$ is isomorphic to the unitary group
$U(N)$, and is therefore characterized by $N^2$ independent
parameters. To acquaint the reader with the flavor of the counting
arguments which will accompany us through this Appendix (and with
the nontrivial aspects contained therein), let us combine the
Williamson and the Euler decompositions to determine the number of
degrees of freedom of a generic $N$-mode Gaussian state (up to first
moments), given by $N+2N^2+N-N=2N^2+N$. The $N$ subtracted from the
sum of the numbers of symplectic eigenvalues and of degrees of
freedom of a symplectic operation takes into account the invariance
under single-mode rotations of the local Williamson forms --- which
`absorbs' one degree of freedom {\em per mode} of the symplectic
operation describing the state according to \eq{willia}. Actually,
the previous result is just the number of degrees of freedom of a
$2N\times2N$ symmetric matrix (in fact, the only constraint $\sig$
has to fulfill to represent a physical state is the semidefinite
$\sig+i\Omega \ge 0$, which compactly expresses the uncertainty
relation for many modes \cite{serafozziprl}).

}

%%%%%%%%%%%%%%%%%%%%%%%%%%%%%%%%%%%%%%%%%%%%%%%%%%%%%%%%%%%

\section{Degrees of freedom of pure Gaussian states}\label{condpuri}

{\sf

Pure Gaussian states are characterized by CMs with Williamson form
equal to the identity. As we have seen (Sec.~\ref{SecWilly}), the
Williamson decomposition provides a mapping from any Gaussian state
into the uncorrelated product of thermal (generally mixed) states:
such states are pure (corresponding to the vacuum), if and only if
all the symplectic eigenvalues are equal to $1$.

The symplectic eigenvalues of a generic CM $\sig$ are determined as
the eigenvalues of the matrix $|i\Omega\sig|$, where $\Omega$ stands
for the symplectic form. Therefore, a Gaussian state of $N$ modes
with CM $\sig$ is pure if and only if \be -\sig\Omega\sig\Omega =
\id_{2N} \; . \label{puri} \ee

It will be convenient here to reorder the CM, and to decompose it in
the three sub-matrices $\sig_q$, $\sig_p$ and $\sig_{qp}$, whose
entries are defined as \be  (\sig_q)_{jk} = \tr[\varrho \hat{q}_j
\hat{q}_k],\quad (\sig_p)_{jk} = \tr[\varrho \hat{p}_j
\hat{p}_k],\quad (\sig_{qp})_{jk} = \tr[\varrho
\{\hat{q}_j,\hat{p}_k\}/2], \ee such that the complete CM $\sig$ is
given in block form by \be \sig = \left(\begin{array}{cc}
\sig_q & \sig_{qp} \\
\sig_{qp}^{\sf T} & \sig_{p}
\end{array}\right) \; . \label{sig}
\ee Let us notice that the matrices $\sig_{q}$ and $\sig_p$ are
always symmetric and strictly positive, while the matrix $\sig_{qp}$
does not obey any general constraint.

Eqs.~{\rm(\ref{puri})} and \pref{sig} straightforwardly lead to the
following set of conditions \bea
\sig_{q}\sig_{p} = \id_N + \sig_{qp}^2 \; , \label{first} \\
\sig_{qp}\sig_{q} - \sig_{q}\sig_{qp}^{\sf T} = 0 \; , \label{second} \\
\sig_{p}\sig_{q} = \id_N + \sig_{qp}^{\sf T\,2} \; , \label{redu1} \\
\sig^{\sf T}_{qp}\sig_{p} - \sig_{p}\sig_{qp} = 0 \; . \label{redu2}
\eea Now, the last two equations are obviously obtained by
transposition of the first two. Moreover, from \pref{first} one gets
\be \sig_{p} = \sig_{q}^{-1}(\id_{N}+\sig_{qp}^2) \; , \label{sigp}
\ee while \eq{second} is equivalent to \be \sig_{q}^{-1}\sig_{qp} -
\sig_{qp}^{\sf T}\sig_{q}^{-1} = 0 \label{secondbis} \ee (the latter
equations hold generally, as $\sig_{q}$ is strictly positive and
thus invertible). \eq{secondbis} allows one to show that any
$\sig_{p}$ determined by \eq{sigp} satisfies the condition
\pref{redu2}. Therefore, only Eqs.~\pref{first} and \pref{second}
constitute independent constraints and fully characterize the CM of
pure Gaussian states.

Given any (strictly positive) matrix $\sig_{q}$ and (generic) matrix
$\sig_{qp}$, the fulfillment of condition \pref{second} allows to
specify the second moments of any pure Gaussian state, whose
sub-matrix $\sig_p$ is determined by \eq{sigp} and does not involve
any additional degree of freedom.

A straightforward counting argument thus yields the number of
degrees of freedom of a generic pure Gaussian state, by adding the
entries of a generic and symmetric $N\times N$ matrix and
subtracting the equations of the antisymmetric condition
\pref{second}: $N^2+N(N+1)/2-N(N-1)/2=N^2+N$, in compliance with the
number dictated by the Euler decomposition of a symplectic
operation: \be \sig = S^{\sf T} \id_{2N} S = O^{\sf T} Z^2 O \; .
\ee Notice that, if either $\sig_{q}$ or $\sig_{qp}$ are kept fixed,
the constraint \pref{second} is just a linear constraint on the
entries of the other matrix, which can be always solved; in fact, it
cannot be overdetermined, since the number of equations $N(N-1)/2$
is always smaller than the number of variables, either $N^2$ or
$N(N+1)/2$.

A preliminary insight into the role of local operations in
determining the number of degrees of freedom of pure CMs is gained
by analyzing the counting of free parameters in the CV version of
the Schmidt decomposition, as introduced in Sec.~\ref{SecSchmidtPS}.
The CM of any pure $(M+N)$-mode Gaussian state is equivalent, up to
local symplectic transformations on the $M$-mode and $N$-mode
subsystems, to the tensor product of $M$ decoupled two-mode squeezed
states (assuming, without loss of generality, $M<N$) and $N-M$
uncorrelated vacua \cite{botero03}. Besides the $M$ two-mode
squeezing parameters, the degrees of freedom of the local symplectic
transformations to be added are $2N^2+N+2M^2+M$. However, a mere
addition of these two values leads to an overestimation with respect
to the number of free parameters of pure CMs determined above. This
is due to the invariance of the CM in `Schmidt form' under specific
classes of local operations. Firstly, the $(N-M)$-mode vacuum (with
CM equal to the identity) is trivially invariant under local
orthogonal symplectics, which account for $(N-M)^2$ parameters.
Furthermore, one parameter is lost for each two-mode squeezed block
with CM $\sig^{2m}$ given by \eq{tms}: this is due to an invariance
under single-mode rotations peculiar to two-mode squeezed states.
For such states, the sub-matrices $\sig^{2m}_{q}$ and
$\sig^{2m}_{p}$ have identical --- and all equal --- diagonal
entries, while the sub-matrix $\sig^{2m}_{qp}$ is null. Local
rotations embody two degrees of freedom --- two local `angles' in
phase space --- in terms of operations. Now, because they act
locally on $2\times2$ identities, rotations on both single modes
cannot affect the diagonals of $\sig^{2m}_{q}$ and $\sig^{2m}_{p}$,
nor the diagonal of $\sig^{2m}_{qp}$, which is still null. In
principle, they could thus lead to two (possibly different)
non-diagonal elements for $\sig^{2m}_{qp}$ and/or to two different
non-diagonal elements for $\sig^{2m}_{q}$ and $\sig^{2m}_{p}$
(which, at the onset, have opposite non-diagonal elements),
resulting in
$$
\sig^{2m}_{q}=\left(\begin{array}{cc}
a & c_1 \\
c_1 & a
\end{array}\right) \, ,\quad
\sig^{2m}_{p}=\left(\begin{array}{cc}
a & c_2 \\
c_2 & a
\end{array}\right) \, ,\quad
\sig^{2m}_{qp}=\left(\begin{array}{cc}
0 & y \\
z & 0
\end{array}\right)\,.
$$
However, elementary considerations, easily worked out for such
$2\times2$ matrices, show that Eqs.~\pref{second} and \pref{sigp}
imply
$$
c_1=-c_2 \;\; ,\;\; y=z \quad {\rm and} \quad a^2-c_1^2 = 1+y^2 \, .
$$
These constraints reduce from five to two the number of free
parameters in the state: {\em the action of local single-mode
rotations --- generally embodying two independent parameters --- on
two-mode squeezed states, allows for only one further independent
degree of freedom}. In other words, all the Gaussian states that can
be achieved by manipulating two-mode squeezed states by local
rotations (``phase-shifters'', in the experimental terminology) can
be obtained by acting on only one of the two modes. One of the two
degrees of freedom is thus lost and the counting argument displayed
above has to be recast as $M+2N^2+N+2M^2+M-(M-N)^2-M=(M+N)^2+(M+N)$,
in compliance with what we had previously established.

As we are about to see, this invariance peculiar to two-mode
squeezed states also accounts for the reduction of locally invariant
free parameters occurring in pure two-mode Gaussian states.

%%%%%%%%%%%%%%%%%%%%%%%%%%%%%%%%%%%%%%%%%%%%%%%%%%%%%%%%%%%
\subsection{Reduction under single-mode operations}\label{redu}

Let us now determine the reduction of degrees of freedom achievable
for pure Gaussian states by applying local single-mode symplectic
transformations. Notice that all the entanglement properties (both
bipartite and multipartite) of the states will solely depend on the
remaining parameters, which cannot be canceled out by local
unitaries.

In general, for $N$-mode systems, local symplectic operations have
$3N$ degrees of freedom, while $N$-mode pure Gaussian states are
specified, as we just saw, by $N^2+N$ quantities. The subtraction of
these two values yields a residual number of parameters equal to
$N^2-2N$. However, this number holds for $N\ge3$, but fails for
single- and two-mode states. Let us analyze the reasons of this
occurrence.

For single-mode systems, the situation is trivial, as one is
allowing for all the possible operations capable, when acting on the
vacuum, to unitarily yield any possible state. The number of free
parameters is then clearly zero (as any state can be reduced into
the vacuum state, with CM equal to the $2\times2$ identity). The
expression derived above would instead give $-1$. The reason of this
mismatch is just to be sought in the invariance of the vacuum under
local rotations: only two of the three parameters entering the Euler
decomposition actually affect the state. On the other hand, one can
also notice that these two latter parameters, characterizing the
squeezing and subsequent last rotation of the Euler decomposition
acting on the vacuum, are apt to completely reproduce any possible
single-mode state. Clearly, this situation is the same as for any
$N$-mode pure Gaussian state under global operations: the first
rotation of the Euler decomposition is always irrelevant, thus
implying a corresponding reduction of the free parameters of the
state with respect to the most general symplectic operation.

As for two-mode states, the above counting argument would give zero
locally invariant parameters. On the other hand, the existence of a
class of states with a continuously varying parameter determining
the amount of bipartite entanglement [the two-mode squeezed states
of \eq{tms}], clearly shows that the number of free parameters
cannot be zero. Actually, local symplectic operations allow one to
bring {\em any} (pure or mixed) two-mode Gaussian state in a
``standard form'' with $\sig_{qp}=0$ and with identical diagonals
for $\sig_{q}$ and $\sig_{p}$. Imposing then \eq{second} on such
matrices, one finds that the only pure states of such a form have to
be two-mode squeezed states. Therefore, we know that the correct
number of locally invariant free parameters has to be one. Even
though local symplectic operations on two-mode states are determined
by $6$ parameters, they can only cancel $5$ of the $6$ parameters of
pure two-mode states. This is, again, due to the particular
transformation properties of two-mode squeezed states under
single-mode rotations, already pointed out in the previous Section
when addressing the counting of degrees of freedom in the
Schmidt-like decomposition: local rotations acting on a two-mode
squeezed state add only one independent parameter. The most general
two-mode pure Gaussian state results from a two-mode squeezed state
by a single local rotation on any of the two modes, followed by two
local squeezings and two further rotations acting on different
modes. Notice that the same issue arises for $(M+N)$-mode states to
be reduced under local $M$- and $N$-mode symplectic operations. A
mere counting of degrees of freedom would give a residual number of
local free parameters equal to $(M+N)^2+M+N-2M^2-2N^2-M-N=
-(M-N)^2$. This result is obviously wrong, again due to a loss of
parameters in the transformations of particular invariant states. We
have already inspected this very case and pointed out such
invariances in our treatment of the Schmidt decomposition (previous
Section): we know that the number of locally irreducible free
parameters is just $\min(M,N)$ in this case, corresponding to the
tensor product of two-mode squeezed states and uncorrelated vacua.

For $N\ge3$, local single-mode operations can fully reduce the
number of degrees of freedom of pure Gaussian states by their total
number of parameters. The issue encountered for two-mode states does
not occur here, as the first single-mode rotations can act on
different non-diagonal blocks of the CM ({\em i.e.}, pertaining to
the correlations between different pairs of modes). The number of
such blocks is clearly equal to $(N^2-N)/2$ while the number of
local rotations is just $N$. Only for $N=1,2$ is the latter value
larger than the former: this is, ultimately, why the simple
subtraction of degrees of freedom only holds for $N\ge3$. To better
clarify this point, let us consider a CM $\sig^{3m}$ in the limiting
instance $N=3$. The general standard form for (mixed) three-mode
states implies the conditions (see Sec.~\ref{sform}) \be {\rm
diag}\,(\sig^{3m}_{q}) = {\rm diag}\,(\sig^{3m}_{p})  \label{diagga}
\ee and \be \sig^{3m}_{qp} = \left(\begin{array}{ccc}
0&0&0 \\
0&0&u \\
s&t&0
\end{array}
\right) \; . \label{sparse} \ee The diagonal of $\sig^{3m}_{q}$
coincides with that of $\sig^{3m}_{p}$ (which always results from
the local single-mode Williamson reductions) while six entries of
$\sig^{3m}_{qp}$ can be set to zero. Let us now specialize to pure
states, imposing the conditions \pref{first} and \pref{second}.
\eq{second} results into a linear system of three equations for the
non-null entries of $\sig^{3m}_{qp}$, with coefficients given by the
entries of $\sig^{3m}_{q}$. The definite positivity of
$\sig^{3m}_{q}$ implies that the sub-system on $s$ and $t$ is
determinate and thus imposes $s=t=0$. This fact already implies
$(\sig^{3m}_{qp})^2=0$ and thus
$\sig_{p}^{3m}=(\sig_{q}^{3m})^{-1}$. As for $u$, the system entails
that, if $a\neq0$, then the entry $(\sig^{3m}_{q})_{13}=0$. But, as
is apparent from \eq{puri} (and from $\sig>0$), the determinant of
the CM of any pure state has to be equal to $1$. Now, working out
the determinant of the global CM $\sig^{3m}$ under the assumptions
\pref{diagga}, \pref{sparse} and $s=t=(\sig^{3m}_{q})_{13}=0$ one
gets $\det{\sig^{3m}}=(\alpha+(u^2-\sig)\beta)/(\alpha-\sig\beta)$
with $\beta>0$ (again from the strict positivity of $\sig^{3m}$),
which is equal to $1$ if and only if $u=0$. Therefore, for pure
three-mode Gaussian states, the matrix $\sig^{3m}_{qp}$ can be made
null by local symplectic operations alone on the individual modes.
The entries of the symmetric positive definite matrix
$\sig^{3m}_{q}$ are constrained by the necessity of
Eqs.~\pref{first} --- which just determines $\sig^{3m}_{p}$ --- and
\pref{diagga}, which is comprised of three independent conditions
and further reduces the degrees of freedom of the state to the
predicted value of three. An alternative proof of this is presented
in Sec.~\ref{secpuri} \cite{3mpra}.

Let us also incidentally remark that the possibility of reducing the
sub-matrix $\sig_{qp}$ to zero by local single-mode operations is
exclusive to two-mode (pure and mixed) and to three-mode pure
states. This is because, for general Gaussian states, the number of
parameters of $\sig_{qp}$ after the local Williamson
diagonalizations is given by $N(N-1)$ (two per pair of modes) and
only $N$ of these can be canceled out by the final local rotations,
so that only for $N<3$ can local operations render $\sig_{qp}$ null.
For pure states and $N>2$ then, further $N(N-1)/2$ constraints on
$\sig_{qp}$ ensue from the antisymmetric condition \pref{second}:
this number turns out to match the number of free parameters in
$\sig_{qp}$ for $N=3$, but it is no longer enough to make
$\sig_{qp}$ null for pure states with $N\ge4$. This is further
discussed in Chapter \ref{ChapGeneric}.

Summing up, we have rigorously determined the number of  ``locally
irreducible'' free parameters of pure Gaussian states
\cite{sformato}, unambiguously showing that the quantification and
qualification of the entanglement (which, by definition, is
preserved under local unitary operations) in such states of $N$
modes is completely determined by $1$ parameter for $N=2$ and
$(N^2-2N)$ parameters for $N>2$, as reported in \eq{npuri}.

}

%\pagenumbering{roman}
%\def\bibname{List of Publications}

%\def\bibname{azz}

\cleardoublepage
\def\bibname{List of Publications}

\label{listofpubs}
%\bibliographystylepub{torabibpub}
%\bibliographypub{mypapers}

\clearpage
\def\bibname{Bibliography}

%\bibliographystyle{torabib}
%\bibliography{tesibib,New,Old,Books}

\end{document}